\documentstyle[12pt]{article}
\newskip\humongous \humongous=0pt plus 1000pt minus 1000pt

\newif\ifdtup

\def\oldrefledge{\hangindent3\parindent}
\def\oldreffmt#1{\rlap{[#1]} \hbox to 2\parindent{}}
\def\oldref#1{\par\noindent\oldrefledge \oldreffmt{#1}
	\ignorespaces}
\def\figledge{\hangindent=1.25in}
\def\figfmt#1{\rlap{Figure {#1}} \hbox to 1in{}}
\def\fig#1{\par\noindent\figledge \figfmt{#1}
	\ignorespaces}
\def\ie{\hbox{\it i.e.}{}}	\def\etc{\hbox{\it etc.}{}}
\def\eg{\hbox{\it e.g.}{}}	\def\cf{\hbox{\it cf.}{}}
\def\etal{\hbox{\it et al.}}
\def\dash{\hbox{---}}
\def\tr{\mathop{\rm tr}}
\def\Tr{\mathop{\rm Tr}}
\def\Im{\mathop{\rm Im}}
\def\Re{\mathop{\rm Re}}
\def\bR{\mathop{\bf R}{}}
\def\bC{\mathop{\bf C}{}}
\def\partder#1#2{{\partial #1\over\partial #2}}
\def\secder#1#2#3{{\partial^2 #1\over\partial #2 \partial #3}}
\def\bra#1{\left\langle #1\right|}
\def\ket#1{\left| #1\right\rangle}
\def\VEV#1{\left\langle #1\right\rangle}
\def\gdot#1{\rlap{$#1$}/}
\def\abs#1{\left| #1\right|}
\def\pr#1{#1^\prime}
\def\ltap{\raisebox{-.4ex}{\rlap{$\sim$}} \raisebox{.4ex}{$<$}}
\def\gtap{\raisebox{-.4ex}{\rlap{$\sim$}} \raisebox{.4ex}{$>$}}
\def\contract{\makebox[1.2em][c]{
	\mbox{\rule{.6em}{.01truein}\rule{.01truein}{.6em}}}}
\def\slash#1{#1\!\!\!/\!\,\,}
\def\beq{\begin{equation}}
\def\eeq{\end{equation}}
\def\bea{\begin{eqnarray}}
\def\eea{\end{eqnarray}}
\def\half{\frac{1}{2}}
\def\aeq{\eeq}
\def\bq{\begin{quote}}
\def\eq{\end{quote}}
\def\pr{{\sl Phys. Rev.~}}
\def\np{{\sl Nucl. Phys.~}}
\def\pl{{\sl Phys. Letters~}}
\def\prl{{\sl Phys. Rev. Letters~}}
\def \Msol {M_\odot}
\def\GeV{\,{\rm GeV}}     
\def\eV {\,{\rm  eV}}     
\def\Mpc{\,{\rm Mpc}}     
\def\pc{\,{\rm pc}}     
\def\half{\frac{1}{2}}     
\def \lta {\mathrel{\vcenter
     {\hbox{$<$}\nointerlineskip\hbox{$\sim$}}}}
\def \gta {\mathrel{\vcenter
     {\hbox{$>$}\nointerlineskip\hbox{$\sim$}}}} 
\def \endpage {\vfill \eject}
\def \endline {\hfill \break}
\def \etal {{\it et al.}\ }
\relax
\def\ts{\thinspace}

\def\ol{\bar}
\def\tr{\hbox{tr}}
\def\be{\begin{equation}} 
\def\ee{\end{equation}} 
\def\bea{\begin{eqnarray}}
\def\eea{\end{eqnarray}}
\def\ba{\begin{array}}
\def\ea{\end{array}}
\def\kslash{\raise.15ex\hbox{/}\kern-.57em k}

\def\gev{{\rm GeV}}

\def\GeV{{\rm GeV}}

\def\ifb{{\rm fb}^{-1}}
\def\half{{\textstyle{ { 1\over { 2 } }}}}

\renewcommand{\theequation}{\thesection.\arabic{equation}}

\newcommand{\bear}{\begin{eqnarray}}
\newcommand{\eear}{\end{eqnarray}}

\newcommand{\lae}{\begin{array}{c}\,\sim\vspace{-21pt}\\< \end{array}}
\newcommand{\gae}{\begin{array}{c}\,\sim\vspace{-21pt}\\> \end{array}}

\newcommand{\down}{\mbox{$N$}}
\newcommand{\inl}{{\scriptscriptstyle L}}
\newcommand{\inr}{{\scriptscriptstyle R}} 
\begin{document}

\begin{titlepage}
\begin{flushright}
FERMI--PUB--02/045-T\\
BUHEP-01-09\\
March, 2003
 \\
\end{flushright}
\par \vskip .05in
 \begin{center}
{\huge \bf
 Strong Dynamics and Electroweak\\ [2mm]
Symmetry Breaking
           } 
 \end{center}
  \par \vskip .1in \noindent
\begin{center}
{\bf  Christopher T. Hill$^{1}$ }\\ [2mm]
and \\ [2mm]
{\bf  Elizabeth H. Simmons$^{2,3}$ }
\end{center}
\begin{center}
  \par \vskip .1in \noindent
{$^1$Fermi National Accelerator Laboratory\\
P.O. Box 500, Batavia, IL, 60510}
  \par \vskip .1in \noindent
{$^2$ Dept. of Physics, Boston University\\
590 Commonwealth Avenue, Boston, MA,  02215}
  \par \vskip .1in \noindent
{$^3$ Radcliffe Institute for Advanced Study and\\
      Department of Physics, Harvard University\\
      Cambridge, MA,  02138}
   \par \vskip .1in \noindent
\end{center}     
\begin{center}{\large Abstract}\end{center}
\begin{quote}

The breaking of electroweak symmetry, and origin of the associated ``weak
scale,'' $v_{weak} = 1/\sqrt{2\sqrt{2}G_F} = 175$ GeV, may be due to a new
strong interaction. Theoretical
developments over the past decade have led to viable
models and mechanisms that are consistent
with current experimental data. 
Many of these schemes feature a privileged role for the top
quark, and third generation, and are natural
in the context of theories of extra space dimensions at the weak scale. 
We review various models and their phenomenological
implications which will be subject 
to definitive tests in future
collider runs at the Tevatron, and the LHC, and 
future linear $e^+e^-$ colliders, as well
as sensitive studies of rare processes. 
\end{quote}
\par \vskip .1in

\vfill
\end{titlepage}

\def\baselinestretch{1.0}
\tiny
\normalsize

\tableofcontents

\newpage

\section{Introduction}

\subsection{Lessons from QCD}

The early days of accelerator-based particle physics were largely
explorations of the strong interaction scale, associated roughly with the
proton mass, of order 1 GeV.  Key elements were the elaboration of the hadron
spectroscopy; the measurements of cross-sections; the elucidation of
spontaneously broken chiral symmetry, with the pion as a Nambu-Goldstone
phenomenon 
\cite{Nambu:1960xd, Goldstone:1961eq, Gell-Mann:1960np}; 
the evolution of the flavor
symmetry $SU(3)$ and the quark model 
\cite{Gell-Mann:1953aa, Gell-Mann:1956aa, Zweig:1964jf};
the discovery of scaling behavior in electroproduction;
\cite{Bjorken:1969dy, Bjorken:1969ja, Feynman:1969ej}; 
and the observation of quarks and gluons as
partons.  Eventually, this work culminated in Quantum Chromodynamics (QCD): a
description of the strong scale based upon the elegant symmetry principle of
local gauge invariance, and the discovery of the Yang-Mills
gauge group $SU(3)$ of color \cite{Fritzsch:1973pi, Greenberg:1964pe}.
Today we recognize that the strong scale is a well-defined quantity,
$\Lambda_{QCD} \sim 100$ MeV, the infrared scale at which the
perturbatively defined running coupling constant of QCD blows up,
and we are beginning to understand how to perform detailed
nonperturbative numerical computations of the associated phenomena. 

There are several lessons in the discovery of QCD which illuminate our
present perspective on nature.  First, it took many years to get from
the proton, neutron, and pions, originally thought to be elementary
systems, to the underlying theory of QCD.  From the perspective of a
physicist of the 1930's, knowing only the lowest-lying states, or one
of the 1950's seeing the unfolding of the resonance spectroscopy and
new quantum numbers, such as strangeness, it would have seemed
astonishing that those disparate elements came from a single
underlying Yang-Mills 
gauge theory \cite{Yang:1954ek}.  Second, the process of solving the problem
of the strong interactions involved a far-ranging circumnavigation of
all of the ideas that we use in theoretical physics today.  For
example, the elaboration of the resonance spectrum and Regge behavior
of QCD led to the discovery of string theory\footnote{Proponents of
concepts such as nuclear democracy, duality, and Veneziano models
(which contain an underlying string theory spectrum) believed this
perspective, at the time, to be fundamental.}! 
QCD embodies a rich list of phenomena, including
confinement, perturbative asymptotic freedom 
\cite{Gross:1973id,  Politzer:1973fx},
topological fluctuations,  and, perhaps
most relevant for our present purposes, a BCS-like
\cite{Bardeen:1957kj, Bardeen:1957mv} 
mechanism leading to
chiral symmetry breaking \cite{Nambu:1960xd, Goldstone:1961eq, Gell-Mann:1960np}.
Finally, the strong interactions are
readily visible in nature for what might be termed ``contingent''
reasons: the nucleon is stable, and atomic nuclei are abundant. If a
process like $p\rightarrow e^+ + \gamma $ occured with a large rate,
so that protons were short-lived, then the strong interactions would
be essentially decoupled from low energy physics, for $\sqrt{s} <
2m_\pi$.  A new strong dynamics, if it exists, presumably does not
have an analogous stable sector (else we would have seen it), and this
dynamics must be largely decoupled below threshold.

 QCD provides direct guidance for this review because of the light it
may shed on the scale of electroweak symmetry breaking (EWSB).  The origin of
the scale $\Lambda_{QCD}$, and the ``hierarchy'' between the strong
and gravitational scales is, in principle, understood within QCD in a
remarkable and compelling way.  If a perturbative input for $\alpha_s
$ is specified at some high energy scale, e.g., the Planck scale, then
the logarithmic renormalization group running of $\alpha_s$ naturally
produces a strong-interaction scale $\Lambda_{QCD}$ which lies far
below $M_{Planck}$.  The value of scale $\Lambda_{QCD}$ does not
derive directly from the Planck scale, e.g. through multiplication by
some ratio of coupling constants.  Rather $\Lambda_{QCD}$ arises
naturally and elegantly from quantum mechanics itself, through
dimensional transmutation \cite{Coleman:1973jx}.  Hence, QCD produces
the large hierarchy of scales without fine-tuning.  The philosophy
underlying theories of dynamical EWSB is that
the ``weak scale'' has a similar dynamical and natural origin.

\subsection{The Weak Scale}

We have mentioned two of the fundamental mass scales in
nature, the strong-interaction scale $\Lambda_{QCD}$ and the scale of
gravity, $M_{Planck} = (\sqrt{G_N})^{-1} \approx 10^{19}$ GeV.  The
mass scale, which will figure most directly in our discussion, is the
scale of weak physics, $v_{weak} = (\sqrt{2\sqrt{2}G_F})^{-1} = 175$
GeV. The weak scale first entered physics, approximately 70 years ago,
when Enrico Fermi constructed the current-current interaction
description of $\beta$-decay and introduced the constant, $G_F$, into
modern physics \cite{Fermi:1934hr}.  The Standard Model
\cite{Glashow:1961tr, Weinberg:1967tq, Salam:1980jd} identifies 
$v_{weak}$ with
the vacuum expectation value (VEV) of a
fundamental, isodoublet, ``Higgs'' scalar field.  Both the gravitational and
weak scales are associated with valid low energy effective theories.
In the case of $M_{Planck}$ we have classical General Relativity;  In
the case of $v_{weak}$ we have the Standard Model.

The Standard Model is predictive and enjoys spectacular success in
almost all applications to the analysis of existing experimental
data. This hinges upon its renormalizability: it is a valid quantum
theory.  Now that current experiments are sensitive to electroweak
loop effects, these two aspects have become closely entwined.  In
almost all channels, experiment continues simply to confirm the
Standard Model's predictions; those processes which are sensitive to
loop contributions are starting to constrain the mass of the Higgs
Boson.  Combining all experimental data sensitive to electroweak loop
corrections yields an upper bound on the Standard Model Higgs boson's
mass of order $200$ GeV \cite{LEPEWWG:2001}.\footnote{At this writing
there are internal inconsistencies in the precision
electroweak data; the $Z^0$-pole data of
leptons alone predicts a Higgs boson mass that is $\sim 20 - 40$ GeV and is
directly ruled out, while the hadronic $A^b_{FB}$ predicts a very
heavy Higgs mass $\sim 300$ GeV; these discrepancies are significant,
at the $3.5\sigma$ level, and it is not clear that the Higgs mass
bound obtained by combining all data is meaningful
\cite{Chanowitz:2001bv}.  }

Beyond this fact, however, we know nothing in detail about the Higgs
boson, or whether or not it actually exists as a fundamental particle
in nature.  If new dynamics {\em were} assumed to be present at the
TeV scale, the Higgs boson could be a bound state and the upper bound on
the composite Higgs mass would rise to $\sim$ 1 TeV
\cite{Lee:1977eg}. Certain models we will describe in this review,
such as the Top Quark Seesaw scheme, predict a composite Higgs boson
with a mass $m_H\sim 1$ TeV and are otherwise in complete agreement
with electroweak constraints (see Section 4).

In considering the Higgs sector, the foremost question is that of motive:
``why should nature provide a unique elementary particle simply for the
purpose of breaking a symmetry?''  Other issues involve naturalness, i.e.,
the degree of fine-tuning required to provide the scale of the mass of the
putative Higgs boson \cite{'tHooft:1980xb, Wilson:1971dh, Wilson:1974jj}.
Certainly, when compared to the completely natural origin of $\Lambda_{QCD}$
relative to $M_{Planck}$, the origin of a Higgs boson mass, $m_H \sim 200$
GeV in the Standard Model is a complete mystery.  These questions hint at the
need for a more general mechanism or some enveloping symmetries that do the
equivalent job of, or provide a rationalization for, the Higgs Boson.
Therefore, it is fair to say that the true mechanism of 
EWSB in nature is unknown.

The paradigm we will explore in the present review is that the EWSB
 physics, in analogy to the strong interactions of QCD,
arises from novel strong dynamics.  We emphasize at the outset that new
strong dynamics (NSD) is incompatable with a completely perturbative view of
physics near the electroweak scale -- but does not exclude the possibility of
a low-scale Supersymmetry (SUSY).  For the most part in this review, our discussion
will focus on the nature and implications of the new strong dynamics itself,
and not on the additional possible presence of SUSY.  

In focusing this review in this direction, we should consider what we
hope to learn by looking beyond the more popular supersymmetric theories.
Certainly, SUSY is an elegant extension of the Lorentz group
that fits naturally into string theory, our best candidate for a
quantum theory of gravity. SUSY, moreover, gives us an
intriguing {\em raison d'etre} for the existence of fundamental scalar
particles:  If we look at the fermionic content of a model, such as
the Minimal Supersymmetric Standard Model (MSSM) we see that Higgs
bosons can be viewed as superpartners of new vector-like leptons
(i.e., a pair of left-handed leptonic isodoublets, one with $Y=1$ and
the other with $Y=-1$ in the MSSM).  As additional rewards we find
that: (i) the hierarchy, while not explained, is protected by the
chiral symmetries of the fermions; (ii) the resulting theory can be
weakly coupled and amenable to perturbative studies; and (iii) there
is reasonably precise unification of the gauge coupling constants.

On the other hand,  SUSY
offers fairly limited insight into why there occurs
EWSB, since the Higgs sector is 
essentially added by hand, just as the original Higgs boson was added
to the Standard Model, to accomodate the phenomenon.  
The
significance of the successful unification of the
gauge coupling constants 
\cite{Georgi:1974sy, Langacker:1991an, Amaldi:1991cn}, which
is certainly one of the more tantalizing aspects of the MSSM, is
nevertheless inconclusive because the unification condition that
obtains in nature remains unknown.  For example, higher dimension
operators associated with the Planck or GUT scales can modify the
naive unification condition, thus permitting unification in models
that might otherwise be rejected
\cite{Hill:1984xh, Shafi:1984gz}.  Moreover, new Strong
Dynamical Models (NSD's) of EWSB can
in principle unify.  However, because the primary dynamical issues in
models of new strong dynamics arise at the weak scale and because
complete NSD models are few in number, unification of NSD models has
not been developed very far.  

We believe that the central problem facing particle
physics today is to explain the origin of the electroweak mass scale,
or equivalently, EWSB: What causes the scale
$v_{weak} \propto (G_F)^{-1/2}$ in nature?  Theories of new strong
dynamics offer new insights into possible mechanisms of electroweak
symmetry breaking.  We recall that before QCD was understood to be a
local gauge theory with its intrinsic rich dynamics, speculation about
what lay beyond the strong interaction scale was systematically
flawed.  Thus, while many of the fundamental symmetries controlling
the known forces in nature are understood, speculation as to what lies
on energy scales well above $v_{weak}$ should be viewed as tentative.
Let us therefore focus on physics at the weak scale.

We will begin by providing an introductory tutorial survey of the
elementary physical ingredients of dynamical symmetry breaking.  In
Section \ref{sec:supcond}, we will consider a sequence 
of ``five easy pieces,'' or 
illustrative models which successively incorporate the key elements of
the electroweak  Standard Model.  This is complemented by a discussion
of the full Standard Model, including oblique radiative corrections,
in 1.4, 3.2,  and Appendix A, and by a discussion of a 
toy model of strong dynamics,
the Nambu--Jona-Lasinio model, in Appendix B.  Issues related to
naturalness in the QCD and electroweak sectors of the Standard Model
are discussed in Section \ref{sec:natural}.  These 
are intended to provide newcomers or non-specialists with the
collected ideas and a common language to make the rest of the review
accessible.  The arrangement of our discussion of modern theories and
phenomenology of dynamical EWSB is given in
Section \ref{sec:synopsis}.

A reader who may wish to ``cut to the chase'' is advised 
to skip directly to Section 1.5.

\subsection{Superconductors, Chiral Symmetries, and Nambu-Goldstone Bosons}
\label{sec:supcond}

A particle physicist's definition of an ordinary electromagnetic
superconductor is a ``vacuum'' or groundstate in which the photon
becomes massive.  For example, when a block of lead (Pb) is cooled to
$3^oK$ in the laboratory, photons impinging on the material acquire a
mass of about $1$ eV, and the associated phenomena of
superconductivity arise (e.g. expulsion of magnetic field lines,
low-resistance flow of electric currents, etc.).  The vacuum of our
Universe, the groundstate of the Standard Model, is likewise an
``electroweak superconductor'' in which the masses of the $W^\pm$ and
$Z^0$ gauge bosons are nonzero, while the photon remains massless.
Moreover there occurs in QCD the phenomenon of ``chiral symmetry
breaking,'' \cite{Nambu:1960xd, Goldstone:1961eq, Gell-Mann:1960np},
analogous to BCS superconductivity 
\cite{Bardeen:1957kj, Bardeen:1957mv} , in which the very
light up, down and strange quarks develop condensates in the vacuum
from which they acquire larger ``constituent quark masses,''
in analogy to the ``mass gap'' of a BCS superconductor.\footnote{In a
superconductor the mass gap is actually a small Majorana-mass,
$\sim \psi\psi + h.c.$
for an electron, an operator which carries net charge $\pm 2$}

In what follows, we will develop an understanding of chiral symmetry
breaking and dynamical mass generation in the more familiar context of
electrodynamics and the Landau-Ginzburg model of superconductivity.
To accomplish this, we examine the following sequence of toy models:
(i) the free superconductor in which the longitudinal photon is a
massless spin-$0$ field and manifest gauge invariance is preserved;
(ii) a massless fermion; (iii) the simple $U(1)_L\times U(1)_R$ fermionic
chiral Lagrangian in which the fermion acquires mass spontaneously and
a massless Nambu-Goldstone boson appears; (iv) the Abelian Higgs
model (also known as the Landau-Ginzburg superconductor); and putting
it all together, (v) the Abelian Higgs model together with the
fermionic chiral Lagrangian in which the Nambu-Goldstone boson has
become the longitudinal photon.  We will then indicate how the
discussion generalizes to our main subject of interest in Section 1.4,
the
electroweak interactions as described by the Standard Model.  

\vskip 0.2cm
\noindent
{\bf 1.3(i) Superconductor $\leftrightarrow$ A Massive Photon}
\vskip 0.2cm

The defining principle of electrodynamics is local $U(1)$ gauge
invariance.  Can a massive photon 
be consistent with the gauge symmetry?  After all, a photon mass term
like $\half M^2 A_\mu A^\mu$ appears superficially not to be invariant
under the gauge transformation $A_\mu \rightarrow A_\mu +
\partial_\mu\chi$.  It can be made manifestly
gauge invariant, however, if we provide an additional ingredient in
the spectrum of the theory: a massless spinless (scalar) mode, $\phi$,
coupled longitudinally to the photon.  The Lagrangian of pure QED
together with such a mode may then be written:
\begin{eqnarray}
\label{one}
{\cal{L}} & =& 
 -\frac{1}{4}F_{\mu\nu}F^{\mu\nu} + \half(\partial_\mu \phi)^2 
 + \half e^2f^2(A_\mu)^2 
- efA_\mu\partial^\mu\phi 
\nonumber \\                      
& =& 
 -\frac{1}{4}F_{\mu\nu}F^{\mu\nu} + \half e^2f^2(A_\mu -\frac{1}{ef} 
                       \partial_\mu \phi)^2
\end{eqnarray}
In the second line we have completed the square of the scalar
terms and we see that
this Lagrangian is gauge invariant, i.e., 
if the transformation $A_\mu \rightarrow
A_\mu + \partial_\mu\chi/ef$ is accompanied by $\phi \rightarrow \phi +
\chi$ then ${\cal{L}}$ is invariant.  The quantity $f$, is called the ``decay constant'' of $\phi$. It
is the direct analogue of $f_\pi$ for the pion of QCD (for a discussion
of normalization conventions for $f_\pi$ see Section 2.1).

We see that the photon vector potential and the massless mode have
combined to form a new field: $ B_\mu = A_\mu -\partial_\mu \phi/ef $.
Physically, $B_\mu$ corresponds to a ``gauge-invariant massive
photon'' of mass $m_\gamma = ef$.  Thus, the Lagrangian can be written
directly in terms of $B_\mu$ as 
\begin{eqnarray}
{\cal{L}} & =& 
 -\frac{1}{4}F_{B\mu\nu}F_B^{\mu\nu} + \half m_\gamma^2 (B_\mu)^2 
\end{eqnarray}
where $F_{B\mu\nu} \equiv \partial_\mu B_\nu - \partial_\nu B_\mu
= F_{\mu\nu}$.
The $\phi$ field has now blended with $A_\mu$ to form the heavy
photon field; we say that the field $\phi$ has been ``eaten'' by
the gauge field to give it mass.

More generally, in order for this mechanism to produce a
superconductor, the Lagrangian for the mode $\phi$ must possess the
symmetry $\phi \rightarrow \phi + \xi(x)$ where $\xi(x)$ can be any
function of space-time ($\phi \rightarrow \phi + \xi$ where $\xi$ is a
constant is the corresponding global symmetry in the absence of the
gauge fields) .  This essentially requires a massless field $\phi$,
with derivative couplings to conserved currents $ \sim j_\mu
\partial^\mu\phi $.  The shift will then change the $\phi$ action
by at most a total divergence, and we can eliminate surface terms by
requiring that all fields be well behaved at infinity.

 Fields like $\phi$, called Nambu-Goldstone bosons (NGB's), always
arise when continuous symmetries are spontaneously broken. 

\vskip 0.2cm
\noindent
{\bf 1.3(ii) A Massless Fermion $\rightarrow$ Chiral Symmetry}
\vskip 0.2cm

Consider now a fermion $\psi (x) $, described by a four component
complex  Dirac
spinor. We define the ``Left-handed'' and ``Right-handed''
projected fields as follows:
\beq
\psi_L = \half(1-\gamma_5)\psi\; ;\qquad
\psi_R = \half(1+\gamma_5)\psi
\eeq
The $(1\pm \gamma_5)/2$ operators are
projections, and the reduced fields are 
equivalent to two independent two-component complex
spinors, each, by itself, forming an irreducible representation of
the Lorentz group.
The Lagrangian of a massless Dirac spinor decomposes
into two independent fields' kinetic terms as:
\beq
{\cal{L}} = \bar{\psi} i\slash{\partial }\psi
=
\bar{\psi}_L i\slash{\partial }\psi_L + \bar{\psi}_R i\slash{\partial }\psi_R
\eeq
This Lagrangian is invariant under
two independent global symmetry transformations,
which we call the {\em ``chiral symmetry''} $U(1)_L \times U(1)_R$:
\beq
\label{onedot5}
\psi_L \rightarrow \exp(-i\theta) \psi_L \; ;\qquad
\psi_R \rightarrow \exp(-i\omega) \psi_R\; ;\qquad
\eeq
The symmetry transformation corresponding to 
the conserved fermion number has
($\theta =\omega$)
while an axial, or $\gamma^5$, 
symmetry transformation has ($\theta = -\omega$).
The corresponding Noether currents are:
\beq
j_{\mu L} \equiv \frac{\delta {\cal{L}}}{\delta\partial_\mu \theta(x)} 
= \half \bar{\psi}\gamma_\mu (1 - \gamma_5) \psi;
\qquad
 j_{\mu R} \equiv \frac{\delta {\cal{L}}}{\delta\partial_\mu \omega(x)} 
= \half \bar{\psi}\gamma_\mu (1 + \gamma_5) \psi;
\eeq
We can form the {\em vector current}, $j_{\mu } = j_{\mu R} + j_{\mu L}
= \bar{\psi}\gamma_\mu\psi $
and the {\em axial vector current}, $j^5_{\mu } = j_{\mu R} - j_{\mu L} 
= \bar{\psi}\gamma_\mu \gamma_5\bar{\psi}$.

If we add a mass term to our
Lagrangian we couple together the
two independent $L$- and $R$-handed fields
and thus break the chiral symmetry:
\beq
\label{onedot7}
{\cal{L}} = \bar{\psi} i\slash{\partial }\psi - m\bar{\psi}\psi
=
\bar{\psi}_L i\slash{\partial }\psi_L + \bar{\psi}_R i\slash{\partial }\psi_R 
- m (\bar{\psi}_L \psi_R  +  \bar{\psi}_R \psi_L )
\eeq
The original $U(1)_L\times U(1)_R$ chiral symmetry of the massless theory has
now broken to a residual $U(1)_{L+R}$, which
is the vectorial symmetry of fermion number
conservation.  We can see explicitly that the vector current is conserved
since the transformation eq.(\ref{onedot5}) with $\theta = \omega$ is still a
symmetry of the Lagrangian eq.(\ref{onedot7}).  The axial current, on the other
hand, is no longer conserved: %
\begin{eqnarray}
\label{onedot8}
\partial_\mu \bar{\psi}\gamma^\mu \gamma_5  \psi
& = & \bar{\psi}\stackrel{\leftarrow}{\slash{\partial}} \gamma_5  \psi
+ \bar{\psi} \gamma_5 \stackrel{\rightarrow}{\slash{\partial}} \psi
\nonumber \\
& = &
-2im\bar{\psi} \gamma_5  \psi
\end{eqnarray}
The Dirac mass term has spoiled the axial 
symmetry ($\theta =
-\omega$).

\vskip 0.2cm
\noindent
{\bf 1.3(iii) Spontaneously Massive Fermion $\rightarrow$ Nambu-Goldstone Boson}
\vskip 0.2cm

Through a sleight of hand, however, we can preserve the full $U(1)_L
\times U(1)_R$ chiral symmetry, and {\em still give the fermion a
mass!}  We introduce a complex scalar field $\Phi$ with a Yukawa
coupling ($g$) to the fermion.  We assume that $\Phi$ transforms under
the $U(1)_L \times U(1)_R$ chiral symmetry as:
\begin{equation}
\label{eq19}
\Phi \rightarrow \exp[-i(\theta -\omega)]\Phi
\end{equation}
that is, $\Phi$ has nonzero charges under both 
the $U(1)_L$ and $U(1)_R$ symmetry groups. 
Then, we write the Lagrangian of the system as: 
\begin{equation}
\label{chira1}
{\cal{L}} = \bar{\psi}_L i\slash{\partial }\psi_L + \bar{\psi}_R i\slash{\partial }\psi_R
 - g(\bar{\psi}_L \psi_R\Phi +  \bar{\psi}_R \psi_L 
\Phi^*) + {\cal{L}}_\Phi 
\end{equation}
where
\begin{equation}
\label{chiral2}
{\cal{L}}_\Phi = |\partial\Phi|^2 - V(|\Phi|) 
\end{equation}
Unlike the previous case where we added the fermion mass term and
broke the symmetry of the Lagrangian, ${\cal{L}} $ remains invariant
under the full $U(1)_L \times U(1)_R$ chiral symmetry
transformations. The vector current remains the same as in the pure
fermion case, but the axial current is now changed to:
\begin{equation}
j_\mu^5 
= \bar{\psi}\gamma_\mu \gamma_5  \psi  + 2i\Phi^*(\stackrel{\rightarrow}{\partial}_\mu 
- 
\stackrel{\leftarrow}{\partial}_\mu)\Phi
\end{equation}
We can now arrange to have 
a ``spontaneous breaking of the chiral symmetry'' to give mass to
the fermion.  Assume the potential for the field $\Phi$ is:
\begin{equation}
\label{pot1}
V(\Phi) = - M^2|\Phi|^2  + \half \lambda |\Phi|^4
\end{equation}
The vacuum built around the field configuration $\VEV{\Phi} = 0 $ is
unstable.  Therefore, let us ask that $\VEV{\Phi} = v/\sqrt{2} $, and without
loss of generality we can take $v$ real.  The potential energy is minimized
for:
\begin{equation}
\frac{v}{\sqrt{2}} = \frac{M}{\sqrt{\lambda}}
\end{equation}
We can parameterize the ``small oscillations'' around the vacuum state
by writing:
\begin{equation}
\label{param1}
{\Phi} = \frac{1}{\sqrt{2}}(v + h(x))\exp(i\phi(x)/f)
\end{equation}
where $\phi(x)$ and $h(x)$ are real fields.  Substituting this anzatz into
the scalar Lagrangian (\ref{chiral2}) we obtain:
\begin{eqnarray}
{\cal{L}}_\Phi & = &
\half (\partial h)^2 -
M^2 h^2 - \sqrt{\frac{\lambda}{2}}M h^3 -\frac{1}{8} \lambda h^4
\nonumber \\
& & +\frac{v^2}{2f^2}(\partial\phi)^2 + \frac{1}{2 f^2} h^2 (\partial\phi)^2
+ \frac{\sqrt{2}M}{\lambda f^2}h (\partial\phi)^2 + \Lambda
\end{eqnarray}
where we have a negative vacuum energy density,
or cosmological
constant, $\Lambda = -M^4/2\lambda$ (of course, 
we can always add a bare cosmological
constant to have any arbitrary vacuum energy we wish).

We see that $\phi(x)$ is a massless field (a Nambu--Goldstone mode). It
couples only derivatively to other fields because of the
symmetry $\phi\rightarrow \phi+\xi $.\footnote{This is a general feature of a
Nambu--Goldstone mode, and implies ``Adler decoupling'': any NGB emission
amplitude tends to zero as the NGB four--momentum is taken to zero.}  The field
$h(x)$, on the other hand, has a positive mass-squared of $m^2 = 2M^2$.  
The proper normalization
of the kinetic term, for $\phi$, i.e., $(v^2/2f^2)(\partial\phi)^2$, requires
that $f = v$. Again, $f$ is the decay constant of the pion--like object
$\phi$. {\em The decay constant $f$ is always equivalent to the vacuum
expectation value} (apart from a possible conventional factor like
$\sqrt{2}$).  

Notice that the mass of $h(x)$ can be formally taken to be arbitrarily
large, i.e., by taking the limit $M\rightarrow \infty$, and
$\lambda\rightarrow \infty$ we can hold $v^2 = f^2 = 2M^2/\lambda$ fixed.
This completely suppresses fluctuations in the
$h$ field, and leaves us with a {\em nonlinear $\sigma$ model}
\cite{Gell-Mann:1960np}.  
In this case
only the Nambu-Goldstone  $\phi$ field is relevant at low energies.  
In the nonlinear $\sigma$ model we can
directly parameterize $\Phi = (f/\sqrt{2}) \exp(i\phi/f ) $.  
The axial current
then becomes:
\begin{equation}
j_\mu^5 = \overline{\psi}\gamma_\mu\gamma^5\psi - 2f\partial_\mu\phi
\end{equation}
where the factor of $2$ in the last term stems from the axial charge $2$ of
$\Phi$ (eq.(\ref{eq19})).  Let us substitute this into the Lagrangian eq.(\ref{chira1})
containing the fermions:
\begin{equation}
{\cal{L}} = \bar{\psi}_L i\slash{\partial }\psi_L + \bar{\psi}_R i\slash{\partial }\psi_R 
+ \half(\partial\phi)^2  
- (gf/\sqrt{2}) (\bar{\psi}_L \psi_R e^{i\phi/v} +  
\bar{\psi}_R \psi_Le^{-i\phi/f} ) 
\end{equation}
If we expand in powers of $\phi/f$ we obtain:
\begin{equation}
{\cal{L}} = \bar{\psi} i\slash{\partial }\psi + \bar{\psi} i\slash{\partial }\psi 
+ \half(\partial\phi)^2   - (gf/\sqrt{2}) \bar{\psi} \psi 
- i(g/\sqrt{2}) \phi \bar{\psi}\gamma^5 \psi + ...
\end{equation}
We see that this Lagrangian describes a Dirac fermion of mass
$m=gf/\sqrt{2}$, and a massless pseudoscalar Nambu-Goldstone boson $\phi$,
which is coupled to $i\bar{\psi}\gamma_5 \psi $ with coupling strength $g =
\sqrt{2}m/f$. This last result is the ``unrenormalized Goldberger-Treiman
relation'' \cite{Goldberger:1958tr}. The Goldberger-Treiman
relation holds experimentally in QCD for the axial coupling
constant of the pion $g_A$ and the nucleon, with $m=m_N$, $f=f_\pi$, and
is one of the indications that the pion is a Nambu-Goldstone boson.
 The Nambu-Goldstone phenomenon is ubiquitous throughout
the physical world, including spin-waves, water-waves, and waves on an
infinite stretched rope.

\vskip 0.2cm
\noindent
{\bf 1.3(iv) Massive Photon $\rightarrow$ Eaten Nambu-Goldstone Boson}
\vskip 0.2cm

We now consider what happens if $\Phi$ is 
a charged scalar field, with charge
$e$, coupled in a gauge invariant way to a vector potential.  
Let us ``switch off'' the fermions for the present.
We construct
the following Lagrangian:
\begin{eqnarray}
\label{abelian1}
{\cal{L}}'_\Phi & = & 
 -\frac{1}{4}F_{\mu\nu}F^{\mu\nu} + 
 |(i\partial_\mu -  e A_\mu)\Phi|^2 - V(\Phi) 
\end{eqnarray}
This is gauge invariant in the usual way, since with $A_\mu \rightarrow A_\mu
+ \partial_\mu\chi$ we can rephase $\Phi$ as $ \Phi \rightarrow e^{-ie\chi}
\Phi$.  The scalar potential $V(\Phi)$ is as given in eq.(\ref{pot1}),
hence, $\Phi$ will again develop a constant VEV $v = \sqrt{2/\lambda}
M$.  We can parameterize the oscillations around the minimum
as in eq.(\ref{param1}) and introduce the new vector potential,
\begin{eqnarray}
\label{manifest}
B_\mu & = &   A_\mu - \frac{1}{e}\partial_\mu\phi  \ .
\end{eqnarray}    
The Lagrangian (\ref{abelian1}) in this reparameterized form becomes:
\begin{eqnarray}
{\cal{L}}'_\Phi & = & 
 -\frac{1}{4}F_{\mu\nu}F^{\mu\nu}  
+ \half e^2 v^2 B_\mu B^\mu + \half (\partial_\mu h)^2 \nonumber \\ 
& & \qquad \qquad  
M^2 h^2 - \sqrt{\frac{\lambda}{2}}M h^3 -\frac{1}{8} \lambda h^4
+\half e^2 \left(h^2 + \frac{\sqrt{2}M}{\lambda}h \right)  B_\mu B^\mu
\end{eqnarray}
where $M$ is defined as in eq.(\ref{pot1}).  

Hence, we have recovered the massive photon
$B_\mu$ together with an electrically neutral field $h$, which we call the
``Higgs boson.'' The Higgs boson has a mass $\sqrt{2} M$ and has both cubic
and quartic self-interactions, as well as linear and bilinear couplings
to pairs of the massive photon. This model is 
essentially a (manifestly Lorentz invariant) Landau-Ginzburg model of
superconductivity, also known as the ``abelian Higgs model.''
We emphasize that it is manifestly gauge invariant because
the gauge field and Nambu-Goldstone mode occur in the 
linear combination of eq.(\ref{manifest}), as in eq.(\ref{one}).
We say that the Nambu-Goldstone boson has been ``eaten'' to become
the longitudinal spin degree of freedom of the photon. 

\vskip 0.2cm
\noindent
{\bf 1.3(v) Massive Photon and Massive Fermions Come Together}
\vskip 0.2cm

Finally, we can put all of these
ingredients together in one grand scheme. For example,
we can combine, e.g., a left-handed, fermion, $\psi_L$, of
electric charge $e$ and a neutral right-handed fermion, $\psi_R$,
with an Abelian Higgs model:
\begin{eqnarray}
{\cal{L}} & =& {\cal{L}}'_\Phi
+ \bar{\psi}_L (i\slash{\partial } - e\slash{A} ) \psi_L 
+ \bar{\psi}_R i\slash{\partial }\psi_R - 
g(\bar{\psi}_L \psi_R\Phi +  \bar{\psi}_R \psi_L 
\Phi^\dagger ) 
\end{eqnarray}
The Lagrangian is completely invariant under the electromagnetic gauge
transformation.  Also, the theory is intrinsically ``chiral'' in that
{\em the left-handed fermion has a different gauge charge than the
right-handed one.}  Now we see that, upon writing $\Phi$ as in
eq.(\ref{param1}), and performing a field redefinition, $\psi_L
\rightarrow
\exp(i\phi/v)\psi_L$, we obtain:
\begin{eqnarray}
{\cal{L}} & =&  -\frac{1}{4}F_{\mu\nu}F^{\mu\nu}  
+ \half e^2 v^2 B_\mu B^\mu + \half (\partial_\mu h)^2 \nonumber \\ 
& & \qquad \qquad  -M^2 h^2 -M h^3 -\frac{1}{8} \lambda h^4 
+ \half e^2 h^2  B_\mu B^\mu
\nonumber \\  
& & \qquad \qquad
+ \bar{\psi}i\slash{\partial }\psi - eB^\mu \bar{\psi}_L\gamma_\mu\psi_L  
- m\bar{\psi}\psi - \frac{1}{\sqrt{2}}gh\bar{\psi}\psi
\end{eqnarray}
Thus we have generated: (1) a dynamical gauge boson mass, $ev$, and 
(2) a dynamical
fermion Dirac mass $ m = g v/\sqrt{2}$. The Dirac mass mixes chiral fermions
carrying different gauge charges, and would superficially
appear to violate the gauge
symmetry and electric charge conservation. However, electric charge
conservation is spontaneously broken by the VEV of $\Phi$.  There remains a
characteristic coupling of the fermion to the Higgs field $h$ proportional to
$g$.\footnote{Unfortunately, this simple model is not consistent at quantum
loop level, since an axial anomaly occurs in the gauge fermionic current.
This can be remedied by, e.g., introducing a second pair of chiral fermions
with opposite charges.}

\subsection{The Standard Model} 
\label{sec:natural}

\noindent
{\bf 1.4(i) Ingredients}
\vskip 0.2cm

Analogues  of all of the above described ingredients 
are incorporated into
the Standard Model of the electroweak interactions.  
In the Standard Model electroweak sector, the gauge group
is $SU(2)\times U(1)$.  The
scalar $\Phi$ is replaced by a spin-$0$, weak isospin-$1/2$, field,
known as the Higgs doublet.  An arbitrary component of $\Phi$ develops a VEV
which defines the neutral direction
in isospin space (it is usually chosen to be the upper component of
$\Phi$ without loss of generality).
Three of the four components of the Higgs isodoublet then become
Nambu-Goldstone bosons, and combine with the gauge fields to make them
massive. The $W^\pm$ and
$Z$ bosons acquire masses, while the photon remains massless. 
The fourth component of $\Phi$ is a left-over, physical, massive object
called the Higgs boson, the  analogue of the $h$ field in our
toy models above.  Because the
Standard Model weak interactions are governed by the non-abelian group
$SU(2)$, tree-level $(h, h^2)\times(W^\pm W^\mp, ZZ)$ interactions occur; the
analogous tree-level $h\gamma\gamma$ coupling is absent in the abelian
Higgs model (and in the Standard Model) since the QED $U(1)$ symmetry
is unbroken.  The electroweak theory is thus, essentially, a
mathematical generalization of a (Lorentz invariant) Landau--Ginzburg
superconductor to a nonabelian gauge group.  We give the explicit
construction of the Standard Model in Appendix A.

We also see that there are striking parallels between
the dynamics of spontaneous symmetry breaking with
an explicit Higgs field, such as $\Phi$, and the dynamical behavior of QCD
near the scale $\Lambda_{QCD}$.  
Consider QCD with two flavors of massless quarks ($\Psi \equiv (u,d)$)
\begin{equation}
{\cal{L}}_{QCD} =  \bar\Psi_L i\slash{D} \Psi_L + \bar\Psi_L i\slash{D} \Psi_L
\end{equation}
where $D_\mu$ is the QCD covariant derivative.  The Lagrangian possesses an
$SU(2)_L \times SU(2)_R$ chiral symmetry.  When the running QCD coupling
constant becomes large at the QCD scale, the strong interactions bind quark
anti-quark pairs into a composite $0^+$ field 
$\bar{\Psi}\Psi$. This develops a non-zero
vacuum expectation value $\langle \bar{\Psi}\Psi \rangle \approx
\Lambda_{QCD}^3$, in analogy to the Higgs mechanism. 
This, in turn, 
spontaneously breaks the chiral symmetry $SU(2)_L \times SU(2)_R$
down to $SU(2)$ of isospin.  The light
quarks then become heavy, developing their ``constituent quark mass'' of
order $m_{Nucleon}/3$.  The pions, the lightest pseudoscalar mesons, are the
Nambu-Goldstone bosons associated with the spontaneous symmetry breaking and
are massless at this level (the pions are not
identically massless because of the fundamental
quark masses,
$m_u\sim 5$ MeV, $m_d\sim 10$ MeV).  
The essence of this dynamics is captured in a
toy model of QCD chiral dynamics known as the Nambu-Jona-Lasinio (NJL) model
\cite{Nambu:1961tp, Nambu:1961fr}. The NJL
model is essentially a transcription to a
particle physics setting of the BCS theory of superconductivity.  In Appendix
B we give a treatment of the NJL model.

If we follow these lines a step further and {\em switch
off} the Higgs mechanism of the electroweak interactions,
then we would have
unbroken electroweak gauge fields coupled to identically massless
quarks and leptons. However,  
it is apparent that the QCD-driven condensate $\langle
\bar{\Psi}\Psi \rangle \neq 0$ will then spontaneously break the
electroweak interactions at a scale of order $\Lambda_{QCD}$.  
The resulting Nambu-Goldstone bosons (the pions) will then be eaten
by the gauge fields to become
the longitudinal modes of the $W^\pm$ and $Z$ bosons.  The chiral
condensate characterized by a quantity $\Lambda_{QCD}\sim f_\pi$ 
would then provide the scale
of the $W^\pm$ and $Z$ masses, i.e., the weak scale in a theory of
this kind is given by $v_{weak}\sim f_\pi$.  Because $f_\pi \approx 93$
MeV is so small compared to $v_{weak}\sim 175$ GeV,
the familiar hadronic strong interactions cannot be the
source of EWSB in nature.
However, it is clear that EWSB could well involve
{\em a new strong dynamics similar to QCD,} with a higher-energy-scale,
$\Lambda \sim v_{weak}$, with
chiral symmetry breaking, and ``pions'' that become the longitudinal $W^\pm$
and $Z$ modes.  This kind of hypothetical new dynamics, known as
Technicolor, was proposed in 1979 (Section 2).

\vskip 0.2cm
\noindent
{\bf 1.4(ii) Naturalness}
\vskip 0.2cm

Various scientific definitions of 
``naturalness'' emerged in the early days
of the Standard Model.  ``Strong naturalness'' is
associated with the dynamical origin of a very small 
physical parameter in a theory in which no initial small 
input parameters occur.  The foremost example is the mechanism that generates
the tiny ratio, $\Lambda_{QCD}/M_{Planck} \sim 10^{-20}$ in QCD.  This
is also the premiere example of the phenomenon of ``dimensional
transmutation,'' in which a dimensionless quantity ($\alpha_s$) becomes a
dimensional one ($\Lambda_{QCD}$) by purely quantum effects.  
Here, the input parameter at very high
energies is $\alpha_s = g_s^2/4\pi$, the gauge coupling constant of QCD, 
which is a
dimensionless number, of order ${\cal{O}}(10^{-1})-{\cal{O}}(10^{-2})$,
not unreasonably far from unity. The renormalization group and
asymptotic freedom of QCD (effects of order $\hbar$) then determine
$\Lambda_{QCD}$ as the low energy scale at which
$\alpha_s(\Lambda)\rightarrow \infty$.\footnote{Perhaps an
enterprising string theorist will one day compute
$\alpha_s(M_{Planck})$, obtaining a plausible result such as
$\alpha_s(M_{Planck}) \sim 1/4\pi^2$ thus completely explaining the
detailed origin of $\Lambda_{QCD}$.}  

More introspectively, the scale
$\Lambda_{QCD}$ arises from the explicit scale breaking in QCD that is
encoded into the ``trace anomaly,'' the divergence of the scale
current $S_\mu$:
\begin{equation}
\label{scale}
\partial^\mu S_\mu= T^\mu_\mu = -\frac{\beta(g_s)}{2g_s}\; G^{\mu\nu}G_{\mu\nu}
\end{equation}
where $\beta(\alpha_s)$ is the QCD-$\beta$ function, arising
from quantum loops, and is of order $\hbar$ (we neglect quark masses
and, indeed, this has nothing to do with the quarks; the phenomenon happens in
pure QCD).
The smallness of $\alpha_s$ at high energies implies that scale invariance is
approximately valid there.  Asymptotic freedom implies that as we descend to
lower energy scales, $\alpha_s$ slowly increases, until the scale breaking
becomes large, finally self-consistently generating the dynamical scale
$\Lambda_{QCD}$ and the hierarchy $\Lambda_{QCD}/M_{Planck}$.  Note that most
of the mass of the nucleon (or constituent quarks)
derives from the nucleon matrix element of the
{\em RHS} of eq.(\ref{scale}).  In a sense, the ``custodial symmetry'' of
this enormous hierarchy is the approximate scale invariance of the theory at
high energies, 
in the ``desert'' where $\partial^\mu S_\mu \approx 0$.  Indeed,
if we were to arrange for $\beta(g) =0$, either by cancellations in the
functional form of $\beta$ or by having a nontrivial fixed point, then the
coupling would not run and $\Lambda_{QCD} \rightarrow 0$!  Strong
naturalness thus underlies one large hierarchy we see
in nature, i.e., how the ratio $\Lambda_{QCD}/M_{Planck} \sim 10^{-20}$ 
can be generated in principle is more-or-less understood! 

A parameter in a physical theory that must be {\em tuned} to a particular
tiny value is said to be ``technically natural'' if  radiative
corrections to this quantity are {\em multiplicative.}  Thus,
the small parameter stays small under radiative corrections.
This happens if setting
the parameter to zero leads the theory to exhibit a symmetry
which forbids radiative corrections from inducing a nonzero
value of the parameter. We then say that the
symmetry  ``protects'' the small
value of the parameter; this symmetry is called a ``custodial
symmetry.''

For example, if the electron mass, $m_e$, is set to
zero in QED, we have an associated chiral symmetry $U(1)_L\times U(1)_R$
which
forbids the electron mass from being regenerated by 
perturbative radiative corrections. 
The chiral  $U(1)_L\times U(1)_R$ in the $m_e=0$ limit is the
custodial symmetry of a small electron mass.
Radiative corrections in QED to the electron mass are a
perturbative power series in $\alpha$, and {\em they multiply a
nonzero bare electron mass}. Multiplicative radiative corrections 
insure that the electron mass,
once set small, remains small to all orders in perturbation
theory.  Now, clearly, technical naturalness begs a deeper, strongly natural,
explanation of the origin of the parameter $m_e$, but no apparent conflict
with any particular small value of $m_e$.

Typically scalar particles, such as the Higgs boson, have
no custodial symmetry, such as chiral symmetry, 
protecting their mass scales.
This makes fundamental scalars, such as the Higgs boson, 
unappealing and unnatural.
The scalar boson mass is
typically subject to large {\em additive} renormalizations, i.e., radiative
corrections generally induce a mass even if the mass is {\em ab initio} set to
zero.\footnote{This point is
actually somewhat more subtle; scale symmetry
can in principle act as a custodial symmetry if
there are no larger mass scales in the problem; see \cite{Bardeen:1995kv}.} 
The important 
exceptions to this are (i) Nambu-Goldstone bosons which can have
technically natural low masses due to their 
spontaneously broken chiral symmetry;
(ii) composite scalars which 
only form at a strong scale such as $\Lambda_{QCD}$
and could receive only additive renormalizations of order $\Lambda_{QCD}$; 
(iii) a technically natural mechanism for having fundamental
low mass scalars is also
provided by SUSY because the scalars are 
then associated with fermionic
superpartners.  The chiral symmetries of these 
superpartner fermions then protect the mass
scale of the scalars so long as SUSY is intact.  Hence, to
use SUSY to technically protect the electroweak mass scale in this way requires that
SUSY be a nearly exact symmetry on scales not far above the weak scale.

In attempting to address the question of naturalness of the Standard
Model, we are thus led to exploit these several exceptional possibilities
in model building to construct a natural symmetry
breaking (Higgs) sector.  In 
SUSY, the Higgs boson(s) are truly fundamental and the theory is
perturbatively coupled.  The SUSY technical naturalness protects the mass
scales of the scalar fields, and one hopes that the strong natural
explanation of the weak scale will be discovered eventually, perhaps
in the origin of SUSY breaking, perhaps incorporating
a trigger mechanism
involving the heavy top quark.  Here one takes the point of
view that there are, indeed, fundamental scalar fields in nature, and
they are governed by the organizing principle of SUSY that
mandates their existence.  This leads to the 
MSSM in which all of the
Standard Model fields are placed in $N=1$ supermultiplets, and are
thus associated with superpartners.  SUSY and the electroweak
symmetry must be broken at similar energy scales to avoid unnaturally
fine-tuning the scalar masses.  

In Technicolor, the scalars are
composites produced by new strong dynamics at the strong scale.  
Pure Technicolor, like
QCD, is an effective nonlinear--$\sigma$ model 
\cite{Gell-Mann:1960np},\footnote{The Higgs boson
is then the analogue of the $\sigma$-meson in QCD, which is a very
wide state, difficult to observe experimentally,
and can be decoupled in the nonlinear $\sigma$-model limit.} and the
longitudinal $W$ and $Z$ are composite NGB's (technipions).  More
recent models, such as the Top Quark Seesaw, feature an observable
composite heavy Higgs boson.  
In theories with composite scalar bosons, one hopes to imitate the
beautiful strong naturalness of QCD. This strategy, first introduced by
Weinberg \cite{Weinberg:1976gm} and Susskind
\cite{Susskind:1978ms} in the late 1970's seems {\em a priori} 
compelling.   Leaving aside the problem of the quark and
lepton masses, one can immediately write down a theory in which there
are new quarks (techniquarks), coupled to the $W$ and $Z$ bosons, and
bound together by new gluons (technigluons) to make technipions.  If
the chiral symmetries of the techniquarks are exact, some of the
technipions become exactly massless, have decay constants, $f \sim
v_{weak}$, and are then ``eaten'' by the $W$ and $Z$ to provide their
longitudinal modes.  We will call this ``pure Technicolor.'' 
Pure TC
can be considered to be a limit of the Standard Model in which all quarks
and leptons are approximately massless and the EWSB is
manifested mainly in the W and Z boson masses.  In this limit the
longitudinal W and Z are the original massless NGB's, or pions, of
Technicolor, and the scale of the new strong dynamics (i.e., the
analogue of $\Lambda_{QCD}$) is essentially $\sim v_{weak}$.  Again, the
scale is set by quantum mechanics itself; one need only specify
$\alpha_{techni}$ at some very high scale, such as the Planck scale,
to be some reasonable number of $O(10^{-2})$, and the renormalization
group produces the scale $v_{weak}$ automatically.  

A third possibility, is that the Higgs boson
is a naturally low-mass pseudo-Nambu-Goldstone boson
\cite{Georgi:1984af}, like the pion in
QCD. This idea, dubbed ``Little Higgs Models,'' 
has recently come back into vogue in the context of
deconstructed space-time dimensions 
\cite{Arkani:2001nc,Hill:2000mu,Arkani-Hamed:2001ca} (see Section 4).
The renaissance of this idea is so recent that, unfortunately, we will
not be able to give it an adequate review. It is currently being examined for
consistency with electroweak constraints and the jury is still out
as to how much available parameter space, and how little fine-tuning, 
will remain in Little Higgs Models. Nonetheless, the basic idea is compelling
and may lead ultimately to a viable scenario.

\subsection{Purpose and Synopsis of the Review}
\label{sec:synopsis}

This review will show the interplay between theory and
experiment that has guided the development of strong
dynamical models of EWSB,
particularly during the last decade.  Because they invoke new
strongly-interacting fields at an (increasingly) accessible energy
scale of order one TeV, dynamical models are eminently testable and
excludable.  To the extent that they attempt to delve into the origins
of flavor physics, they become vulnerable to a plethora of low-energy
precision measurements.  This has forced model-builders to be
creative, to seek out a greater understanding of phase transitions in
strongly-coupled systems, to seek out connections with other
model-building trends such as SUSY, and to re-examine ideas
about flavor physics.  Because experiment has played such a key role
in guiding the development of these theories, we choose to present the
phenomenological analysis in parallel with the theoretical.  Each set
of experimental issues is introduced at the point in the theoretical
story where it has had the greatest intellectual impact.

Chapter 2 explores the development of pure Technicolor theories.  As
already introduced in this chapter and further discussed in 2.1, pure
Technicolor (an asymptotically free gauge theory which spontaneously
breaks the chiral symmetries of the new fermions to which it couples)
can explain the origins of EWSB and the
masses of the $W$ and $Z$ bosons.  Section 2.2 discusses the
mathematical implementation of these ideas in the minimal two-flavor
model and the resulting spectrum of strongly-coupled techni-hadron
resonances.  The phenomenology of these resonances and the prospects
for discovering new strong dynamics in studies of vector boson
scattering at future colliders are are also explored.  The one-family
TC model and its rich phenomenology are the subject of section 2.3.

A more realistic Technicolor model must include a mechanism for
transmitting EWSB to the ordinary quarks and
leptons, thereby generating their masses and mixing angles.  
The original suggestion of an
Extended Technicolor (ETC) gauge interaction involving both ordinary and
techni-fermions alike is the classic physical realization of that
mechanism.  As discussed in sections 3.1 and 3.2, the extended
interactions can cause the strong Technicolor dynamics to affect
well-studied quantities such as oblique electroweak corrections or the
rates of flavor-changing neutral curent processes. Moreover, the
extended interactions require more symmetry breaking at higher energy
scales, so that the merits of the weak-scale theory are, as with
SUSY, entwined with mechanisms operating at higher energies.
These issues have had a profound influence on model-building.  Section
3.3 describes some of the explicit ETC scenarios designed to address
questions of flavor physics, further symmetry breaking, and
unification.

As one moves beyond the minimal TC and ETC theories, the conflict
inherent in a theory of flavor dynamics become sharper: creating large
quark masses requires a low ETC scale, while avoiding large
flavor-changing neutral currents mandates a high one.  An intriguing
resolution is provided by ``Walking'' Technicolor dynamics
(section 3.4). This departs radically from the QCD analogy: 
the dynamics remains strong far above the TC scale, up to the ETC scale, 
because the
$\beta$-function is approximately zero.  
This, in turn, has led to 
multi-scale and low-scale theories of Technicolor (section 3.5), which
predict many low-lying resonances with striking experimental signatures at
LEP, the Fermilab Tevatron and LHC.  As discussed in section 3.6, first
searches for these resonances have been made and extensive
explorations are planned for Run II.  Finally, while the initial
motivation for TC theories was the avoidance of fundamental
scalars, several variants of model-building have led to low-energy
effective theories that incorporate light scalars along with
TC; these are the subject of section 3.7.

Chapter 4 explores an idea that has taken hold as it became clear that the
top quark's mass is of order the EWSB scale
($v_{weak} \sim 175$ GeV): it is likely that the top quark
plays a special role in any complete model of strong electroweak symmetry
breaking.  In some sense, the top-quark may be a bona-fide techniquark with
dynamical mass generation of its own.  The first attempts at models along
these lines, known as top-quark condensation (section 4.1), demonstrate the idea
in principle, but are ultimately unacceptably fine-tuned theories.  However,
by generalizing the idea of top-quark condensation, and building realistic models
of the new ``Topcolor'' forces that underpin the dynamics, one is led back to
acceptable schemes under the rubric of Topcolor-Assisted Technicolor (TC2).
The TC2 models incorporate the best features of the TC and Topcolor
ideas in order to explain the full spectrum of fermion masses, while avoiding
the classic isospin violation and FCNC dilemmas that plague traditional
ETC models.  The Topcolor theory, its relationship to
TC, and associated phenomenology are the focus of sections 4.2
and 4.3.  Further insights into the dynamics of mass generation have arisen
in the context of Top-Seesaw models (section 4.4), in which the top quark's
large mass arises partly through mixing with strongly-coupled exotic quarks.

Most recently, as discussed in section 4.5, 
Topcolor is a forerunner of and has a natural setting
in latticized or ``deconstructed'' extra dimensions 
\cite{Hill:2000mu,Arkani-Hamed:2001ca}. Topcolor may represent
a  connection between the phenomenlogy of EWSB and the possible
presence of extra-dimensions of space-time at the $\sim$ TeV scale.  All in all, new
information about the top quark and new ideas about the structure of
space-time have fostered a mini--renaissance in the arena of new strong
dynamics and EWSB.

\newpage
\section{Technicolor}



Motivations underlying Technicolor have been described
for the reader in Section 1. We
wish to mention a few of the many detailed 
earlier reviews.  The review of Farhi and
Susskind \cite{Farhi:1981xs} and vintage
lectures by various authors \cite{Sikivie:1980fm,
Dimopoulos:1980tn, Lane:1980wc, Kaul:1983uk, Kane:1981ee,
Ellis:1981gx, King:1989ec} remain
useful introductions. There also 
exists a collection of reprints \cite{Farhi:1982vt}
tracing the early developments.  To our knowledge there is no
comprehensive review of the ``medieval'' period of TC, ca. late
1980's to early 1990's.  For more recent surveys, the reader should consult
the reviews of K. Lane, \cite{Lane:2000pa, Lane:1993wz} and
S. Chivukula \cite{Chivukula:1996uy,
Chivukula:1993nj, Chivukula:1998if}.
S. King has also written a more recent review \cite{King:1995yr} 
which develops some specific models, particularly of ETC (see Section 3).

Certain aspects of TC model-building
will not be addressed in the present discussion and we refer the interested
reader to the literature.  We will not discuss gauge coupling unification (see, e.g., \cite{Rubakov:1982vi, King:1985si, 
 Decker:1982ci, Decker:1982ae, Anselm:1981pz, Elias:1980ej}), 
 nor will we discuss cosmological implications (see,
e.g., \cite{Chivukula:1990qb}).  We will only briefly mention the 
idea of Supersymmetric TC
\cite{Witten:1981nf, Dine:1981za,
Dine:1982qj}, in the section on SUSY and
EWSB in Section 3.7.   
While TC has largely
evolved in directions somewhat orthogonal 
to Supersymmetry, the
overlap of these approaches may blossom 
in coming years should evidence for
NSD should emerge at the weak scale.

We presently begin with a description of the essential elements of
TC theories, addressing the problem of generating the $W$ and $Z$
masses.  We postpone to Section 3 the more involved details of ETC
and the problems and constraints associated with creating fermion
masses.  Accordingly, we will discuss the core phenomenology of TC
models in this section, and additional phenomenological discussion will
appear in Section 3 as the more detailed schemes unfold.


\subsection{Dynamics of Technicolor}

Technicolor (TC) was introduced by Weinberg \cite{Weinberg:1976gm} and Susskind
\cite{Susskind:1978ms} in the late 1970's.  The heaviest known fermion at
that time was the $b$-quark, with a mass of $\sim 5$ GeV and the top quark
was widely expected to weigh in around $15$ GeV.  The predicted Standard Model
gauge sector, on the other hand, 
was composed of the massless photon and gluon, 
and the 
anticipated, heavy gauge bosons, $W$ and $Z$, 
with $M_W\sim 80$ GeV and $M_Z\sim 90$ GeV.  
Since the matter sector appeared to contain
only relatively light fermions, it was useful
to contemplate a limit in which all of the elementary fermions are
approximately massless, and seek a mechanism to
provide only the heavy gauge boson
masses.  TC was a natural solution to this
problem.

\subsubsection{The TC $\leftrightarrow$ QCD Analogy}

TC is a gauge theory with properties similar to those of QCD.  For
concreteness, consider a TC gauge group
$G_{T} = SU(N_{T})$, having $N_{T}^2-1$ gauge bosons, called ``technigluons.''  
We introduce {\em identically massless} chiral ``techniquarks'' subject to this
new gauge force: $Q_L^{ai}$ and
$Q_R^{ai}$, where $a$ refers to TC and $i$ is a flavor index.  We
will assume that the $Q$'s fall into the fundamental, $\bf{N_{T}}$
representation of $SU(N_{T})$.  We further
assume that we have $N_{Tf}$ flavors of the $Q$'s. 
This then implies that we have an overall global chiral symmetry:
$SU(N_{Tf})_L\times SU(N_{Tf})_R\times [U(1)_A]\times U(1)_Q$
(where the $U(1)_A$ is broken by the axial anomaly
and is thus written in the square brackets $[...]$).  We will call
this the ``chiral group'' of the TC theory.  

TC, like QCD, is assumed to be a confining
theory\footnote{\addtolength{\baselineskip}{-.4\baselineskip} Note
that this is not the case in other schemes, such as Topcolor, and
a spontaneously broken nonconfining TC has been 
considered \cite{Hill:1993ev}.}
and has an intrinsic (confinement) mass scale $\Lambda_{T}$, which
must be of order the weak scale $v_{weak}$. Hence,
the physical spectrum will consist of TC singlets that are
either technimesons, composed of $\overline{Q}{Q}$, or technibaryons
composed of $N_{T}$ techniquarks. We expect various resonances above
the lowest lying states, showing Regge behavior, precocious scaling,
and ultimately even technijet phenomena
\cite{Kolb:1984dn}.  Since these objects are not found as stable states
in nature, the complete theory must provide for the decay of techniquarks
into the light observed quarks and leptons.  This is part of the function of
a necessary extension of the theory, 
called Extended Technicolor (ETC), which will serve
the role of giving the light quarks and leptons their masses as well.  We
postpone a detailed discussion of ETC until Section 3.

In complete analogy to QCD, the theory
produces a dynamical chiral condensate of fermion bilinears in the
vacuum, i.e., if $Q^i_{L,R}$ are techniquarks of flavor $i$, then 
TC yields:
\be
\label{cond1}
\VEV{\overline{Q}_{iL} Q_{jR} } \approx \Lambda_{TC}^3\delta_{ij}
\ee
This phenomenon occurs in QCD, in analogy to
the Cooper pair condensate in a BCS superconductor,
and gives rise to the large
nucleon and/or constituent quark masses.
The general implication of eq.(\ref{cond1}) is the
occurence of a ``mass gap'':
the techniquarks acquire constituent masses
of order $m_0 \sim \Lambda_{T}$.
Technibaryons
composed of $N_{T}$ techniquarks will be heavy with
masses of order $\sim N_{T}\Lambda_{T}$.  There must also
occur $N_{Tf}^2-1$ massless
Nambu-Goldstone bosons, 
with a common
decay constant $F_T \sim \Lambda_{T}$.  

By analogy, in QCD, if we consider 
the two flavors of up and down quarks to be
massless, then there is a global chiral symmetry of the Lagrangian of the
form $SU(2)_L\times SU(2)_R\times [U(1)_A] \times U(1)_B$, (where ``$A$'' stands
for axial, and ``$B$'' for baryon number). 
Within an approximation to the chiral dynamics of
QCD, known as the ``chiral constituent quark
model'' (e.g., see \cite{Bijnens:1993uz}), based upon
the Nambu--Jona-Lasinio (NJL) model,
(Appendix B) we can give a description of the dynamical chiral
symmetry breaking in QCD or TC 
(we'll refer to this as the NJL model below). 
 In this approximation, we obtain the Georgi-Manohar
\cite{Manohar:1984md} chiral Lagrangian for constituents
quarks and mesons as the low energy solution to the
model. In this NJL approximation there is a cut-off scale $M$, 
which is of order  $\sim m_\rho$, 
and we can relate $f_\pi$ 
to the dynamically generated ``mass gap'' of
the theory, i.e., the ``constituent quark mass'' of QCD.  If $N_f$ fermion
flavors condense, each having $N_c$ colors, then the quarks will have a
common dynamically generated constituent mass $m_0$ and produce a common
decay constant,
$f_\pi$ for the $(N_f^2-1)$ Nambu-Goldstone bosons given by the Pagels-Stokar
relation \cite{Pagels:1979hd} (see Appendix B; in the next section we discuss
normalization conventions for $f_\pi$):
\be
f_\pi^2 = \frac{N_c}{4\pi^2} m_0^2 \ln(M^2/m_0^2)\ .
\ee
In the NJL approximation 
we also obtain an explicit formula for the quark condensate bilinear:
\be
\label{cond2}
\VEV{\overline{Q}_{iL} Q_{jR}}  = \delta_{ij}\frac{N_c}{8\pi^2}m_0 M^2
\ee
Improvements can be made to the NJL model by softening the
four-fermion interaction and treating technigluon exchange in the ladder
approximation (see, e.g., \cite{Appelquist:1988yc, Lane:1988en, Mahanta:1989sn}).
Alternatively, lattice gauge theory techniques can be
brought to bear upon TC as well (see, e.g., \cite{Kogut:1983sm, Kogut:1982fn}).

The TC condensate is diagonal in an arbitrary basis of
techniquarks, $Q_i$'s, where the chiral subgroup, $SU(N_{Tf})_L\times
SU(N_{Tf})_R\times U(1)_Q$, is an exact symmetry ($ U(1)_A $ is broken
by instantons, and the Techni-$\eta$' is heavy like the $\eta$' of QCD).  
The
Standard Model gauge interactions, $SU(3)\times SU(2)_L\times U(1)_Y$ will
be a gauged subgroup of this exact TC chiral subgroup.  Indeed,
since we want to dynamically break electroweak symmetries, then
$SU(2)_L\times U(1)_Y$ {\em must always be} a subgroup of the chiral group.
In the minimal model described in the next section, QCD is not a subgroup of
the chiral group, while in the Farhi-Susskind model, both QCD and electroweak
gauge groups are subgroups of the chiral group.

When the $SU(3)\times SU(2)_L\times U(1)_Y$ interactions are turned on, a
particular basis for the $Q_{L,R}^{ai}$ has
thus been selected (the general models of Q's
contains various $SU(3)\times SU(2)_L\times U(1)_Y$ representations). Thus,
an ``alignment'' occurs in the 
dynamical condensate pairing of $\overline{Q}_L^i$ with $Q_R^i$.  In
general it is not obvious {\em ab initio} that this alignment preserves
the exact gauge symmetries (like electromagnism and QCD; i.e., the 
electric charge and color generators must commute with
the condensate to preserve these symmetries), and breaks yet
other symmetries (electroweak) in the desired way.  This is one
example of the
``vacuum alignment'' problem \cite{Peskin:1980gc, Preskill:1981mz,
Binetruy:1982uf, Leon:1983dg, Luty:1992xz, Georgi:1994at, Chivukula:1998uf,
Lane:2000es}.    In the simplest TC representations the
desired vacuum alignment is manifest.

\subsubsection{Estimating in TC by Rescaling QCD; $f_\pi$, $F_T$, $v_{weak}$}

Since TC is based upon an analogy with the dynamics
of QCD, 
we can use QCD as an ``analogue
computer'' to determine, by appropriate rescalings, the properties of the
pure TC theory .  A
convenient set of scaling rules due originally to `t Hooft
\cite{'tHooft:1977am, 'tHooft:1978yv, 'tHooft:1983wm} (see,
furthermore, e.g., \cite{Manohar:1984md, Georgi:1986kr}), characterize
the behavior of QCD.  These rules have been extensively applied to TC
\cite{Dimopoulos:1980sp}.  The main scaling rules are:
\be
\label{cond3}
f_\pi \sim   \sqrt{N_c }\Lambda_{QCD} 
\qquad
\VEV{\overline{Q}_i Q_j }  \sim \delta_{ij}{N_c}\Lambda_{QCD}^3
\qquad
m_0 \sim \Lambda_{QCD}
\ee
These rules follow from the NJL approximation with the identification
$m_0\sim M\sim \Lambda_{QCD}$.
When we discuss TC models we will use
the notation, $F_T$, to refer to the corresponding NGB, or
technipion, decay constant.

We typically
refer to the weak scale $v_{weak} =2^{-3/4}G_F^{-1/2} = 175$ GeV,
which is related to the usual Higgs VEV as $v_{weak} =v_0/\sqrt{2}$,
and $v_0 =246$ GeV. Hence, in the spontaneously
broken phase of the Standard Model
we can parameterize the Higgs field with its VEV
as: 
\be
H = \exp(i\pi^a\tau^a/v_0) \left( \begin{array}{c} {v_0}/{\sqrt{2}} + {h_0}/{\sqrt{2}}
\\ 0 \end{array} \right).
\ee
This gives the kinetic terms for the $\pi^a$ (and $h^0$) the proper 
canonical normalizations in the limit of switching off the gauge fields.
>From the Higgs boson's kinetic terms we extract,
where the electroweak covariant derivative $D_\mu$ 
is defined in eq.(A.1):
\be
\label{Hkin}
D_\mu H^\dagger D^\mu H \rightarrow \frac{g_2}{2}v_0 W^+_\mu\partial^\mu \pi^- +
\frac{g_2}{2}v_0  W^-_\mu\partial^\mu \pi^+ +
v_0  (\frac{g_2}{2}W^0_\mu + \frac{g_1}{2}B_\mu  )\partial^\mu \pi^0 + ...
\ee
Now, in QCD $f_\pi$ is defined by:
\be
\label{pion}
<0|j^{a5}_\mu|\pi^b> = if_\pi p_\mu \delta_{ab} \qquad F_\pi\approx 93 \;MeV
\ee
where $j^{a5}_\mu = \overline{\psi}\gamma_\mu\gamma^5\frac{\tau^a}{2} \psi$
where $\psi =(u, d)$ in QCD  
(Note: another definition in common use involves the 
matrix elements of the charged currents
and differs by a factor of $\sqrt{2}$, i.e., $F_\pi = \sqrt{2}f_\pi$).
When pions (Nambu-Goldstone bosons) or technipions are introduced through
chiral Lagrangians, we have typically a nonlinear-$\sigma$ model
field $U$ that transforms under $G_L\times G_R$ as
$U\rightarrow L U R^\dagger$, and its kinetic term is of the form:
\be
U = \exp(i \pi^a \tau^a/f)\qquad {\cal{L}} = \frac{f^2}{4}\Tr (\partial^\mu U^\dagger
\partial_\mu U)
\ee
Then the normalization is $f = f_\pi = 93$ MeV, which can be seen
by working out the axial current, 
$j_\mu^5 = \delta {\cal{L}}/\delta \partial_\mu \pi^a$ 
and comparing with eq.(\ref{pion}).

We will similarly
define $F_T$ as the techni-pion, $\tilde{\pi}$, to vacuum matrix
element for the corresponding techniquark axial current,
involving a single doublet of techniquarks, in TC models,
i.e., $\tilde{j}^{a5}_\mu = \overline{Q}\gamma_\mu\gamma^5\frac{\tau^a}{2} Q$
where $Q =(T, B)$ are techniquarks:
\be
\label{TCpion}
<0|\tilde{j}^{a5}_\mu|\tilde{\pi}^b> = iF_T p_\mu \delta_{ab} \qquad F_T\propto v_{weak}
\ee
Including electroweak gauge interactions the 
techniquark kinetic terms take
the form:
\be
\label{tquarks}
\overline{Q_L} i\slash{D}Q_L + \overline{Q_R} i\slash{D}Q_R
\longrightarrow 
\frac{F_T^2}{4}\Tr ((D^\mu U)^\dagger
(D_\mu U))
\ee
where $D_\mu $ is defined in eq.(\ref{a1}).  We have also written
the corresponding chiral Lagrangian describing the technipions
with a nonlinear-$\sigma$ model, or chiral field 
$U = \exp(i\pi^a\tau^a/F_T)$ (in the chiral Lagrangian 
the left-handed electroweak generators act on the left side of $U$, while
vectorial generators act on both left and right,
and are commutators with $U$).
We can thus form matrix elements of
the techniquark kinetic terms between vacuum and technipion states,
or expand the chiral Lagrangian to first order in $\pi^a$.
>From eqs.(\ref{TCpion},\ref{tquarks}) we obtain the effective Lagrangian
describing the longitudinal coupling of $(W^\pm, Z)$
to $(\pi^\pm, \pi^0)$:
\be
\label{pikin}
\frac{g_2}{2}F_T W^+_\mu\partial^\mu \pi^- + 
\frac{g_2}{2}F_T W^-_\mu\partial^\mu \pi^+ +
F_T (\frac{g_2}{2}W^0_\mu + \frac{g_1}{2}B_\mu  )\partial^\mu \pi^0 
\ee
Hence, comparing
eq.(\ref{Hkin}) to eq.(\ref{pikin}) 
we see that the Higgs VEV $v_0 = F_T$ 
when we have a single doublet of technipions.  If $N_D$ doublets
carry weak charges then eq.(\ref{tquarks}) contains $N_D$
terms.  $F_T$ remains the same, but the weak scale becomes
$v_0 = \sqrt{N_D}F_T$.\footnote{If we could switch off nature's EWSB,
the preceding discussion indicates precisely how QCD itself
would then break the electroweak interactions.
Indeed, (see e.g. \cite{Lane:2000pa}) if there is no EWSB
(no Higgs boson), then the up and down quarks are
identically massless.  Then we would have the QCD chiral condensate, 
generating constituent quark masses for up and down of order
$300$ MeV, and massless composite pions in the absence of electroweak
gauge interactions.  In the presence of electroweak gauge interactions
the pions are mathematically ``eaten'' to become $W_L$ and $Z_L$.
Now, however the $W$ and $Z$ masses given by $M'_W = f_\pi M_W/v_{0}
\sim 29$ MeV and $M'_Z = f_\pi M_Z/v_{0} \sim 33$ MeV. The
longitudinal W and Z would thus be the ordinary $\pi$'s of QCD. Thus,
QCD misses the observed masses by $\sim 4$ orders of magnitude, but it
gets the ratio of $M_W/M_Z$ correct! } 

Consider a TC gauge group $SU(N_{T})$ with $N_D$
electroweak left-handed doublets and $2N_D$ singlets of right-handed
techniquarks, each in in the $\bf{N_{T}}$ representation.  The strong
$SU(N_{T})$ gauge group will form a chiral condensate pairing the left-handed
fermions with the right-handed fermions.  This produces $(2N_D)^2-1$
Nambu-Goldstone bosons (technipions, $\pi_T$, and the
singlet $\eta_T'$) each with decay constant $F_T$. Hence, we can estimate $F_T$
from the QCD analogue $f_\pi$ using the scaling rules:
\be
F_T \sim  \sqrt{\frac{N_{T}}{3}}\left(
\frac{\Lambda_{T}}{\Lambda_{QCD}} \right) f_\pi;
\qquad v_0 = \sqrt{N_D}F_T \sim \sqrt{\frac{N_D N_{T}}{3}}\left(
\frac{\Lambda_{T}}{\Lambda_{QCD}} \right) f_\pi
\ee
As stated above,  $v_0^2$ receives contributions from $N_D$ copies
of the electroweak condensate. 

Now, in the above example of scaling, we have taken the point of view that
$\Lambda_{T}$ is fixed and, e.g., $v_0$ varies as $\sim\sqrt{N_{T} N_D}
$ as we vary $N_{T}$ and $N_D$.  This is an unnatural way to define the
TC theory, since $v_0$ is an {\em input parameter} whose value is
fixed by $G_F$. Hence, in discussing
TC models from now on we will use the ``{\em TC Scaling}''
scheme in which we hold $v_0$ fixed, and vary $\Lambda_{T}$ together with
$N_{T}$ and $N_D$.  Hence, the same example rewritten in the language of
TC Scaling is:
\be
\label{eqscale}
F_T \sim  v_0\sqrt{\frac{1}{N_D}};\qquad 
\Lambda_{T} = \Lambda_{QCD}\frac{v_0\sqrt{3}}{f_\pi \sqrt{N_DN_{T}}};
\qquad
v_0 = 246 \;\makebox{GeV.}
\ee
With these scaling rules in hand, we turn to 
the key properties of the main classes of TC models.

\noindent
\subsection{The Minimal TC Model of Susskind and Weinberg}

\subsubsection{Structure}

In the minimal TC scheme, 
introduced by Weinberg \cite{Weinberg:1976gm} and
Susskind \cite{Susskind:1978ms}, the gauge group is $SU(N_{T})
\times SU(3) \times SU(2)_L\times U(1)_Y$. In addition to the
ordinary Standard Model fermions, 
we include at least
one flavor doublet of color singlet technifermions, $(T,B)$. These
form two chiral weak doublets, $(T,B)_L$ and $(T,B)_R$,
and the chiral group is therefore $SU(2)_L\times SU(2)_R\times
[U(1)_A]\times U(1)_B$. The
left-handed weak doublet $(T,B)_L$ will have $I=\half$ electroweak
$SU(2)_L$ gauge couplings, while the $(T,B)_R$ form
a pair of singlets.  
The gauge anomalies of the Standard Model and TC
vanish if we take ``vectorlike'' assignments under the weak
hypercharge $Y$:
\be
Q^a_L = \left( \begin{array}{c} T \\ B \end{array} \right)^a_L\qquad (Y=0);
\qquad \qquad
Q^a_R = ( T_R, \; B_R)^a \qquad \frac{Y}{2} = ( \frac{1}{2}, -\frac{1}{2})\ .
\ee
As ``$a$'' is the TC index, we have $N_{T}$
TC copies of these objects. 
Anomalies involving $Y$ are absent because 
we have introduced
a vector-like pair $(T_R,B_R)$, each element with opposite $Y$
(the Witten global $SU(2)$ anomaly \cite{Witten:1982fp} vanishes 
provided $N_{T}$ is even for any $N_D$).  We can readily generalize the
model to include arbitrary flavors,
$N_D>1 $ doublets, of the same color-singlet technifermions.
With these assignments the techniquarks have electric charges as defined
by $Q=I_3+Y/2$,  of $+1/2$ for $T$ and $-1/2$ for $B$.

Without developing a detailed treatment of grand unification, we can see that
the scale $\Lambda_{T}$ plausibly corresponds to the electroweak scale, and
can be generated naturally by choosing the contents of the model
appropriately.  If we assume that the TC gauge coupling constant is
given by a high energy theory (e.g., a GUT or string theory) with the
unification condition,
\be
\alpha_{T}(M_{GUT}) = \alpha_3(M_{GUT}),
\ee
then evolving down by
the renormalization group, we obtain for the TC scale
$\Lambda_{T}$ to one-loop precision:
\be
\frac{\Lambda_{T}}{\Lambda_{QCD}}  = \exp\left[
\frac{2\pi (b_0' - b_0 ) }{b_0 b_0'\alpha_{3}(M_{GUT})}\right]
\ee
where, 
\be
b_0 = 11 -\frac{2}{3} n_f \qquad \makebox{and} \qquad b_0' = \frac{11N_{T}}{3} 
-\frac{4}{3}N_{D}
\ee
$b_0$ and $b_0'$ are the one-loop $\beta$-function coefficients of QCD and TC
respectively.  Putting in some ``typical'' numbers, we
find that $N_{T}=4$, $n_f= 6$, $N_{D} = 4$ and $\alpha_3^{-1}(M_{GUT})
\approx 30$ implies that ${\Lambda_{T}}/{\Lambda_{QCD}} \approx 8.2\times 10^2$ (for
smaller $N_D$ the ratio rapidly increases).  Hence, using $\Lambda_{QCD}=200$
MeV, we find $\Lambda_{T}\approx 165$ GeV, and $F_T\approx 95$ GeV, and we
predict $v_{weak} \approx 190$ GeV, not far from the known $175$ GeV.  
The minimal model thus crudely exhibits how TC, in
principle, can generate the electroweak hierarchy, $v_0/M_{Planck} \sim
10^{-17}$, thus generate a TC mass scale $\Lambda_{T} \sim 10^2 - 10^3$ GeV, 
in a natural way from the scale anomaly (renormalization group
running) and grand unification.  

Let us turn this around and assume that we have naturally obtained the
hierarchy $\Lambda_{T}/M_{Planck} \sim 10^{-17}$ in some particular and
complete TC theory. We will not bother further with unknown
details of unification, and take, rather, a bottom-up approach.  The
QCD/TC scaling laws allow us to make estimates of the chiral
dynamics if we know $\Lambda_{T}$, and to conversely derive $\Lambda_{T}$
from the fixed electroweak parameter $v_0$. 
In a minimal TC model with $N_{T} = 4$ and $N_{D} =1$,
$\Lambda_{QCD} \sim 200$ MeV, $f_\pi \sim 93$ MeV,
we obtain the result from eq.(\ref{eqscale}) that $\Lambda_{T} \sim 458$ GeV 
(whereas we obtain
$\Lambda_{T} \sim 205$ GeV with $N_D=5$, which is closer to
consistency with the top-down estimate given above).

\subsubsection{Spectroscopy of the Minimal Model}
\vskip .1in
\noindent
{\bf 2.2.2(i) Techniquarks}
\vskip .1in

The spectrum of the minimal model 
follows QCD as well.  The mass gap of the
theory can be estimated by scaling from QCD.  From
eq.(\ref{eqscale}) the techniquarks
will acquire dynamical constituent masses of order:
$m_{TQ} \sim m_0 v_0\sqrt{3}/f_\pi\sqrt{N_{T}N_D}$ 
where $m_0\sim m_N/3 \sim 300 $ MeV is the constituent
quark mass in QCD. This gives $m_{TQ} \sim 690$ GeV for
the  minimal model with $N_{T}=4$, $N_D=1$.
Hence, there will be a spectrum of ``baryons'' composed
of $QQQQ$ with a mass scale of order $\sim 3/\sqrt{N_D} $ TeV. At the
pre-ETC level the lightest of these
objects is stable, and cosmologically
undesireable \cite{Chivukula:1990qb}. 
However, in the presence
of the requisite ETC interactions there will be $Q\rightarrow
q+X$, $Q\rightarrow\ell+X$  transitions
and the baryons will become unstable to decay
into high multiplicity final states of light
quarks and leptons, e.g., final states 
containing $12$ top-- or bottom-- quarks! Note
that at higher energies
the minimal model is asymptotically free. At very high
energies, $E >> 10$ TeV we would expect the formation of ``techni-jets'' 
\cite{Kolb:1984dn}.  

\vskip .1in
\noindent
{\bf 2.2.2.(ii) Nambu-Goldstone Bosons}
\vskip .1in

The spectrum of the lowest lying mesons is predictable in the QCD-like
TC models by analogy with QCD.  It is controlled by the chiral group
of the model.  With $N_D=1$ we have a chiral group of $SU(2)_L\times
SU(2)_R\times [U(1)_A]\times U(1)_V$, and (switching off the gauge
interactions) there are three identically massless isovector pseudoscalar
Nambu-Goldstone bosons, $\pi^\pm_T, \pi^0_T$, which are dubbed technipions.
The decay constant of the technipions is $F_T = v_0/\sqrt{N_{D}} 
= 246/\sqrt{N_{D}}$ GeV,
so that the pion decay constant and $v_0$ coincide in models with a single
weak-doublet of left-handed techniquarks.  When we gauge the $SU(2)_L\times
U(1)_Y$ subgroup of the chiral group, the technipions become the longitudinal
weak gauge bosons $W^\pm_L$ and $Z_L$.  following the conventional
Higgs/superconductor mechanism as described in Section 1.3.  Indeed, it is a
general aspect of all dynamical symmetry breaking schemes that the $W^\pm_L$
and $Z^0_L$ are composite NGB's.

With $N_D=1$ the only remnant Nambu-Goldstone boson, corresponding to the
spontaneous breaking of the ungauged $U(1)_A$ subgroup of the chiral group,
is an isosinglet Techni-$\eta'$, or $\eta_{T}'$.
This is the analogue of the $\eta'$ in QCD.  
The $\eta_{T}'$ acquires mass through TC instantons.
The original analysis of this state in TC is due to
deVecchia and Veneziano \cite{DiVecchia:1980xq}, and a very nice and more
recent treatment, which we follow presently, is given by Tandean
\cite{Tandean:1995ci}.

The symmetry, $U(1)_A$, is broken by the axial vector anomaly, i.e. the
triangle diagrams with the emission of two technigluons. The anomaly is
actually suppressed in the large $N_{T}$ limit, as $\sim 3/N_{T}$,
\cite{Witten:1979vv, Witten:1980sp, Coleman:1980mx}, and an
estimate of the $\eta_{T}'$ mass is
therefore\footnote{\addtolength{\baselineskip}{-.4\baselineskip} An
overall $\sqrt{6}/2$ factor is associated with the present
techni-flavor group $SU(2)$ relative to the flavor group $SU(3)$ of
QCD in which the $\eta'$ lives
\cite{Tandean:1995ci}. } $m_{\eta' TC} \sim (3\sqrt{6}/2N_{T})
\sqrt{3/N_{T}N_D}(v_0/f_\pi) m_{\eta'} \sim 2 $ TeV for $N_{T}=4$ and
$N_D=1$.  
Thus the $\eta_{T}'$ is a relatively heavy state;
in the minimal $N_D=1$ model, it TC is fairly well hidden from
direct searches at low energies $E \lta 1$ TeV.

The decay of $\eta_{T}'$ parallels, in part, that of the $\eta'$ in QCD and
involves anomalies. Note that while the
$U(1)_A$ anomaly is suppressed by $1/N_{T}$, the flavor anomalies relevant to
these decays (these are the analogues of the $\pi^0\rightarrow 2\gamma$
anomaly in QCD) are generally proportional to $N_{T}$.  We expect the
principle decay modes $\eta_{T}'\rightarrow ZZ,\gamma Z,\gamma\gamma $ and
$\eta_{T}' \rightarrow W^+W^-Z, ZZZ$ as analogues to the multi-pion and
photonic decays of the $\eta'$ in QCD.  However, more novel decays directly
into $gg$ ($g\sim$ gluon) or into top quark pairs can also be possible.
These decay modes depend on the existence and details of a coupling to
ordinary quarks and leptons, which is the subject of ETC,
discussed in Section 3 (statements about ETC are 
more model dependent, especially regarding the top quark).
Typical estimates for decay widths can be
gleaned from \cite{Tandean:1995ci}: 
\bea 
\Gamma(\eta_{T} \rightarrow
t\overline{t}) & \sim & \left(\frac{3}{N_{T}}\right) 107 \lambda^2
/\sqrt{N_D} \;\makebox{MeV}.  \nonumber \\ \Gamma(\eta_{T} \rightarrow gg) &
\sim & \left(\frac{3}{N_{T}}\right) 56 /\sqrt{N_D} \;\makebox{MeV}.
\nonumber \\ \Gamma(\eta_{T} \rightarrow W^+W^-) & \sim &
\left(\frac{3}{N_{T}}\right) 26/\sqrt{N_D} \;\makebox{MeV}.  
\eea 
where the parameter $\lambda$ describes the ETC coupling of
$\eta_{T}$ to the top quark.  Note that the $\overline{t}t$ mode,
though model dependent, can
dominate provided that $\lambda \sim M^2_{T}/M_{ETC}^2$ is not too
small.  See \cite{Tandean:1995ci} for more
details.

Next, let us consider increasing the number of technidoublets.  For general
$N_D > 1$, we first turn off the Standard Model gauge interactions.  Then our
$SU(N_{T})$ theory has a global chiral group of $ SU(2N_D)_L\times SU(2N_D)_R
\times [U(1)_A ]\times U(1) $.  
This leads to $(2N_D)^2-1$ Nambu-Goldstone bosons, $\pi^a_T$ in the
adjoint representation of $SU(2N_D)$, and the singlet, $\eta_{T}'$ described
above.  These are analogues of the pseudoscalar octet, $\pi,K,\eta$ in QCD,
and the singlet $\eta'$.  Note that because our techniquarks carry no QCD
color, the NGB's will likewise be colorless. In the absence of the
electroweak interactions all of the remaining NGB's are massless (except
$\eta_T'$ as described above).  As such low mass objects are not seen in the
spectrum, sources of NGB masses will be required to avoid conflict with
experiment.  

Switching on the gauge interactions, the $SU(2)_L\times U(1)_Y$ becomes 
a gauged subgroup of the full 
$ SU(2N_D)_L\times SU(2N_D)_R \times [U(1)_A ]\times U(1)$ 
chiral group. When we restore the electroweak interactions, specific linear
combinations of triplets of the NGB's are eaten to give masses to $W$ and
$Z$. Of the NGB's we thus have several classes of objects.  Some of the
remaining $\pi_T$'s carry electroweak charges, but form linear
representations of $SU(2)_L$.  These are the NGB's that correspond to
generators of $ SU(2N_D)_L\times SU(2N_D)_R $ that do not commute with the
$SU(2)_L\times U(1)$ subgroup. They are analogues of $K$ mesons in $SU(3)$
that form linear isodoublets.  In addition there are generalizations of the
$\eta$ in QCD, which is an isosinglet.  These objects correspond to
generators that commute with $SU(2)_L\times U(1)_Y$, and do not acquire
masses via gauge couplings.

To count and to classify the various technipions, it is useful 
to use matrix direct product notation.  First ignore
$U(1)_Y$. Our particular model has an $SU(2N_D)$ vector flavor
subgroup of the chiral group. This group has $4N_D^2-1$ generators.  We can
arrange all of the $N_D$ doublets into a column vector with respect to this
group, i.e., the fundamental ${\bf 2N_D}$ representation of $SU(2N_D)$.  On
the other hand, the horizontal flavor subgroup, which acts on weak doublets, is
$SU(N_D)$.  

Hence, we can classify the generators associated with the NGBs
according to their transformation properties under the direct product of the
$SU(2)_L$ and $SU(N_D)$ as follows:  

\begin{itemize}
\item The three generators of $SU(2)_L$ can be written as the direct
product $\tau^a \otimes I_d$ where $I_d$ is the $d$-dimensional unit matrix
acting on the left-handed doublets.  Therefore, the three NGBs corresponding to
the generators $\tau^a \otimes I_d$ are the states that become the $W^\pm_L$ and
$Z^0_L$. 

\item There are $3N_D^2-3$ matrices of the form $\tau^a
\otimes\lambda^A $, where $\lambda^A$ are the $N_D^2-1$ generators of
$SU(N_D)$.  Since these matrices do not commute with the $\tau^a
\otimes I_d$ $SU(2)_L$ charges, the corresponding $3N_D^2-3$ PNGB's carry
$SU(2)_L$ charge.  These non-neutral PNGB's acquire mass when we switch on
the gauge interactions, just as the $\pi^\pm$ acquire a mass splitting
relative to $\pi^0$ in QCD due to the coupling to electromagnetism.

\item The matrices of the form $I_2\otimes \lambda^A $ commute with
$SU(2)_L$, so there are $N_D^2-1$ objects which are sterile under
$SU(2)_L$. These objects are PNGB's that remain identically massless (modulo
electroweak instanton effects), and are dubbed ``techni-axions.''

\item The final element $I_2\otimes I_d$ corresponds to the $\eta_{T}'$.

\end{itemize}

\begin{figure}[t]
\vspace{7cm}
\includegraphics{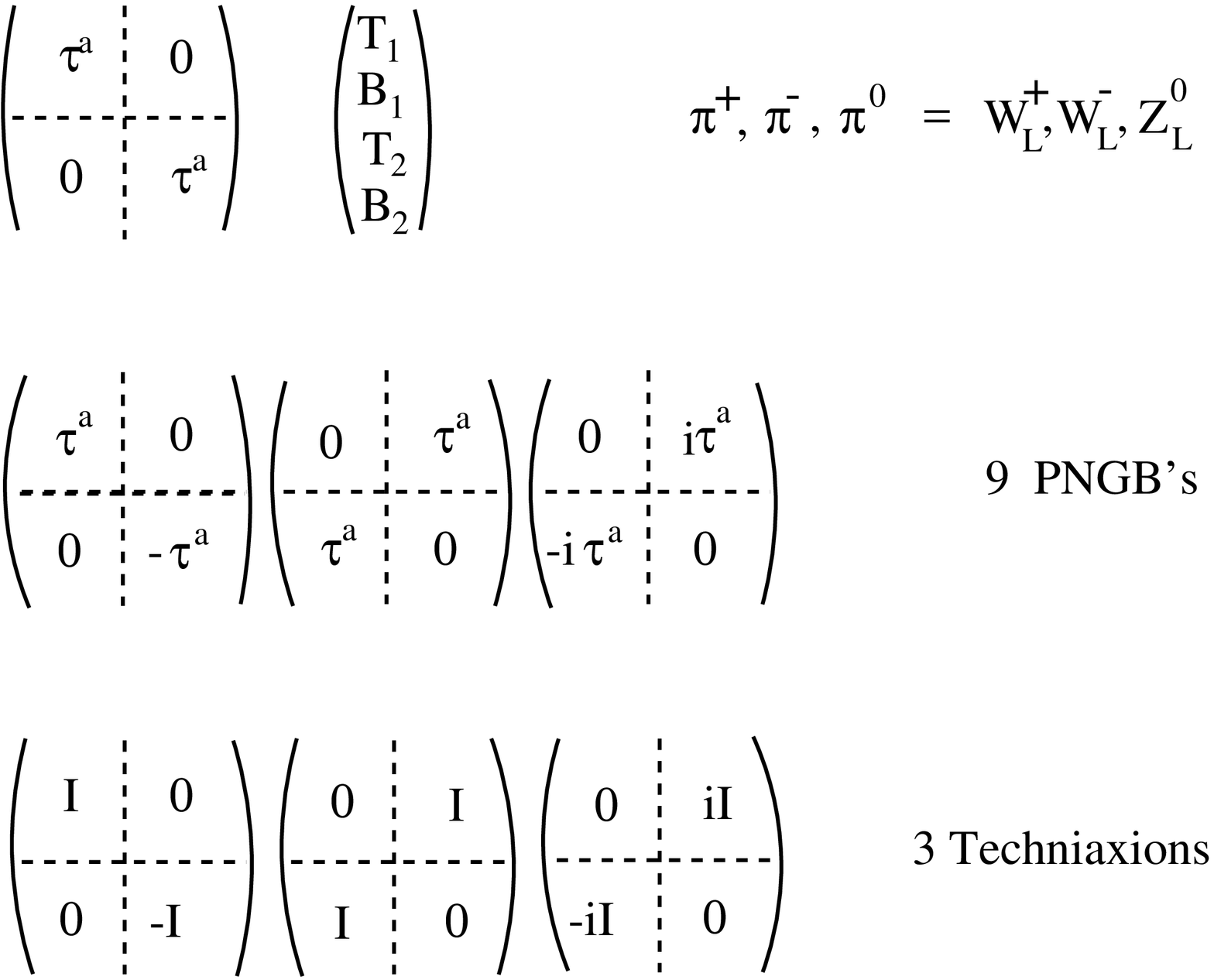}
\vspace{5cm}
\caption[]{\small \addtolength{\baselineskip}{-.4\baselineskip} The 
$(2N_D)^2-1$ Nambu-Goldstone bosons  corresponding to an $N_D=2$
extension of the minimal model. The $\eta'_{T}$
corresponds to the unit matrix.}
\label{cth1}
\end{figure}

\noindent We illustrate the above decomposition explicitly in
Fig.(\ref{cth1}) for the special case of $N_D=2$.

A similar analysis in the presence of $U(1)_Y$ adds a little more
information.  The PNGB's that can carry charges under $U(1)_Y$ are given by
the generators that do not commute with $Y \otimes I_d$.  These are the
electrically charged technipions -- hence, it is easier to examine the
properties under $U(1)_{EM}$.  In contrast, the techniaxions are neutral
under $U(1)_Y$, rendering them sterile under all of the gauge interactions.

To estimate the masses of the charged PNGB's, we again use QCD as an analogue
computer, and rescale the $\pi^\pm - \pi^0$ mass-squared difference.  The
mass-squared's of the non-neutral weak-charged PNGB's are therefore estimated
to be of order $m^2_{\pi T}\sim \alpha_2\Lambda^2_{T}$ or $m_{\pi T} \sim 6$
GeV. \cite{Dimopoulos:1979qi}) There are further small corrections when we
switch on $U(1)_Y$.  Unfortunately, charged scalars with such low masses are
ruled out experimentally (see section 3), another difficulty for the
minimal model.  

All electrically neutral PNGB's remain massless at this level of
model-building.  In particularly, the techniaxions remain perturbatively
massless, since they correspond to a residual exact global symmetry
(spontaneously broken).  Their associated axial currents have electroweak
anomalies like the $\pi^0$, but no QCD anomaly, and this can lead to a
miniscule electroweak instanton source of mass, $\propto
\exp(-4\pi^2/g_2^2)v_0$. These objects behave like axions, with decay
constant $F_T$.  They would be problematic, since axion-like
objects have restricted $F_T$'s, typically $\geq 10^8$ to $10^{10}$ GeV by
the usual astrophysical arguments.  As we will see in Section 3, these problems
are ameliorated, in part, by the effects of ETC, which provide a stronger
source of gauge mass for technipions.

Incidently, there is expected to be a $\theta$-angle in TC, and it
is ultimately interesting as a potential novel source of CP-violation 
(see e.g., \cite{Dimopoulos:1979qi}).  

\vskip .1in
\noindent
{\bf 2.2.2(iii) Vector Mesons}
\vskip .1in

In TC there will generally occur  isovector and isosinglet s-wave
vector mesons, the analogues of $\rho(770)$ and $\omega(782) $ in QCD, which
we denote: $\rho^\pm_{T}, \rho^0_{T}$ and $\omega^0_{T}$.  The vector mesons
are particularly important phenomenologically, because of their decays to
weak gauge bosons and technipions (e.g., $\rho_T\rightarrow WW$ is the direct
analogue of the QCD process $\rho\rightarrow \pi\pi$).  The vector mesons
provide potentially visible resonance structures in processes like $pp$ or
$e^+e^- \rightarrow W^+W^-$ and, more generally, make large contributions to
technipion pair production.  The masses of vector mesons can be estimated by
TC scaling, $m_{\rho} \sim m_{\omega TC} \sim m_{\rho } (F_T/f_\pi)
\sim m_{\rho } (v_0/f_\pi)\sqrt{3/N_{T}N_D}$ which yields the approximate
value of $m_{\rho,\omega TC}\sim 1.8/\sqrt{N_D}$ TeV for $N_T=4$.
 

Let us
follow the conventional discussion of vector mesons
in QCD by introducing the dimensionless
phenomenological ``decay constants'' $f_{\rho TC} $ and $f_{\omega TC}$:
\be
\label{rhoomeg}
<\rho^a_{T}| j^b_\mu |0>  = \epsilon_\mu \delta^{ab}\frac{m_{\rho TC}^2}{f_{\rho TC}}
\qquad \qquad
<\omega_{T}| j^0_\mu |0>  = \epsilon_\mu \frac{m_{\omega TC}^2}{f_{\rho TC}}
\ee
where $j^{0}_\mu = \overline{Q}\gamma_\mu  Q /\sqrt{2}$, 
 $j^{0,a}_\mu = \overline{Q}\gamma_\mu (\tau^a/2) Q$,
and $Q$
is a techniquark doublet. It is difficult to extract $f_\omega$ from QCD due
to $\omega-\phi $ mixing, so one typically assumes ``nonet'' symmetry, $f_\rho =
f_\omega$. We then determine the decay constants from the partial width,
$\Gamma(\rho^0 \rightarrow e^+e^-) = 4\pi\alpha m_\rho/3f_\rho^2$; this
implies $f_\rho \approx f_\omega \approx 5.0$. 

 We must determine
how $f_\rho$ and $f_\omega$ undergo TC scaling.  Note that the
current matrix elements of eq.(\ref{rhoomeg}) involve a TC singlet
combination of techniquarks, and a normalized initial state, and might
therefore be expected to scale from QCD as $\sim \sqrt{N_{T}/3}
(\Lambda_{T}/\Lambda_{QCD})^2 $.  However, for fixed $v_0$ we write this as
$\sim (v_0^2/f_\pi^2)\sqrt{3/N_{T}N^2_D} $.  so the amplitude $ m_{\rho
TC}^2/f_{\rho TC} $ must scale as $\sim \sqrt{3/N_{T}N_D^2}$. Since $ m_{\rho
TC} \sim \sqrt{3/N_{T}N_D}$ we see that \cite{Chivukula:1990rn}:
\be
f_{\rho TC} \sim \sqrt{3/N_{T}} f_\rho 
\sim 4.3  \qquad \makebox{for}\;\; N_T=4
\ee
This result makes intuitive sense only if one keeps in mind that,
in TC scaling, we 
are holding $v_0$ fixed!

The decay modes of the vector mesons have been considered in
the literature  \cite{Dimopoulos:1981yf, Dimopoulos:1982fj,
Eichten:1986eq, Chivukula:1990rn}.  
Some can be treated by scaling from the principle QCD decay
modes $\rho^\pm \rightarrow \pi^\pm \pi^0$,
$\rho^0 \rightarrow 2\pi^0$,
$\rho^0 \rightarrow \pi^+ \pi^-$,
and $\omega\rightarrow \pi^+\pi^0\pi^-$,
$\omega\rightarrow \pi^0\gamma $. Note that the $\omega$ 
decay modes
are associated with anomalies in a chiral
Lagrangian description, and have relatively 
tricky scaling properties \cite{Chivukula:1990rn}.
In TC, by invoking the ``equivalent Nambu-Goldstone
boson rule'' for the longitudinal gauge bosons,
one finds the analogue modes $\rho_{T}^\pm \rightarrow W_L^\pm Z_L^0$,
$\rho_{T}^0 \rightarrow W_L^+W_L^-, 2Z_L^0$,
and $\omega_{T}\rightarrow W_L^+ Z_L^0 W_L^-$,
$\omega_{T}\rightarrow Z \gamma, ZZ, WW $.
Scaling from QCD yields:
\bea
\Gamma(\rho_{T}^0 \rightarrow W^+W^- + 2Z^0)
& \sim & \left(\frac{3k_1}{N_{T}}\right)\frac{m_{\rho T}}{m_\rho}
\Gamma(\rho\rightarrow \pi^+\pi^-) \sim
\left(\frac{3}{N_{T}}\right)^{3/2}
\frac{280}{\sqrt{N_D}} \;\makebox{GeV}.
\nonumber \\
\Gamma(\rho_{T}^\pm \rightarrow W^\pm Z^0)
& \sim & \left(\frac{3k_1}{N_{T}}\right)\frac{m_{\rho T}}{m_\rho}
\Gamma(\rho\rightarrow \pi^+\pi^-) \sim
\left(\frac{3}{N_{T}}\right)^{3/2}
\frac{280}{\sqrt{N_D}} \;\makebox{GeV}.
\nonumber \\
\Gamma(\omega_{T}\rightarrow W_L^+ Z_L^0 W_L^-)
& \sim & k_2\left(\frac{3}{N_{T}}\right)^{5/2}\frac{m_{\omega T}}{m_\omega}
\Gamma(\omega\rightarrow \pi^+\pi^0\pi^-) \sim
\frac{35}{\sqrt{N_D}} \;\makebox{GeV}.
\eea
where $k_1\sim 1.2$ and
$k_2\sim 4$ are compensation factors arising from
the $\rho$ and $\omega$ decay having phase-space suppression
owing to the finite pion mass \cite{Chivukula:1990rn}.
Other decays of the $\omega$ are treated in \cite{Chivukula:1990rn}:
\bea
\Gamma(\omega_{T}\rightarrow Z^0 \gamma) & \sim &
\left (\frac{3}{N_{T}}\right)^{1/2}
2.3/\sqrt{N_D} \;\makebox{GeV},
\nonumber \\
\Gamma(\omega_{T}\rightarrow Z^0 Z^0)  & \sim &
\left (\frac{3}{N_{T}}\right)^{1/2}
1.1/\sqrt{N_D} \;\makebox{GeV}'
\nonumber \\
\Gamma(\omega_{T}\rightarrow W^+W^-)  & \sim &
\left (\frac{3}{N_{T}}\right)^{1/2}
5.2/\sqrt{N_D} \;\makebox{GeV}. \qquad
\eea

\vskip .1in
\noindent
{\bf (iv) Higher p-wave Resonances}
\vskip .1in

The parity partners of the $\rho_T$ and $\omega_T$ are the the p-wave axial 
vector mesons, $a_{1T}$ and $f_{1T}$.  If the chiral symmetry breaking of QCD or
TC were somehow switched off 
\footnote{\addtolength{\baselineskip}{-.4\baselineskip} There
really is no conceptual limit of the true theory that
can do this. It would be analogous to taking the coupling
constant on the NJL model to be sub-critical;
see \cite{Bardeen:1994ae}} states and their parity partners would be
degenerate.  Given the presence of spontaneous chiral symmetry breaking, we
must estimate the masses of the axial vector mesons by scaling from QCD: 
 $ m_{a_{1T}} \approx
m_{f_{1T}} \sim m_{a_{1},f_1}(1260) (v_0/f_\pi)\sqrt{3/N_{T}N_D} 
\sim 2.9/\sqrt{N_D}$ TeV.

The spectrum should also include p-wave parity partners of the $\pi_T$ and
$\eta_T$: the isotriplet and isosinglet multiplets of $0^+$ mesons which
analogues of the QCD states $a_0^\pm, a_0^0$, and $f_0$. Their masses are
roughly given by $ m_{a_{0T}} \approx m_{f_{0T}} \sim m_{a_{0},f_0}(980)
(v_0/f_\pi)\sqrt{3/N_{T}N_D} \sim 2.2/\sqrt{N_D}$ TeV.  A chiral Lagrangian
approach to estimating the masses of the $0^+$ states more carefully is
beyond the scope of this discussion.  Note that instanton effects can
also be substantial in a chiral quark model for these states. Typically the
$a_0$ nonet in the NJL approximation to QCD has a low mass of order $2m_Q\sim
600$ MeV, while contributions from the 't Hooft determinant can roughly
double this result, bringing it into consistency with the experimental
values.

\vskip .1in
\noindent
{\bf (iv) Summary of Static Properties}
\vskip .1in

Table \ref{tab:cth:technipi} summarizes the spectroscopy and decays of the
main components of the minimal model of Susskind and Weinberg.  These
static properties, as well as  production
cross-sections and observable processes, are extensively
discussed in EHLQ
\cite{Eichten:1986eq}, \cite{Eichten:1984eu}, and 
\cite{Dimopoulos:1981yf, Dimopoulos:1982fj, Chivukula:1990rn}.  
We briefly review
production and detection of the 
techni-vector mesons of the minimal model in  the
next section.

{\small
\begin{table}
\begin{tabular}{|c||c|c|c|c|}
\hline
state &  $ I (J^{PC}) $  & mass (TeV) & decay width (GeV)
\\ 
\hline\hline
$ \pi^\pm_T , \pi^0_T \; \rightarrow \; W_L^\pm , Z_L $ &  $ 1(0^{-+}) $ 
& $M_W, M_Z$ (eaten) &  $ \Gamma_{W}\;, \Gamma_Z $  
\\ 
\hline
$ \eta^0_T $ (a) &  $ 0(0^{-+}) $ 
& $ \sim r m_{\eta} \sim 0.4 \rightarrow 0.8 $   &  $ 
\Gamma_{t\overline{t}}
\sim  8.0 - 64.0 $, $ \Gamma_{gg}\sim 0.3- 3.0 $  
\\ 
\hline
$ \rho^\pm_T $, $ \rho^0_T $  &  $ 1(1^{--})$ 
& $ q m_\rho/s \sim 1.2 $  &  
$ \Gamma_{\rho_T}(WW) \sim  
\Gamma_\rho(\pi\pi)/sq^2
\sim  350  $ 
 \\ \hline
$ \omega^0_T $  &  $ 0(1^{--})$ 
& $ \sim   q m_\omega/s \sim 1.2 $   &  
$ \Gamma_{\omega_T}(WWZ) \sim  \Gamma_\omega(\pi\pi\pi)/sq^2   
\sim 80 $ 
\\ 
\hline
$ {a_{0T}}^\pm , {a_{0T}}^0 $ 
&  $ 1(0^{++}) $ 
& $r m_{a_0} \sim 1.5$   
& $\eta_T W^\pm$,  $\eta_T Z^0$;
$\Gamma \sim \Gamma_{a_0}/s \sim 100 $ 
\\ 
\hline
$ f_{0T}^0 \sim \sigma_T $ 
&  $ 0(0^{++}) $ 
& $qf_0/s \sim 2 $  &  $\Gamma\sim \Gamma_{f_0}/s \sim 1000$ 
\\ 
\hline 
${a_{1T}}^\pm$, ${a_{1T}}^0 $  &  $ 1(1^{++}) $ 
& $ q m_{a_1}/s \sim 2 $  &  
$\Gamma(WW) \sim \Gamma_{a_1}/s \sim 700 $  
 \\ 
 \hline
$f_{1T} $  &  $ 0(1^{++})$ 
& $qm_{f1}/s \sim 2$  & 
$\Gamma(4W) \sim \Gamma_{f_1}/s \sim 100 $  
\\ 
 \hline
\end{tabular}
\caption[tab1]{\small \addtolength{\baselineskip}{-.4\baselineskip} Estimated properties of 
lowest-lying (pseudo-) scalar and 
(axial-) vector 
mesons in the minimal TC model with a single electroweak doublet 
of techniquarks  $N_D=1$, $N_{T}=4$, and $q=\sqrt{3/N_{T}} = 0.86$.  We take 
$r\equiv \Lambda_{T}/\Lambda_{QCD}
= 1.5\times 10^3$, and $\Lambda_{QCD} = 200$ MeV, and 
$s \equiv f_\pi/ F_T  = 5.7\times 10^{-4}$,  where 
$f_\pi =100$MeV, $F_T = 175$ GeV. The combination
$r^3s^2 = 1.1\times 10^3 $ frequently occurs.
(a) These are estimates from the discussion
of \cite{Eichten:1994nc} .
} 
\label{tab:cth:technipi}
\end{table}
}

\vskip .2in

\subsubsection{Non-Resonant Production and Longitudinal
Gauge Boson Scattering}

Since the longitudinal $W$ and $Z$ are technipions, the minimal
TC model predicts that high energy $W_L-W_L$, $Z_L-Z_L$ or
$W_L-Z_L$ scattering will be a strong-interaction phenomenon.  Studying the
pair-production and scattering of the longitudinal $W$ and $Z$ states
thus provides a potential window on new strong dynamics.  As
Nambu--Goldstone bosons, the longitudinal $W$ and $Z$ are described by
a nonlinear $\sigma$-model chiral Lagrangians, \cite{Weinberg:1968de,
Coleman:1969sm, Callan:1969sn, Appelquist:1980vg,
Longhitano:1980iz, Longhitano:1981tm, Renken:1983ap, Gasser:1984yg,
Gasser:1985gg, Golden:1991ig, Holdom:1990tc, Dobado:1991zh}.
This is called ``the equivalence theorem,''
\cite{Weinberg:1966kf, Chanowitz:1998wi, Chanowitz:1996si,
Chanowitz:1987vj, Chanowitz:1986hu, Chanowitz:1985hj,
Chanowitz:1984ne, Berger:1992tv, Chanowitz:1994zh,
Cornwall:1974km,Yao:1988aj,Bagger:1990fc,He:1992ng,He:1994yd,
He:1997cm,He:1995br,He:1994qa}, and often this is viewed
as an abstract approach, without specific reference to
TC. However, to the extent that we can use QCD as an analogue
computer for TC, we expect that many of the familiar low
energy $\pi -\pi$ theorems of QCD transcribe into the ``low energy''
TC regime, $\lta 1 $ TeV.  Therefore, in models with a
strongly--coupled EWSB sector, certain
``low-energy-theorem'' or ``non-resonant'' contributions to the
production and scattering of the Nambu--Goldstone bosons are present.
In theories where the EWSB sector also
includes resonances, such as the techni-$\rho$, that couple strongly
to the Nambu--Goldstone bosons, the scattering contributions from the
resonances may also be present and even dominate.  The longitudinal
$W-W$ scattering processes are therefore a {\it minimal} requirement of new
strong dynamics.

There are several important mechanisms for producing vector boson pairs at
future hadronic or leptonic colliders.  The first is annihilation of a light
fermion/anti-fermion pair.  This process yields vector boson pairs that are
mostly transversely polarized and will usually be a background to the
processes of interest here.  A key exception is the production of
longitudinally polarized vector bosons in a $J=1$ state (see
\cite{Golden:1995xv} and references therein), which renders this production
channel sensitive to new physics with a vector resonance like a techni-$\rho$.

A second mechanism applicable to hadronic colliders is gluon fusion
\cite{Georgi:1978gs, Dicus:1987dj, Glover:1989rg, Kao:1991tt}, in which the
initial gluons turn into two vector bosons via an intermediate state (e.g.
top quarks, colored techihadrons) that couples to both gluons and electroweak
gauge bosons.  In this case, only chargeless $V_L V_L$ pairs can be produced,
and thus this channel is particularly sensitive to new physics with a scalar
resonance like a heavy Higgs boson.  Finally, the vector-boson fusion
processes \cite{Jones:1979bq, Dicus:1985zg, Hioki:1983yz, Han:1985zn,
  Cahn:1984ip}, $V_L^{}V_L^{} \to V_L^{}V_L^{}$, are important because they
involve all possible spin and isospin channels simultaneously, with scalar
and vector resonances as well as non-resonant channels.

The review article of Golden, Han and Valencia \cite{Golden:1995xv} 
examines the possibilities of making relatively model-independent searches
for the physics of EWSB in $V_L V_L$ scattering at
the LC (in $e^+ e^-$ or $\gamma\gamma$ modes) and LHC.  Their discussion
compares three basic scenarios: (i) No resonances present in the
experimentally accessible region ($\sim 1.0 - 1.5$ TeV) so that PNGB
production is dominated by the nonresonant low energy theorems;  
(ii) Production physics dominated by a spin-zero
isospin-zero resonance like the Higgs boson or a techni-sigma;  (iii) 
physics  dominated by a new spin-one isospin-one resonance like a
techni-$\rho$.  We summarize their results, along with updates from other
sources (see, e.g. 
\cite{Boos:1999kj,Boos:1998gw,Accomando:1998wt,He:1997rb,He:1996nm}), 
here and in Table \ref{golden-lhclum}.

\begin{figure}[tbh]
 \vspace{9.0cm}
\includegraphics{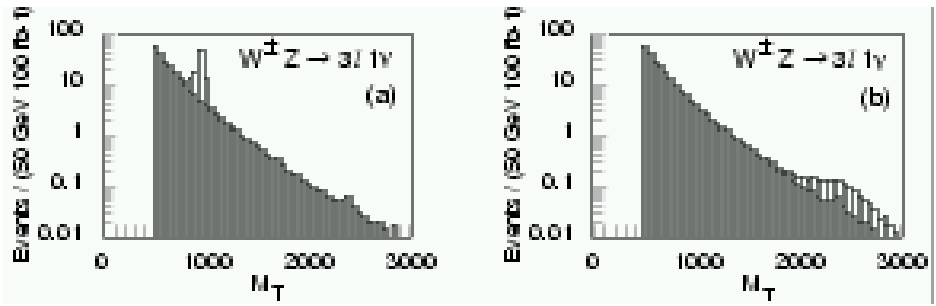}
\vskip-4.0truecm
 \caption{\small \addtolength{\baselineskip}{-.4\baselineskip}
      Event yields at the LHC for $\rho_T \to W^\pm Z^0 \to \ell^\pm
      \nu_\ell \ell^+ \ell^-$ for $M_{\rho_T} = 1.0$, $2.5$ TeV; from
      Ref.\protect\cite{Golden:1995xv}. A
conventional techni-$\rho$ resonance of mass much above 1 TeV would be invisible
in the channel $\rho_T \to W^\pm Z^0 \to \ell^\pm \nu_\ell \ell^+ \ell^-$,.
    \label{ieri-lhc} }
\end{figure}

The LHC can detect strongly interacting
EWSB physics in di--boson production.  Large 
Standard Model (SM)
backgrounds for pair-production of leptonic $W$'s can be
suppressed if one imposes stringent leptonic cuts,
forward-jet-tagging and central-jet-vetoing.  
Refs.~\cite{Golden:1995xv, Chanowitz:1998wi} explored
complementarity for  $W^\pm Z$ and $W^\pm W^\pm$ channels in
studying a vector-dominance model or a non-resonant model. A
systematic comparison of the different final states allows one to
distinguish between different models to a degree.  A
statistically significant signal can be obtained for every model
(scalar, vector, or non-resonant) in at least one channel with a few
years of running at an annual luminosity of 100 fb$^{-1}$.
Detector simulations demonstrate that the
semileptonic decays of a heavy Higgs boson, $H \to W^+W^- \to l \nu
jj$ and $H \to ZZ \to l^+ l^- jj$, can provide statistically
significant signals for $m_H=1$ TeV, after several years of running at
the high luminosity.

\begin{table}[htbp]
\centering
\bigskip
\begin{tabular}{l|cccccccc}
\hline\hline
& \multicolumn{8}{c}{Model}\\
\cline{2-9}
Channel& SM& Scal& $O(2N)$& V1.0& V2.5& CG&
LET-K& Dly-K\\
\hline
$ZZ(4\ell)$ &
1.0  & 2.5  & 3.2  & ~    & ~   & ~    & ~     & ~    \\
$ZZ(2\ell2\nu)$ &
0.5  & 0.75 & 1.0  & 3.7  & 4.2 & 3.5  & 4.0   & 5.7  \\
$W^+W^-$ &
0.75 & 1.5  & 2.5  & 8.5  & ~   & 9.5  & ~     & ~    \\
$W^{\pm} Z$ &
 ~   & ~    & ~    & 7.5  & ~   & ~    & ~     & ~    \\
$W^{\pm}W^{\pm}$ &
4.5  & 3.0  & 4.2  & 1.5  & 1.5 & 1.2  & 1.2   & 2.2  \\
\hline\hline
\end{tabular}
\caption{\small \addtolength{\baselineskip}{-.4\baselineskip}
\label{golden-lhclum}%
Number of years at LHC with annual luminosity 100 fb$^{-1}$
required for a 99\% confidence level signal.  The
models considered are: the Standard Model (SM), strongly-coupled models
with new scalar resonances (Scal, O(2N)), strongly-coupled models with new
vector resonances of mass 1 TeV (V1.0) or 2.5 TeV (V2.5), and strongly-coupled
models with non-resonant scattering following the low-energy theorems (CG,
LET-K, Dly-K). From 
ref. \cite{Golden:1995xv}.}
\end{table}

The LHC's power to use di-boson production to see vector
resonances associated with a strong EWSB
sector has limited reach in mass \cite{Golden:1995xv,
Chanowitz:1998wi}.  For example, a conventional techni-$\rho$
resonance of mass much above 1 TeV would be invisible in the channel
$\rho_T \to W^\pm Z^0 \to \ell^\pm \nu_\ell \ell^+ \ell^-$, as shown
in Fig.(\ref{ieri-lhc}).  A heavier techni-$\rho$ would,
instead, make itself felt in the complementary $W^\pm W^\pm$ channel
\cite{Golden:1995xv, Chanowitz:1998wi}.  Models of ``low-scale''
TC with lighter vector resonances more visible in the $WZ$
channel at the LHC will be discussed in Section 3.5 and 3.6.

\subsubsection{Techni-Vector Meson Production
and  Vector Meson Dominance}

\begin{figure}
\vspace{6.0cm}
\includegraphics{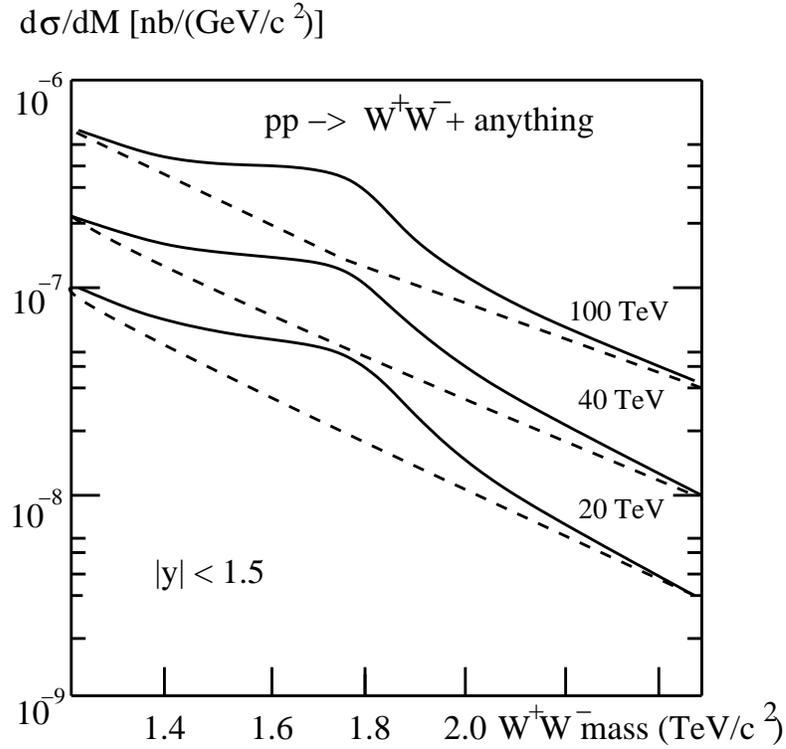}
\vspace{2.5cm}
\caption[cap]{\small \addtolength{\baselineskip}{-.4\baselineskip}
 Vector Meson Dominance production of
techni-$\rho$ with subsequent decay to
$W^+Z$ in $pp$ collider with center-of-mass
energies, $20$, $40$ and $100$ TeV (from EHLQ \cite{Eichten:1998kn}).}
\label{fig:ehlq1}
\end{figure}

\begin{figure}
\vspace{6.0cm}
\includegraphics{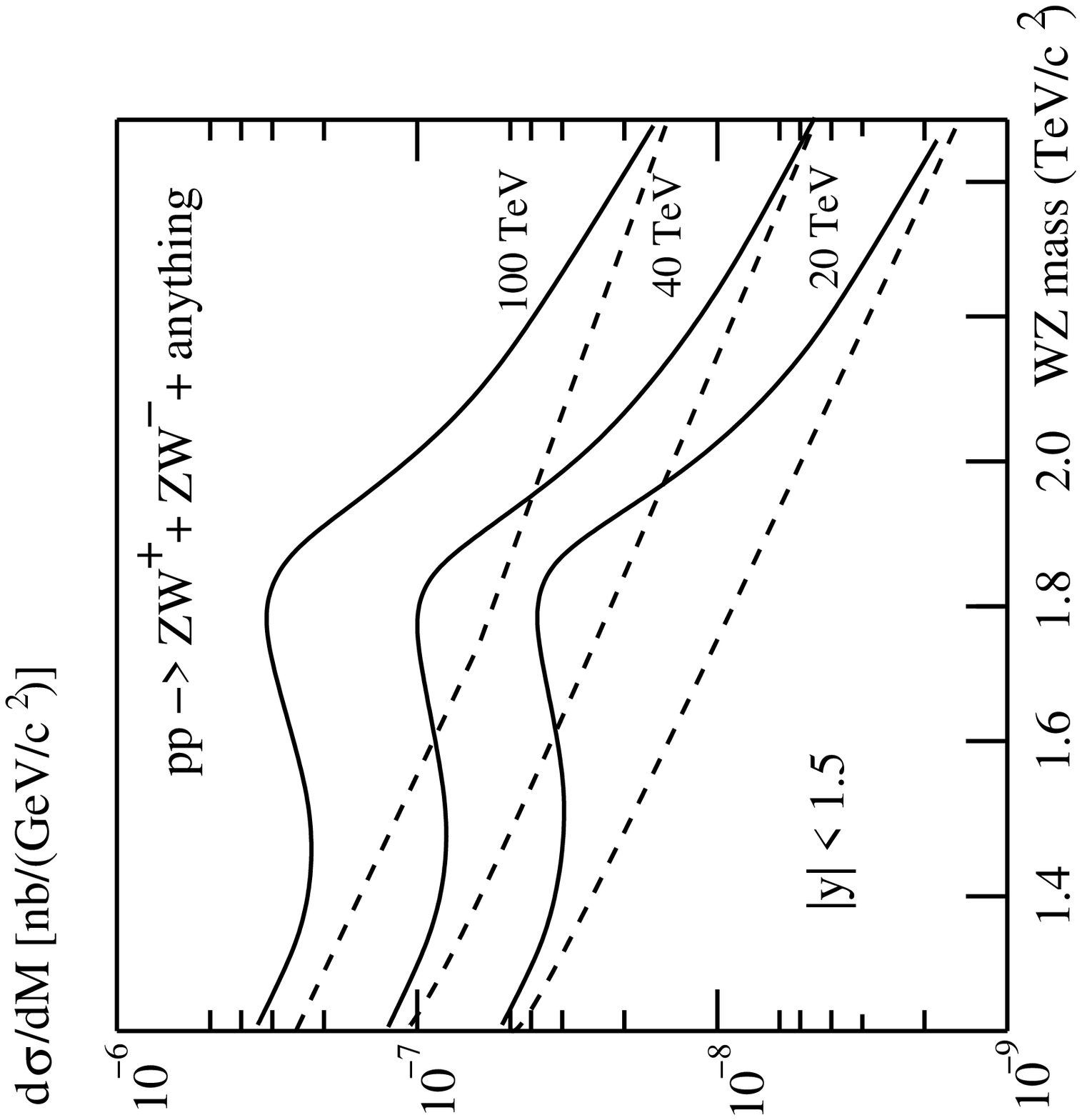}
\vspace{2.5cm}
\caption[cap]{\small \addtolength{\baselineskip}{-.4\baselineskip} 
Vector Meson Dominance production of
techni-$\rho$ with subsequent decay to
$W^\pm Z$  in $pp$ collider with center-of-mass
energies, $20$, $40$ and $100$ TeV (from EHLQ \cite{Eichten:1998kn}). }
\label{fig:ehlq2}
\end{figure}

In the minimal model with $N_D=1$ the lowest-lying resonances that can
provide an obvious signal of physics beyond the Standard Model are the
techni-vector mesons.  The first accessible process involves the
annihilation (at scales of order a few TeV) of a fermion and
antifermion into a techni-vector meson such as $\rho_T$ and its
subsequent decay into gauge boson pairs, known as the ``Susskind process''
\cite{Susskind:1978ms}.  The vector boson
pairs produced in this way are mostly transversely polarized (the case
of longitudinal polarizaion was discussed in the previous section).
Production and detection of techni-$\rho$ states in $f \bar{f} \to
\rho_T \to V V$ processes at various present and future colliders have
been discussed extensively in the literature, beginning with EHLQ
\cite{Eichten:1986eq, Eichten:1984eu}, and \cite{Chanowitz:1985hj,
Chivukula:1990rn, Rosenfeld:1988th, Rosenfeld:1989im, Carone:1994vg,
Casalbuoni:1995as, Bagger:1994zf, Chanowitz:1995ap, 
Casalbuoni:1993su, Lane:1991qh}.
Examples of the calculated cross-sections for $\rho_T$ producion and decay at
an LHC or VLHC are given in Figures \ref{fig:ehlq1} and
\ref{fig:ehlq2}. Detailed search strategies and limits for the ``low-scale''
variants of the minimal model are discussed in Section 3.5.

Absent a direct coupling between ordinary and techni-fermions such as
ETC  can provide (Section 3), how are the techni-vector
mesons of the pure Minimal TC model to be produced in $q\bar{q}$ or
$e^+ e^-$ annihilation?  The answer is that Vector Meson Dominance (VMD) 
enables
the techni-vector mesons to couple to currents of ordinary fermions.  Most
relevant to a discussion of $\rho_T$ production are vector dominance mixing
of the $\rho_T^0$ with $\gamma$ and $Z$ and the $\rho_T^\pm$ with $W^\pm$,
which we discuss below.  The production and detection of the techni-$\omega$
is considered in \cite{Chivukula:1990rn}.

Let us briefly review the theory of VMD.  Consider
a schematic effective Lagrangian in which we introduce the photon,
$A_\mu$ together with a single neutral vector
meson\footnote{\addtolength{\baselineskip}{-.4\baselineskip} This can
be directly generalized to electroweak gauge fields and an isotriplet
$\rho^a$, or to gluons and a color octet $\rho_8^a$, but see
\cite{Zerwekh:2001uq,SekharChivukula:2001gv}.}  $\rho_\mu$:
\be
L = -\frac{1}{4}(F_A^{\mu\nu})^2 -\frac{1}{4}(F_\rho^{\mu\nu})^2
-\half\epsilon F_A^{\mu\nu}F_{\rho\;\mu\nu}
+ \half m_\rho^2(\rho_\mu)^2 -e A_\mu J^\mu -\kappa \rho_\mu J^\mu_{had}
\ee 
$J^\mu $ is the ordinary electromagnetic current, and we define $F_X^{\mu\nu}
= \partial^\mu X^\nu - \partial^\nu X^\mu$.  $J_{had}^\mu $ is a vector
hadronic current which describes the strong interactions of the $\rho$;
$J_{had}^\mu $ contains, for example, $i\pi^+\partial_\mu \pi^-$, and we 
would typically fit the parameter
$\kappa$ to describe the strong interaction decay $\rho \rightarrow
\pi^+\pi^-$.  The $\epsilon$ term represents mixing between the photon and
$\rho$, and can be viewed as arising from the 
nonzero amplitude $<0| T J^\mu(0) \;
J_{had}^\nu(x)|0>$; we will work to order $\epsilon$.  
Note
that the vector meson, $\rho$,  can
be viewed as a gauge field that has acquired mass through spontaneous
symmetry breaking \cite{Sakurai:1960ju} (Indeed, Bando, Kugo, Yamawaki and
others \cite{Bando:1988br, Bando:1988ym, Bando:1985rf, Bando:1985ej} 
have argued that vector meson effective Lagrangians
always contain a hidden local symmetry). This is why we choose the $\rho$
kinetic term to be in the form of the photon kinetic term, and it implies
that we are always free to choose a gauge, such as $\partial_\mu\rho^\mu=0$.

Upon integrating by parts, we can rewrite the $\epsilon$ term
as $\epsilon A_\nu\partial_\mu F_{\rho\;\mu\nu}$. Using equations
of motion for the $\rho$ to order $\epsilon$ we obtain: 
\be
L = -\frac{1}{4}(F_A^{\mu\nu})^2 -\frac{1}{4}(F_\rho^{\mu\nu})^2
 +
\epsilon m_\rho^2 A_\nu \rho^\nu + \epsilon \kappa A_\nu J_{had}^\nu
+ \half m_\rho^2(\rho_\mu)^2 -e A_\mu J^\mu -\kappa \rho_\mu J^\mu_{had}
\ee 
The first $\epsilon m_\rho^2 A_\nu \rho^\nu $ term can be viewed as arising
from the matrix element of the electromagnetic interaction, $ eA_\mu <\rho|
J^\mu |0> $, and by comparison with eq.(\ref{rhoomeg}) we identify $\epsilon
= e/f_\rho $.  In this form we have an induced mass mixing between the
$\rho$ and photon.\footnote{\addtolength{\baselineskip}{-.4\baselineskip} This does not violate gauge invariance.  While a
shift $A_\mu \rightarrow A_\mu + \partial_\mu\theta$ leads to $\epsilon
m_\rho^2 \partial_\nu\theta \rho^\nu$, we can integrate by parts $-\epsilon
m_\rho^2 \theta \partial_\nu\rho^\nu$, and the gauge invariance of the $\rho$
allows this term to be set to zero. That is, $\rho_\mu$ behaves like a
conserved current if it is a gauge field with hidden local symmetry
\cite{Bando:1985ej}.)}  

The $\rho$ and photon can now redefined as
$\rho \rightarrow \rho - \epsilon A$, $A \rightarrow A + \epsilon
\rho_\mu$. Thus we obtain:
\be
L = -\frac{1}{4}(F_A^{\mu\nu})^2 -\frac{1}{4}(F_\rho^{\mu\nu})^2
+ \epsilon \rho_\mu J^\mu
+ \half m_\rho^2(\rho_\mu)^2 -e A_\mu J^\mu -\kappa \rho_\mu J^\mu_{had}
\ee 
This removes the 
mass mixing term, and
the
$\epsilon A_\nu J_{had}^\nu $ term, but leads to an induced
$\epsilon \rho_\mu J^\mu$ term.
Thus, we can view the $\rho$ as having
an induced direct coupling to the full electromagnetic current
of strength $\epsilon$!

Alternatively, upon integrating by parts, we could
have  written the $\epsilon$ term
as $\epsilon \rho_\nu\partial_\mu F_{A}^{\mu\nu}$, and using equations
of motion for the $\rho$ to order $\epsilon$ we have 
$\epsilon e\rho_\nu J^\nu  = e^2 \rho_\nu J^\nu /f_\rho$.
Thus, we can view the effect of the $\epsilon$ term as
directly inducing
the coupling of the $\rho$ to any electromagnetic current with
strength $e^2/f_\rho$, e.g., the $\rho$
will
couple directly to the electron's electromagnetic current.  
While this is a small coupling, the $\rho$
is generally a narrow state.  On-resonance the production rate can be
substantial. 

An equivalent non-Lagrangian description of this treats the
$\epsilon$ term as a mixing effect in the propagator
of the photon. The propagator becomes a matrix, allowing 
the photon to mix with the
$\rho$ and couple directly to $J_{had}$.  The propagator connecting
an electromagnetic current to the hadronic current
becomes:
\be
\frac{-i}{q^2} \times -i\epsilon\; q^2 \times \frac{-i}{q^2 -m_\rho^2}
\ee

In the context of VMD, we can summarize our expectations
for the minimal model with $N_D = 1$.  We have already estimated the 
the techni-$\rho$ mass to be of order $\sim 1.8$ TeV in this case.  The 
dominant expected production and decay modes of the $\rho_T$ are then:
\be
f\bar{f} \rightarrow (\gamma, Z^0) \rightarrow \rho^0_T \rightarrow
W^+W^- 
\ee
and
\be
f\bar{f} \rightarrow (W^\pm ) \rightarrow \rho^\pm_T \rightarrow
W^\pm Z^0
\ee
The annihilation of a fermion and anti--fermion a $\rho_T$ and its
subsequent decay into technipion $+$ gauge boson, or technipions can
also be significant.  With $N_D=1$ we would expect
$f\bar{f}\rightarrow
\rho_T\rightarrow \eta_T' + Z $ to be of interest.\footnote{\addtolength{\baselineskip}{-.4\baselineskip}  Tandean
\cite{Tandean:1995ci} has also considered the signature and detectability of the
$\eta'_T$ produced in a TeV scale $\gamma\gamma$ collider.}. 

In models with more doublets, $N_D>1$, there are more vector meson
states which can be enumerated and classified according to the scheme
discussed above for technipions (see Fig.(\ref{cth1})).  However, the
only vector mesons which participate in the VMD mixing
are those transforming like the Standard Model gauge fields, i.e. as
$\tau^a \otimes I_d$.  These vector mesons can, in turn, decay to any
pair of technipions for which the $V$ and technipions together form an
overall $I\otimes I$
combination\footnote{\addtolength{\baselineskip}{-.4\baselineskip}
Recall that under multiplcation of direct product representations, we
have $(X\otimes Y)\times (W\otimes U) = XW\otimes YU$ }. Hence, decays
to gauge boson pairs are allowed since $(\tau^a \otimes I_d)\times
(\tau^b \otimes I_d)\times (\tau^c \otimes I_d)$ contains the singlet
$\epsilon^{abc}(I\otimes I)$.  Moreover, we see that the decays to the
$3(N_D^2-1)$ PNGB's of the form $\tau^a \otimes \lambda^a$ are also
possible.  Decays to the techniaxions of the form $I\otimes
\lambda^a$, however, are disallowed.

Because we expect the techni-$\rho$ to have $M_{\rho_T} > 2 M_{\pi_T}$ in the
minimal model with $N_D>1$, the dominant decay of the techni-$\rho$ should be
to a pair of technipions.  The technipions will, in turn, decay to $WW$,
$WZ$, and so forth, through direct couplings or anomalies.  Variant models
may include decays to gluon pairs; for models incorporating ETC interactions,
final states with fermion pairs are also expected.  We defer further
discussion of searches and phenomenology for now, and move on to sketch out a
more general TC model.


\newpage
\noindent
\subsection{Farhi--Susskind Model}

\subsubsection{Structure}

The minimal model is neither a unique prescription for the
construction of a TC theory, nor is it likely to contain
sufficient richness to ultimately allow the generation of the observed
range of fermion masses and mixing angles.  We anticipate that a more
complete model would need to include both a ``quark'' sector of
color-triplet techniquarks, and a ``leptonic'' sector of color-singlet
technifermions (like those in the minimal model).  These states could
ultimately act under extended TC interactions\footnote{See
section 3.} to give mass terms to the ordinary quarks and leptons.
Toward this end, we turn now to describing the Farhi-Susskind
TC model \cite{Farhi:1979zx} which contains a much richer
spectrum of technifermions imitating the anomaly free representations
of the Standard Model.

The Farhi-Susskind model extends the flavor content of the minimal
model to imitate one full generation of quarks and leptons with the
usual anomaly free isospin and $Y$ assignments:
\begin{eqnarray}
Q_L & = & \left( \begin{array}{c} T \\ B \end{array} \right)^i_L
\;\; Y =
y;\; \qquad \qquad
\left( \begin{array}{c} N \\ E \end{array} \right)_L\;\; Y =
-3y;
\\ \nonumber
Q_R & = & ( T^i_R, \; B^i_R  \;\;
 N_R, \; L_R ) \;\; Y =  \left( y+1,\; y-1,\; -3y+1,\; -3y-1 \right)
\end{eqnarray}
where the $SU(3)$ color index $i$ takes the values $i = 1,2,3$ for
the techniquarks.  This is an anomaly free representation for
any choice of the parameter $y$.  For the particular
standard choice of $y=1/3$, the techniquarks and technileptons
have electric charges identical to those of the quarks and leptons.

In the present model the techniquarks and technileptons
couple to the full $SU(3)\times SU(2)_L\times U(1)_Y$ gauge group in
the usual way.  Now we postulate that each $Q$ multiplet carries, 
in addition, an $SU(N_{TC})$ quantum number in
the $N_{TC}$ representation of the strong
TC gauge group. Note that 
the existence of a right-handed technineutrino, $N_R$, is required to provide 
anomaly cancellation for the TC
gauge interaction.  The $SU(N_{TC})$ is essentially a gauged
horizontal generation symmetry.

Unlike the minimal model with $N_D=1$, 
in which all three NGB's
are absorbed into the longitudinal modes of the electroweak gauge
bosons, the Farhi-Susskind model has a low energy spectrum containing
numerous pseudo-Nambu-Goldstone bosons (PNGB's). Their quantum numbers
may be ascertained by observing that, in the limit
of vanishing $SU(3)\times
SU(2)_L\times U(1)$ couplings, there is an $SU(8)_L\times
SU(8)_R\times U(1)_A\times U(1)$ global chiral group.  The full
Standard Model
$SU(3)\times SU(2)_L\times U(1)_Y$ interactions are a gauged subgroup of
this chiral group. At the scale $\Lambda_{TC}$, the TC gauge
coupling is strong, and causes a degenerate chiral condensate to form:
\be
\VEV{\overline{T}_{Li}T_{Ri}}
=
\VEV{\overline{B}_{Li}B_{Ri}}
=
\VEV{\overline{N}_{L}N_{R}}
=
\VEV{\overline{E}_{L}E_{R}} \sim \Lambda_{TC}^3
\ee
where $i$ is a (unsummed) color index ranging from
$i=1,2,3$.
The chiral group  is thus
broken spontaneously 
to an approximate $SU(8)\times U(1)$ vectorial 
symmetry, producing
$63+1$ NBB's.  

>From the previous remarks about condensates, we see that there exist four
composite electroweak doublets. This is similar to the structure of a
4-Higgs-doublet model in which each Higgs boson gets a
common VEV $F_T$. The electroweak scale is thus related to the common
VEV's of the four Higgs bosons, $F_{T}$, as $v_{0}^2 = 4 F_T^2$, and thus
$F_T= 123$ GeV.  One combination of the NGB's will become the longitudinal
$W$ and $Z$, while the orthogonal states remain in the spectrum, as we will
describe in the next subsection.


\subsubsection{Spectroscopy}

\vskip .1in
\noindent
{\bf (i) {Color $\bf \{1\}$, $\bf \{3\}$, and $\bf \{8\}$, 
Pseudo-Nambu-Goldstone Bosons}}
\vskip .1in

Let us examine the content of the low lying $(8\times 8)$
PNGB's of the
Farhi-Susskind model.  As might be expected, the enhanced variety of
technifermions yields a larger selection of PNGB states.  Their properties
are summarized in Table \ref{tab:cth:fstechnipi}.

We will begin with the eight color-singlet states.
By symmetry, three linear combinations are identically massless and become the
longitudinal $W^\pm $ and $Z^0 $:
\be
\overline{T}_iB^i + \overline{N}E \sim \pi^- \qquad
\overline{B}_iT^i + \overline{E}N \sim \pi^+ \qquad
\overline{T}_iT^i-\overline{B}_iB^i+ \overline{N}N-\overline{E}E
\sim \pi^0_L
\ee 
There remain $5$ orthogonal color singlet objects, two
with non-zero electric charge (we follow the nomenclature and normalization
conventions of Eichten, Hinchliffe, Lane and Quigg (EHLQ) \cite{Eichten:1984eu} 
\cite{Eichten:1986eq}):
\be
\overline{T}_iB^i - 3\overline{N}E \sim P^-\qquad
\overline{B}_iT^i - 3\overline{E}N \sim P^+ \qquad
\ee
and three which are electrically  neutral:
\be
\overline{T}_iT^i-\overline{B}_iB^i- 3(\overline{N}N - \overline{E}E) \sim
P^0
\qquad
\overline{T}_iT^i+\overline{B}_iB^i- 3(\overline{N}N + \overline{E}E) \sim P^0{}'
\qquad
\ee
and
\be
\overline{T}_iT^i+\overline{B}_iB^i+ 
\overline{N}N+\overline{E}E \sim \eta'_T
\ee 

The PNGB which is neutral under all gauge interactions receive mass
via instantons.  The $\eta_T'$, receives mass from the instantons of
TC, and is expected to be heavy, as in the case of the
Weinberg-Susskind model. From the
discussion of section 2.2.2(ii) we estimate $m_{\eta' TC} \sim
(\sqrt{6}/N_{T})\sqrt{3/N_{T}N_D}(v_0/f_\pi) m_{\eta'} \sim 700 $ GeV
for $N_{T}=4$ (where $N_D=4$ in the Farhi-Susskind model).  The
$P^0{}'$ receives a mass only of order 1 GeV from QCD instantons, but
will receive a larger contribution from ETC effects (see Section 3).
The $P^0$, likewise, receives its mass from ETC.

\begin{table} 
\begin{tabular}{|c||c|c|c||}
\hline
state &  $I(J^{P})$, color, $[Q]$   & mass (GeV) 
\\ \hline\hline
$ \pi^{-}_T \sim {\tiny (\overline{T}_iB^i + \overline{N}E)
} $ &  $ 1[0^{-},1^- ]\;0[-1]$ 
& $M_W$       
\\ \hline 
$ 
\pi^{0}_T \sim {\tiny (\overline{T}_i
T^i-\overline{B}_iB^i +\overline{N}N-\overline{E}E) }   $ 
&  $ 1[0^{-},1^- ]\;0\; [0]$ 
& $M_Z$   
\\ \hline 
$ \eta'_T  \sim {\tiny (\overline{T}_i
T^i+\overline{B}_iB^i +\overline{N}N+\overline{E}E) }   $ 
&  $ 1[0^{-}, 1^- ]\;0\; [0]$ 
& $\sim 10^3$     
\\ \hline\hline
$  P^+ \sim{\tiny (\overline{B}_iT^i - 3 \overline{E}N)
 } $ &  $ 1[0^{-},1^- ]\;0\;[1]$ 
& $\sim 100(4/N_{TC})^{1/2} $       
\\ \hline 
$ P^0 \sim{\tiny (\overline{T}_i
T^i-\overline{B}_iB^i -3(\overline{N}N-\overline{E}E)) }  
$ &  $ 1[0^{-},1^- ]\;0\;[0]$ 
& $ \sim 100 $[ETC]     
\\ \hline 
$ P^0{}'\sim{\tiny (\overline{T}_i
T^i+\overline{B}_iB^i -3(\overline{N}N+\overline{E}E)) }   
$ &  $ 0[0^{-},1^- ]\;1\;[0]$ 
& $ \sim 100 $[ETC]      
\\ \hline 
$P_3^1 \sim  \overline{E}T  $  &  $ 1(0^-)\;3\;[5/3] $ 
&  $ \sim 160\; (4/N_{TC})^{1/2}\;$
 \\ \hline
$P_3^0 \sim \overline{N}T -\overline{E}B  $  &  $ 1(0^-)\;3\;[2/3] $ 
&   $ \sim 160\;(4/N_{TC})^{1/2}\; $
\\ \hline
$P_3^{-1}\sim \overline{N}B $  &  $ 1(0^-)\;3\; [-1/3]$ 
&   $\sim 160\; (4/N_{TC})^{1/2}\; $ 
 \\ \hline
$P_3{}' \sim \overline{N}T+ \overline{E}B   $  &  $ 0(0^-)\;3\;[2/3] $ 
&  $\sim 160\; (4/N_{TC})^{1/2}\; $
\\ \hline  
$P_8^{+} \sim \overline{B}T   $  &  $ 1(0^-)\;8\;[1] $ 
&  $\sim 240\; (4/N_{TC})^{1/2}\; $
\\ \hline 
$P_8^{0} \sim \overline{T}T - \overline{B}B  $  &  $ 1(0^-)\;8\;[0] $
&  $\sim 240\; (4/N_{TC})^{1/2}\; $
\\ \hline 
$P_8^{0}{}' \sim \overline{T}T + \overline{B}B  $  &  $ 0(0^-)\;8\;[0] $
&  $\sim 240\; (4/N_{TC})^{1/2}\; $
\\ \hline\hline
\end{tabular}
\caption[tab2]{\small \addtolength{\baselineskip}{-.4\baselineskip} Properties of scalar states 
in the Farhi-Susskind TC model following
\cite{Eichten:1984eu},
\cite{Eichten:1986eq}; see also \cite{Dimopoulos:1980sp}.
To each listed spin-0 state there is a corresponding
$s$-wave spin-1 analogue; there will also be $p$-wave analogue
resonance states.
Nonself-conjugate states have corresponding (unlisted) antiparticles.
Quoted masses are crude estimates, 
quoted in GeV; their theoretical
values are very model dependent, modulo walking ETC, etc.
For analogue vector and resonance masses 
see the discussion of the text.}
\label{tab:cth:fstechnipi}
\end{table}

The PNGB's with  electroweak gauge charges,
but no color, $P_0^Q$,  receive masses from the
gauge interactions in analogy to the electromagnetic mass splitting of
the ordinary $\pi^+$ and $\pi^0$ of QCD. In QCD we have $\delta
m^2_\pi = {m_{\pi^+}^2 - m_{\pi^0}^2} \approx (35\; MeV)^2$. In the
chiral limit (current masses $m_u=m_d=0$) we have $m^2_{\pi^0} = 0$, and
thus $\delta m^2_\pi $ roughly reflects the electromagnetic mass
contribution of the $\pi^\pm$.  Scaling from this, one finds that the
full electroweak induced mass for a technipion is of order
$m^{2}_{P_0^\pm} \sim
F^2_T\delta m^2_\pi/f_{\pi}^2\sin^2\theta_W $.
Hence numerically, $m_{P_0^{\pm}} \sim
100(4/N_{TC})^{1/2}$ GeV.
The $P_0^0$ and $P_0{}'$ masses arise from ETC alone.
More detailed estimates for the PNGB mass splittings in TC
are discussed in Dimopoulos \cite{Dimopoulos:1980sp} and Eichten and
Lane \cite{Eichten:1980ah}.

The spectrum of the Farhi-Susskind model also includes technipions with 
non-zero color charge: technileptoquark PNGB's, 
$P_3\sim {\overline{L}Q}$, and color
octet PNGB's, $P_8\sim \overline{Q}(\lambda^A/2)Q$.  
These states acquire masses principally
through the
$SU(3)_{QCD}$ interactions \cite{Chadha:1981yt} 
\cite{Chadha:1981rw} \cite{Preskill:1981mz}.
If the technipion carries net color in the $R$ representation, 
then the QCD contribution to the
mass is of order $m^2_{\pi T} \sim ({C_2(R)\alpha_3(F_T)/\alpha(F_T)})
F^2_T\delta m^2_\pi/f_{\pi }^2$, giving us:
\be
M(P_3) \sim 160\; (4/N_{TC})^{1/2}\;\makebox{GeV}.
\ee
\be
M(P_8) \sim 240\; (4/N_{TC})^{1/2}\;\makebox{GeV}.
\ee 
(here we use $\alpha_3(F_T)=0.1$, $\alpha(F_T) =1/128$, $C_2(3) =4/3$,
$C_2(8) = 3$).
More generally, the various electroweak and strong contributions to
the mass of a colored PNGB are added in quadrature to form the full
mass, $m^2 = m_c^2 + m_{EW}^2$.  In a more complete model, there can
also be Extended TC contributions to the masses, but these
are expected to be smaller than the QCD masses given above.

The reader is advised to consider the spectrum of states, and
reasonable decay and production estimates (which follow), 
but not take literally
the model estimates of masses at this stage. Any TC model we review
from the 1980's is, at best, incomplete, and can only serve as a guide to what
may be contained in more modern reincarnations of the models.
Moreover, in subsequent sections on ETC, Walking TC and hybrid models like TC2
many mechanisms will surface that can rearrange the masses of states
in these models. 

Color-triplet (leptoquark) technipions decay
via Extended TC
as $P_3\rightarrow q \overline{\ell}$ (without
the standard choice of $y=1/3$ these objects would be
stable).  The color octet technipions decay
into $P_8\rightarrow \overline{q}q + ... $ final states.
The rates for these PNGB decay processes depend upon the
scale and details of ETC, but are expected typically to
be of order $\Gamma \lta M^3/\Lambda_{TC}^2$ (e.g, in
older ETC models $\Lambda_{TC} \sim 1$ TeV in this estimate is
replaced by  $\Lambda_{ETC} \sim 100$ TeV, while in
``Walking ETC'' (Section 3.4) we expect  $\Lambda_{TC} \sim 1$ TeV is
replaced by  $\sqrt{\Lambda_{TC} \Lambda_{ETC}} \sim 10$ TeV).
We review the phenomenology of the colored technipions further in
Sections \ref{ssec:pdhc} and \ref{ssec:epemcoll}.

\vskip .1in
\noindent
{\bf (ii) Vector Mesons}
\vskip .1in 
 
As was the case in the minimal TC models, the Farhi-Susskind model
includes s-wave vector (J=1) states which we refer to collectively as
$\rho_T$ (the literature also refers to these states as $V_T$).  Their
masses follow from the estimate in section 2.2.2(iii) if we set $N_D = 4$.
Neglecting the QCD corrections, which are expected to be less than $\sim
15\%$, we find
\be
M(\rho_T\equiv V_T) \sim 700\; (4/N_{TC})^{1/2}\;\makebox{GeV}
\ee
Note that there now exist color-octet $V_8$ states which have the
quantum numbers of the gluon, and act like a multiplet of heavy
degenerate
gluons\footnote{\addtolength{\baselineskip}{-.4\baselineskip} Similar
objects will crop up as gauge particles (colorons) in Topcolor models
(Section 4.2) or as KK modes in extra-dimensional models (Section
4.6)}.  These will exhibit vector-dominance-like mixing with gluons in
processes like $
\overline{q}q \rightarrow G
\rightarrow V_8 \rightarrow P_A P_B$.

There are also, as in the minimal model, p-wave parity partners of the
PNGB's (the techni-$a_0$'s and techni-$f_0$'s) and 
parity partners of the vector mesons, the
axial-vector mesons $a_{1T}$'s and $f_{1T}$.
Following the discussion of section 2.4, we find them to have masses of order:
\be
M(a_{1T}, f_{1T} ) \sim 1700\; (4/N_{TC})^{1/2}\;\makebox{GeV}
\ee
\be
M(a_{0T}, f_{0T} ) \sim 1300\; (4/N_{TC})^{1/2}\;\makebox{GeV}
\ee
Generally speaking, the parity-partner states form identical
representations of the symmetry groups and have
identical charges, but are significantly heavier. 

\subsubsection{Production and Detection at Hadron Colliders}\label{ssec:pdhc}

\smallskip
\noindent{\bf (i) Color-singlet PNGB: $P^0$ and $P^0{}'$}
\smallskip

\begin{figure}
\vspace{7.0cm}
\includegraphics{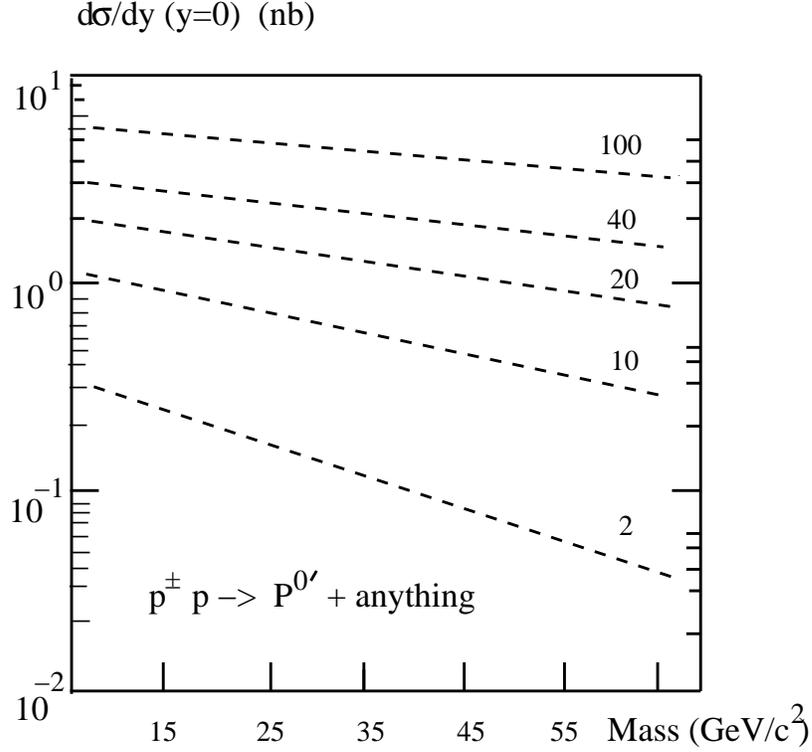}
\vspace{2.0cm}
\caption[cap21]{\small \addtolength{\baselineskip}{-.4\baselineskip} Differential cross section
for production of $P^0{}'$ at
$y=0$ in $pp$ and $p\bar{p}$
collisions with indicated center-of-mass energy in
TeV (reproduced from EHLQ \cite{Eichten:1998kn}).}
\label{fig:ehlq4}
\end{figure}

EHLQ 
\cite{Eichten:1986eq}, \cite{Eichten:1984eu}, 
provided the original discussion of production and detection of neutral
PNGBs in hadron colliders at various energies.  The differential
cross-section for production of $P^0{}'$ is shown in Figure
\ref{fig:ehlq4}.  The relevant decay widths for 
for $P^0{}'$ in the Farhi-Susskind
model are (including possible ETC contributions)
\cite{Chivukula:1995dt}:
\begin{eqnarray}
\Gamma (P^{0 \prime} \rightarrow l \bar{l} (q \bar{q})) &=& \frac{(3)}{8 \pi}
\frac{m_{l(q)}^2}{F^2} m_P
\left( 1 - 4 m_{l(q)}^2/m_P^2 \right)^{3/2}  \\
\Gamma (P^{0 \prime} \rightarrow gg) &=& \frac{ \alpha_s^2}{6 \pi^3}
\left( \frac{N}{4} \right)^2 \frac{m_P^3}{F^2}  \\
\Gamma (P^{0 \prime} \rightarrow \gamma \gamma) &=&
\frac{ \alpha^2}{27 \pi^3}
\left( \frac{N}{4} \right)^2 \frac{m_P^3}{F^2}   \; .
\end{eqnarray}
If $m_{P^{0'}} < m_t/2$, then the best hope of finding
$P^{0\prime}$ at a hadron collider is through the decay modes
$P^{0 \prime} \rightarrow \gamma\gamma$, $P^{0
\prime} \rightarrow \overline{b}b, \tau^+ \tau^-$, since 
the dijet decay modes will be invisible
against the large QCD background.  The signal in the two-photon channel
will resemble that of an intermediate mass Higgs boson; the small
branching ratio (of order 0.001) is compensated by the large production
rate.  The signal in the $\tau^+ \tau^-$ final state has as background 
from the
corresponding Drell-Yan process.  According to ref.~\cite{Eichten:1984eu}
the {\it effective} integrated luminosity (i.e. luminosity times
identification efficiency) required to find the $P^{0
\prime}$ in this channel would be in the range $3 \times 10^{35}$ --
$5 \times 10^{36}\ {\rm cm}^{-2}$ for colliders with center-of-mass energies
in the range $2$ -- $20$ TeV.

\begin{figure}[tb]
\vspace{5cm}
\begin{center}
\includegraphics{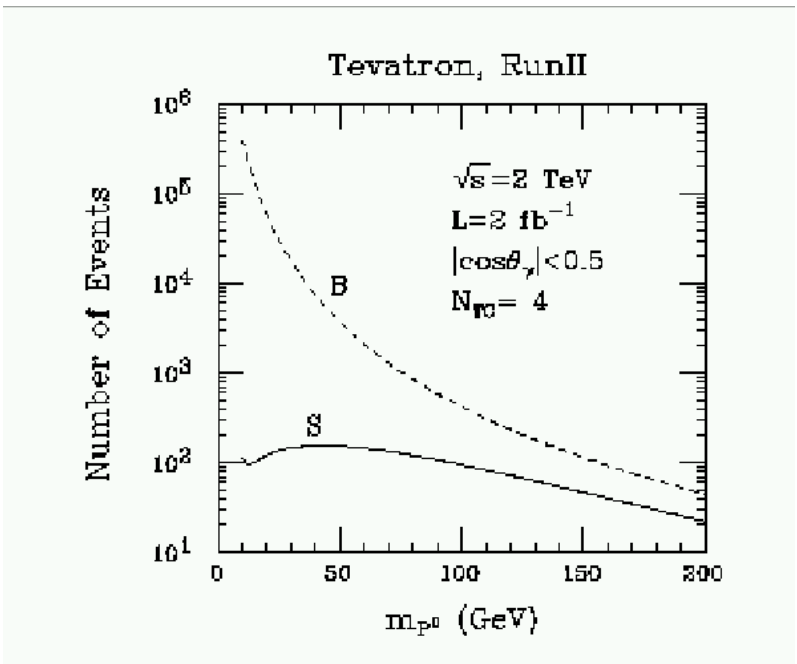}
\end{center}
\vspace{3cm}
\caption[cap22]{\small \addtolength{\baselineskip}{-.4\baselineskip} Projected $gg \to P^0 \to \gamma\gamma$ signal rate $S$ and 
  irreducible background rate for an integrated luminosity of 2 $fb^{-1}$ at
  Tevatron Run II; from \protect\cite{Casalbuoni:1998fs}.}
\label{fig:casal-hadr}
\end{figure}

Recently, Casalbuoni \etal
\cite{Casalbuoni:1998fs} (see also \cite{Casalbuoni:1992nw}) 
have looked in detail at the possibility of
finding the $P^0$ state at the Tevatron and LHC.  The $P^0$ could be visible in
the process $gg \to P^0 \to \gamma\gamma$ at Run II 
of the Tevatron in the mass range $60$ GeV
$\leq M \leq$ $200$ GeV and at the LHC 
in the mass range $30$ GeV $\leq M \leq$
$200$ GeV.  Moreover, a very precise
measurement of the signal rate would be possible at the LHC, enabling the
determination of some model parameters.  Figure \ref{fig:casal-hadr} shows
the projected signal and irreducible background rates at 
Tevatron Run II.

\smallskip
\noindent{\bf (ii) Color-triplet PNGB's: the $P_3$ Leptoquarks}
\smallskip

As we have seen, the spectrum of the 
Farhi-Susskind TC model includes
color-triplet technipions, with leptoquark quantum numbers
\cite{Farhi:1981xs,Eichten:1986eq,Lane:1991qh}. 
These objects, $P_3$'s, would be predominantly resonantly pair-produced
through a color-octet techni-$\rho$ ($V_8$) coupled either to gluons, or
through octet vector dominance mixing (VMD) with the ordinary gluon.  The latter
case assures a signal in $q\bar{q}$ annihilation provided the $V_8$ can be
excited.

$P_3$'s decay (via ETC
\cite{Dimopoulos:1979es} \cite{Eichten:1979ah}) preferentially to third-generation
quarks and leptons.  At leading order, the leptoquark pair production cross
section depends only on the masses ($M_{V_8}, M_{P_3}$) and the $V_8$ decay
width ($\Gamma_{V_8}$).  The latter depends, in turn, on masses, on the
size of the TC group $N_{TC}$ and on the mass-splitting between
the color-octet and color-triplet technipions ($\Delta M$).  While we would
expect that $V_8$ VMD with the gluon would provide the
largest contribution to the hadronic production of colored PNGB's, provided
the $V_8$ pole is within reach of the machine energy, at this writing there
is no study of this contribution and its effects.

EHLQ  \cite{Eichten:1986eq}, and more recently
W. Skiba \cite{Skiba:1996ne}, have studied of hadron collider signatures of
colored PNGB's, without the $V_8$ resonant enhancement.
Possible processes in which 
colored PNGBs can be produced in hadron
colliders are: (i) Gluon-gluon and quark--anti-quark annihilations 
as sources of
PNGB pair production; (ii) Quark-gluon fusion 
producing single PNGB's; (iii) anomalous couplings to two gluons
producing single PNGB's.
Direct PNGB couplings
to fermions (through ETC effects) are typically
too small to give significant cross sections. 
The $V_8$ VMD is
neglected in the present discussion, and we know
of no reference in the literature in which
its effects are included.

\begin{figure}[t]
\vspace{7cm}
\includegraphics{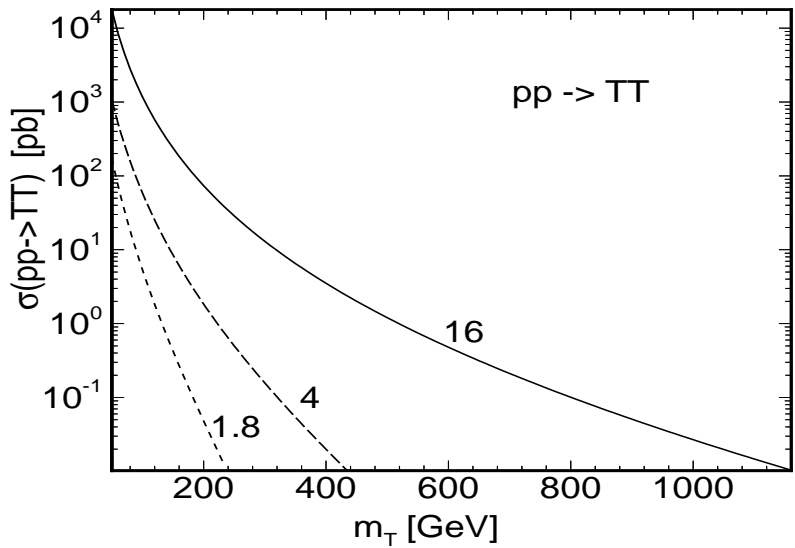}
\vspace{2cm}
\caption[cap23]{\small \addtolength{\baselineskip}{-.4\baselineskip} The cross section for pair production of leptoquarks in pp 
           collisions, for $\protect\sqrt{s}=$
           1.8, 4 and 16 TeV, (from W. Skiba
           \cite{Skiba:1996ne};  $T=P_3$ in Skiba's notation).}
\label{fig:lq}
\end{figure}

The cross section for the pair production of 
colored PNGB's has been calculated 
in Ref. \cite{Eichten:1984eu} for the general case of 
pseudo(scalar) particles in any
representation of $SU(3)_{color}$. Quark anti-quark
annihilation yields
\begin{equation}
  \frac{d \sigma}{d \hat{t}}(q \bar{q} \rightarrow P P) = 
  \frac{2 \pi \alpha_s^2}{9 \hat{s}^2} \, k_d \beta^2 (1-z^2)
\end{equation}
and $g-g$ annihilation gives 
\begin{equation}
  \frac{d \sigma}{d \hat{t}}( g g  \rightarrow P P) =
  \frac{2 \pi \alpha_s^2}{\hat{s}^2} \, k_d 
  \left( \frac{k_d}{d} -\frac{3}{32} (1-\beta^2 z^2) \right)
  \left( 1 -2 V + 2 V^2 \right) .
\end{equation}
where $k_d$ is the ``Dynkin index'' of the $d$-dimensional 
representation ($k_3=\frac{1}{2}$, $k_8=3$), $z$ the cosine of parton 
scattering angle in the center of mass system, 
\begin{displaymath}
V=1-\frac{1-\beta^2}{1-\beta^2 z^2} \ \ {\rm and} \ \ 
\beta^2=1-\frac{4 m_P^2}{\hat{s}},
\end{displaymath}
$\hat{s}$ and $\hat{t}$ are Mandelstam variables at the parton level.
Using these formulae and parton distributions from
Ref.~\cite{Devoto:1983sh} (set 1), Skiba obtains
\cite{Skiba:1996ne} production rates for leptoquarks, $P_3$, and 
octets, $P_8$, which agree with the results of
Refs.~\cite{Eichten:1986eq} (see also Hewett \etal ,
\cite{Hewett:1993ks}).  As the cross-section for pair production of
leptoquarks at LHC energies is sizeable (see Fig.(\ref{fig:lq})), the LHC
can potentially observe leptoquarks with masses up to approximately
$\sim 1$ TeV.

\begin{figure}[t]
\vspace{7cm}
\includegraphics{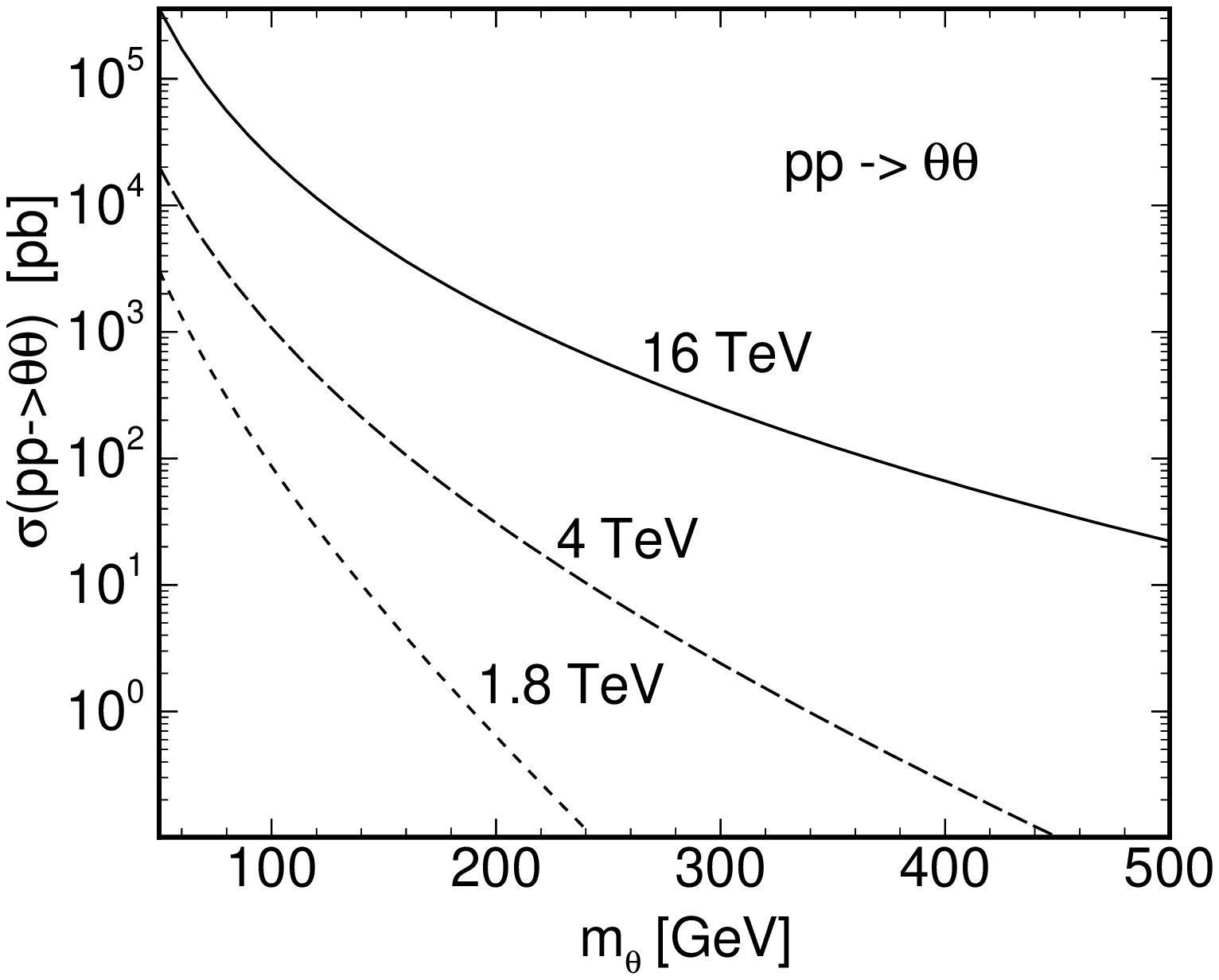}
\vspace{2cm}
\caption[]{\small The cross section for pair production of color-octet PNGBs in pp
           collisions, for $\protect\sqrt{s}=$
           1.8, 4 and 16 TeV, (from W. Skiba
          \cite{Skiba:1996ne}; $\theta=P_8$ in Skiba's notation).}
\label{fig:octet}
\end{figure}

\smallskip
\noindent{\bf (iii) Color-octet PNGB's, $P_8$, and vector mesons $\rho_{T8}$}
\smallskip

Due to color factors, the production cross sections for color octet
$P_8$'s are an order of magnitude larger than for leptoquarks, as a
comparison of Figures \ref{fig:lq} and \ref{fig:octet} reveals.  The
detection of $P_8$'s is, on the other hand, far more difficult.
$P_8$'s typically decay into two hadronic jets, so that pair-produced
$P_8$'s yield four-jet final states.  QCD four-jet production is the
main source of background, and it is overwhelmingly large.  We mention
here two suggestions from the literature as to how $P_8$ states might
be detected.  One involves using special cuts and kinematic variables
to bring out the $P_8$ signal in four-jet final states.  The other
turns to a single-production of $P_8$ and a rarer $P_8$ decay mode
with lower backgrounds.

Chivukula, Golden and Simmons Ref.~\cite{Chivukula:1991zk,
Chivukula:1991di} have evaluated the possibility of detecting new
colored particles at the Tevatron and LHC in multi-jet final states.
The analysis begins with estimating the QCD multi-jet background
\cite{Parke:1986gb, Kunszt:1988it, Mangano:1989rp, Mangano:1990by,
Berends:1989ie, Berends:1990hf} and calculating the signal from heavy
particle decays.  Isolation and centrality cuts must be applied to
ensure all jets are detectable.  Then appropriate kinematic
variables must be chosen to make the signal stand out cleanly above
background.

\begin{figure}[htb]
\begin{center}
\vspace{7cm}
\includegraphics{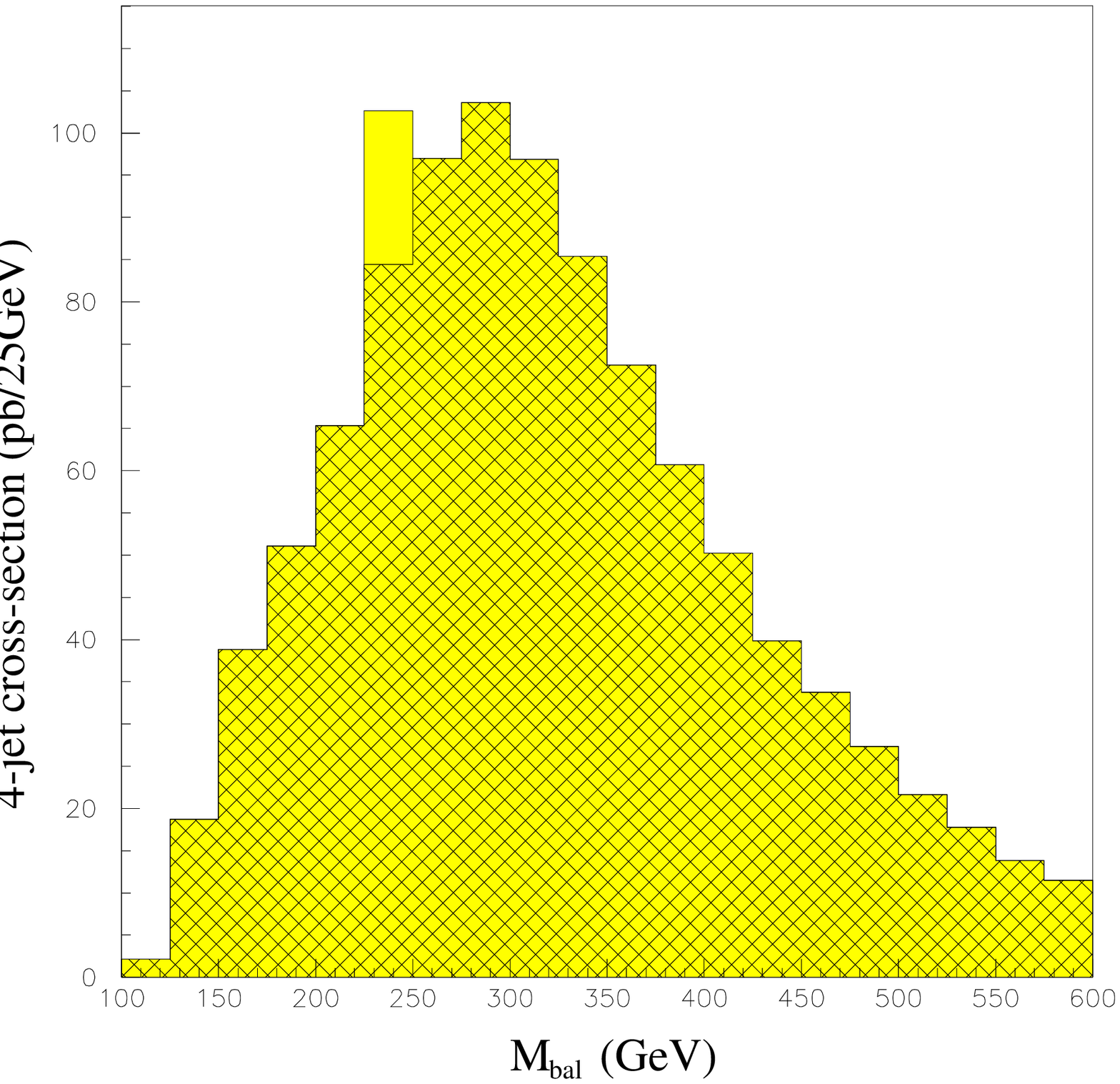}
\vspace{2cm}
\caption[cap24]{\small \addtolength{\baselineskip}{-.4\baselineskip} Four-jet rate $\frac{d \sigma}{ d m_{bal}}$ at a
$17$ TeV collider with $p_T^{min}$ of $100$ GeV.
QCD background (hatched) and $240$ GeV
technipion signal are shown.  No resolution effects included
(from ref. \protect\cite{Chivukula:1991zk}).}
\label{fig:p88jets}
\end{center}
\end{figure}

The following strategy allows one to pull a $P_8$ signal out of the
large QCD background \cite{Chivukula:1991zk}, as illustrated in Figure
\ref{fig:p88jets}.  For each four-jet event, consider all possible
pairwise partitions of the jets.  Choose the partition for which the
two pairs are closest in invariant mass, and define the ``balanced
pair mass,'' $m_{bal}$, as the average of the two masses. The signal
cross-section will cluster about $m_{bal} = m_{new\ particle}$ while
the background will not.  Imposing a large minimum-$p_T$ cut on the
jets further enhances the signal; the background is peaked at low
$p_T$ due to infrared QCD singularities.

Analyses of this kind indicate that real scalar color-octet particles with
masses as high as $325$ GeV should be visible at the LHC if a $p_T$ cut of
about $\sim 170$ GeV is employed \cite{Chivukula:1995dt}.  The lower end of the
visible mass range depends strongly on just how energetic ($p_T^{min}$) and
well-separated ($\Delta R$) jets must be in order for an event to be
identified as containing four distinct jets.  Discovery of color-octet
scalars at the Tevatron is likely to be possible, albeit
difficult, in a reduced mass range  \cite{Chivukula:1991zk}.

\begin{figure}[htb]
\begin{center}
\vspace{7cm}
\includegraphics{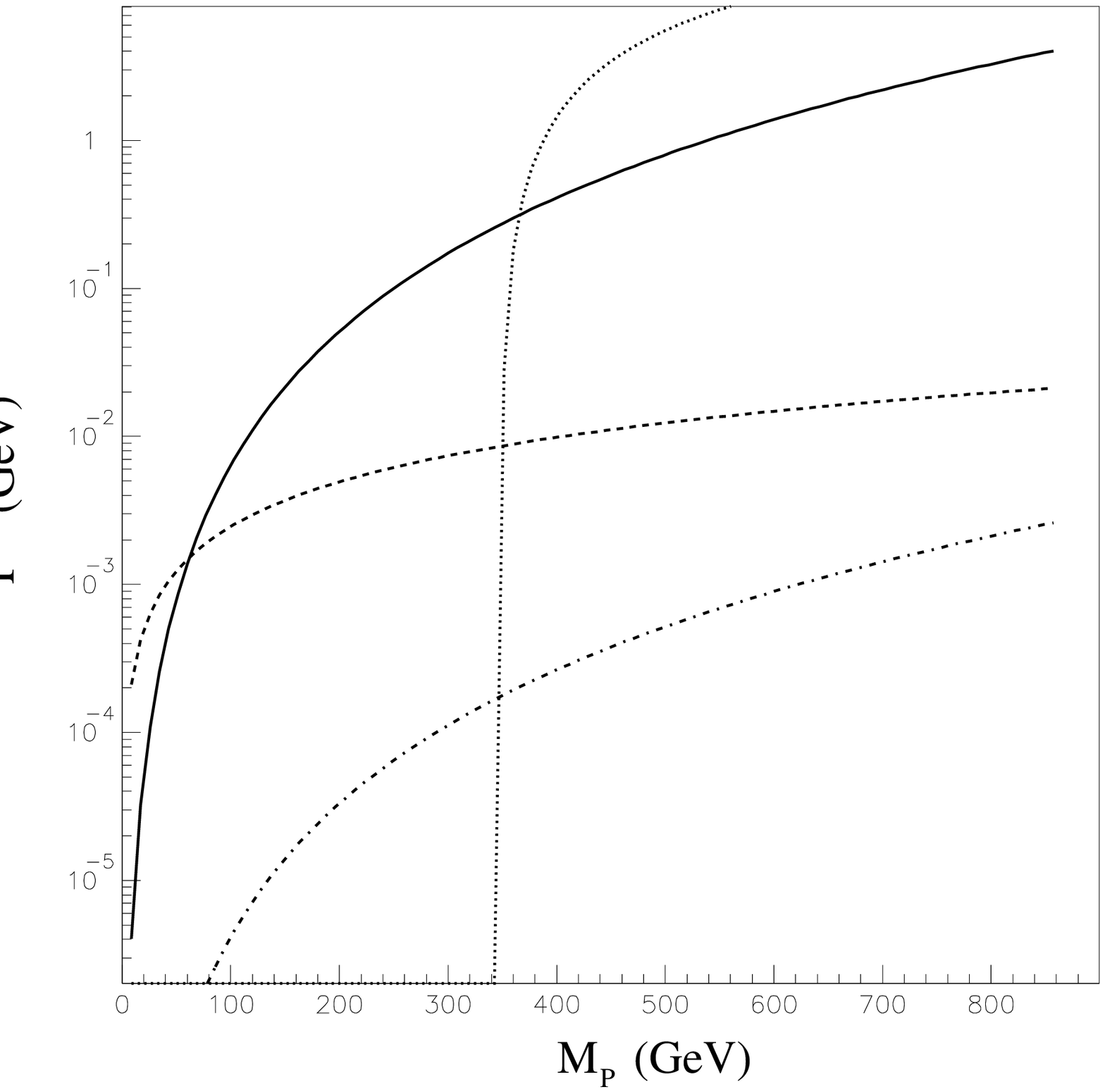}
\vspace{2cm}
\caption[cap25]{\small \addtolength{\baselineskip}{-.4\baselineskip} 
Partial widths for the decay of $P_8^{0 \prime}$
into gluon-gluon (solid line),
$\bar{b} b$ (dashed line), $\bar{t} t$ (dotted line) and $Z$-gluon
(dot-dashed line) in the Farhi-Susskind model \protect\cite{Chivukula:1995dt}.}
\label{fig:p8widths}
\end{center}
\end{figure}

\begin{figure}[htb]
\begin{center}
\vspace{5cm}
\includegraphics{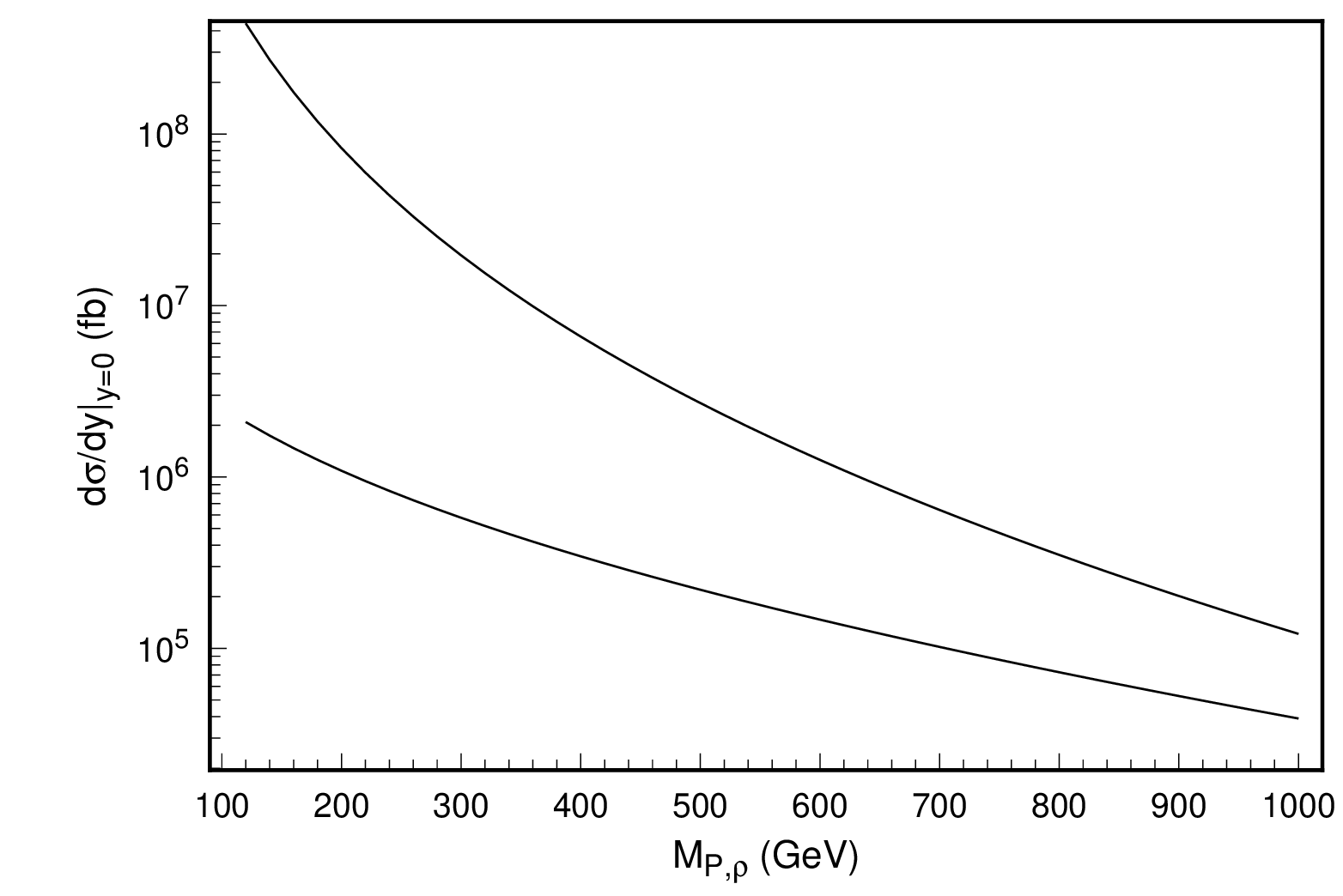}
\vspace{2cm}
\caption[cap26]{\small \addtolength{\baselineskip}{-.4\baselineskip} Differential cross section at $y = 0$ for single production
of $\rho_{T8}^{0 \prime}$ (solid line) and $P_8^{0 \prime}$ (dashed line) at
the LHC in femtobarns \protect\cite{Chivukula:1995dt}.}
\label{fig:p8cross}
\end{center}
\end{figure}

An alternative strategy is to use processes with smaller backgrounds, such as single-production followed by a rarer decay mode.  
At hadron colliders, the widths relevant for single production and decay of
the $P_8^{0 \prime}$ of the Farhi-Susskind model are
\cite{Chivukula:1995dt, Belyaev:1999xe} :
\begin{eqnarray}
\Gamma (P_8^{0 \prime} \rightarrow q \bar{q}) &=& \frac{3}{16 \pi}
\frac{m_q^2}{F^2} m_P \left( 1 - 4 m_q^2/m_P^2 \right)^{3/2} \nonumber \\
\Gamma (P_8^{0 \prime} \rightarrow gg) &=& \frac{5 \alpha_s^2}{24 \pi^3}
\left( \frac{N}{4} \right)^2 \frac{m_P^3}{F^2} \;   \nonumber  \\
\Gamma (P_8^{0 \prime} \rightarrow g\gamma) &=& 
\left( \frac{N e g_s}{4\pi F} \right)^2 \frac{m_P^3}{576\pi} \; .
\end{eqnarray}
Figure \ref{fig:p8widths} gives the decay widths of the $P_8^{0 \prime}$ in
the Farhi-Susskind model.  
Figure \ref{fig:p8cross} illustrates gluon fusion production at the LHC
and we see that the rate of single $P_8^{0 \prime}$ production is 
high at a hadron collider.  
However, the signal in the primary decay channels (gluon or b-quark pairs) is
swamped by QCD background.  If the PNGB mass is above the $t
\bar{t}$ threshold, the decay mode $P \rightarrow t \bar{t}$ becomes dominant
and may alter the standard QCD value of the $t \bar{t}$ cross section 
and $t\bar{t}$ spin correlations.  The decay channel $P \to
g\gamma$ \cite{Hayot:1981gg, Belyaev:1999xe} holds some promise 
for the models of ``low-scale'' TC discussed 
in Section 3.5; 
however the PNGB's of the Farhi-Susskind TC 
model are not likely to be visible in this mode.

The $\rho_{T8}^{0 \prime}$ coupling to $gg$ and $q \bar{q}$ can be
estimated by assuming that it mixes with the gluon under a generalized
vector meson dominance (VMD). VMD applies to the 
$gg\rightarrow \rho_{T8}^{0 \prime}$
process as well as the $q \bar{q} \rightarrow \rho_{T8}^{0 \prime}$ process
ref. \cite{SekharChivukula:2001gv}.  The partial
widths relevant for $\rho_{T8}^{0 \prime}$ production at hadron
colliders in the narrow width approximation are \cite{Chivukula:1995dt}:
\begin{eqnarray}
\Gamma(\rho_{T8}^{0 \prime} \rightarrow gg) &=&
\frac{\alpha_s^2}{2 \alpha_\rho} m_\rho   \\
\Gamma(\rho_{T8}^{0 \prime} \rightarrow q \bar{q}) &=&
\frac{5 \alpha_s^2}{6 \alpha_\rho} m_\rho
\label{eq:rho8wid}
\end{eqnarray}
Figure \ref{fig:p8cross} compares the single production cross section of the
$\rho_{T8}^{0 \prime}$ and $P_{8}^{0 \prime}$ at the LHC.

If kinematically allowed the dominant decay mode of $\rho_{8T}^{0 \prime}$
is into two colored PNGB \cite{Chivukula:1995dt}:
\begin{equation}
\Gamma( \rho_{8T}^{0 \prime} \rightarrow P_8 P_8) =
\frac{\alpha_{\rho_T}}{4}\; m_\rho \;
\left( 1 - 4 m_P^2/m_\rho^2 \right)^{3/2} .
\end{equation}
In this case the $\rho_{T8}^{0 \prime}$ contributes strongly to the cross
section for color-octet PNGB pair production discussed in above and
improves the signal.

If the $\rho_{T8}$ are light, e.g. in low-scale models (Section 3.5),
there can be a sizeable cross section for their {\it pair} production.
N\"aively, well above the pair-production threshold one expects:
\cite{Chivukula:1995dt} :
\begin{equation}
\frac{\sigma (pp \rightarrow \rho_{T8} + X)} {\sigma (pp \rightarrow
\rho_{T8} \rho_{T8} + X) } \simeq \frac{1}{g_{\rho_T}^2} \simeq \frac{1}{40}  .
\end{equation}
Note that {\it all} types of colored vector
resonances can be pair-produced, whereas the isosinglet $\rho_{T8}^{0
\prime}$ dominates single production.  Hence, pair production, unlike
single production, can result in interesting decays to longitudinal
electroweak gauge bosons.  One may also expect spectacular $8$--jet events
whose kinematics distinguish them from the QCD background, as in the case
of color-octet PNGB pair production. 

The $\rho_{T8}$ can significantly affect the
invariant mass distribution of
$\bar{t}t$ final states \cite{Eichten:1994nc}.
Through gluonic vector dominance processes like 
$\bar{q}q\rightarrow g+\rho_{T8}
\rightarrow \bar{t}t$ the gluon and $\rho_{T8}$
add coherently.  Significant mass imits can be placed
from the Tevatron in Run II (this process
has a direct analogue in Topcolor, where $\rho_{T8}$
becomes the coloron \cite{Hill:1994hs}).

\vskip .1in
\noindent
{\bf (iv) Non-Resonant Production and Longitudinal
Gauge Boson Scattering}
\vskip .1in

At the next level, below resonance production, we expect gauge boson
fusion through techniquark loops (including anomalous loops) into
techni-vector mesons and/or techni-pions.  This can include gluon
fusion \cite{Georgi:1978gs}, \cite{Dicus:1987dj},
\cite{Glover:1989rg}, \cite{Kao:1991tt} at machines such as the LHC,
or $W-W$ fusion at an LHC or LC (see section \ref{ssec:epemcoll})
\cite{Jones:1979bq}, \cite{Dicus:1985zg}, \cite{Hioki:1983yz},
\cite{Han:1985zn}, \cite{Cahn:1984ip}.  Scattering of the longitudinal
electroweak modes, $V_L^{}V_L^{} \to V_L^{}V_L^{}$, is especially
important (see Section 2.2.3). These processes involve all possible
spin and isospin channels simultaneously, and can proceed through
scalar and vector resonances or non-resonant channels.

\smallskip
\noindent{\bf (v) Rescattering}
\smallskip

An alternative way of finding color-octet technipions is to detect
them after they have re-scattered into pairs of $W$ or $Z$ particles.
For example, in the Farhi-Susskind TC model the production of
gauge boson pairs through gluon fusion includes a potentially sizeable
\cite{Bagger:1991qx} contribution from loops of colored technipions
through the process $P_8^{0,\pm} P_8^{0,\pm} \to WW$ or $ZZ$.  This
may prove to be a very useful diagonostic.  If colored technipions are
first detected in 4-jet final states, their association with
electroweak symmetry breaking will not be obvious.  However, the {\it
combination} of their discovery with the observation of a large number
of gauge-boson pairs may permit us to deduce that the colored scalars
are PNGB's of the symmetry breaking sector
\cite{Chivukula:1992bk}.

\begin{figure}[htb]
\begin{center}
\vspace{7cm}
\includegraphics{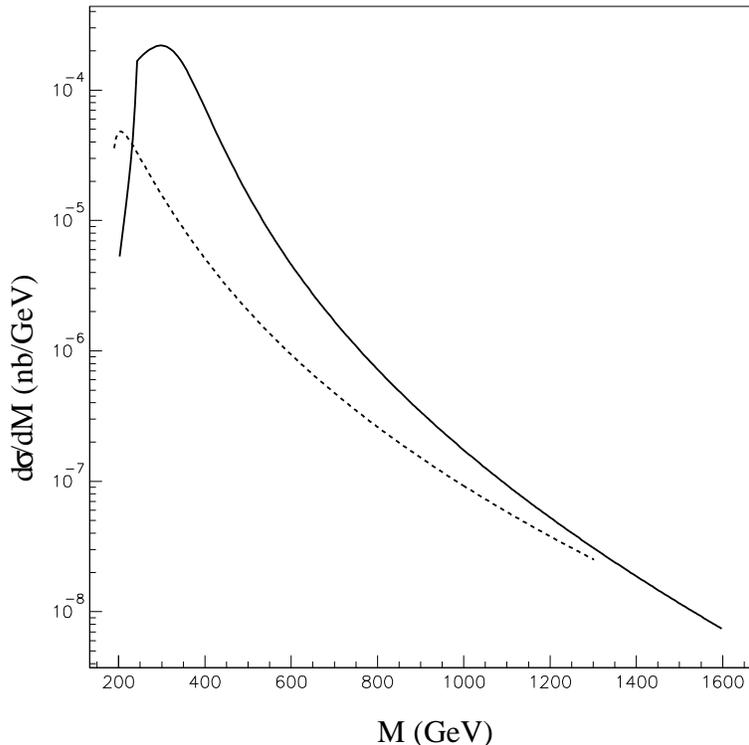}
\vspace{3cm}
\caption[cap27]{\small \addtolength{\baselineskip}{-.4\baselineskip} The $ZZ$ differential cross section in nb/GeV vs.
$M_{ZZ}$ in a toy $O(N)$ scalar-model
with three color-octet PNGBs (upper curve), and
the continuum $q\bar{q}$ annihilation background (lower curve).
A pseudo-rapidity cut $|\eta|<2.5$ is imposed on the final state
$Z$'s. From \protect\cite{Chivukula:1992bk} }
\label{fig:ssekhar}
\end{center}
\end{figure}

The contribution of loops of colored technipions to the production of
gauge-boson pairs through gluon fusion was calculated to leading order in
chiral perturbation theory in ref.~\cite{Bagger:1991qx}.  Unfortunately,
general considerations \cite{Soldate:1990fh,Chivukula:1993gi} show that in
theories with many Nambu--Goldstone bosons, chiral perturbation theory breaks down
at very low energies. In the Farhi-Susskind model, for example, chiral
perturbation theory breaks down at a scale of order 440 GeV!  Hence only a
qualitative estimate of the signal size is possible.  Figure
\ref{fig:ssekhar} shows the $ZZ$ differential cross section as a function
of $ZZ$ invariant mass at the LHC in a toy $O(N)$ scalar-model
\cite{Chivukula:1992bk}; model parameters were chosen so the signal size is
representative of Farhi-Susskind TC.  Note that there are almost an
order of magnitude more events due to gluon fusion than due to the
continuum $q\bar{q}$ annihilation for $ZZ$ invariant masses between 300 GeV
and 1 TeV. The observation of such a large two gauge-boson pair rate at a
hadron collider would be compelling evidence that the EWSB sector couples
to color.

\subsubsection{Production and Detection at $e^+e^-$ Colliders}
\label{ssec:epemcoll}

\vskip .1in
\noindent{\bf (i) pair-production of EW bosons}
\vskip .1in

\begin{figure}[htb]
\vspace{7.0cm}
\includegraphics{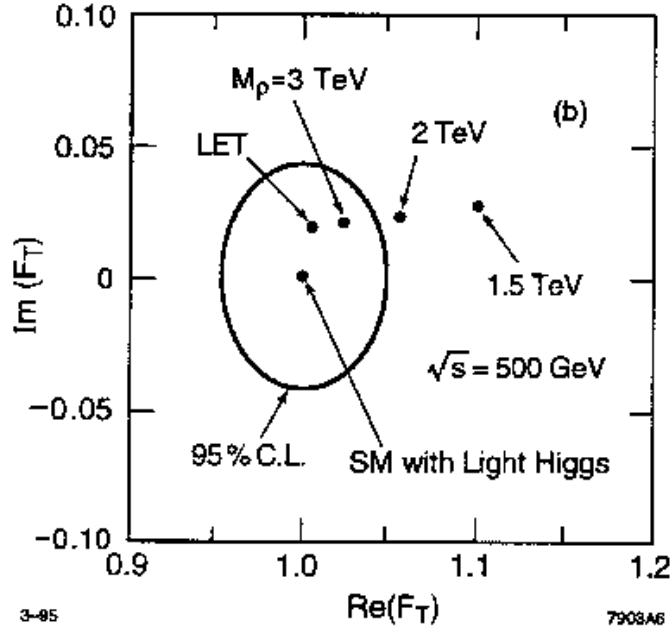}
\vskip2.2truecm
 \caption[cap28]{\small \addtolength{\baselineskip}{-.4\baselineskip}
      Sensitivity of a 500~GeV LC with 80~$fb^{-1}$ of data to
      $M_{\rho_T}$ via the $W$--boson form factor; from
      Ref.\protect\cite{Barklow:1995sk}.  The predicted values for a
      3-TeV $\rho_T$ or non-resonant scattering (LET) lie within the
      95\% c.l.  curve for the prediction of the Standard Model; these
      cases cannot, then be reliably distinguished.  Lighter $\rho_T$
      give predictions which differ significantly from those of the
      Standard Model.}  
\label{LC-ww-fig}
\end{figure}

\begin{figure}[tbh] 
\vspace{10.0cm}
\includegraphics{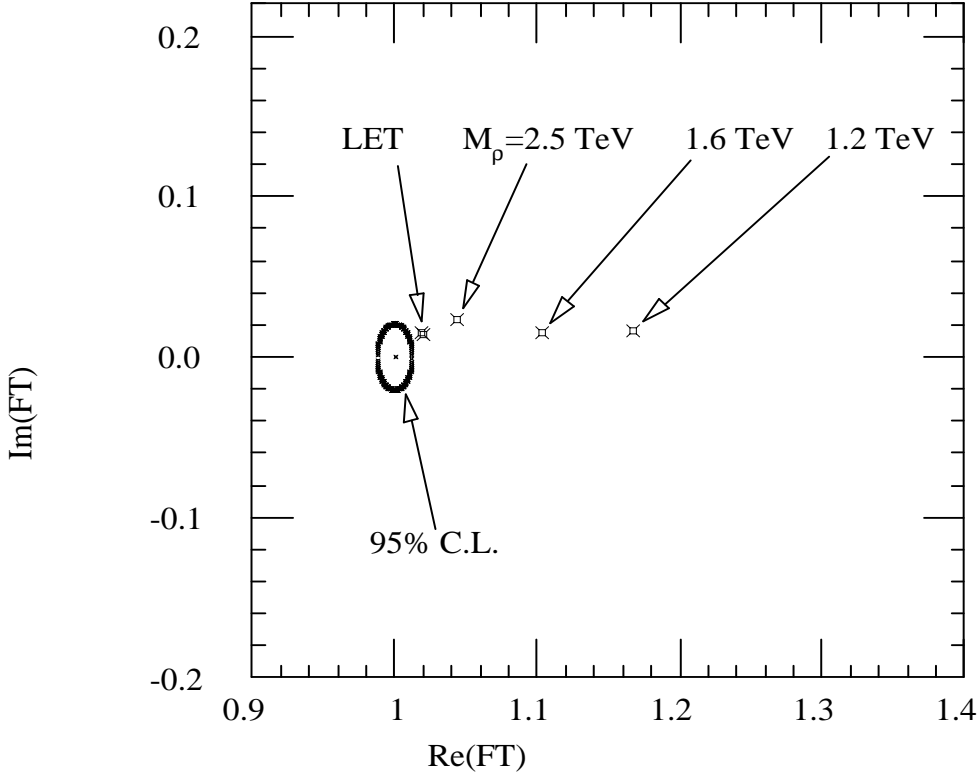}
\vspace{10pt}
\caption{\small \addtolength{\baselineskip}{-.4\baselineskip}
Sensitivity of a 500~GeV LC with 500~$fb^{-1}$ of data to
      $M_{\rho_T}$ via the $W$--boson form factor; from
      Ref.\protect\cite{Barklow:2002su}.  The predicted values for a
      2.5-TeV $\rho_T$ or non-resonant scattering (LET) now lie outside
      95\% c.l.  curve for the prediction of the Standard Model.}
\label{fig:fteight}
\end{figure}

The $s$-channel process $e^+e^- \to W W$ is an effective probe of strong
electroweak symmetry breaking, especially for physics with a vector resonance
\cite{Barklow:1994uf, Barklow:1995sk, Golden:1995xv, Barklow:2002su}.  
A $500$ GeV linear collider with only 80 fb$^{-1}$ of integrated
luminosity would already be sensitive to radiative corrections induced
by vector resonances with masses up to about 2 TeV, but would not be
able to observe off-resonance contributions to $W W$ production (see
Figure \ref{LC-ww-fig}).  With 500 fb$^{-1}$ of integrated luminosity,
even the non-resonant or low-energy-theorem contributions would become
distinguishable from Standard Model expectations, as illustrated in
Figure \ref{fig:fteight}.  A higher-energy $e^+e^-$ collider would have even
greater search potential.  With $\sqrt {s}$=1.5 TeV and an integrated
luminosity of 190 fb$^{-1}$, it should be possible to distinguish the
radiative effects of a very heavy techni-$\rho$ or a non-resonant
amplitude from those of the standard model with a light Higgs boson;
the 4~TeV (6 TeV) techni-$\rho$ corresponds to a 6.5$\sigma$ (4.8
$\sigma$) signal.  At a slightly higher integrated luminosity of 225
fb$^{-1}$, it would be possible to obtain 7.1$\sigma$, 5.3$\sigma$ and
5.0$\sigma$ signals for a 4~TeV techni-$\rho$, a 6~TeV techni-$\rho$,
and non-resonant contributions, respectively.  

The $WW$ fusion processes are complementary to the $s$-channel $W^+W^-$ mode
since they involve more spin-isospin channels (the $I=0$ and $I=2$ channels).
For an $e^+e^-$ collider at $\sqrt s=1.5$~TeV with 190 fb$^{-1}$ of data, the
$W^+W^-/ZZ$ event ratio can be a sensitive probe of a strongly-interacting
electroweak sector \cite{Golden:1995xv, Barger:1995cn, Kurihara:1993jv,
  Barger:1994wa, Kurihara:1993jg} as illustrated in
Table~\ref{golden-tableii}.  Statistically significant signals are found for
a 1 TeV scalar or vector particle.  There is also about a 6$\sigma$ signal
for non-resonant (low-energy-theorem) amplitudes via the $W^+W^- \to ZZ$
channel alone.  For an $e^-e^-$ collider with the same energy and luminosity,
the non-resonant-amplitude signal rate for the $\nu\nu W^-W^-$ ($I=2$)
channel \cite{Barger:1994wa} is similar to that for $e^+e^-\to\bar\nu\nu ZZ$,
as anticipated from symmetry arguments, while the background rate is
higher \cite{Golden:1995xv}.  The signals are doubled for an $e^-_L$ polarized
beam (or quadrupled for two $e^-_L$ beams), whereas the backgrounds increase
by smaller factors.  A 2 TeV $e^+e^-$ linear collider would increase the
signal rates by roughly a factor of 2--2.5.  

\begin{table}[htbp]
\centering
\bigskip
\begin{tabular}{|l|c|c|c|c|c|} \hline
channels & SM  & Scalar & Vector   & LET  \\
\noalign{\vskip-1ex}
& $m_H=1$ TeV & $M_S=1$ TeV & $M_V=1$ TeV &\\
\hline
$S(e^+ e^- \to \bar \nu \nu W^+ W^-)$
& 160   & 160   & 46  & 31  \\
$B$(backgrounds)
& 170    & 170   & 4.5  & 170  \\
$S/\sqrt B$ & 12 & 12 & 22 & 2.4 \\
\hline
$S(e^+ e^- \to \bar\nu \nu ZZ)$
&  120  & 130  & 36  & 45   \\
$B$(backgrounds)
& 63    & 63   & 63  & 63  \\
$S/\sqrt B$ & 15& 17& 4.5& 5.7\\
\hline
\hline
$S(e^- e^- \to \nu \nu W^- W^-)$
& 27  & 35  & 36  & 42  \\
$B$(backgrounds)
& 230  & 230   & 230  & 230  \\
$S/\sqrt B$ & 1.8 & 2.3 & 2.4 & 2.8 \\
\hline
\end{tabular}
\caption{\small \addtolength{\baselineskip}{-.4\baselineskip} \label{golden-tableii}%
Total numbers of $W^+W^- \to 4$-jet and  $ZZ \to  4$-jet
signal $S$ and background $B$ events calculated for  a 1.5~TeV
LC with  integrated luminosity 200~fb$^{-1}$.  Results are shown for the
SM, for strongly-coupled models with scalar or vector resonances and for
low-energy-theorem (LET) scattering.  Events are summed
over the mass range $0.5 < M_{WW} < 1.5$~TeV except for the $W^+W^-$ channel
with  a narrow vector resonance in which $0.9 < M_{WW} < 1.1$~TeV. The
statistical significance $S/\protect\sqrt B$ is also given.
For comparison, results for $e^-e^- \to \nu \nu W^-W^-$
are also presented, for the same energy and luminosity. The hadronic 
branching fractions of $WW$ decays and the $W^\pm/Z$
identification/misidentification are included. From
ref. \protect\cite{Golden:1995xv}. }
\end{table}

\begin{figure}[tb] 
\vspace{12.0cm}
\includegraphics{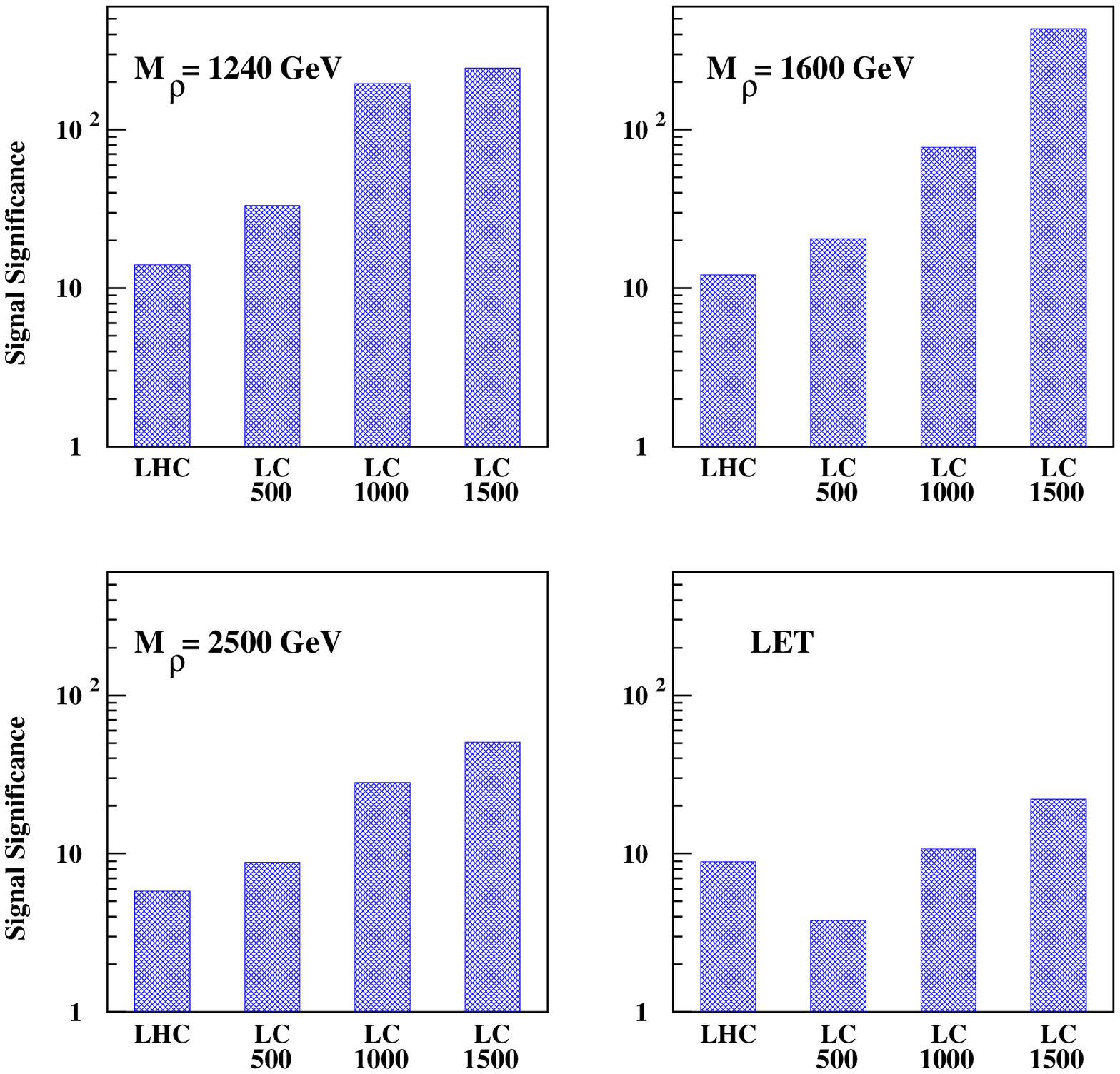}
\vspace{1pt}
\caption{\small \addtolength{\baselineskip}{-.4\baselineskip} 
Direct strong symmetry breaking signal significance 
in $\sigma$'s for various masses $M_\rho$ of a vector resonance 
in $W_L W_L$ scattering \protect\cite{Barklow:2002su}.
The numbers below the ``LC'' labels refer to the
center-of-mass energy of the linear collider in GeV.
The luminosity of the LHC is assumed to be 300~$fb^{-1}$, 
while the luminosities
of the linear colliders are assumed to be  500, 1000, and 
1000~$fb^{-1}$ for 
$\sqrt{s}$=500, 1000, and 1500~GeV respectively.  
The lower right hand plot ``LET'' refers to the
case where no vector resonance exists at any mass 
in strong $W_L W_L$ scattering.}
\label{fig:strong_lc_lhc}
\end{figure}

When both the both the pair production and gauge boson fusion
processes are taken into account, the direct signal of a
strongly-coupled symmetry-breaking sector is generally stronger at a
high-energy linear collider than at the LHC; 
Figure \ref{fig:strong_lc_lhc}.  

A strongly--coupled electroweak symmetry breaking sector can also be
investigated \cite{Golden:1995xv} at photon--photon colliders in the modes
$\gamma \gamma \to ZZ$, $\gamma \gamma \to W^+W^-$, $\gamma \gamma \to
W^+W^-W^+W^-$, and $\gamma\gamma \to W^+W^-ZZ$.  While irreducible backgrounds in the first two cases
are severe, the four-boson final states seem more
promising \cite{Brodsky:1993xp}.  To be sensitive to a heavy scalar resonance
(e.g. heavy Higgs) with a mass up to 1 TeV would require a 2 TeV $e^+e^-$
collider with luminosity of 200 fb$^{-1}$ \cite{Jikia:1995vz},
\cite{Cheung:1994sa}.
A $1.5$ TeV linear collider running in the $\gamma \gamma $ mode is found to
be roughly equivalent to a $2$ TeV $e^+e^-$ collider for the purpose of
studying strongly-interacting ESB physics \cite{Jikia:1995vz, Cheung:1994sa}
(the correspondence scales roughly as $\sqrt{s_{\gamma \gamma}}\sim
0.8\sqrt{s_{ee}}$).

\vskip .1in
\noindent{\bf (ii) colorless neutral PNGBs, $P^0, P^0{}'$} 
\vskip .1in

A light enough neutral PNGB can be singly produced at an $e^+e^-$
collider via the axial-vector anomaly \cite{Bell:1969ts, Adler:1969gk} 
which couple the PNGB to
pairs of electroweak gauge bosons.  At LEP I the PNGB, $P^a$, would
have been
primarily produced \cite{Manohar:1990eg, Randall:1992gp,Rupak:1995kg}
through an anomalous coupling to the $Z$ boson and either a photon ($Z
\to \gamma P^a$) or a second, off-shell $Z$ boson ($Z \to Z^* P^a$).
At the higher center of mass energies of LEP II, PNGBs over a wider
range of masses could have been 
produced through $s$-channel $\gamma^*/Z^*$
exchange and through a $2 \to 3$ production mechanism~\cite{Lubicz:1996xi,
Casalbuoni:1998fs}.

If the TC group is $SU(N_{T})$, the
anomalous coupling between the PNGB and the gauge bosons
$G_1$ and $G_2$ is given, as for the QCD pion,
by~\cite{Dimopoulos:1981yf, Ellis:1981hz, Holdom:1981bg}
\begin{equation}
  N_{T} A_{G_1 G_2} \frac{g_1 g_2}{2 \pi^2 F_T}
  \epsilon_{\mu\nu\lambda\sigma} k^\mu_1 k^\nu_2 \varepsilon^\lambda_1
  \varepsilon^\sigma_2\ ,
\end{equation}
where $N_{T}$ is the number of technicolors, $A_{G_1 G_2}$ is the
anomaly factor (see below), the $g_i$ are the gauge couplings of the
gauge bosons, and the $k_i$ and $\varepsilon_i$ are the four-momenta
and polarizations of the gauge bosons.  

The model-dependent value of the anomaly factor is calculated
in ~\cite{Dimopoulos:1981yf}, \cite{ Ellis:1981hz}, \cite{Holdom:1981bg}.
The dominant production mode for a PNGB at LEP I was generally the $Z \to
\gamma  P^a$ process \cite{Manohar:1990eg} which has a branching ratio of order
$10^{-5}$  
\begin{equation}
  \label{eqn:pzpwidth}
  \Gamma_{Z^0 \to \gamma P^a} = 2.3\times10^{-5}{\rm GeV}
  \left(\frac{123{\rm GeV}}{F_{T}}\right)^2 \left(N_{T} 
  A_{\gamma Z^0}\right)^2
  \left(\frac{M_{Z^0}^2 - M_{P^a}^2}{M_{Z^0}^2}\right)^3\ .
\end{equation}
The final states contain a hard, mono-energetic photon, along with the
decay products of $P^a$.  Production in combination with an off-shell $Z^0$
is harder to observe.  An upper bound on the decay width of the
process $Z^0 \to Z^{0*}P^a \to P^a f\bar{f}$ is given
by~ \cite{Randall:1992gp, Lynch:2000hi}.  One expects branching ratios of
order $10^{-7}$ to $10^{-6}$, depending on the process of
interest.  At LEP II and an NLC, production modes include $e^+ e^- \to
\gamma^*, Z^* \to \gamma^{(*)}P^a, Z^{(*)} P^a$ and the 2-to-3 process, $e^+
e^- \to e^+ e^- P^a$ \cite{Lubicz:1996xi}.

In the minimal models, the PNGB's decay back into electroweak bosons.
Which decays are allowed or dominant depends on the values of the anomaly
factors.  For example, a neutral colorless PNGB produced by $Z$-decay can
certainly decay to an off-shell $Z$ plus another electroweak gauge boson
(photon or $Z$).  It may also be able to decay to a pair of photons; if
allowed, this mode dominates over decays via off-shell $Z$'s.  In any given
non-minimal TC model, the dominant decay mode of the neutral
PNGB's depends both on the gauge couplings of the technifermions and on any
interactions coupling technifermions to ordinary fermions.  If some
technifermions are colored, the PNGB may have an anomaly coupling allowing
it to decay to gluons.  If the PNGB gets its mass from effective
four-fermion interactions (e.g. due to extended TC), then it will
be able to decay to an $f\bar f$ pair.  Finally, in some models, the PNGB
may decay dominantly to particles in an invisible sector.  

Searches for the neutral PNGB's at LEP I and LEP II were sensitive only
to PNGB's with small technipion decay constants and large anomaly
factors. The $P^a$ of the Farhi-Susskind model are invisible at these
colliders.  However, data from the LEP experiments is able to test the
properties of the PNGB's of various extended models, such as the
Appelquist-Terning model \cite{Appelquist:1993gi}, the Manohar-Randall weak
isotriplet model \cite{Manohar:1990eg}, Lane's low-scale TC model 
\cite{Lane:1999uh, Lane:1999uk}, and the Lane-Ramana multiscale model 
\cite{Lane:1991qh}.
Details about all of the constraints are available in
Ref. \cite{Lynch:2000hi}; a summary for a few walking TC
models will be given in Section 3.  For now, we turn to the
possibility of direct searches at higher-energy linear colliders.

\begin{figure}[tbh]
\vspace{6cm}
\begin{center}
\includegraphics{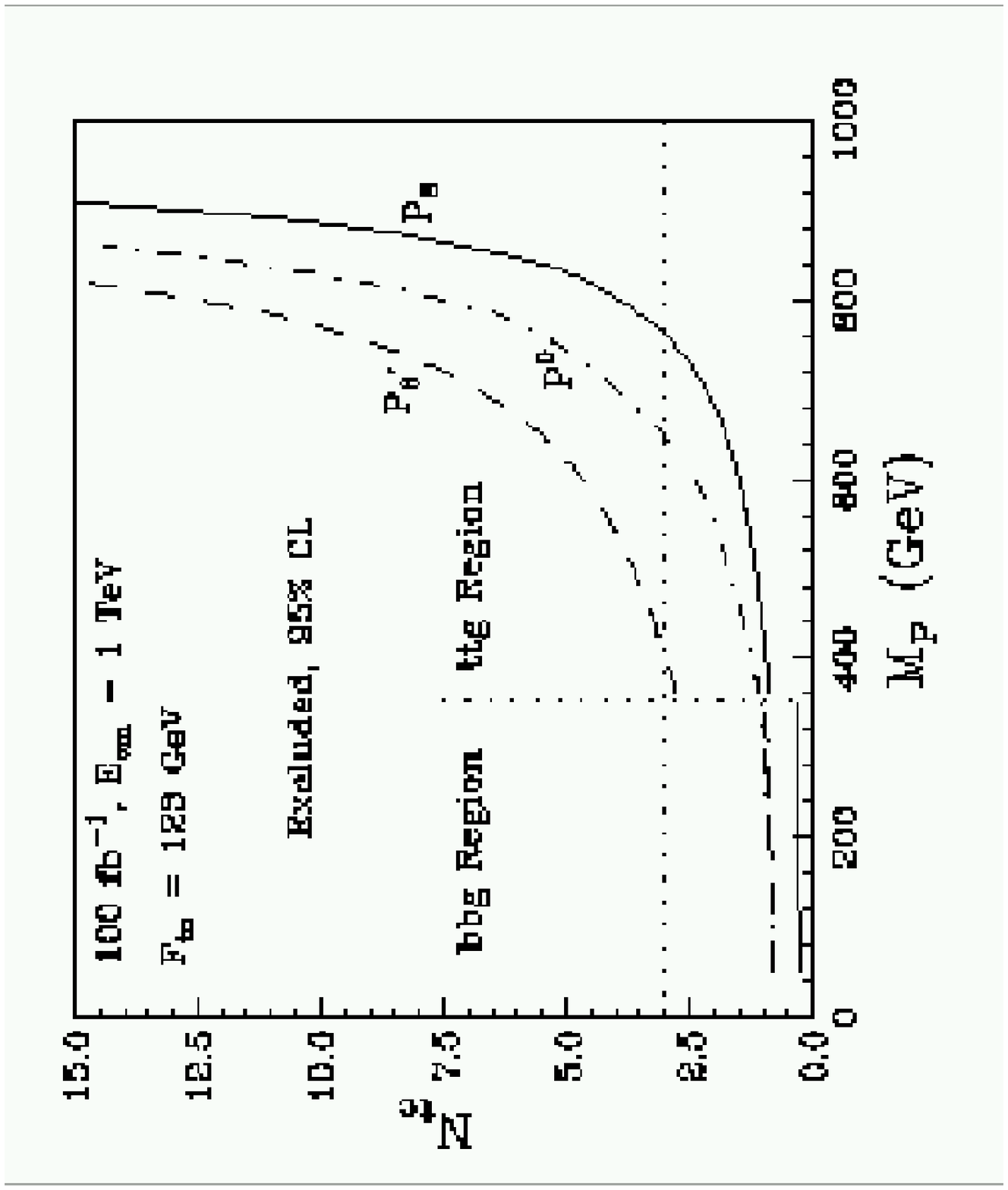}
\end{center}
\vspace{2cm}
\caption[28a]{\small \addtolength{\baselineskip}{-.4\baselineskip} Projected 95\% c.l. exclusion reach for electrically 
neutral PNGB
 as a function of mass and number of technicolors at a 1 TeV LC
\protect\cite{Swartz:1996}.}
\label{fig:swartz-p8-e}
\end{figure}

\begin{figure}[tbh]
\vspace{6cm}
\begin{center}
\includegraphics{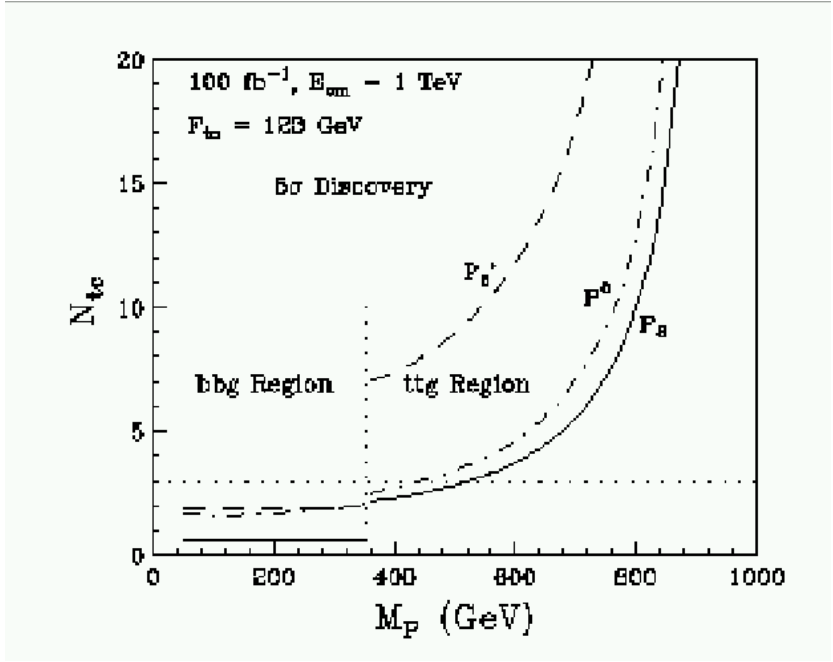}
\end{center}
\vspace{2cm}
\caption[29a]{\small \addtolength{\baselineskip}{-.4\baselineskip} 
Projected 5-sigma discovery reach for electrically neutral PNGB
as a function of mass and number of Technicolors at a 1 TeV LC
  \protect\cite{Swartz:1996}.}
\label{fig:swartz-p8-d}
\end{figure}

The possibility of searching at a $1$ TeV Linear Collider 
for a generic neutral
$P^{0 (')}$ decaying predominantly to the heaviest available fermions
\begin{eqnarray}
e^+e^- \to P^0_{T} \gamma &\to& t \bar{t} \gamma 
 \to b\bar{b} W^+W^- \gamma \qquad M_{P}
                         > 2 m_t \nonumber \\
                         &\to& b \bar{b} \gamma \qquad\qquad M_{\Pi} < 2 m_t.
\end{eqnarray}
was studied in ref. \cite{Swartz:1996}.
The signal events will stand out from the background due to the hard
monochromatic photon recoiling against the PNGB decay products. Based on a
Monte Carlo simulation of signal and background, ref. \cite{Swartz:1996}
finds that for models with $N_{T} = 3$ and $f_{T} = 123$, the
$P^0{}'$ can be excluded at the 95\% c.l. up to a mass of 646 GeV and
could be discovered at the 5-sigma level up to a mass of 452 GeV.  The
$P^{0}$ production rate appears too small to allow discovery in the models
examined.  These results are illustrated in Figures \ref{fig:swartz-p8-e} and
\ref{fig:swartz-p8-d}.

  The electrically neutral color-octet $P_{8}^{0 (')}$ will 
  be singly produced in 
  association with a gluon through the anomalous coupling of the PNGB to the
  gluon and a photon or $Z$ boson.  Provided that the PNGB decays to the
  heaviest available fermion pair, the production and decay chain will look
  like \cite{Swartz:1996}
\begin{eqnarray}
e^+e^- \to P^0_{T 8} g &\to& t \bar{t} g \to b\bar{b} W^+W^- g \qquad M_{P8}
                         > 2 m_t \nonumber \\
                         &\to& b \bar{b} g \qquad\qquad M_{P8} < 2 m_t.
\end{eqnarray}
Characteristics that will help distinguish the signal events from those of
the $q\bar{q} g$ background are monochromaticity of the gluons in the signal
events are monochromatic and their large spatial separation from the PNGB
decay products.  The left-right asymmetries and $t\bar{t}$ spin correlations
will also differ from those of the background distributions.  Based on a
Monte Carlo simulation of signal and background, ref. \cite{Swartz:1996}
asserts that the electrically neutral color-octet PNGB can be excluded or
discovered for any realistic values of $N_{TC}$ provided that $M_{P} < 2
m_t$.  For higher-mass PNGB, finding the $P_{T 8}^{0'}$ will require
including events with hadronic $W$ decays for $N_{TC} < 7$.  For the specific
case $N_{TC} = 3$, it will be possible for a 1 TeV LC to exclude a $P_{T
  8}^{0}\ (P_{T 8}^{0'})$ of mass less than 761 (404) GeV at 95\% c.l. or
to discover a $P_{T 8}^{0}$ with five-sigma significance for $M < 524$ GeV
\cite{Swartz:1996}.  The exclusion and discovery limits are shown in Figures
\ref{fig:swartz-p8-e} and \ref{fig:swartz-p8-d}.

The ability of an LC to study the lightest PNGB  of a
particular model \cite{Casalbuoni:1992nw} has been probed in 
ref.\cite{Casalbuoni:1998fs}.  They find that the process $e^+ e^- \to P^0
\gamma$ would allow discovery so long as the PNGB mass was not close to $M_Z$
and for $N_{T} \geq 3$.  Moreover, if running in $\gamma\gamma$ mode, the
process $\gamma\gamma \to P^0 \to b\bar{b}$ could not only allow detection of
a PNGB with mass between 10\% and 70\% of $\sqrt{s}$, 
but would also provide more precise information about
model parameters.  
A muon collider could in principle help determine the
couplings of a PNGB in greater detail.
\begin{figure}[tb]
\vspace{5cm}
\begin{center}
\includegraphics{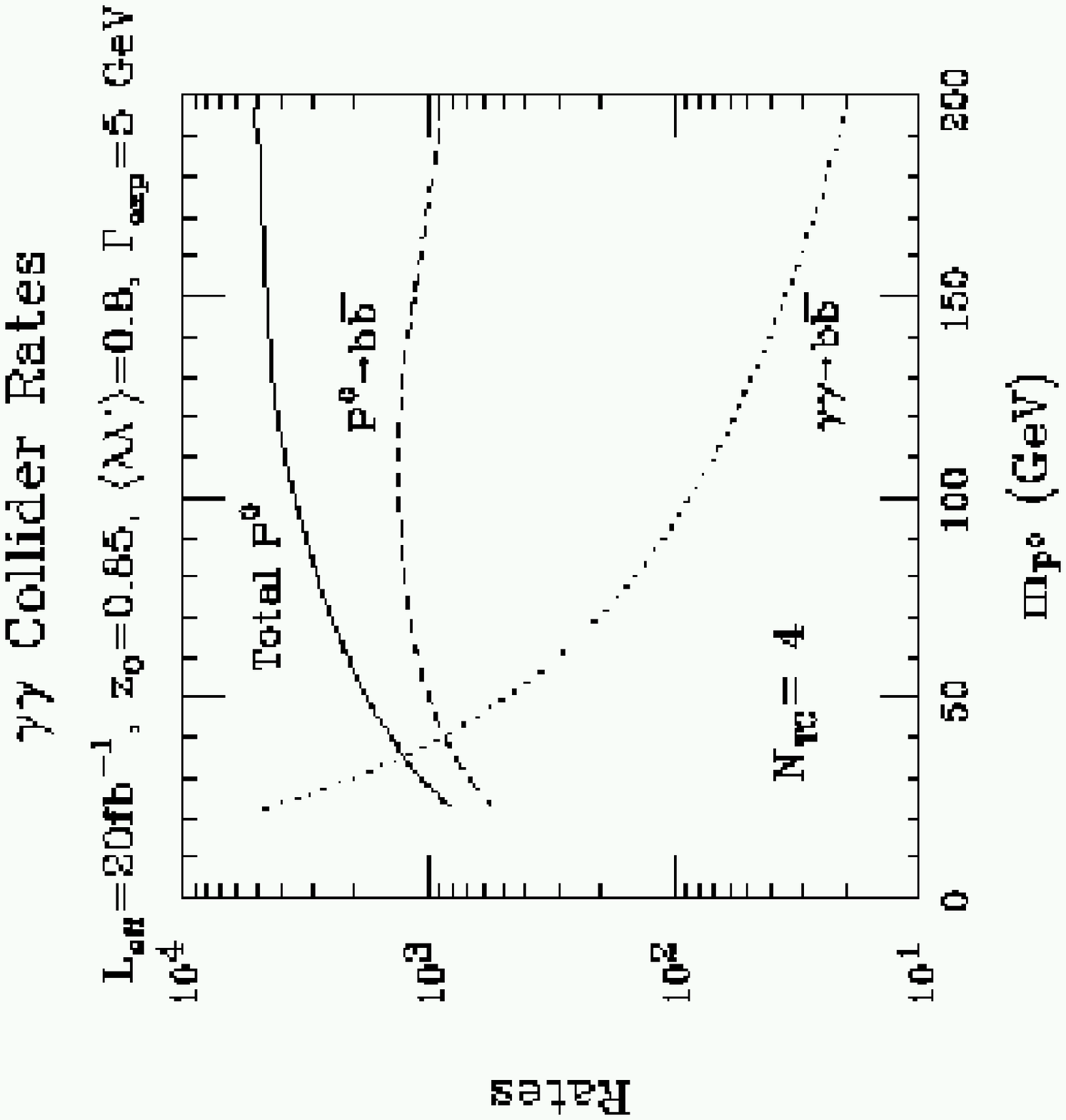}
\end{center}
\vspace{3cm}
\caption[cap29b]{\small \addtolength{\baselineskip}{-.4\baselineskip} Projected  $P^0$ production rate (solid), rate 
including PNGB decay
  to $b\bar{b}$ (dashed), and irreducible $\gamma\gamma \to b\bar{b}$
  background (dotted) for
  $\gamma\gamma$ collisions at $\sqrt{s} = 500$ GeV with integrated
  luminosity of 20 $fb^{-1}$ \protect\cite{Casalbuoni:1998fs}.}
\label{fig:casal-lept}
\end{figure}

The production and detection of
the $\eta_T'$ state has been studied in \cite{DiVecchia:1980xq} and
\cite{Tandean:1995ci}, the latter with emphasis on the laser
back-scattering technique at an $e^+e^-$ LC
\cite{Ginzburg:1983vm}.

\vskip .1in
\noindent{\bf (iii) colorless charged PNGBs, $P^\pm$} 
\vskip .1in

Electrically charged PNGB's would be abundantly pair-produced at a
high-energy, high-luminosity linear electron-positron collider.  In
terms of the electromagnetic point cross section $\sigma_o = 4 \pi
\alpha^2(s)/ 3 s$, the continuum cross-section for producing $P_0^+ P_0^-$
pairs at an NLC is $0.30\beta^3
\sigma_o$, where $\beta$ is the c.m. frame velocity of $P_0^\pm$
\cite{Swartz:1996}.  Assuming the $P_0^\pm$ decay dominantly to $tb$ pairs,
the final state contains a pair of on-shell W bosons and four b quarks.
The leading background comes from $t\bar{t} g^*$ events where the off-shell
gluon materializes as a $b\bar{b}$ pair; the cross-section for this process
is less than a tenth of the $t\bar{t}$ production cross-section of
1.9$\sigma_o$.  Given a detector with good angular coverage and high
efficiency for detection of the b-jets, it should be possible to detect
$P^\pm_0$ with a mass below about 0.45$\sqrt{s}$ at the five-sigma
significance level \cite{Swartz:1996}.  Similarly, the large
electromagnetic pair-production cross-section and moderate backgrounds for
pair-production of $P^{\pm}_{T 8}$ should make 5-sigma significance
discovery possible out to a mass of approximately 0.49 $\sqrt{s}$
\cite{Swartz:1996}.

As seen from the discussion of section 2.2.4, the $\rho_T^0$ will
couple via VMD to any electromagnetic current (or
to the weak neutral current, which can include $\omega_T$).  
VMD production on the narrow resonances will produce dramatic
signatures in the conventional final states as in $e^+e^- \rightarrow
(\rho_T,\omega_T) \rightarrow (WW+ZZ, W\pi+ZP,
PP)$. Such processes have been discussed
in the context of muon colliders, where the
better beam energy resolution is an advantage,
 in ref.\cite{Eichten:1998kn}.

In contrast, the work of ref \cite{Wang:1998us} suggests that it will be
difficult to detect the effects of charged technipions on single top
production at an $e\gamma$ collider in the process $e\gamma \to t b \nu$.

\newpage
\section{Extended Technicolor}

%
%
%

\subsection{The General Structure of ETC}

A realistic Technicolor model must address the flavor problem.  That is, it
must incorporate a mechanism for generating quark and lepton masses and the
various weak mixing angles, together with CP-violation.  This implies that
the ordinary quarks and leptons of the Standard Model need to couple to the
technifermion condensate that breaks the electroweak symmetry.  There must
also be a mechanism for violating the conserved techni-baryon quantum
number: techniquarks must be able to decay, since stable techni-baryonic
states are cosmologically problematic \cite{Chivukula:1990qb}.  

The classic way to fulfill
these requirements is to extend the Technicolor gauge interactions to
include additional gauge bosons coupling both to ordinary and Technicolored
fermions.  The extended interactions are part of a large gauge group which
necessarily breaks down to its Technicolor subgroup at an energy above the
scale at which the Technicolor coupling becomes strong,
$\Lambda_{TC}$.

The construction of the first theories of ``Extended Technicolor'' (ETC) is
due to Eichten \& Lane \cite{Eichten:1979ah} and Dimopoulos \& Susskind
\cite{Dimopoulos:1979es}.  Their work, and the papers which followed in the
early 1980's, especially
\cite{Ellis:1981hz}, \cite{Dimopoulos:1982fj}, \cite{Dimopoulos:1981kp}
\cite{Dimopoulos:1981yf}, \cite{Peskin:1980gc}, \cite{Sikivie:1980hm},
flesh out the general scheme and identify the challenges.  We will
summarize the major issues, indicate trends in modern model-building, and
refer the reader to the literature for further detail.

\subsubsection{Master Gauge Group ${\cal{G}}_{ETC}$}

The kinds of new interactions that are required for ETC involve transition
couplings of technifermions $Q_{L,R}$ into the ordinary quarks and leptons,
$\psi_{L,R}$.  Currents of the form $\bar{Q}_{L,R}\gamma_\mu \psi_{L,R}$
are therefore required, coupling to the new ETC gauge bosons.  In a full
theory we assume a large gauge group ${\cal{G}}_{ETC}$ which contains all
of the desired currents of the form $\bar{Q}Q$, $\bar{Q}\psi$ and
$\bar{\psi}\psi$.  A simple generalization is a blockwise imbedding of
$SU(N_{TC})$ into $SU(N_{ETC})$ with $N_{ETC}> N_{TC}$.  Clearly $N_{ETC}$
must be large enough to accomodate representations containing both $\psi$'s
and $Q$'s.

For example, consider how this imbedding might work in the
minimal model.  The minimal model includes one weak
doublet of technifermions ($N_D$ = 1) and a typical choice of Technicolor
group $N_{TC}=4$.  ETC must couple the technifermions to all $12$
left-handed electroweak doublets of ordinary quarks and leptons and all
$12$ singlets.  Then we should think of the ordinary fermions of each
generation as falling into 3 doublets of quarks and one of leptons under
the gauged $SU(2)_L$ 
and an analogous set of doublets under a global $SU(2)_R$.  This yields
an ETC scheme with a gauge group ${\cal{G}}_{ETC}
= SU(16)\times SU(2)_L\times
SU(2)_R \times U(1)$. Both $SU(N_{TC})$ and $SU(3)_c$ are imbedded
into 
$SU(16)$; 
hypercharge arises as $Y/2 = I_{3R} + (B-L)/2$,
where $B-L$ is a diagonal generator in $SU(16)$.  The
fermions form two fundamental { \bf{16}} multiplets; one of them:
\begin{eqnarray}
&\ &(Q_c, Q_k, Q_m, Q_y, \psi_r^1, \psi_g^1, \psi_b^1, \psi_r^2, \psi_g^2,
\psi_b^2, \psi_r^3, \psi_g^3, \psi_b^3, \psi_{lep}^1, \psi_{lep}^2,
\psi_{lep}^3)_L \\
&\ &{\rm c, k, m, y = Technicolors; \ \ \ r, g, b = colors} \nonumber
\end{eqnarray}
is a doublet under the electroweak group $ SU(2)_L$, and a singlet under $
SU(2)_R$, a $(2,1)$; the other is a $(1,2)$.  The Technicolor condensate
will therefore break both the $SU(2)_L$ and the $SU(2)_R$. 
For the
Farhi-Susskind model, an analogous ETC extension might be to a gigantic
gauge group placing all of the quarks and leptons and techniquarks
(including each Technicolor copy) into a single multiplet of a compact
gauge group, e.g., $SU(56)$.  Here the full Standard Model gauge
interactions themselves would be subgroups of the master group $SU(56)$.
Clearly a large variety of models are possible, and we will describe
several general approaches to building consistent models in this Section
(several variant schemes are also reviewed in \cite{King:1995yr}).

Starting from a high-energy theory based on a master gauge group
${\cal{G}}_{ETC}$, it is necessary to arrive at a low-energy theory in which
the only surviving gauge groups are those of Technicolor and the Standard
Model.  The group ${\cal{G}}_{ETC}$ has generators, $T^a$, which
form a Lie Group,
\begin{equation}
[T^a, T^b] = if^{abc}T^c
\end{equation}
The Technicolor gauge group ${\cal{G}}_{TC}$ 
must be a subgroup of ${\cal{G}}_{ETC}$,
since the $\psi$'s do not carry Technicolor, while the $Q$'s do.  Then 
${\cal{G}}_{ETC}$ must undergo symmetry breaking at a
scale $\Lambda_{ETC}$ down to its subgroup ${\cal{G}}_{TC}$:
\begin{equation}
{\cal{G}}_{ETC} \rightarrow {\cal{G}}_{TC}\times ...  \qquad \makebox{at $\Lambda_{ETC}$}.
\end{equation}
where the ellipsis denotes other factor groups, including 
perhaps the full Standard Model
$SU(3)\times SU(2)\times U(1)$.
This leaves the Technicolor gauge bosons (generators denoted $\tilde{T}^a$)
massless, and elevates all of the coset ETC gauge bosons to masses of order
$\Lambda_{ETC}$.  Indeed, the breaking may proceed in a more complicated
way, e.g., it may occur in $n$ steps ${\cal{G}}_{ETC} \rightarrow
{\cal{G}}_{1} \rightarrow {\cal{G}}_{2} \rightarrow ...  {\cal{G}}_{n-1}
\rightarrow {\cal{G}}_{TC} $. A class of models in which this
symmetry-breaking sequence is achieved dynamically is known as ``Tumbling
Gauge Theories'' \cite{Raby:1980my, Georgi:1981mh, Dimopoulos:1983gc,
Veneziano:1981yz, Eichten:1982mu, Kobayashi:1986fz, Martin:1992aq}.
Typically, at each step, the subgroup will evolve and become strong,
permitting new condensates to form, which further break the theory into the
next subgroup. This occurs repetitively at different scales, and may in
principle be a way of generating the mass hierarchy for quarks and leptons
of different flavors.  Another possibility is that there are fundamental
scalars associated with supersymmetry at some high energy scale, and the
breaking of ${\cal{G}}_{ETC}$ (whether single or sequential) can be driven by
a Higgs mechanism (see Section 3.7).  

\subsubsection{Low Energy Relic Interactions}

While the only elements of the original ETC gauge group that survive at low
energies are the generators of the Technicolor and Standard Model gauge
groups, the low-energy phenomenology of these models includes additional
effects caused by the broken ETC generators.  On energy scales $\mu
\lta \Lambda_{ETC}$ exchange of the heavy  ETC bosons corresponding to
those broken generators produces three types of effective contact
interactions among the ordinary and technifermions:
\begin{equation}
\label{contact1}
\bar\alpha_{ab}\frac{\bar{Q}\gamma_\mu \bar{T}^aQ \bar{Q}\gamma^\mu \bar{T}^bQ}{\Lambda_{ETC}^2} 
+ 
\bar\beta_{ab}\frac{\bar{Q}\gamma_\mu \bar{T}^a\psi \bar{\psi}\gamma^\mu
\bar{T}^bQ}{\Lambda_{ETC}^2}
+ 
\bar\gamma_{ab}\frac{\bar{\psi}\gamma_\mu \bar{T}^a\psi \bar{\psi}\gamma^\mu \bar{T}^b\psi}{\Lambda_{ETC}^2}
\end{equation}
Here the $\alpha$, $\beta$ and $\gamma$ are coefficents that are
contracted with generator indices and their
structure depends upon the details of the parent
ETC theory.  In the $\bar{T}^a$ we include chiral factors such as
$(1\pm\gamma^5)/2$ (i.e., the theory is flavor-chiral and the ETC
generators have different actions on left- and right-handed fermions).
We can now Fierz rearrange these operators to bring them into
the form of products of scalar and pseudo--scalar densities.
Upon Fierz rearrangement, we can pick out the generic terms of
greatest phenomenological relevance:
\begin{equation}
\label{contact2}
\alpha_{ab} \frac{\bar{Q}T^aQ \bar{Q}T^b Q }{\Lambda_{ETC}^2} 
+ 
\beta_{ab} \frac{\bar{Q}_LT^aQ_R \bar{\psi}_RT^b \psi_L}{\Lambda_{ETC}^2}
+ 
\gamma_{ab} \frac{\bar{\psi}_LT^a\psi_R \bar{\psi}_RT^b \psi_L}{\Lambda_{ETC}^2}
+ ...
\end{equation}
Note that after Fierzing we must include the identity matrix among the
generators, which we do by extending the range of the generator indices to
include zero: $\bf{1} \equiv T^0$.

\begin{figure}[t]
\vspace{4cm}
\includegraphics{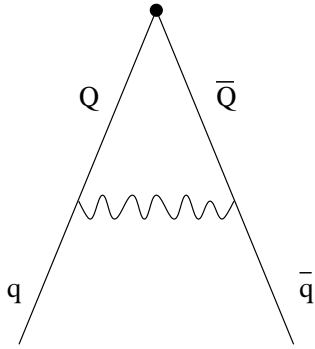}
\vspace{1cm}
\caption[]{\small \addtolength{\baselineskip}{-.4\baselineskip}
The exchange of an ETC gauge boson allows
a standard quark (or lepton) to communicate 
with the techniquark condensate, $\VEV{\overline{Q}Q}$.}
\label{fig:cth:etc-beta}
\end{figure}

As a consequence of these generic terms we see that the physical
effects of ETC go beyond generating the quark and lepton masses and
mixings.  On the positive side, ETC interactions, such as the
$\alpha$-terms, can elevate the masses of some of the light
Nambu-Goldstone bosons, such as the techniaxions, to finite nonzero
values, essential to rendering models consistent with experiment.  On
the negative side, Extended Technicolor produces the $\gamma$-terms,
four-fermion contact interactions amongst ordinary fermions of the
same Standard Model gauge charges.  This leads generally to
flavor-changing neutral current effects in low energy hadronic
systems, e.g., it can lead to dangerously large contributions to the
$K_L K_S$ mass difference.  Lepton number violation will also
generally occur, leading to enhanced rates for $\mu\rightarrow
e+\gamma$ and related processes.  It is a difficult model-building
challenge to limit these dangerous effects while generating adequately
large quark and lepton masses to accomodate the heavier fermions,
e.g., charm, $\tau$, $b$, and especially $t$.  It is, moreover,
important that the oblique electroweak corrections involving TC and  ETC
effects be under control \cite{Georgi:1989pt}, as will be discussed in
Section 3.2.  These and other\footnote{There has recently been a great
deal of interest in the possibility of a small departure of the
observed $g-2$ of the muon from the Standard Model expectation
\cite{Brown:2001mg,Onderwater:2001ie}.  Some authors claim that it is difficult
for TC (or Topcolor) to yield comparable effects \cite{Xiong:2001rt},
while others claim that it is not \cite{Yue:2001db}.} phenomenological
considerations have had a large influence on ETC model-building.  We
will look at explicit examples of models constructed to address such
questions in Section 3.3.

\subsubsection{The $\alpha$-terms: Techniaxion Masses.}

\begin{figure}[t]
\label{etcpngbfig}
\vspace{4cm}
\includegraphics{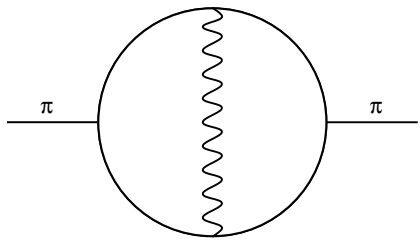}
\vspace{0cm}
\caption[]{\small \addtolength{\baselineskip}{-.4\baselineskip}
ETC gauge boson exchange across a fermion loop.
This is an $\alpha$ term insertion leading to PNGB mass.}
\end{figure}

The four-technifermion terms (with coefficients $\alpha$) can potentially
solve a problem we encountered in the previous discussion of the minimal and
Farhi-Susskind models in Section 2.  Loops involving $\alpha$ term
insertions, which represent ETC gauge boson exchange across a technifermion
loop, with external PNGB's (Fig.(\ref{etcpngbfig})) generally induce masses
for the PNGB's.  This mechanism can elevate the masses of the undesireably
light PNGB's to larger values more consistent with experiment.

To see this explicitly, we can make use of an NJL or Georgi-Manohar model
of the techniquark condensate, which instructs us to replace the
techniquark bilinears in eq.(\ref{contact2}) with the corresponding chiral
fields (a discussion of the chiral dynamics of ETC may be found in
\cite{Chadha:1981rw}, \cite{DiVecchia:1980wy} ).  Specifically one makes
the replacement, motivated from the NJL approximation
or QCD:  
\begin{equation}
\bar{Q}_{aR}Q^{\dot{b}}_{L} \rightarrow 
cN_{TC}\Lambda_{TC}^3\Sigma_a^{\dot{b}}
\qquad\qquad
\Sigma_a^{\dot{b}} \equiv \exp(i\pi^c \tilde{T}^c/F_T)_a^{\dot{b}}
\label{eqn:cth:bos}
\end{equation}
where dotted indices are for $SU(2)_L$ and undotted are for $SU(2)_R$ and
the $ \tilde{T}^a$ are generators of the Technicolor subgroup of
${\cal{G}}_{ETC}$.  This leads, through the diagram of Fig.(20),
 to a PNGB interaction
term in the effective Lagrangian:
\begin{equation}
\sim \frac{\alpha_{ab}c^2 N_{TC}^2\Lambda_{TC}^6}{\Lambda_{ETC}^2} 
\Tr ( \Sigma T^a \Sigma^\dagger T^b )
\label{eqn:cth:a-leff}
\end{equation}
Expanding eq.(\ref{eqn:cth:a-leff}) in the $\pi^a\tilde{T^a}$ we see that
the induced technipion mass terms are:
\begin{equation}
\sim -\frac{\alpha_{ab}c^2 N_{TC}^2\Lambda_{TC}^6}{\Lambda_{ETC}^2F_T^2} 
\Tr ( [\pi^c \tilde{T}^c, T^a][T^b,\pi^d \tilde{T}^d] 
\end{equation}
Those technipions associated with the Technicolor generators,
$\tilde{T}^a$ that commute with ETC, i.e., for which:
\begin{equation}
[\tilde{T^a}, T^b] = 0\ ,
\end{equation}
will have vanishing mass contributions from ETC.  Technipions associated
with the noncommuting generators, on the other hand, will generally receive
nonzero ETC contributions to their masses of order 
$\sim N_{TC}\Lambda^2_{TC}/\Lambda_{ETC}$.

This situation is akin to that of QCD: the ${T^a}$ are analogues of the
electric charge operator, and $\tilde{T^a}$ are the generators of nuclear
isospin. The neutral pion of QCD is associated with a generator, $I_3$ that
commutes with the electric charge, $Q = I_3 + Y/2$, hence the neutral pion
receives no contribution to its mass from electromagnetism.  The charged
pions, associated with generators $I_1 \pm iI_2$, which do not commute with
$Q$, receive nonzero electromagnetic contributions to their masses
\cite{Chadha:1981yt}.

Among the PNGB's that arise in the Minimal and the Farhi-Susskind models,
the techniaxions can receive masses, at best, {\em only} from ETC.  In
classic ETC the prospects for mass generation have been treated in detail in
\cite{Dimopoulos:1981yf}; a typical result is: 
\begin{equation}
M_{axion}^2 \approx \frac{1}{N_{TC}} ( \makebox{a few GeV})^2\ .
\end{equation}
We will see
subsequently that electrically charged PNGB's 
receive masses from ETC that are small
compared to their electromagnetic contributions.  Similarly the colored
states, $P_3$ and $P_8$ of the Farhi-Susskind model receive larger masses
from QCD than from conventional ETC.  

We will discuss below, however, an
additional dynamical ingredient of modern ETC theories known as
``Walking.''  In Walking Technicolor, any effective operator involving the
techniquarks can be significantly amplified by the effects of Technicolor.
This tends to enhance the PNGB masses by factors of order
$\Lambda_{ETC}/\Lambda_{TC}
\sim 10^3$. This in principle remedies the problem of elevating
the masses of the techniaxions to evade experimental bounds.

\subsubsection{The $\beta$ terms: Quark and Lepton Masses}

The terms in eq.(\ref{contact2}), with coefficients $\beta_{ab}$, will
generally give masses and mixing angles to the ordinary quarks and
leptons. Technicolor condenses the technifermions, 
$\VEV{\bar{Q}{Q}} \sim {N_{TC}}\Lambda_{TC}^3$ at the
TC scale.  The natural scale
for the ETC-induced quark and lepton masses is then of order:
\begin{equation}
\label{etcmass1}
m_{q,\ell} \sim \beta \frac{{N_{TC}}\Lambda_{TC}^3}{\Lambda_{ETC}^2}
\end{equation}
This would seem to liberally allow a fit of
ordinary quark or lepton masses to a scale of order
$\sim \Lambda_{TC}$.  For example, with $m_{q} \sim m_{charm} \sim
1 $ GeV and $\Lambda_{TC}\sim 100$ GeV, we have for $\beta \sim 1$,
the result,
$\Lambda_{ETC}\lta $ (few)$\times 10^3 $ GeV.  Thus, the ordinary quark and
lepton masses place an {\em upper bound} on the effective value of
$\Lambda_{ETC}$.  The pattern and scale of masses and mixing angles is
sensitive to the pattern of breaking of ETC.  It is not unreasonable to
expect to obtain hierarchical mass patterns between generations.

Higher-order effects of the $\beta$ terms also yield observable
consequences.  The key example for ETC model building is $R_b$.  In a
classic ETC model, $m_t$ is generated by the exchange of an
electroweak-singlet ETC gauge boson of mass $M_{ETC}\sim g_{ETC}\Lambda_{ETC}$ 
coupling with
strength $g_{ETC}$.  At energies below $M_{ETC}$, ETC gauge boson exchange
may be approximated by local four-fermion operators, and $m_t$ arises from
an operator coupling the left- and right-handed currents:
\begin{equation}
\label{mass111}
   - {g_{ETC}^2 \over  M_{ETC}^2}  \left({\bar\psi}_L^i \gamma^\mu
T_L^{iw}\right) \left( {\bar U^w}_R \gamma_\mu t_R \right) + {\rm h.c.}\ 
\label{eqn:ehs:topff}
\end{equation}
where $T=(U,D)$ are technifermions,
and $i$ and $k$ are weak and Technicolor
indices. Assuming there is only a single weak doublet of
technifermions:
\begin{equation}
\label{labelcs1}
   m_t\ = {g_{ETC}^2 \over M_{ETC}^2}
   \langle{\bar U}U\rangle\ \approx\ {\Lambda_{TC}^3 \over \Lambda_{ETC}^2} \ .
\label{eqn:ehs:topmass}
\end{equation}

\begin{figure}[tb]
\vspace{-2.0cm}
\includegraphics{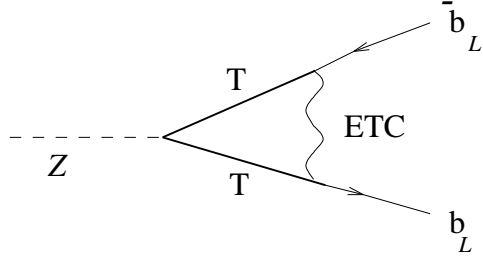}
\vspace{5.5cm}
\caption{\small \addtolength{\baselineskip}{-.4\baselineskip}
 Direct correction to the $Zb\bar b$ vertex
from exchange of the ETC gauge boson that gives rise to the top quark
mass.  Technifermions are denoted by `T'. }
\label{fig:ehs:zbb-diag}
\end{figure}

The ETC boson which produces $m_t$ through
eq.(\ref{mass111}) also necessarily induces the related operator \cite{Chivukula:1992ap}:
\begin{equation}
\label{zbb111}
   - {g_{ETC}^2 \over  M_{ETC}^2}  \left({\bar\psi}_L^i \gamma^\mu
T_L^{iw}\right)\left({\bar T}_L^i \gamma^\mu
\psi_L^{iw}\right)  + {\rm h.c.}\ 
\label{eqn:ehs:topffd}
\end{equation}
This affects the $Zbb$ vertex as shown in Figure \ref{fig:ehs:zbb-diag},
which indicates the ETC boson being exchanged between the left-handed
fermion currents (with 
$T \equiv D_L$ since the ETC boson is a
weak singlet).
This process alters 
the $Z$-boson's tree-level coupling to left-handed bottom quarks 
$g_L =
(-\frac{1}{2} + {1\over 3}\sin^2\theta_W)(e/\sin\theta_W \cos\theta_W)$ 
by \cite{Chivukula:1992ap}:
\begin{equation}
\label{labelcs2}
\delta g_L^{ETC}\ = \ -{\xi^2 \over 2} {\Lambda_{TC}\over \Lambda_{ETC}^2} 
{e\over{\sin\theta \cos\theta}}(I_3)
\ =\ {1\over 4} {\xi^2} {m_t\over{\Lambda_{TC}}}
\cdot {e\over{\sin\theta \cos\theta}} \label{tb}
\end{equation}
where the right-most expression follows from applying eq. 
(\ref{eqn:ehs:topmass}). Here $\xi $
is a mixing parameter angle between the $W$ and $Z$ 
and the ETC bosons of the theory.

Such shifts in $g_L$ alter the ratio of $Z$ boson decay widths
\begin{equation}
R_b \equiv {\Gamma(Z\to b\bar b) \over {\Gamma(Z \to {\rm hadrons})}}.
\end{equation}
This is a convenient observable to work with since oblique
\cite{Lynn:1985fg} \cite{Golden:1991ig}
\cite{Holdom:1990tc} \cite{Peskin:1990zt}
\cite{Dobado:1991zh} \cite{Peskin:1992sw} 
and QCD corrections largely cancel in
the ratio, making it sensitive to the direct vertex correction.
One finds:
\begin{equation}
{\delta R_b \over R_b} \approx -5.1\% \xi^2 \left({m_t
\over 175{\rm GeV}} \right).
\label{eqn:ehs:rbb}
\end{equation}
Such a large shift in $R_b$ would be readily detectable in current
electroweak data.  In fact, the experimental value of $R_b = 0.2179\pm
0.0012$ lies close enough to the Standard Model prediction
\cite{Langacker:1994ah} \cite{Blondel:1994ke} (.2158) that a 5\% reduction
in $R_b$ is excluded at better than the 10$\sigma$ level.  Alternatively we
require a suppression of $\xi \lta 0.2$ at the 2$\sigma$ level.

It should be noted that there are discrepancies in the LEP data
with the Standard Model, in particular
in the Forward-Backward asymmetry of the $b$-quark \cite{Chanowitz:2001bv},
though, from the present discussion,
it is unclear that they can be reconciled with
a Technicolor-like dynamics.
ETC models in which the ETC and weak gauge groups commute are 
evidently ruled out.  However, models which can separate third generation and
first and second generation dynamics, such as non-commuting ETC (Section
3.3.2), can produce acceptable $R_b$ corrections.  Low-Scale Technicolor
(Section 3.5) and Topcolor or Topcolor-Assisted Technicolor models (Section
4) generally also predict values of $R_b$ in accord with the data 
\cite{Yue:2000ay, Burdman:1997pf}.

\subsubsection{The $\gamma$ terms: Flavor-Changing Neutral Currents}

Severe constraints from flavor-changing neutral currents turn out to
exclude the possibility of generating large fermion masses in classic ETC
models.  The interactions of the third class of terms in
eq.(\ref{contact2}), associated with coefficients $\gamma_{ab}$, cause
these problems.  Because ETC must couple differently to fermions of
identical Standard Model gauge charges (e.g. $e, \mu, \tau$) in order to
provide the observed range of fermion masses, flavor-changing neutral
current interactions amongst quarks and leptons generally result. Processes
like:
\begin{equation}
\label{etcmass2}
\frac{(\bar{s}\gamma^5 d)\, (\bar{s}\gamma^5 d)}{\Lambda_{ETC}^2}
+
\frac{(\bar{\mu}\gamma^5 e)\, (\bar{e}\gamma^5 e)}{\Lambda_{ETC}^2}
+ ...
\end{equation}
are generally induced, and these give new contributions to experimentally
well-constrained quantities \cite{Buras:1983ff}.  For example, the first
term causes $\Delta S = 2$ flavor-changing neutral current interactions
which give a contribution to the experimentally well-measured $K_L K_S$
mass difference. The matrix element of the operator between $K^0$ and
$\bar{K}^0$ yields \cite{Eichten:1979ah}: 
\begin{equation}
\label{bound1}
\delta m^2/m_K^2 \sim \gamma \frac{f_K^2 m_K^2}{\Lambda_{ETC}^2} \lta 10^{-14} 
\end{equation}
where we might expect $\gamma \sim \sin^2\theta_c \sim 10^{-2}$ in any
realistic model. Hence, we obtain:
\begin{equation}
\Lambda_{ETC} \gta 10^3 \; \makebox{TeV}
\label{eqn:cth:strlim}
\end{equation}
The second term of eq.(\ref{etcmass2}) induces the lepton-flavor-changing
process $\mu \rightarrow e\bar{e}e,; e\gamma $; from this, we estimate a
somewhat weaker bound, $\Lambda_{ETC} \gta 10^1 $ TeV \cite{Eichten:1979ah}. 

Applying the bound eq.(\ref{bound1}) to our expression for
ETC-generated fermion masses (\ref{etcmass1}), and assuming
$\alpha \sim \beta \sim \gamma$, 
yields an upper bound on
the masses of ordinary quarks and leptons that a generic ETC model can
produce (we use $\Lambda_{TC}\lta 1 $ TeV, $\beta \lta 10$,
$N_{TC}\lta 10$):
\begin{equation}
m_{q,\ell} \lta N_{TC}\frac{\Lambda_{TC}^3}{\Lambda_{ETC}^2} \lta 100 \; \makebox{MeV}
\end{equation}
Hence, producing the mass of the charm quark is already
problematic for a classic ETC model.  

A remedy for production of the charm and, marginally, 
the $b$-quark masses is ``Walking
Technicolor,'' as we discuss below (Section 3.4).  We remark that another
way to suppress the $\gamma_{ab}$ effects is to construct theories based
upon $SO(N)$ groups containing ETC and Technicolor, with fermions in the
real $\bf{N}$ representation. These can avoid flavor-changing neutral
current-like interactions in the tree approximation \cite{Giudice:1992sz}
\cite{Raby:1990pq}, a consequence of the representations that can be
generated by products of the fundamental $\bf{N}$ representation at that
order.  An operator such as eq.(\ref{etcmass2}) will be generated in loops,
and one still obtains a limit of order $1$ GeV on the allowed fermion
masses.  Ultimately addressing the heavy top quark requires new dynamical
mechanisms such as non-commuting ETC (Section 3.3.2) or ``Topcolor'' (section
4.2).

These $\gamma$-term problems, it should be noted, have analogues in the
Minimal Supersymmetric Standard Model, under the rubric of 
``The SUSY
Flavor Problem.'' The mass matrices of
the squarks and sleptons may in general require diagonalization by
different flavor rotations than those of the quarks and leptons. This leads
to similar unwanted FCNC and lepton number violating processes.  The
typical constraints upon mass differences between first and second
generations sfermions are $\delta m^2/M^2 \lta 10^{-3}$, where $M^2$ is the
center-of-mass of a multiplet.  The need to keep $M \sim 1 $ TeV in SUSY
places constraints on the models that may be hard to realize. It is
intriguing that both SUSY models, and Technicolor models focus our
attention upon nagging flavor physics issues at the multi-TeV scale.

\subsection{Oblique Radiative Corrections}

Precision electroweak measurements have matured considerably throughout the
LEP and Tevatron era, with the copious data on the $Z$--pole, the discovery
of the top quark, and precise $W$ mass measurements.  Taken together with
low energy data such as atomic parity violation and neutrino scattering,
these experiments test the Standard Model at the level of multi-loop
radiative corrections.  The resulting constraints are often troublesome for
classic Extended Technicolor.  Yet these constraints can generally be
avoided or ameliorated in the presence of Walking ETC (Section 3.4) or new
strong top quark dynamics (section 4).

The conventional parlance for discussing the ``oblique'' radiative
corrections to the Standard Model is to use the $S$, $T$, and $U$
parameters.  For definitions of these parameters, we follow the formalism
of Peskin and Takeuchi \cite{Peskin:1990zt}
\cite{Peskin:1992sw}. The $S$, $T$, and $U$ parameters are defined
and computed for a fermion bubble in Appendix A (see eqs.(\ref{vstu})).

\begin{figure}[t]
\vspace{7cm}
\includegraphics{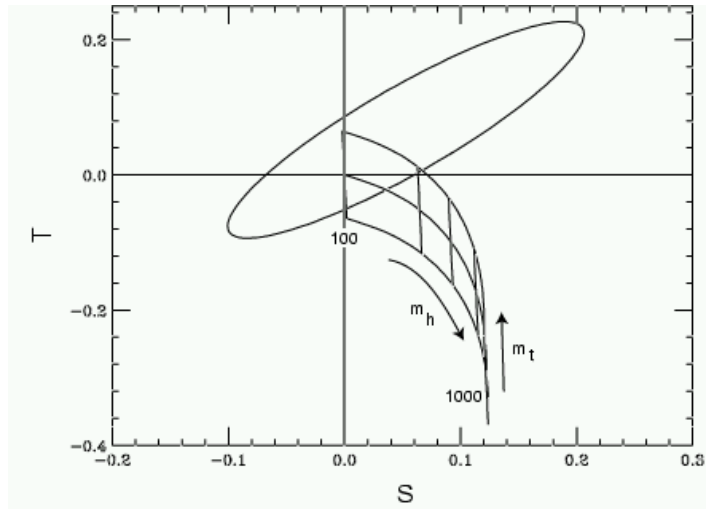}
\vspace{2cm}
\caption[]{\small \addtolength{\baselineskip}{-.4\baselineskip}
Fit of the precision electroweak data to the Standard
Model plus the $S$, $T$ parameters described in the text (from Peskin and
Wells \cite{Peskin:2001rw}). The ellipse is the fit; the warped grid is the
Standard Model predictions. The figure uses only the values of the three
best electroweak observables, $M_W$, the value of the weak mixing angle
which appears in $Z^0$ decay asymmetries, and $\Gamma_\ell$ (the leptonic
width of the $Z^0$. All data is combined in a plot of M. Swartz at the 1999
Lepton-Photon Conference \cite{Swartz:1999xv}).  }
\label{STfit}
\end{figure}

Precision electroweak observables are linear functions of $S$ and $T$.
Thus, each measurement picks out an allowed band in the $S-T$ plane, and
measurement of several processes restricts one to a bounded region in this
plane, the $S-T$ error ellipse.  By convention, offsets are added to $S$
and $T$ so the point $S = T = 0$ corresponds to the prediction of the
Standard Model for a set of fixed ``reference values'' of the top quark and
Higgs boson masses.  A recent review by Peskin and Wells
\cite{Peskin:2001rw} addresses general issues relevant to theories beyond
the Standard Model, including Technicolor, new strong top quark dynamics,
and extra-dimensions.  Peskin and Wells take the reference values to be
$m_t = 174.3$ GeV and $m_H = 100$ GeV. In Fig.~\ref{STfit}, we present the
68\% confidence contour (1.51 $\sigma$) for a current $S-T$ fit from
Peskin and Wells \cite{Peskin:2001rw}.  The overlay of the perturbative
Standard Model prediction with $m_t = 174.3 \pm 5.1$ GeV and $m_H$ running
from 100 GeV to 1000 GeV, with $m_H = 200, \;300, \; 500$ GeV is indicated
with vertical bands in the warped grid in the Figure.

While Figure \ref{STfit} favors a low mass Higgs boson, $M_h \lta 200$
GeV, we emphasize that this {\em does not imply that a low mass Higgs boson
is compelled to exist}.  The warped grid is only a {\em perturbative}
Higgs boson mass plot.  If the Higgs boson is indeed heavy, then there must
be associated new strong dynamics, and there will occur additional new
physics contributing to $S$ and $T$. One cannot, therefore, rely upon the
perturbative mass plot overlay for large Higgs boson masses.  An example of
a model in which the Higgs boson mass is $\sim 1$ TeV, but is naturally
consistent with the $S$-$T$ error ellipse is the Top Quark Seesaw theory
\cite{Dobrescu:1998nm} \cite{Chivukula:1998wd}, described in 
Section 4.4.

In general, however, one may still use the $S-T$ error ellipse itself
to assess the agreement of a given theory with the data.  
 The contributions to $S$ from classic Technicolor are large
and positive.  As seen in Appendix A, the contribution of a single
weak isodoublet with $N_T$ Technicolors is, in a QCD-like dynamics:
\begin{equation}
\Delta S \approx   \frac{N_T}{6\pi}
\end{equation}
since the techniquarks are massless and have degenerate constituent quark
masses. Thus, with $N_T=4$ the minimal model with a single weak isodoublet
will give a correction to the $S=T=0$ reference point of $\Delta S = +0.2$
and corresponds roughly to the case of a $\sim 1$ TeV Higgs boson mass with
$\Delta T \sim -0.3$.  The Farhi-Susskind model has a full generation of
$3$ colors each of techniquarks and a ``fourth color'' of leptons, and thus
yields $\Delta S \sim 0.8$ with $\Delta T \sim -0.3$, and would appear to
be ruled out.  In the minimal model we might be able to pull ourselves from
the point $(0.2, -0.3)$ back into the experimentally favored ellipse by
adding a positive contribution to $\Delta T \sim 0.4$ from additional new
physics.  To return to the ellipse in a Farhi--Susskind--like model would
require large negative sources of $S$ as well. 

There are possible physical sources of the additional contributions to $S$
and $T$ which are required to make Technicolor consistent with the measured
oblique parameters.  First, strong dynamics models that address the flavor
problem must usually include new higher-dimension operators from physics at
higher scales that will affect the oblique parameters; a general analysis
has been made by a number of authors \cite{Golden:1991ig}, 
\cite{Chivukula:1992nw}, \cite{Chivukula:1991bx} \cite{Chivukula:1999az}.
The possible existence of such operators obviates any claim that new
dynamical models are ruled out by the $S$-$T$ constraints.  Second, the
presence of precocious ($\sim$ TeV scale) extra dimensions can play an
important role, as recently discussed by Hall and Kolda,
\cite{Hall:1999fe}.  Third, Appelquist and Sannino have shown that 
the ordering pattern for vector-axial hadronic states in $SU(N)$ 
vector-like gauge theories close to a conformal transition 
need not be the same as predicted in QCD-like theories.  
As a consequence, the $S$-parameter in near-critical 
technicolor theories can be greatly reduced relative 
to QCD-like theories
\cite{Appelquist:1998xf}.
Fourth, specific particle content that can produce overall modest
contributions to $S$ and $T$ has been identified.  We will mention
some of the proposed models shortly.

The contribution to the $S$
parameter from a scalar field with exotic weak charge was considered in
\cite{Dugan:1991ck}, \cite{Georgi:1991ci}.  Both the ordinary weak charge
under $SU(2)_L$ and the charge under
custodial\footnote{\addtolength{\baselineskip}{-.4\baselineskip} The
custodial $SU(2)$ is essentially the global $SU(2)_R$ acting on
right-handed fermions, and broken by $U(1)_Y$ and the Higgs-Yukawa
couplings.  It insures that when the electroweak symmetries are
broken, there remains an approximate $SU(2)$ global symmetry.  This
global $SU(2)$ symmetry implies, in a chiral Lagrangian, that
$f_{\pi^+}=f_{\pi^0}$. } $SU(2)_R$ play a role.
Building on this, Peskin and Wells \cite{Peskin:2001rw} label a
field's weak isospin $j_L$ and assign it the appropriate quantum
number $j_R$ under the custodial $SU(2)$; they further
define: $j_+ = j_L + j_R$ and $j_- = |j_L - j_R|$.  If one assumes
a particle with $J = j_-$ has the lowest mass $m$, and all other
particles in the multiplet have a common mass $M$, then $\Delta S$ is
roughly:
\begin{equation}
    \Delta S \sim  \frac{\kappa'}{3\pi} \log {M^2\over m^2} \, 
\end{equation}
where: 
\begin{equation}
    \kappa' = - \frac{1}{12}\left[ \left( {(j_+ + 1)\over (j_-+1)} \right)^2 - 1
                   \right] j_-(j_-+1)(2j_-+1)\; .
\end{equation}
If the particles with the smallest $j_-$ are the lightest, then the
multiplet can yield negative contributions to $S$.  Large values of
$|\Delta S|$ can be obtained from multiplets with large weak isospin.

Several avenues for producing small or even negative contributions to
$S$ have been discussed in the literature.  
Models with Majorana neutrino condensates 
naturally produce negative contributions to $S$ \cite{Gates:1991uu}, 
which can in principle
compensate large positive contributions.  Luty and Sundrum
\cite{Luty:1992fe} constructed models in which the PNGB's give
negative contributions to $S$.  To obtain $\Delta S \simeq - 0.1$ from
this source, one needs technifermions with $j_L = 2$ and PNGB's with
masses of order $200$ GeV.  More generally, other models may include
arbitrary numbers of vectorlike pairs of electroweak charged particles
\cite{Maekawa:1995yd} (e.g., both a left-handed and right-handed
gauged doublet with $j_-=0$ and a Dirac mass coming from other
external dynamics) with no cost in $S$.  The Top--Seesaw model
(Section 4.4) contains additional vectorlike fermions, allowing both
$S$ and $T$ to be completely consistent with present data.  Likewise,
the Kaluzsa-Klein recurrences of ordinary quarks and leptons in
extra-dimensional models are vectorlike pairs, and are largely
unconstrained by $S$.  These examples
are discussed in Section 4.6.

It is straightforward to engineer large positive contributions to
$\Delta T$ \cite{Einhorn:1981cy}.  Particles with masses much larger
than $\sim 1$ TeV can contribute to $\Delta T$ if their masses have an
up-down flavor asymmetry.  The contribution is of order:
\beq
          |\Delta T| \sim  \left| \frac{m_U^2 - m_D^2}{ m_U^2 + m_D^2} \right|
	  \; .
\eeq
$|m_U - m_D|$ can typically be at of order $100$ GeV.  One class of
theories that generate this kind of contribution to $\Delta T$ are the
Technicolor models in which ETC gauge interactions are replaced by
exchange of weak-doublet techni-signlet scalars \cite{Simmons:1989pu,
Carone:1993rh, Carone:1994xc} (see Section 3.5).

The effects of new $Z'$ bosons on oblique corrections were studied by
many authors (see \cite{Peskin:2001rw} and references therein).  Such
particles occur in new strong dynamical models, e.g. as in Topcolor
Models.  Peskin and Wells investigate the region of parameters for any
$Z^{0\prime}$ model in which the shifts due to the $Z^{0\prime}$
compensate those of a heavy Higgs boson.  Models with extra dimensions
can also exhibit ``compensation'' between effects of new vector bosons
and heavy Higgs bosons.  Rizzo and Wells \cite{Rizzo:1999br} have
shown that the 95\% C.L. bound on the Higgs boson mass could reach
$\sim 300$ to $\sim 500$ GeV for Kaluzsa-Klein masses $M_{KK}$ in the
range $\sim 3$ to $\sim 5$ TeV.  The coupling of the new sector is
typically strong (see Section 4).
In this vein, Chankowski \etal\ \cite{Chankowski:2000an} has argued
that two-Higgs-doublet models (which can be viewed as
effective Lagrangians for composite Higgs models) 
can be made consistent with the
electroweak fits for a Higgs boson mass of $\sim 500$ GeV. Technicolor
models can also be made consistent with the precision electroweak fits
through these kinds of effects.

A modification of the dynamics of the Technicolor gauge theory itself
can produce smaller oblique corrections than naively estimated above.
In Walking Technicolor models (Section 3.4), where the Technicolor
gauge coupling remains strong between the TC and ETC
scales\footnote{\addtolength{\baselineskip}{-.4\baselineskip} This
could be engineered, for instance, by including many vectorial
techniquarks which would affect the $\beta$ function without adding to
$S$.}, the theory is intrinsically non-QCD-like and conventional
chiral lagrangian, or chiral constituent techniquark estimates are
expected to fail. Models of this type would not have a visible Higgs
boson and may have no new particles below the first techi-vector
mesons, or open techniquark/anti-techniquark threshold: it is
difficult to anticipate the spectroscopy in a reliable way since QCD
no longer functions as an an analogue computer.  There is
calculational evidence that dangerous contributions to $\Delta S$ are
suppressed by walking.  For example, models have been proposed in
which the Technicolor enhancements to $\Delta S$ are of order $0.1$
\cite{Sundrum:1993xy},
\cite{Appelquist:1997fp}, \cite{Appelquist:1993gi}.
It is possible to construct a Technicolor model that is consistent
with the electroweak data by including enough weak isospin breaking to
provide a small positive correction to $\Delta T$.  Such a model might have,
for example $\Delta S \sim 0.2$, $\Delta T \sim 0.2$.  

To summarize, the potential conflicts between models of new strong
dynamics and the tight experimental constraints on the size of the
oblique electroweak parameters can be averted in several ways.
Whether it is particle content, the details of Technicolor dynamics,
or the behavior of higher-dimension operators from higher-energy
physics which comes to the rescue, it is clear that strong dynamics
models are not {\em a priori} ruled out.  The challenge for
model-builders is to produce concrete models that are consistent with
the constraints.

\subsection{Some Explicit ETC Models}

Having introduced the broad theoretical outlines and main experimental
challenges of ETC, we now present some explicit models.  Our summary
emphasizes strategies employed in model-building and does not represent a
complete list of all models.  This discussion will anticipate some of the
ideas of Walking ETC which will be more fully treated in Section 3.4.

\subsubsection{Techni-GIM}
\vskip .1in

An essential problem in ETC is to find a way to suppress the flavor
changing neutral processes arising from the dangerous $\gamma$ terms. We 
seek a mechanism to  naturally make $\gamma << \beta$. One
possibility is to include a GIM mechanism in the ETC sector (TC-GIM).
Here, the aim is to construct a detailed model in which the $\beta$
coefficients are of order unity (to make the fermion masses as large as
possible) and a GIM cancellation causes $\gamma << \beta$,
protecting against the unwanted $\Delta S =
2$ flavor-changing neutral current interactions.  With such a GIM mechanism
we might hope to suppress the flavor-changing current-current interactions
of eq.(\ref{etcmass2}) by a further factor of $\sim
\Lambda_{TC}^2/\Lambda_{ETC}^2$.  The lower bound on $\Lambda_{ETC}$ would then be
weakened to $\Lambda_{ETC} \gta 10$ TeV, allowing ETC to generate a generic quark
mass as large as $10$ GeV.  This would accomodate the $c$ and $b$-quark,
but would remain insufficient to produce a top quark mass of order
$\Lambda_{TC}$.

A candidate mechanism for achieving a GIM mechanism is to introduce
separate ETC gauge groups for each weak hypercharge fermion species
\cite{Dimopoulos:1983gc} \cite{Randall:1993vt}
\cite{Chivukula:1987fw}  \cite{Chivukula:1987py}.
Hence one has a separate ETC gauge group for each the left-handed electroweak
doublets, right-handed up-type singlets, and down-type singlets.
Weak $SU(2)_L$ commutes with all of these gauge groups.  The ETC groups are
broken in such a way that approximate global symmetries hold, broken only
by terms necessary to generate quark and lepton masses, and by gauge
symmetries.  This implies that the $(\overline{\psi}\psi)^2$ operators must
be approximately invariant under these global symmetries, and thus, in
general no $\Delta S=2$ operators can occur.

\begin{figure}[t]
\vspace{6cm}
\includegraphics{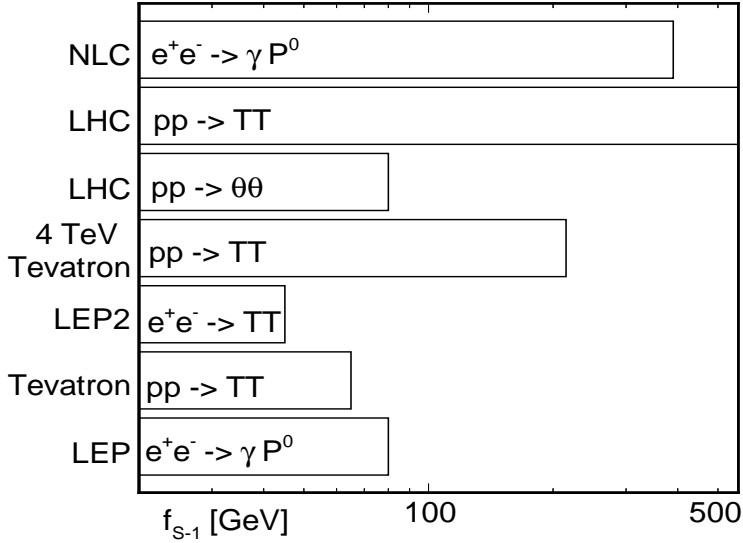}
\vspace{2cm}
\caption[]{\small \addtolength{\baselineskip}{-.4\baselineskip}
 The potential for probing the scale of TC-GIM interactions at
           present and future colliders (the high-scale models). The most
           important reaction(s) for each collider is(are) indicated inside
           the bars (from W. Skiba
	  \cite{Skiba:1996ne}; $T=P_3$, $\theta=P_8$
	   in Skiba's notation).}
\label{fig:scales}
\end{figure}

This is akin to the Glashow-Weinberg discrete flavor symmetry
\cite{Glashow:1977nt} that was introduced long ago into the multi-Higgs boson
extensions of the Standard Model.  Indeed the TC-GIM models essentially
have Glashow-Weinberg-natural multi-Higgs models as low energy effective
Lagrangians.  These models are fairly complicated and suffer from the usual
limitations on the fermion masses that can be generated in even a Walking
TC scheme, e.g., the top quark mass requires an ETC boson mass of order
$\Lambda_{TC}$ within this framework.  Nonetheless, models which achieve
the natural GIM suppression may be possible.  The third generation is then
clearly special, and part of a sort of inverted hierarchy.

A noticeable feature of these models is the presence of leptoquark PNGB's,
akin to those in the Farhi-Susskind model.  Leptoquarks, in most models,
typically decay into a quark and a lepton of the same generation. In the
TC-GIM models, the leptoquarks can carry the lepton and quark numbers of
any generation.  This means, for example, that existing limits for
leptoquarks decaying only to first-generation or 2nd-generation fermions
can be applied directly to TC-GIM models.  For example, there are limits on
``down-type'' first-generation leptoquarks \cite{Abachi:1994hr} (those
decaying to $e + \bar{d}$) and on leptoquarks decaying to 2nd generation
fermions\cite{Abe:1995fj}.  These essentially
rule out versions of TC-GIM models in which the dynamical scale is
relatively low.  The color-octet PNGBs also expected in these models may be
detectable at the LHC (see Section 2.3.3).

Skiba has given a detailed survey of signatures of TC-GIM models
\cite{Skiba:1996ne}, as summarized in Figure \ref{fig:scales}.  One of most
promising TC-GIM signatures at an NLC is pair production of leptoquarks:
$e^+e^- \rightarrow \gamma,\,Z \rightarrow P\bar{P}$, where $ P\rightarrow
\ell \bar{q}$.  Each leptoquark decays into a hadronic jet and an isolated
lepton. The events would have two opposite-sign leptons, two hadronic jets
and no missing energy, a signature easily disentangled from the
backgrounds. It is expected that leptoquarks can be discovered provided
that their masses are at least a few GeV below the NLC's kinematic limit.

\subsubsection{Non-Commuting ETC Models}

Another approach to ETC model building, which addresses the deviation in
$R_b$ caused by the $\beta$ terms discussed in 3.1.4, is to drop
the assumption that the ETC and weak gauge groups commute
\cite{Chivukula:1994mn}.  A heavy top quark must then receive its mass from
ETC dynamics at low energy scales; if the ETC bosons responsible for
$m_t$ carry weak charges, we can embed the weak group $SU(2)_{heavy}$
under which $(t,b)_L$ is a doublet into an ETC group, or more
precisely, a subgroup of ETC which is broken at low scales.  To retain
small masses, the light quarks and leptons should not be charged under
this low-scale ETC group; hence the weak $SU(2)_{light}$ group for the
light quarks and leptons must be distinct from $SU(2)_{heavy}$.  The
resulting symmetry-breaking pattern is:
\begin{eqnarray}
ETC & \times& SU(2)_{light} \times U(1)' \Rightarrow\ \ (f)\nonumber\\ 
TC \times SU(2)_{heavy} & \times& SU(2)_{light} \times U(1)_Y \Rightarrow\
\ (u) \nonumber\\ 
TC & \times& SU(2)_W \times U(1)_Y \Rightarrow\ \ (v_0)\nonumber \\
TC & \times& U(1)_{EM}, \nonumber  
\end{eqnarray}
where $f$, $u$, and $v_0 = 246$ GeV are the VEV's of the order parameters for
the three different symmetry breakings.  Note that, since we are interested
in the physics associated with top-quark mass generation, only $t_L$, $b_L$
and $t_R$ {\bf must} transform non-trivially under $ETC$.  However, to
ensure anomaly cancelation we take both $(t,b)_L$ and $(\nu_\tau,\tau)$ to
be doublets under $SU(2)_{heavy}$ and singlets under $SU(2)_{light}$, while
all other left-handed ordinary fermions have the opposite $SU(2)$
assignment.

The two simplest possibilities for the $SU(2)_{heavy} \times SU(2)_{light}$
transformation properties of the order parameters that mix and break the
extended electroweak gauge groups are:
\begin{equation}
\langle \varphi \rangle \sim (2,1)_{1/2},\ \ \ \ \langle
\sigma\rangle \sim (2,2)_0 ~,\ \ \ \ \ \ \ ``{\rm heavy\ case}"~,
\end{equation}
\begin{equation}
\langle \varphi \rangle \sim (1,2)_{1/2},\ \ \ \ \langle
\sigma\rangle \sim (2,2)_0 ~,\ \ \ \ \ \ \ ``{\rm light\ case}"~.
\end{equation}
The order parameter $\langle\varphi\rangle$ breaks $SU(2)_L$ while
$\langle\sigma\rangle$ mixes $SU(2)_{heavy}$ with $SU(2)_{light}$.  In the
``heavy'' case \cite{Chivukula:1994mn}, the technifermion condensation that
provides mass for the third generation of quarks and leptons is also
responsible for the bulk of electroweak symmetry breaking (as measured by
the contribution to the $W$ and $Z$ masses).  The ``light'' case is just
the opposite: the physics that provides mass for the third generation {\it
does not} provide the bulk of electroweak symmetry breaking.  While this
light case is counter-intuitive (after all, the third generation is the
heaviest!), it may in fact provide a resolution to the issue of how large
isospin breaking can exist in the fermion mass spectrum (and, hence, the
technifermion spectrum) without leaking into the $W$ and $Z$ masses.  This
is essentially what happens in multiscale models
\cite{Lane:1989ej,Terning:1995sc} and in Topcolor Assisted Technicolor
\cite{Hill:1995hp}.  Such hierarchies of technifermion masses are also
useful for reducing the predicted value of $S$ in Technicolor
models\cite{Appelquist:1993gi}.
  
Below the scale of ETC breaking, ETC boson exchange
yields four-fermion operators among
third-generation quarks and technifermions \cite{Chivukula:1994mn}:
\begin{equation}
{\cal L}_{4f} \sim - {1\over f_{ETC}^2}\left(\xi\bar\psi_L
\gamma^\mu U_L + {1\over \xi}\bar t_R \gamma^\mu T_R\right) \left(\xi
\bar U_L \gamma_\mu
\psi_L + {1\over \xi}\bar T_R \gamma_\mu t_R \right) ~,
\label{eqn:ehs:4f}
\end{equation}
where $\xi$ is a model-dependent coefficient.  When the techniquark
condensate forms in the $LR$
cross-terms\footnote{\addtolength{\baselineskip}{-.4\baselineskip} The
$LR$-interactions become enhanced in strong-ETC models, in which the
ETC coupling is fine-tuned to be close to the critical value necessary
for the ETC interactions to produce chiral symmetry breaking.  
Physically, this is due to the presence of a composite scalar
\cite{Chivukula:1990bc, Appelquist:1991kn} which is light compared to
$\Lambda_{TC}$ and communicates electroweak symmetry breaking to the
top quark.}  in these operators we obtain the top quark mass of order
eq.(\ref{labelcs2}) \cite{Dimopoulos:1979es,Eichten:1979ah}.  In the
heavy case the technifermions responsible for giving rise to the
third-generation masses also provide the bulk of the $W$ and $Z$
masses, and we expect $F_{T} \approx 125$ GeV (which, for $m_t
\approx 175$ GeV, implies $f_{ETC} \approx 375$ GeV)
\cite{Chivukula:1996gu}.  Even in the light case there must also be some
$SU(2)_{heavy}$ breaking VEV, in order to give the top quark a mass.

The spectrum of non-commuting ETC models includes an extra set of W
and Z bosons which affect weak-interaction physics at accessible
energies.  Mixing between the two sets of weak gauge bosons alters the
$Zff$ couplings.  In addition, the one-loop diagram involving exchange
of the top-mass-generating boson shifts the coupling of $b_L$ to the
$Z$ as in eq.(\ref{labelcs2}); here, the techniquarks in the loop are
up-type, reversing the sign of the final answer.  The two physical
effects shift $R_b$ in opposite directions by similar amounts, leaving
$R_b$ consistent with experiment \cite{Chivukula:1996gu}.  More
generally, a recent \cite{ehsrsc:2002} fit of precision electroweak
data to the predictions of the NCETC models yields limits on the
masses and couplings of the heavy W and Z bosons.  Essentially, if the
only modifications to the electroweak observables come from the extra
electroweak and ETC bosons of non-commuting ETC, then the heavy W and
Z bosons must weigh at least 2.4 TeV in the ``light'' case and at
least 3.3 TeV in the ``heavy'' case.  More general tests of the idea
of an extended weak gauge sector at current and future colliders are
discussed in Section 3.6.

\subsubsection{Tumbling and Triggering}

Any successful ETC scheme must provide a dynamical explanation for the
breaking of the ETC group to its TC subgroup, as well as for the breaking
of the electroweak symmetry.  Some interesting ideas fall under the rubrics
of ``tumbling'' and ``triggering''.

As mentioned earlier, some ETC models seek to explain the flavor hierarchy
of fermion masses by tying it to the sequential breaking of a large ETC
gauge group into a succession of subgroups containing, ultimately, the
Technicolor interactions.  This is called ``Tumbling '' \cite{Raby:1980my,
Georgi:1981mh, Veneziano:1981yz, Eichten:1982mu, Kobayashi:1986fz,
Martin:1992aq}.  The breaking of the full ETC group down to the Technicolor
group proceeds in $n$ steps ${\cal{G}}_{ETC} \rightarrow {\cal{G}}_{1}
\rightarrow {\cal{G}}_{2} \rightarrow ...  {\cal{G}}_{n-1} \rightarrow
{\cal{G}}_{TC} $.  At each step, the subgroup's gauge coupling will evolve
and become strong, permitting new condensates to form, which further break
the theory into the subsequent subgroup.  Eventually a low energy theory
described by the Technicolor and Standard Model gauge groups is achieved.  For
discussions of the detailed dynamics of Tumbling, we refer the reader to
\cite{Gusynin:1983kp} and more recently \cite{Kikukawa:1992ig,
Kitazawa:1994pd}.  Ref. \cite{Sualp:1983bq} analyzes all possible tumbling
schemes based upon $SU(N)$.

While the tumbling mechanism mentioned above is based on the idea of a
group's breaking itself, some interesting models involve the alternative
idea of triggering, in which one gauge theory produces breaking in another.
Appelquist \etal \cite{Appelquist:1997fp},
\cite{Appelquist:1994sg}, \cite{Appelquist:1993gi}
explore Technicolor models featuring a non-trivial infrared fixed point of
the Technicolor gauge coupling, in which QCD is used to play a triggering
role for EWSB.  These models, thus, predict a relation between the
electroweak scale and the QCD confinement scale. They also predict exotic
leptoquarks with masses of $\sim 200$ GeV.  Since the oblique corrections
are a perennial problem in ETC theories, we mention that isospin splitting
and techniquark/technilepton mass splitting in one-family models of this
kind have been found reduce the predicted value of the oblique 
parameter $\Delta S$ to acceptable levels, without giving
large contributions to $\Delta T$
\cite{Appelquist:1993gi}.

\subsubsection{Grand Unification}

While the dynamics of pure TC models is tied strongly to the
electroweak scale, ETC models inevitably involve physics at higher energy
scales.  We have already seen that avoiding FCNC in the 
neutral-$K$ system
requires pushing the symmetry-breaking scale of classic ETC models up to at
least 1000 TeV.  As a result, it is natural for one to ask whether ETC
theories can be embedded into a grand-unified gauge theory at still higher
energies.  A proof-of-principle for the possibility of ETC unification was
mentioned earlier: the Farhi-Susskind extension to an SU(56) under which
all quarks, leptons, and technifermions transform within one multiplet.  We
mention some additional unified models in which some of the key theoretical and
phenomenological challenges of ETC are also addressed.

Giudice and Raby \cite{Giudice:1992sz} \cite{Raby:1990pq} have considered a
unified model that starts with an ETC group of $SO(4)$ with a full standard
model family of fermions in the $\bf{4}$ vector representation. This sector
of the model is, in spirit, similar to the Farhi-Susskind model with
$SO(4)$ replacing $SU(N_{TC})$ and containing the ETC gauge interactions.
At the ETC breaking scale $\lambda_{ETC}\sim 10^3$ TeV the group $SO(4)$
breaks to $SO(3)$ which is the true gauge Technicolor group.  As mentioned
above, theories based upon $SO(N)$ groups containing ETC and Technicolor,
with fermions in the $\bf{N}$ representation, can avoid flavor-changing
neutral current-like interactions ($\gamma$ terms) in the tree
approximation.  This leads to gauge bosons with masses of order $5$ TeV for
third generation,and $\sim 10^3$ TeV for first and second, that mediate the
ETC interactions.  The model exhibits a fixed point in the strong coupling
of the Technicolor group, leading the coupling to run slowly (see Section
3.4 on Walking TC) above the TC scale; a key side-effect of walking, the
non-calculability of the $S$-parameter, is discussed.  The model also
invokes four fermion effective operators, and shares features with Top
Condensation models (see Section 4).

The main idea is to imbed this ETC theory into a large gauge group such as
$SO(18)$ and attempt to achieve a grand unified theory.  The breaking
chains required for this GUT, curiously, can be generated by Majorana
condensates of neutrinos (see e.g. \cite{Hill:1991ge}), thus incorporating
the Gell-Mann--Ramond--Slansky seesaw mechanism \cite{Gell-Mann:1978pg}
more intimately into the dynamics.  Note that neutrino condensates have
been invoked elsewhere to break electroweak symmetries, or in concert with
fourth generation Farhi-Susskind Technicolor,
\cite{Hill:1990vn, Hill:1991ge, Akhmedov:1996vm, Martin:1991xw,
Evans:1993yj}.  The model of Giudice and Raby leaves open the question of
breaking the ETC gauge group at $\Lambda_{ETC}$.  The exact nature of the
Technicolor fixed point (walking) with the additional four-fermion
operators is also left unclear.

King and Mannan \cite{King:1992ev}, \cite{King:1995yr} give a schematic
model based upon the three sequential breakings $SO(10)\rightarrow
SO(9)\rightarrow SO(8)\rightarrow SO(8) $, where the ETC bosons for $n$th
generation of fermions is produced at the $n$th breaking scale.  This
attempts to explain the generational hierarchy of masses and mixings of the
quarks and leptons.  The model incorporates grand unification and neutrino
masses, but does not address the dynamics behind the sequential breakings
as in the Giudice--Raby model.

\subsection{Walking Technicolor}

\subsubsection{Schematic Walking}

Extended Technicolor
(ETC) has difficulties in producing the observed heavy quark
and lepton masses.  Even the charm quark is heavy enough to cause problems
in building models with sufficiently suppressed flavor-changing
neutral-current interactions.  In the preceeding section, we surveyed
various attempts to deal with this difficulty from a structural point of
view.  We turn now to an intriguing dynamical possibility which emerges
from a closer examination of the full TC
strong--dynamics.

Let us consider the TC radiative corrections to the operators from
ETC that generate the quark and lepton masses. These operators appear as the
``$\beta$'' contact terms of eq.(\ref{contact1},
\ref{contact2}) at the scales $\mu \lta \Lambda_{ETC}$. Since $\Lambda_{ETC} >>
\Lambda_{TC}$, these operators are subject to renormalization
effects by TC,
\begin{equation}
\VEV{\bar{Q}Q_{ETC}} = \exp\left( \int_{\Lambda_{TC}}^{\Lambda_{ETC}}
{d\ln(\mu)} \;\gamma_m(\alpha(\mu)) \right) \VEV{\bar{Q}Q_{TC}}
\label{eq:ehs:radcf}
\end{equation}
where $\gamma_m$ is the operator's anomalous dimension.
 
If TC is QCD-like, then the TC coupling constant $\alpha(\mu)$ is
asymptotically free, and falls logarithmically as 
$\alpha(\mu) \propto 1/\ln(\mu)$ above the scale
$\Lambda_{TC}$.  With the anomalous dimension $\gamma_m \propto
\alpha(\mu)$ we see that the radiative correction is proportional to
$\exp[\gamma_m\ln(\ln(\mu)] \sim 
(\ln(\Lambda_{ETC}/\Lambda_{TC}))^{\gamma_m}$.
  Hence the
radiative corrections are power-logarithmic factors,
 similar to the behavior of QCD
radiative corrections to the nonleptonic weak interactions in the Standard
Model.

If, however, $\alpha(\mu)$ is approximately constant, 
i.e., if the TC theory
exists approximately at a ``conformal fixed point,'' 
$\alpha(\mu)= \alpha^\star\neq 0$, where $\beta(\alpha^\star)=0$, then
the radiative correction is converted into a power
law, proportional to $\exp[\gamma_m(\alpha^\star)\ln(\mu)]
\sim (\Lambda_{ETC}/\Lambda_{TC})^{\gamma_m(\alpha^\star)}$, which is
a substantially larger renormalization effect.

Such a theory is not QCD-like, but is an {\em a priori} possible behavior
of a Yang-Mills gauge theory.  This behavior can provide significant
{\em amplification} to both
the $\alpha$ and $\beta$ terms of ETC which involve the
technifermion bilinears, but does not alter the 
dangerous $\gamma$ terms, which
involve only the Standard Model fermions. This holds out 
the possibility of
enhancing ETC-generated fermion and PNGB masses 
without increasing the rate of neutral
flavor-changing processes.   A TC theory with an
approximately constant coupling $\alpha(\mu)=\alpha^\star$  in the range
$\Lambda_{TC}\lta \mu \lta \Lambda_{ETC}$ is said to be 
 ``Walking Technicolor'' (WTC),  
an idea which was first
proposed by Holdom \cite{Holdom:1981rm}; some early implications were
discussed in refs. 
\cite{Yamawaki:1986zg}, \cite{Bando:1986bg}, \cite{Appelquist:1986an},  
\cite{Appelquist:1987tr}.

\subsubsection{Schwinger-Dyson Analysis}

As demonstrated by Yamawaki, Bando and  Matumuto,
\cite{Yamawaki:1986zg}, 
and further elaborated by  Appelquist, Karabali and Wijewardhana
\cite{Appelquist:1986an}, for Walking TC it suffices to have 
an {\em approximate} fixed point,
$\beta(\alpha^\star) << 1$ with $\alpha^\star
\sim \alpha_c$ near $\Lambda_{TC}$, 
for the relevant scales $\Lambda_{TC} \lta \mu\lta
\Lambda_{ETC}$. Here $\alpha^\star $ is near the the critical
coupling $\alpha_c$ for the formation of chiral condensates.  An analysis
of the Schwinger-Dyson equations for the mass-gap of the theory 
\cite{Yamawaki:1986zg}, \cite{Appelquist:1988yc} 
then shows that the radiative correction 
of eq.(\ref{eq:ehs:radcf})
can enhance the techniquark bilinear operator by a factor of order
$\Lambda_{ETC}/\Lambda_{TC}$.  
Essentially, the solution of the Schwinger-Dyson
equation with fixed coupling gives 
large anomalous dimensions $ \simeq 1$ near
the critical coupling, $\alpha^*  \sim \alpha_c$. This
result obtains in the chiral broken phase as well as the symmetric phase,
\cite{Yamawaki:1986zg}
near critical coupling 
and hence resolves the difficulties
in TC (which is supposed to be in the chiral broken 
phase near $\alpha_c$).  
We recapitulate the argument below.

The Euclideanized Schwinger-Dyson equation for the self-energy of a fermion
in Landau gauge is given by \cite{Johnson:1964da}:
\begin{equation}
\Sigma(p^2) = 3C_2(R) \int \frac{d^4 k}{(2\pi)^4}
\frac{\alpha((k-p)^2)}{(k-p)^2}\frac{ \Sigma(k^2)}{Z(k^2) k^2 + \Sigma^2(k^2)}
\end{equation}
Typically we approximate $Z(k^2) \approx 1$, 
and linearize the equation by neglecting the
$\Sigma^2(k^2)$ denominator term.
We assume $\alpha(\mu) \approx \alpha_c$
is slowly varying.  Two solutions are then found:
\begin{equation}
\Sigma(p^2) = \Sigma(\mu)\left(\frac{\mu^2}{p^2}\right)^{b_\pm} \qquad
b_\pm = \half (1 \pm (1 -\alpha(\mu)/\alpha_c)^{1/2}) 
\end{equation}
where the critical coupling constant is 
$\alpha_c = \pi/3C_2(R)$, and $C_2(R)$ is the quadratic Casimir 
of the complex technifermion representation $R$ (recall $C_2 = (N^2-1)/2N$
for the fundamental representation). 
 
The normal perturbative anomalous dimension of the $\overline{Q}{Q}$ 
operator is
\begin{equation}
\gamma_m = 1 -(1 -\alpha(\mu)/\alpha_c)^{1/2} \sim \frac{
3C_2(R)\alpha(\mu)}{2\pi}
\end{equation}
 Hence, the solution with $b_{-}$ corresponds to the running of a
 normal mass term of nondynamical origin. The solution with $b_{+}\sim
 1$ corresponds to the high momentum tail of a dynamically generated
 mass having the softer $\sim 1/p^2$ behavior at high energies
 \footnote{\addtolength{\baselineskip}{-.4\baselineskip} Jackiw and
 Johnson \cite{Jackiw:1973tr} showed long ago that this solution also
 forms the Nambu-Goldstone pole, confirming it is a dynamically
 generated mass.}.  Note that, at the critical coupling
 $\alpha(\mu)=\alpha_c$, the two solutions coincide, which is believed
 to be a generic phenomenon, \cite{Appelquist:1988yc},
 \cite{Cohen:1989sq}, \cite{Mahanta:1989rb}.  Moreover, if we suppose
 that $\alpha(\mu) = \alpha^* > \alpha_c$ for $0\leq \mu \leq
 \Lambda^*$, then we find that a true dynamical symmetry breaking
 solution exists where:
\begin{equation}
\Sigma(0) \sim 
\Lambda^* \exp \left( -\pi / \sqrt{ \frac{\alpha^* }{\alpha_c } -1} \right)
\end{equation}
In the energy range $\Lambda_{TC} \leq \mu \leq
\Lambda_{ETC}$, the large value of $\alpha(\mu)\approx \alpha^*$
corresponds to an anomalous dimension of order 1, making the radiative
correction factor for the technifermion bilinear (\ref{eq:ehs:radcf}) of order
 $\Lambda_{ETC}/\Lambda_{TC}$.

What are the implications of Walking Extended TC for the ordinary
fermion masses?  In classic ETC, we have seen that fermion masses typically
scale as $\Lambda_{TC}^3/\Lambda_{ETC}^2$.  Since $\Lambda_{TC} \approx 1$ TeV,
the phenomenological
constraint on $\Lambda_{ETC}\gta 100$ TeV implies $m_{q,\ell} \lta 100$
MeV.  Walking ETC brings a large renormalization
enhancement of the techniquark bilinear by
a factor of order $\sim \Lambda_{ETC}/\Lambda_{TC}$, so that we now have
\cite{Yamawaki:1986zg}:
\begin{equation}
m_{q,\ell} \sim \Lambda_{TC}^2/\Lambda_{ETC} \sim 1\; \makebox{GeV}\ ,
\end{equation}
which is large enough to accomodate the strange and charm quarks,
and the $\tau$ lepton. This is born out by more
detailed studies which include the full ETC boson exchange
in the gap equations \cite{Holdom:1981rm},
 \cite{Holdom:1985sk}, \cite{Yamawaki:1986zg},  \cite{Bando:1986bg}
 \cite{Appelquist:1986an}, \cite{Appelquist:1987tr}
 \cite{Appelquist:1987fc}.

On the other hand, this is barely large enough to 
accomodate the bottom quark and, certainly not the top quark
masses\footnote{\addtolength{\baselineskip}{-.4\baselineskip} For an
heroic attempt see, e.g.,
\cite{Appelquist:1996kp}.  Such models lead either to gross
violations of precision electroweak constraints, or to excessive fine-tuning.}.
Consider that, if TC has QCD-like dynamics, 
the value of $\Lambda_{ETC}$ required to fit the top mass is given by (see
\cite{Chivukula:1996uy}):
\begin{equation}
\Lambda_{ETC} \sim 1\;\makebox{TeV}\;\left(\frac{175
\makebox{GeV}}{m_t}\right)^{1/2}\,.
\label{eqn:cth:walk-lo}
\end{equation}
The measured value of the top quark mass therefore implies $\Lambda_{ETC}
\sim \Lambda_{TC}$.  This leaves too little ``distance'' between
energy scales for walking to make a difference
\cite{Chivukula:1993tz}.  It therefore appears that ETC alone, even in the
presence of walking, can only contribute a fraction of the observed $m_t$. 
 We will address alternative mechanisms for producing the masses
of the $t$ and $b$ quarks in Section 4.  Some of these include new strong
gauge dynamics peculiar to the third generation, while others invoke
instantons \cite{Hill:1991at} \cite{Baluni:1981ty}. 

What is the origin of the small $\beta$ function and what other
effects may arise as a  consequence? 
Consider the one-loop $\beta$
function of an $SU(N)_{TC}$ gauge theory with $N_f$
techniquarks in the fundamental $N_{TC}$  representation:
\begin{equation}
\beta_{TC} = - \frac{g_{TC}^3}{16\pi^2} \left( \frac{11}{3}
  N_{TC} - \frac{8}{3} N_f \right) + \cdots
\label{eq:ehs:betaone}
\end{equation}
 Clearly, for $\alpha_{TC}(\mu) =g_{TC}^2/4\pi$ 
to walk requires having many technifermions
active 
between the scales $\Lambda_{TC}$ and
$\Lambda_{ETC}$.  These need {\em not} all be electroweak doublets, e.g., 
they
may be singlets or vectorlike doublets 
with respect to $SU(2)_L$, which
can help suppress contributions to $S$.  
Higher tensor representations of
the $SU(N_{TC})$ gauge group are also possible
\cite{Eichten:1980ah}.  
In general, after the ETC breaking, the fermions in the lower energy theory
fall into subsets carrying [i] only (TC); [ii] (TC)$\times$(color); [iii]
(TC)$\times$(flavor), and so on, including both the fundamental and higher
representations of TC.  The technifermions in different representations may
condense at different scales, as will be discussed in the next section.  

Whether walking is caused by the presence of many technifermions in the
fundamental TC representation or technifermions in higher
TC representations, the chiral symmetry-breaking sector is
enlarged relative to that of minimal TC models.  As a result, one
expects a proliferation of technipions and small technipion decay constants
$F_{T} \ll v_0$.  At first glance, it appears that the models will suffer
from unacceptably large contributions to S (because of the large number of
technifermions) and from the presence of many light pseudo-Nambu-Goldstone
bosons (PNGBs) which have not been observed.  However, the effects of the
strong walking-TC dynamics on both issues must be taken into
account.

As emphasized by Lane \cite{Lane:1993wz}, the ingredients that enter
conventional QCD-inspired estimations of $S$ are not applicable to a
walking ETC theory.  Given the altered pattern of resonance masses
(possibly a tower of $\rho_T$ and $\omega_T$ states), the proliferation of
flavors, and the presence of fermions in non-fundamental gauge representations
\cite{Lane:2000pa}, it is unclear how to estimate $S$ reliably in a
walking model.  We note that when estimated naively in a QCD-like
TC theory, the S-parameter comes out larger than anticipated,
roughly twice the naive result in the fermion bubble approximation.  The
fermion bubble approximation appears, in spirit, to be closer to the
situation in walking ETC.  Indeed, existing estimates of $S$ in walking
models \cite{Appelquist:1992is} \cite{Sundrum:1993rf}, approach the fermion
loop estimate.  However, in the absence of compelling evaluations of S in
walking gauge theories, S does not provide a decisive test of these models.

In contrast, the effect of walking dynamics on the masses of the PNGB's is
unequivocal: the enhanced condensate raises these masses.  Previously we
found $ m_P^2 \sim
\VEV{\bar{Q}Q}^2/\Lambda_{ETC}^2F_T^2
\sim N_{TC}^2\Lambda_{TC}^6/\Lambda_{ETC}^2F_T^2 $, which could be
dangerously small.  With the enhancements on the pair of bilinear
techniquark operators this becomes:
\begin{equation}
 m_P^2 \sim 
 N_{TC}^2 \Lambda_{TC}^4/F_T^2 \sim N_{TC}^2 \Lambda_{TC}^2
\end{equation}
and the PNGB's are now safely elevated out of harm's way from current
experiments, but left potentially accessible to the Tevatron and LHC.

More generally, Walking TC is actually an illustration of the
physics of chiral dynamics in the large $N_{flavor}$ limit.  Recent
interest in the phase structure of chiral gauge theories has been inspired
in part by duality arguments in SUSY theories where there exist exact
results for the phase structure of an $SU(N)$ gauge theory with $N_f$
flavors (see the review
\cite{Intriligator:1996au}).  The presence of infrared fixed points of the
gauge coupling appears to be fairly generic in theories with a large number
of flavors.

 The infrared fixed point of the strong $SU(N)_{TC}$ gauge theory 
 can also arise from
 the interplay of the first two terms of the $\beta$-function
 (\ref{eq:ehs:betaone}). The size of the fixed point coupling constant,
 $\alpha^*$, can be controlled by adjusting $N_f$ and $N_{TC}$.  By judicious
 choice of $N_f\sim N_{TC}$ one can make $\alpha^*$ small and a perturbative
 analysis should be valid.  The weak coupling fixed point, truncating
 $\beta$ on the first two terms, is called the Banks-Zaks fixed point
\cite{Banks:1982nn}. 
The ladder-approximation gap equation can be used to probe the chiral
breaking transition with fixed point $\alpha^*$
\cite{Appelquist:1998rb}, which appears to occur for large $N_f \lta
4N \pm 20\% $.\footnote{\addtolength{\baselineskip}{-.4\baselineskip}
See \cite{Terning:1997jj} for a discussion of the
interplay of chiral breaking and confinement in the SUSY case.}.

\subsection{Multi-Scale and Low-Scale TC}

 Eichten and Lane 
\cite{Lane:1989ej} \cite{Eichten:1997yq} have suggested
that, if TC dynamically generates the weak scale,
there may be distinct sectors of the full theory that contribute
components to the full electroweak scale, so that $v_{0}^2 = v_1^2 +
v_2^2 +...$.  This general idea is known as ``Multi-Scale TC.''
The sector of the theory with the smallest
$v_i$, the ``Low-Scale TC''
sector, may produce visible phenomenological consequences 
at the lowest energy scales in current colliders. As we have seen
in the previous section, a ``Low-Scale'' sector is 
generally expected in
a walking TC theory.

Low-Scale TC produces a rich phenomenology, much of which is directly
accessible to the Tevatron in Run II or to a low energy LC; a plethora
of new states should also be visible at the LHC.  Processes of
interest in a $f\bar{f}$-collider are then the familiar $f\bar{f}
\rightarrow (\rho_T,\omega_T)
\rightarrow (WW+ZZ, WP_T + ZP_T, P_TP_T)$ as in the Farhi-Susskind
model, but transplanted now to the lower scale $v_i$.  In this section, we
will first discuss how the Low-Scale TC idea fits within the
general ETC framework and then sketch the phenomenological consequences.
Current limits on these theories are discussed in the next section.

The motivation for Low-Scale TC arises directly from the ideas
of WTC discussed in the previous section.  We have already
seen that a WTC coupling is desirable because it offers
some remedies for the questions of fermion and PNGB masses.  Suppose we
follow the lead of Eichten and Lane \cite{Lane:1989ej} in constructing a
walking model by including technifermions in higher-dimensional TC
representations.  As the theory evolves downward towards the TC scale,
chiral condensates form.  However, they do not all form at a single scale:
the higher representations will condense out at higher energies since their
binding is controlled by the Casimir coefficients, $C_2(R)$ of their
representations \cite{Marciano:1980zf}.  Moreover, since we have seen how
walking dynamics enhances technifermion bilinears, the expected separation
of the various condensate scales may be large.

In the most phenomenologically interesting cases of Low-Scale TC, the
sector of the theory with the smallest $v_i$ includes states light
enough to be accessible to current experiments.  One caveat is that a
very light Low-Scale sector may include charged technipions $P^\pm$
with masses less than $m_{top}$.  If top quark decays are to remain
dominated by the conventional Standard Model channel $t \rightarrow
W+b$, as consistent with Tevatron Run I data, it is necessary to
suppress a fast $t \rightarrow P_T +b$, decay.  One possibility is to
require that $m_P\gta 160$ GeV \cite{Balaji:1997va}.  Alternatively,
one can decouple the top quark from the Low-Scale sector to suppress
the $t\bar{b}P$ vertex \cite{Eichten:1994nc}. This is generally
accomplished by using separate strong top dynamics to generate the top
quark (and b-quark) masses and Low-Scale TC for the bulk of the EWSB
\cite{Hill:1995hp} \cite{Lane:1995gw}.  A scheme of this type called
``Topcolor Assisted TC,'' or TC2, and the phenomenological
implications of the new strong top dynamics will be discussed in
Section 4.2.

In general, the spectroscopy of a Low-Scale TC model can
accomodate everything from a minimal model through the Farhi-Susskind
structure.  The Low-Scale spectrum will, at the very least, include light
PNGB's (technipions) and techni-vector mesons.  For example, with $F_T
\sim 60$ GeV one expects $M_{P_T} \sim 100$ GeV and $M_{\rho_T} \sim 200$
GeV \cite{Lane:2000pa}. The technipions will be resonantly produced
via techni-$\rho$ vector meson dominance (VMD)
\footnote{\addtolength{\baselineskip}{-.4\baselineskip} Per the
discussion of Section 2.3.3, there is an induced coupling of the
$\rho_T$ to {\em any} electromagnetic current, so this also applies at
$e^+e^-$ and $\mu^+ \mu^-$ colliders
\cite{Eichten:1998kn}). See \cite{SekharChivukula:2001gv, Zerwekh:2001uq} about the caveats that apply when
estimating production of color-octet $\rho_T$ by vector meson dominance.}
with large rates at the Tevatron, LHC, and a linear collder
\cite{Eichten:1996dx}. The technivector mesons are expected to be,
in analogy to the minimal model, an isotriplet, color-singlet $\rho_T$, and
the isoscalar partner $\omega_T$.  Isospin is likely to be a good
approximate symmetry, so $\rho_T$ and $\omega_T$ should be approximately
degenerate in mass, as is the $I=1$ multiplet of technipions.  The
enhancement of technipion masses due to walking suggests that the decay
channels $\rho_T \rightarrow P_TP_T$ and $\omega_T
\rightarrow P_TP_TP_T$ are probably closed. 
Thus, the decay modes $\rho_T \rightarrow W_L P_T$ and
$Z_L P_T$, where $W_L$, $Z_L$ are longitudinal weak bosons, and $\rho_T,
\omega_T
\rightarrow \gamma P_T$ may dominate. 
Because technipion
couplings to fermions, like those of scalars, are proportional to
mass, one expects the most important decay modes to be
\begin{equation}
\label{eq:singdecay}
\ba{ll}
P_T^0 &\rightarrow b \bar{b} \\
P_T^+ &\rightarrow c \bar{b} \ts\ts\ts {\rm or} \ts\ts\ts c \ol s, \ts\ts
\tau^+ \nu_\tau \ts.
\ea
\end{equation}
Heavy-quark jet tagging is, then, important in searches for Low-Scale
TC.

Eichten, Lane and Womersley \cite{Eichten:1997yq} have performed an
extensive analysis of the phenomenological signatures of Low-Scale
TC at the Tevatron.  They present simulations of $\bar{p} p
\rightarrow \rho_T^\pm \rightarrow W^\pm_L P_T^0$ and $\omega_T
\rightarrow \gamma P_T^0$ for the Tevatron collider with an integrated
luminosity of $1\,\ifb$. For $M_{\rho_T} \simeq 200$ GeV and $M_{P_T}
\simeq 100\,$ GeV, the cross sections at the Tevatron are expected to be of
order a few picobarns.  The narrowness of the $\rho_T$ and $\omega_T$
suggests that appropriate cuts, e.g., on the final state invariant mass,
can significantly enhance the signal$/$background ratio.  Furthermore, the
final states include $b$-quark jets from technipion decay and either a
photon or the isolated lepton from weak boson decay.  This results in some
dramatic signals which stand out well above background when
$b$-tagging is required.  Cross sections for the LHC are an order of
magnitude larger than at the Tevatron, so detection of the light
technihadrons should be straightforward there as well.

\begin{figure}
\vspace{9cm}
\includegraphics{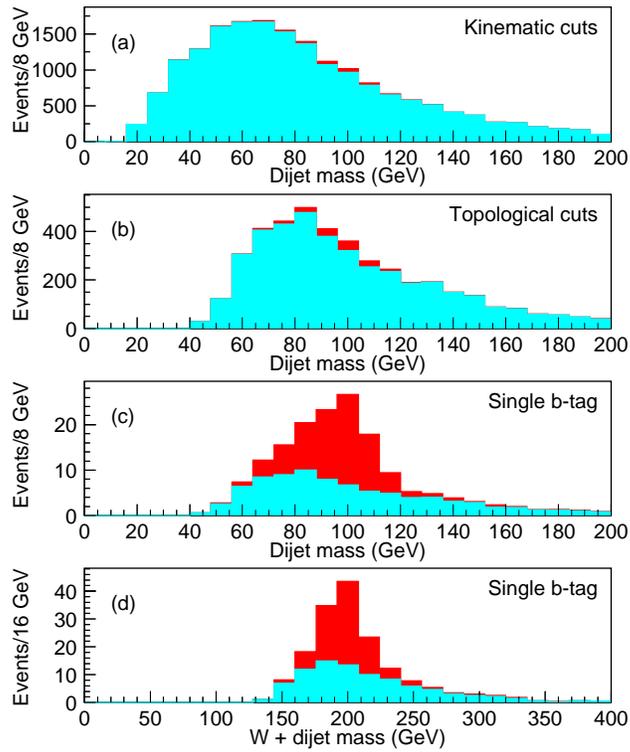}
\vspace{3cm}
\caption[]{\small \addtolength{\baselineskip}{-.4\baselineskip} 
 Predicted invariant mass distributions at Tevatron Run II for
$\rho_T$ signal (black) and $Wjj$ background (grey); vertical scale is
events per bin in 1~fb$^{-1}$ of integrated luminosity from
Ref.(\cite{Eichten:1997yq}).  Dijet mass distributions (a) with
kinematic selections only, (b) with the addition of topological
selections, and (c) with the addition of single $b$-tagging; (d)
$W+$dijet invariant mass distribution for the same sample as (c).
\label{tech_rho1}}
\end{figure}

More specifically, ref.\cite{Eichten:1997yq} considers one light isotriplet
and isoscalar of color-singlet technihadrons and uses VDM for techni-$\rho$
production and decay to determine the rates for:
\begin{equation}
\label{eq:singlet}
\begin{array}{lll}
q \ol q' & \rightarrow W^\pm + \rho_T^\pm &
\rightarrow \ts\ts W_L^\pm Z_L^0;  \quad W_L^\pm
P_T^0, \ts\ts P_T^\pm Z_L^0;  \quad P_T^\pm P_T^0 \\ \\
q \ol q & \rightarrow \gamma + Z^0 + \rho_T^0 &
\rightarrow \ts\ts W_L^+ W_L^-; \quad
W_L^\pm P_T^\mp; \quad P_T^+ P_T^- \;
\end{array}
\end{equation}
For $M_{\rho_T} \sim 200\,\gev > 2M_{P_T}$ and $M_{P_T} \sim 100\,\gev$, the
dominant processes have cross sections of $1-10$~pb at the Tevatron and
$\sim 10 - 100$ ~pb at the LHC.  The modes with the best 
signal-to-background are
$\rho_T \rightarrow W_L P_T$ or $Z_L P_T$ and $\rho_T\omega_T 
\rightarrow \gamma P_T$.

Figure~\ref{tech_rho1}(d) shows invariant mass distributions for the $Wjj$
system after kinematic and topological cuts and $b$-tagging have been
imposed as discussed in Ref.\cite{Eichten:1997yq}.  A clear peak is visible
just below the mass of the $\rho_T$, and the peaks in the dijet mass and
the $Wjj$ mass are correlated.  If the $\rho_T$ and $P_T$ exist in the
mass range favored by the Low-Scale TC models they can be easily
found in Run~II of the Tevatron.  The $\omega_T$ is likewise produced in
hadron collisions via vector-meson-dominance coupling through $\gamma$ and
$Z^0$.  We expect $\omega_T \rightarrow \gamma P_T^0$, $Z^0 P_T^0$ will
dominate $\omega_T \rightarrow P_TP_T$.

\begin{figure}[t]
\vspace{7cm}
\includegraphics{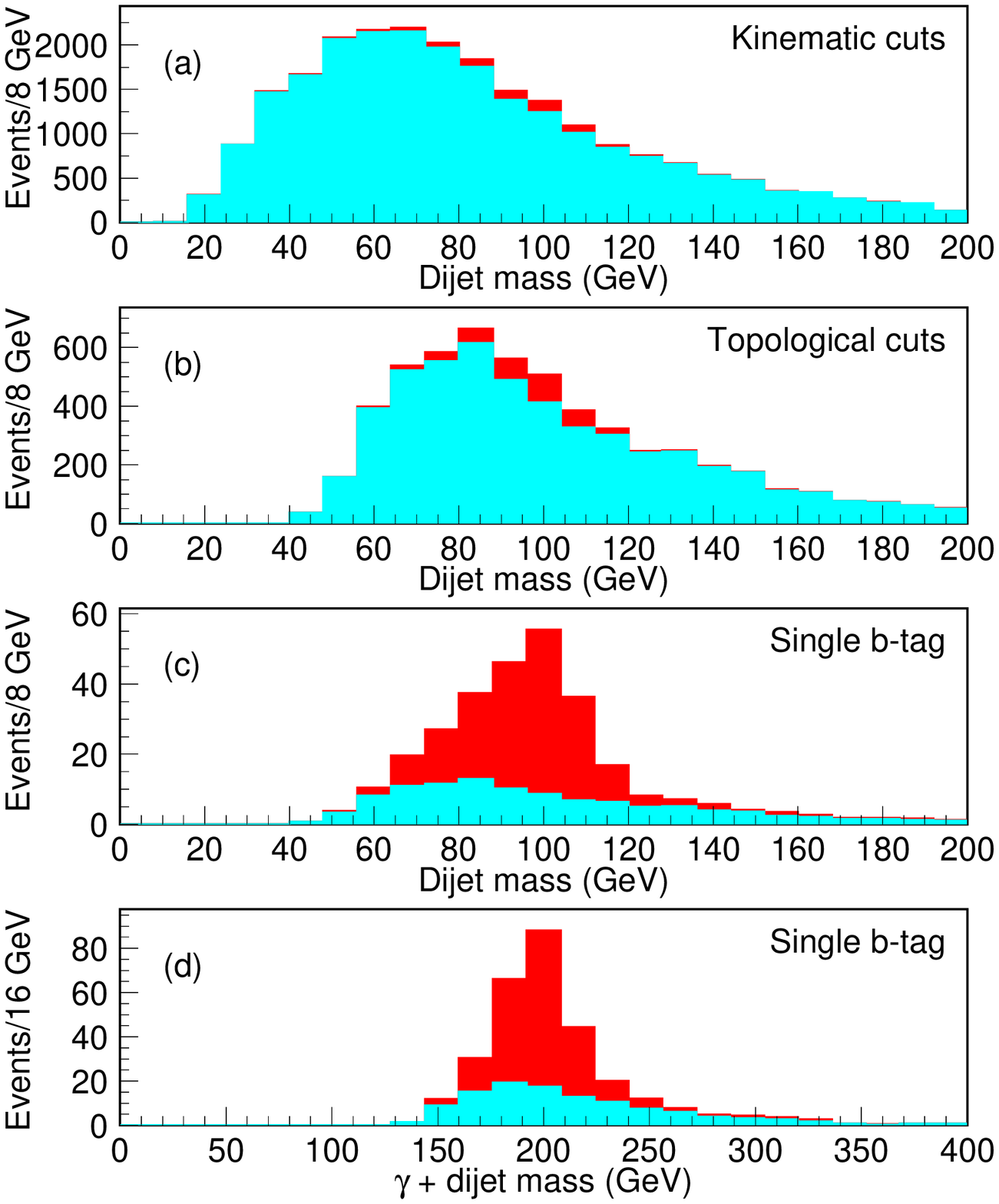}
\vspace{3cm}
\caption[]{\small \addtolength{\baselineskip}{-.4\baselineskip} 
Predicted invariant mass distributions at Tevatron Run II for
$\omega_T $ signal (black) and $\gamma jj$ background (grey);
vertical scale is events per bin in 1~fb$^{-1}$ of
integrated luminosity.
Dijet mass distributions (a) with 
kinematic selections only, (b) with the addition 
of topological selections,
and (c) with the addition of single $b$-tagging; 
(d) $\gamma+$dijet invariant mass distribution for the same
sample as (c).}
\label{restrho}
\end{figure}

Figure~\ref{restrho}(a) shows the invariant mass distribution of the two
highest-$E_T$ jets for signal (black) and background (grey) events, that
pass certain kinematic criteria, 
for an integrated luminosity of $1\,\ifb$.  The
effect of topological cuts is seen in Fig.~\ref{restrho}(b). Tagging one $b$-jet
significantly improves the signal/background as in Fig.~\ref{restrho}(c),
and a peak below the $P_T$ mass can be seen.  Figure~\ref{restrho}(d)
shows the photon$+$dijet invariant mass after various topological cuts and
$b$-tagging are implemented \cite{Eichten:1997yq}.

The ultimate virtue of Low-Scale TC is that the new strong
dynamics, presumably the key to other fundamental questions, such as those
of fermion masses and flavor physics, would
commence at relatively lower
accessible energies.  As we have seen, Low-Scale TC signatures
$\rho_T \rightarrow W P_T$ and $\omega_T \rightarrow \gamma P_T$ can be
discovered easily in Run~II of the Tevatron for production rates as low as
a few picobarns.  In the next section, we will examine existing
experimental constraints on these models.

\subsection{Direct Experimental Limits and Constraints
on TC}

Experiments performed in the past few years have had the large data
samples, high energies, and heavy-flavor-tagging capabilities required
to begin direct searches for the new phenomena predicted by models of
new strong dynamics.  We have just examined some of the theoretical
implications for preferred channels of Low-Scale TC models.  We now
summarize the status of experimental limits on the accessible scalar
mesons, vector mesons, and gauge bosons, that are generically
predicted by these theories.  Our focus here is on the direct searches
performed largely by the LEP and Tevatron experiments, and we present
many of their original exclusion plots.  Tables summarizing many of
these results may be found in
refs. \cite{Groom:2000in,Barklow:2002su}.  As appropriate, we will
also comment on more indirect searches for new physics via
measurements of precision electroweak observables and on the prospects
for the LHC and future lepton colliders.  For further discussion of
technicolor searches at LHC and even higher-energy hadron colliders,
see ref. \cite{Barklow:2002su}.


\subsubsection{Searches for Low-Scale Color-singlet 
Techni--$\rho$'s and Techni--$\omega$'s  
(and associated Technipions)}

In light of the Low-Scale TC model, and the generic phenomenology of
TC, it is useful to examine the present-day constraints that exist,
mostly from LEP and the Tevatron, on the direct observables, i.e., the
masses and production signatures of technivector mesons and
techni-$\pi$'s.  Note that a number of the searches have taken
advantage of the fact that the $\omega_T$ of Low-Scale models can be
visible in collider experiments \cite{Lane:1989ej, Eichten:1996dx,
Eichten:1997yq, Lane:2000pa}.  Enhancement of technipion masses by WTC
can quench the decay $\omega_T \to P_T P_T P_T$, resulting in the
dominant mode $\omega_T \to \gamma P_T$ \cite{Lane:1989ej}.
Techni-$\omega$ decays to $q\bar{q}$, $\ell^+ \ell^-$ and
$\nu\bar{\nu}$ can also be significant, but the decay $\omega_T
\to Z P_T$ is phase-space suppressed.

CDF has published two searches for color-singlet techni-$\rho$'s
\cite{Affolder:1999du, Affolder:2000hp}.  One study assumes that the
channel $\rho_T \to P_T P_T$ is closed and the other that it is
open.  If $M_{\rho_T} < M_W + M_{P_T}$ then the techni-$\rho$ (and
its isoscalar partner, $\omega_T$) decays to pairs of ordinary quarks
or leptons.  D0 has looked for light $\rho_T, \omega_T$ decaying to
$e^+ e^-$ \cite{Handa:1999gx}.  The LEP collaborations L3
\cite{Kounine:1999tp}, DELPHI \cite{Borisov:2000tp}, 
and OPAL \cite{opal:2001prelim} have released preliminary results on
multi-channel searches for light $\rho_T$ and $\omega_T$. 

\begin{figure}
\vspace{5.0cm}
\includegraphics{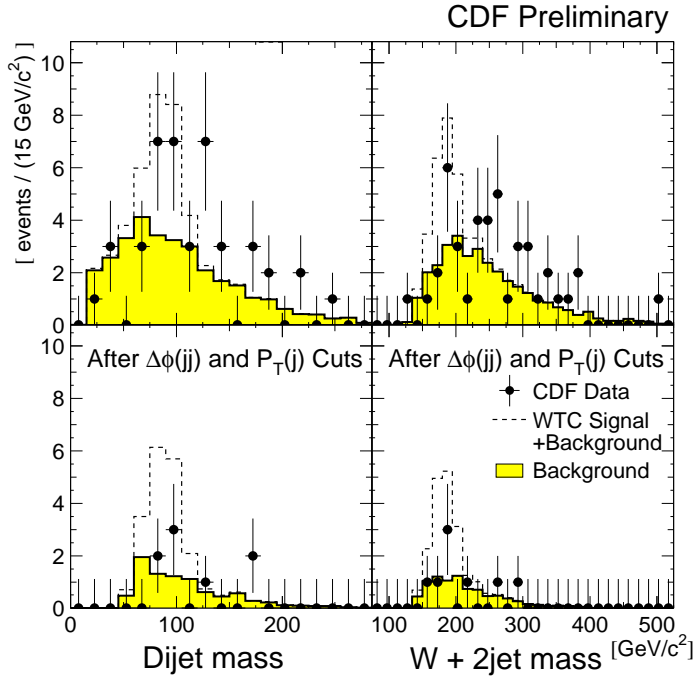}
\vspace{2.5cm}
\caption[cap]{\small \addtolength{\baselineskip}{-.4\baselineskip} 
Invariant mass of the dijet system and of the W + 2 jet
  system (with a leptonically decaying W) in the CDF search for $\rho_T \to
  W^\pm P_T$ \protect\cite{Affolder:2000hp}.  The points are data; the
  solid histogram is background; the dashed histogram shows background plus
  the signal from a walking TC model with $M_{\rho_T} = 180$ GeV and
  $M_{P_T} = 90$ GeV.  The topological cuts leading to the lower figures
  are described in \protect\cite{Eichten:1998kn}. }
\label{fig:rho1-peak}
\end{figure}

\begin{figure}
\vspace{5.50cm}
\includegraphics{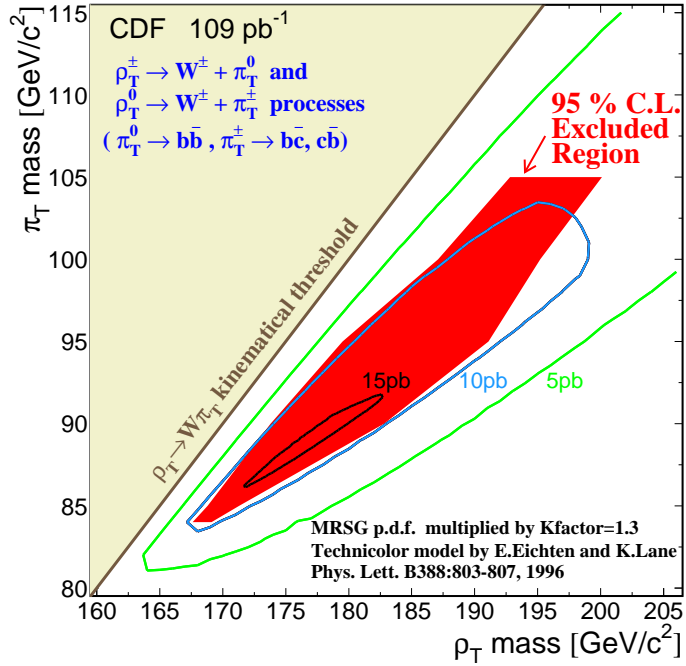}
\vspace{3.0cm}
\caption[cap]{\small \addtolength{\baselineskip}{-.4\baselineskip} 
Excluded region for the CDF search for color singlet
  techni-$\rho$ search in the mode $\rho_T \to W^\pm P_T$.
\protect\cite{Affolder:2000hp}}
\label{fig:rho1-count}
\end{figure}

\begin{figure}
\vspace{5.0cm}
\includegraphics{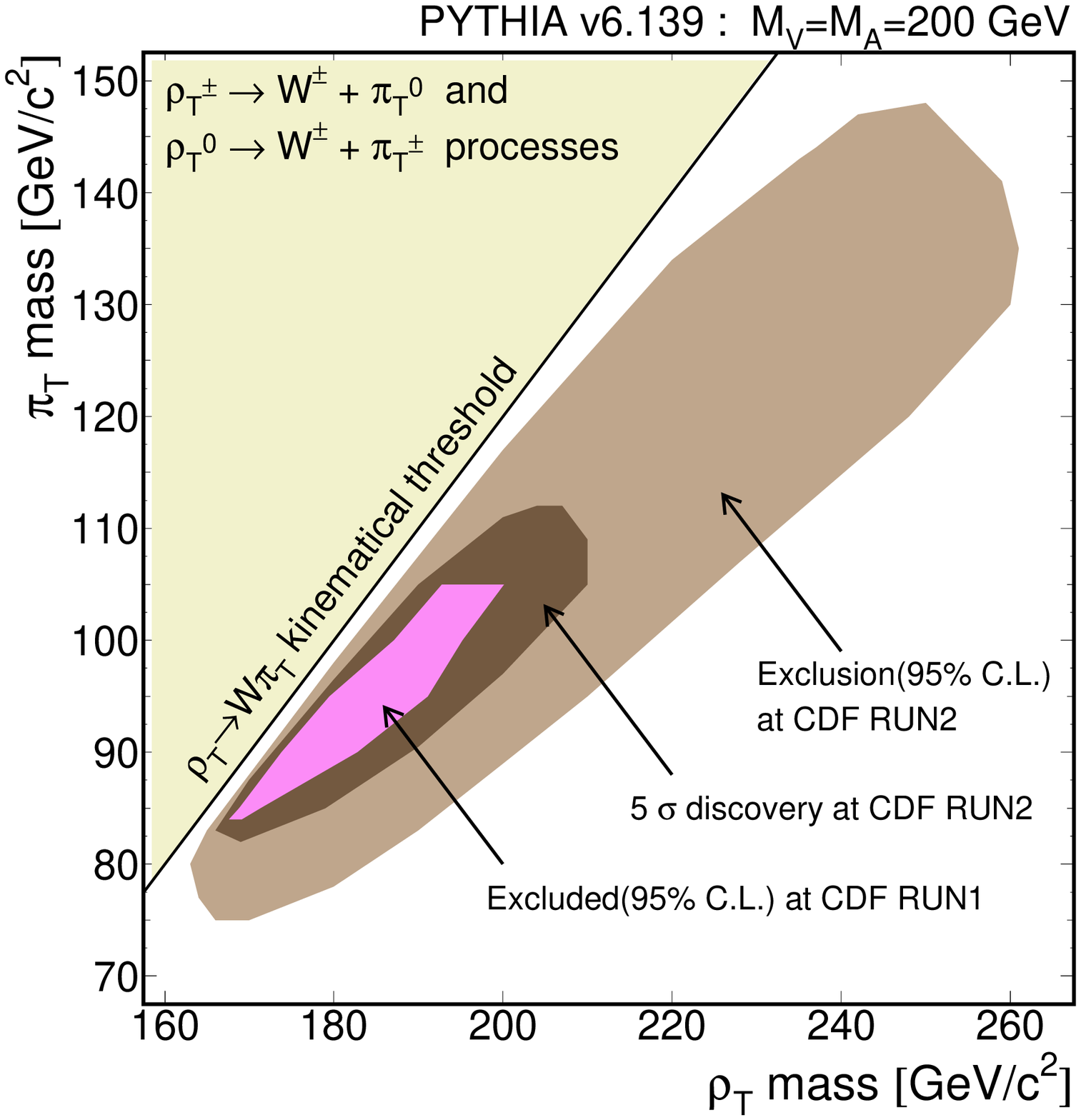}
\vspace{3.0cm}
\caption[cap]{\small \addtolength{\baselineskip}{-.4\baselineskip} 
Predicted reach of CDF in Run II for $\rho_T \to W \pi_T$
  assuming $M_V = M_A = 200$ GeV \protect\cite{Lane:2000pa}
  (note the notational variance, $P_T=\pi_T$).}
\label{fig:rho2-count}
\end{figure}

CDF performed a counting experiment looking for $\rho_T \to W P_T \to \ell
\nu b\bar{b}, \ell \nu c \bar{b}$ in 
 109 pb$^{-1}$ of Run I data
\cite{Affolder:1999du, Affolder:2000hp}.  
They selected candidate lepton plus 
two-jet events with at least one jet $b$-tagged. 
The presence of peaks in the $M_{b,jet}$
and $M_{W,b,jet}$ distributions would signal the presence of the
$P_T$ and $\rho_T$, respectively, as indicated in figure
\ref{fig:rho1-peak}.  No deviation
from Standard Model expectations was observed.  CDF therefore 
set upper limits on the techni-$\rho$ production cross-section for
specific pair of techni-$\rho$ and technipion masses as indicated in Figure
\ref{fig:rho1-count}.  In Run II, with a larger data sample and a doubled
signal efficiency, CDF expects to explore a significantly larger range of
$\rho_T$ masses using the same selection criteria, as indicated in Figure
\ref{fig:rho2-count}.

CDF also searched \cite{Affolder:1999du, Affolder:2000hp} for 
technipions in the shape of the two $b$-jet mass distribution
in $\ell + 2$-jet and $4$-jet events,  using 91 pb$^{-1}$ of Run I
data.  The former topologies
arise as described above; the latter can result from
either $\rho_T \to W P_T$ or $\rho_T \to
P_T P_T$ where the technipions decay to heavy flavors and the $W$
decays hadronically.  
 CDF notes that the upper limits ($\sim
100$ pb) this search sets on production of $\sim 200$ GeV
techni-$\rho$ decaying to $\sim 100$ GeV technipions provides no
immediate improvement in the constraints on TC models.  
Run II should provide significantly improved sensitivity to these modes.

\begin{figure}[tb]
\vspace{5.0cm}
\includegraphics{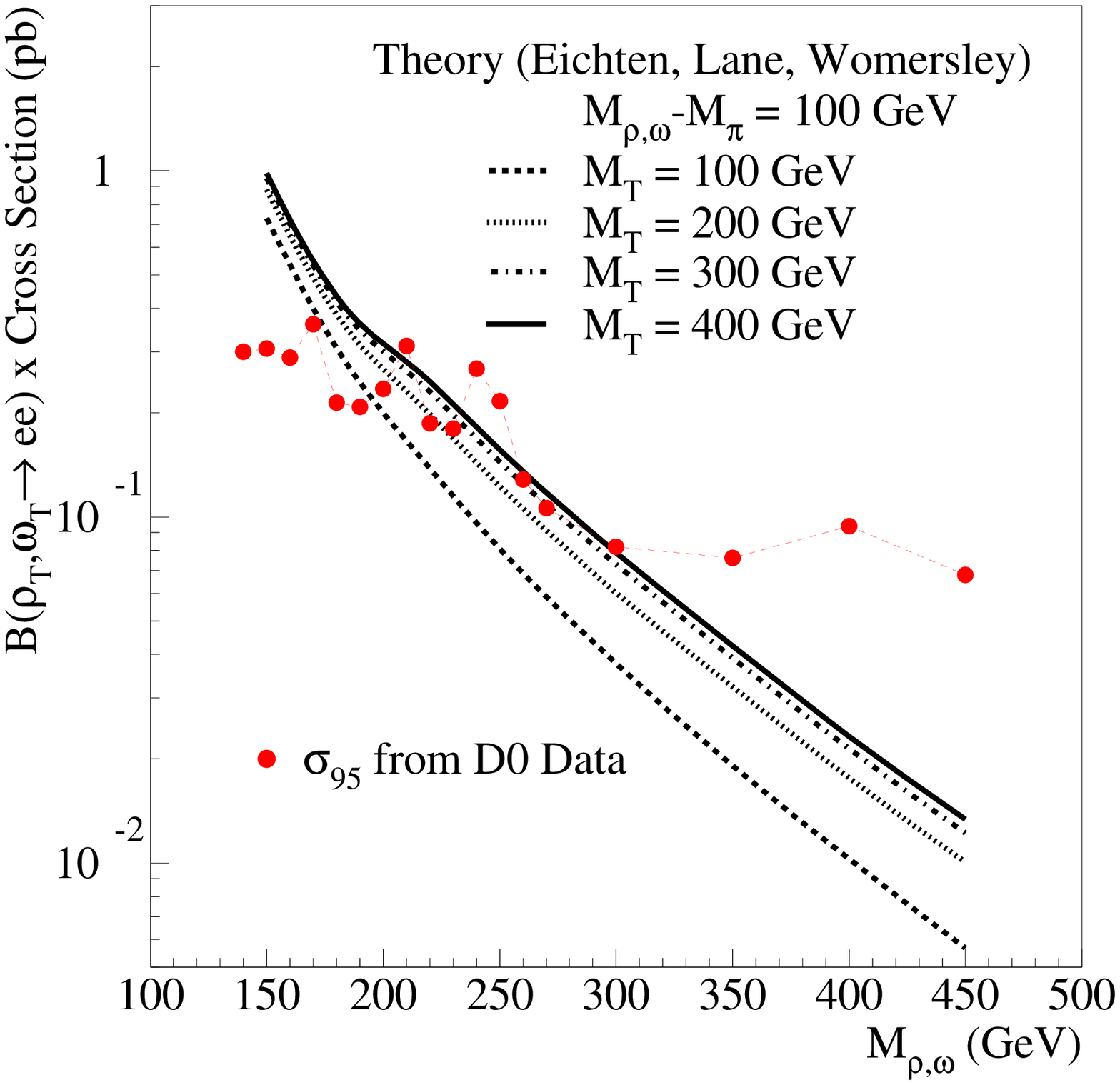}
\vspace{3.5cm}
\caption[cap]{\small \addtolength{\baselineskip}{-.4\baselineskip} 
Excluded regions for the D0 search for $\rho_T,
  \omega_T \to e^+e^-$ \protect\cite{Abazov:2001qd}.}
\label{fig:d0-ee}
\end{figure}

\begin{figure}[tb]
\vspace{5.0cm}
\includegraphics{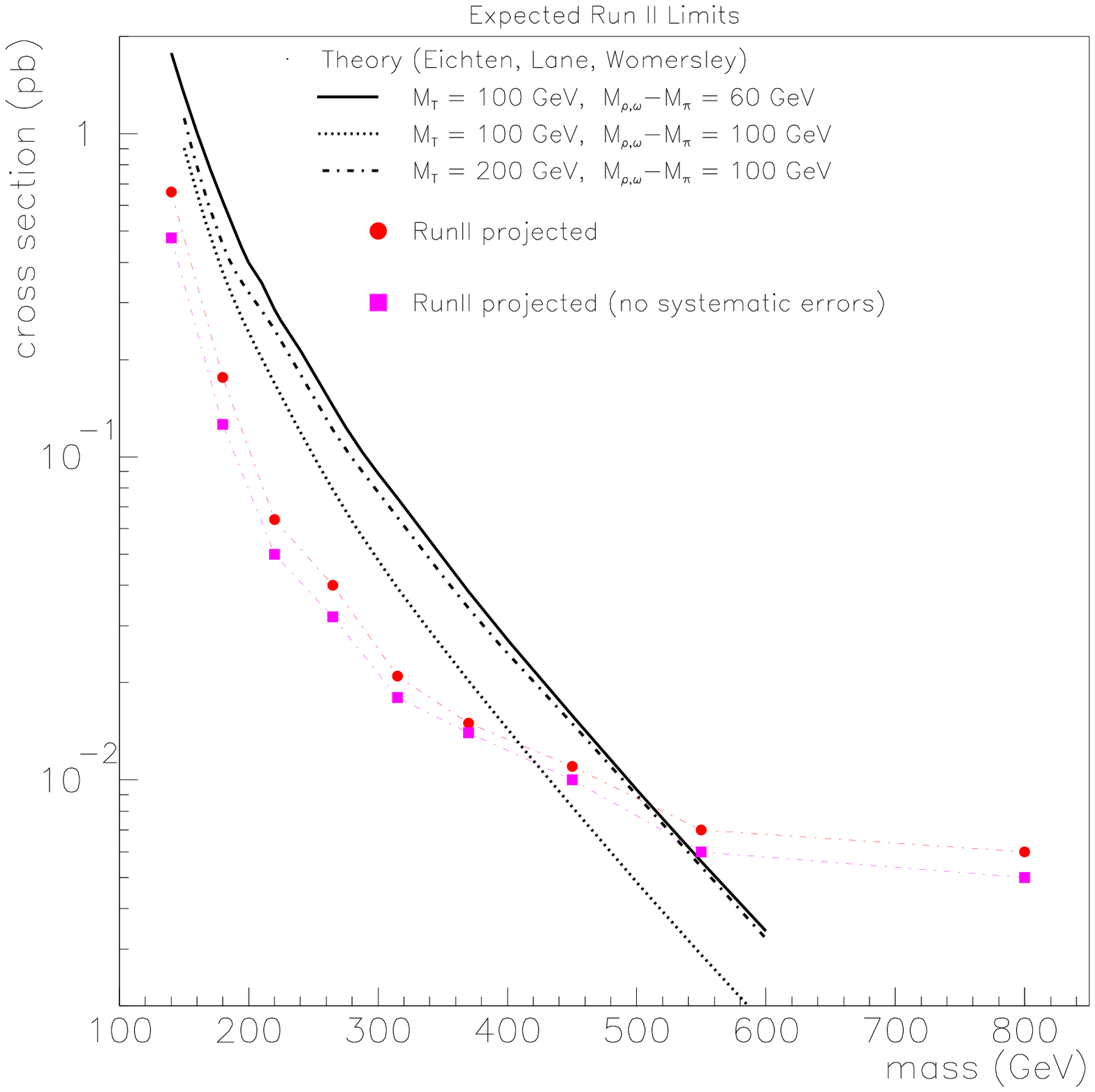}
\vspace{3.50cm}
\caption[cap]{\small \addtolength{\baselineskip}{-.4\baselineskip} 
Projected reach of the D0 detector in Tevatron Run IIa
  for $\rho_T, \omega_T \to e^+e^-$ \protect\cite{Lane:2000pa}. }
\label{fig:d0-ee2}
\end{figure}

D0 has looked for the light $\rho_T$ and $\omega_T$ in WTC in which
the vector mesons are unable to decay to $W+ P_T$
\cite{Handa:1999gx,Abazov:2001qd}.  To study the decays $\rho_T,
\omega_T \to e^+ e^-$, they selected events with two isolated
high-$E_T$ electrons, one required to be central, from $\sim 120$
pb$^{-1}$ of Run I data.  No excess above expected backgrounds was
seen.  For a model with $N_{TC} = 4$, techniquark electric charges of
$Q_D = Q_U - 1 = -1/3$ and $M_T = 100$ GeV (this mass parameter is the
Low-Scale TC scale, of order the weak interaction scale)
\cite{Eichten:1997yq}, $\rho_T$ and $\omega_T$ mesons with masses
below $200$ GeV were ruled out at 95\% {\em c.l.}  provided that
$M_{\rho_T} - M_{P_T} < M_W$ (see Figure \ref{fig:d0-ee}).  Projected
improvements in the reach of this search in Run II are shown in Figure
\ref{fig:d0-ee2}.

\begin{figure}[tb]
\vspace{5.0cm}
\includegraphics{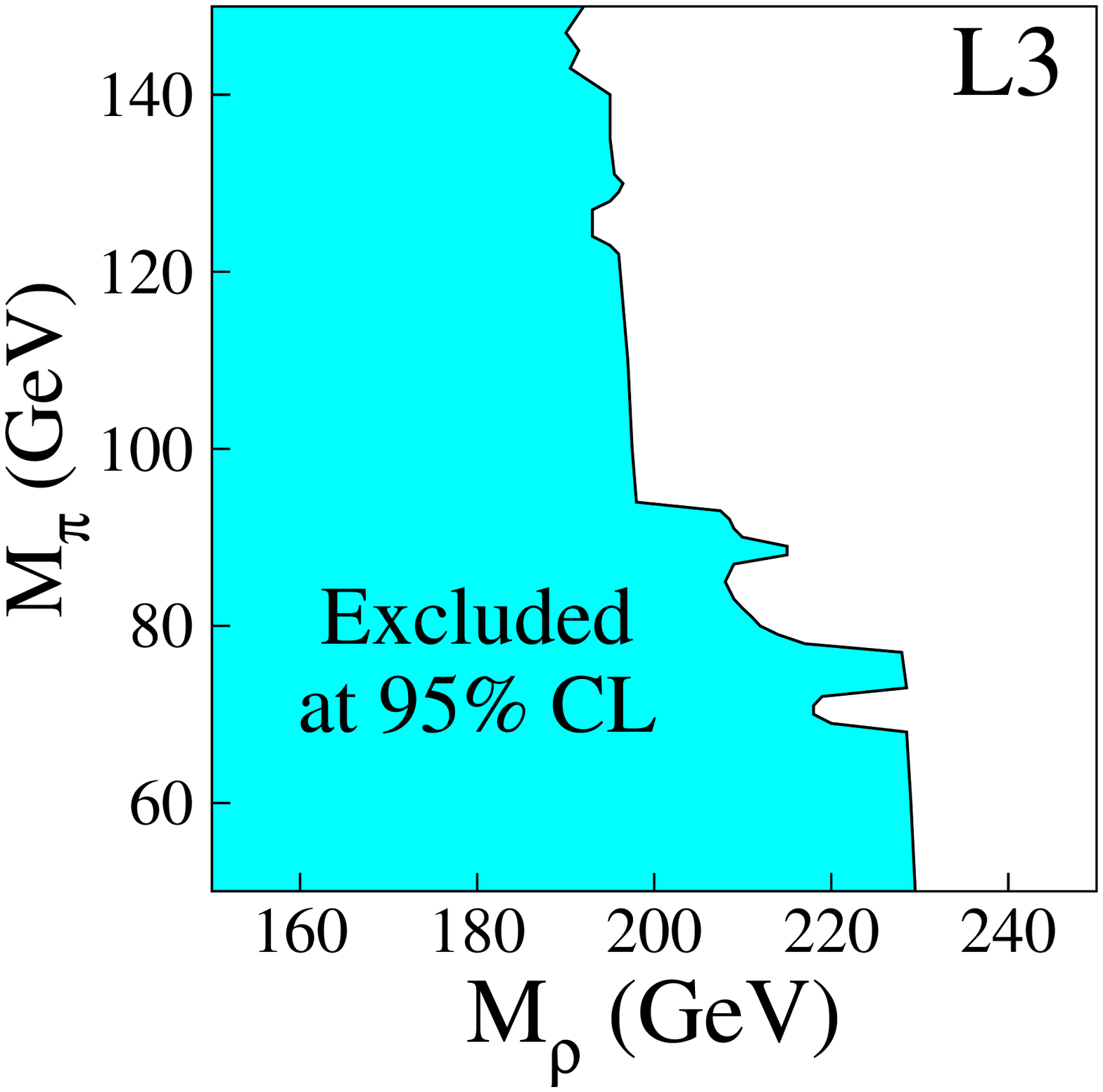}
\vspace{3.50cm}
\caption[cap]{\small \addtolength{\baselineskip}{-.4\baselineskip} 
The $M_{\rho_T} - M_{P_T}$ region excluded by L3 at
  95\%c.l. \protect\cite{Kounine:1999tp}. }
\label{fig:l3-prelim}
\end{figure}

L3 used $176.4$ pb$^{-1}$ of data collected at an average center of mass
energy of $188.6$ GeV to search for color-singlet $\rho_T$.
\cite{Kounine:1999tp}.  The search took into account the four major
techni-$\rho$ decay modes
\begin{equation}
\rho_T \to W_L W_L, W_LP_T, P_TP_T, \gamma P^{(')}_T
\end{equation}
for the following range of techni-$\rho$ and technipion masses
\begin{equation}
50 \;GeV < M_{P_T} < 150\; GeV\ \ \ \ \ \ \ \ \ \ 150 \;GeV < M_{\rho_T} < 250
\; GeV
\end{equation}
In  the $WW$ decay channel, all decay modes of the W bosons are
included. The result
is that an upper limit of 0.47 pb was set at 95\% {\em c.l.} on the
possible increase of the $e^+ e^- \to WW$ cross-section due to
contributions from TC.  

In other decay channels, the technipions decay
predominantly to $b\bar{b}$ or $b\bar{c}$ (the calculated branching
ratios ranged from 50\% to 90\%)
no statistically
significant excess of techni-$\rho$-like events was observed.  L3
found the following approximate mass ranges for technipions and
techni-$\rho$'s were excluded at 95\% {\em c.l.}, for $150< M_{\rho_T} < 200$
GeV, technipion masses $50 < M_{P_T} < 150$ GeV were excluded and
for $50 < M_{P_T} < 80$ GeV, techni-$\rho$ masses $150 < M_{\rho_T}
<$ 230 GeV were excluded.  Figure \ref{fig:l3-prelim} shows the
boundaries of the excluded region.

\begin{figure}[tb]
\vspace{5.0cm}
\includegraphics{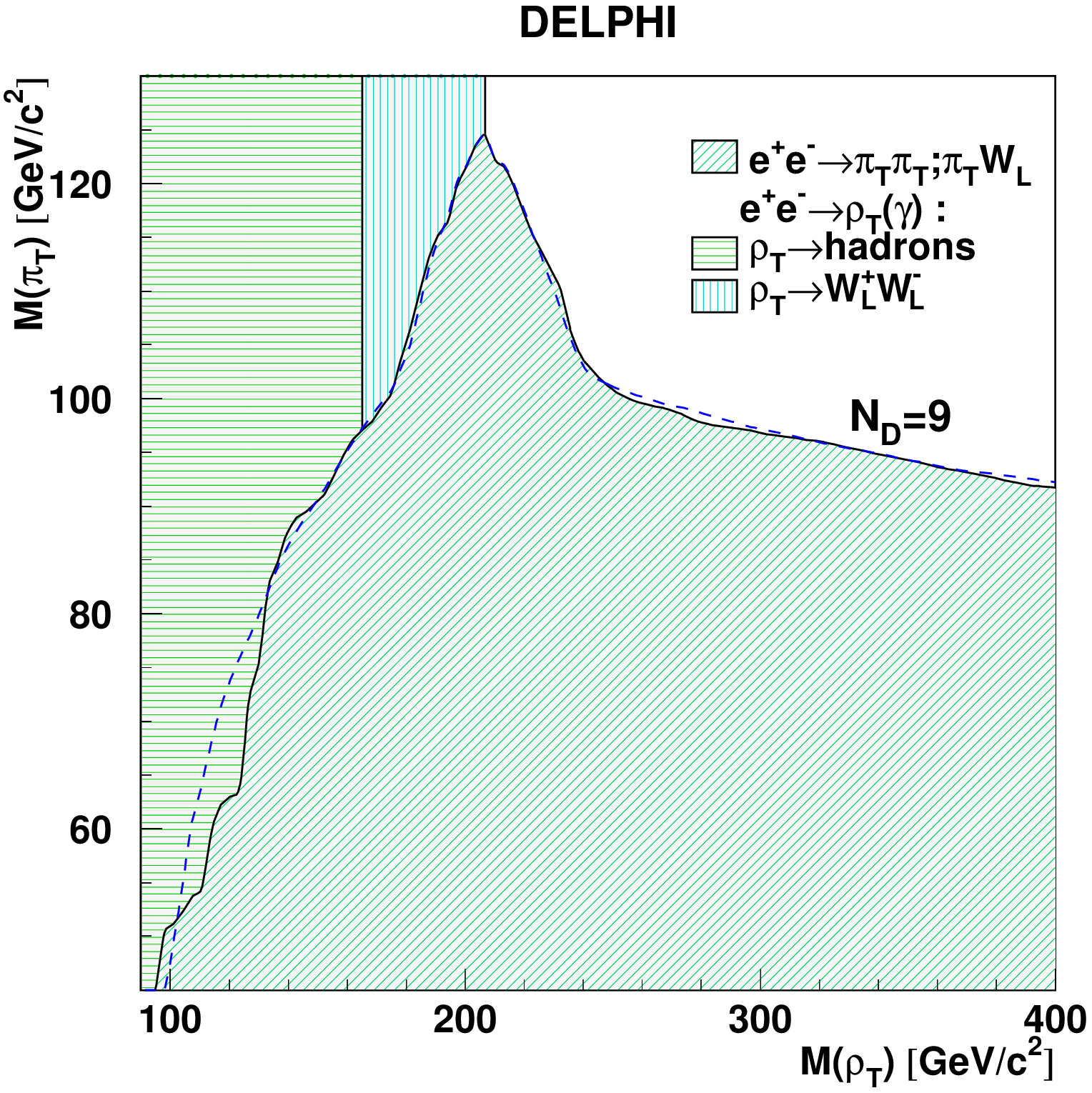}
\vspace{3.5cm}
\caption[cap]{\small \addtolength{\baselineskip}{-.4\baselineskip} 
 $M_{\rho_T} - M_{P_T}$ region excluded by DELPHI at
  95\%c.l.  \protect\cite{Abdallah:2001ft}}
\label{fig:delphi-prelim}
\end{figure}

The DELPHI collaboration searched for technipion production in the
final states $W_L P_T$ and $P_T P_T$ 
in four-jet final states \cite{Abdallah:2001ft}.  
No significant
contribution of technipion production was observed, either in the total
rate, or in the $M_{jj}$ distribution.  DELPHI thus excludes at 95\%
{\em c.l.} technipions of mass between $70$ and $130$ GeV 
when the techni-$\rho$
mass is between $120$ and $200$ GeV.  For heavier $M_{\rho_T}$, the range
of excluded technipion masses shrinks, as shown in Figure
\ref{fig:delphi-prelim}, e.g., 
for $M_{\rho_T} = 250$ GeV, $70 < M_{P_T} <
100$ GeV is ruled out.

\begin{figure}[tb]
\vspace{5.0cm}
\includegraphics{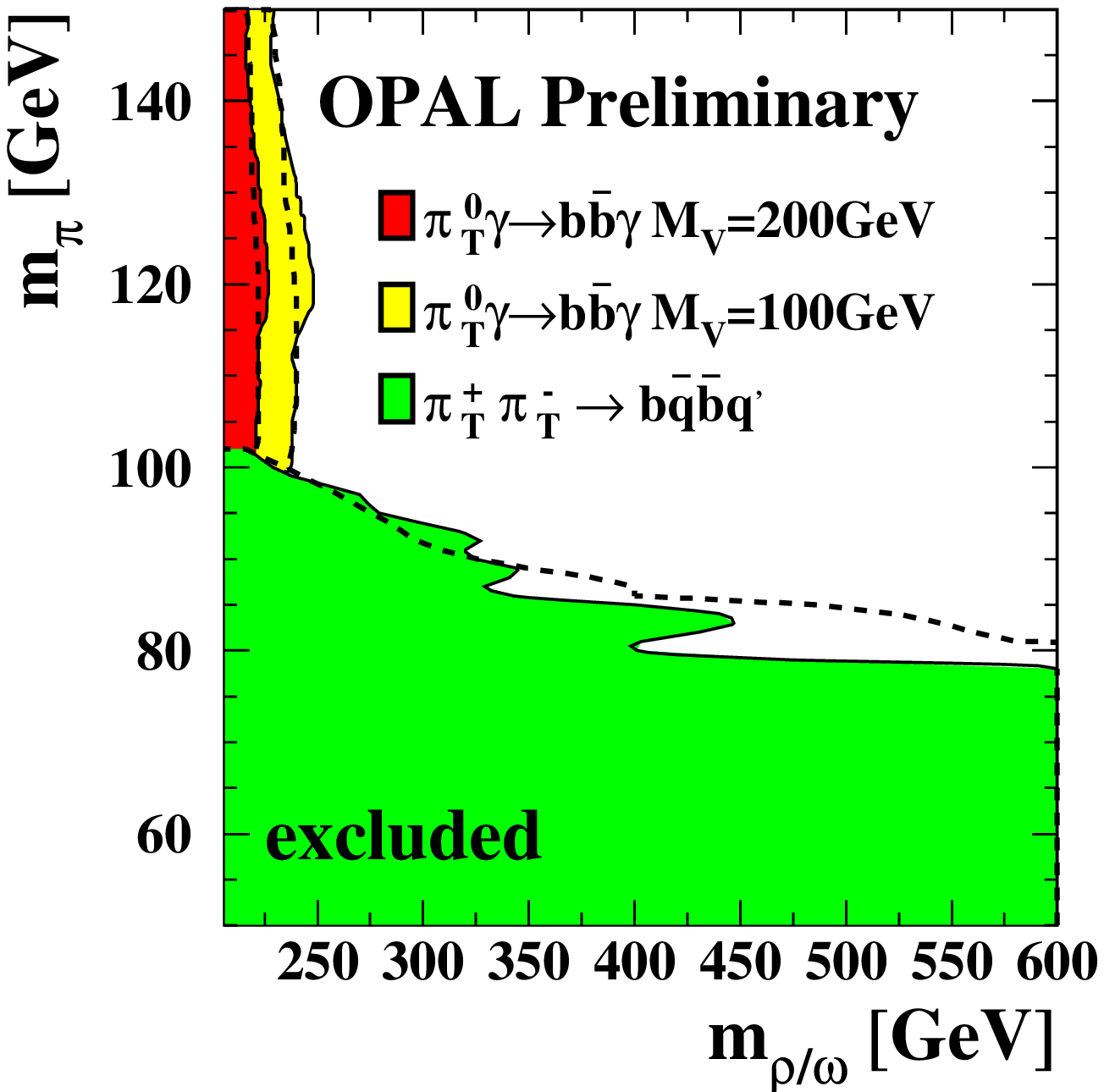}
\vspace{3.5cm}
\caption[cap]{\small \addtolength{\baselineskip}{-.4\baselineskip} 
The 95\% c.l. excluded region in the ($M_{\rho_T,\omega_T}$) plane
  from the combination of the $P^+_T P^-_T \to b\bar{q}\bar{b}q'$
  search and the $P^0_T\gamma \to b\bar{b}\gamma$ search.  The dashed
  lines show the corresponding median expected exclusions for the
  background only hypothesis.  \protect\cite{opal:2001prelim}}
\label{fig:opalprelimmpit}
\end{figure}

The OPAL collaboration \cite{opal:2001prelim} searched for
technihadrons in the channels $e^+e^- \to \rho_T, \omega_T \to P^+_T
P^-_T \to b \bar{q} \bar{b} q'$ and $e^+e^- \to \rho_T, \omega_T \to
P^0 \gamma \to b\bar{b}\gamma$ at $\sqrt{s} \approx 200-209$ GeV.  
No significant excess over
SM background was observed in the total number of events or the dijet
mass distribution.  Cross-section $\times$ branching ratio upper
bounds of 60 - 200 fb were set at 95\% c.l., depending on the
$P^\pm+T$ mass.  In the search for neutral technipions, events with a
pair of b-jets and an energetic isolated photon were chosen.  After
kinematic and topological cuts, no excess over the SM background was
observed.  An upper limit of approximately 50 fb (except when the
technipion mass is nearly $M_Z$) on the cross-section $\times$
branching ratio was established at 95\% c.l. The combined results of
these searches are shown in Figure \ref{fig:opalprelimmpit}.

CDF searched  for $\omega_T \to \gamma P_T$,
assuming that the $P_T \rightarrow b\bar{b}$ \cite{Abe:1998jc}.  
About 200 
events with a photon, a $b$-jet and at least one additional jet were
selected from $85$ pb$^{-1}$ of data.  This was found to be consistent
with the Standard Model.  The distributions of
$M_{b,jet}$ and $M_{\gamma,b,jet} - M_{b,jet}$ show no evidence of
resonance production.  Thus, CDF obtains an upper limit on
(cross-section)$\times$(branching ratio), which it compared to the
predictions of the TC models of ref. \cite{Eichten:1997yq}.  The range of
excluded masses of
$P_T$ and $\omega_T$ are shown in Figure \ref{fig:omega-t} (the exclusion
region boundary is ragged because of statistical fluctuations in
the data).  CDF noted that, if the
channels $\omega_T \to P_T P_T P_T$ or $\omega_T \to Z P_T$
were open, the excluded region would be reduced as shown in the
Figure \cite{Abe:1998jc}.   
Had their Monte Carlo included $\rho_T \to
\gamma P_T$ and $\rho_T, \omega_T \to \gamma \pi^\prime_T$, their
excluded region would increase.

\begin{figure}[tb]
\vspace{8.0cm}
\includegraphics{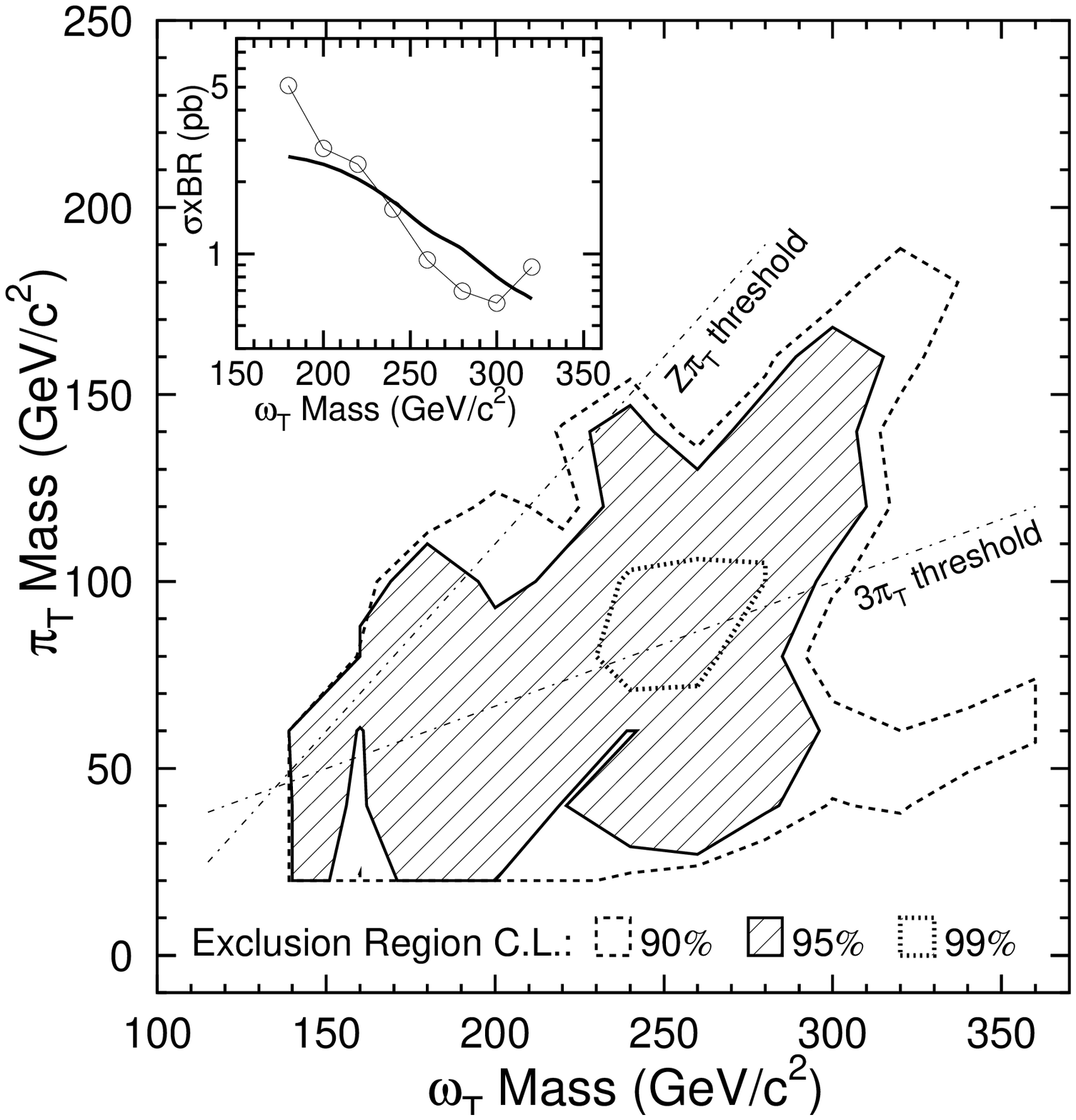}
\vspace{3.0cm}
\caption[cap]{\small \addtolength{\baselineskip}{-.4\baselineskip} 
The 90\%, 95\% and 99\% {\em c.l.} 
exclusion regions for the
CDF search for $\omega \to \gamma P_T$ \cite{Abe:1998jc}.  The inset 
shows the limit on $\sigma\cdot B$ for $M_{P_T} = 120$ GeV;  
the circles 
indicate the limit and the solid line shows 
the prediction from \protect{\cite{Eichten:1997yq}}.}
\label{fig:omega-t}
\end{figure}

Future experiments should provide 
definitive tests of Low-Scale TC 
\cite{Eichten:1997yq}. 
The cross-section for $\omega_T$ production is
$\sim 1-10$ pb for Tevatron, and  an order of
magnitude larger at the LHC. Simulations by the ATLAS
collaboration \cite{ATLAS:TDR} show that the relatively light
techni--vector resonances of Low-Scale TC
\cite{Lane:1999uh} are well within reach 
at LHC (Figure \ref{ieri-atlas}).

\begin{figure}[tbhp]
 \vspace{9.0cm}
\includegraphics{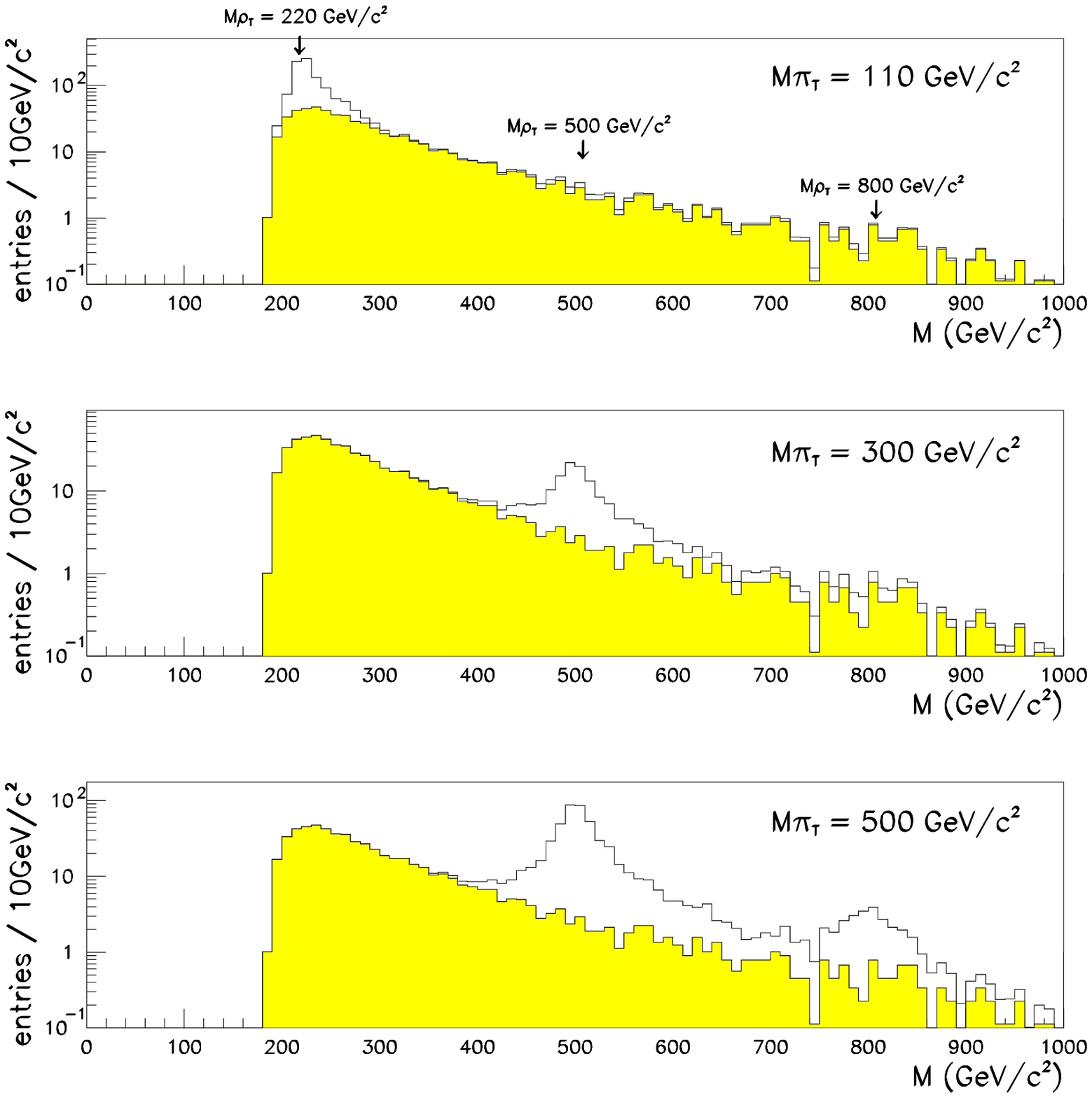}
\vskip5.0truecm
 \caption{\small \addtolength{\baselineskip}{-.4\baselineskip}
   Simulated event and background rates in the ATLAS detector 
   for $\rho_T \to
   W^\pm Z \to \ell^\pm\nu_\ell \ell^+\ell^-$ 
   for various $M_{\rho_T}$ and
   $M_{P_T}$ in low-scale TC models \protect\cite{Lane:1999uh}; from
     Ref.~\protect\cite{ATLAS:TDR}.
    \label{ieri-atlas} }
\end{figure}

Experiments at an $e \gamma$ collider have potential to
discover and study an $\omega_T$ with a mass up to about $\sim 1$ TeV in the
processes $e^- \gamma \to e^-\omega_T \to e^-\gamma Z, e^-W^+ W^- Z$
\cite{Godfrey:1998pm}.  As shown in Figure \ref{fig:god-omega}, at the stage
where only detector acceptance cuts have been applied, the
cross-section for an $\omega_T$ decaying to $WWZ$ can be of order
$\sim 10-100$ fb,
 and as much as an order of magnitude above background 
 ($\omega_T\rightarrow Z\gamma$ is below
background).   Applying various kinematic cuts,
the $\omega_T$ could be visible for a 
range of masses and decay widths.  For example, with 200 $fb^{-1}$ 
collected at $\sqrt{s} =$ 1.5 TeV, an $\omega_T$ of width $100$ GeV can
be detected at the $3\sigma$ level up to a mass of about $\sim 1.3$ TeV.

\begin{figure}[htbp]
\begin{center}
\vspace{8cm}
\includegraphics{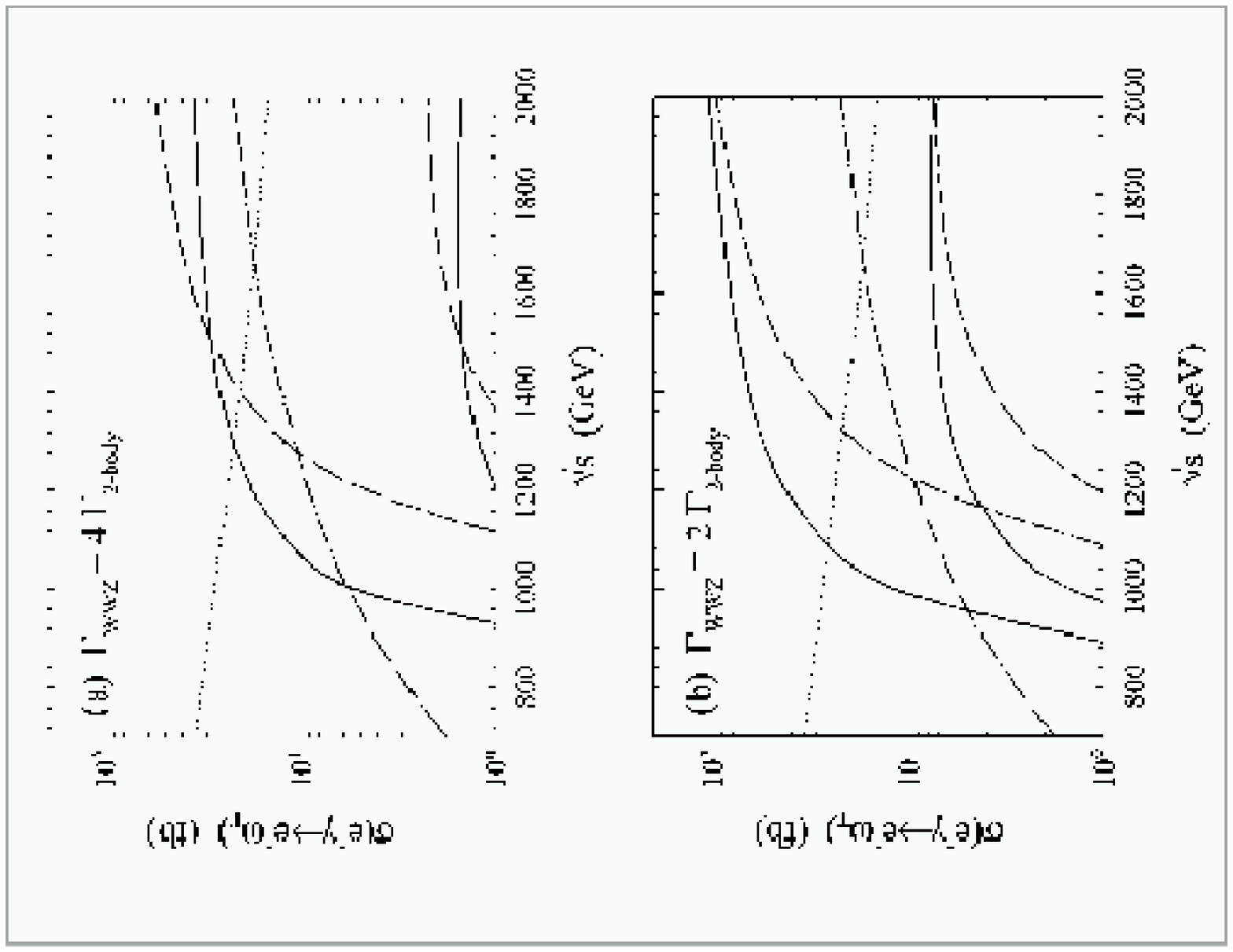}
\vspace{1cm}
\caption{\small \addtolength{\baselineskip}{-.4\baselineskip} 
Cross-section vs $e^+e^-$ CM energy for $e^-\gamma \to e^-\omega_T$
  with $\omega_T \to WWZ$ and $Z\gamma$ and the SM backgrounds thereto.
  Solid (long-dashed) lines are for $M_{\omega_T} = 0.8$ (1.0) TeV.  In each
  case, the upper curve is for $WWZ$ and the lower is for $Z\gamma$.  The
  dash-dot (dotted) line is for the SM $e^- WWZ$ ($e^- Z\gamma$) background.
  In (a), $0.25 \Gamma_{WWZ} = \Gamma_{Z\gamma}$ = 5 GeV (solid) or 20 GeV
  (long-dashed); in (b), $0.5 \Gamma_{WWZ} = \Gamma_{Z\gamma}$ = 15 GeV
  (solid) or 40 GeV (long-dashed). (From \protect\cite{Godfrey:1998pm})}
\label{fig:god-omega}
\end{center}
\end{figure}

%


\subsubsection{Separate Searches for color-singlet  
$P^0, P^0{}^\prime$}

\begin{table}
  \begin{center} 
\label{tab:tcsm}
\vspace{\baselineskip}
    \begin{tabular}{|c|c|c|}   \hline
      \multicolumn{1}{|c|}{$M_{P_T^{0'}} \leq$}
      &\multicolumn{2}{c|}{$N_{TC}\sqrt{N_D} \leq$}  \\ 
      &\multicolumn{1}{c}{${P_T^{0'}} \to g g $} &
      \multicolumn{1}{c|}{${P_T^{0'}} \to \bar{b}b$}\\ \hline
      30 GeV & $28^{(b)}$ & $24^{(b)}$ \\ 
      60 GeV & $67^{(b)}$ & $70^{(b)}$ \\ 
      80 GeV & $283^{(b)}$ & $25^{(a)}$ \\ 
      100 GeV & --- & $40^{(a)}$ \\ 
      120 GeV & --- & $42^{(a)}$ \\ 
      140 GeV & --- & $49^{(a)}$ \\ 
      160 GeV & --- & $68^{(a)}$ \\ \hline
    \end{tabular}
    \caption{\small \addtolength{\baselineskip}{-.4\baselineskip} 
Limits (from ref. \protect\cite{Lynch:2000md} 
on the number of technicolors, $N_{TC}$,
    and weak doublets of technifermions, $N_D$, for hadronically
    decaying PNGBs in TCSM~\protect\cite{Lane:1999uh, Lane:1999uk}
    models as a function of the upper bound on the PNGB mass.  
    The superscripted labels indicate the
    data used to calculate the limits: (a) means ${\cal A}_{\gamma\gamma P^a}$ 
    ; (b) means ${\cal A}_{\gamma Z P^a}$.} 
  \end{center}
\end{table}

As discussed in Sections 3.4 and 4, many TC models 
require a large number $N_D$ of
weak doublets of technifermions.  For a given TC
gauge group $SU(N_{TC})$, the number of doublets required to make the gauge
coupling $g_{TC}$ ``walk'' is $N_D
\approx 10$,
as in the models of refs.~\cite{Lane:1995gw, Lane:1996ua, Lane:1998qi}.  
Topcolor-Assisted TC models (see
Section 4) also tend to require many doublets of technifermions
~\cite{Lane:1999uh, Lane:1995gw}.  One phenomenologically
interesting consequence of the large number of doublets is the
presence of PNGB states with small technipion decay constants: $F_T
\approx v_0/\sqrt{N_D}$. Such states will be lighter, and have
generically longer lifetimes.

As outlined in Section 2.3.4, data from LEP can place limits on
single production of light neutral PNGB's from a variety of TC models.
One benchmark example is the Lane's low-scale TC ``Straw Man Model''
(TCSM)~\cite{Lane:1999uh, Lane:1999uk}, in which the lightest
technifermion doublet, composed of technileptons $T_U$ and $T_D$ can
be considered in isolation.  The result is two, nearly degenerate
neutral mass eigenstates, whose generators are given by $P_T^0 \sim
\bar{T}_U \gamma_5 T_U - \bar{T}_D \gamma_5 T_D$ and $P_T^0{}' \sim
\bar{T}_U \gamma_5 T_U + \bar{T}_D \gamma_5 T_D$.  As shown in 
ref. \cite{Lynch:2000md} LEP searches for
hadronically-decaying scalars produced in association with a Z or
$\gamma$ may be used to place an upper bound on the product
$N_{TC}\sqrt{N_D} {\cal A}$ (where ${\cal A}$ is the relevant anomaly
factor).  Inserting the value of ${\cal A}$ appropriate to a
particular PNGB yields an upper on $N_{TC}\sqrt{N_D}$.
Table~\ref{tab:tcsm} gives these bounds as a function of PNGB mass for
the cases where either 2-gluon or $\bar{b}b$ decays dominate.

Consider the case where $M_{P_T^{0}{}'} \leq 30$ GeV and $b$-quark
decays dominate; the limit $N_{TC} \sqrt{N_D} \leq 24$ applies.  As a
result, for $N_{TC} = (4, 6, 8, 10, 12)$ the largest number of
electroweak doublets of technifermions allowed by the LEP data is,
respectively, $N_{D} = (36, 16, 9, 5, 4)$.  The results are very
similar if the two-gluon decays of the PNGB dominate instead.  How do
these results accord with the requirements of Walking TC?  Requiring
the one-loop TC beta function to satisfy $\beta_{TC}\approx 0$,
implies that $11 N_{TC} / 4$ weak doublets of technifermions are
needed, according to eq. (\ref{eq:ehs:betaone}).  The analysis of
ref. \cite{Lynch:2000md} thus shows that WTC and a very light
$P_T^{0}{}'$ can coexist only in models with $N_{TC}$ = 4, 6.
Similarly, the size of the TC group is restricted to $N_{TC} \le 6
[12]$ if the PNGB mass is 80 GeV [160 GeV].  The results are similar
if the 2-loop $\beta$-function is used.

As a second example, we mention the results of
ref. \cite{Lynch:2000md} for the WTC model of Lane and
Ramana~\cite{Lane:1991qh} whose LEP--II and NLC phenomenology was
studied by Lubicz and Santorelli~\cite{Lubicz:1996xi}. The model has
$N_D = 9$: one color-triplet of techniquarks ($N_Q = 1$) and six
color-singlets of technileptons ($N_L = 6$).  Of the several neutral
PNGBs in this model, the one whose relatively large anomaly factors
and small decay constant ($F_T \approx 40$ GeV) makes it easiest to
produce is $P_L^3 \sim \bar{N}_\ell \gamma_5 N_\ell -
\bar{E}_\ell \gamma_5 E_\ell $ where the subscript implies a sum over
technilepton doublets.  This PNGB is expected to have a mass in the
range $\sim 100 - 350$ GeV \cite{Lubicz:1996xi}.  
Depending on the value of the ETC
coupling between the PNGB and fermions, the dominant decay of this
PNGB may be into a photon pair or $\bar{b}b$.  Ref. \cite{Lynch:2000md} 
finds that if the
two-photon decays dominate, the PNGB must have a mass in excess of 160
GeV; if the $\bar{b}b$ decay is preferred, the mass range 80 GeV $\leq
M_P \leq$ 120 GeV is excluded.


\subsubsection{Separate searches for color-singlet $P_T^\pm$}

Models with more than the minimal two
flavors of new fermions (e.g. TC with more than one weak
doublet) typically contain electrically charged PNGB
states ($P_T^{\pm}$).  Experimental limits on charged scalars are often
phrased in the language of a two-higgs-doublet model, i.e., in
terms of the ratio of the vacuum expectation values of the two doublets
($\tan\beta$) and the mass of the charged scalar ($M_H^+$).

\smallskip
\noindent
{\bf (i) LEP limits}
\smallskip

\begin{figure}[tbh]
\vspace{7cm}
\begin{center}
\includegraphics{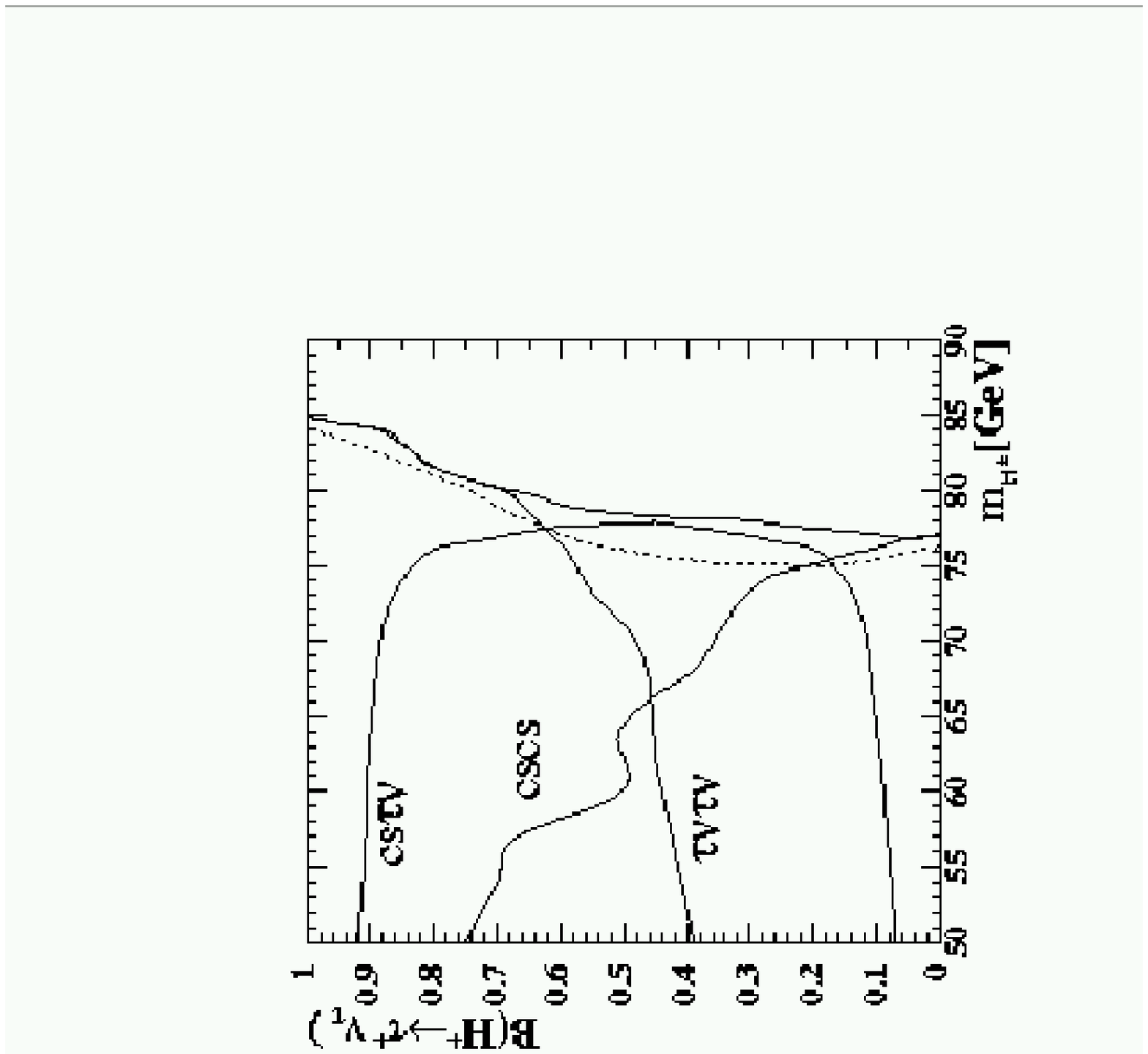}
\end{center}
\caption[a]{\small \addtolength{\baselineskip}{-.4\baselineskip} 
LEP lower bounds on $M_{H^\pm}$ as a function of $B(H^+ \to
  \tau\nu)$.  From \protect\cite{Groom:2000in}.}
\label{fig:chgddhlep}
\end{figure}

Color-singlet electrically-charged technipions $P_T^{\pm}$ with the quantum
numbers of the charged scalars in two-higgs-doublet models are directly
constrained by the limits on pair-production of $H^\pm$ derived from LEP
data.  When $\tan\beta$ is large, the charged scalars of two-higgs-doublet
models decay mostly to $\tau\nu$ ; if $\tan\beta$ is small, light charged
scalars decay to $c\bar{s}$, but for $M_{H^\pm}$ heavier than about 130 GeV,
the channel $H^+ \rightarrow t\bar{b} 
\rightarrow W b\bar{b}$ dominates \cite{Djouadi:1996gv,
  Ma:1998up}.  The LEP searches assume that $H^+ \to \tau^+\nu_\tau,
c\bar{s}$ (as consistent with the mass range they can probe), and derive
limits on the rate of $H^+H^-$ as a function of the branching ratio $B(H^+
\to \tau^+\nu_\tau)$.  As shown in Figure \ref{fig:chgddhlep}, the lower
limit is at least $77$ GeV for any value of the branching ratio.

\begin{figure}[tbh]
\vspace{6cm}
\begin{center}
\includegraphics{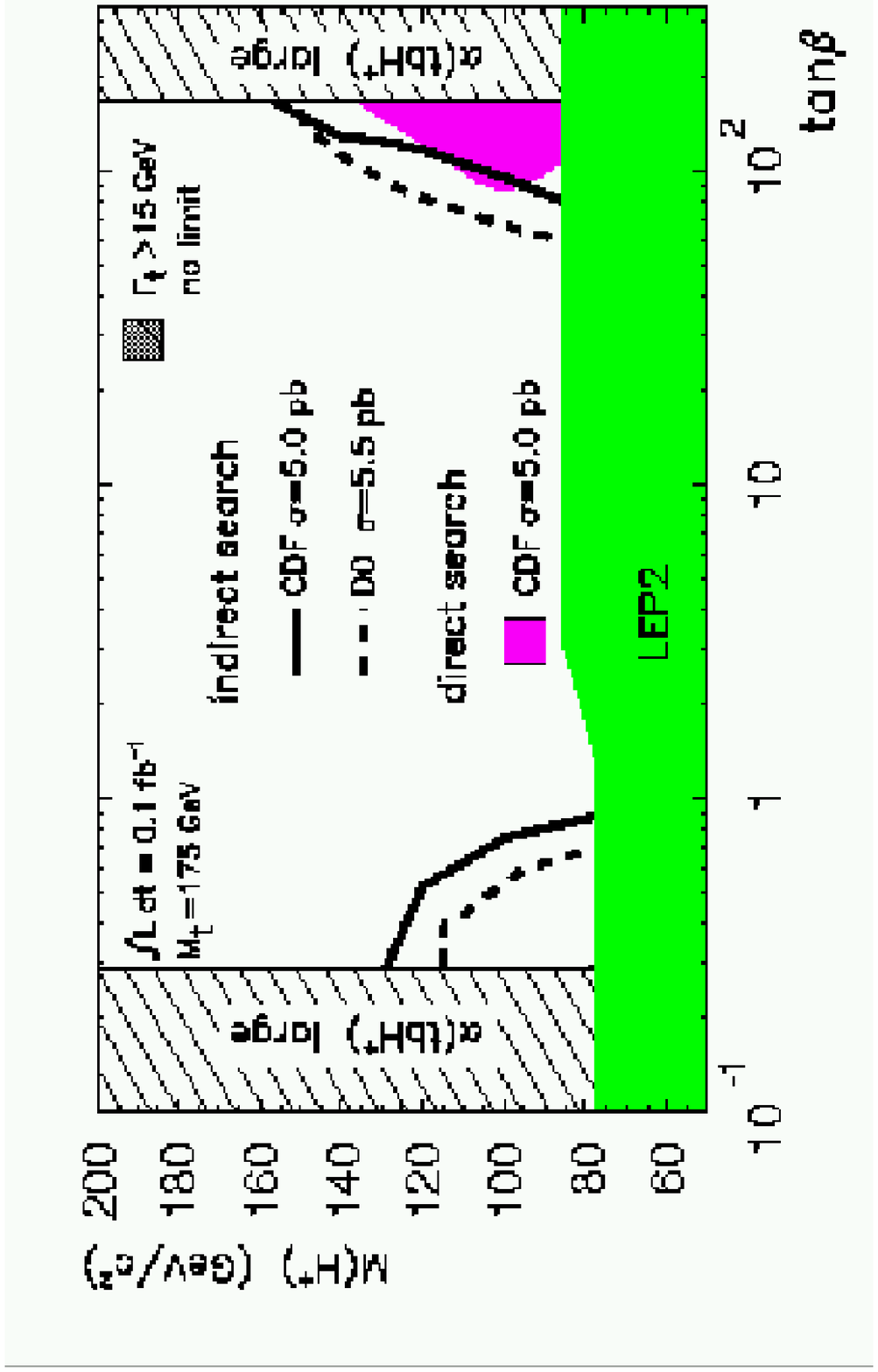}
\end{center}
\caption[a]{\small \addtolength{\baselineskip}{-.4\baselineskip} 
Limits on charged scalar mass as a funciton of $\tan\beta$. The
  95\% exclusion bounds from CDF and D0 studies of top decays are strong
  functions of $\tan\beta$. LEP limits from figure \ref{fig:chgddhlep} are
  also shown.  From \protect\cite{Groom:2000in}.}
\label{fig:chgddh}
\end{figure}

If the mass of the charged scalar is less than $m_t - m_b$, then the decay
$t \to H^+ b$ can compete with the standard top decay mode $t \to W b$.
Since the $t b H^\pm$ coupling can be parameterized in terms of $\tan\beta$
as \cite{higgshunt}
\begin{equation}
g_{tbH^+} \propto m_t\cot\beta(1+\gamma_5) + m_b\tan\beta(1-\gamma_5)\ ,
\end{equation}  
we see that the additional decay mode for the top is significant for either
large or small values of $\tan\beta$.  The charged scalar, in turn, decays
as $H^\pm \to c s$ or $H^\pm \to t^*b \to Wbb$ if $\tan\beta$ is small and
as $H^\pm \to \tau \nu_\tau$ if $\tan\beta$ is large.  In any case, the
final state reached through an intermediate $H^\pm$ will cause the original
$t\bar{t}$ event to fail the usual cuts for the lepton + jets channel.  A
reduced rate in this channel can therefore signal the presence of a light
charged scalar.  As shown in Figure
\ref{fig:chgddh}, D0 and CDF have each set a limit \cite{Groom:2000in} on
$M_{H^\pm}$ as a function of $\tan\beta$ and $\sigma_{tt}$.  In Run II the
limits should span a wider range of $\tan\beta$ and reach nearly to the
kinematic limit as shown in Figure~\ref{fig:chgddh-3}.

\begin{figure}[bt]
\vspace{8cm}
\begin{center}
\includegraphics{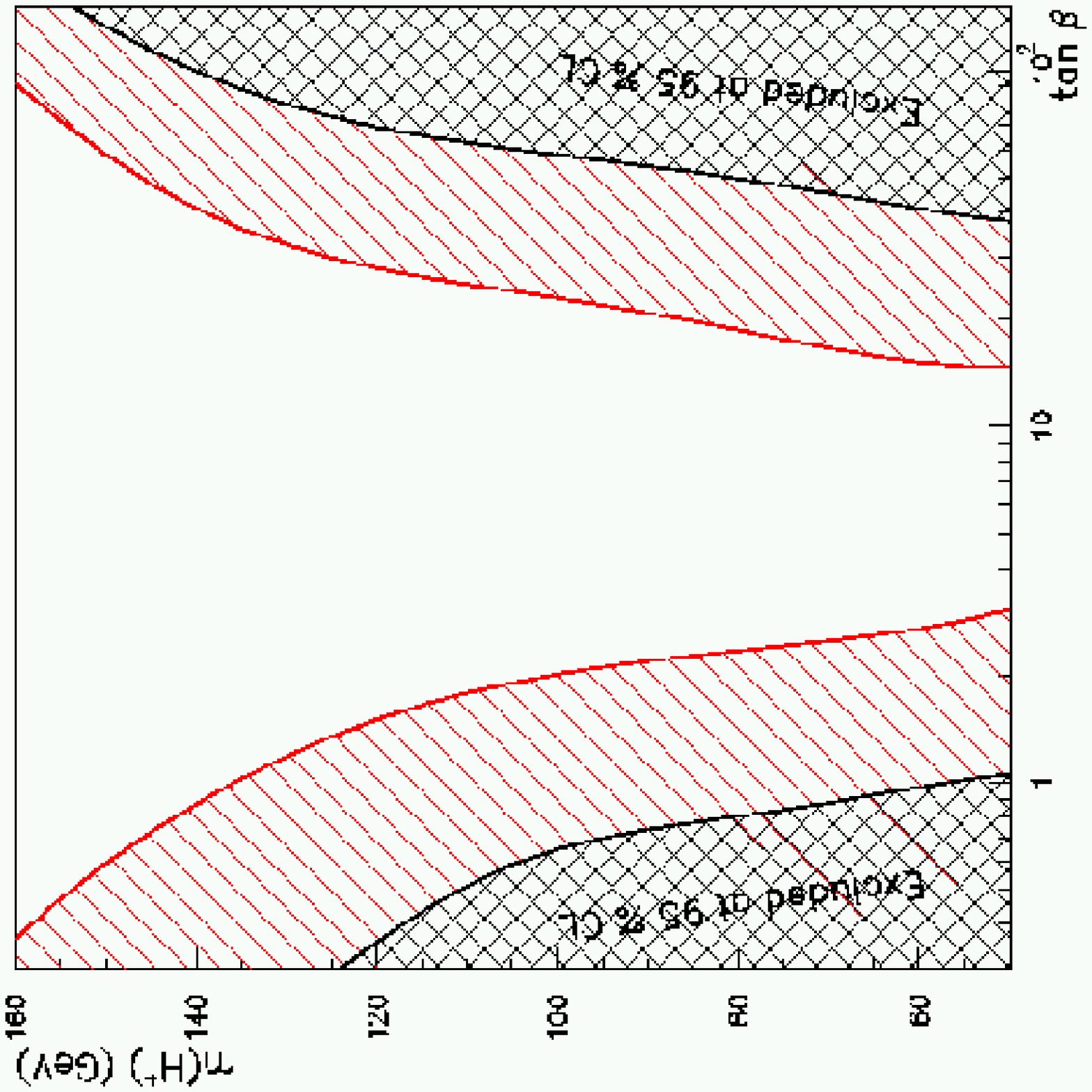}
\end{center}
\caption[b]{\small \addtolength{\baselineskip}{-.4\baselineskip} 
Projected Run II reach
of D0 charged
scalar search in $t\to H^\pm b$ assuming
$\sqrt{s} = 2$ TeV, $\int{\cal L} dt = 2 {\rm fb}^{-1}$, and
$\sigma(t\bar{t})$ = 7pb.}
\label{fig:chgddh-3}
\end{figure}

Limits on the mass and coupling of $P_T^{\pm}$ are also implied by the
experimental 95\% {\em c.l.} upper limit $B(b \to s \gamma) < 4.5\times 10^{-4}$
obtained by the CLEO Collaboration \cite{Alam:1995aw, Ahmed:1999fh}.
Radiative corrections due to the $P_T$ tend to increase \cite{Ellis:1986xy,
  Barger:1990fj} the branching ratio above the Standard Model prediction of
$(3.28\pm 0.33) \times 10^{-4}$ \cite{Chetyrkin:1997vx, Kagan:1998ym}, as
would be true in a model with an extended Higgs sector. To the extent that
this is the only new physics contribution to $b\to s \gamma$, it implies an
upper bound of order 300 GeV on the mass of the charged scalar.
However, the contributions of other new particles or non-standard gauge
couplings can also affect the branching ratio, making the exact limit quite
model-dependent.  Weaker, and also model-dependent, bounds can also be
derived from measurements of $b\to s\gamma$ and $b\to \tau \nu_\tau X$ and
from $\tau$-lepton decays at LEP \cite{Adriani:1993yn, Buskulic:1995gj,
  Adam:1996ts, Barate:1998vz, Ackerstaff:1998yk}.


\subsubsection{Searches for Low-Scale Color-octet Techni--$\rho$'s 
(and associated Leptoquark Technipionss)}

In models of Walking TC, in which enhancement of the technipion
masses prevents the decay $V_8 \to \bar{P}_T P_T$
($V_8$ is the color-octet techni--$\rho$), decay to dijets can
dominate \cite{Holdom:1981rm, Eichten:1996dx}.  The CDF Collaboration has
used 87 pb$^{-1}$ of Run I data to search for the effects of $V_8$'s
on the dijet and $b$-tagged dijet invariant mass spectra
\cite{Abe:1998uz}. No deviations from the Standard Model backgrounds were
observed.  A narrow $V_8$ with mass $350 < M_{V_{8}} <
440$ GeV is excluded at 95\% c.l. using the $b$-tagged distribution, as may
be seen in figure \ref{fig:harris-topgluon}. The mass range $260 <
M_{V_{8}} < 470$ GeV is ruled out using the untagged sample.

In searches for leptoquark
techni-$\pi$'s $P_{3}$, CDF sets $N_{TC} = 4$ and allows
a relevant parameter, $\Delta M$, to
take on the expected value of 50 GeV  and the
limiting values of 0 and $\infty$. 
More precisely, $\Delta M$ is the mass difference between $P_8$ and
$P_3$, and enters the calculation of
the partial width for $V_8\rightarrow \bar{P}_3P_3$
\cite{Lane:1991qh}.
CDF reports joint limits
on the masses of the $P_{3}$ and $V_8$.

\begin{figure}
\vspace{5.0cm}
\includegraphics{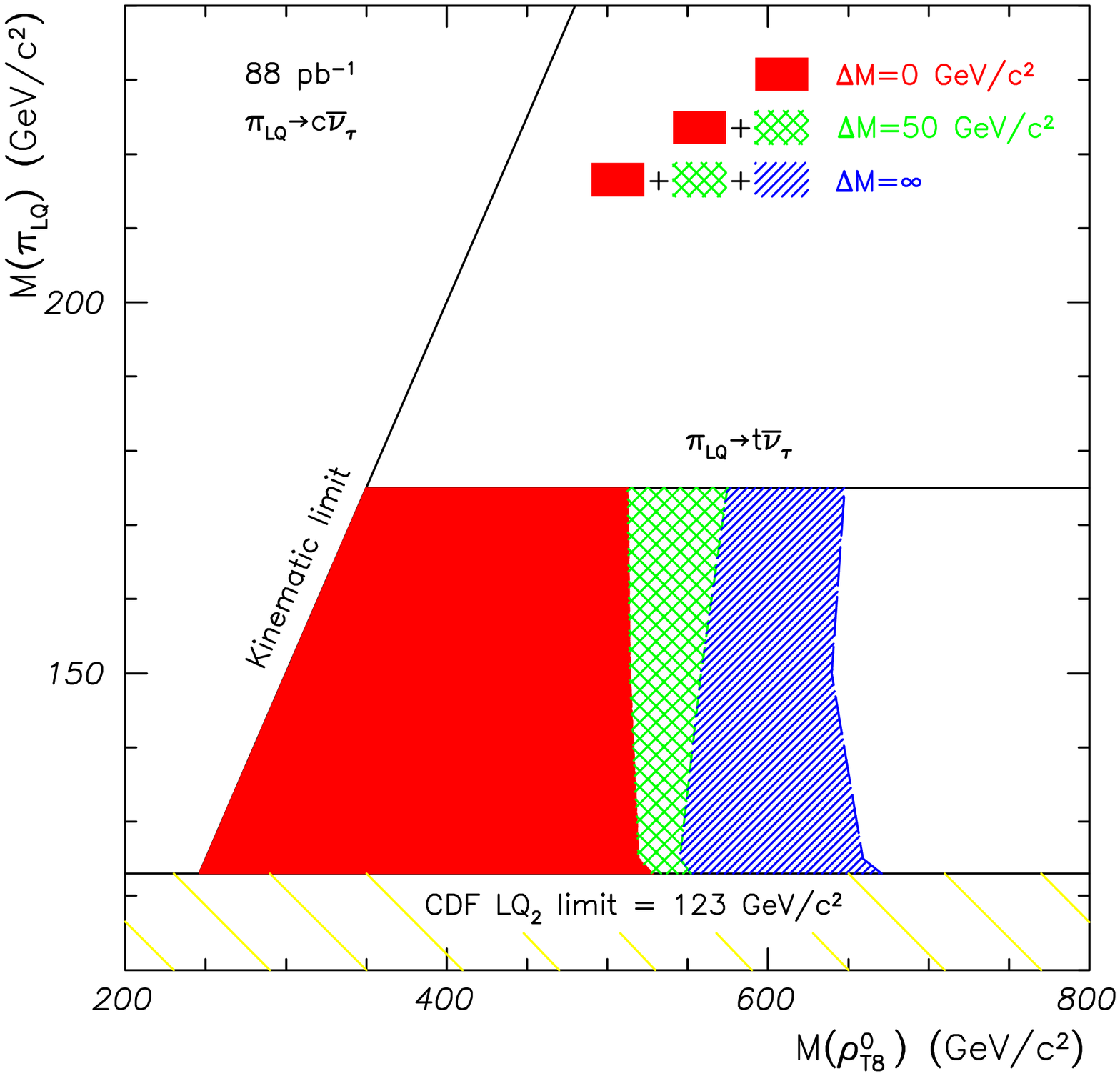}
\vspace{3cm}
\caption[cap]{\small  \addtolength{\baselineskip}{-.4\baselineskip} 
CDF's 95\% {\em c.l.} exclusion region for 
$\rho_8 \to \bar{P}_{3} P_{3} \to c c \nu_\tau \nu_\tau$ 
\protect\cite{Affolder:2000ny} for several 
limiting values of $\Delta M
= M(P_8) - M(P_3)$ which affect the $V_8$ decay partial width.}
\label{fig:lq-c-nu}
\end{figure}

In the first search \cite{Abe:1998iq} CDF considered the decay path
$P_{3} \to \bar{b} \tau^-$.  The observed yield was consistent with
Standard Model backgrounds (dominated by $Z \to \tau\tau$ plus jets,
diboson, and $\bar{t}t$ production).  CDF was able to exclude
techni--$\pi$ masses up to ($M_{V_8} / 2$) for $V_8$ masses up to
about $\sim 450$ $(500, 620)$ GeV assuming $\Delta M = 0$ $(50,
\infty)$, as illustrated in Figure \ref{fig:lq-b-nu}.  This extends
a continuum leptoquark analysis
\cite{Abe:1997dn} which had previously set the limit $M_{P_{3}} \ge 99$ GeV.

In a second set of searches \cite{Affolder:2000ny}, CDF considered
the decay paths $P_{3} \to c \bar{\nu}_\tau$ and $P_{3} \to b
\bar{\nu}_\tau$. No excess of observed events
over Standard Model backgrounds (dominated by W + jets) was found.
For $P_{3}$ decaying to charm, CDF rules out (Figure
\ref{fig:lq-c-nu}) technipion masses up to $m_t$ for techni-$\rho$
masses up to about $\sim 450$, $(500, 650)$ GeV assuming $\Delta M = 0$,
$(50,
\infty)$; heavier $P_{3}$ would decay to $t
\nu_\tau$. The lower bound from the continuum search in this channel is
$M_{P_{3}} \ge 122 GeV$ at 95\% {\em c.l.}  For $P_{3}$ decaying to $b
\bar{\nu}_\tau$, CDF's continuum search set the 95\% lower bound
$M_{P_{3}} \ge 149 GeV$ and the technipion search excludes $M_{P_{3}}$
up to the kinematic limit ($M_{V_8} / 2$) for techni-$V_8$ masses up to about
$\sim 600 $ (650, 700) GeV assuming $\Delta M = 0$ (50, $\infty$), as 
shown in Figure
\ref{fig:lq-b-nu}.

\begin{figure}
\vspace{5.0cm}
\includegraphics{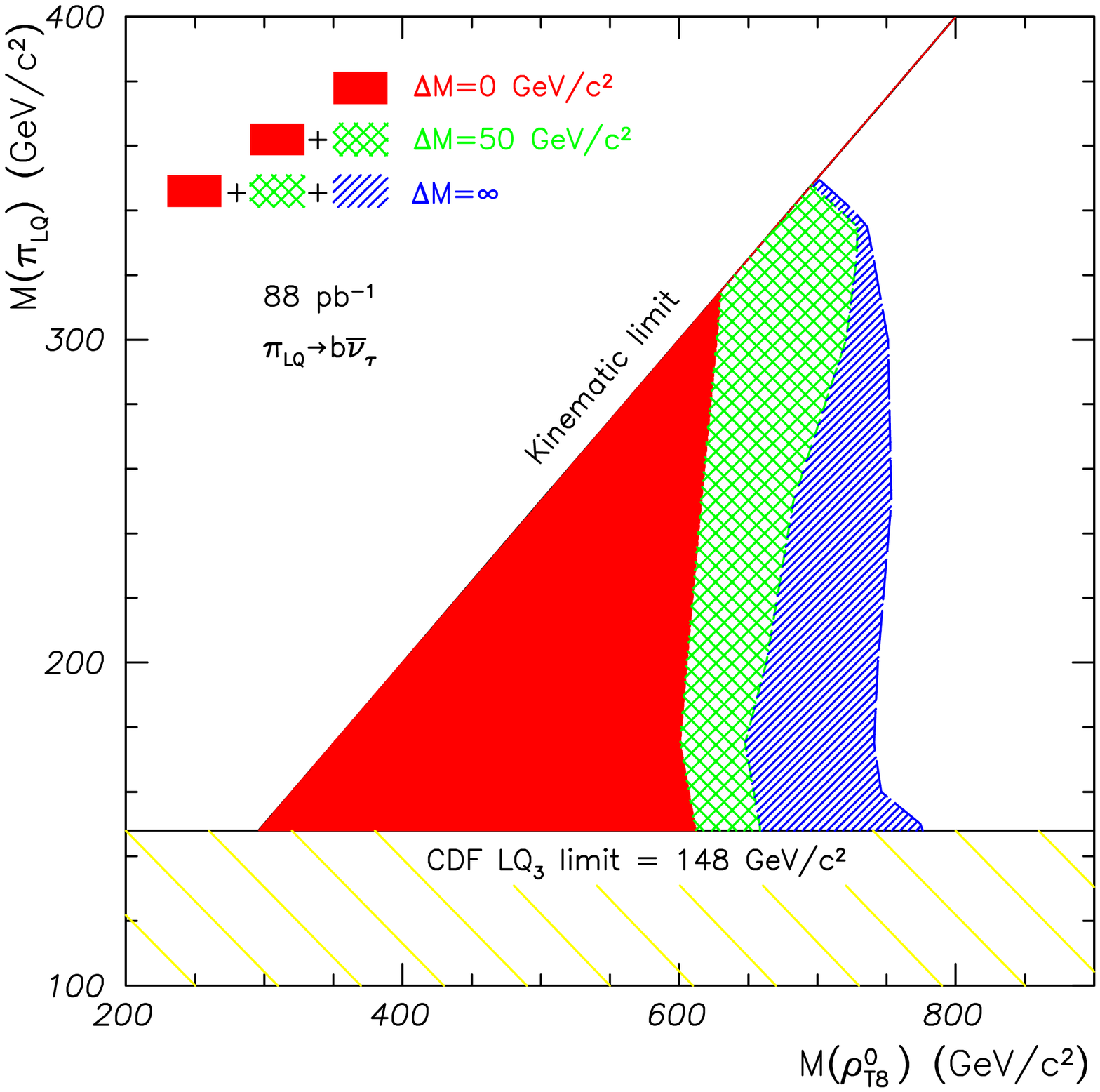}
\vspace{3.0cm}
\caption[cap]{\small \addtolength{\baselineskip}{-.4\baselineskip} 
CDF's 95\% c.l. exclusion region for 
$V_8 \to P_{3} P_{3} \to b b \nu_\tau \nu_\tau$ 
\protect\cite{Affolder:2000ny} for several values of $\Delta M
= M(P_8) - M(P_3)$.}
\label{fig:lq-b-nu}
\end{figure}

Leptoquarks typically decay into a quark and a lepton
of the same generation. In the TC-GIM models, the $P_3$'s
carry lepton and quark numbers of any generation.  A
leptoquark decaying into an $e$ and $d$--quark has the signature of a
``first generation leptoquark.'' The first generation leptoquarks
decays into $e+\bar{d}$ with branching ratio $\epsilon$, or into a $\nu +
\bar{u}$ with branching ratio $1-\epsilon$, and experimental limits are
sensitive to the unknown ratio $\epsilon$.  In TC-GIM models we have
down-type leptoquarks with $\epsilon=1$, and distinct up-types have
$\epsilon=0$.  We note that the D0 Collaboration ~\cite{Abachi:1994hr}
has excluded down-type leptoquarks up to $\sim 130$ GeV, while second
generation leptoquarks are excluded by CDF up to $\sim 133$
GeV~\cite{Abe:1995fj}.  This essentially rules out low energy scale
versions of the TC-GIM models.  

Searches for color-octet technipions will present a greater challenge.
It is difficult to pull color-octet technipions, $P_8$'s, out of the
multijet backgrounds at LHC. However, TC-GIM models can have many
different kinds of $P_8$'s; if these are nearly degenerate in mass
they may give a correspondingly stronger signal.  Run II searches for
$P_8$'s should also consider the rare decay channel decay channel $P_8
\to g\gamma$
\cite{Hayot:1981gg, Belyaev:1999xe}.  Although the rate for this mode
is down from the two-gluon channel by a factor of about $\sim 100$, a
smaller background makes the signal potentially visible, as
illustrated in Figure
\ref{fig:p8ggam}.  By employing a PYTHIA-level simulation the authors of ref.
\cite{Belyaev:1999xe} were able to identify cuts on the transverse momentum
and invariant mass of the photon + jet system that significantly enhance the
signal relative to the background.  They find that 
the Tevatron Run II can exclude 
$P_8^{0 \prime}$ in Low-Scale TC models
where $F_T$ is reduced to about $\sim 40$ GeV,  up to 
$\sim 350$ GeV, or achieve a $5\sigma$ discovery up to $\sim 270$ GeV.  
The larger value of $F_T$ in the Farhi-Susskind TC model
relative to Low-Scale models renders its PNGB invisible in this mode.
\begin{figure}[tb]
\begin{center}
\vspace{8cm}
\includegraphics{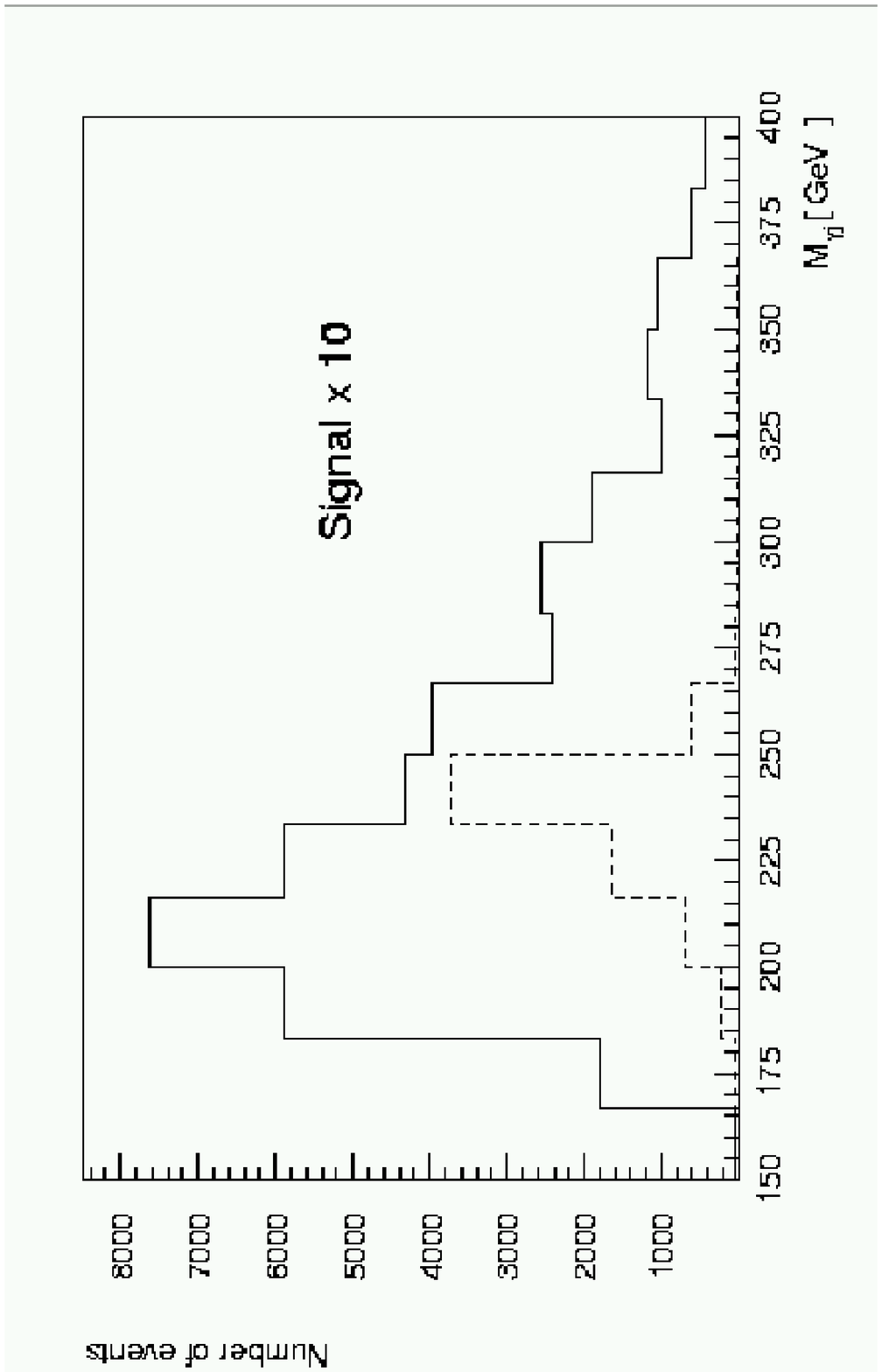}
\vspace{1cm}
\caption{\small \addtolength{\baselineskip}{-.4\baselineskip} 
Invariant mass distribution for the $p\bar p \to P_8^{0\prime} \to g
  \gamma$ signal (dashed) and background
  (solid) in multi-scale TC with $M_{P}$ = 250 GeV and 
$F$ = 40 GeV \protect\cite{Belyaev:1999xe}.}
\label{fig:p8ggam}
\end{center}
\end{figure}


\subsubsection{Searches for $W'$ and $Z'$ bosons from $SU(2) \times SU(2)$} 

As discussed in Section 3.3.2, an integral part of some dynamical
theories is an extended $SU(2)_h \times SU(2)_\ell$ structure for the
weak interactions in which the first two generations of fermions are
charged under the weaker $SU(2)_\ell$ and the third generation feels
the stronger $SU(2)_h$ gauge force.  Examples include
the non-commuting extended TC (NCETC) models
\cite{Chivukula:1994mn} and the related topflavor models
\cite{Muller:1996dj,Malkawi:1996fs,Muller:1997eg,Muller:1996qs}.  
The low-energy spectrum of the models includes massive W' and Z'
bosons that couple differently to the third-generation fermions.  The
more strongly suppressed the new gauge bosons' couplings to first and
second generation fermions, the less effective traditional searches
for new electroweak weak bosons become.  It is also notable that the
new gauge bosons couple, at leading order, only to left-handed
fermions (through weak isospin).  This section discusses search
techniques that exploit the flavor non-universal couplings of the W'
and Z'.

The LEP experiments have studied the possibility that new physics is
contributing to electron-positron scattering via four-fermion contact
interactions.  This allows them to set a lower bound on the mass of a
Z' boson, because at energies well below the mass of the Z' boson its
exchange may be approximated by a four-fermion contact interaction.
Their strongest limits on exchange of a Z' boson arising from extended
weak interactions come from processes involving pair production of
third-generation fermions: $e^+_L e^-_L \to \tau^+_L \tau^-_L$ and
$e^+_L e^-_L \to b_L \bar{b}_L$.  As discussed in
ref. \cite{Lynch:2000md}, the LEP limits on contact-interaction scale
$\Lambda$ translate into the following bounds from $\tau\tau$
production
\begin{equation} 
M_{Z'} = \Lambda \sqrt{\alpha_{em} / 4\sin^2\theta} 
      \  >\  \left\{\ {365\ {\rm GeV}\ \ \  {\rm ALEPH}} \atop {355\ {\rm GeV}\ \ \  {\rm OPAL}} \right\} \ .
\end{equation}
and from $bb$ production
\begin{equation} 
M_{Z'} = \Lambda \sqrt{\alpha_{em} / 4\sin^2\theta} 
      \  >\  \left\{\ {523\ {\rm GeV}\ \ \  {\rm ALEPH}} \atop {325\ {\rm GeV}\ \ \  {\rm OPAL}} \right\} \ .
\end{equation}
While these are weaker than the current limits on non-commuting ETC
 \cite{ehsrsc:2002} or topflavor \cite{Muller:1996dj,Muller:1996qs} Z'
 bosons from precision electroweak data (section 3.3.2), they are
 complementary in the following sense.  The limits from a fit to
 electroweak data assume that the only new relevant physics comes from
 the extra weak and ETC gauge bosons (or Higgs bosons in topflavor);
 the contact-interaction limits are lower bounds on the Z' mass
 regardless of the other particle content of the theory.

The Fermilab Tevatron experiments have, likewise, searched for new contact
interactions contributing to dijet and dilepton production.  Because their
searches involve only first and second generation fermions, the implied
bounds on the mass of a Z' primarily coupled to the third generation are
significantly weaker than those from the LEP data \cite{Lynch:2000md}.

A W' boson preferentially coupled to the third generation fermions should be
detectable in single top-quark production at Run IIb \cite{Simmons:1997ws} if
its mass is less than about 1-1.5 TeV.  The ratio of cross-sections $R_\sigma
\equiv \sigma(\bar{p}p \to t b) / \sigma(\bar{p}p \to l \nu)$ can be measured
(and calculated) to an accuracy 
\cite{Smith:1996ij,Heinson:1996pi,Heinson:1997zm} of at least
$\pm 8\%$.  The extra weak gauge bosons present in a non-commuting
Extended TC model can change $R_\sigma$ in several
ways\footnote{\addtolength{\baselineskip}{-.4\baselineskip} Exchange
of the ETC boson that generates $m_t$ does {\bf not} modify the $Wtb$
vertex, because the boson does not couple to all of the required
fermions: ($t_R, b_R, U_R, D_R$).}.  Mixing of the $W_h$ and $W_\ell$
bosons alters the light $W$'s couplings to the final state fermions.
Exchange of both $W$ and $W'$ mass eigenstate bosons contributes to
the cross-sections.  As a result, a visible {\bf increase} in
$R_\sigma$ is predicted.

A direct search for Z' bosons primarily coupled to the third generation
can be made by looking at heavy flavor production at the Tevatron.  No
searches for these bosons have been made thus far.  A recent study
\cite{Lynch:2000md} (using PYTHIA and a simple model of the D0 detector)
indicates that the channel $p \bar{p} \to Z' \to \tau\tau \to e \mu \nu
\bar{\nu}$ is promising.  By requiring large leptonic transverse momenta, low
jet multiplicity, and a sufficiently large opening angle between the
$e$ and $\mu$, the Run II experiments should be able to overcome the
Standard Model backgrounds from $Z^0$, $t\bar{t}$ and $W W$
production. A Z' boson with a mass up to 700 GeV (depending on mixing
angle) should be visible, as indicated in figure \ref{fig:ehs:zprime}.

\begin{figure}[tb]
\vspace{5.0cm}
\includegraphics{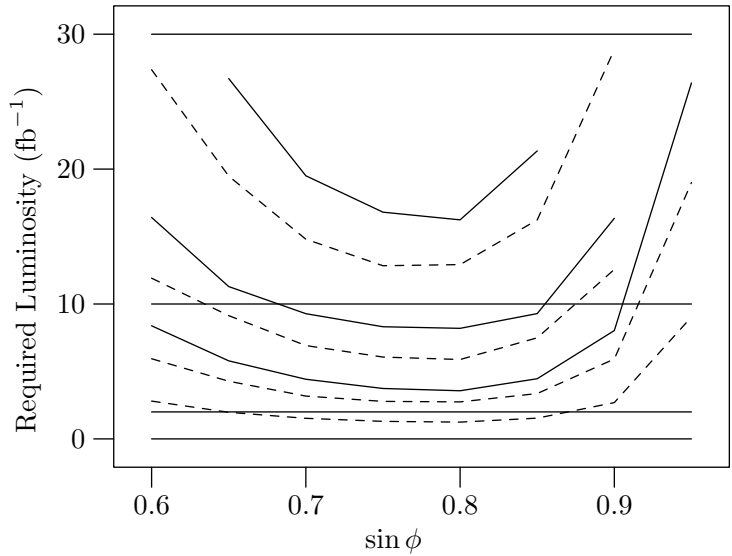}
\vspace{2.5cm}
\caption{\small \addtolength{\baselineskip}{-.4\baselineskip} 
Luminosity required to discover SU(2) Z' bosons 
preferentially coupled to the third generation \protect\cite{Lynch:2000md}.  
  Dashed curves are $3\sigma$ discovery curves for a fixed mass,
  while solid curves are $5\sigma$ discovery curves.  From bottom to
  top, $3\sigma$ curves are displayed for Z' masses of
  550 GeV, 600 GeV, 650 GeV, and
  700 GeV.  From bottom to top, $5\sigma$ curves are
  displayed for Z' masses of 550 GeV, 600 GeV,
  and 650 GeV.  The horizontal lines indicate luminosity
  targets for Run II.}
\label{fig:ehs:zprime}
\end{figure}

Various methods of finding $W'$ and $Z'$ at the LHC, NLC or FMC have been
suggested in the context of topflavor models
\cite{Muller:1996dj,Malkawi:1996fs,Muller:1997eg,Muller:1996qs}; in principle
these should work equally well for dynamical models.  Calculations of
expected production rates indicate that a number of processes are
worth further study.  The presence of a $W'$ could boost single top
production at the LHC to a rate rivaling that of $t\bar{t}$; the
distinctive final state would include a pair of b-jets and a
high-$p_T$ lepton.  Both single-lepton and dilepton production at the
LHC could be visibly modified by new weak bosons.  A $Z'$ boson might
visibly alter the rate of di-muon production at an NLC or that of
$t\bar{t}$ production at an NLC, FMC or LHC.  Since only the
left-handed couplings of the top quark would be affected, top angular
distributions might also be be altered.  Finally, if flavor-changing
mixing between second and third-generation leptons were large,
production of unlike-sign $\mu\tau$ pairs might be seen.  Because the
$V_H V_L V_L$ triple gauge boson coupling vanishes to leading order,
di-boson production will not reveal the presence of a $W'$ or $Z'$.


\subsection{Supersymmetric and Bosonic Technicolor}


\subsubsection{Supersymmetry and Technicolor }

  Shortly after the introduction of Technicolor, Witten described
the general aspects of a {\em Supersymmetric Technicolor} theory
\cite{Witten:1981nf}, while
Dine, Fischler, and Srednicki \cite{Dine:1981za} \cite{Dine:1982qj}
constructed explicit models.  Several interesting possibilities arise when
the ideas of Supersymmetry (SUSY) and TC are combined.  Notably, it
becomes possible to raise the scale of ETC
\cite{Buras:1983nb},  and thus suppress the dangerous $\gamma$ terms.
One might even be able to achieve dynamical electroweak symmetry breaking
in concert with Supersymmetry breaking.  These models are unfashionable at
present, relative to the MSSM which guarantees a low-mass Higgs boson
accessible to Run IIb or the LHC.  Nonetheless, the general idea has some
theoretical merit and the models have interesting and accessible
phenomenology (including the potential for light fundamental scalars).

In ordinary QCD we know that a chiral condensate,
$\VEV{\overline{q}q}\neq 0$ is dynamically generated when the gauge
coupling becomes strong. In Supersymmetric QCD a quark $q$
possesses a superpartner, denoted $\tilde{q}$, the squark. The
supercharge, $Q$,
generates a transformation 
on these fields of the form:
\be
\{ Q, \tilde{q}^\dagger q + {q}^\dagger \tilde{q} \}  = \overline{q}q
\ee 
Hence, the existence of the fermionic condensate 
$\VEV{ \overline{q}q}\neq 0$ implies that:
\be
Q|0> \neq 0 
\ee
Whenever $Q$ does not annihilate the vacuum it
means that SUSY is broken. The field $ \tilde{q}^\dagger q +
{q}^\dagger \tilde{q}$ then 
becomes a massless Goldstone fermion, a Goldstino.  
Thus, SUSY breaking and
dynamical EWSB can have an intimate connection 
through fermion condensation.

In addition to the key papers mentioned above, we refer the interested
reader to futher works on Supersymmetric Technicolor, 
refs.\cite{Murayama:2000dw, Dobrescu:1995gz, Buras:1983yg, Nilles:1982my,
Dine:1982qj}, and to ref.\cite{Liu:1999ah}, 
which discusses the combination of Topcolor with SUSY.

\subsubsection{Scalars and Technicolor: Bosonic Technicolor }
\vskip .1in

The key advantage of introducing fundamental scalars into TC
is to provide an alternative to ETC.
Most of the constraints and problems of ETC can be dismissed 
by assuming general masses and couplings of the fundamental
scalar sector.  We thus consider
theories in which the additional fields that 
communicate the TC condensates
to the ordinary quarks and leptons are fundamental
scalars.  We view these scalars as
ultimately associated with Supersymmetry,
where their masses can be
protected by fermionic chiral symmetries 
\cite{Samuel:1990dq, Dine:1990jd, Kagan:1991gh, Kagan:1990az, Kagan:1992gi,
  Kagan:1991ng, Dobrescu:1995gz}. 
Alternatively, these scalars may, in principle, be bound
states arising within a high energy strongly coupled theory
\cite{Chivukula:1990bc, Appelquist:1991kn}. 
Conceivably, 
they can also be viewed
as relics from extra dimensions, such as Wilson lines
(Section 4.6).   In what follows, we momentarily
disregard the details of the higher-energy physics that enables the scalars
to have masses of order the weak scale and consider the key features and
phenomenology of TC models with fundamental or effective scalars.
We classify the models according to the weak and TC charges of the
scalar states.

\vskip .1in
\noindent
{\bf(i) Weak-doublet Techni-singlet Scalars}
\vskip .1in

We begin by discussing TC models whose spectrum includes one or more
weak-doublet TC-singlet scalars.   These correspond to a natural
low-energy limit of strongly-coupled ETC models
\cite{Chivukula:1990bc}.  Moreover, the presence of a weak-doublet
techni-singlet scalar has a sufficiently large effect on the vacuum
alignment of the technifermion condensate to make an $SU(2)$ TC
gauge group viable \cite{Dobrescu:1998ci}.

In the minimal model of this type \cite{Simmons:1989pu}, one adds to the
Standard Model gauge and fermion sectors a simple $SU(N)$ TC
sector, with two techniflavors that transform as a left-handed doublet
$\Upsilon_L = (p_L, m_L)$ and two right-handed singlets, $p_R$ and $m_R$,
under $SU(2)_W$, with weak hypercharge assignments $Y(\Upsilon _L)=0$,
$Y(p_R)=1/2$, and $Y(m_R)=-1/2$.  The technifermions and ordinary fermions
each couple to a weak scalar doublet $\phi= (\phi^+
\phi^0)^T $ which has the quantum numbers of the Higgs doublet of the
Standard Model.  Unlike the Standard Model Higgs doublet, $\phi$ has a {\it
 nontachyonic} mass ($M_\phi^2 \geq 0$) 
 and is not the primary source of electroweak
  symmetry breaking.

The scalar has Yukawa couplings to the technifermions:
\begin{equation}
{\cal L}_{\phi T} = \bar\Upsilon_L \tilde\phi\thinspace \lambda_+ p_R\ +\
    \bar\Upsilon_L \phi\thinspace \lambda_- m_R\ +\ h.c.
\label{eeq:ytc}\end{equation}
When the technifermions condense, these couplings
cause $\phi$ to acquire an effective VEV, $\langle\phi\rangle = f' \approx
{4 \pi \lambda_T f^3 / M_\phi^2}$.  
Because the scalar couples to ordinary fermions as well
as technifermions, the ordinary fermions obtain masses from the diagram
sketched in Figure
\ref{fig:ehs:bosonictc}
\begin{equation}
m_{f} \approx \lambda_{f} h {4 \pi f^3 \over M_\phi^2 }.
\end{equation}
where
$h$ is the scalar coupling to the
ordinary fermions.  
The coupling matrices $\lambda_{f}$ are proportional to the mass matrices
$m_f$ and drive flavor symmetry breaking.  The quark flavor
symmetries are broken according to the same pattern as in the Standard
Model so that the quarks mix via the usual CKM matrix and the standard GIM
mechanism prevails.

\begin{figure}
\includegraphics{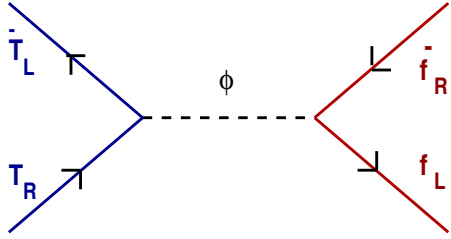}
\vspace{4.5cm}
\caption[cap]{\small \addtolength{\baselineskip}{-.4\baselineskip} 
Fermion-technifermion interaction through scalar
  exchange.  This diagram is responsible for fermion mass generation
once the technifermions condense.}
\label{fig:ehs:bosonictc}
\end{figure}

Both the
technipion decay constant and the scalar VEV contribute to the 
electroweak scale:
$f^2+f'^2=v_0^2$.  The technipions and the isotriplet components of $\phi$
mix.  One linear combination becomes the longitudinal component of the $W$
and $Z$; the orthogonal one remains in the low-energy theory as an
isotriplet of physical scalars \cite{Carone:1995mx}. The coupling of the
charged physical scalars to quarks has the same form as in a type-I
two-Higgs doublet model
\cite{Carone:1995mx}.    Imposing the requirement that the isoscalar
component of $\phi$ have no VEV enables one to eliminate $f$ and $f'$ and
calculate observables in terms of $M_\phi$, $h$ and $\lambda$.  This model
has been studied in the limits {\em (i)} in which $\lambda$ is negligible
and one works in terms of ($M_\phi, h$)
\cite{Simmons:1989pu, Carone:1993rh, Carone:1995mx}, and {\em (ii)} in
which $M_\phi$ is negligible and one works in terms of $(\lambda, h)$
\cite{Carone:1994xc, Carone:1995mx}.  The leading Coleman-Weinberg
corrections to the scalar potential have been included in studies of
both limits.

\begin{figure}[tb]
\vspace{10.0cm}
\includegraphics{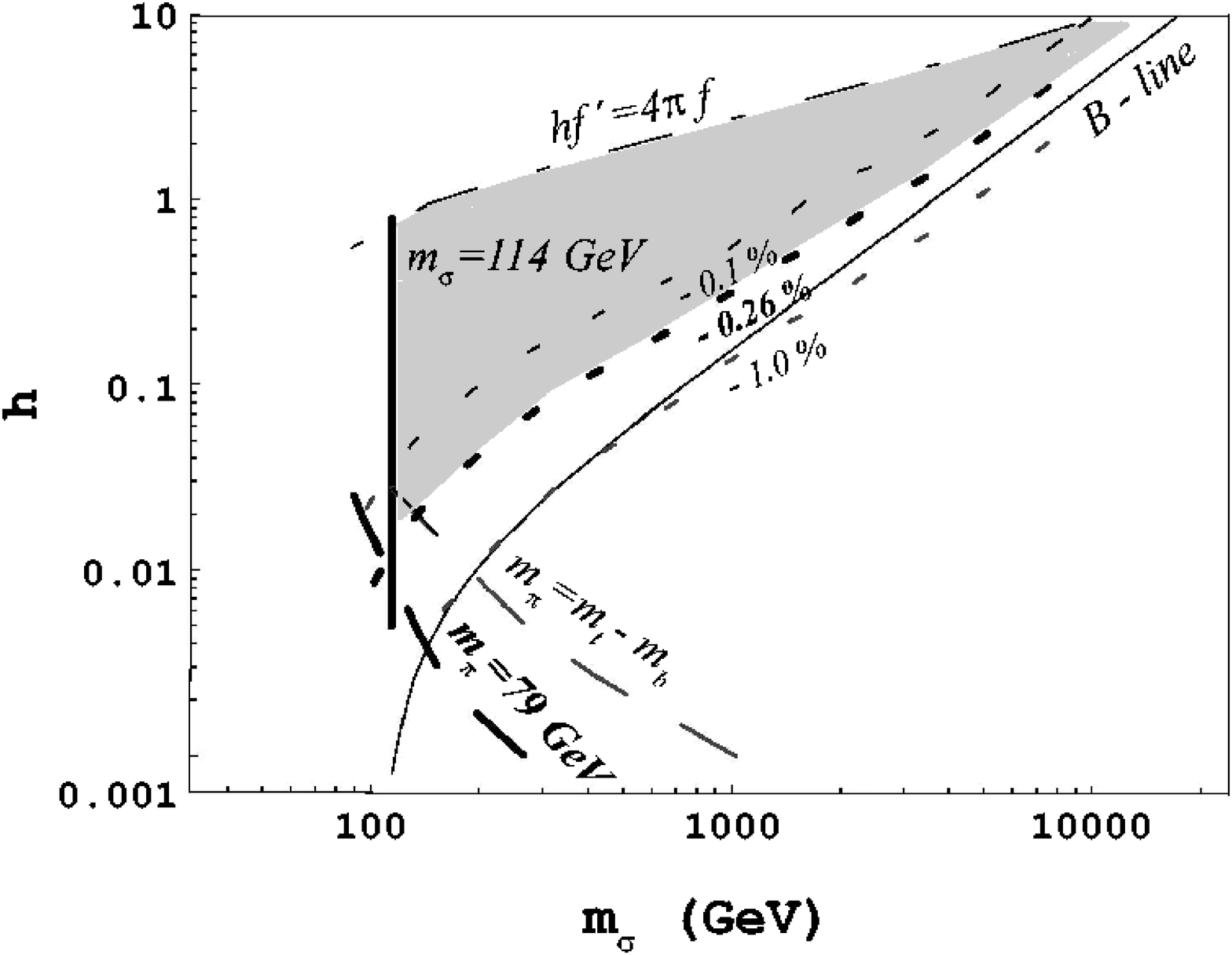}
\vspace{-1.5cm}
\caption{\small \addtolength{\baselineskip}{-.4\baselineskip} Constraints on technicolor with scalars in limit [i], where
the scalar self-coupling is negligible, plotted in the physical basis
($m_\sigma$, h). The allowed region of parameter space (shaded) is
bounded by the contours $m_\sigma = 114$ GeV (solid), $R_b - R_b^{SM}
= 0.26\%$ (dashes) and $h f' = 4 \pi f$ (dot-dash).  Other contours of
constant $R_b$ are shown for reference. The current bound from
searches for charged scalars $m_{\pi^\pm_p} = 79$ GeV is shown (long
dashes) along with the reference curve $m_{\pi^\pm_p} = m_t - m_b$.
The constraint from $B^0\bar{B}^0$ mixing is labeled
``B-line''. \protect\cite{Hemmige:2001vq}}
\label{fig:ehs:tc+phi-i}
\end{figure}

The phenomenology of TC models with weak-doublet scalars has
been found to be in agreement with experiment.  Models of this kind do
not produce unacceptably large contributions to $K^0$-$\overline{K^0}$
or $B^0$-$\overline{B^0}$ mixing, nor to the electroweak $S$ and $T$
parameters \cite{Simmons:1989pu, Carone:1993rh, Carone:1994xc}.  In
addition, the new scalars in the model can be made heavy enough to
evade detection, even in the limit where the scalar doublet is assumed
to have a vanishing $SU(2) \times U(1)$ invariant mass
\cite{Carone:1994xc}.  There are negative corrections to $R_b$, but they
never exceed $-1$\% in the regions of parameter space allowed by other
constraints; likewise, the rate for $b \to s \gamma$ is less than in the
Standard Model, but not so altered as to conflict with experiment 
\cite{Carone:1995mx}.  Current bounds on the parameter space of the model
\cite{Hemmige:2001vq} 
are shown in Figures \ref{fig:ehs:tc+phi-i} and \ref{fig:ehs:tc+phi-ii};
in creating the plots, the parameters $M_\phi$ (limit i) and $\lambda_\phi$
(limit ii) have been eliminated in favor of mass of the physical isoscalar
state, $m_\sigma$.

\begin{figure}[tb]
\vspace{11.0cm}
\includegraphics{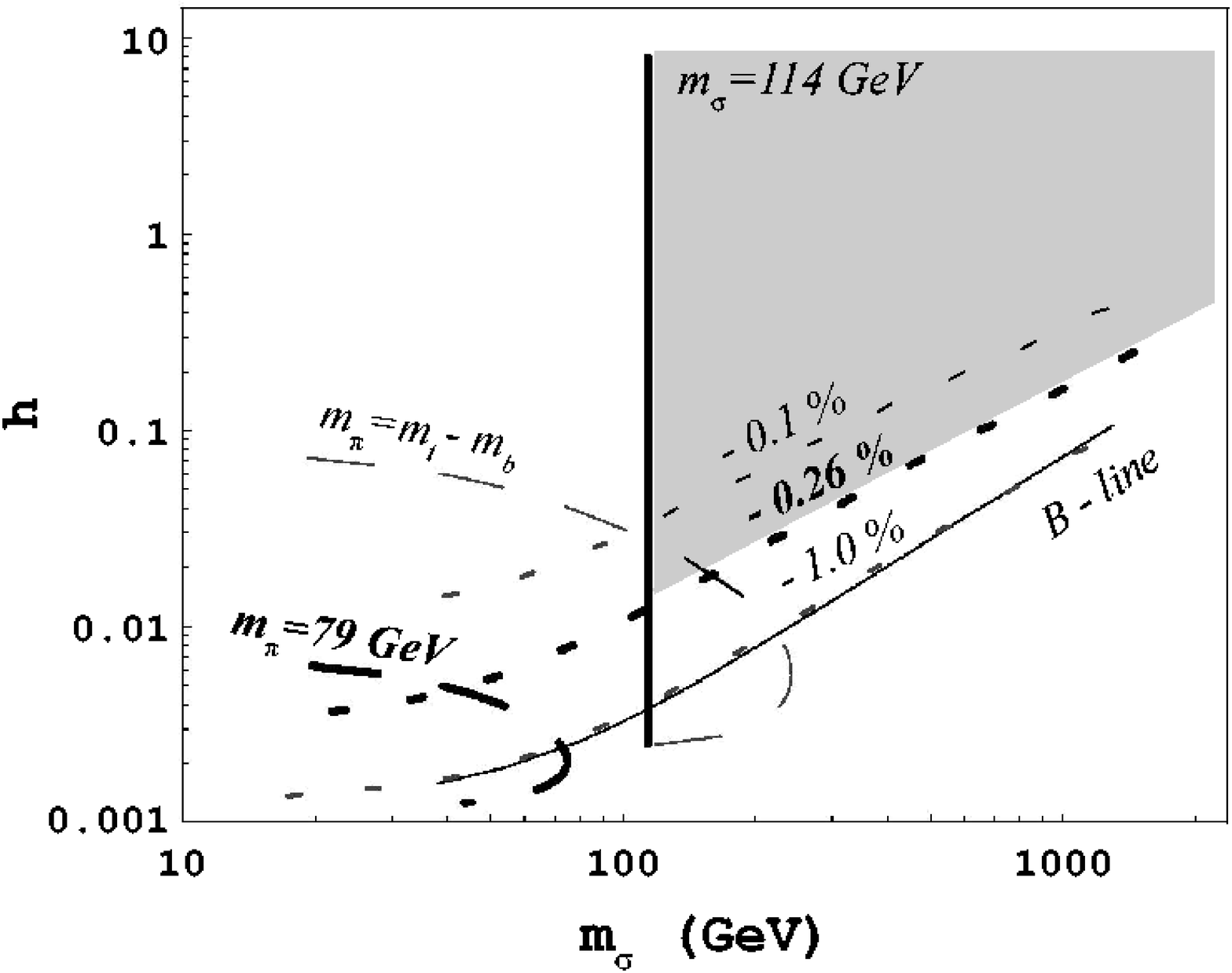}
\vspace{-2cm}
\caption{\small\addtolength{\baselineskip}{-.4\baselineskip} 
  Constraints on technicolor with scalars in limit
[ii], where the scalar mass is negligible, plotted in the physical
basis ($m_\sigma$, h). Curves labeled as in Figure
\protect\ref{fig:ehs:tc+phi-i}.  From 
\protect\cite{Hemmige:2001vq}.  }
\label{fig:ehs:tc+phi-ii}
\end{figure}

Supersymmetrized models with weak-doublet techni-singlet scalars were
introduced in refs.\cite{Samuel:1990dq, Dine:1990jd} and aspects of flavor
physics and renormalization group evolution were studied in refs.\cite{
Kagan:1991gh, Kagan:1990az, Kagan:1992gi, Kagan:1991ng}.  The minimal such
model \cite{Samuel:1990dq, Dine:1990jd} contains all the fields of the
Minimal Supersymmetric Standard Model (MSSM), 
a set of $SU(N)_{TC}$ gauge bosons
and their superpartners, and color-singlet technifermion superfields
transforming under $SU(N)_{TC} \times SU(2)_W \times U(1)_Y$ as
\begin{equation}
T_{U_R} \equiv (N_{TC}, 1, 1/2),\ \ \ T_{D_R} \equiv (N_{TC}, 1, -1/2),\ \
\ T_L \equiv (N_{TC}, 1, 0).
\label{eqn:ehs:oneset}
\end{equation}
The superpotential includes $W_{SSM}$ from the MSSM and an additional part
$W_{HTC}$ from Yukawa couplings of the two Higgs superfields $H_U$ and
$H_D$ to the technifermion superfields $T$.
\begin{equation}
W_{HTC} = g_U H_U T_{U_R} T_L + g_D H_D T_{D_R} T_L
\end{equation}
R-parity is imposed, just as in the MSSM; 
the technifermions transform like
matter.

The Higgs fields have positive squared masses in the perturbative vacuum.
The superpotential terms $W_{HTC}$, however, produce terms linear in
the Higgs fields in the Lagrangian when the technifermions
condense \cite{Samuel:1990dq,Dine:1990jd}.  This causes the Higgs fields to
acquire VEV's:
\begin{equation}
\langle H_U \rangle = g_U \langle T_{U_R} T_{U_L} \rangle / m_{H_U}^2\,,
\ \ \ \ \ 
\langle H_D \rangle = g_D \langle T_{D_R} T_{D_L} \rangle / m_{H_D}^2\, .
\end{equation}
Thus for $\langle T T \rangle \sim (600 \rm{GeV})^3$, $g_U \sim 1$ and
$m_{H_U} \sim 1$ TeV, one obtains $\langle H_U\rangle \sim 100$ GeV,
yielding a realistic top quark mass.  If $m_{H_D} > m_{H_U}$, the VEV
of $H_D$ will be much smaller than that of $H_U$, producing the
required top-bottom mass splitting.  In variant models with larger
technifermion content, 
$W_{HTC}$ generates technifermion current-algebra
masses, yielding masses of order $200$ GeV -- $1$
TeV \cite{Samuel:1990dq,Dine:1990jd} for the technipions in the spectrum.

These supersymmetrized TC models minimize the FCNC problems that
usually affect SUSY and TC models \cite{Kagan:1991gh, Kagan:1990az,
Kagan:1992gi, Kagan:1991ng}.  In the minimal version, scalar exchange among
quarks and leptons generates no tree-level FCNC.  If additional Higgs
multiplets are introduced to explain the hierarchy of fermion masses and
mixings, tree-level FCNC may be re-introduced; scalar masses of order $\sim 10$
TeV suffice to suppress them.  Furthermore, because the lightest Higgs bare
mass $m_{H_U}$ is of order $\sim 1$ TeV, superpartner masses of order 
$\sim 1-10$ TeV
become natural, reducing the degree of low-energy squark and slepton mass
degeneracy required to avoid undue FCNC from loops.

\vskip .1in
\noindent
{\bf(ii) Weak-singlet Technicolored Scalars}
\vskip .1in

Theories which include weak-singlet technicolored scalars address the
intergenerational fermion mass hierarchy more directly.  In the early model
of ref.\cite{Kaplan:1991dc}, exchange of technicolored weak-singlet
scalars induces four-fermion interactions between a trio of technifermions
and one fermion.  Ordinary fermions then mix with technibaryons and become
massive.  Unfortunately, this model predicts unacceptably large FCNC and
tree-level contributions to the $T$ parameter.  A more natural account of
the mass hierarchy is provided by models \cite{Kagan:1991ng, Kagan:1995qg,
Dobrescu:1995gz, Dobrescu:1997kt} in which exchange of technicolored
weak-singlet scalars induces four-fermion interactions between pairs of
fermions and technifermions.  The small CKM elements associated with the
third-generation quarks arise because the $t_L$ and $b_L$ mass eigenstates
are automatically aligned.  However, because the scalar can couple to both
$t_R$ and $b_R$, the top-bottom mass ratio must be provided by a ratio of
Yukawa couplings.  We will discuss three models with interesting features
in more detail.

\begin{figure}
\includegraphics{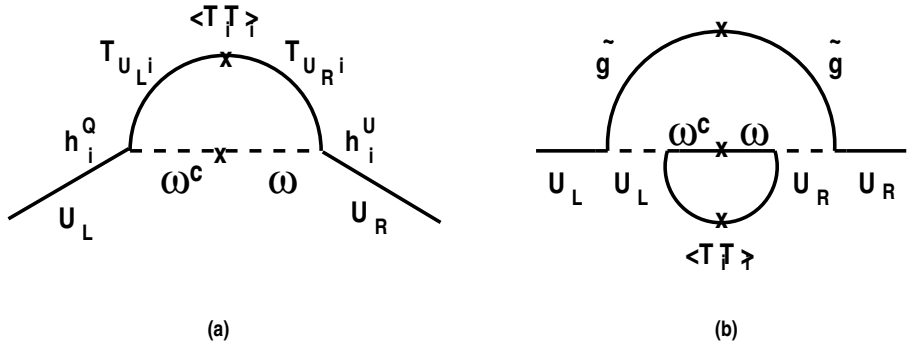}
\vspace{4.5cm}
\caption[cap]{\small\addtolength{\baselineskip}{-.4\baselineskip} 
  (a) Tree-level mass for two generations of fermions 
due to technifermion condensation.  (b) Radiative mass contribution from 
squark and gluino exchange.  Analogous contributions to charged lepton 
masses arise from exchange of $\xi$ and electroweak gauginos.}
\label{fig:ehs:kaganmass}
\end{figure}

A supersymmetric model with technicolored weak-singlet scalars and some
interesting mass phenomenology was introduced by Kagan \cite{Kagan:1991ng}.
The gauge group is $SU(N_{TC})\times SU(3)\times SU(2)\times U(1)$; three
families of quark and lepton superfields and two sets ($T_1, T_2$) of the
color-singlet technifermion superfields as in equation
(\ref{eqn:ehs:oneset}) are present.  The Higgs sector, consists of a
vector-like pair of color triplet Higgs fields transforming under
$SU(N_{TC})\times SU(3)\times SU(2)\times U(1)$ as $\omega^c \equiv
(N_{TC}, \bar{3}, 1, -1/6)$ and $\omega \equiv (N_{TC}, 3, 1, 1/6)$ and a
vector-like pair of color singlet fields $\xi^c \equiv (N_{TC}, 1, 1, 1/2)$
and $\xi \equiv (N_{TC}, 1, 1, -1/2)$.  While only two generations of
quarks receive masses at tree-level in this model, gaugino and sfermion
exchange enable the first generation fermions to receive radiative masses,
as indicated in Figure \ref{fig:ehs:kaganmass}.  SUSY breaking masses can
be in the $1-10$ TeV range and little or no squark and slepton mass
degeneracy is needed.  For $N_{TC} = 3$, adding a pair of technivector,
color, weak- and hypercharge singlet Higgs superfields induces
four-technifermion interactions of the form $T_L T_L T_{U_R} T_{D_R}$.
When technifermion condensation occurs, this generates large current masses
for the technifermions, usefully enhancing the masses of the technipions.
Phenomenology associated with techniscalar-generated chromomagnetic moments
for the quarks and several scenarios for quark mass generation in
non-supersymmetric versions of the model are discussed in
\cite{Kagan:1991ng, Atwood:1995vm}.

Dobrescu has studied a related supersymmetric TC model with
only one doublet of technifermions \cite{Dobrescu:1995gz}.  Only
the third generation fermions acquire large masses and the other two
generations' smaller masses are generated radiatively by sfermion and
gaugino exchange.  The predicted rates of neutral $K$ and $B$ meson mixing,
CP violation in $K_L$ and $B$ meson decays, and CP asymmetries in $B$ decays or
$\Delta S = 1$ transitions, are all consistent with experiment.

Dobrescu and Terning \cite{Dobrescu:1997kt} have also shown  that a
TC model with technicolored weak-singlet scalars can produce
a negative contribution to the S parameter.  This model includes the
Standard Model fields, an $SU(N_{TC})$ technicolor gauge group, two
flavors of technifermions $\Psi_R \equiv (P_R, N_R) \equiv (N_{TC},
1,2,0)$, $P_L \equiv (N_{TC}, 1,1,1)$, and  $N_L \equiv (N_{TC}, 1,1,-1)$
and three technicolored scalars transforming as
$\phi \equiv (N_{TC}, \bar{3}, 1, -1/3)$, $\omega_t \equiv (N_{TC},
\bar{3}, 1, -7/3)$, and $\omega_b \equiv (N_{TC}, \bar{3}, 1, 5/3)$.
The couplings of the $Z$ boson to the $(t,b)_L$ doublet ($\delta g_L$) and the
$t_R$ and $b_R$ quarks ($\delta g_R^t,\, \delta g_R^b$) are found to
be shifted by amounts depending on the Yukawa couplings and the scalar
masses.  The model therefore predicts that the value of the $S$ parameter is
reduced from the typical one-doublet techifermion contribution by an amount
depending on those couplings:
\begin{equation}
S \approx 0.1 N_{TC} - 1.02 \left( 3 \delta g_L - 2 \delta g_R^t + \delta
g_R^b\right)\,.
\end{equation}

\vskip .1in
\noindent
{\bf(iii) Weak-doublet Technicolored Scalars}
\vskip .1in

Models incorporating weak-doublet technicolored scalars
\cite{Dobrescu:1998ci} can explain not only the intergenerational fermion
mass heirarchy, but also some elements of the intragenerational mass
hierarchies.  These models
\cite{Dobrescu:1998ci} include the Standard Model gauge and fermion sectors
together with a minimal TC sector. This takes
the form of an asymptotically free $SU(N_{\rm
  TC})$ gauge group, which becomes strong at a scale of order 1 TeV, and one
doublet of technfermions which transform under the $SU(N_{\rm TC}) \times
SU(3)_{\rm C} \times SU(2)_{\rm W} \times U(1)_{\rm Y}$ gauge group as
$\Psi_L \equiv (P_L, N_L) = (N_{TC}, 1, 2)_0$, $P_R = (N_{TC}, 1,1)_{+1}$
  and $N_R = (N_{TC}, 1,1)_{-1}$.
The large top quark mass is generated by a scalar
multiplet, $\chi^t$, which transforms under the $SU(N)_{\rm TC} \times
SU(3)_{\rm C} \times SU(2)_{\rm W} \times U(1)_{\rm Y}$ gauge group as:
$(\overline{N_{\rm TC}}, 3, 2)_{4/3}$. Its 
Yukawa interactions may be written without loss of generality as
\begin{equation}
{\cal L}_{t} = C_q\, \overline{q^3_{\inl}}\, \down_{\inr}\, \chi^t 
\, +\, C_t\, \overline \Psi_{\!\!\inl}\, t_{\inr}\, i
\sigma_2\,\chi^{t \dagger} 
\, +\, {\rm h.c.} ~,
\label{eqn:ehs:tyuk}
\end{equation}
where $q_{\inl}^3 \equiv (t_{\inl}, b_{\inl})^\top$ is the left-handed weak
eigenstate $t-b$ quark doublet, and the Yukawa coupling constants, $C_q$
and $C_t$, are defined to be positive.  Below the scale of technifermion
condensation, scalar exchange generates a top quark mass.  Because the
hypercharge of $\chi^t$ allows it to couple to $t_R$ but not to $b_R$, the
model as described gives mass only to the top quark.  If other fermion
masses arise from physics above the TC scale, it will be natural
for the top to be the heaviest fermion. Such physics could be a
weak-doublet techni-singlet scalar like that described earlier, or a set of
technicolored scalars.  In any case, the oblique corrections and Z-pole
observables predicted by this class of models are found to be consistent
with experiment
\cite{Dobrescu:1998ci}.

\newpage
\section{Top Quark Condensation and Topcolor}


The large top quark mass is suggestive of new dynamics, potentially
associated with electroweak symmetry breaking.  The early papers
attempt to identify all of the EWSB with the
formation of a top quark condensate, and thus a bound-state Higgs
composed of $\bar{t}t$.  From this perspective, in
contrast to TC, the $W$ and $Z$ {\em and the top
quark} are the first order massive particles; all other quarks and
leptons viewed as {\em a priori} massless.  We first
discuss these pure top condensation scenarios and then review more
realistic models incorporating top condensation into a larger
dynamical framework.

\subsection{Top Quark Condensation in NJL Approximation}

\subsubsection{The Top Yukawa Quasi-Infrared Fixed Point}

Top quark condensation is related to the
quasi-infrared fixed point of the top quark Higgs--Yukawa coupling,
first discussed by Pendleton and Ross \cite{Pendleton:1981as} and
Hill, \cite{Hill:1981yr}, \cite{Hill:1981sq}, with implications for
the Higgs boson mass, \cite{Hill:1985tg}. At issue are
particular properties of
the solutions to the Renormalization Group (RG) equations
for the top Yukawa coupling constant.  The Pendleton-Ross fixed
point corresponds to a fixed ratio
$g_t/g_{QCD}$\cite{Pendleton:1981as} and is the basis of
the idea of ``reduction of coupling constants'' schemes (e.g., see
\cite{Oehme:1985jy}).  It defines a critical trajectory above which
$g_t$ has a Landau pole, and below which $g_t$ is asymptotically
free. It predicted $m_t \sim 120$ GeV.

On the other hand,
the formation of a Higgs bound-state composed of $\bar{t}t$ implies a
very large top quark Higgs-Yukawa coupling constant at the scale of
the binding interaction $M$. If the top quark Higgs--Yukawa
coupling is sufficiently large at a high energy scale $M$ (e.g., if
$g_t(M)$ is at least of order unity) then one obtains a 
robust low energy prediction of $m_{t} = g_t(m_t)v_{wk}$,
e.g. for $M\sim 10^{15} GeV$ of order $m_t\sim 220$  GeV 
in the Standard
Model \cite{Hill:1981yr}, \cite{Hill:1981sq}.
(or $m_t \sim 200 \sin\beta$ GeV, in the MSSM \cite{Bardeen:1994rv}).
For arbitrary but sufficiently large initial $g_t(M)$
the low energy value of $g_t(v_{wk})$  is determined by
the RG equations alone. It remains logarithmically sensitive to $M$, but  
becomes insensitive to the initial
$g_t(M)$ boundary condition.
The term ``quasi-infrared fixed point'' refers
to this solution for the top-Yukawa coupling, and 
is the RG improved solution
of top quark condensation schemes
\cite{Bardeen:1990ds}.

\subsubsection{The NJL Approximation}

Nambu discussed the idea of a $\bar{t}t$ condensate in the
context of generic chiral dynamics,
\cite{Nambu:1988mr}. This
was independently, and more concretely, developed by Miransky, 
Tanabashi, Yamawaki, \etal \cite{Miransky:1989xi,Miransky:1989ds}, 
who placed the idea
firmly in the context of an NJL model and derive the 
relationship between the top quark mass and electroweak scale
through the Pagel's-Stokar formula.
It was subsequently elaborated
by Marciano \cite{Marciano:1990mj},
\cite{Marciano:1989xd},
and Bardeen, Hill and Lindner \etal \cite{Bardeen:1990ds}.  The latter
paper provides a detailed analysis of the scheme with the connection
to, and improvement by, the full renormalization group. The Higgs
boson is shown to be a ``deeply bound state'' \cite{Miransky:1991gz}
composed of $\bar{t}t$, where the top and Higgs masses are predicted
by the quasi-infrared fixed point \cite{Hill:1981sq, Hill:1985tg}.

The NJL model for
top quark condensation must be considered as an approximation
to some supposed new strong dynamics. 
Indeed, the discussion of models only becomes complete when a
concrete proposal for the new dynamics is given,
e.g., a new gauge interaction, ``Topcolor,'' \cite{Hill:1991at} 
which we discuss in the next section. 
Technically, the Standard Model can always be rewritten as an
NJL model {\em for any fermion}, by a sufficiently 
arbitrary and wide range of choices of
higher dimension operators at the composite scale, 
as emphasized by Hasenfratz \etal 
\cite{Hasenfratz:1991it}.  
For example, combining certain $d=8$ operators with the $d=6$ NJL
interaction, restricting oneself to the fermion bubble approximation,
and choosing coefficients for these operators to be absurdly large, of
order $\sim 10^6$, one can argue that the Higgs is composed of the
electron and positron!  In doing this, however, one is perversely
tuning cancellations of the coupling of composite Higgs boson to its
constituents in the infrared, a phenomenon one does not expect in any
natural or realistic dynamics.

In Topcolor the additional operators are under control, 
and one can estimate the coefficients to be
small, $\lta {\cal{O}}(1)$.  Indeed, Topcolor was invented 
\cite{Hill:1991at} to address
the criticism raised in ref.\cite{Hasenfratz:1991it}. There are
proposals other than Topcolor, based upon strongly coupled $U(1)$ or
non--gauge interactions, which also aim to provide a concrete basis
for the new strong interaction, see refs. \cite{Lindner:1993ah},
\cite{Lindner:1992bs}, and \cite{Blumhofer:1993kv}.
As these interactions are not asymptotically free, it is difficult to
understand how they become strong in the infrared
or unified at high energies. Note that $U(1)$ interactions,
moreover, are typically
subleading in $N_c$ and the NJL model
is a large-$N_c$ approximation. 
Topcolor thus has an advantage;  we will have
a new dynamics such as Topcolor in mind throughout the
following discussion. 

We can implement the notion of top quark condensation by adapting the
Nambu--Jona-Lasinio (NJL) model following 
ref.\cite{Miransky:1989xi,Miransky:1989ds}.
A new fundamental interaction associated with a high energy scale, $M
$, involving principally the top quark, is postulated as a
four-fermion interaction potential:
\beq
\label{top1}
V \sim -\frac{g^2}{M^2} (\bar{\psi}^a_Lt_{Ra})_i(\bar{t}_{Rb}\psi^b)^i + ...
\eeq
where $(a,b)$ are color indices and $(i)$ is an $SU(2)_L$ index.  This
is viewed as a cut-off theory at the scale $M$, (i.e., the interaction
presumeably softens due to topgluon exchange above this scale).  For
any $g^2$ this interaction is attractive
and will form a bound-state boson, $H \sim
\bar{\psi}_L t_R$.  With sufficiently large (supercritical) $g^2 >
g_c^2 $ the interaction will trigger the formation of a low energy
condensate, $\VEV{H} \sim \VEV{\bar{t}t}$. The condensate has the
requisite $I=1/2$ and $Y=-1$ quantum numbers of the Standard Model
Higgs boson condensate, enabling it to break $SU(2)\times
U(1)\rightarrow U(1)$ in the usual way.

The subsequent
analysis is a straightforward application of the NJL model, as
discussed in Appendix B.  For supercritical $g^2 > g_c^2$ the theory
undergoes spontaneous symmetry breaking.  The associated
Nambu--Goldstone modes become the longitudinal $W$ and $Z$.  In the
fermion--loop approximation, one obtains the Pagels-Stokar formula
which connects the Nambu--Goldstone boson decay constant, $f_\pi = v_{wk}$,
to the constituent quark mass,
\cite{Miransky:1989xi,Miransky:1989ds,Bardeen:1990ds}:
\beq
\label{ps1}
f_\pi^2 = v^2_{wk} = \frac{N_c}{16\pi^2} m^{2}_{t} (
\log\frac{M^2}{m_t^{2}} + k).
\eeq
Here $m_t$ is the top quark mass, which is
now the dynamical mass gap of the theory.  The constant $k$
is associated with the precise matching onto, and definition of, the high
energy theory; it comes from the ellipsis of eq.(\ref{top1}),  
and in Topcolor models $k \lta {\cal{O}}(1)$.  

Now, if we take the cut--off to be $M\sim 1$ TeV and $k\approx 1$ we
predict too large a top quark mass, $m_t\sim 600 $ GeV.  On the other
hand, with very large $M\sim 10^{15}$ GeV and $k\approx 1$ we find
remarkably that $m_t\sim 160 $ GeV. Moreover, note that for large $M$
we become systematically less sensitive to $k$.

Unfortunately, in the limit of very large $M >> v_{wk}$ the model is 
extremely fine-tuned.    This is seen from the quadratic running
of the (unrenormalized) composite Higgs boson mass in
the NJL model, e.g., 
eq.(B.5):
\be
m^2_H (\mu) =  \frac{M^2}{g^2} - \frac{2N_c}{(4\pi)^{2}} (M^2 - \mu^2)
\ee
Taking $\mu \rightarrow v_{wk}$, 
we see that, in order to have  $M >> v_{wk} \sim m_H$, 
the coupling constant of the 
NJL theory must be extremely fine-tuned:
\be
\frac{g^2N_c}{8\pi^{2}} = 1 + {\cal{O}}(v_{wk}^2/M^2)
\ee
This implies extreme proximity of $g^2$ to the critical value $g_c^2 =
8\pi^2/N_c$.  If $M\sim 10^{15}$ GeV then the coupling must
be fine-tuned to within $\sim 1:10^{-30}$ of its critical value.

Critical coupling corresponds to a scale invariant
limit of the low energy theory, as in the case of critical coupling in
a second order phase transition in condensed matter physics.  Choosing
a near-critical coupling essentially tunes the hierarchy between the
large scale $\sim M$ and the weak scale $v_{weak}$ by
having approximate scale invariance over a large ``desert.''

In the limit of $M >> v_{wk}$ one can reliably use the renormalization
group to improve the predictions for the low-energy top and Higgs
masses. The full QCD and Standard Model contributions can be included
to arbitrary loop order.  For the top quark, the main effect is
already seen in the one-loop RG equation:
\be
\label{rgtop}
16\pi^2 \frac{\partial g_t}{\partial \ln\mu }=
\left(N_c +\frac{3}{2}\right)g_t^3 - (N_c^2-1)g_tg_{QCD}^2
\ee
with the ``compositeness boundary condition'':
\be
\frac{1}{g_t^2} \rightarrow \frac{N_c}{(4\pi)^{2}} \ln(M^2/\mu^2)
\qquad
\mu \rightarrow M
\qquad
\ee
The boundary condition states that if the Higgs doublet is a pure
$\bar{t}t$ bound state, then the Higgs-Yukawa coupling to the
top-quark must have a Landau pole at the composite scale $M$.
The low energy value of $g_t$, as given by the solution to eq.(\ref{rgtop}),
is the quasi-infrared fixed point, given
by the approximate vanishing of the {\em rhs} of eq.(\ref{rgtop}). The
fixed point is only $\ln(\ln(M))$
sensitive to the UV scale $M$.  The low energy quasi-infrared fixed
point prediction in the Standard Model of $m_t \approx 220$ GeV with
$M\sim 10^{15}$ GeV \cite{Hill:1981yr,Hill:1981sq} is large in
comparison to the observed $m_t = 176\pm 3$ GeV.  The result does,
however, depend on the exact structure (e.g., the particle content) of
the high energy theory, and one comes
sufficiently close to the physical top mass
that perhaps new dynamics can be introduced into the
Standard Model to fix the prediction.
 
In the MSSM, for example, one obtains $m_t
\sim 200\sin\beta$ GeV, which determines a predicted $\tan\beta$ when
compared to the experimental $m_t$ \cite{Bardeen:1994rv},
\cite{Carena:1992ky}.
Top quark condensation has been adapted to supersymmetric models
\cite{Clark:1990tq}, but a problem arises. Essentially, the
nonrenormalization of the superpotential implies that the effective
bound state Higgs mass does not run quadratically; it runs only
logarithmically owing to the Kahler potential, or kinetic term,
renormalization effects.  Hence, the requisite new strong interaction
scale $M$ must be of order $v_{weak}$, and cannot be placed in any
sensible way at $\sim M_{GUT}$.  This may be acceptable in the context
of top-seesaw models (Section 4.4), but the general viability of
Supersymmetry in these schemes has not been examined.  While the
quasi-infrared fixed point does play an important role in the MSSM
\cite{Bardeen:1994rv}, \cite{Carena:1992ky}, it appears that the naive 
compositeness interpretation is difficult to maintain in SUSY.

There have been many applications of these ideas to other specific
schemes; we mention a few variants here.  M. Luty demonstrated how to
produce multi-Higgs boson models in the context of top and bottom
condensation \cite{Luty:1990bg} (see also \cite{Andrianov:1999xd},
\cite{Andrianov:1996ag}).  The problem of generating the masses of all third
generation fermions has been attacked in this context by various authors
\cite{Babu:1991vx}, \cite{Froggatt:1990wa}.

Yet another application of these ideas is to the neutrino spectrum,
and the formation of neutrino condensates; Hill, Luty and Paschos have
considered the dynamical formation of right-handed neutrino Majorana
condensates and the see-saw mechanism, \cite{Hill:1991ge}
\cite{Hill:1991ge} (see also: \cite{Martin:1991xw}).  Some of these
models include additional exotic matter such as a sequential fourth
generation or leptoquark bound states \cite{Siegemund-Broka:1992ei}.  

Suzuki has discussed the reinterpretation of the compositeness
condition as indicating formation of a composite right-handed top
quark \cite{Lebed:1992qv}.  Such composites in the NJL context are
necessarily Dirac particles, and the Suzuki model affords an
interesting variation on the NJL model in which a composite Dirac
fermion arises (this of course happens automatically in the composite
SUSY models).  For related comprehensive reviews of top condensation
which describe other variant schema see: \cite{Cvetic:1997eb},
\cite{King:1995yr} and \cite{Hill:1990mt}.

The key problem with top
condensation models is that either {\em (i)} the new dynamics lies at a very
high energy scale, and the top mass is predicted, but there is
an enormous amount of fine--tuning, or {\em (ii)} the new dynamics
lies at the TeV scale, hence less fine-tuning,
yet the predicted top quark mass is then too
large.  We turn presently to a discussion of more
natural models of class {\em (ii)}, based upon
Topcolor, which can provide remedies for these problems.

\subsection{Topcolor}

%
\def\a{\alpha}
\def\b{\beta}
\def\c{\chi}
\def\d{\delta}
\def\D{\Delta}
\def\e{\epsilon}
\def\f{\phi}
\def\F{\Phi}
\def\vf{\varphi}
\def\g{\gamma}
\def\G{\Gamma}
\def\h{\eta}
\def\i{\iota}
\def\j{\psi}
\def\J{\Psi}
\def\k{\kappa}
\def\l{M}
\def\L{M_0}
\def\m{\mu}
\def\n{\nu}
\def\o{\omega}
\def\O{\Omega}
\def\p{\pi}
\def\P{\Pi}
\def\q{\theta}
\def\Q{\Theta}
\def\r{\rho}
\def\s{\sigma}
\def\S{\Sigma}
\def\t{\tau}
\def\u{\upsilon}
\def\U{\Upsilon}
\def\x{\xi}
\def\X{\Xi}
\def\z{\zeta}
\def\sl{l}
\def\sf{f}
\def\fr{\frac}
\def\ba{\begin{array}}
\def\ea{\end{array}}
\def\bz{\begin{equation}}
\def\ez{\end{equation}}
\def\by{\begin{eqnarray}}
\def\ey{\end{eqnarray}}
\def\ma{\matrix}
\def\nn{\nonumber}
\newcommand{\ls}[1]
   {\dimen0=\fontdimen6\the\font 
    \lineskip=#1\dimen0
    \advance\lineskip.5\fontdimen5\the\font
    \advance\lineskip-\dimen0
    \lineskiplimit=.9\lineskip
    \baselineskip=\lineskip
    \advance\baselineskip\dimen0
    \normallineskip\lineskip
    \normallineskiplimit\lineskiplimit
    \normalbaselineskip\baselineskip
    \ignorespaces}

\def\oldrefledge{\hangindent3\parindent}
\def\oldreffmt#1{\rlap{[#1]} \hbox to 2\parindent{}}
\def\oldref#1{\par\noindent\oldrefledge \oldreffmt{#1}
	\ignorespaces}
\def\figledge{\hangindent=1.25in}
\def\figfmt#1{\rlap{Figure {#1}} \hbox to 1in{}}
\def\fig#1{\par\noindent\figledge \figfmt{#1}
	\ignorespaces}
%
\def\ie{\hbox{\it i.e.}{}}	\def\etc{\hbox{\it etc.}{}}
\def\eg{\hbox{\it e.g.}{}}	\def\cf{\hbox{\it cf.}{}}
\def\etal{\hbox{\it et al.}}
\def\dash{\hbox{---}}
\def\tr{\mathop{\rm tr}}
\def\Tr{\mathop{\rm Tr}}
\def\Im{\mathop{\rm Im}}
\def\Re{\mathop{\rm Re}}
\def\bR{\mathop{\bf R}{}}
\def\bC{\mathop{\bf C}{}}
\def\partder#1#2{{\partial #1\over\partial #2}}
\def\secder#1#2#3{{\partial^2 #1\over\partial #2 \partial #3}}
\def\bra#1{\left\langle #1\right|}
\def\ket#1{\left| #1\right\rangle}
\def\VEV#1{\left\langle #1\right\rangle}
\def\gdot#1{\rlap{$#1$}/}
\def\abs#1{\left| #1\right|}
\def\pr#1{#1^\prime}
\def\ltap{\raisebox{-.4ex}{\rlap{$\sim$}} \raisebox{.4ex}{$<$}}
\def\gtap{\raisebox{-.4ex}{\rlap{$\sim$}} \raisebox{.4ex}{$>$}}
\def\contract{\makebox[1.2em][c]{
	\mbox{\rule{.6em}{.01truein}\rule{.01truein}{.6em}}}}
\def\slash#1{#1\!\!\!/\!\,\,}
\def\beq{\begin{equation}}
\def\eeq{\end{equation}}
\def\bea{\begin{eqnarray}}
\def\eea{\end{eqnarray}}
\def\half{\frac{1}{2}}
\def\aeq{\eeq}
\def\bq{\begin{quote}}
\def\eq{\end{quote}}
\def\pr{{\sl Phys. Rev.~}}
\def\np{{\sl Nucl. Phys.~}}
\def\pl{{\sl Phys. Letters~}}
\def\prl{{\sl Phys. Rev. Letters~}}
\def \Msol {M_\odot}
\def\GeV{\,{\rm GeV}}
\def\eV {\,{\rm  eV}}
\def\Mpc{\,{\rm Mpc}}
\def\pc{\,{\rm pc}}
\def\half{\frac{1}{2}}
\def \lta {\mathrel{\vcenter
     {\hbox{$<$}\nointerlineskip\hbox{$\sim$}}}}
\def \gta {\mathrel{\vcenter
     {\hbox{$>$}\nointerlineskip\hbox{$\sim$}}}}
\def \endpage {\vfill \eject}
\def \endline {\hfill \break}
\def \etal {{\it et al.}\ }

\newcommand{\bdm}{\begin{displaymath}}
\newcommand{\edm}{\end{displaymath}}
\def\simlt{\mathrel{\lower2.5pt\vbox{\lineskip=0pt\baselineskip=0pt
           \hbox{$<$}\hbox{$\sim$}}}}
\def\simgt{\mathrel{\lower2.5pt\vbox{\lineskip=0pt\baselineskip=0pt
           \hbox{$>$}\hbox{$\sim$}}}}
\catcode`@=11
\newcount\@tempcntc
\def\@citex[#1]#2{\if@filesw\immediate\write\@auxout{\string\citation{#2}}\fi
  \@tempcnta\z@\@tempcntb\m@ne\def\@citea{}\@cite{\@for\@citeb:=#2\do
    {\@ifundefined
       {b@\@citeb}{\@citeo\@tempcntb\m@ne\@citea\def\@citea{,}{\bf ?}\@warning
       {Citation `\@citeb' on page \thepage \space undefined}}%
    {\setbox\z@\hbox{\global\@tempcntc0\csname b@\@citeb\endcsname\relax}%
     \ifnum\@tempcntc=\z@ \@citeo\@tempcntb\m@ne
       \@citea\def\@citea{,}\hbox{\csname b@\@citeb\endcsname}%
     \else
      \advance\@tempcntb\@ne
      \ifnum\@tempcntb=\@tempcntc
      \else\advance\@tempcntb\m@ne\@citeo
      \@tempcnta\@tempcntc\@tempcntb\@tempcntc\fi\fi}}\@citeo}{#1}}
\def\@citeo{\ifnum\@tempcnta>\@tempcntb\else\@citea\def\@citea{,}%
  \ifnum\@tempcnta=\@tempcntb\the\@tempcnta\else
   {\advance\@tempcnta\@ne\ifnum\@tempcnta=\@tempcntb \else \def\@citea{--}\fi
    \advance\@tempcnta\m@ne\the\@tempcnta\@citea\the\@tempcntb}\fi\fi}
\catcode`@=12
\newcommand{\SM}{Standard Model}
\newcommand{\TDM}{Two-Doublet Model}
\newcommand{\THDM}{Two-Higgs-Doublet Model}
\newcommand{\tdm}{two-doublet model}
\newcommand{\br}{branching ratio}
\newcommand{\Ma}{M_{A^{0}}}
\newcommand{\Mg}{M_{H^{\pm}}}
\newcommand{\bsg}{b \rightarrow s \gamma }
\newcommand{\la}{M}
\newcommand{\La}{M}
\newcommand{\bP}{\bar{\Psi}_L}
\newcommand{\cD}{{\cal D}}
\newcommand{\cL}{{\cal L}}

%
%
%


\subsubsection{Gauging Top Condensation}

\noindent

Our previous discussion of top condensation noted that the
relevant interaction Lagrangian, eq.(\ref{top1}), must be viewed
as an effective description of a more fundamental theory. Let
us presently consider what that theory might be.
A key observation is
that a Fierz rearrangement of the interaction term leads to
\cite{Hill:1991at}:
\beq
  -\frac{g^2}{M^2}(\bar{\psi}^a_Lt_{Ra})_i(\bar{t}_{Rb}\psi^b)^i
=
\frac{g^2}{M^2}(\bar{\psi}_{iL} \gamma_\mu \frac{\lambda^A}{2} \psi^i_L )
(\bar{t}_{R} \gamma^\mu \frac{\lambda^A}{2} t_{R}) + O(1/N_c)
\eeq
where $N_c=3$ is the number of colors.  This is exactly
the form (including the sign) induced
by a massive color octet vector boson exchange, and suggests 
a new gauge theory with certain properties: 
{\em (i)} it must be spontaneously broken at a scale of
order $M$; {\em (ii)} it must be strongly coupled at the scale $M$
to produce deeply bound composite Higgs bosons and trigger chiral
condensates; {\em (iii)} it must involve the color degrees of freedom of
the top quark, analogous to QCD.  The relevant models therefore
involve embedding of QCD into some large group $G$ which is
sensitive to the flavor structure of the Standard Model 
\cite{Hill:1991at,Hill:1995hp}.

>From this point
of view the embedding of QCD into a minimal $SU(3)_1\times SU(3)_2$
gauge group at higher energies seems
a plausible 
scenario\footnote{\addtolength{\baselineskip}{-.4\baselineskip} 
We will see in Section 4.5 that such a structure
occurs in theories with extra dimensions, and 
anticipates the idea of ``deconstruction.''}. The second (weaker)
$SU(3)_2$ gauge interactions act upon the first and second
generation quarks while the first (stronger) $SU(3)_1$
interaction acts upon the third generation. This
drives the formation
of the top condensate $\langle \bar{t}t\rangle =0$. 
Clearly some additional dynamics is then required to
suppress the formation of the 
$b$-quark consdensate, 
$\langle \bar{b}b\rangle =0$.

We note that a different class of models based upon a strong $U(1)$
gauge interaction has also been proposed 
by Bonische, \cite{Bonisch:1991hm}, Giudice and Raby,
\cite{Giudice:1992sz}, and Lindner and Ross, \cite{Lindner:1992bs}
in which the top quark carries the extra, strong $U(1)$ charge. 
The desired NJL interaction term of eq.(\ref{top1})
occurs, but with a $1/N_c$ suppression.

\begin{figure}[t]
\label{fierz}
\vspace{4cm}
\includegraphics{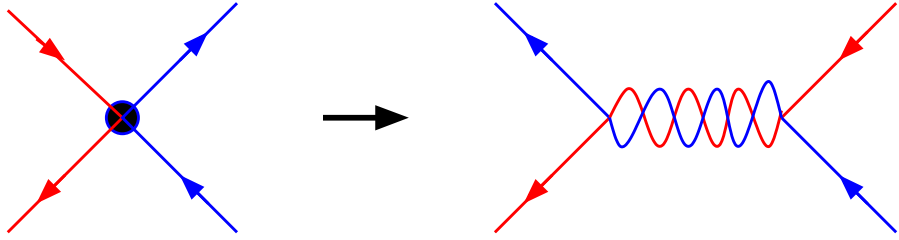}
\vspace{1cm}
\caption[]{\small \addtolength{\baselineskip}{-.4\baselineskip} 
Fierz rearrangement of the attractive
Nambu-Jona-Lasinio interaction
contains the color current-current
interaction to leading order in $1/N_c$.
}
\end{figure}

Topcolor Assisted Technicolor (TC2)
\cite{Hill:1991at,Hill:1995hp}
postulates that the
top quark mass is large because  it is a combination of 
{\em (i)} a small {\em
fundamental component}, $\epsilon m_t << m_t$ generated by, e.g., ETC
or a fundamental Higgs boson, plus {\em (ii)} a large
{\em dynamical quark mass
component,} $(1-\epsilon)m_t\approx m_t$, generated by Topcolor dynamics at
the scale $M \sim 1$ TeV, which is coupled preferentially to the third
generation.  In a pure top condensation model we would
have $\epsilon = 0$, and we produce the usual
three NGB's that become longitudinal
$W^\pm_L$ and $Z_L$. With nonzero $\epsilon$ we are relaxing the
requirement that the $\VEV{\bar{t}t}$ condensate account for all of
the electroweak symmetry breaking.  

The ETC component of the top quark mass,
the $\epsilon
m_t$ term, is expected to be  $\sim m_b$. Hence we 
assume $\epsilon \lta 0.1$.
Furthermore, the $b$-quark can receive mass contributions
from instantons in $SU(3)_{1}$ \cite{'tHooft:1976up,Hill:1991at}, 
so that the fundamental EWSB sector (e.g., ETC) 
needs to provide only a very small, or possibly none of the, fundamental 
contribution to $m_b$.  Remarkably
the strong CP--$\theta_1$ angle in the Topcolor $SU(3)_1$ can provide
the origin of CKM CP-violation \cite{Buchalla:1996dp}.

After all of the dynamical symmetry breaking there are
three NGB's from the TC sector, and three NGB's from top condensation
sector. One linear combination of these, mostly favoring the TC
NGB's, will become the longitudinal $W_L^\pm$ and $Z_L$.  The orthogonal
linear combination will appear in the spectrum as an isovector multiplet of
PNGB's, $\tilde{\pi}^a$. These objects
acquire mass as a consequence of the interference between the dynamical and
ETC masses of the top quark, i.e., the masses of the $\tilde{\pi}^a$
will be proportional to 
$\epsilon$.\footnote{\addtolength{\baselineskip}{-.4\baselineskip} 
Technicolor itself can be a spontaneously broken theory with NJL-like
dynamics \cite{Hill:1993ev}.  This has the benefit of suppressing
resonant contributions to the $S$ and $T$ parameters, and may be part
of the TC2 scenario described below.}   
We refer to the $\tilde{\pi}^a$ as {\em
top--pions} (note that in the minimal pairing of the $\psi_L =(t,b)_L$
doublet with the $t_R$ singlet there is no singlet top-$\eta'$).
For $\epsilon\lta 0.05 - 0.10$, we will find that the
top--pions have masses of order $\sim 200$ GeV.   
They are phenomenologically forbidden from occuring much below $\sim
165$ GeV \cite{Balaji:1997va} due to the absence of the decay
mode $t \to \tilde{\pi}{}^+ + b$. 

Because Topcolor and low-scale TC address complementary issues, it is
natural to combine these schemes into TC2.
\cite{Lane:1995gw,Lane:2000pa}.  Happily, each half of 
the model appears to solve many difficulties experienced by the other.
For example, since the top quark couples only weakly to ETC in TC2
models, the problem of preventing $t \to P^+ + b$ in Low-Scale TC is
avoided.  Moreover, since ETC need not now provide the full top or
bottom quark masses, the devastating FCNC constraints on ETC are
alleviated (perhaps including some ``walking''), 
producing a viable Technicolor component
of the TC2 model.  Even the $S$ parameter is likely to be less
problematic.  We do, however, have to deal with FCNC's induced by mass
mixing of $t$ and $b$ with light quarks, and off-diagonal light PNGB
couplings.  These can evidently be controlled by systematic choices of
fermion mass matrix textures, and the model again appears viable.  For
discussion of these flavor physics questions, we refer the reader to
Section 4.2.4, to \cite{Buchalla:1996dp}, which elaborates on the
issues raised in \cite{Kominis:1995fj}, and also to
\cite{Burdman:2000in} and \cite{Simmons:2001va}.

We mention some variant applications and elaborations of Topcolor.
``Bosonic Topcolor,'' \cite{Aranda:2000vk,Aranda:2000xy}, is
based on maintaining the fundamental sector as an elementary
Higgs boson, similar in spirit to Bosonic Technicolor (section 3.7.2)
 and is consistent
with current phenomenological constraints.  We view Bosonic Topcolor
or TC as a useful framework for identifying acceptable
model-building directions, with the eventual aim of replacing the
fundamental Higgs boson with something dynamical or Supersymmetric
\cite{Carone:1997kc}. Bosonic Topcolor anticipates a class of
models in which SUSY arrives, not near the EWSB scale, but rather
closer to the multi-TeV scale. 
``Topcolor Assisted Supersymmetry,'' 
provides a mechanism for solving the flavor
problems in the Sfermion sector
\cite{Liu:1999ah}.  Some authors \cite{Triantaphyllou:2000uu} have
argued for a natural origin of Topcolor-like structures in
grand-unified models, and models with mirror fermions
\cite{Lindner:1998er,Triantaphyllou:1998ke}. 
Applications to left-right symmetry breaking have also been considered
in \cite{Akhmedov:1996vm}.  Finally, we note that related models
involving the idea of a horizontal replication of $SU(2)_L
\rightarrow G$, where $G$ contains
some strong interactions, have been explored under the rubrics of
non-commuting extended technicolor \cite{Chivukula:1994mn}, (see
section 3.3.2) and top-flavor
\cite{Muller:1996dj,Malkawi:1996fs,He:1999vp}.  A potent
constraint, often overlooked in these schemes, is the presence of
potentially dangerous large instanton effects that violate $B+L$ over
some portion of the parameter space of the model
\cite{'tHooft:1976up,Hill:1991at}).

\subsubsection{Gauge Groups and the Tilting Mechanism}

In Topcolor models, a new strong dynamics occurs primarily in
interactions that involve $\overline{t}t\overline{t}t$,
$\overline{t}t\overline{b}b$, and $\overline{b}b\overline{b}b$. 
The dynamics must cause the top
quark to condense, $\VEV{\overline{t}t} \neq 0$, while simultaneously
suppressing the formation of a large bottom quark condensate,
$\VEV{\overline{b}b} \approx 0$.  This requires
``tilting'' the vacuum.

Common solutions involve introducing additional strong $U(1)'$
interactions that are attractive in the ${\overline{t}t}$ channel and
repulsive in the $\VEV{\overline{b}b}$ channel \cite{Hill:1991at}.
This can take the form of imbedding the weak
hypercharge, $U(1)_Y\rightarrow U(1)_{Y1}\times U(1)_{Y2}$,
since $Y$ has the desired properties.
There are then constraints on the running of the strong $U(1)$
couplings and the demands of tilting.
Effective field theory analysis of tilting indicates it is not an
unreasonable possibility, but may require some fine-tuning at the few
percent level \cite{Chivukula:1998vd}.

\vskip 0.1cm
\noindent
{\bf (i) Classic Topcolor}
\vskip 0.2cm

Topcolor 
with a $U(1)'$ tilting mechanism
 leads to the following gauge group structure
at high energies
\cite{Hill:1991at}:
\beq
SU(3)_1\times SU(3)_2
\times U(1)_{Y1}\times U(1)_{Y2}
\times SU(2)_L \rightarrow
SU(3)_{QCD}\times U(1)_{EM}
\eeq
where $SU(3)_1\times U(1)_{Y1}$ ($SU(3)_2\times U(1)_{Y2}$) couples
preferentially to the third (first and second) generations.  The
$U(1)_{Yi}$ are just rescaled versions of electroweak $U(1)_{Y}$ The
fermions are then assigned to $(SU(3)_1, SU(3)_2, {Y_1}, {Y_2}$) as
follows:
\bea
(t,b)_L \;\;   &\sim  & (3,1,{1}/{3},0) \qquad \qquad
(t,b)_R \sim \left(3,1,({4}/{3},-{2}/{3}),0\right) \\ \nonumber
(\nu_\tau,\tau)_L &\sim & (1,1,-1,0) \qquad \qquad
\tau_R \sim \left(1,1,-2,0\right) \\ \nonumber
  & & \\ \nonumber
(u,d)_L,\;\;  (c,s)_L & \sim & (1,3,0,{1}/{3}) \qquad \qquad
(u,d)_R, \;\; (c,s)_R \sim \left(1,3,0,({4}/{3},-{2}/{3})\right) \\  
\nonumber
(\nu, \ell)_L\;\; \ell = e,\mu & \sim & (1,1,0,-1) \qquad \qquad
\ell_R \sim \left(1,1,0,-2\right)
\eea
The extended color interactions must be broken to the diagonal
subgroup which can be identified with QCD. We assume this is
accomplished through an (effective) scalar field:
\beq
\Phi \sim  (3,\bar{3}, y, -y) \label{phi_q}
\eeq
In fact, when $\Phi$ develops a VEV
the theory spontaneously breaks down to ordinary QCD
$\times U(1)_{Y}$ at a scale assumed to be
$\sim 1$~TeV, before Topcolor becomes
confining.  Nonetheless, the $SU(3)_1$ is assumed to
be strong enough to form chiral condensates.  
The vacuum will be tilted by the $U(1)_{Y1}$ couplings
which permits the formation of a $\VEV{\overline{t}t}$
condensate but disallows the $\VEV{\overline{b}b}$ condensate is due
to the $U(1)_{Yi}$ couplings.

The symmetry breaking will gives rise to three top-pion states near
the top mass scale.  If the Topcolor scale is of the order of 1~TeV,
the top-pions will have a decay constant, estimated from the
Pagel's-Stokar relation to be $f_\pi \approx 60$ GeV, and a strong
coupling constant given by, $g_{tb\pi} \approx
m_t/\sqrt{2}f_\pi\approx 2.5$.  If $m_{\tilde{\pi}} > m_t + m_b$, the
top-pions may be observable in $\tilde{\pi}^+\rightarrow t +
\overline{b}$.

Let us define
the coupling constants (gauge fields) of $SU(3)_1\times SU(3)_2$ to be,
respectively, $h_1$ and $h_2$ ($A^A_{1\mu}$ and $A^A_{2\mu}$) while for
$U(1)_{Y1}\times U(1)_{Y2}$ they are respectively ${q}_1$ and $q_2$,
$(B_{1\mu}, B_{2\mu})$.  The $U(1)_{Yi}$ fermion couplings are then
$q_i\frac{Yi}{2}$, where $Y1, Y2$ are the charges of the fermions
under $U(1)_{Y1}, U(1)_{Y2}$ respectively.
A techni--condensate or explicit Higgs can break $SU(3)_1\times SU(3)_2
\times U(1)_{Y1}\times U(1)_{Y2}
\rightarrow SU(3)_{QCD}\times  U(1)_Y$
at a scale $M \gta 240$ GeV, or it can fully break $SU(3)_1\times
SU(3)_2
\times U(1)_{Y1}\times U(1)_{Y2}\times SU(2)_L
\rightarrow SU(3)_{QCD}\times  U(1)_{EM}$
at the scale $M_{TC}= 240$ GeV.  Either scenario typically leaves a
{\em residual global symmetry} implying a
degenerate, massive color octet of ``topgluons,'' $B_\mu^A$, and a
singlet heavy $Z'_{\mu}$.  The gluon $A_\mu^A$ and topgluon $B_\mu^A$
(the SM $U(1)_Y$ field $B_\mu$ and the $U(1)'$ field $Z'_\mu$), are
then defined by orthogonal rotations with mixing angle $\theta$
($\theta'$):
\bea
& &
\cot\theta = h_1/h_2;\qquad\qquad
\frac{1}{g_3^2} = \frac{1}{h_1^2} +
\frac{1}{h_2^2} 
\eea
and:
\bea
& &
\cot\theta' = q_1/q_2;\qquad\qquad 
\frac{1}{g_1^2} = \frac{1}{q_1^2} +
 \frac{1}{q_2^2} ;
\eea
where $g_3$ ($g_1$) is the QCD ($U(1)_Y$)
coupling constant at $M_{TC}$.
We demand $\cot\theta \gg 1$ and $\cot\theta' \gg 1$ to tilt the
strongest interactions and to select the top quark direction for
condensation.  The masses of the degenerate octet of topgluons and
$Z'$ are given by $M_B\approx g_3M/\sin\theta\cos\theta$ $ M_{Z'}
\approx g_1M/\sin\theta'\cos\theta'$.  The usual QCD  ($U(1)_Y$
electroweak) interactions are obtained for any quarks that carry
either $SU(3)_1$ or $SU(3)_2$ triplet quantum numbers (or $U(1)_{Yi}$
charges).  

Integrating out the heavy bosons $Z'$ and $B$ gives  
rise to effective low energy four fermion interactions. 
The effective Topcolor interaction mediated by $B$
is strongest in the third generation and takes the form:
\bea
{\cal{L}}'_{TopC} & =  & -\frac{2\pi\kappa}{M_B^2}\left(
\bar{t}\gamma_\mu \frac{\lambda^A}{2} t +
 \bar{b}\gamma_\mu \frac{\lambda^A}{2} b
\right)
\left(
\bar{t}\gamma^\mu \frac{\lambda^A}{2} t +
 \bar{b}\gamma^\mu \frac{\lambda^A}{2} b
\right) .\label{topc_in}
\eea
where:
\begin{equation}\label{kk1def}
\kappa=\frac{g^2_3\cot^2\theta}{4\pi}
\end{equation}
This interaction is attractive in the color-singlet $\bar{t}t$ and $\bar{b}b$
channels and
invariant under color $SU(3)$
and $SU(2)_L\times SU(2)_R
\times U(1) \times U(1)$ where $SU(2)_R$ is the custodial
symmetry of the electroweak interactions.

In addition  we have the $U(1)_{Y1}$
interaction Lagrangian (which breaks custodial $SU(2)_R$):
\be
{\cal{L}}'_{Y1} =   -\frac{2\pi\kappa_1}{M^2_{Z'}}\left(
\frac{1}{6}\bar{\psi}_L\gamma_\mu \psi_L +
\frac{2}{3}\bar{t}_R\gamma_\mu t_R
-\frac{1}{3}\bar{b}_R\gamma_\mu  b_R -
\frac{1}{2}\bar{\ell}_L\gamma_\mu \ell_L -
\bar{\tau}_R\gamma_\mu \tau_R \right)^2
\label{u1_in}
\ee
where:
\begin{equation}\label{kk2def}
\kappa_1=\frac{g^2_1\cot^2\theta'}{4\pi} \qquad
\end{equation}
and where $\psi_L = (t,b)_L$, $\ell_L = (\nu_\tau,\tau)_L$ and
$\kappa_1$ is assumed to be $O(1)$.  Note that while too small a value
for $\kappa_1$ signifies fine-tuning in the model, phenomenological
constraints limit how large $\kappa_1$ can be.

For sufficiently large $\kappa$,
we trigger the formation of a low energy condensate,
$\VEV{\overline{t}t + \overline{b}b}$, which breaks
$SU(2)_L\times SU(2)_R\times U(1)_Y
\rightarrow U(1)\times SU(2)_{c}$, where $SU(2)_{c}$ is
a global custodial symmetry. The $U(1)_{Y1}$ force is attractive in the  
${\overline{t}t}$
channel and repulsive in the ${\overline{b}b}$ channel. 
We find, in concert, 
the critical and subcritical values of the combinations:
\beq
\kappa + \frac{2\,\kappa_1}{9N_c} > \kappa_{crit} ;
\qquad
\kappa_{crit} > \kappa - \frac{\kappa_1}{9N_c}
\label{crit_con}
\eeq
by using the large-$N_c$ NJL (fermion loop)
approximation (similar criticality constraints
were discussed previously in the NJL model of top
condensation of Miransky, 
Tanabashi, Yamawaki, \etal \cite{Miransky:1989xi,Miransky:1989ds}). 

\begin{figure}[t]
\vspace{10cm}
\includegraphics{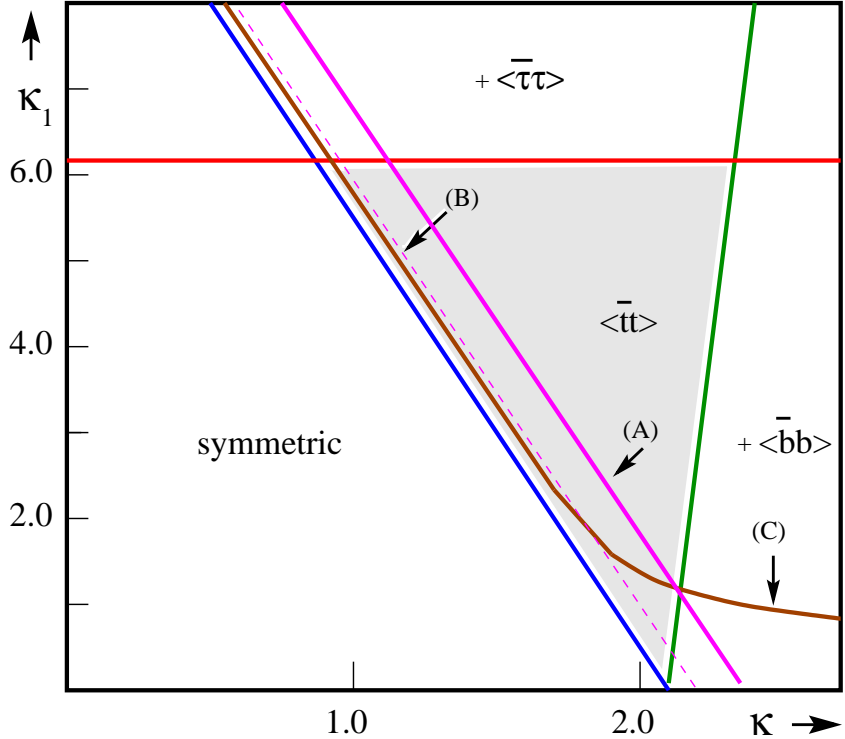}
\vspace{1cm}
\caption[]{\small \addtolength{\baselineskip}{-.4\baselineskip} 
The full phase diagram of the Topcolor model.  The top quark alone
condenses for values of $\kappa$ and $\kappa_1$ corresponding to
hatched region. In other regions additional unwanted condensates turn
on (for still larger $\kappa_1$ a $\VEV{\bar{\tau} b}$ condensate
forms. (A) $M_B=1.0$ TeV; (B) $M_B =3.0$ TeV; (C) upper bound from
$Z\rightarrow \bar{\tau}\tau$ (figure from \cite{Buchalla:1996dp}).}
\label{fig:phase-dg}
\end{figure}

The phase diagram of the model is shown in Fig.(\ref{fig:phase-dg})
from \cite{Buchalla:1996dp}.  The criticality conditions
(\ref{crit_con}) define the allowed region in the $\kappa_1$--$\kappa$
plane in the form of the two straight solid lines intersecting at
$(\kappa_1=0,\kappa=\kappa_{crit})$.  To the left of these lines lies
the symmetric phase, in between them the region where only a
$\VEV{\bar{t}t}$ condensate forms and to the right of them the phase
where both $\VEV{\bar{t}t}$ and $\VEV{\bar{b}b}$ condensates
arise. The horizontal line marks the region above which $\kappa_1$
makes the $U(1)_{Y1}$ interaction strong enough to produce a
$\VEV{\bar{\tau}\tau}$ condensate. This line is meant only as a rough
indication, as the fermion-bubble (large-$N_c$) approximation, which
we have used, fails for leptons.  There is an additional constraint
from the measurement of $\Gamma(Z\to\tau^+\tau^-)$, confining the
allowed region to the one below the solid curve. This curve
corresponds to a $2\sigma$ discrepancy between the Topcolor prediction
and the measured value of this width.  In the allowed region a top
condensate alone forms. The constraints favor a strong $SU(3)_{1}$
coupling and allow a range of $U(1)_{Y1}$ couplings.

How
does Technicolor know to break Topcolor when the latter
triggers chiral condensates?
This may involve self-breaking
schemes, as considered by Martin \cite{Martin:1992aq},
\cite{Martin:1992aq}, \cite{Martin:1993mj}. 
Topcolor interactions may themselves produce the
condensate of $\Phi$ as a dynamical boundstate, leading to
self-breaking at the Topcolor scale.
Other such scenarios require further study.

We can also construct a model with no strong
$U(1)$ tilting interaction \cite{Hill:1991at,Hill:1995hp}
called ``Type II Topcolor.''  The model 
requires additional heavy vectorlike $b$' quarks,
but  there is no $Z$'
required for tilting because 
of the particular gauge charge assignments
under Topcolor.  These models are intriguing, but have received
little attention in the literature, and we refer the interested
reader to the original references  
\cite{Hill:1991at,Hill:1995hp,Buchalla:1996dp}.

\vskip 0.1cm
\noindent
{\bf (ii) Flavor-Universal Topcolor}
\vskip 0.2cm

The flavor-universal variant of Topcolor
\cite{Chivukula:1996yr,Lane:1998qi,Popovic:1998vb} is 
based on the same gauge group as above, but assigns 
all quarks to be triplets under the
strongly-coupled $SU(3)_1$ gauge group and singlets of $SU(3)_2$.  As a
result, the heavy color-octet of gauge bosons that results from
breaking of the extended color group to its QCD subgroup are
``flavor-universal colorons'' coupling to all quark flavors with equal
strength \cite{Chivukula:1996yr}:
\beq
{\cal L}_{C} = g_3 \cot\theta (C^A \cdot J^A_C)
\eeq
where
\beq
J_C^{\mu\,A} = \sum_q \bar{q} \gamma^\mu \frac{\lambda^A}{2} q
\eeq 
The weak and hypercharge assignments
of the quarks and leptons are as in Classic Topcolor, so that the
properties and couplings of the Z' boson are unchanged.

Once again, the values of $\kappa_i = g_i^2/4\pi$ are jointly
constrained by several different pieces of physics.  Above all, the
top quark should be the only fermion to condense and acquire a large
mass.  Applying the gauged NJL analysis
\cite{Bardeen:1986sm,Leung:1986sn,Appelquist:1988fm,Kondo:1989qd} 
to all the Standard Model fermions,
one can seek solutions with $\langle\,\bar{t}t\,\rangle\neq 0$ and
$\langle\,\bar{f}f\,\rangle = 0$ for $f
\neq t$.  Top condensation occurs provided 
that\footnote{\addtolength{\baselineskip}{-.4\baselineskip} 
Likewise, if one applies the 
fermion bubble approximation to leptons, The strong U(1) interaction will
not cause leptons to condense if \cite{Popovic:1998vb,Buchalla:1996dp}
$\kappa_{1}\ < 2 \pi - 6 \alpha_Y$.  This approximate condition is
superceded by other constraints.}
\cite{Popovic:1998vb,Buchalla:1996dp}
\begin{equation}
\kappa_{3} + {2\over 27}\kappa_{1} \geq {{2 \pi}\over 3} -
{4\over3}\alpha_s - {4\over 9}\alpha_Y
\end{equation}
where we again take $\kappa_{crit, NJL} \approx 2\pi/3$
\cite{Nambu:1961tp,Nambu:1961fr}. 
Quarks other than the top quark will not condense provided that
additional limits on $\kappa_1$ and $\kappa_3$ are satisfied.  In
Classic TC2 models, the limit comes from requiring
$\langle\,\bar{b}b\,\rangle = 0$, while
in flavor-universal TC2 models, an even stronger limit comes 
from ensuring
$\langle\,\bar{c}c\,\rangle= 0$ \cite{Popovic:1998vb}:
\begin{equation}        
\kappa_{3} + {2\over 27}{\alpha_{Y}^{2}\over \kappa_{1}} < 
{{2 \pi}\over 3} - {4\over3}\alpha_s - {4\over 9}\alpha_Y
\label{eq-yess}
\end{equation}
As shown in Figure (\ref{fig:kappakappa-plane}), the phase diagram of
the flavor-universal model includes a region in which only top
condensation occurs; the region is similar to that of classic TC2.
(see also discussion of phase plane in \cite{Appelquist:1988fm,Kondo:1989qd}).
\begin{figure}[tb]
\vspace{6cm}
\includegraphics{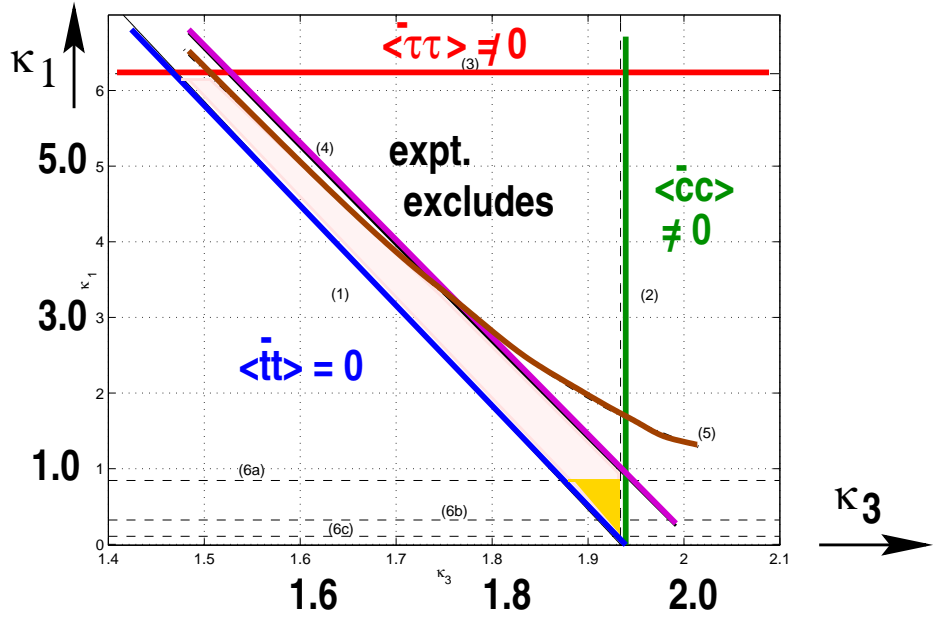}
\vspace{1cm}
\caption[two]{\small \addtolength{\baselineskip}{-.4\baselineskip} 
Joint constraints on coloron and Z' couplings 
(from \protect\cite{Popovic:1998vb}).  Curves
  (1), (2), (3) outline the `gap triangle' where only $\langle\,
  \bar{t} t \,\rangle \neq 0$ in flavor-universal TC2 models; in
  ordinary TC2 models, the triangle is roughly symmetric about the
  lowest point.  The region above curve (4) is excluded by data on
  $\Delta\rho_{*}$; the region above curve (5) is excluded by data on
  $Z\rightarrow\tau^{+}\tau^{-}$. Lines (6a-6c) are upper bounds on
  $\kappa_1$ that hold if the U(1) Landau pole lies one, two, or five
  orders of magnitude above $\Lambda$.
  The right-hand bound is steep because it
  comes from the $\langle\,\bar{c}c\,\rangle= 0$
constraint, eq. (\ref{eq-yess}).}
\label{fig:kappakappa-plane}
\end{figure}

Several types of physics further constrain the allowed region of the
$\kappa_1-\kappa_3$ plane.  As already mentioned for Classic TC2, ,
mixing between the Z and Z' bosons alters the predicted value of the Z
decay width to tau leptons from the standard model value
\cite{Buchalla:1996dp,Popovic:1998vb}.  The $\rho$ or $T$ parameter is also
sensitive both to $Z-Z'$ mixing \cite{Chivukula:1996gu} and to single
coloron exchange across the top and bottom quark loops of $W$ and $Z$
vacuum polarization diagrams \cite{Chivukula:1995dc,Popovic:1998vb}.
Finally, if the Landau pole of the strong $U(1)$ interaction is to lie
at least an order of magnitude above the symmetry-breaking scale
$\Lambda$, then $\kappa_1 \lta 1$ 
\cite{Popovic:1998vb}.  Figure \ref{fig:kappakappa-plane} summarizes
these constraints on the $\kappa_1 - \kappa_3$ plane for
flavor-universal TC2 models.  The constraints from $\Delta\rho$ and
the Landau pole also apply to the phase diagram in figure
\ref{fig:phase-dg}.  In the end, both kinds of Topcolor models are confined to a
region of the $\kappa_1 - \kappa_3$ plane in which $\kappa_1$ is small
and $\kappa_3 \approx \kappa_{crit, NJL} \approx 2\pi/3$
\cite{Popovic:1998vb}.

\subsubsection{Top-pion Masses; Instantons; The b-quark mass}

In TC2 a multiplet of PNGB top-pions is naturally
present in the spectrum. 
Top-pion masses are
estimated from the loop of Figure(\ref{epsilon-fig}), which feels
both the dynamical mass $\sim m_t$ and the explicit mass
$\sim \epsilon m_t$
\cite{Hill:1991at,Hill:1995hp}.
The top--pion decay constant is
estimated from the Pagels-Stokar
formula; using $M = M_{B}\sim 1.5$ TeV and $m_t = 175$ GeV, it is $f_\pi
\approx 60$ GeV.  The Lagrangian for the coupling of the top-pions to
quarks takes the form:
\beq
 g_{bt\tilde{\pi}} 
\left[ {i}\overline{t} 
\gamma^5 t \tilde{\pi}^0 
+\frac{i}{\sqrt{2} }
\overline{t} (1-\gamma^5) b \tilde{\pi}^+ 
+ \frac{i}{\sqrt{2} }
\overline{b} (1+\gamma^5)t  \tilde{\pi}^- 
\right]
\eeq
and the coupling strength is 
$g_{bt\tilde{\pi}} \approx m_t/\sqrt{2}f_\pi$ \cite{Hill:1995hp}.
This Lagrangian, written above in the current basis, will 
in general contain generational mixing
when one passes to the mass-matrix eigenbasis.

\begin{figure}[t]
\label{epsilon-fig}
\vspace{4cm}
\includegraphics{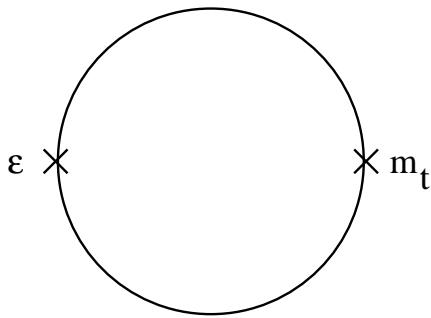}
\vspace{1cm}
\caption[]{\small \addtolength{\baselineskip}{-.4\baselineskip} 
The interference of the explicit breaking tewrm $\propto \epsilon$,
and the dynamical mass $\sim m_t$ yields an estimate of the
top-pion mass.}
\end{figure}

 Estimating the
induced top-pion mass from the fermion loop yields:
\beq
m_{\tilde{\pi}}^2 = 
\frac{N \epsilon m_t^2 M_B^2 }{8\pi^2 f_\pi^2} 
= \frac{\epsilon M_B^2 }{\log(M_B/m_t)}
\eeq
where the Pagels-Stokar formula 
is used for $f_\pi^2$
(with $k=0$) in the
last expression. For 
$\epsilon = (0.03,\; 0.1)$, 
$M_B\approx (1.5,\; 1.0) $ TeV, 
and $m_t=175$ GeV  this 
predicts $m_{\tilde{\pi}}= (180,\; 240)$ GeV.
We would
expect that $\epsilon$ is subject to
large renormaliztion effects 
and, even a bare value of $\epsilon_0 \sim 0.005$
consistent with ETC, can produce
sizeable $m_{\tilde{\pi}} > m_t$.  

Charged top-pions as light as $\sim 165$ GeV, would provide a
detectable decay mode for top quarks \cite{Balaji:1997va}.
Burdman has discussed potentially dangerous effects in $Z\rightarrow
b\bar{b}$ resulting from low mass top-pions and decay constants as small as
$\sim 60 $ GeV \cite{Burdman:1997pf}.  A comfortable phenomenological
range is slightly larger than our estimates: $m_{\tilde{\pi}}
\gta 300$ GeV and $f_\pi \gta 100$ GeV. 
These values remain  subject to large uncertainties.

The $b$ receives a mass of $\sim O(1)$
GeV from ETC. Remarkably, however, it also obtains an induced mass from
instantons in $SU(3)_{1}$.   The instanton
effective Lagrangian may be approximated by the `t Hooft flavor
determinant\footnote{\addtolength{\baselineskip}{-.4\baselineskip} 
Note that the $SU(3)_1$ CP-angle, $\theta_1$, 
cannot be eliminated from the full
quark mass matrix because of the ETC contribution to the $t$ and $b$
masses.  Indeed, it can lead to induced scalar couplings of the
neutral top--pion, and an induced CKM CP--phase,
\cite{Buchalla:1996dp}.  }, which is the effective
Lagrangian generated by zero-modes when instantons are integrated out.
The instanton 
induced $b$-quark mass can then be estimated as:
\beq
m^\star_b \approx 
\frac{3 k m_t}{8\pi^2} \sim 6.6\; \hat{k}\; GeV 
\eeq
where we generally
expect $\hat{k}\sim 1$ to $10^{-1}$ as in QCD, from fitting
the QCD 't Hooft determinant to the $\eta'$ mass.
This yields a reasonable estimate of the observed $b$ quark mass,
and for $\hat{k}\sim 1$ it comes very close to the observed value.
There also occurs an induced top--pion
coupling to $b_R$ coming from instantons.
Many of the features of the theory we have just outlined imitate
the successful chiral-constituent quark model approximation to
QCD (see, e.g., \cite{Bijnens:1993uz}), thus yielding 
a reliable picture of Topcolor dynamics.

\subsubsection{Flavor Physics: Mass Matrices, CKM and CP-violation}

Topcolor has an obvious challenge: it violates the GIM symmetry by
treating the interactions of third generation differently than those of the
first and second. Indeed, it does so with a new strong interaction.
This contrasts with, e.g., the flavor universal coloron model in which GIM
is not violated by the strong $SU(3)$ spectator interaction
\cite{Chivukula:1996yr,Simmons:2001va}.  Of course, the Higgs--Yukawa 
couplings of the Standard Model violate GIM as well, and so too must
ETC interactions in Technicolor if they are to provide the observed
nondegenerate fermion masses.  Clearly, when we seek a deeper theory
of the origin of fermion masses we must face the true dynamical origin
of GIM violation.  Topcolor faces this issue head-on at accessible
energy scales.  In consequence, it confronts significant constraints
and potential observable violations of Standard Model predictions.

We can study these questions in the context of Topcolor by examining
the textures of the fermion mass matrices that arise in these models.
The textures are controlled by the breaking patterns of horizontal
global (flavor) symmetries \cite{Buchalla:1996dp}.  The models
sketched above include a Topcolor symmetry at energies above the weak
scale, presumably subsequent to some initial breaking of a larger
group structure at a much higher energy scale $M_0$.  The third
generation fermions have different Topcolor assignments than do the
second and first generation fermions.  Thus the mass matrix texture,
particularly the mixing of the light quarks and leptons with the third
generation, will depend quite strongly on the way in which Topcolor is
broken.  Flavor physics phenomenology associated with mixing of the
third and first generations, particularly $B^0-\bar{B}^0$ mixing, will
be seen to constrain model-building significantly
\cite{Burdman:2000in,Simmons:2001va}, 
while $K$ physics turns out to be less significant a
probe.

Let us parameterize the electroweak symmetry breaking in TC2 by a
fundamental Higgs boson, which ultimately breaks $SU(2)_L\times
U(1)_Y$.  This is simply shorthand for a techniquark bilinear operator
which receives a VEV of order $v_{weak}$.  Similarly, we consider an
effective field $\Phi$ which breaks Topcolor.
We specify the full Topcolor charges of these fields, e.g., under
$SU(3)_1\times SU(3)_2 \times U(1)_{Y1}\times U(1)_{Y2}\times SU(2)_L$
we choose:
\beq
\Phi \sim (3,\bar{3}, \frac{1}{3}, -\frac{1}{3}, 0)
\qquad
H \sim (1,1,0, -1, \half)
\eeq
Also, let $\Sigma =\exp(i\pi^a T^a/f_{\pi})$ be 
the nonlinear chiral field composed
of the (bare) top-pions.
Then, the effective Lagrangian
couplings to fermions that generate mass terms in the
up quark sector are of the form:
\begin{eqnarray}
{\cal L}_{{\cal M}_U} & = & m_0 \bar{T}_L\Sigma P T_R +c_{33}\bar{T}_Lt_R  
H\fr{\det{\Phi^\dagger}}{M_0^3}+
c_{32}\bar{T}_L c_R H\fr{\F}{\L} + c_{31}\bar{T}_L u_R  
H\fr{\F}{M_0} \nn\\
& &+c_{23}\bar{C}_L t_R H \F^\dagger  
\fr{\det{\F^\dagger}}{M_0^4}+c_{22}\bar{C}_L c_R
H + c_{21}\bar{C}_L u_R H  \label{lag_mass}\\
& & + c_{13}\bar{U}_L t_R H \F^\dagger  
\fr{\det{\F^\dagger}}{M_0^4}+c_{12}\bar{U}_L c_RH
+c_{11}\bar{U}_L u_R H  + {\rm h.c.} \nn
\end{eqnarray}
Here $T=(t,b)$, $C=(c,s)$ and $U=(u,d)$.  The mass $m_0$ is
the dynamical Topcolor condensate top mass.
Furthermore $\det{\Phi}$ is defined by
\bz
\det{\Phi} \equiv \frac{1}{6}\e_{ijk}\e_{lmn}\Phi_{il}\Phi_{jm}\Phi_{kn}
\label{phi_det}
\ez
where in $\Phi_{rs}$ the first(second) index refers to $SU(3)_1$ ($SU(3)_2$).  
The matrix elements require these factors
of $\Phi$ to connect the third with the
first or second generation color indices. The down quark
and lepton mass matrices are generated by couplings analogous to
(\ref{lag_mass}).

To explore what kinds of textures arise naturally,
let us assume that the ratio $\Phi/M_0$ is small, O($\epsilon$).
The field $H$ acquires a VEV of $v$.
Then the resulting mass  matrix is:
\bea
&& \left( \begin{array}{ccc}
c_{11}v &  c_{12}v & \sim 0    \cr 
 c_{21}v & c_{22}v  &\sim 0    \cr
c_{31}O(\epsilon)v & c_{32}O(\epsilon)v & \sim m_0 + O(\epsilon^3)v    
\cr
 \end{array}\right)\label{m_trian}
\eea
where we have kept only terms of $\cal O (\e)$ or larger.

This is a {\em triangular matrix} (up to the $c_{12}$ term).  When it is
written in the form $U_L {\cal D} U^{\dagger}_R$ with $U_L$ and $U_R$
unitary and ${\cal D}$ positive diagonal, restrictions on $U_L$ and
$U_R$ may be inferred.  In the present case, the elements $U^{3,i}_L$
and $U^{i,3}_L$ are vanishing for $i\neq 3$ , while the elements of
$U_R$ are not constrained by triangularity.  Analogously, in the down
quark sector $D^{i,3}_L=D^{3,i}_L=0$ for $i\neq 3$ with $D_{R}$
unrestricted.  The situation is reversed when the opposite corner
elements are small, which can be achieved by choosing $H \sim (1,1,
-1, 0, \half)$.  The full CKM matrix is, as usual, given by $K =
U_LD^\dagger_L$.

Triangularity, thus implies that either the matrix $U_L$ has large off
diagonal elements, while $U_R$ has small off diagonal elements (we'll
denote this case $(U:10)$), or vice versa, $(U:01)$. Without
triangularity $(U:11)$ is allowed; with exact flavor symmetry we have
$(U:00)$.  These are not fine-tunings, but rather systematic choices
we make in implementing the symmetry.  They will ultimately have a
deeper dynamical origin.  For an example application, we can invoke
triangularity to cause the CKM matrix $K = U_LD^\dagger_L$ to be
generated by pure $U_L$ rotations, with no contribution from $D_L$,
$(U:10,D:01)$ or $(U:10,D:00)$, or vice versa $(U:01,D:10)$ or
$(U:00,D:10)$.

The restrictions on the quark mass rotation matrices owing to their
triangular structure have important phenomenological consequences
\cite{Buchalla:1996dp}.  For instance, in the process $B^0\rightarrow
\overline{B^0}$ there are potentially large contributions from
top-pion and coloron exchange.

\begin{figure}[t]
\label{bbarpi}
\vspace{5cm}
\includegraphics{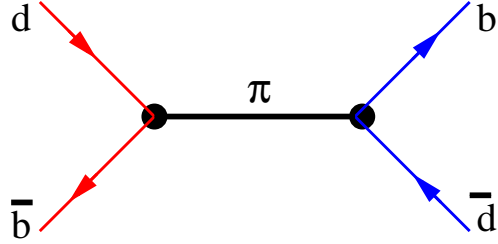}
\vspace{1cm}
\caption[]{\small \addtolength{\baselineskip}{-.4\baselineskip} 
Top-pion contribution to $B^0-\bar{B}^0$ mixing.
}
\end{figure}

In the top-pion graph of Fig.(\ref{bbarpi}) we find a dangerously
large contribution to $\delta m^2/m^2$, about two orders of magnitude
above the experimental limit if normal CKM angles are assumed in a
$(D:11)$ solution.  However, owing to chirality the graph is
proportional to the product $D_{L,3,1} D^\dagger_{R,3,1}$.  Hence, we
can systematically suppress this effect if we allow one of the three
solutions $(D:10)$, or $(D:01)$, or $(D:00)$, while $(D:11)$ is
disallowed.  However, the topgluon graph yields a result that is
similarly large, and proportional to $|D^{3,1}_L|^2 + |D^{3,1}_R|^2$
\cite{Burdman:2000in}.  Hence, we conclude that only the symmetric solution,
$(D:00)$, is allowed, \ie , the off-diagonal mass matrix elements
mixing $(1,3)$ in the down quark sector are all small.  
This suppresses Topcolor $b\rightarrow s\gamma$ effects as well.

\begin{figure}[t]
\label{bbarglu}
\vspace{5cm}
\includegraphics{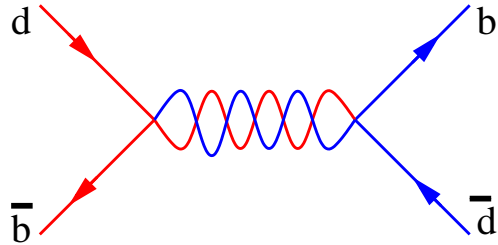}
\vspace{1cm}
\caption[]{\small \addtolength{\baselineskip}{-.4\baselineskip} 
Top-gluon contribution to
$B^0-\bar{B}^0$ mixing.}
\end{figure}

Thus, the CKM matrix elements mixing to the third generation must be
controlled by $U_L$, and we thus require a $(U:10)$ or $(U:11)$
solution.  The same effects occur in $D^0-\bar{D}^0$ mixing, where
top-pion graphs are $\propto U_L U^\dagger_R$ off-diagonal elements,
and topgluon graphs yield an expression $\propto |U_L|^2 + |U_R|^2$.
Therefore, the latter result implies that $D^0-\bar{D}^0$ must be in
excess of the Standard Model prediction by about two orders of
magnitude.  This is close to the current experimental limit.  Thus,
$D^0-\bar{D}^0$ mixing emerges as an intriguing probe of
topcolor. Data showing that the rate of $D^0-\bar{D}^0$ mixing is
close to the SM prediction would pose a severe difficulty for topcolor
models.

There are various other effects and limits on the models that can be
obtained in mostly B and D, and to a lesser degree in K physics.  Many
of these are sensitive to the $Z'$ which is more model dependent.  We
refer the interested reader to some of the relevant literature,
beginning with \cite{Buchalla:1996dp}, \cite{Burdman:1997vn} and 
\cite{Burdman:1997qn}. 

The measurement of $R_b$ in $Z\rightarrow b\bar{b}$ also yields an 
important constraint, \cite{Hill:1995di},
\cite{Burdman:1997pf}, \cite{Loinaz:1998jg}. This is subject to larger
uncertainties because subleading $1/N_c$ top-pion loops dominate the
leading large $N_c$ top-gluon loops.  The degree of uncertainty can be
seen by examining the analysis of \cite{Burdman:1997pf} and noting
that in their exclusion figure, with the Topcolor $f_{\pi} \sim 60$
GeV the model appears to be ruled out for all reasonable top-pion
masses, while for $f_{\pi} \sim 100$ GeV there is no constraint.  We
take the difference of these two results to be within the errors of
the subleading $N_c$ calculation.  Similar uncertainties plague
efforts to evaluate the oblique electroweak constraints on these
models
\cite{Loinaz:1998jg}. Also, diagrams that
generate off-shell longitudinal mixing
of the $Z$ with $\pi^0$ have not been included in 
any of these analyses.

Studying questions related to flavor physics in the context of
Topcolor models has proven to be quite instructive.  We now
turn to other phenomenological questions and evaluate the status of
limits on Topcolor scenarios.

\subsection{Topcolor Phenomenology}



\subsubsection{Top-pions}


If Topcolor produces a $\bar{t}t$ condensate
to elevate the top quark mass, it necessarily 
produces an isovector of PNGB top-pions $\sim$ 
$t\bar{b}, t\bar{t}-b\bar{b}$
and $b\bar{t}$. There may also be a heavier
top-Higgs  $\sim t\bar{t}+b\bar{b}$. In TC2 this multiplet
of top-pions is uneaten by gauge bosons and  the masses of these objects
have been estimated in minimal schemes to be of order
$\sim 200$ GeV, though this is fairly uncertain. The
top-pions are strongly coupled to $\bar{t}t$, $\bar{b}b$ and 
$\bar{t}b$.

 Top-pion vertex corrections therefore  can 
noticeably decrease $R_b$ \cite{Burdman:1997pf,Loinaz:1998jg}, 
though the estimates are very sensitive to 
choice of $f_\pi$ (the effect becomes
negligible with $f_\pi \sim 100$ GeV). 
Additional physics can in principle cancel the effect 
\cite{Yue:2000ay} (e.g., topgluons, for example, can increase $R_b$
\cite{Hill:1995di}), but ref. \cite{Burdman:1997pf} surveyed a number of
possible sources of cancellation and found none producing an effect of
the requisite size.  
Exchange of the light composite
pseudoscalars in top seesaw models likewise contributes 
to the $T$ parameter. The
interplay of the resulting limits is discussed in
section 4.5.  Several searches for top--pion and top--higgs ($\sigma$)
production  have been proposed; we now review these.

\vskip .1in
\noindent
{\bf (i) Neutral Top--pions, Top--Higgs}
\vskip .1in

\begin{figure}[tb]
\vspace{5cm}
\begin{center}
\includegraphics{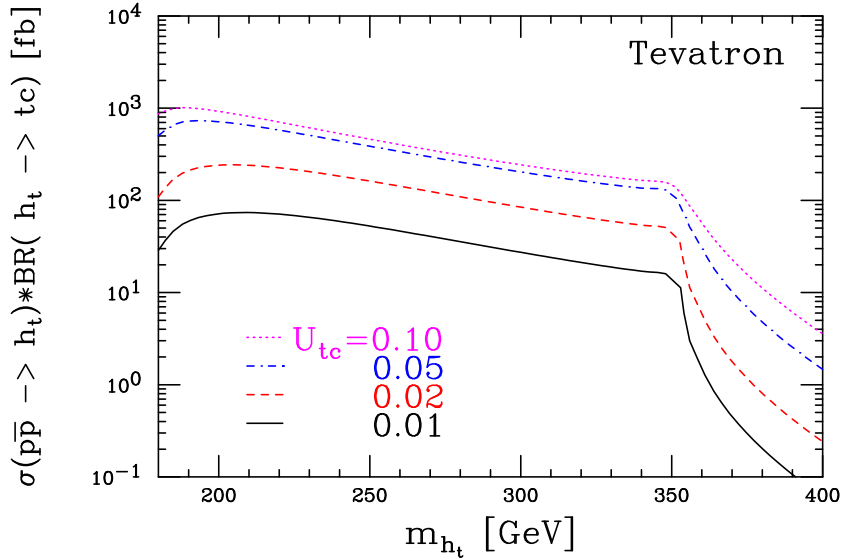}
\end{center}
\caption{\small \addtolength{\baselineskip}{-.4\baselineskip} 
Top-higgs production cross-section via gluon fusion times
  branching ratio to $\bar{t}c$ at $\sqrt{s} =$ 
  2 TeV. \protect\cite{Burdman:1999sr}}
\label{fig:toppia}
\end{figure}

A singly-produced neutral $\hat{\pi}^0$ 
that dominantly decays to $\bar{t}c$, via flavor mixing,
could possibly be
detected \cite{Burdman:1999sr} at Run IIb or the LHC.  
 The strength of the $\hat{\pi}^0$ coupling to $\bar{t}c$ is
governed by the model-dependent parameter $U_{tc} \equiv \sqrt{\vert
  U^L_{tc}\vert^2 + \vert U^R_{tc} \vert ^2}$ where $U^L$ and $U^R$ are the
matrices that diagonalize the up-quark mass matrix.  In order for the
flavor-changing decay to dominate, the decays to $t\bar t$
must be energetically
disallowed.  Decays to the electroweak gauge bosons  is more
model-dependent.  In TC2 models, where the Yukawa coupling of scalars coupling
to $t_R$ is enhanced by $r^2 \approx (m_t / f_{\pi_t})^2 \approx 10$, the
top-scalar coupling to vector boson pairs is likewise suppressed by $1/r^2$
\cite{Burdman:1996iv}.  In top seesaw models (Section 4.4), $r=1$ 
so that the CP-even scalar decays mainly to $VV$; however, there can
also be a light CP-odd scalar that is unable to decay to $VV$ and
therefore decays to $\bar{t}c$ instead
\cite{Chivukula:1998wd}.

\begin{figure}[tb]
\vspace{5cm}
\begin{center}
\includegraphics{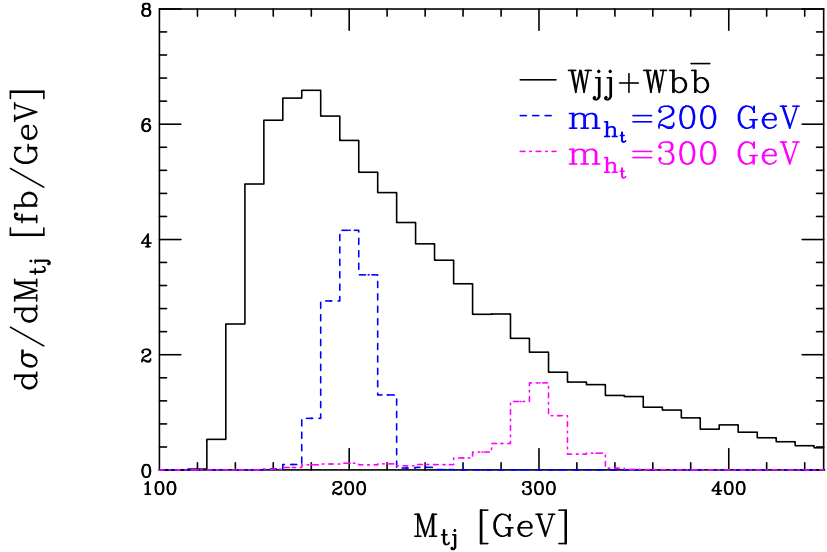}
\end{center}
\caption{\small \addtolength{\baselineskip}{-.4\baselineskip} 
Mass distribution for the $\bar{t}c$ jet system 
  for $m_{h_t} = 200,
  300$ GeV and for the $Wjj$ and $Wbb$ backgrounds at $\sqrt{s} =$
  $2$ TeV \protect\cite{Burdman:1999sr}.}
\vspace{2cm}
\label{fig:toppi1} 
\end{figure}

As shown in Figure \ref{fig:toppia}, the production cross-section times
flavor-changing branching ratio at Run IIb can provide several hundred
top-higgs events even for $U_{tc} \sim 0.02$.  Since the top-higgs is
expected to be rather narrow, the signal peak should be visible for scalar
masses up to the top threshold (see figure \ref{fig:toppi1}).  At LHC, the
production rate is dramatically enhanced since the primary production mode is
gluon-gluon fusion; the cross-section times branching ratio is about 100
times that shown in Figure \ref{fig:toppia} for the same values of $U_{tc}$.
Recent studies are optimistic for this process  \cite{Cao:2002af},
but detailed background studies are required.

Recent work \cite{Yue:2000fe,Yue:2000xa,Yue:2001uv,Huang:2001pm} 
also suggests that 
$\hat{\pi}^0$'s may be visible through 
their $\bar{t}c$ decays at a LC running in
$e^+e^-$ or $\gamma\gamma$ collision mode.  The cross-section for $e^+e^- \to
\bar{t}c$ or $\gamma\gamma \to \bar{t}c$ at a linear collider receives a contribution of
order $20$ fb from neutral top-pions; an integrated luminosity of 
$50$ fb$^{-1}$
might make the top-pion visible \cite{Yue:2000xa}.  On the other hand, the
processes $e^+e^- \to \gamma \pi_t, Z \pi_t$, with $\pi_t \to \bar{t}c$, have
cross-sections of order $1.5$ pb and $0.3$ pb, respectively for top-pion masses
below $350$ GeV \cite{Yue:2000fe}.  A LC with $\sqrt{s} = 1$ TeV would produce
of order $400$ $(100)$ $\gamma \bar{t} c$ ($Z \bar{t}c$) events due to 
top-pions, potentially
making both channels observable.

\vskip .1in
\noindent
{\bf (ii) Charged Top--pions}
\vskip .1in

In the presence of sizeable flavor-changing couplings $\bar{t} c $ of the
$\hat{\pi}_t^0$, a large flavor-mixing coupling for the charged top-pions $
\bar{c}
b \pi_t^+$ can be induced \cite{He:1998ie}.  This enables charged top-pions
to be singly produced at sizeable rates in the s-channel at 
hadron colliders \cite{Balazs:1998sb}
or photon-photon linear colliders \cite{He:2002fd}.  
Ref.\cite{He:1998ie} has calculated the
production cross-sections for several charged top-pion masses at a variety of
colliders using the following benchmark set of couplings (typical for
topcolor models): $U^L_{tb} = U^L_{cb} = 0$, $U^R_{tb} = 5 U^R_{cb}$,
$g_{\pi tb} = 3
\sqrt{2} m_t / v$.  As shown in Figure \ref{fig:toppi2}, the cross-sections
for Run II and the LHC are 4.0 pb and 0.55pb, respectively; those at
photon-photon colliders are also sizeable.  Hence, the authors of
\cite{He:1998ie} suggest that charged top-pions would be visible up to masses
of 300-350 GeV at Run II and $~$1 TeV at LHC, and that light enough top-pions
would also be detectable at a linear collider running in the photon-photon
mode.

In contrast, the work of ref \cite{Wang:1998us} suggests that it will be
difficult to detect the effects of charged top-pions on single top production
at an $e\gamma$ collider in the process $e\gamma \to \bar{t} b \nu$.

\begin{figure}[tbh]
\vspace{7cm}
\begin{center}
\includegraphics{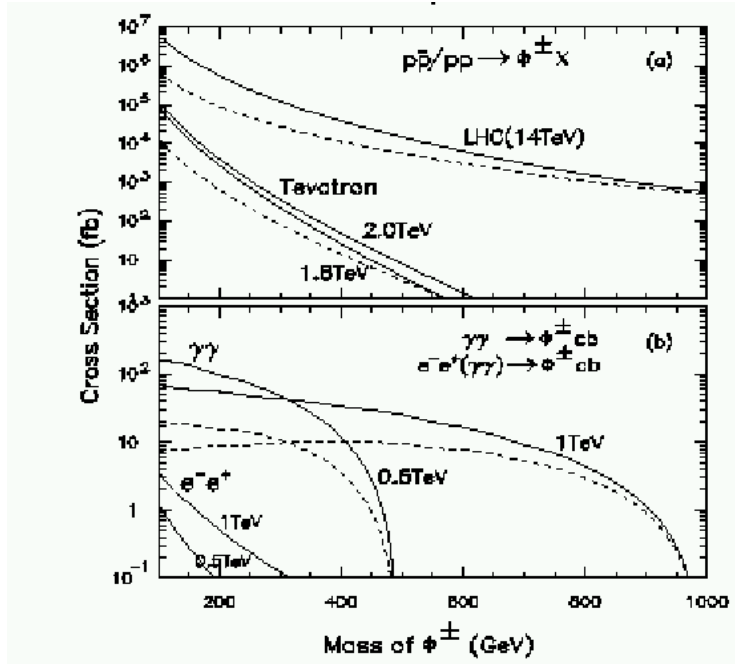}
\end{center}
\vspace{2cm}
\caption{\small \addtolength{\baselineskip}{-.4\baselineskip} 
$s$-channel charged top-pion production with the benchmark Yukawa
  couplings given in the text.  For reference, the dashed curves show the
  results for Yukawa couplings satisfying the 3-sigma $R_b$ bound \protect\cite{He:1998ie}.}
\label{fig:toppi2}
\end{figure}

We mention that in titling scenarios and variant
schemes, there will be analogous bottom-pions which couple
strongly to $\bar{b}b$. For a discussion of the associated phenomenology of
these objects see
\cite{Balazs:1998nt,Diaz-Cruz:1998qc,He:1999vp}. 


\subsubsection{Colorons: New Colored Gauge Bosons}

Several of the dynamical models we have discussed include 
extended strong interactions $SU(3)_1 \times SU(3)_2$, with
coupling constants $h_2 \gg h_1$.  A general prediction of such models
is the existence of a color-octet of massive gauge bosons (colorons).  
The best search strategy for colorons depends on how they couple to the
different quark flavors.  We will discuss the phenomenology
resulting from the two most common choices for the quarks' charge
assignments.

\vskip .1in
\noindent
{\bf (i) Topgluons}
\vskip .1in 

In models such as Topcolor \cite{Hill:1991at}, TC2
\cite{Hill:1995hp,Buchalla:1996dp}, or Top Seesaw \cite{Dobrescu:1998nm},
only the third-generation quarks transform 
principally under the stronger $SU(3)_2$ group.
As a result, the massive topgluons ($B^{\mu a}$) couple predominantly to
third-generation quarks \cite{Hill:1991at}.  
The topgluons are expected to be heavy ($M ~ 0.5 - 2.0$ TeV)
and broad ($\Gamma/M ~ 0.3 - 0.7$) resonances.  In
production at, e.g., the Tevatron, in the dominant process $\bar{q}q
\rightarrow (g,B) \rightarrow \bar{t} t$ the amplitude involves
$(g_3\tan\theta)\times (-g_3\cot\theta)$ so that the factors involving
$\theta$ cancel (and there is characteristic destructive
interference {\em above threshold}).  Thus, $\theta$ affects the
rates only through the decay width of $B^{\mu a}$.

Topgluons of moderate mass may be produced directly at the Tevatron
\cite{Hill:1994hs,Han:2003pu}.
CDF has recently used their measured upper limit on cross-section time
branching ratio of new resonances decaying to $b\bar{b}$ to place a
limit on topgluons \cite{Abe:1998uz} .  They exclude topgluons of
width $\Gamma_B = 0.3 M$ in the mass range $280 < M < 670$ GeV, of width 0.5 M in
the range $340 < M < 640$ GeV, and of width $\Gamma_B = 0.7 M$ 
in the range $375 <
M < 560$ GeV.  A simulation of topgluon production and decay combined
with an extrapolation of the CDF $b$-tagged dijet mass data from Run I
\cite{tracking} indicates that in Run IIb, the topgluon discovery mass reach in
$b\bar{b}$ final states should be 0.77-0.95 TeV with 2 $fb^{-1}$ of
integrated luminosity and 1.0 - 1.2 TeV with 30fb$^{-1}$.

\begin{figure}[tb]
\vspace{5.0cm}
\includegraphics{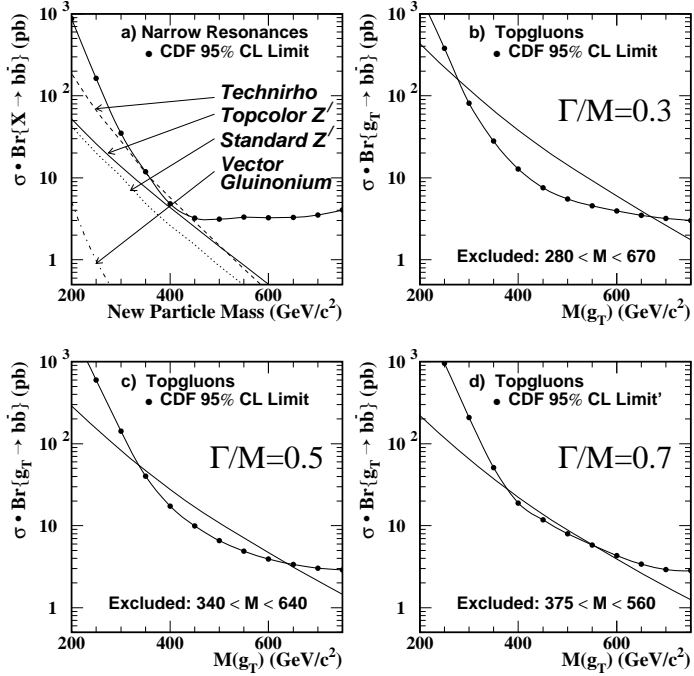}
\vspace{2.5cm}
\caption[topglus]{\small \addtolength{\baselineskip}{-.4\baselineskip} 
CDF search for topgluons in $\bf b\bar{b}$ \protect\cite{Abe:1998uz}}
\label{fig:harris-topgluon}
\end{figure}

\begin{figure}[tb]
\vspace{4.0cm}
\includegraphics{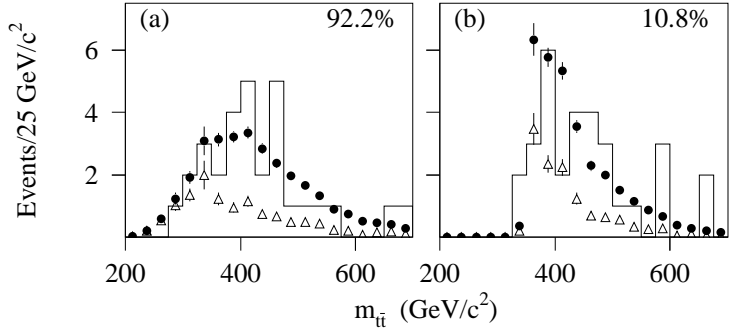}
\vspace{2cm}
\caption[abcf]{\small \addtolength{\baselineskip}{-.4\baselineskip} 
Invariant mass distribution for top pairs: 
pairs: D0 data (histogram), simulated background (triangles), simulated
S+B (dots).  In (a) $m_t$ unconstrained; 
in (b) $m_t = 173$ GeV.\protect\cite{Abachi:1997jv,Abbott:1998fv,Abbott:1998dc,Abbott:1998dn}}
\label{fig:d0ttdist}
\end{figure}

\begin{figure}[tb]
\vspace{4.0cm}
\includegraphics{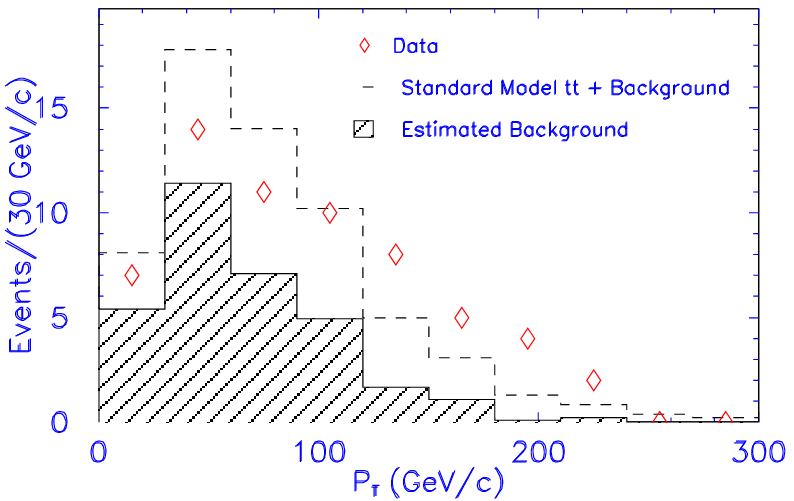}
\vspace{2cm}
\caption[abb]{\small \addtolength{\baselineskip}{-.4\baselineskip} 
$P_T$ distribution for hadronically-decaying tops in
lepton+jets events from CDF.\protect\cite{Affolder:2000dt}}
\label{fig:hiptcdf}
\end{figure}

Lower backgrounds make topgluons easier to find in their decays
to top quarks \cite{Hill:1994hs,Han:2003pu}.  
Initial measurements of the invariant mass ($M_{tt}$)
and transverse momentum ($p_T$) distributions of the produced top
quarks have been made, as shown in Figure \ref{fig:d0ttdist}.  While a
comparison with the measured $M_{jj}$ distribution for QCD dijets
\cite{Abbott:1998wh} illustrates how statistics-limited the Run I top
sample is, some preliminary limits on new physics are being
extracted\footnote{\addtolength{\baselineskip}{-.4\baselineskip} 
It has been noted, e.g. that a narrow 500 GeV Z'
boson is inconsistent with the observed shape of the high-mass end of
CDF's $M_{tt}$ distribution.\cite{Affolder:2000eu}}.  The $p_T$
distribution for the hadronically-decaying top in fully-reconstructed
lepton + jets events (Figure
\ref{fig:hiptcdf}) constrains any non-SM physics which increases the
number of high-$p_T$ events.  The fraction $R_4 =
0.000^{+0.031}_{-0.000} (stat)^{+0.024}_{-0.000}(sys)$ of events in
the highest $p_T$ bin ($225 \leq p_T \leq 300$ GeV) implies
\cite{Affolder:2000dt} a 95\% c.l. upper bound $R_4 \leq 0.16$ as
compared with the SM prediction $R_4$ = 0.025.

In Run IIb, the $\sigma_{tt}$ measurement will be dominated by
systematic uncertainties; the collaborations will use the large data
sample to reduce reliance on simulations \cite{Amidei:1996dt}.
Acceptance issues such as initial state radiation, the jet energy
scale, and the $b$-tagging efficiency will be studied directly in the
data.  It is anticipated \cite{Amidei:1996dt} that an integrated
luminosity of $1$ $(10, 100)$ fb$^{-1}$ will enable $\sigma_{tt}$ to
be measured to $\pm$ $11$ $(6, 5)$ \%.  Topgluons of mass up to
$1.0$-$1.1$ TeV ($1.3 - 1.4$ TeV) would be visible in 2 fb$^{-1}$ (30
fb$^{-1}$) at Run IIb, either in a search for a new broad resonance, or
through their effects on the magnitude of the $t\bar{t}$ total
cross-section.  Projected Run IIb limits on $\sigma\cdot B$ for new
resonances decaying to $t\bar{t}$ are illustrated in Figure
\ref{fig:sigbxtt}.  For further details on current and 
future topgluon searches at the Tevatron, see Table III of
ref. \cite{Barklow:2002su}.
The situation at the LHC and VLHC has been recently 
studied in considerable  in \cite{Han:2003pu},
conclude that $\bar{t}t$ is overwhelmed by QCD backgrounds.

\vskip .1in
\noindent
{\bf (ii) Flavor-universal Colorons}
\vskip .1in

In theories in which all quarks carry only $SU(3)_2$ charge
\cite{Chivukula:1996yr,Lane:1998qi,Popovic:1998vb}, 
the massive colorons couple with equal strength to
all quark flavors (section 4.1.2(ii)).  
As a result, they will appear in
dijet production and heavy flavor production at hadron colliders.  A
comparison of these processes \cite{Simmons:1997fz} has indicated that
the limits on colorons from dijet production should be the more
stringent ones.

Existing limits on flavor-universal colorons from several sources are
summaried in Figure \ref{fig:bertram-coloron}.  Exchange of light
strongly-coupled colorons across quark loops would cause large
contributions to the $T$ parameter; accordingly, the region $(M /
\cot\theta) < 450$ GeV is 
excluded.\footnote{\addtolength{\baselineskip}{-.4\baselineskip} 
A fit to the full set of
electroweak data gives a slightly stronger bound at large coloron mass
\cite{Burdman:1999us}.}  Light narrow colorons would have been seen by
a Run I CDF search for new particles decaying to dijets (see
\cite{Bertram:1998wf}); this excludes the cross-hatched region at low
$\cot^2\theta$. The light-shaded region $(M / \cot\theta < 759$ GeV)
is excluded by the shape of the dijet angular distribution measured by
D0 \cite{Abbott:1998yy}.  Finally, the shape of the dijet mass
spectrum measured by D0
\cite{Abbott:1998wh} sets the strongest limit: $M/\cot\theta > 837$
GeV \cite{Bertram:1998wf}.

\begin{figure}[tb]
\vspace{5.0cm}
\includegraphics{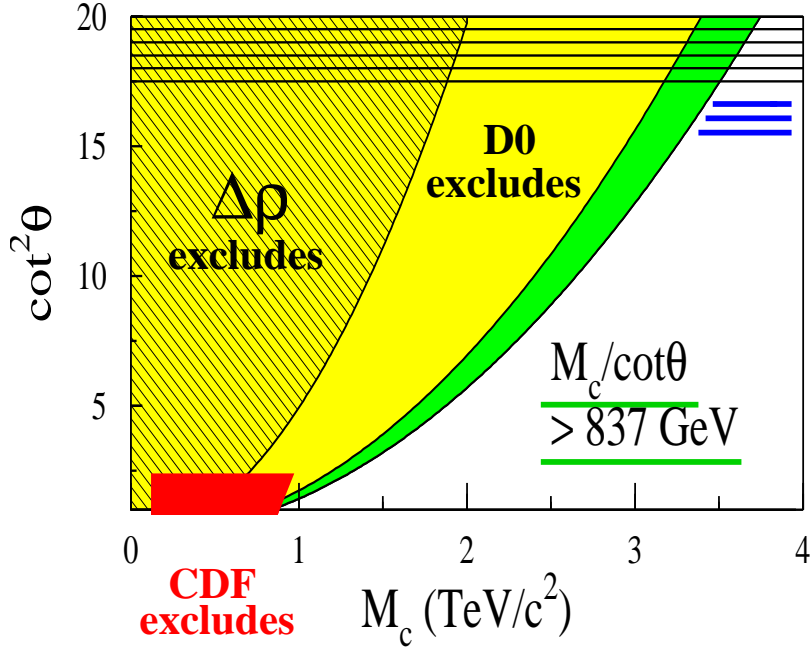}
\vspace{2.5cm}
\caption{\small \addtolength{\baselineskip}{-.4\baselineskip} 
Experimental limits on flavor-universal colorons from sources 
  described in the text \protect{\cite{Bertram:1998wf}}.  The horizontally
  hatched region at large $\cot\theta$ is not part of the Higgs phase of the
  model \protect{\cite{Chivukula:1996yr}}.}
\label{fig:bertram-coloron}
\end{figure}

In the context of flavor-universal TC2 models, the
value of $\kappa_3$ must be approximately $2$ in order for the colorons to help
cause the top quark to condense as discussed earlier \cite{Popovic:1998vb};
this is equivalent to $\cot\theta \approx 4$.  Hence, in these models, the
limit set by D0 on the coloron mass is $M_c > 3.4$ TeV.

Run IIb will, naturally, be sensitive to somewhat heavier colorons.  According
to the TeV 2000 report the limit on $\sigma\cdot B$  will improve as shown in
Figure \ref{fig:tev2000-coloron}.  The predicted $\sigma\cdot B$ for a
coloron of $\cot\theta = 1$ is equivalent to that for an axigluon (shown);
larger values of $\cot\theta$ increase the rate.
\begin{figure}[tb]
\vspace{10.0cm}
\includegraphics{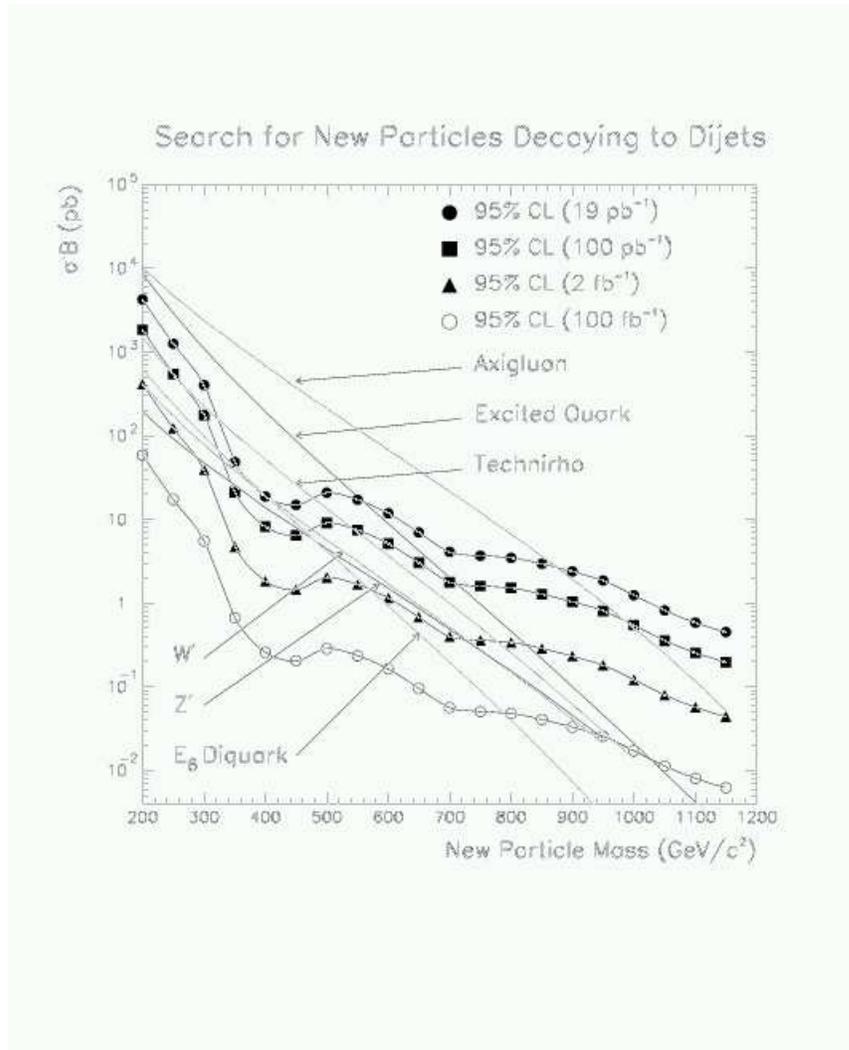}
\vspace{2.5cm}
\caption{\small \addtolength{\baselineskip}{-.4\baselineskip} 
Anticipated 95\% c.l. upper limit on $\sigma\cdot B$ for new
  particles decaying to dijet as a function of the new particle mass for
  various integrated luminosities at the Tevatron.  Axigluon curve
  corresponds to a coloron of $\cot\theta = 1$\protect{\cite{Amidei:1996dt}}.}
\label{fig:tev2000-coloron} 
\end{figure}

\subsubsection{New $Z'$ Bosons}

As discussed in section 4.1, both Classic TC2
\cite{Hill:1995hp,Buchalla:1996dp} and Flavor-Universal TC2
\cite{Lane:1998qi,Popovic:1998vb} models include an extra $U(1)$ group
and predict the presence of a massive $Z'$ boson.  The couplings of this
$Z'$ to fermions are generally not flavor-universal, and the more
strongly suppressed the $Z'$ couplings to first and second generation
fermions are, the less effective traditional searches for $Z'$
become. This section discusses experimental searches and limits that
exploit the flavor non-universal couplings of the $Z'$.

TC2 models include an extended electroweak sector of the form
\begin{equation}
SU(2)_W \times U(1)_h \times U(1)_\ell \ .
\end{equation}
where the coupling of the first $U(1)$ group is the stronger one, $g_h >>
g_\ell$.  At a scale above the weak scale, the two hypercharge groups break
to a subgroup identified as $U(1)_Y$.  As a result, a $Z'$ boson that is a
linear combination of the original two hypercharge bosons becomes massive.
This heavy $Z'$ boson couples to a fermion as 
\begin{equation} 
-i\frac{e}{\cos\theta} \left( \frac{s_\chi}{c_\chi} Y_\ell 
-\frac{c_\chi}{s_\chi} Y_h \right) 
\label{eqn:ehs:zpru1}
\end{equation}
where $Y_h$ ($Y_\ell$) is the fermion's hypercharge under the $U(1)_h$
($U(1)_\ell$) group and $\chi$ is the mixing angle between the two
original hypercharge sectors $\cot{\chi} = (g_h / g_\ell)^2$.
Compared with the couplings of the $Z'$ boson from an extended
weak group (Section 3.6.5), the significant physical differences are
that the overall coupling is of hypercharge rather than weak strength,
and the $Z'$ couples to both left-handed and right-handed fermions at
leading order.

\begin{figure}[tb]
\vspace{2.0cm}
\vbox{ 
      \includegraphics{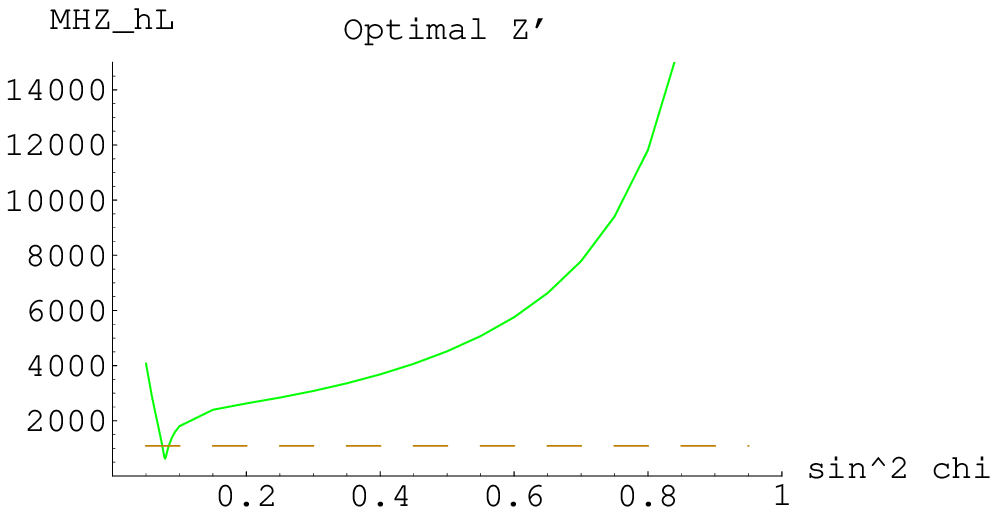} 
      \vspace{7cm} \includegraphics{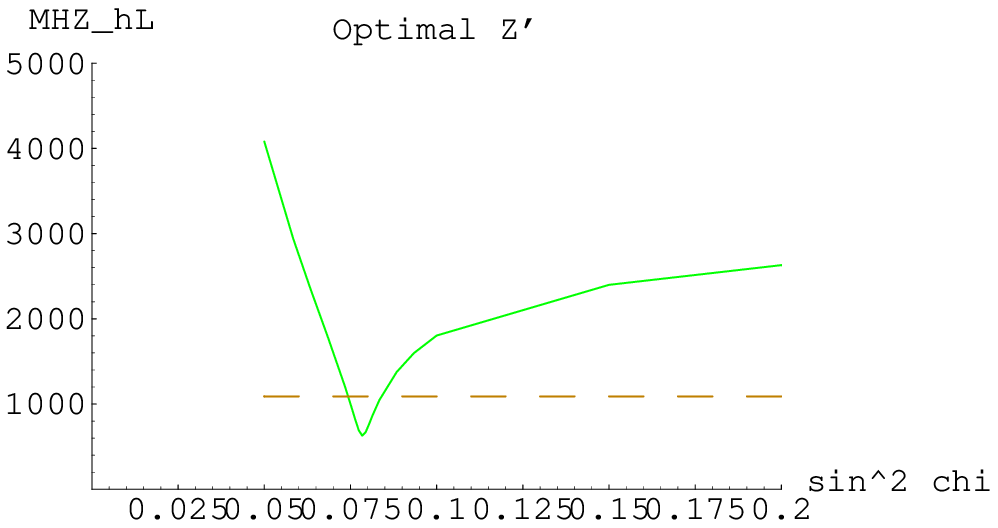}}
\vspace{2.5cm}
\caption{\small \addtolength{\baselineskip}{-.4\baselineskip}
 Lower bounds at 95\% c.l. on mass of ``optimal'' non-universal Z' boson. \protect\cite{ehsrsc:2002}.  Solid curve is from a global fit to precision electroweak data; dashed line is from LEP II contact interaction limits. The bottom plot is a close-up of the region where the smallest Z' mass is allowed.}
\label{fig:ehs:optimal}
\end{figure}

\vskip .1in
\noindent
{\bf (i) $Z'$ and Precision Tests}
\vskip .1in 

Studies of precision electroweak limits on the $Z'$ bosons
\cite{ehsrsc:2002,Chivukula:1996cc} in these models have obtained 
lower bounds on $M_{Z'}$ as a function of $\sin^2\chi$.  The lower bound
is found to depend quite strongly on the hypercharge assignments of
the fermions because the shift in the Z boson's coupling to fermions
depends directly on $Y_h$:
\begin{equation}
\delta g^f \approx - \left( \frac{e \sin\theta}{x \cos\theta \cos^2\chi}\right)
\left( 1 + \frac{\epsilon}{\sin^2\chi}\right) 
\left[ Y_h^f - \sin^2\chi Y^f \right]
\end{equation}
where $\epsilon \equiv 2 f_{\pi_t}^2/v^2 (Y_h^{t_L} - Y_h^{t_R})$.  In
models with a so-called ``optimal'' $Z'$, in which the third generation
fermions have $Y_h = Y,\, Y_\ell = 0$ and those of the other two
generations have $Y_h = 0,\, Y_\ell = Y$, the $Z'$ can be as light as
$630$ GeV (see figure
\ref{fig:ehs:optimal}) at 95\% c.l. for $\sin^2\chi \approx 0.0784$; a
$Z'$ mass less than a TeV is allowed for $0.0744
\lae \sin^2\chi \lae 0.0844$ \cite{ehsrsc:2002}.  In the TC2 model of
Lane \cite{Lane:1996ua}, where the fermions have considerably larger $Y_h$
charges (see table), the lower bound on the $Z'$ mass is correspondingly
higher, about $20$ TeV.  The strong constraint here comes from sensitivity to
atomic parity violation in $Cs$ \cite{Lane:1996ua}.  A variant of Lane's model
\cite{kdl:temp} in which the lepton's $U(1)_h$ couplings are vectorial has a
$Z'$ limit much closer to that of the ``optimal'' scenario (though
lacking the low-mass region near $\sin^2\chi = .0784$)
\cite{Chivukula:1996cc}.

\begin{table}[tb]
\begin{center}
\begin{tabular}{|l|l||l|l|}
\hline\hline
1st, 2nd & $Y_h$ & 3rd & $Y_h$\\
\hline\hline
$(u,d)_L,\ (c,s)_L$ & -10.5833 & $(t,b)_L$ & 8.7666\\
\hline
$u_R,\ c_R$& -5.78333 & $t_R$&11.4166 \\
\hline
$d_R,\ s_R$& -6.78333 & $b_R$&10.4166 \\
\hline
$(\nu_e,e)_L,\ (\nu_\mu,\mu)_L$&-1.54&$(\nu_\tau,\tau)_L$&-1.54\\
\hline
$e_R,\ \mu_R$ & 2.26 & $\tau_R$ & 2.26\\
\hline\hline
\end{tabular}
\caption{\small \addtolength{\baselineskip}{-.4\baselineskip} 
Fermion charges for Lane's model \protect{\cite{Lane:1996ua}}}
\end{center}
\end{table}

The non-universal couplings of the $Z'$ boson to fermions cause it to
induce lepton number violating processes.  A study in
ref. \cite{Rador:1998is} has shown that $\mu-e$ conversion in nuclei
is about an order of magnitude better than the decay $\mu\to 3 e$
\cite{Yue:2000fh} 
for constraining the magnitudes of the lepton mixing
angles.  The decay $\mu \to e\gamma$ yields weaker bounds.  Present
data allows the $Z'$ mass to be as large as 1 TeV and the magnitudes
of the lepton mixing angles lie roughly between the analogous CKM
entries and their square-roots \cite{Rador:1998is}.  Looking to the future, 
ref. \cite{Murakami:2001cs} finds that the MECO and PRIME experiments
will probe lepton flavor violating $Z'$ bosons out to a mass of order $10$
TeV, far better than the MEG ($\mu \to e \gamma$) experiment can do.

At energies well below the mass of the $Z'$ boson, its exchange in the
process $e^+ e^- \to f \bar{f}$ where $f$ is a $\tau$ lepton or $b$
quark may be approximated by four-fermion contact interactions.  As
discussed in section 3.6.5, limits the ALEPH and OPAL experiments have
given on the scale $\Lambda$ of new contact interactions may be
translated into lower bounds on the mass of the $Z'$ boson.  The
strongest such limits on an ``optimal'' $Z'$ boson are for the process
$e^+_R e^-_R
\to \tau^+_R \tau^-_R$ \cite{Lynch:2000md} 
\begin{equation} 
M_{Z'} = \Lambda \sqrt{\alpha_{em} / \cos^2\theta} 
      \  >\  \left\{\ {370\ {\rm GeV}\ \ \  {\rm ALEPH}} \atop {370\ {\rm GeV}\ \ \  {\rm OPAL}} \right\} \ .
\end{equation}
which improves a bit on the limit from precision electroweak data
\cite{Chivukula:1996cc}.

\begin{figure}[tb]
\vspace{9.0cm}
\includegraphics{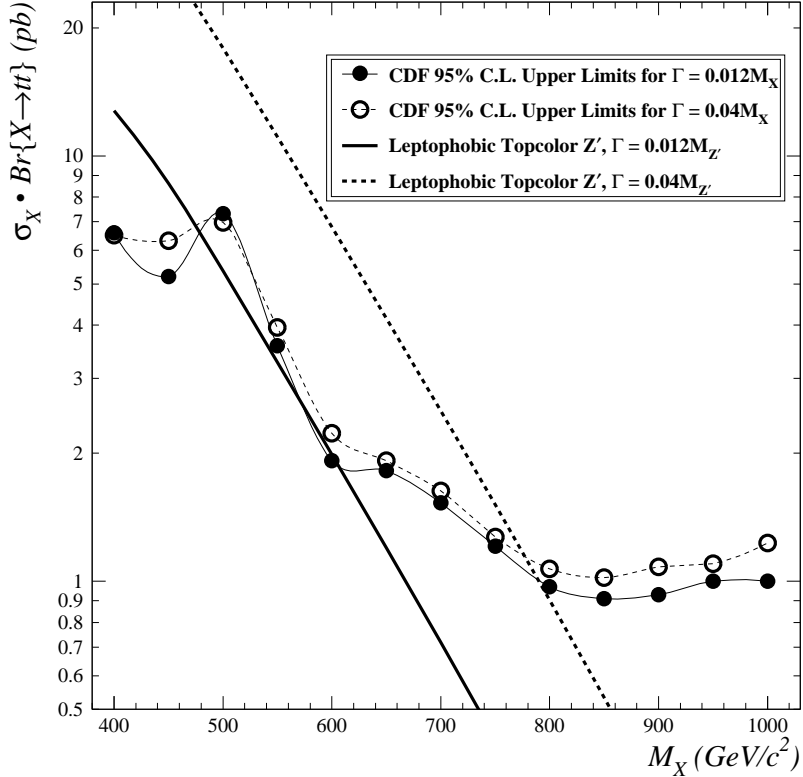}
\vskip2.3truecm
\caption{\small \addtolength{\baselineskip}{-.4\baselineskip} 
The 95\% c.l. upper limits on 
 $\sigma_X \cdot BR{X \to t\bar{t}} $ as a function of mass 
(solid and open points) compared to the cross section for a 
leptophobic topcolor $Z'$ (thick solid and dashed curves) for 
two resonance widths \protect\cite{Affolder:2000eu}.}
\label{fig:ehs:cdf-zprime}
\end{figure}

\begin{figure}[tb]
\vspace{5.0cm}
\includegraphics{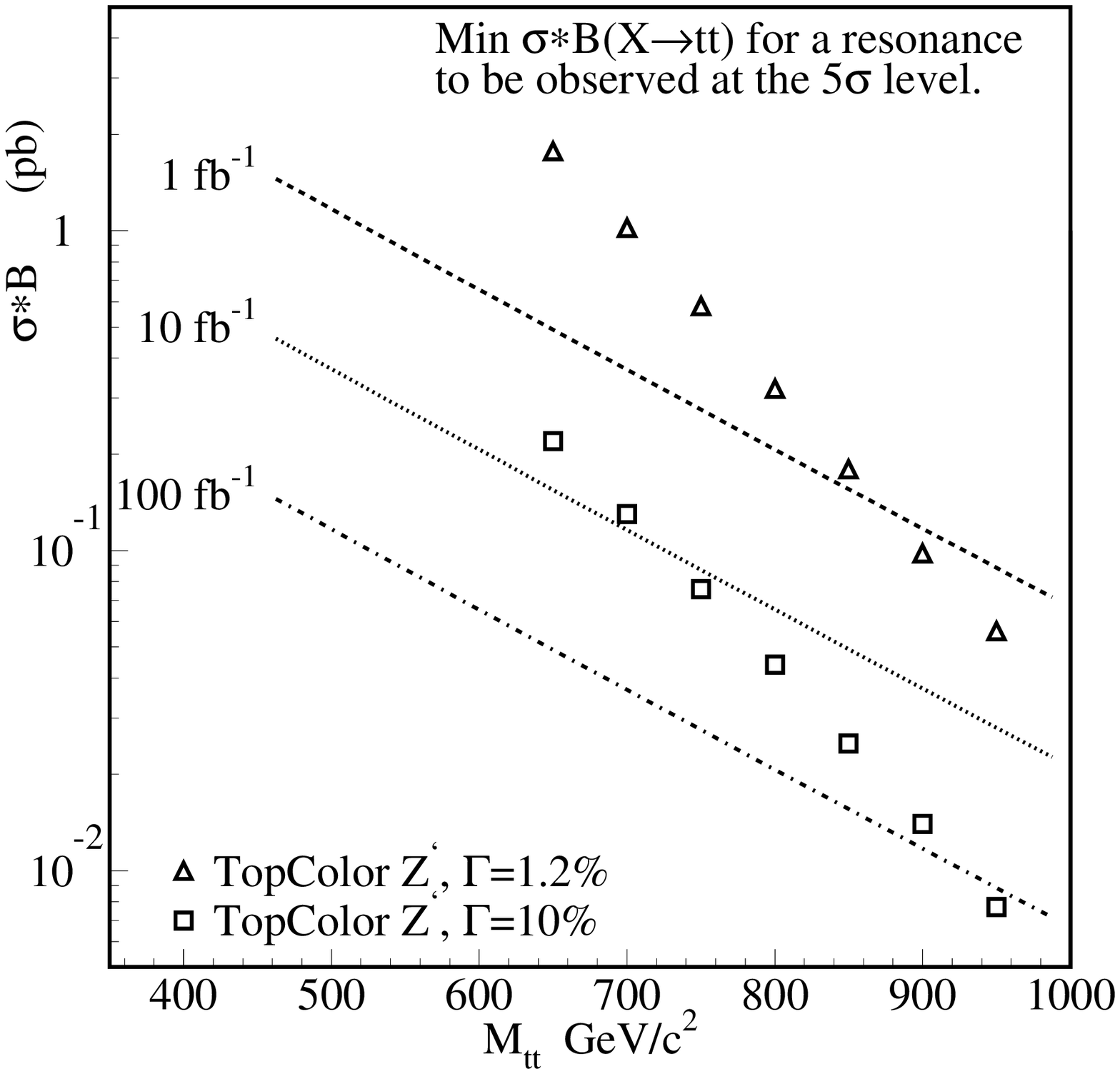}
\vspace{3.5cm}
\caption[ab]{\small \addtolength{\baselineskip}{-.4\baselineskip} 
Anticipated\protect\cite{Amidei:1996dt} Run IIb limits on
$\sigma\cdot B(X \to t\bar{t})$ and predictions for a topcolor Z'.}
\label{fig:sigbxtt}
\end{figure}

Dijet and Drell-Yan data from CDF and D0 can, likewise, be interpreted
as setting limits on the mass of a new $Z'$ boson.  The lower bounds on an
``optimal'' $Z'$ derived from limits on quark-lepton contact interactions are
significantly weaker than those from the LEP data \cite{Lynch:2000md}.  The
constraints on the $Z'$ of Lane's TC2 model \cite{Lane:1996ua} from dijet and
Drell-Yan data are sufficient to indicate that the model requires significant
fine-tuning \cite{Su:1997nn}; as mentioned earlier, however, the bounds from
precision electroweak data are even stronger.

\vskip .1in
\noindent
{\bf (ii) $Z'$ Production Searches}
\vskip .1in

Evidence of a $Z'$ boson coupled primarily to the third generation can more
profitably be sought in heavy flavor production at the Tevatron.  The CDF
collaboration has searched in $t\bar{t}$ events \cite{Affolder:2000eu} for a
narrow leptophobic $Z'$  boson present in some models of topcolor-assisted
technicolor \cite{Harris:1999ya}.  Finding no evidence of a new narrow
residence, CDF sets the limits $M_{Z'} > 780$ GeV for such a $Z'$  of natural
width $\Gamma_{Z'} = 0.04 M_{Z'}$ and $M_{Z'} > 480$ GeV for $\Gamma_{Z'} =
0.012 M_{Z'}$ as shown in figure \ref{fig:ehs:cdf-zprime}.  Simulations of
$Z' \to t\bar{t} \to l \nu b\bar{b}jj$ for Run II indicate that with
2$fb^{-1}$ (30 fb${-1}$) of data a narrow topcolor $Z'$ of mass up to 0.92 TeV
(1.15 TeV) would be visible \cite{Amidei:1996dt}; figure
\ref{fig:sigbxtt} summarizes the anticipated Run II sensitivity to
$\sigma\cdot B$ in the $M_{tt}$ distribution, and the corresponding
predictions for topcolor $Z'$  bosons..  The search by the CDF
Collaboration for new resonances decaying to $b\bar{b}$
\cite{Abe:1998uz}, which was previously mentioned (section 4.3.2,
figure \ref{fig:harris-topgluon}) as setting limits on technirhos, falls slightly short
of setting a limit on topcolor $Z'$ bosons.  In addition to pursuing Z'
bosons in these heavy quark channels, the Run IIb will also be able to
look for $Z'$ bosons decaying to $\tau$ leptons as described in section
3.6.5. In the case with the lowest standard model background, $Z' \to
\tau\tau \to e \mu \nu \bar{\nu}$, a $Z'$ boson with a mass up to about
600 GeV should be accessible (depending on mixing angle)
\cite{Lynch:2000md}.  For further details on current and 
future Topcolor $Z'$ boson searches at the Tevatron, see Table IV of
ref. \cite{Barklow:2002su}.

A $Z'$ boson would also be visible in the process $e^+ e^- \to \tau \tau$ at an
NLC \cite{Popovic:1998vb}.  Assuming a $50$\% efficiency for idenitfying 
$\tau$--pairs and requiring an excess over standard model backgrounds of
$(N^{\tau\tau} - N^{\tau\tau}_{SM}) \geq 5 \sqrt{N^{\tau\tau}_{SM}}$, it
appears that the effects of a 2.7 TeV $Z'$ boson with $\alpha_1\tan^2\chi \leq
1$ would be visible in a 50 fb$^{-1}$ sample taken at $\sqrt{s} = 500$ GeV.
A $1.5$  TeV NLC with 200 fb$^{-1}$ of data would be sensitive to $Z'$ bosons as
heavy as 6.6 TeV.  The other channels suggested earlier in the context of
topflavor models might also be useful for finding this $Z'$ at a LC, LHC, or
FMC.

\subsection{Top Seesaw} 



The 
electroweak mass gap, i.e., the dynamical fermion
mass associated with the generation of $v_{weak}$
through the Pagels-Stokar relation eq.(\ref{ps1}), is $\sim 600$ GeV
when the scale of Topcolor is taken
to be a natural scale of order $\sim 1$ TeV.
If the top quark mass were this large, our problems would
be solved, and EWSB would be identified naturally with a
$\bar{t}t$ condensate. By itself, however,
the top quark is too light
to produce the full electroweak condensate
in this way.

We can, however, write a viable model in which
the $I=\half$ top quark mass term, the term associated
with electroweak symmetry breaking, is indeed
$\sim 600$ GeV, and exploit a seesaw mechanism
to tune the physical mass of the top quark to $m_t =175$ GeV
\cite{Dobrescu:1998nm,Chivukula:1998wd}. This gives
up the predictivity of the top quark mass, since
the tuning is done through a new mixing angle. Because the top quark
is heavy, however,
the mixing angle need not be chosen artificially small.

The Top Seesaw mechanism can be implemented with the introduction of a pair of
iso-singlet, vectorlike quarks, $\chi_L$ and $\chi_R$ which both have $Y =
4/3$, analogues of the $t_R$.  This model produces a bound-state Higgs boson,
primarily composed of $\bar{t}_L\chi_R$, with the Higgs mass $\sim 1$ TeV,
saturating the unitarity bound of the Standard Model.  Such a large Higgs
mass would seemingly be {\em a priori} 
ruled out by the $S-T$ parameter constraints.
Remarkably, however, the Top Seesaw theory 
has the added bonus of supplying a rather large
$T$-parameter contribution associated with the presence of the $\chi$
fermions \cite{Dobrescu:1998nm,Collins:1999rz}.  

In 1998 when the Top Seesaw theory was proposed, it was in poor
agreement with the $S-T$ bounds.  With the most recent compilation of
LEP and worldwide data, including a refined initial state radiation
and $W$-mass determination, the Top Seesaw theory lies within the
$95\%$ c.l. error ellipse (see Section 4.5). Indeed the theory lies
within the $S-T$ plot for expected natural values of the $\chi$
masses.  Hence, the theory has already scored a predictive success:
the $S-T$ ellipse has moved to accomodate it.
We may, in fact, view the measured error ellipse as a determination of the
$\chi$ mass in this scheme; we obtain roughly $M_\chi \sim
4$ TeV.  In this picture, the worldwide electroweak precision measurements
are probing the mass of a heavy new particle, the $\chi$, significantly above
the electroweak scale.

%
%

Note that the Top Seesaw model, unlike TC2,  {\em 
does not invoke Technicolor}; rather, it replaces Technicolor
entirely with Topcolor.  It
offers novel model building possibilities for attacking the flavor
problem
\cite{Georgi:2000wt,Burdman:1998vw,He:1999vp,Popovic:2001cj}.
Extensions have been constructed in which the
$W$, $Z$, $t$ and $b$ all receive dynamical masses. Construction
of an explicit model of all quark and lepton masses seems
plausible \cite{Georgi:2000wt,Burdman:1998vw}.   
Because the $\chi$ quarks
need not carry weak-isospin quantum numbers, and
enter in vectorlike pairs\footnote{See
refs. \cite{He:1999vp,Popovic:2001cj} for models where the $\chi$'s 
are weak isodoublets.}, the constraints on
the number of techniquarks from the $S$ parameter are essentially
irrelevant for the Top Seesaw. 
The Top Seesaw also
finds an attractive setting in extra dimensional models,
as discussed in Section 4.6. The $\chi$ quarks and Topcolor
itself may be interpreted
as Kaluza-Klein modes. Because the mass-gap is $600$ GeV rather than $m_t=175$
GeV, the masses of all the
colorons, and any additional heavy gauge bosons, are
naturally moved to slightly larger mass scales  than in TC2.
One then has more model-building elbow room: phenomenological
exploration of the Top Seesaw dynamics will rely on the VLHC rather than
the LHC.   

\subsubsection{The Minimal Model}

In the minimal Top Seesaw scheme the full EWSB occurs via the condensation of
the left-handed top quark with a new, right-handed weak-singlet quark, which
we refer to as a $\chi$-quark. The $\chi_R$ quark has hypercharge $Y= 4/3$
and is thus indistinguishable from the $t_R$.  The dynamics which leads to
this condensate is Topcolor, as discussed below, and no tilting $U(1)'$ is
required. The fermionic mass scale of this weak-isospin $I=1/2$ condensate is
$\sim 0.6$ TeV.  This corresponds to the formation of a dynamical bound-state
weak-doublet Higgs field, $H \sim (\overline{\chi_R} t_L, \overline{\chi_R}
b_L)$.  To leading order in $1/N_c$ this yields a VEV for the Higgs boson,
via the Pagels-Stokar formula, of the appropriate electroweak scale,
$v_{weak} = 175$ GeV, and the top quark 
acquires an $I=\half$ dynamical mass $\mu$.
The ($I=0$, $Y=4/3$) $\chi$-quarks have an allowed Dirac mass
$M_\chi$ and a mass term $m_0$ that mixes $t_R$ and $\chi_L$.
Thus,
the full fermionic mass matrix takes the form:
\be 
\overline{\left( 
\begin{array}{c}
t_{L} \\ 
\protect\chi _{L}
\end{array}
\right) }\left( 
\begin{array}{cc}
0 & \protect\mu  \\ 
m_{0} & M_{\protect\chi }
\end{array}
\right) 
\left( 
\begin{array}{c}
t_{R} \\ 
\protect\chi _{R}
\end{array}
\right)
\label{ehs-mm-m}
\ee
where the vanishing entry is a term forbidden by the
Topcolor assignments (see 4.4.2 below).
We diagonalize the mass matrix:
\be
{\cal M}=U_{L}^{\dagger }(\phi_{L}) {\cal M}_{D} U_{R}(\phi_{R})
\label{ehs-diag-m}
\ee
leading to eigenvalues:
\be
{\cal M}_{D}=\left( 
\begin{array}{cc}
m_{1} & 0 \\ 
0 & m_{2}
\end{array}
\right) 
\ee
where:
\bea
m_{1}^{2} & = & \frac{1}{2}\left[
{m_{0}^{2}+M_{\chi }^{2}+\mu ^{2}}-{\sqrt{\left(
m_{0}^{2}+M_{\chi }^{2}+\mu ^{2}\right) ^{2}-4\mu ^{2}m_{0}^{2}}}\right] 
\bigskip 
\nonumber \\
& \approx & 
\frac{ m_0^2 \mu^2 }{ M_\chi^2 + m_0^2 + \mu^2 } + {\cal O} 
( m_0^4 \mu^4 / M_\chi^6)
\eea 
and 
\bea m_{2}^{2} & = & \frac{1}{2}\left[ {m_{0}^{2}+M_{\chi }^{2}+\mu
    ^{2}}+{\sqrt{\left(
        m_{0}^{2}+M_{\chi }^{2}+\mu ^{2}\right) ^{2}-4\mu
      ^{2}m_{0}^{2}}}\right] \bigskip
\nonumber \\
& \approx & 
M_\chi^2 + m_0^2 + \mu^2 +  {\cal O} ( m_0^2 \mu^2 / M_\chi^2 )
\eea 
where limits for large $M_\chi$ are indicated.

The fermionic mass matrix thus
admits a conventional seesaw mechanism, yielding the physical top quark mass
as an eigenvalue that is $\sim m_0\mu/M_\chi << \mu \approx 600$ GeV.  The
top quark mass can be adjusted to its experimental value. The diagonalization
of the fermionic mass matrix does not affect the VEV ($v_{weak} = 175$ GeV)
of the composite Higgs doublet.  Indeed, the Pagels-Stokar formula is now
modified to read: \beq
\label{ps2}
v_{weak}^2 \equiv f_\pi^2 = \frac{N_c}{16\pi^2}
\frac{m_{t}^2}{\sin^2\phi_L} ( \log\frac{M^2}{\overline{M}^2} + k) \eeq
where $m_t$ is the physical top mass, and $\phi_R$ the mass matrix
diagonalizing mixing angle.  The Pagels-Stokar formula differs from that
obtained (in large $N_c$ approximation) for top quark condensation models by
the large enhancement factor $1/\sin ^{2}\phi_L$. This is a direct
consequence of the seesaw mechanism.  The Top Seesaw employs $\psi_L=
(t_L,b_L)$ as the source of the weak $I=1/2$ quantum number of the composite
Higgs boson, and thus the origin of the EWSB vacuum condensate.  This neatly
separates \cite{Simmons:1989pu} the problem of EWSB from that of the new
weak-isosinglet states in the $\chi_{L,R}$ and $t_R$ sector, a distinct
advantage since the electroweak constraints on new physics are not very
restrictive of isosinglets.

\subsubsection{Dynamical Issues}

How does Topcolor
produce the $\mu$ mass term?  We introduce an embedding of QCD
into the gauge groups $SU(3)_{1}\times SU(3)_{2}$, with coupling constants
$h_{1}$ and $h_{2}$ respectively. These symmetry groups are broken down to
$SU(3)_{QCD}$ at a high mass scale $M$. The assignment of the elementary
fermions to representations under the full set of gauge groups
$SU(3)_{1}\times SU(3)_{2}\times SU(2)_{W}\times U(1)_{Y}$ is as follows:
\begin{equation}
\psi_{L} : \, ({\bf 3,1,2,} \, +1/3) \;\;\; ,  \;\;\
\chi_{R} : \, ({\bf 3,1,1,} \, +4/3) \;\;\; ,  \;\;\
t_{R}, \chi_{L} : \, ({\bf 1,3,1,} \, +4/3) ~.
\end{equation}
This set of fermions is incomplete: the representation specified has
$[SU(3)_1]^3$, $[SU(3)_2]^3$, and $U(1)_Y [SU(3)_{1,2}]^2$ gauge anomalies.
These anomalies will be canceled by fermions associated with either the
dynamical breaking of $SU(3)_1 \times SU(3)_2$, or with the $b$-quark mass
generation (a specific example of the latter case is given at the end of this
section).  

\noindent
Schematically things look like
this:
\begin{center}
\underline{$\ \ \ \ \ \ \ \ \ \ SU(3)_{1}$ \ \ \ \ \ \ \ $\ \ \ \   
SU(3)_{2}$ \ \ \ \ \ \ \ }\\
$\left( 
\begin{array}{c}
\left( 
\begin{array}{c}
t \\ 
b
\end{array}
\right) _{L} \\ 
\left( \chi \right) _{R} \\ 
...
\end{array}
\right) ${\small \ \ \ \ \ }$\left( 
\begin{array}{c}
\left( 
\begin{array}{c}
t \\ 
b
\end{array}
\right) _{R} \\ 
\left( \chi \right) _{L} \\ 
...
\end{array}
\right) $
\end{center}
The dynamics of EWSB and top-quark mass generation will not depend
on the details of the additional fermions.

We introduce a scalar Higgs field, $\Phi$, transforming as $({\bf
\overline{3},3,1,} \, 0)$, which develops a diagonal VEV, $\langle
\Phi{}^i_j \rangle = {\cal V}\delta_{j}^{i}$.  This field is presumably
yet another dynamical condensate, or it can be interpreted as a relic
of compactification of an extra dimension.  Topcolor is then broken
to QCD,
\begin{equation}
SU(3)_{1}\times
SU(3)_{2}\longrightarrow SU(3)_{QCD}
\end{equation}
yielding massless gluons and an octet of degenerate colorons 
with mass $M$ given by 
\begin{equation}
M^{2}= (h_{1}^{2}+h_{2}^{2})\,{\cal V}^{2}\,.
\end{equation}

We can also exploit $\Phi$ to
provide the requisite $M_\chi$
by introducing a Yukawa coupling of the fermions $\chi_{L,R}$ of the form:
$
-\xi\,\overline{\chi _{R}}\,\Phi \, \chi _{L}+{\rm h.c.}
\longrightarrow -M_{\chi}\,
\overline{\chi }\chi
$.
We emphasize that this is an electroweak singlet mass term.
$\xi$ can be a perturbative 
coupling constant so ${\cal V} \gg \xi{\cal V} =  M_{\chi }$.
Finally, since both $t_{R}$ and $\chi _{L}$ carry identical Topcolor and
$U(1)_{Y}$ quantum numbers we are free to include the explicit mass term,
also an electroweak singlet, of
the form $m_0\overline{\chi _{L}}\,t_{R}+{\rm h.c.}$.

The Lagrangian of the model at scales below the
coloron mass is $SU(3)_C\times SU(2)_W\times U(1)$ invariant and
becomes:
\begin{equation}
{\cal {L}}_{0}= {\cal {L}}_{\rm kinetic}-(M_{\chi}\,\overline{\chi _{L}}
\,\chi_{R}
+m_0\,\overline{\chi _{L}}\,t_{R}+{\rm h.c.})+{\cal {L}}_{\rm int}
\label{njll}
\end{equation}
${\cal {L}}_{\rm int}$ contains the residual Topcolor interactions from the
exchange of the massive colorons:
\begin{equation}
{\cal {L}}_{\rm int}=  -\frac{h^{2}_{1}}{M^{2}}\left( \overline{\psi _{L}}\,
\gamma^{\mu }\frac{M ^{A}}{2}\psi _{L}\right)
\left( \overline{\chi _{R}}\,
\gamma_{\mu }\frac{M ^{A}}{2}\chi _{R}\right) +LL+RR+...\;
\label{njl0}
\end{equation}
where $LL$ $(RR)$ refers to purely left-handed (right-handed) current-current
interactions.  It suffices to retain in the low
energy theory only the effects of the operators shown in eq.~(\ref{njl0})
even though higher dimension operators may be present.
To leading order in $1/N_c$, upon performing the
familiar Fierz rearrangement, we have:
\begin{equation}
{\cal {L}}_{\rm int}= \frac{h_{1}^{2}}{M^{2}}(\overline{\psi_{L}}\,
\chi_{R})\,(\overline{\chi_{R}}\,\psi _{L}) ~.
\label{njl1}
\end{equation}
It is convenient to pass to a mass eigenbasis with the following
redefinitions:
\begin{equation}
\chi _{R} \rightarrow \cos \phi_R \;\chi _{R} - \sin \phi_R\;
t_{R}\,;
\quad t_{R} \rightarrow \cos \phi_R \;t_{R} + \sin \phi_R \;\chi _{R}
\end{equation}
where
\begin{equation}
\tan \phi_R = \frac{m_0}{ M_{\chi }}
\end{equation}
In this basis, the NJL Lagrangian takes the form:
\bear
{\cal {L}}_{0} & = & {\cal {L}}_{\rm kinetic}-
\overline{M}\,\overline{\chi_{R}}\,\chi_{L}+{\rm h.c.}
\nonumber \\ [2mm]
& & + \frac{h_{1}^{2}}{M^{2}}
\left[ \overline{\psi _{L}}\, \left(\cos \phi_R \;\chi_{R}
-\sin \phi_R \;t_{R}\right)\right] 
\left[ \left(\cos\phi_R \; \overline{\chi_{R}}
-\sin\phi_R \; \overline{t_{R}} \right) \,\psi _{L}\right] 
\label{njl2}
\eear
where 
\be
\overline{M}=\sqrt{M_{\chi}^{2}+ m_0^{2}} ~.
\label{barM}
\ee

At this stage we have the choice of using the renormalization group, or
looking at the mass gap equations for $\mu$.  A rationale 
for studying the gap
equations is that they in principle allow one to explore limits, such as
$\overline{M} > M$ which are conceptually more difficult with the
renormalization group\footnote{For instance, the $d=6$ operator makes no
  sense above the scale $M$ in the renormalization group, but the cut-off
  theory can still be expressed in the gap equation language.}.  
For a more complete
analysis of the dynamics see \cite{He:2001fz}.

\begin{figure}[t]
\vspace{7cm}
\includegraphics{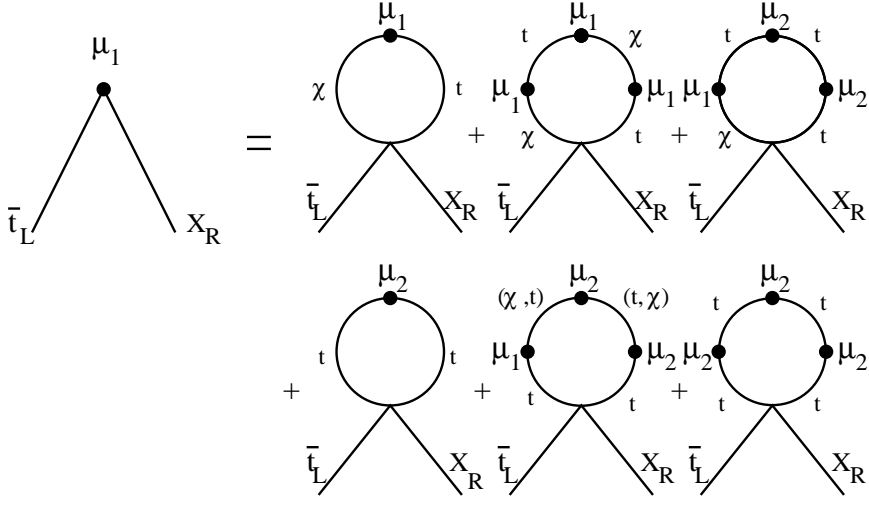}
\vspace{1cm}
\caption[]{\small \addtolength{\baselineskip}{-.4\baselineskip} 
Gap equations for $\mu_1$.  A similar set
of terms is obtained for $\mu_2$ \cite{He:2001fz}.
}
\label{diagramone}
\end{figure}

Let us summarize the gap equation analysis.  
We assume two dynamical mass
terms:
\be
-\mu_1\overline{t}_L \chi_R -\mu_2\overline{t}_L t_R 
\ee
and we compute the gap equations to order ${\cal{O}}(\mu^3)$.
This will produce no IR divergences in terms to order $\cos^2\phi_R$,
but there is an IR log-divergence in the last term of Fig.(\ref{diagramone}) of
order $\sin^2\phi_R$, and we take this to be $\ln (m_t^2)$. 
Fig.(\ref{diagramone}) 
produces the coupled system of gap equations for $\mu_1$:
\begin{eqnarray}
\mu_1 & = & \frac{h_1^2N}{8\pi^2M^2}
\mu_1\cos^2\phi_R\left(M^2 -
\overline{M}^2\ln\left(\frac{M^2+\overline{M}^2}{\overline{M}^2}\right)\right)
\nonumber \\
& &
+ \frac{h_1^2N}{8\pi^2M^2}
\mu_1\cos^2\phi_R\left(
-\mu_1^2\ln\left(\frac{M^2+\overline{M}^2}{\overline{M}^2}\right)
+\frac{\mu_1^2M^2}{M^2 + \overline{M}^2}
-\mu_2^2\ln\left(\frac{M^2+\overline{M}^2}{\overline{M}^2}\right)
\right)
\nonumber \\ 
& & 
+ \frac{h_1^2N}{8\pi^2M^2}
\mu_2\cos\phi_R\sin\phi_R \left(
M^2 -
\mu_1^2\ln\left(\frac{M^2+\overline{M}^2}{\overline{M}^2}\right)
- \mu_2^2\ln\left(\frac{M^2}{m_t^2}\right)
\right)
\end{eqnarray}
and a similar set for $\mu_2$.
With the
substitutions $
\mu_1 = \mu \cos\phi_R $ and
$\mu_2 =\mu\sin\phi_R $
the two independent gap equations reduce to a single mass gap
equation:
\begin{eqnarray}
\label{gapts}
\mu & = & \frac{h_1^2N}{8\pi^2M^2}
\mu \left(M^2 -
\overline{M}^2\cos^2\phi_R
\ln\left(\frac{M^2+\overline{M}^2}{\overline{M}^2}\right)
\right.
\nonumber \\
& &
\left.
-\mu^2\ln\left(\frac{M^2+\overline{M}^2}{\overline{M}^2}\right)
+\frac{\mu^2M^2}{M^2 + \overline{M}^2}
-\mu^2\sin^4\phi_R \ln\left(\frac{M^2(M^2+\overline{M}^2)}{m_t^2
\overline{M}^2}\right)
\right)
\end{eqnarray}
The gap equation eq.(\ref{gapts}) shows 
that we require supercritical coupling
as the mass $\overline{M}$ becomes large. 
Moreover, for fixed supercritical coupling, $h_1^2/4\pi $,
as we raise the scale $\overline{M}$ the condensate turns off
like a second order phase transition. 
\begin{figure}[t]
\vspace{7cm}
\includegraphics{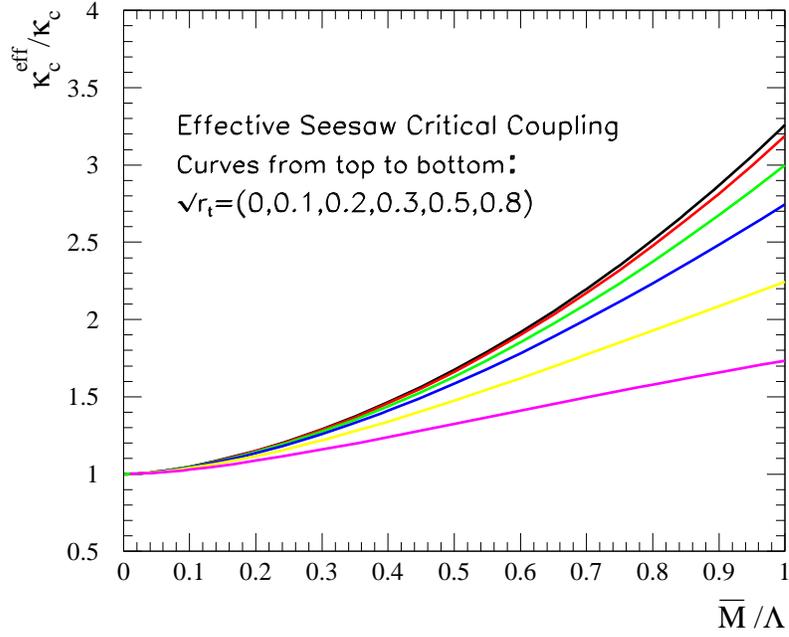}
\vspace{1cm}
\caption[]{\small \addtolength{\baselineskip}{-.4\baselineskip} 
Gap equation
with  coupling constant $\kappa = h_1^2/4\pi $
(scaled by constant $\k_c \equiv 2\pi/3$) as a function
of $\overline{M}/M $, for
$\sqrt{\tan\phi_R} =(0,\,0.1,\,0.2,\,0.3,\,0.5,\,0.8)$.
from Hill, He, and Tait \cite{He:2001fz}}
\end{figure}

We can also see that these reproduce normal
top condensation in the decoupling limit. For example,
choose $\overline{M}\rightarrow \infty$ for fixed $M$,
and we find:
\begin{eqnarray}
\label{gaptsred}
\mu & = & \frac{h_1^2N}{8\pi^2M^2}
\mu \left(M^2(1-\cos^2\phi_R) 
-\mu^2\sin^2\phi_R \ln\left(\frac{M^2}{m_t^2}\right)
\right)
\end{eqnarray}
and redefining $\tilde{g}=h_1\sin\phi_R$, and $m_t=\mu\sin\phi_R$
we have:
\begin{eqnarray}
m_t & = & \frac{\tilde{g}^2N}{8\pi^2M^2}
\mu \left(M^2 
- m_t^2 \ln\left(\frac{M^2}{m_t^2}\right)
\right)
\end{eqnarray}
which is the top condensation gap equation. Here we have decoupled $\chi_L$
and $\chi_R$ with $\overline{M}\rightarrow \infty$.  We can also
obtain top condensation by setting $\sin^2\phi_R = 0$ and
$\overline{M}\rightarrow 0$, which decouples $\chi_L$ and  $t_R$, and causes
$\chi_R$ to play the role of $t_R$.

\begin{figure}[t]
\vspace{7cm}
\includegraphics{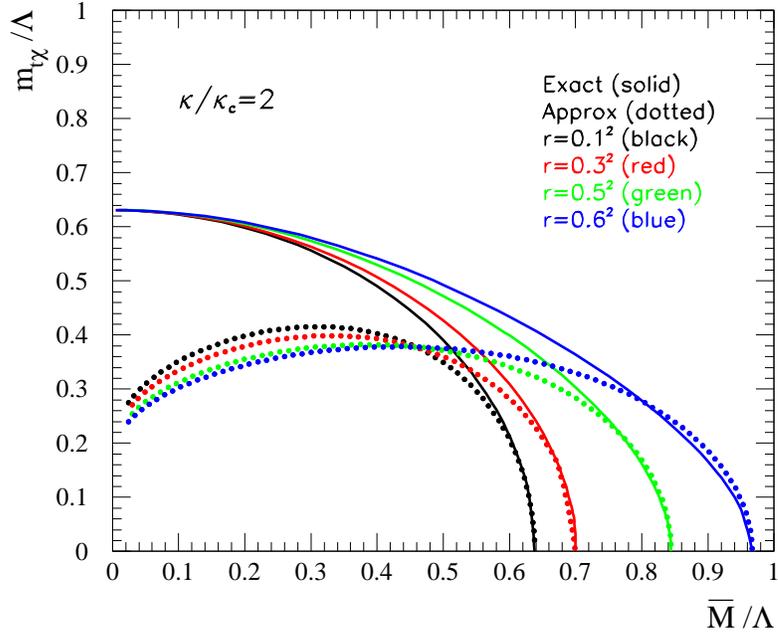}
\vspace{1cm}
\caption[]{\small \addtolength{\baselineskip}{-.4\baselineskip}
Behavior of gap equation for fixed
supercritical coupling
$h_1^2/4\pi\k_c = 2.0 $ as $\overline{M}/M$
is increased. In the notation of the figure,
$m_{t\chi} = \mu$. 
$\sqrt{\tan\phi_R} =(0,\,0.1,\,0.2,\,0.3,\,0.5,\,0.8)$.
The figure shows results for our gap equation (solid)
and an approximation (dotted) that treats $\overline{M}$
as a mass insertion (see \cite{He:2001fz}).
}
\end{figure}

A separate question of interest is the structure of the electroweak
corrections in the Top Seesaw theory. This is most
easily determined from an effective
Lagrangian for the composite Higgs boson. The most complete
study to date is that of  Hill, He and Tait \cite{He:2001fz}.
The composite Higgs boson mass satisfies the approximate
NJL result, $m_{Higgs} \approx 2\mu \sim 1.2$ TeV.  Experience
with Topcolor suggests that radiative corrections from $SU(2)_2$
will reduce this. The composite
Higgs mass is consistent with the unitarity limit of
the Standard Model.  By itself, this pulls the theory to 
large negative $T$. However,
there are $\chi-t$ mixing corrections to the T parameter
as well.  We obtain for T:
\be
T=\frac{Nm_{t}^{2}}{32\pi ^{2}v_{weak}^{2}\alpha _{Z}}\left[ \frac{\mu ^{2}}{%
m_{0}^{2}}+\frac{2\mu ^{2}}{\overline{M}^{2}}\left( \ln \left( \frac{%
\overline{M}^{4}}{\mu ^{2}m_{0}^{2}}\right) -1\right) \right]
\ee
Putting in typical numbers, such as  $\mu = 0.6$ TeV,
 $m_0 = m_t\overline{M}/\mu \approx 1.2$ TeV,
$\alpha_Z=1/128$, one observes a large positive correction
to $T$, and one recovers consistency with the error ellipse
constraint for $\overline{M} \approx 4$ TeV.
Note that the $S$-parameter
is vanishing, since
the additional $\chi$ quarks are weak isosinglets.

We note that, using the freedom to adjust $\sin\phi_R$, we can
in principle dynamically 
accommodate any fermion mass lighter than $\sim 600$ GeV -- at the
price of some fine-tuning.  This freedom may be useful in constructing
more complete models involving all three generations
\cite{Burdman:1998vw,Georgi:2000wt}. The top quark is unique, however,
in that it is very difficult to accommodate such a heavy quark in any
other way, and there is less apparent fine-tuning.  We therefore
believe it is generic, in any model of this kind, that the top quark
receives the bulk of its mass through this seesaw mechanism.  In a
more general theory that includes the seesaw mechanism there are more
composite scalars, and one of the neutral Higgs bosons may be as light
as ${\cal O}(100$ GeV).

\subsubsection{Including the $b$-quark}

Inclusion of the $b$-quark is straightforward, and two distinct
schemes immediately suggest themselves.  We include additional
fermionic fields of the form $\omega _{L}$, $\omega _{R}$, and $b_{R}$
with the assignments:
\begin{equation}
b_{R}, \omega _{L} : \, ({\bf 1,3,1,} -2/3) \;\;\; , \;\;\; 
\omega _{R}   : \, ({\bf 3,1,1,} -2/3) ~.
\label{omega}
\end{equation}
These fermion gauge assignments cancel the anomalies noted above.
We further allow $\overline{\omega _{L}}\omega _{R}$ and  $\overline{\omega
_{L}}b_{R}$ mass terms, in direct analogy to the $\chi $ and $t$ mass terms:
\begin{equation}
{\cal {L}}_{0}\supset -(M_{\omega }\overline{\omega _{L}}\omega
_{R}+m_{\omega }\overline{\omega _{L}}b_{R}+{\rm h.c.})
\end{equation}
With the previous assignments
for the $\chi$ quarks,
schematically, this looks like the following:

\noindent
{\bf Inclusion of b-quark I:$\protect\bigskip $}
\begin{center}
\underline{$\ \ \ \ \ \ \ \ \ \ \ \ SU(3)_{1}$ \ \ \ \ \ \ \ \ \ \ \ \ \ \ \
\ \ \ \ $\ \ \ \ \ \ \ \ SU(3)_{2}$ \ \ \ \ \ \ \ \ \ \ \ }\\
$\left( 
\begin{array}{c}
\left( 
\begin{array}{c}
t \\ 
b
\end{array}
\right) _{L}I=%
{\frac12}%
\\ 
\left( 
\begin{array}{c}
\chi \\ 
\omega
\end{array}
\right) _{R}I=0
\end{array}
\right) ${\small \ \ \ \ \ }$\left( 
\begin{array}{c}
\left( 
\begin{array}{c}
t \\ 
b
\end{array}
\right) _{R}I=0 \\ 
\left( 
\begin{array}{c}
\chi \\ 
\omega
\end{array}
\right) _{L}I=0
\end{array}
\right) $
\end{center}

Alternatively, we
may define
the $\chi$ and $\omega$ fields
to
form isodoublets, with assignments as:
\begin{equation}
b_{R},  : \, ({\bf 1,3,1,} -2/3) \;\;\; , \;\;\; 
(\chi, \omega)_{R},   : \, ({\bf 3,1,2,} 1/3) 
\;\;\; , \;\;\; 
, (\chi, \omega)_{L} : \, ({\bf 3,1,2,} 1/3)
~.
\label{omega2}
\end{equation}

\noindent
{\bf Inclusion of b-quark II: $\protect\bigskip $}
\begin{center}
\underline{\ \ \ \ \ \ $ SU(3)_{1}$ \ \ \ \ \ \ \ \ \ \ \ \ 
\ \ \ \ \ $\ \ \ \ \  SU(3)_{2}$ 
\ \ \ \ \ \ \ \ \ \ \ }\\
$\left( 
\begin{array}{c}
\left( 
\begin{array}{c}
t \\ 
b
\end{array}
\right) _{L}I=%
{\frac12}%
\\ 
\left( 
\begin{array}{c}
\chi \\ 
\omega
\end{array}
\right) _{R}I=%
{\frac12}%
\end{array}
\right) ${\small \ \ \ \ \ }$\left( 
\begin{array}{c}
\left( 
\begin{array}{c}
t \\ 
b
\end{array}
\right) _{R}I=0 \\ 
\left( 
\begin{array}{c}
\chi \\ 
\omega
\end{array}
\right) _{L}I=%
{\frac12}%
\end{array}
\right) $
\end{center}

The
schematic model affords a simple way to suppress the formation of a $b$-quark
mass comparable to the top quark mass. 
We can suppress the formation of the $\overline{\omega _{L}}b_{R}$
condensate altogether by choosing $\overline{M}_{\omega } =
\sqrt{\mu^2_{\omega\omega}+\mu^2_{\omega b}}\sim M$. In this limit we do
not produce a $b$-quark mass.  However, by allowing $\mu_{\omega
\omega }\leq M$ and $\mu_{\omega b}/\mu_{\omega \omega }\ll 1$ we
  can form an acceptable $b$-quark mass in the presence of a small
  $\overline{\omega_{L}}b_{R}$ condensate.  
  
Yet another possibility arises, one which seems to be phenomenologically
favored \cite{He:2001fz}, which
is to exploit instantons. If we suppress the
formation of the $\overline{\omega _{L}}b_{R}$ condensate by
choosing $\overline{M}_{\omega }\sim M,$ there will be a
$\overline{\omega _{L}}b_{R}$ condensate induced via the 't Hooft
determinant when the $t$ and $\chi$ are integrated out.  We then
estimate the scale of the induced $\overline{\omega _{L}}b_{R}$ mass
term to be about $\sim 20$ GeV, and the $b$-quark mass then emerges
as $\sim 20\mu_{\omega b}/\mu_{\omega \omega }$ GeV. We will not
further elaborate the $b$-quark mass in the present discussion,
since its precise origin depends critically upon the structure of
the complete theory including all light quarks and leptons.
Including partners for the $b$-quark,
the $T$ parameter is given by a more general formula
\cite{Collins:1999rz,He:2001fz}.

The Top Seesaw Model offers new possibilities for a dynamical
scheme explaining both EWSB and the origin of flavor masses
and mixing angles.  We focused here in some detail on the
third generation and the EWSB dynamics.  A fully extended
model for light quark and lepton masses has not yet been
analyzed, but it would seem that the Top Seesaw affords
interesting new directions and possibilities that should be
examined.  Some varying attempts in this direction
can be found in \cite{Burdman:1998vw,Georgi:2000wt}.
Remarkably, the vectorlike fermions
introduced here to provide
the seesaw can also help to remedy the discrepancies 
between lepton and $b$-quark forward-backward
and left-right asymmetries
in the LEP data \cite{Chanowitz:2001bv}.

\subsection{Top Seesaw Phenomenology}

%


\subsubsection{Seesaw Quarks}

In the Top Seesaw and related models, the third generation quarks
\cite{Dobrescu:1998nm,Chivukula:1998wd,Collins:1999rz},
and possibly all Standard Model
fermions \cite{Burdman:1998vw,Simmons:1989pu}, acquire mass through seesaw
mixing with exotic, weak-singlet fermions.  As a result, there are several
new types of states which can affect the phenomenology: the new fermions
themselves, and, composite scalars formed at least in part from the new
fermions.  We begin by surveying current bounds on the mixing angles
$\phi_f$ between ordinary and weak-singlet fermions.  We then look at
low-energy limits on the masses of the heavy fermionic seesaw partner,
$f^H$, states and the composite
scalars (these results apply
generally
to Kaluza-Klein modes as well). 
 We comment on variant models in which the mixing is with
weak-doublet fermions instead of weak-singlets.  And to conclude, we discuss
Tevatron limits on the masses of the $f^H$ states.

Suppose the mass matrix of a Standard Model
top quark mixing with a vectorial isosinglet quark $\chi$ has the seesaw
form of eq.(\ref{ehs-mm-m}).  The matrix is diagonalized by performing
separate rotations on the left-handed and right-handed fermion fields (eq.
\ref{ehs-diag-m}). Most of the 
phenomenology is sensitive to the mixing among
the left-handed fermions. Of the two mass eigenstates, the 
physical top quark is the lighter one
$\tilde{t}_L$ and is mostly weak-isodoublet:
\begin{equation}
\tilde{t}_L = \cos\phi_t t_L - \sin\phi_t \chi_L \ ,
\label{eqn:ehs:liferm}
\end{equation}
and has a mass of order $m_0\mu / M_\chi$; the
heavier ($\tilde{\chi}$) 
is mostly weak-isosinglet:
\begin{equation}
T_L^H \equiv \tilde{\chi}_L = \sin\phi_t t_L + \cos\phi_t \chi_L
\label{eqn:ehs:heferm}
\end{equation}
with a mass of order $\sim M_\chi$.  This stucture is readily
generalized to models in which more than one ordinary fermion mixes with
weak singlets \cite{Langacker:1988ur}.

Recent limits on the mixing angles between ordinary and 
weak-isosinglet fermions
from precision electroweak data were obtained in \cite{Popovic:2000dx}.
Separate limits at 95\% c.l. are given for the case in which each fermion
flavor has its own weak partner \cite{Burdman:1998vw,Simmons:1989pu}
\begin{eqnarray}
\sin^2\phi_e \leq  0.0024 \, ,\ \ \ \ \ \ \sin^2\phi_\mu 
\!\!&\leq&\!\! 0.0030\, ,\ \ \ \ \ \sin^2\phi\tau \leq 0.0030\\
\sin^2\phi_d \leq  0.015\, ,\, \ \ \ \ \ \ \ \sin^2\phi_s 
\!\!&\leq&\!\! 0.015 \, ,\ \ \ \ \ \sin^2\phi_b \leq 0.0025\\
\sin^2\phi_u \leq  0.013\, ,\, \ \ \ \ \ \ \ \sin^2\phi_c
\!\!&\leq&\!\! 0.020 \, .
\label{eqn:ehs:casea}
\end{eqnarray}
and the case in which only the third-generation quarks have weak partners
\cite{Dobrescu:1998nm,Chivukula:1998wd,Collins:1999rz}
\begin{equation}
\sin^2\phi_b \leq 0.0013\ .
\end{equation}


The most important precison electroweak constraints on the Top Seesaw
come from the $S$ and $T$ parameters
\cite{Collins:1999rz,Popovic:2000dx,He:2001fz}.  It is
remarkable that the minimal Top Seesaw model, which typically includes
a heavy composite Higgs boson around $1$\,TeV (see
Fig.\,\ref{fig:Mhnew}), is non-trivially compatible with the $S-T$
bounds.  The composite Higgs boson's contributions will drive $T$ in
the negative direction relative to a light SM Higgs.  However, the Top
Seesaw sector has generic weak-isospin violation from the $t$-$\chi$
mixing which will significantly contribute to $T$ in the positive
direction.  Extended models with bottom seesaw are more complex
because of the $b$-$\omega $ mixing and the two composite Higgs
doublets.  Corrections to the $Z\rightarrow \bar{b}b$ vertex are also
relevant in this case.  We summarize the essentials of the $S-T$
analysis of Top Seesaw models and refer the interested reader to the
detailed discussion in ref. \cite{He:2001fz}.

\begin{figure}[t]
\vspace{8cm}
\includegraphics{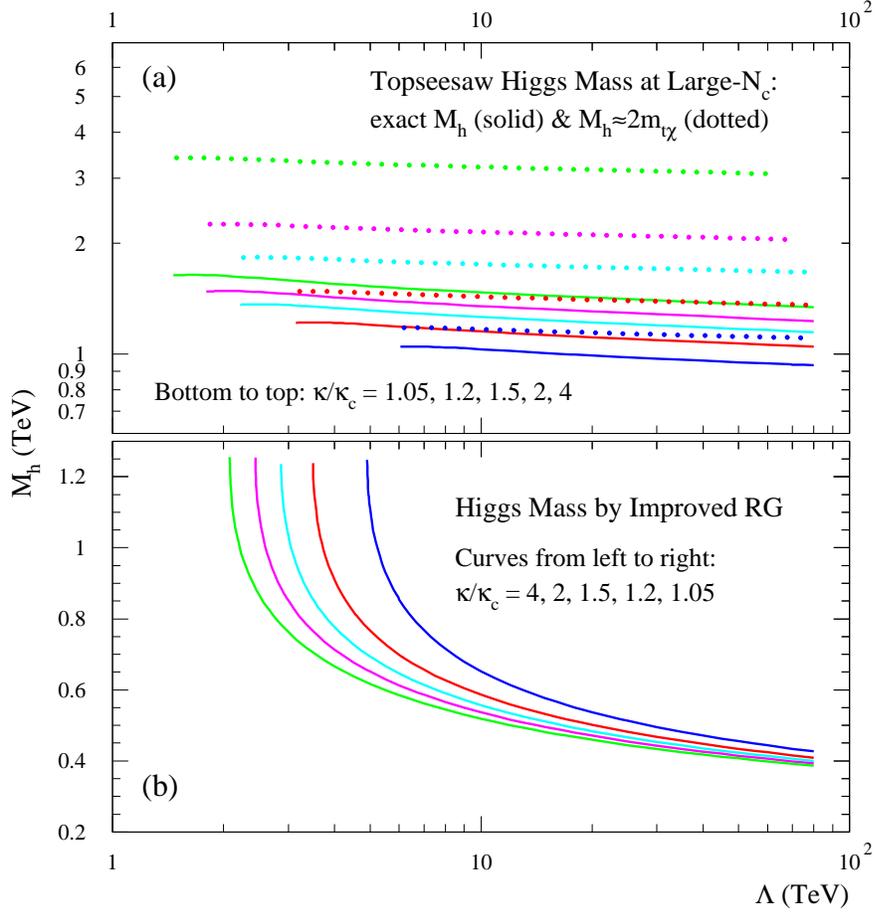}
\vspace{2.5cm}
\caption[]{\small \addtolength{\baselineskip}{-.4\baselineskip}
The predicted mass spectrum of the Top Seesaw Higgs
boson: (a) by the large-$N_c$ fermion-bubble calculation;
and (b) by an improved RG analysis 
including the Higgs self-coupling evolution.
}
\label{fig:Mhnew}
\end{figure}

In Fig.\,\ref{stnew}, 
we give the complete $S$ and $T$
contributions from the minimal Top Seesaw model, including 
corrections from both the composite Higgs boson and the seesaw quarks, 
and compare them with
the 95\%\,C.L. contour for $S-T$.  
Each figure corresponds
to a different choice of critical
coupling, $\kappa / \kappa_c$, and shows the trajectory
in the $S-T$ plane as the $\chi$ mass varies.  
The results are based on both
the large-$N_c$ fermion-bubble calculation 
and an improved RG in ref.\cite{He:2001fz}. 
The improved RG approach gives
lower Higgs mass values (around $400-500$\,GeV) so that the
curves are slightly shifted towards the upper left
(the reality is somewhere between these two trajectories). 
The figure clearly illustrates that the Top Seesaw model is consistent
with the electroweak precision data provided $M_\chi$ is in the appropriate
mass range.
For instance, when the Topcolor force is slightly super-critical, we see that
precision data are effectively probing $M_\chi \sim 4$\,TeV.
In Fig.\,\ref{stmh}, we display the same 
$S-T$ trajectories as in Fig.\,\ref{stnew}, 
but with the corresponding Higgs mass ($M_h$) values marked 
 \cite{He:2001fz}.

\begin{figure}[t]
\vspace{10 cm}
\includegraphics{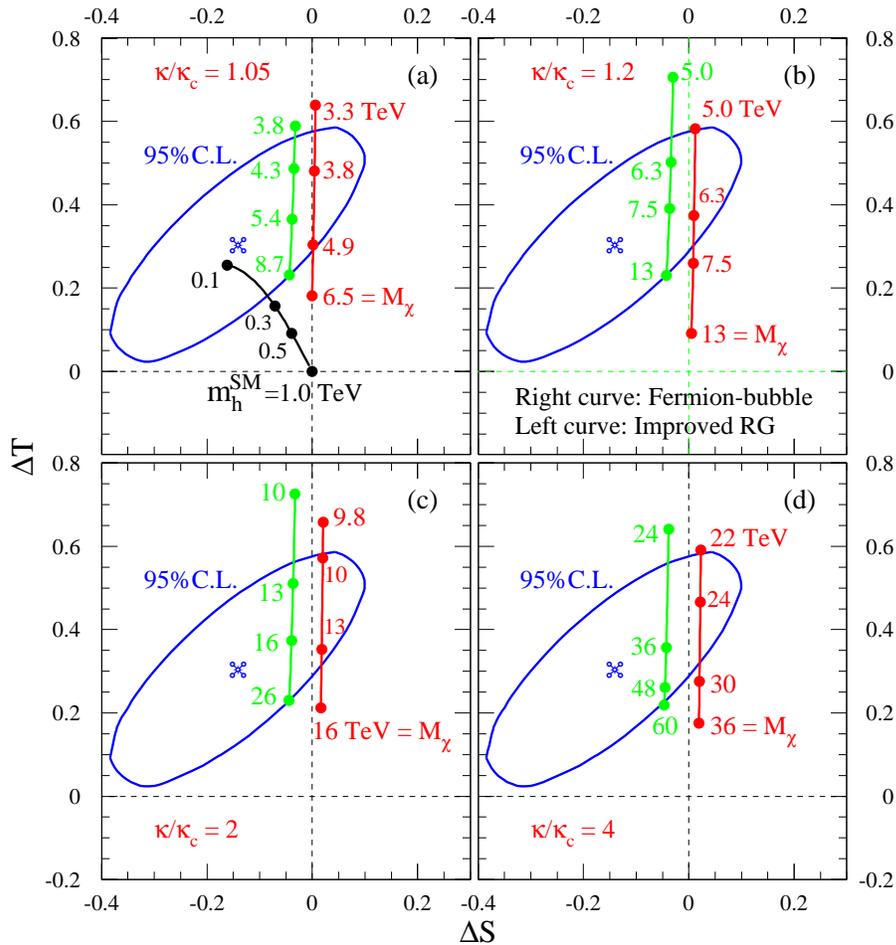}
\vspace{1 cm}
\caption[]{\small \addtolength{\baselineskip}{-.4\baselineskip}
Top Seesaw contributions to $S$ and $T$ are
compared with the 95\% c.l. error ellipse (with $m_h^{\rm ref} = 1$ TeV)
for $\k/\k_c=1.05,\,1.2,\,2,\,4$,
shown as a function of $M_\chi$.  
In each plot, the curve on the right is derived from
the large-$N_c$ fermion bubble calculation, 
the curve on the left is deduced by an 
improved RG approach. 
For reference, the SM Higgs corrections to ($S$,\,$T$), relative to
$m_h^{\rm ref} =1$\,TeV, are given for $m_h^{\rm SM}$ varying from 
100\,GeV up to 1.0\,TeV in plot (a). (from Hill, He and Tait
\cite{He:2001fz}).
}
\label{stnew}
\end{figure}

\begin{figure}[t]
\vspace{10cm}
\includegraphics{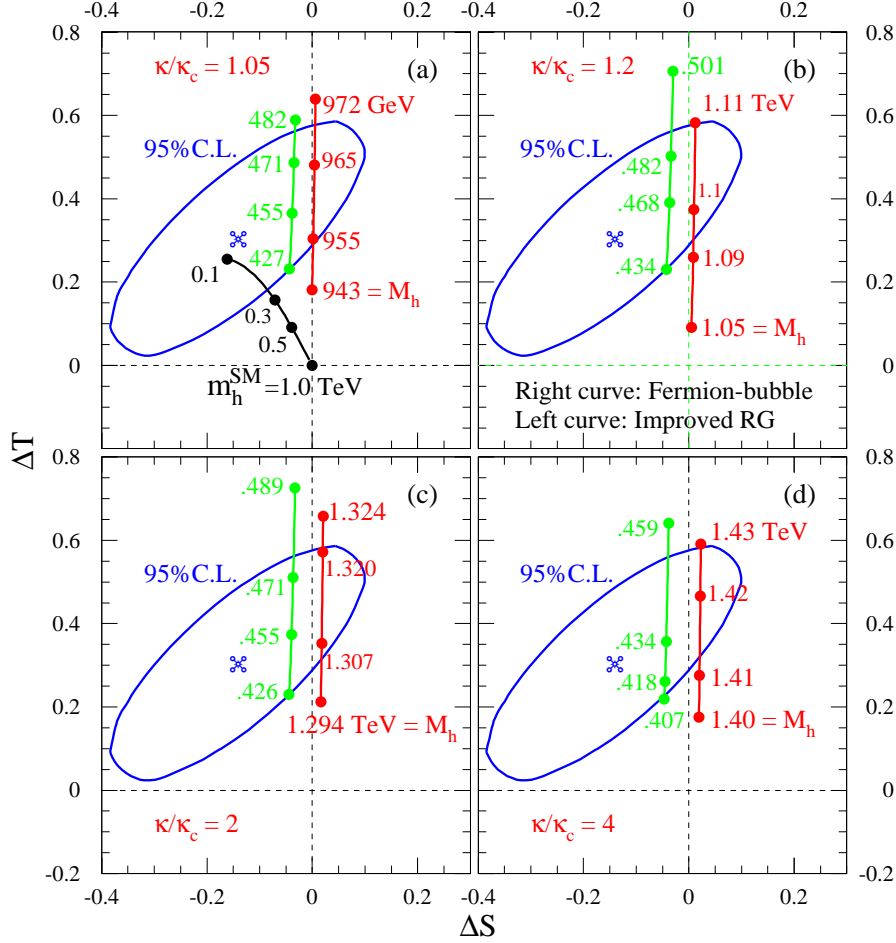}
\vspace{1 cm}
\caption[]{\small \addtolength{\baselineskip}{-.4\baselineskip}
Same as Fig.\,\ref{stnew}, 
but with the corresponding $M_h$ values
marked on the  $S$-$T$ trajectories instead.
(from Hill, He and Tait
\cite{He:2001fz}).
}
\label{stmh}
\end{figure}

\begin{figure}[t]
\vspace{10cm}
\includegraphics{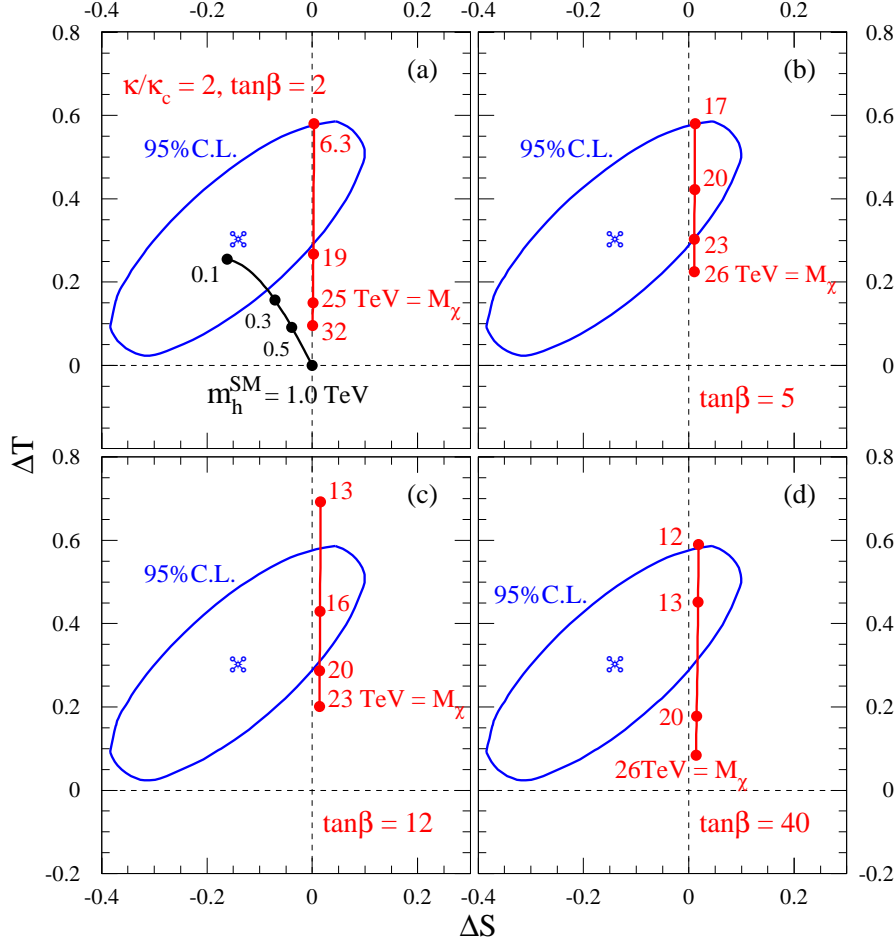}
\vspace{1 cm}
\caption[]{\small \addtolength{\baselineskip}{-.4\baselineskip}
Top and bottom seesaw contributions to $S$ and $T$ are
compared with the 95\% C.L. error ellipse (with $m_h^{\rm ref} = 1$ TeV)
for $\k/\k_c=2$ and $\chi = 3 \times 10^{-3}$ 
with a variety of values of $\tan\beta$.
The $S$-$T$ trajectories (including both Higgs and
quark contributions) are shown as a function of $M_\chi$.
For reference, the SM Higgs corrections to ($S$,\,$T$), relative to
$m_h^{\rm ref} =1$\,TeV, are depicted for $m_h^{\rm SM}$ varying from 
100\,GeV up to 1.0\,TeV in plot (a).
}
\label{fig:tbss_stx}
\end{figure}

\begin{figure}[t]
\vspace{10cm}
\includegraphics{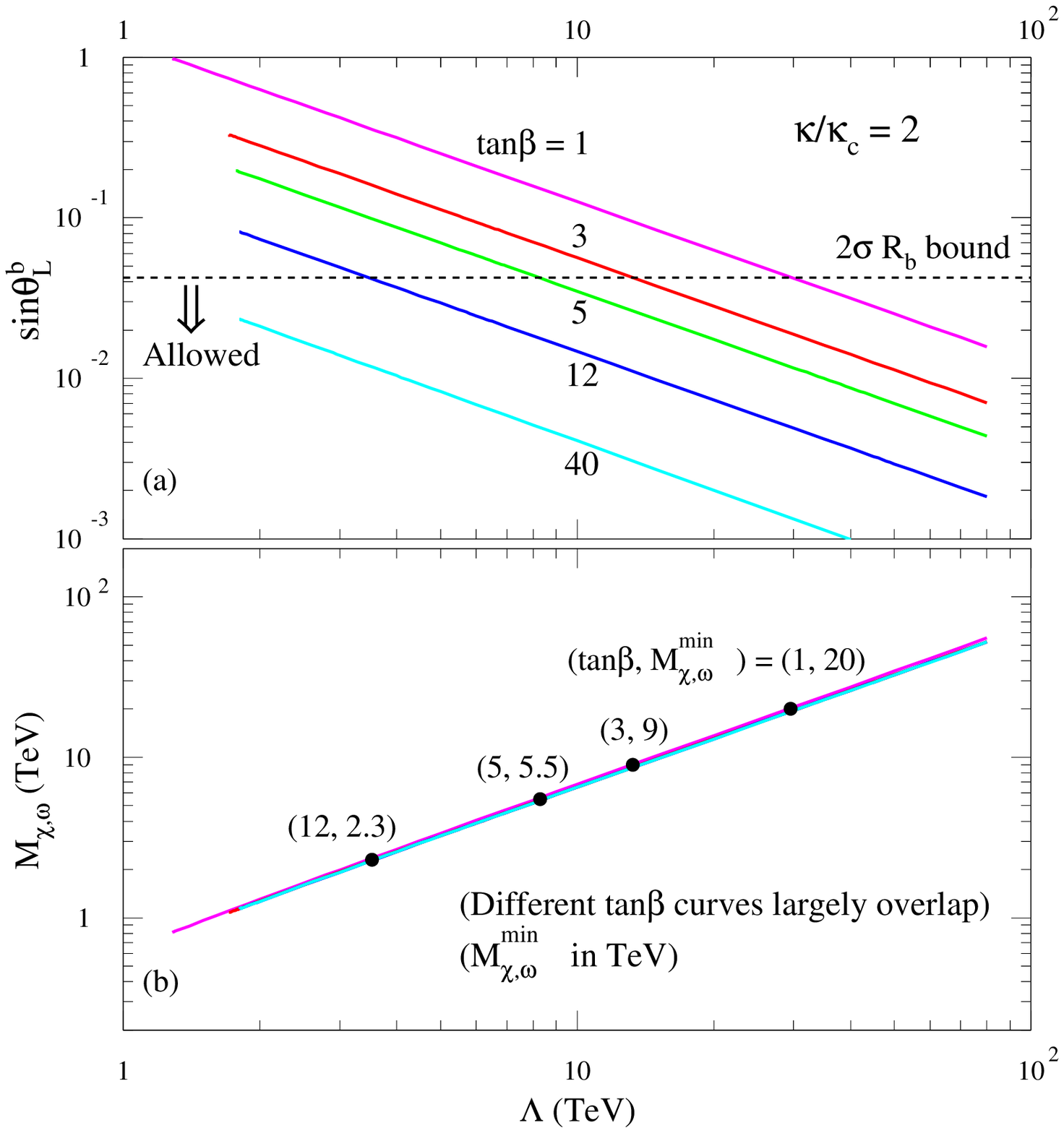}
\vspace{1 cm}
\caption[]{\small \addtolength{\baselineskip}{-.4\baselineskip}
The $R_b$ limits are shown for
the $b$ seesaw angle $s_L^b=\sin\theta_L^b$ in plot (a) and
for the mass $M_\omega(\simeq M_\chi)$ in plot (b).
Here, we choose  $\kappa / \kappa_c = 2$ and a wide range of
$\tan\beta$ values.  
}
\label{fig:tbssZbb}
\end{figure}

The inclusion of a bottom seesaw generates additional $b$-$\omega $ mixing
($\omega\equiv b^H$ is the seesaw partner of $b$)
which makes nontrivial contributions to the $S$ and $T$ parameters
and also to the $Z\rightarrow \bar{b}b$ vertex. 
Furthermore, the composite Higgs
sector now contains two doublets and thus
provides additional corrections to the precision observables.
The $b-\omega$ mixing induces a positive shift in
the left-handed $Z$-$b$-$\bar{b}$ coupling,
\beq
\delta g_L^b =  +\frac{e}{2\sin\!\theta_W\cos\!\theta_W}
                      (\sin\phi_b)^2  \,,
\eeq
which results in a decrease of 
$R_b=\Gamma[Z\to b\bar{b}]/\Gamma[Z\to {\rm hadrons}]$,
i.e., $R_b \simeq R_b^{\rm SM} - 0.39 (s^b_L)^2$, as 
obtained 
in Ref.\,\cite{Collins:1999rz,He:2001fz}. 
This puts an upper bound on the $b$-seesaw angle,  
\beq
\sin\phi_b~\simeq~ \frac{m_b/\mu_{\omega}}{\sqrt{1+r_b}} 
     ~\simeq~ \frac{m_b}{M_\omega\sqrt{r_b}} \,,
\eeq
and correspondingly a lower bound on the mass 
$M_\omega$ ($\simeq M_\chi$), as summarized  
in Fig.\,\ref{fig:tbssZbb}.
The $R_b$ bound will mainly constrain
the low $\tan\beta$ region of
the effective composite two doublet model. 

Variant models \cite{He:1999vp,Popovic:2001cj} in which vectorial
weak-doublet partners exist for both top and bottom give similar
contributions to $T$ as the models with a weak-singlet partner for
top, while the contributions to $R_b$ are suppressed by a small mixing
angle.  The lower bound on these exotic quarks is, then, of order a
few TeV, assuming they are degenerate.  In principle, one could
experimentally distinguish between the models with weak-singlet and
weak-doublet mixings by measuring $A^t_{LR}$ at an NLC
\cite{Popovic:2001cj}.  The predicted shifts relative to the SM value
would be of similar size but opposite sign, as the $Zt_L t_L$ coupling
is altered in the weak-singlet models while the $Z t_R t_R$ coupling
is altered in the weak-doublet models.

As discussed in \cite{Popovic:2000dx}, it is possible to use existing
Tevatron data to set limits on direct production of the $\chi$-like states.  
New,
mostly-singlet, quarks decaying via mixing to an ordinary quark plus a W boson
would contribute to the dilepton events used by the CDF \cite{Abe:1998iz} and
D0 \cite{Abachi:1997re} experiments to measure the top quark production
cross-section.  Since the weak-singlet quarks are color triplets, they would
be produced with the same cross-section as sequential quarks of identical
mass.  However the weak-singlet quarks can decay via neutral-currents (e.g.
$d^H \to Z d^L$) as well as charged-currents (e.g.  $d^H \to W u^L$), and
this lowers the branching fraction of the produced quarks to the final states
to which the search is sensitive.  In fact, the decay width $b^H \to c^L W$
is so strongly suppressed both by Cabbibo factors and the large rate of $b^H
\to b^L Z$, that the Tevatron data do not provide a lower bound on $M^H_b$.
For models in which all quarks have weak-singlet partners, the limits
\begin{equation}
M^H_{d,s}  > \left\{\ {153\ {\rm GeV}\ \ \  {\rm CDF}} \atop 
            {143\ {\rm GeV}\ \ \  {\rm D0}} \right\} 
\end{equation}
 
Similarly, mass limits on new mostly-singlet leptons (for flavor-universal
mixing models) can be extracted from the results of LEP II searches for new
sequential lepton doublets.  In the relevant searches, the new neutral lepton
$N$ is assumed to be heavier than its charged partner $L$ and $L$ is assumed
to decay only via charged-current mixing with a Standard Model lepton (i.e.
$BR(L \to \nu_\ell W^*) = 1.0$).  The OPAL \cite{Ackerstaff:1998cz} and
DELPHI \cite{Abreu:1998jw} experiments have each set a 95\% c.l. lower bound
of order 80 GeV on the mass of a sequential charged lepton.  As detailed in
\cite{Popovic:2000dx}, when one adjusts for the increased production rate and
decreased charged-current branching fraction of the mostly-singlet $\ell^H$
states, the resulting limit is
\begin{equation}
M^H_\ell \geq 84.9 {\rm GeV}\ .
\end{equation}
is slightly stronger.

Finally, CDF limits \cite{Affolder:1999bs} on heavy $b^H$ quarks pair-produced
through QCD processes and decaying via neutral currents can be applied to
the weak-singlet fermions so long as the value of $B(b^H \to b^L Z^0)$ is
included.  CDF finds 95\% c.l. upper limits on the product of cross-section
and squared branching fraction as shown in figure \ref{fig:cdf-nc-b}.  For
$B(b^H \to b^L Z^0) = 100\%$, heavy quarks with masses between 100 and 199
GeV are excluded.  But if the neutral-current branching fraction decreases for
heavier $b'$ quarks, as in the models discussed in \cite{Popovic:2000dx},
$b'$ quarks as light as 160 GeV may still be allowed by the data.

\begin{figure}[tb]
\vspace{5.0cm}
\includegraphics{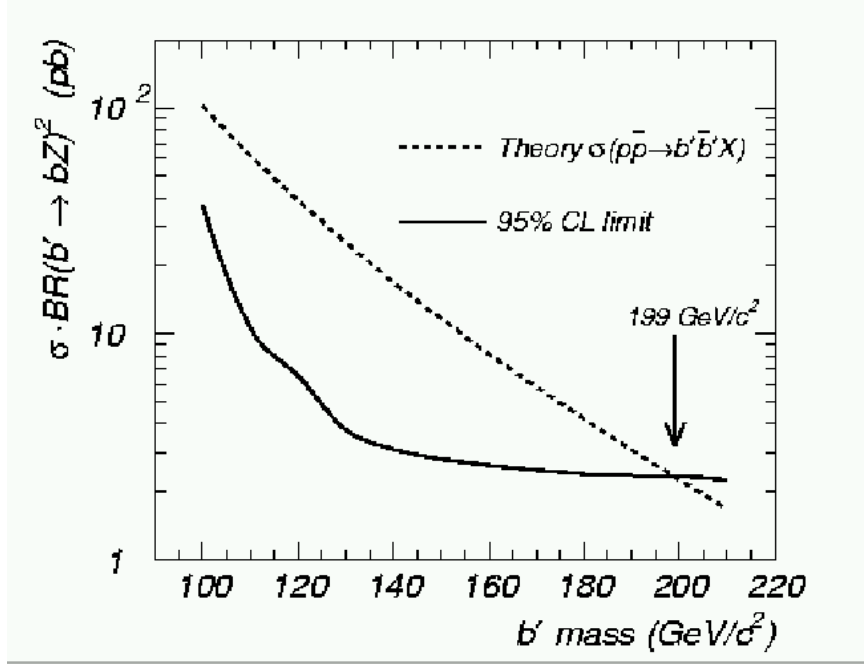}
\vspace{2cm}
\caption{\small \addtolength{\baselineskip}{-.4\baselineskip}
The 95\% c.l. upper limit on $p\bar{p} \to b' \bar{b}' X$ production
  cross section times the $b' \to b Z$ branching ratio squared (solid).  The
  dashed curve shows the value predicted in a theory where the branching
  ratio is 100\% 
  \protect\cite{Affolder:1999bs}.}
\label{fig:cdf-nc-b}
\end{figure}

Ultimately, at higher-energy hadron colliders, direct pair-production of
the $\chi$ states and their subsequent decay via $\chi\rightarrow
ht\rightarrow t\bar{t}t$ could yield \cite{He:2001fz}, a spectacular
$6t$ final state.  As shown in Fig.~{\ref{fig:timt}}, the cross
section for this process with $m_\chi=1$~TeV would be $\sim 10$~pb.

\begin{figure}[tb] %
\vspace{7.0cm}
\includegraphics{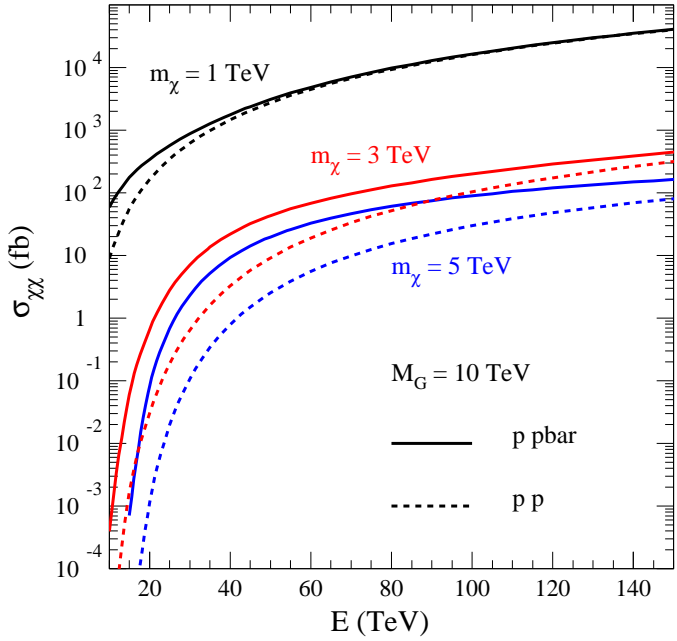}
\vspace{10pt}
\caption{Cross-section for pair production of $\chi$ states 
at high energy hadron colliders, leading to 6-top events
\protect\cite{Barklow:2002su}. }
\label{fig:timt}
\end{figure}


\subsubsection{Flavorons}

As an extension of the Top Seesaw mechanism, models with flavor
universal dynamical symmetry breaking have been proposed
\cite{Burdman:1998vw,King:1992qb}.  In these models, the dynamics are driven by family or
large flavor gauge symmetries; when these symmetries break, multiplets of
heavy ``flavoron'' bosons remain in the specrum.  Their masses, like the
symmetry breaking scale, are expected to be of order a few TeV.  The
flavorons' couplings to fermions must be strong enough to generate
condensates that break the electroweak symmetry and provide the ordinary
fermions with masses.  As a result, the flavorons tend to be readily produced
at hadron colliders like the Tevatron -- but also to have large decay widths
that make their detection in dijet invariant mass spectra a challenge
\cite{Burdman:1998vw}.  

For the moment, some of the best limits on flavor gauge bosons come
from precision electroweak data \cite{Burdman:1999us}.  Flavorons
coupling to ordinary fermions can give rise to direct corrections to
$Zff$ vertices; generally speaking, if the $Zff$ coupling is $g_f$,
the correction is \cite{Hill:1995di}
\begin{equation}
\Delta g_f = g_f \frac{G \kappa_F}{6\pi} \frac{M_Z^2}{M_F^2} \ln
\left(\frac{M_F^2}{M_Z^2} \right)
\end{equation}
where $G$, $\kappa \equiv g_F^2/4\pi$, and $M_F$ are the group theory
factor, coupling, and mass appropriate to the particular flavouron.  Flavoron
exchange across $t$ and $b$ quark loops will generally contribute to the $T$
parameter an amount (in leading log)
\begin{equation}
T = N_c (G^\prime_{LL} + 2 G^\prime_{RR}) 
\frac{\kappa m_t^4}{32\pi^2 \sin^2\theta_W
  \cos^2\theta_W M_Z^2 M_F^2} \left[ \ln (M_F^2 / m_t^2) \right]^2
\end{equation}
where $G^\prime_{LL,RR}$ are group theory factors for flavoron couplings to
  left-handed and right-handed quarks.
  In addition, in models in which some generator of the flavor
group mixes with hypercharge or the diagonal generator of $SU(2)_W$, the
associated $Z - Z'$ mixing also contributes to $T$.

For each of the three representative flavoron models mentioned here, we
assume that flavoron exchange at low energies may be treated in the NJL
approximation (with four-fermion coupling $4\pi\kappa/2\!M_F^2$).  The critical coupling required for chiral
symmetry breaking is 
\begin{equation}
\kappa_{crit} = \frac{2 N \pi}{N^2-1}.
\end{equation}
The minimal model
\cite{Burdman:1998vw,King:1992qb,Georgi:1990ej,Georgi:1994at} has a gauged
SU(3) family symmetry acting on the left-handed quark doublets. This family
symmetry group does not mix with the Standard Model gauge groups and the
model leaves the GIM mechanism intact.  The critical coupling is
$\kappa_{crit} = 2.36$.  Models in which the $SU(9)$ symmetry of color and
family multiplicity of the left-handed quarks is gauged
\cite{Burdman:1998vw,Randall:1993vt} also preserve GIM.  A proto-color group
acting on the right-handed quarks is also present; obtaining the correct
value for $\alpha_s$ in its presence requires
\begin{equation}
\kappa \geq 3 \alpha_s(2 \rm{TeV}) \approx 0.3
\label{eqn:ehs-kappa}
\end{equation}
Ignoring mixing between the flavor and proto-color group generators,
the critical coupling for chiral symmetry breaking is $\kappa_{crit} =
0.71$.  If the full $SU(12)$ flavor symmetry of all the left-handed
quark and lepton doublets is gauged
\cite{Burdman:1998vw,Randall:1993vt}, both a proto-color group and a
proto-hypercharge group must be included for the right-handed
fermions, and the constraint (\ref{eqn:ehs-kappa})\ still applies.  In
this case, $\kappa_{crit} = 0.53$ -- not far above the lower limit.
The 95\% c.l. lower bound on $M_F$ from precision electroweak data for
$\kappa$ at its critical value is approximately 2 TeV for all three
models \cite{Burdman:1999us}.

Comparable limits on $M_F$ (for critical coupling $\kappa$) in the
$SU(3)$ and $SU(9)$ models have been obtained from Tevatron Run I data
\cite{Burdman:2000yq}.  Run II should be sensitive to $M_F$ up to 
2.5 - 3 TeV with 2 $fb^{-1}$ of data.  For the $SU(3)$ model, both
dijet and anomalous single top production are likely to yield large
signals; in the SU(9) model, the dijet signal is relatively enhanced
and the single top signal, relatively reduced in size, enabling the
two models to be distinguished if both channels are studied.

A significantly stronger bound on the flavorons of the $SU(12)$ model
derives from atomic parity violation: $M_F \gae 10$ TeV \cite{Burdman:2000yq}.
Flavorons from the $SU(12)$ model light enough to produce visible
signals at the Tevatron are already excluded by this limit.

\subsection{ Extra Dimensions at the TeV Scale}


Recently there has been considerable interest in theories of extra
dimensions of space which emerge not far from the weak scale.  
We cannot give a comprehensive review of this burgeoning field, 
but we will
outline some key ideas relating to electroweak symmetry
breaking and the physics of new strong dynamics.  The ideas largely
stem from the view that string theories admit an arbitrary hierarchy
between the compactification scale of the extra dimensions, and the
fundamental string scale. Moreover, with gravity in a higher
dimensional bulk the ``true'' Planck scale can emerge as a $\sim 100$
TeV scale, but is subject to rapid power-law renormalization; grand
unification remains viable in these theories, and one is led to a long
list of novel phenomena.

Some of the key early works are
those of Antoniadis, \etal and Lykken (weak scale superstrings
\cite{Antoniadis:1990ew,Antoniadis:1997hk},\cite{Lykken:1996fj});\footnote{
Possibly the earliest proposal for ``weak scale extra dimensions''  
was an heroic effort by Darrell R. Jackson to identify the
$W$ and $Z$ as Kaluza-Klein modes, (private communication to CTH, 
Caltech, {\em ca.} 1975.)
} Arkani-Hamed, Dimopoulos,
and Dvali, (millimeter scale gravity 
\cite{Arkani-Hamed:1998rs,Antoniadis:1998ig});
Dienes, Dudas and Gherghetta, (gauge fields
in the bulk and power-law unification \cite{Dienes:1998vh,Dienes:1998vg});
Randall, Sundrum (DeSitter space in the
bulk and natural hierarchy \cite{Randall:1999ee,Randall:1999vf}).
It is also essential that chiral fermions
emerge in the compactified theory, and this is possible
by implementing orbifold compactification {\em ala} Horava and Witten,
\cite{Horava:1996ma}, 
or domain walls {\em ala}  Arkani-Hamed and Schmaltz
\cite{Arkani-Hamed:1999dc}.
In the latter case 
dynamical mechanisms arise for the origin
of generational hierarchies, and the CKM matrix. 
\cite{Nussinov:2001ps,Nussinov:2001rb}  Many phenomenological constraints
have been placed upon the scales of the new extra
dimension(s), and many variations
and developments of this theme now
exist, too numerous to review here.
A compactification
scale lower limit of order  $\sim 1$ TeV seems to be
indicated phenomenologically.

Why not consider extra space-time dimensions at the weak scale?
After all, supersymmetry is an extra-dimensional theory in
which the extra dimensions are fermionic, or
Grassmanian, variables.
Supersymmetry leads
naturally, upon ``integrating out'' the extra fermionic dimensions
(i.e., descending from a superspace action to a space-time action),
to a perturbative extension of the Standard Model, e.g., the MSSM.  In 
such a scheme
the Higgs sector is at least a two-doublet model,
and the lightest Higgs boson is expected to be in a range determined
by the {\em perturbative} electroweak constraints, $\lta 140$ GeV.
>From a ``bottom-up'' perspective the key lesson
from Supersymmetry is that an  {\em organizing principle} for
physics beyond the Standard Model emerges
from hidden extra dimensions
which are then integrated out.
Upon specifying the algebraic properties of the extra dimensions
one is led to a particular symmetry
structure and class of dynamics for EWSB.   The particular extra dimensions in which we will be interested are ordinary c-numbers, 
but one can contemplate other possibilities 
\cite{Alishahiha:2001nb,Dai:2001bx,Adams:2001ne}.

Extra dimensions do not show up as new ``real estate,'' but rather
they appear in accelerator experiments as new particles \footnote{For
a discussion of one possibility for extra dimensions of time, see
\cite{Nambu:1973qe}.}. These are the excited modes of existing
fields, e.g., quarks, leptons, gauge bosons, the graviton, that
propagate in the compact extra dimensions, and are known as
Kaluza-Klein (KK) excitations.  The KK modes typically appear as a
ladder spectrum of discrete resonances with the same quantum numbers
as the zero-mode field itself.  For example, if we consider the
possibility that QCD gauge fields (gluons) can propagate in the bulk
then the KK modes show up as a sequence of massive colorons.

The idea of extra c-number dimensions near the weak scale
has led to the proposal of dynamical schemes
that can imitate the strong dynamics we have discussed
in this review.  Dobrescu first observed that
the main ingredients of Topcolor, i.e., the colorons,
appear naturally as KK modes of the gluon
\cite{Dobrescu:1998dg}
and a number of detailed models
have been discussed 
\cite{Cheng:1999rq,Cheng:1999bg,Cheng:1999fu,Arkani-Hamed:2000hv}. 
These essentially follow the idea of top condensation
as discussed in Section 4. The other key ingredient of the Top Seesaw
model, the seesaw partner $\chi$ quarks, can emerge
as KK modes as well.

\subsubsection{Deconstruction} 
 
As KK-modes begin to appear in accelerator
experiments, we can ask: ``how are they to be described
phenomenologically, as if we have no
knowledge of the extra dimensions
themselves?'' or, equivalently, ``What is the
effective low energy Lagrangian of the KK modes?''  
This has led to a new twist in the development
of these models known as ``deconstruction.''
Independently, the ``Harvard'' group of Arkani-Hamed,
Cohen and Georgi, 
\cite{Arkani-Hamed:2001ca,Arkani-Hamed:2001nc,Arkani-Hamed:2001vr}
and  the ``Fermilab'' group Hill, Pokorski, Wang
and Cheng \cite{Hill:2000mu,Cheng:2001vd,Cheng:2001nh},
have proposed using a lattice to describe the extra dimensions.
The $3+1$ dimensions of space-time are continuous, while
the extra compact dimensions are latticized (this is known as
a ``transverse lattice'' \cite{Bardeen:1980xx}).
By using
the lattice technique,  one can
 ``integrate out'' the extra dimensions,
preserving local gauge invariance, 
and arrive at a manifestly gauge invariant 
effective Lagrangian including
Kaluza-Klein modes (in a sense the KK modes are analogues of superpartners).

To get a flavor for deconstruction, let us consider the first KK-mode
excitation of the gluon in a compactified theory of $D=4+1$
dimensions. We can write its effective Lagrangian knowing only the
$SU(3)$ symmetry of QCD.  The lowest KK mode appears as a ``coloron,''
a heavy linear octet representation of the gauge group $SU(3)$; its
mass is the only parameter determined by compactification.  A heavy
vector meson which forms a multiplet under a (local or global)
symmetry group $G$, can always be 
described 
as a gauge field \cite{Bando:1985rf}, with a
spontaneously broken gauge group that is an identical copy of $G$. 
 For the single coloron case we have $G =
SU(3)$.  Hence, we may take our theory of a gluon plus its first
excitation to be based upon the gauge group $G' =SU(3)_1\times
SU(3)_2$ and break this to $G=SU(3)$ with a Higgs field $\Phi$
transforming as a $(3,\bar{3})$.
  
By constructing a Higgs potential: 
\be
V(\Phi) =-M^2\Tr(\Phi^2) + \lambda \Tr(\Phi^4) + \lambda \Tr(\Phi^2)^2
+ M'\det(\Phi) 
\ee
we can arrange for $\Phi $ to develop
a vacuum expectation value $V$, by which
all unwanted Higgs degrees of freedom
are elevated to the large
mass scale $\propto M$.  Note that in trying to discard
the extra $10$ degrees of freedom in $\Phi$ as a Higgs field,
we encounter a unitarity bound 
\cite{Lee:1977eg,Chivukula:2002ej,SehkarChivukula:2001hz} 
which predicts the string scale
from $M$.

The Lagrangian containing the gluon
and the first KK mode is therefore:
\be
{\cal{L}} = 
-\frac{1}{4} F^A_{1\mu\nu}F_1^{A\mu\nu}
-\frac{1}{4} F^A_{2\mu\nu}F_2^{A\mu\nu} + |D_\mu\Phi|^2 -V(\Phi).
\ee
The field $\Phi$, after developing a VEV, and decoupling
the Higgs degrees of freedom, can
be parameterized by a chiral field:
\be
\Phi \rightarrow V\exp(i\phi^AT^A/\sqrt{2}V)
\ee
This is exactly the structure that would emerge from a Wilson lattice
with two transverse lattice points (which are now ``branes, each
carrying a $3+1$ theory).  We also see that this is identical to the
gauge sector of Topcolor models.  With the breaking in place,
the theory has two mass eigenstates, the gluon and coloron:
\be
G_\mu^A = \cos\theta A_{1\mu}^A + \sin\theta A_{2\mu}^A;\qquad\qquad
B_\mu^A = -\sin\theta A_{1\mu}^A + \cos\theta A_{2\mu}^A;
\ee
where $A_{i\mu}$ belongs to $SU(3)_i$.  If we allow
$SU(3)_i$ to have a coupling constant $g_i$ then we
find:
\be
\tan\theta = \frac{g_1}{g_2}
\ee
The mass of the coloron is given by 
\be
M_K = V\sqrt{g_1^2 + g_2^2} = \frac{1}{R}
\ee
where $R$ is identified with the radius of the compactified extra
dimension. 
We also see that the low energy QCD-coupling is related
to $g_i$ as:
\be
\frac{1}{g_{QCD}^2} = \frac{1}{g^2_1} + \frac{1}{g^2_2}
\ee
With $g_1=g_2$ we have $g_1^2=2g_{QCD}^2$
and this is the onset of ``power-law
running'' of the coupling constant. 
 The model indeed represents a single KK mode, descending
from higher dimensions.  Conversely, observation of a coloron in the
particle spectrum could be taken to suggest that a new extra dimension
is opening up.

\begin{figure}[t]
\vspace{6cm}
\includegraphics{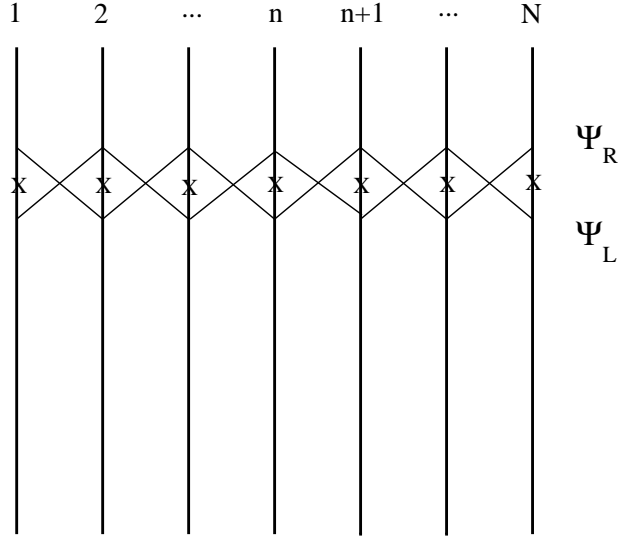}
\vspace{1cm}
\caption[]{\small \addtolength{\baselineskip}{-.4\baselineskip}
Dirac fermion corresponding to constant $\phi$
has both chiral modes on all branes. The $\times$'s denote the
$\phi$ couplings on each brane, and the links are the 
latticized fermion kinetic terms which become Wilson
links when gauge fields are present. (from \cite{HillHeTait:2000xx})
The spectrum here has a singlet lowest massive
mode, and doubled KK modes; by adding a 
Wilson term one can remove one of the
two cross-bars between adjacent branes, and elminate 
second Brillouin zone doubling
in the spectrum.}
\label{dirac1}
\end{figure}

\begin{figure}[bt]
\vspace{6cm}
\includegraphics{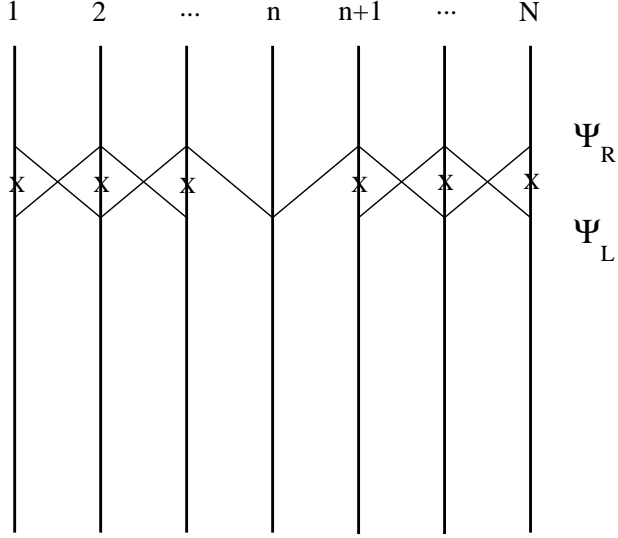}
\vspace{1cm}
\caption[]{\small \addtolength{\baselineskip}{-.4\baselineskip}
A chiral fermion occurs on brane $n$ where 
$\phi(x^5)$ swings rapidly through zero. The chiral fermion has
kinetic term (Wilson links) connecting to adjoining branes. 
(from \cite{HillHeTait:2000xx})}
\label{chiral1}
\end{figure}

\begin{figure}[t]
\vspace{6cm}
\includegraphics{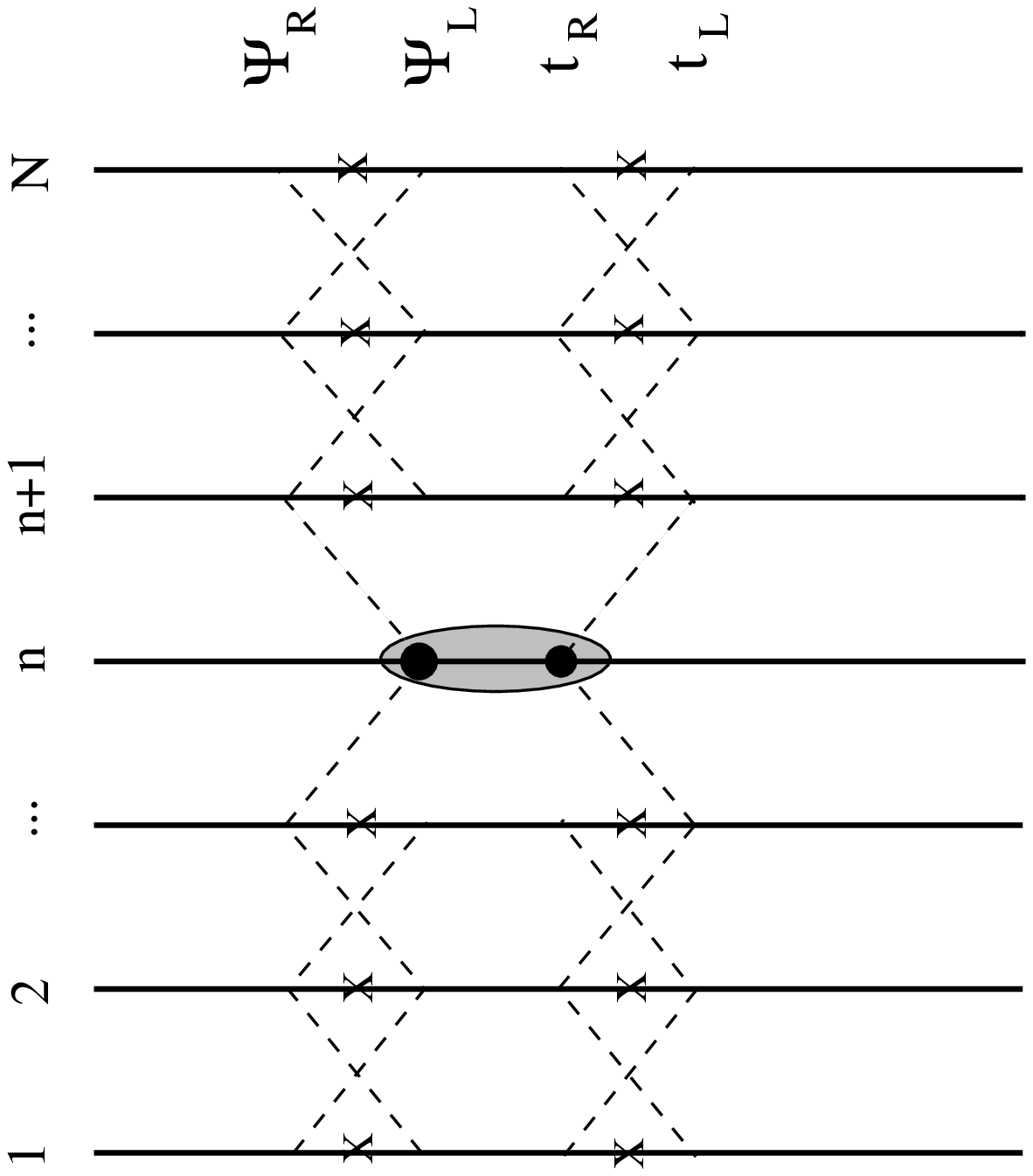}
\vspace{1cm}
\caption[]{\small \addtolength{\baselineskip}{-.4\baselineskip}
Pure top quark condensation by Topcolor is obtained in the limit
of critical coupling on brane $n$ and
decoupling to the nearest neighbors.  Decoupling corresponds to taking the
compactification mass scale large; the links are then denoted by dashed lines.
(from \cite{HillHeTait:2000xx})}
\label{critferm1}
\end{figure}

The full lattice description of QCD
is a straightforward generalization
of this model \cite{Hill:2000mu}. 
If we consider the full lattice gauge theory in $4+1$ dimensions
with a Wilson action for the extra dimension, we obtain
the Lagrangian:

\begin{figure}[t]
\vspace{6cm}
\includegraphics{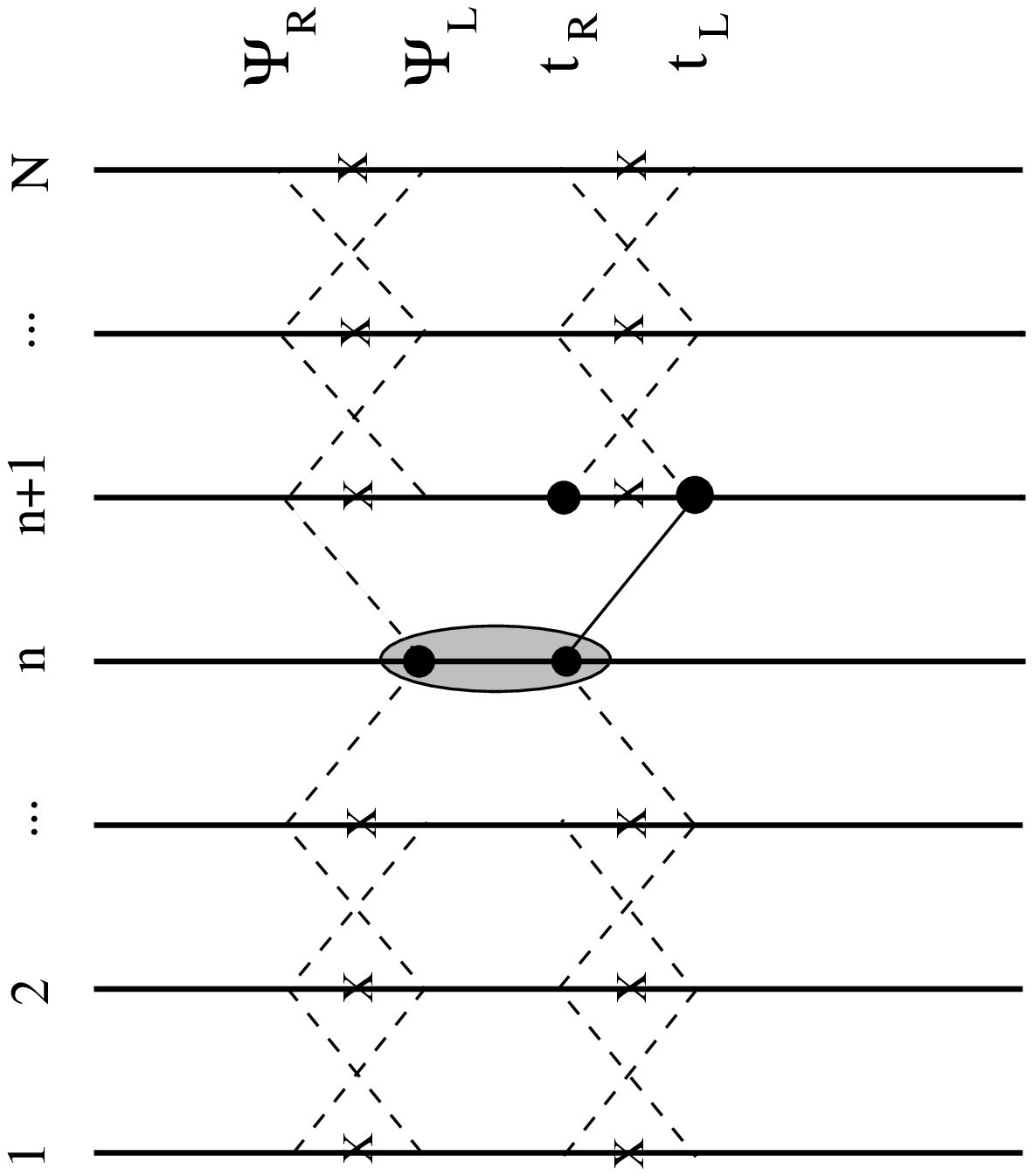}
\vspace{1cm}
\caption[]{\small \addtolength{\baselineskip}{-.4\baselineskip}
Top Seesaw Model arises when the effects of nearest
neighbor vectorlike fermions are retained, i.e., when these
heavier states are only partially decoupled. Keeping more
links maintains the seesaw. Usually we denote $t_{Rn}\sim \chi_R$,
 $t_{Ln+1}\sim \chi_L$, $t_{Rn+1}\sim t_R$.   
(from \cite{HillHeTait:2000xx})}
\label{critferm2}
\end{figure}

\be
{\cal{L}} = 
-\frac{1}{4} \sum_{j=1}^{N} F^A_{j\mu\nu}F_j^{A\mu\nu}
+ \sum_{k=1}^{N-1}\Tr (D_\mu\Phi_k)^2 
\ee
Here there are $N-1$ $\Phi $ fields, and the $kth$ field
transforms as an $(3_k, \bar{3}_{k+1})$ representation, straddling the
nearest neighbor $SU(3)_k$ and $SU(3)_{k+1}$ gauge groups.
$\Phi_k$ is a unitary matrix,
and represents the
Wilson line:
\be
\Phi_k = V\exp \left(ig \int_{x_k}^{x_{k+1}} dx^5 A_5^aT^a   \right)
\ee
$\Phi_k$ may be parameterized by:
\be
\Phi_k = V\exp(i\phi_k^aT^a/\sqrt{2}V)
\ee
where $T^a=\lambda^a/2$.  This clearly matches
our previous discussion of a single KK mode.

The $\Phi_k$ kinetic terms imply a gauge field mass term of the form:
\be
\half g^2V^2 \sum_{k=1}^{N-1}(A^a_{k\mu} - A^a_{(k+1)\mu})^2
\ee
Performing the mass spectrum analysis of the model is equivalent
to analyzing the phonon spectrum of a one-dimensional 
crystal lattice.  The mass term is
readily diagonalized and
we obtain eigenvalues reflecting
a ladder of states with:
\be
M_n = \sqrt{2}gV \sin(\pi n/2N)
\ee
the lowest energy masses given by:
\be
M_n \approx \pi g Vn/N\sqrt{2} \equiv \frac{1}{R}
\ee
This is the spectrum assuming free boundary conditions
on the lattice and corresponds to orbifolding.  With
periodic boundary conditions instead, one gets the $A_5^a$
as low energy modes as well, which act like pseudo-scalars,
and the KK mode levels are doubled. 

The net result is that the effective Lagrangian for, e.g., QCD
propagating in the extra dimensional bulk, is a chain of gauge groups
of the form: $SU(3)\times SU(3)\times SU(3)\times ...$, one gauge
group per lattice brane.  The high energy
coupling constant common to each group
is $g \sim \sqrt{N}g_{QCD}$,
consistent with the power-law running. 
>From a ``bottom-up'' perspective, one writes
down this gauge structure with arbitrary relevant operators with
arbitrary coefficients to describe the most general low energy
effective theory for extra-dimensional physics.

Various additional dynamical issues in
deconstructed theories are currently under 
investigation, such as chiral dynamics \cite{Hashimoto:2000uk,Gusynin:2002cu},
electroweak observables \cite{Muck:2001yv,Kim:2001gk}, unification
\cite{Arkani-Hamed:2001vr,Chankowski:2001hz} and GUT breaking
\cite{Cheng:2001qp}.
A number of authors have applied deconstruction to 
supersymmetric schemes and SUSY breaking 
 \cite{Csaki:2001em,Cheng:2001an} \cite{Csaki:2001qm,Kobayashi:2001fr}. 
There are interesting and novel topological issues in the
presence of many gauge factor groups, e.g., 
\cite{Hill:2001bt,Csaki:2001zx}. 
Duality and supersymmetry 
can have an intriguing interplay \cite{Arkani-Hamed:2001ie} in these
schemes via topology.
Deconstructing gravity is in need of
elaboration, but Bander has made
an interesting first attempt in the language of vierbeins 
\cite{Bander:2001qk}.

Fermions can be accomodated, and chirality
can be engineered.
Chiral fermions can
be localized in the fifth dimension by background fields
\cite{Arkani-Hamed:1999dc,Skiba:2002nx,Kaplan:2001ga}.   A free fermion has
the continuum and latticized
action (we neglect the gauge interactions here,
and do not include Wilson terms):
\beq
\int d^5x \; \bar{\Psi}(i\slash{\partial}
-\partial_5\gamma^5 - \phi(x^5) )\Psi
\rightarrow
\sum_{n=1}^N\int d^4x \; \bar{\Psi}_n(i\slash{\partial}
 - \phi_n )\Psi_n + 
 V\bar{\Psi}_n \gamma_5 \Psi_{n+1} + h.c.
 \label{latferm}
\eeq

If the background  field
is approximately constant
both chiral components of the fermion
appear on each lattice brane,
as depicted in Fig.(\ref{dirac1}). 
The cross bars are the latticized  links
allowing hopping of the fermion from $n$ to $n+1$
in eq.(\ref{latferm})
If $\phi(x^5)$
swings through zero rapidly in the vicinity of brane $n$,
i.e., $\phi_n=0$ then only
a single chiral component is normalizeable
in the vicinity of $n$ and one
gets a dislocation in the lattice 
as shown in Fig.(\ref{chiral1}).

The coupling strength of 
$SU(3)_n$ on the $n$-th brane
will generally be renormalized
by the dislocation and can become  supercritical,
Fig.(\ref{critferm1}).
It would, therefore,  not be coincidental to
expect this to happen; indeed a variety of
effects are expected to occur near the
dislocation, e.g., the chiral fermions themselves
can feed-back onto the gauge fields to produce such renormalization
effects. 
The result is a chiral condensate on brane $n$ forming between
chiral fermions. Identify $\Psi=(t,b)_L$  and $t_R$ as
the chiral zero-modes on brane $n$ of two independent
Dirac fields in the bulk. 
In the limit that we take the compact extra dimension very small,
the nearest neighbor links decouple at low energies. In this limit
we recover a Topcolor model with pure top quark condensation. 

In Fig.(\ref{critferm2}) we illustrate the case that some of the 
links to nearest neighbors are not completely decoupled.
Again, this can arise from renormalizations due to background fields, or
to warping \cite{Cheng:2001vd,Cheng:2001nh}. 
Thus the mixing with heavy vectorlike
fermions occurs in addition to the chiral dynamics
on brane $n$. In this limit we obtain the
Top Quark Seesaw Model.

\subsubsection{Little Higgs Theories}

Recently another approach to electroweak symmetry breaking
has been revived,  
inspired by 
deconstruction, in which the Higgs itself is a 
PNGB, 
\cite{Arkani-Hamed:2001nc}.   This is actually an
old idea, due to Georgi, Kaplan, and others (see, e.g.
\cite{Kaplan:1984fs}\cite{Georgi:1984ef}\cite{Kaplan:1984sm}).  
The novelty presently
is that delocalization in an extra dimension
(or the minimal deconstructed description of such)
leads to a potentially softer scheme in which
dangerous gauge loop quadratic divergences
contributing to the Higgs boson (mass)$^2$ 
can be cancelled in one loop order.  The resulting
models bear some resemblance to the MSSM Higgs
structure.

There are now a number of schemes, most notably the
``minimal moose model,'' \cite{Arkani-Hamed:2002qx}
and $SU(5)/SO(5)$ model
\cite{Arkani-Hamed:2002qy,Arkani-Hamed:2002pa}, and the $SU(6)/SO(6)$
scheme \cite{Low:2002ws}.
We briefly describe the ``minimal moose'' scheme. 

When we periodically compactify a higher dimension we
find that there is, in the absence of loops, a massless mode
corresponding to the zero-mode of $A_5^a$.  For QCD this would appear
in the effective lagrangian as a color octet of pseudoscalars.  
However, since these modes carry color, they will acquire
mass $\sim gV$.  Arkani-Hamed, Cohen and Georgi show how this leads to
a Coleman-Weinberg potential for the modes, and is generally finite.
The authors then construct a scheme \cite{Arkani-Hamed:2001nc} in
which this mode can arise as an electroweak isodoublet.

The idea, from a lattice point of view, is to latticize two
compact dimensions with toroidal boundary
conditions ($T_2$, the surface of a doughnut).  The minimal
plaquette action is then a generalized Eguchi-Kawai model
\cite{Eguchi:1982nm}, but with
the additional links associated with the periodicity.  At three of the
four sites of the plaquette one attaches $SU(3)$ gauge groups.  On the
fourth cite one attaches $SU(2)\times U(1)$.  The Wilson links
(linking Higgs) which attach the $SU(2)\times U(1)$ gauge group to
nearest neighbor $SU(3)$'s then contain components which transform as
$I=\half$ Higgs fields. The Coleman-Weinberg potential produces an
unstable vacuum. The theory leads to an effective two-Higgs doublet
model. The gauge couplings would
be expected to raise the mass  of the lightest Higgs
to order $\sim g_2^2\Lambda^2$, but the discrete symmetries
of the model, associated with the delocalized structure,
lead to a GIM-like cancellation of this contribution.
The dangerous term does occur at two-loop level
$\sim g_2^4\Lambda^2$. With $\Lambda\sim 10$ TeV
the scale of the ``compactification dynamics''  
 the lightest $0^+$ Higgs can be of low mass,
 $\sim 120$ GeV.  Hence a ``little'' hierarchy,
 $10$ TeV $\lta \mu\lta$ $v_{weak}$ is
 protected in this scheme.  

These models are receiving fuller elaboration at present.
For a critical discussion 
of this scheme see K. Lane \cite{Lane:2002pe}.
Two recent works attack several aspects of the models
from the perspective of consistency with electroweak 
radiative corrections
\cite{Csaki:2002qg}. The models contain
ingredients, such as $I=0$ PNGB's, that develop tadpole
VEV's and  may violate $T$ cosntraints. Also, the
models such as $SU(5)/SO(5)$ employ a top seesaw to
give to the top quark mass, and the
extended fermion structure can be
problematic in loops \cite{Hewett:2002px}.
Some recent papers examine collider
phenomenology of these schemes \cite{Burdman:2002ns,Han:2003wu}.

Much remains to be
done with Little Higgs schemes, in particular
in establishing their phenomenological naturalness.

\newpage

\section{Outlook and Conclusions}

The next decade will bring great discoveries in the field of
elementary particle physics.  The Tevatron Run-II program and the LHC
will begin to reveal the mechanism(s) responsible for the origin of mass.
The Large Hadron Collider in concert with
a Linear Collider can begin
detailed studies of the fields involved in electroweak symmetry
breaking.  In the longer term, a Very Large Hadron Collider will have
the power to probe the corollary mechanisms that underlie the physics of
flavor. 

This review has focused on the possibility that the origin of mass 
involves a new strong dynamics near the TeV
scale.  As we have argued here, dynamical electroweak symmetry
breaking, whether it comes in the guise of Technicolor, Topcolor, or
some variant scheme, can provide a natural explanation for
the weak scale.  It can also provide tantalizing clues about the 
magnitudes and origins of
the fermion masses and mixings, and may be able to explain CP
violation.  Indeed, many 
issues of flavor physics that are relegated to the Planck or Gut scale
in perturbative theories, such as the MSSM, become accessible
to sufficiently energetic accelerators in the dynamical framework.  

Technicolor was created in the era in which the $W$ and $Z$ bosons are heavy
and all known fermions are comparatively light.  
In the first approximation, therefore,
the fermions are massless and Technicolor is essentially a pure QCD-like theory
which naturally generates the weak scale, in analogy to the
way the strong scale is
generated by QCD through its ``running coupling constant.''  
By itself, Technicolor requires a new gauge group and additional
fermions, i.e., techniquarks. 
Technicolor is, however, an incomplete theory.  
Extended Technicolor was introduced to
accomodate light fermion masses, but even the charm quark begins to push
the limits on ETC from rare decay processes.  Hence, one is led
to various schemes to accomodate heavier fermion
masses, such as Walking Technicolor
and Multi-Scale Technicolor.  Combinations
with Supersymmetry and Bosonic Technicolor have been considered.

In the post--1990 era, in addition to the heavy $W$ and $Z$, we
have the very heavy top quark. This leads to an alternate point of departure
for dynamical models of EWSB, based on the idea of Topcolor.
Topcolor by itself is a fine-tuned theory in which
the top quark alone condenses, predicting
$m_t\sim 220$ GeV, and this first--order theory can be ruled out.  
However, in concert
with Technicolor, we arrive at viable models in which a new Techni-dynamics
can coexist and the top quark acquires a dynamical mass through
Topcolor. Such models, 
``Topcolor Assisted Technicolor'' or TC2, predict a rich phenomenology
that may be accessible to the Tevatron, and certainly to the LHC.
In this view, the top quark is playing the role of the first techniquark. 

Topcolor by itself, with a small extension,  
can naturally yield a completely viable theory of dynamical
EWSB: the ``Top Quark Seesaw.'' 
In this scheme the top quark condenses,
through strong Topcolor, but there are additional vectorlike fermions that
mix to give the physical mass of the top quark through a
seesaw.  The Top Quark Seesaw, despite an effective boundstate Higgs boson
with a mass of $\sim 1$  TeV, is in agreement with present oblique radiative
corrections (the $S-T$ plot).  
It explains the $W$ and $Z$ masses, as well as $m_t$ and 
$m_b$, the latter through instantons or a simple extension. 
Moreover, it involves the fairly robust
gauge group imbedding of QCD to $SU(3)\times SU(3)$ 
and additional vectorlike fermions, and is therefore
as economical as Technicolor.  Finally, the $SU(3)\times SU(3)$ structure,
and the new vectorlike fermions, are completely natural ingredients 
expected from extra dimensions of space near the weak scale.  The model
may be extendable to generate all known fermion masses and mixing angles, 
though no complete theory has yet been written down. 

Thus, strong dynamical models of EWSB have evolved and remain
viable and consistent with all known experimental limits.  
Of course, regardless of the theoretical beauty of any given
scenario, its validity will ultimately be ascertained through
experiment.  Should the slight discrepancies in the electroweak
precision tests, e.g., the forward-backward
asymmetries of leptons and hadrons, open into a new
phenomenology, then novel TeV-scale dynamics may be in the
offing.  Should the hadron colliders find a spectrum of new states
with unexpectedly large production cross-sections, or
should the Higgs boson be shown to
be very heavy through exclusion, 
then new strong dynamics will be indicated. 

We have seen that new strong dynamics
can be consistent with the presence of (even weak-scale)
Supersymmetry. Another exciting candidate system of
organizing principles  is the existence of c-number extra
dimensions, which we have found to be naturally associated with
several types of strong dynamics, such as the Top Seesaw
or the Little Higgs models.  
Hence, we will likely uncover new fundamental organizing principles
in our quest to unravel EWSB, whether the dynamics is perturbative
or strongly coupled.

The task of theorists is to prepare to understand whatever 
results emerge from the upcoming experiments
aimed at the electroweak scale.  In this review we have focused upon a
set of ideas complementary to the most popular and well-studied
perturbative MSSM scenarios.  We have seen that, when considered in
equivalent detail, these dynamical theories are no less viable, given
current experimental bounds.  They hint at connections to
deeper ideas about the structure and interplay of space, time
and matter.
They remain, in our opinion, exciting
prospects for discovery in the near future.

\section{Acknowledgements}

We wish to thank the following individuals for
useful input and commentary to this review:  Tom Appelquist,
William Bardeen, Gustavo Burdman,
Sekhar Chivukula, Bogdan Dobrescu, Estia Eichten, Keith Ellis,
Hong-Jian He, Ken Lane, Chris Quigg, and Koichi Yamawaki.

\newpage

\section*{Appendix A: The Standard Model}
       \renewcommand{\theequation}{A.\arabic{equation}}
       \setcounter{equation}{0}
\addcontentsline{toc}{section}{Appendix A: The Standard Model }


\noindent
{\bf A(a) Summary of Standard Model Structure}
\noindent
\vskip .1in

The Standard Model is a unified description of
the electromagnetic and weak interactions  based upon
the gauge group $SU(2)\times U(1)$.
The fermions are chiral, i.e., the left-handed fermions are
doublets with respect to the $SU(2)$ gauge
group, while the right-handed fermions are
singlets. Left-- and right--handed fermions also carry different
charges under the $U(1)$ gauge group.

The concept of spontaneous symmetry breaking, to provide both the
gauge boson and fermion masses, is necessarily imbedded into the
Standard Model.  As in the Landau-Ginzburg superconductor, a scalar
field, the ``Higgs field,'' is introduced explicitly, ``by hand,'' to
provide the symmetry breaking condensate.  Recalling the toy model example of
Section 1.3(iii), the fermions can only acquire Dirac masses 
(or neutrino Majorana masses) 
through coupling to the condensing Higgs boson, since explicit fermion
mass terms are not singlets under the gauge group, owing to the fermionic
chiral charge assignments.  The gauge fields also
acquire their masses through the condensing Higgs boson, as per the toy
example of Section 1.3(iv).

We begin by considering an $SU(2)\times U(1)$ gauge theory
with one scalar field that transforms as an isodoublet under
the $SU(2)$ group. The covariant derivative 
for the theory is given by:
\bea
\label{a1}
iD_\mu & = &  i\partial_\mu - g_2W_\mu^a {Q^a} - g_1 B_\mu \frac{Y}{2}
\nonumber \\
& = & i\partial_\mu - g_2{W}_\mu^+Q^- - g_2{W}_\mu^-Q^+ 
-g_2{W}^3_\mu Q^3 - g_1{B}_\mu \frac{Y}{2}
\eea
where the ``charge basis'' expressions
are $Q^\pm = (Q^1 \pm iQ^2)/\sqrt{2}$, and
$W_\mu^\pm = (W_\mu^1 \pm iW_\mu^2)/\sqrt{2}$. 
Note there are  two 
gauge coupling constants, $g_2$ and $g_1$,
one for each simple subgroup of the theory. The
$Q^a$  are the $SU(2)$ weak charges and $Y$ is the $U(1)$ 
hypercharge.  The
$Q^a$ satisfy the $SU(2)$ Lie algebra:
\beq
[Q^a,Q^b] =i\epsilon^{abc}Q^c.
\eeq
These charges  are at this stage {\em abstract operators},
the generators of the associated groups. 
We specialize to particular {\em representations}
for the charges, 
choosing an $I=\half$ weak isospin representation for the
left-handed fermions, and  
Higgs boson, in which
$Q^a = \tau^a/2$ where the $\tau^a$ are Pauli matrices.
The right-handed fermions are singlets, annihilated
by the $Q^a$.
We further define the weak hypercharge for any representation 
by its eigenvalue, $Y_r$, or the abstract operator $Y$. 

In defining the Standard Model we 
choose a  specific generator to
correspond to  the electric
charge operator:
\beq
Q_{EM} = Q^3 + \frac{Y}{2}
\eeq
Note that the choice of $Q^3$ is a choice of basis, i.e.,
we could use any normalized linear combination of
the $Q^a$ in place of $Q^3$, but by a gauge rotation we
can always align that linear combination
with $Q^3$.
Given the electric charges of particles, 
this defines the $Y_r$
for any representation. 

The Standard Model left-handed fermionic matter doublets are then
assigned the following structure and
weak hypercharges:
\beq
\left( \begin{array}{c} u^{2/3} \\ d^{-1/3} \end{array}\right)_L,
Y_r=\frac{1}{3};\qquad\;
\left( \begin{array}{c} \nu^{0} \\ e^{-1} \end{array}\right)_L,
Y_r=-1;\qquad\;
H^c = \left( \begin{array}{c} \phi^+ \\ -\bar{\phi}^0 \end{array}\right)\;
Y_r={+1};
\eeq
where the last entry is the ``charge--conjugated Higgs field''
defined by $H^c = i\tau^2 H^*$.  
The right--handed fermions are $SU(2)$ singlets, 
i.e., $[Q^a, \psi_R] =0$,
hence,
\beq
u_R^{2/3},\;\;Y_r=\frac{4}{3};\qquad\;
d_R^{-1/3},\;\;Y_r=-\frac{2}{3};\qquad\;
e_R^{-1},\;\;Y_r={-2};\qquad\;
\nu_R^{0},\;\;Y_r=0.
\eeq
A right--handed neutrino is sterile, and
usually omitted in the definition of the
``Standard Model.''

Returning to the gauge bosons, 
if we write the linear combinations:
\bea
\label{lcWZ}
W_\mu^3  &=& Z^0_\mu \cos\theta + A_\mu\sin\theta
\nonumber \\
B_{\mu} &=& - Z^0_\mu \sin\theta + A_\mu\cos\theta
\eea
where $Z^0_\mu$ ($A_\mu$) is the physical
$Z$-boson (photon), then we see, 
\beq
g_2\sin\theta = e;\qquad\qquad g_1\cos\theta =e;\qquad\qquad
\eeq
The photon thus couples to $eQ_{EM}$ with 
strength $e$ where:
\beq
\frac{1}{e^2} =  \frac{1}{g_2^2} + \frac{1}{g_1^2}
\eeq
The weak mixing angle is defined by:
\beq
\tan\theta  = \frac{g_1}{g_2} 
\eeq
The gauge covariant field strengths are defined by
commutators of the covariant derivative:
\bea
F_{\mu\nu}^a  &=& 
-\frac{i}{g_2^2}\Tr\left({\tau^a}[D_\mu,D_\nu]\right)
=\partial_\mu A^a_\nu-\partial_\nu A^a_\mu + \epsilon^{abc}A_\mu^bA_\nu^c
\nonumber \\
F_{\mu\nu} &=& -\frac{i}{g_1^2}Y_r^{-1}\Tr\left([D_\mu,D_\nu]\right)
=\partial_\mu B_\nu-\partial_\nu B_\mu
\eea
Then the gauge field kinetic terms are:
\beq
{\cal{L}}_{G.B.\;kinetic} = -\frac{1}{4} F_{\mu\nu}^a F_{\mu\nu}^a 
-\frac{1}{4} F_{\mu\nu} F_{\mu\nu} 
\eeq

Consider the complex doublet scalar Higgs-boson
with $Y_r=-1$:
\beq
H = \left( \begin{array}{c} \phi^0 \\ \phi^- \end{array}\right)
\eeq
The Lagrangian for $H$ takes the form:
\bea
\label{higgs00}
{\cal{L}} & = & (D_\mu H)^\dagger (D^\mu H) - V(H)
\eea
The masses of the gauge bosons are generated by spontaneous
symmetry breaking, in strict analogy to the generation
of mass in a superconductor as discussed in Section 1.3(iv).  
We assume that $V(H)$ has
an unstable extremum for $H=0$ and a nontrivial
minimum, e.g., we can write:
\be
V(H) = \frac{\lambda}{2}(H^\dagger H - v_{weak}^2)^2
\ee 
The Higgs boson then
develops a vacuum expectation value.  This may, without
loss of generality be taken in the upper component (which
is, again, the choice of orientation within the internal $SU(2)$
space that defines the electric charge to be that
combination of generators that annihilates $\VEV{H}$):
\beq
\VEV{H} = \left( \begin{array}{c} v_{weak}  \\ 0 \end{array}\right)
\qquad
\makebox{or:}
\qquad
{H} = 
\exp\left(i\frac{\pi^a\tau^a}{\sqrt{2}v_{weak}}\right)
\left( \begin{array}{c} v_{weak} + \frac{h}{\sqrt{2}} \\ 0 \end{array}\right)
\label{ansazz}
\eeq
In the present normalization $v_{weak} = (2\sqrt{2}G_F)^{-1/2} = 175$
GeV.  Note that $Q_{EM}$ acting on the Higgs VEV is zero, which
implies that the photon remains a massless gauge boson $\gamma$.  

When we substitute the anzatz (\ref{ansazz}) into the Higgs boson kinetic
term of eq. (\ref{higgs00}), the masses of the gauge bosons are
generated (as in the Abelian Higgs model of Section 1.3(iv)):
\bea
\label{higgs1}
{\cal{L}} 
&= & \half(\partial h)^2 +
\half M_W^2 W_\mu^+W^\mu{}^- + \half M_Z^2 Z_\mu Z^\mu 
-\half M_H^2h^2 -\frac{\sqrt{\lambda}}{{2}}M_Hh^3 -\frac{1}{8}\lambda h^4
\nonumber \\
& & 
+ \half \left({h^2} + \frac{ M_H }{\lambda}h  \right)(g_2^2 W_\mu^+W^\mu{}^- + 
(g_1^2 + g_2^2) Z_\mu Z^\mu)
\eea
where $M_H = v_{weak}\sqrt{2\lambda}$.  This Lagrangian also
 exhibits the coupling
of the Higgs $h$ field to itself and to the
physical $W$ and $Z$ fields.
%
%
%
Inverting eq.(\ref{lcWZ}):
\begin{eqnarray}
{A}_\mu & = & \sin\theta\; {W}^3_\mu + \cos\theta\; {B}_\mu
\nonumber \\
{Z}_\mu & = & \cos\theta\;  {W}_\mu^3 - \sin\theta\;  {B}_\mu
= \frac{(g_2  {W}_\mu^3 
- g_1  {B}_\mu)}{\sqrt{g_1^2 + g_2^2}}
\end{eqnarray}
allows us to rewrite the covariant derivative as:
\bea
iD_\mu & = & i\partial_\mu - g_2{W}_\mu^+Q^- - g_2{W}_\mu^-Q^+ 
-e{A}_\mu \left( Q^3 + \frac{Y}{2}\right)
-{Z}_\mu \widetilde{Q}
\label{ehs-covder}
\eea
where the $Z$-boson couples to the neutral current charge:
\bea
\widetilde{Q} &=&
\sqrt{g_1^2 + g_2^2}
\left(\cos^2\theta\; \frac{\tau^3}{2} - \sin^2\theta\; \frac{Y}{2}
\right) \nonumber \\
& = & e
\left(\cot\theta\; \frac{\tau^3}{2} - \tan\theta\; \frac{Y}{2}
\right) 
\eea
We can also extract the masses, upon comparing eq.(\ref{higgs1}) with
eq.(\ref{higgs00}) using the full covariant
derivative (\ref{ehs-covder}):
\beq
\label{gbmasses}
M_W^2 = \half g_2^2v_{weak}^2; \qquad M_Z^2 = \frac{1}{2} v_{weak}^2 (g_1^2 + g_2^2).
\eeq
>From this, it follows:
\beq
\frac{M_W^2}{M_Z^2} = \cos^2\theta_W\qquad\qquad v_{weak}^2 = \frac{1}{2\sqrt{2}G_F}
\eeq

The coupling of the matter fields to gauge fields is then
determined through the gauge--invariant kinetic terms.  For example,
let us consider the left--handed top and bottom quark doublet
$\Psi_L = (t,\; b)_L$.  The kinetic term is:
\bea
\bar\Psi_L i\slash{D} \Psi_L & = &
\bar\Psi_L i\slash{\partial } \Psi_L - 
\frac{1}{\sqrt{2}} \bar{t}\gamma_\mu L b W^{\mu\;+}
-
\frac{1}{\sqrt{2}} \bar{b}\gamma_\mu L t W^{\mu\;-} \nonumber \\
& & - \frac{2e}{3} \bar{t}\gamma_\mu  L t {A}_\mu
+ \frac{e}{3} \bar{b} L b  {A}_\mu 
- \bar\Psi_L \widetilde{Q} \gamma_\mu\Psi_L {Z}_\mu
\eea
where $L=\half(1-\gamma^5)$.
Dirac mass terms would be of the form $\bar{\Psi}_L\psi_R$ and
are therefore isospin$-\half$; we cannot
directly add such nonsinglet terms into the Lagrangian.  
We can, however,
assume there are terms of the form:
\beq
g_t \bar{\Psi}_L\cdot H t_R + g_b \bar{\Psi}_L\cdot H^c b_R 
\eeq
which couple the left- and right-handed fermions to the Higgs field,
and which are invariant under $SU(2)\times U(1)$.  When $H$ develops
its VEV, we see that we obtain masses $m_t = g_tv_{weak}$ and $m_b =
g_b v_{weak}$ for the top and bottom quarks respectively.  In general,
of course, there will occur mixing effects when we include the light
generations; these lead to the Cabibbo-Kobayashi-Maskawa (CKM) matrix
relating the gauge and mass eigenbases for left-handed quarks.

For completeness, we mention that to minimally
generate a neutrino Majorana mass, we would write for
a leptonic doublet $\Psi_L = (\nu, \ell)_L$:
\beq
\frac{g_\nu}{M} (\bar{\Psi}_L H) (H^C{\Psi}^C_L) 
\eeq
where ${\Psi}^C_L$ is the charge-conjugated spinor.
This term produces a Majorana mass for $\nu_L$
of $g_\nu v_{weak}^2/M$. 

The 
interplay between {\em gauge symmetries},
and {\em chiral symmetries}, both of which 
are broken spontaneously, is fundamental
to the Standard Model.  The left-handed fermions carry
the electroweak $SU(2)$ quantum numbers, while the right-handed do not.
All of the 
mathematical features
of the symmetric Lagrangian remain intact, 
but the 
spectrum of the theory does not retain the original obvious
symmetry properties. 
When a massive gauge boson was discovered, such as
the $W^\pm$ or $Z$ of the Standard Model, we also discovered
an extra piece of physics: 
the longitudinal component, i.e.,
the NGB which comes from the symmetry breaking
sector.

\vskip .1in
\noindent
{\bf A(b) Summary of One--Loop 
Oblique Radiative Corrections}
\noindent
\vskip .1in

We will now summarize a particular subclass
of electroweak radiative corrections, the
so-called ``oblique'' corrections. To begin, we 
generalize slightly our definition of $v_{weak}$ in
the masses of the $W$ and $Z$. Let us 
generalize eq.(\ref{gbmasses}) and write:
\beq
M_W^2 = \half v_W^2g_2^2; \qquad M_Z^2 = \frac{1}{2} v_Z^2 (g_1^2 + g_2^2)
\eeq
where $v_W$ ($v_Z$) is the Higgs VEV ``as seen by'' the
$W$-boson ($Z$=boson). In tree level $v_W^2 = v_Z^2 = v_{weak}^2$,
and we want to include radiative effects that split these
quantities and lead to their $q^2$ eveolution.  
The radiative corrections
to the spontaneously broken Standard Model  thus
involve keeping track of four underlying functions of the
momentum scale, $q^2$ 
%
%
%
which are:
$g_1^2(q^2)$, $g_2^2(q^2)$ and $v_W^2(q^2)$ and $v_Z^2(q^2)$.
For the running
of the coupling constants,
to a good approximation, we really only need to know $\alpha(M_Z)
\approx \alpha(M_W)$ in most applications. 

Let us be more precise about $v_W$ and $v_Z$.
First, we scale away the coupling constants by
defining $\tilde{A}^a_\mu = g_2{A}^a_\mu$
and $\tilde{B}_\mu = g_2{B}_\mu$.
Loop corrections 
to correlation functions (or inverse propagators)
for any two gauge fields can be written as:
\be
F.T.<0| T \tilde{A}^A_\mu(0) \tilde{A}^B_\mu(x) |0>
=g_{\mu\nu}\Pi_{AB} - q_{\mu}q_\nu\Pi^T_{AB}
\ee
(subscript $T$ stands for ``transverse''). We then make the specific
definitions of $v_W$ and $v_Z$
\bea 
\half v_Z^2 & = & \half 
v_{weak}^2  - \Pi_{3B} \nonumber \\
& = & \half v_{weak}^2  - \Pi_{3Q} + \Pi_{33} 
\nonumber \\
 \half {v}_W^2 & = & 
\half v_{weak}^2 + \Pi_{WW} - \Pi_{3Q}  
\eea
and of the couplings
\bea
\frac{1}{g_2^2} & = & \frac{1}{g_{2un}^2} - \Pi_{33}^T - \Pi_{3B}^T \nonumber \\
& = &\frac{1}{g_{2un}^2}- \Pi_{3Q}^T
\nonumber \\
\frac{1}{g_1^2} & = & \frac{1}{g_{1un}^2} -\Pi_{BB}^T - \Pi_{3B}^T\nonumber \\
& = & \frac{1}{g_{1un}^2} + \Pi_{3Q}^T -\Pi_{QQ}^T 
\eea
where $g_{i\;un}$ is an unrenormalized coupling constant.
The dominant Standard Model contributions to $v_W^2 $
and $v_Z^2$ at $q^2=0$ come from the third generation 
and a putative Higgs boson. The difference 
$v_W^2 - v_Z^2$ is, upon computing the loops, finite:
\bea
v_W^2 - v_Z^2 &=&  \frac{N_c}{32\pi^{2}}
\left[(m_t^2+m_b^2)
-\frac{2m_t^2m_b^2}{(m_t^2-m_b^2)}\log(m_t^2/m_b^2)
\right.
\nonumber \\
& &\left. + \frac{M_W^2m_H^2}{m_H^2-M_W^2}\ln(m_H^2/M_W^2)
- \frac{M_Z^2 m_H^2}{m_H^2-M_Z^2}\ln(m_H^2/M_Z^2)
\right]
\eea
The $\rho$ parameter of Veltman is:
\bea
\rho & = & \frac{v_W^2}{v_Z^2}
\nonumber \\
& = & 1  
+ \frac{N_c}{32 v_0^2\pi^{2}}
\left[(m_t^2+m_b^2) 
-\frac{2m_t^2m_b^2}{(m_t^2-m_b^2)}\log(m_t^2/m_b^2)
\right.
\nonumber \\
&  &   
+ 
\left. \frac{M_W^2m_H^2}{m_H^2-M_W^2}\ln(m_H^2/M_W^2)
- \frac{M_Z^2 m_H^2}{m_H^2-M_Z^2}\ln(m_H^2/M_Z^2) \right]
\eea

In general, $v_W^2$ and $v_Z^2$ have small, nonzero, $q^2$ dependence.
Moreover, the effect of new physics at the weak scale, $|q| \sim 1$
TeV will generally be to induce additional $q^2$ dependence into
$v_W^2$ and $v_Z^2$ and additional splitting into $ v_W^2 - v_Z^2$
beyond the Standard Model contributions.  For example, new sequentially heavier
chiral fermions will contribute through loops, and their contributions
do not decouple as their masses are taken large.  The squared
couplings $g_1^2$ and $g_2^2$ are less susceptible to effects from new
physics We seek, therefore, a convenient parameterization of the effects of new physics that can be used for comparison with a variety of experiments.  The
resulting interpolation between different experiments may involve
nonstandard values of these new parameters and yield evidence for new
physics.

A simple set of parameters can
be generated by expanding  $v_W^2$ and $v_Z^2$ in a Taylor
series in $q^2$ as follows:
\bea
\label{vstu}
v_W^2(q^2) &=& v_{weak}^2 +  \sigma q^2 + \tau v_r^2  + \omega q^2 
\\
v_Z^2(q^2) &=& v_{weak}^2 +  \sigma q^2 - \tau v_r^2  - \omega q^2 
\eea
Thus, $v^2_{weak}$ is the average $(I=0)$ 
of the two decay constants and contains most
of the physics of symmetry breaking, while $\tau$ $(I=1)$ 
is just a rewriting
of the splitting between these at
zero momentum (i.e., the $\rho$ parameter). The
parameters $\sigma$ and $\omega$  are respectively $(I=0)$ and $(I=1)$
measures of physics contributing to the $q^2$ 
evolution in the effective low energy theory.

A similar parameterization exists in the literature and is due to
Kennedy and Lynn, \cite{Kennedy:1989sn}, Altarelli and Barbieri
\cite{Altarelli:1991zd} and Peskin and Takeuchi \cite{Peskin:1992sw}.
These authors consider the coefficients of the full vacuum
polarization tensors $\Pi_{XY}$.  Peskin and Takeuchi define:
\bea
S &= & 16\pi\left[ \frac{\partial}{\partial      q^2}\left.\Pi_{33}\right|_{q^2=0}  
 -          \frac{\partial}{\partial q^2}\left.\Pi_{3Q}\right|_{q^2=0} 
\right]
\\
T &= & \frac{4\pi}{\sin^2\theta\cos^2\theta M_Z^2}\left[ \left.\Pi_{WW}\right|_{q^2=0}  
 -          \left.\Pi_{33}\right|_{q^2=0} 
\right]
\\
U &= & 16\pi\left[ \frac{\partial}{\partial      q^2}\left.\Pi_{WW}\right|_{q^2=0}  
 -          \frac{\partial}{\partial q^2}\left.\Pi_{33}\right|_{q^2=0} 
\right]
\eea
Recalling our definitions of $v_W^2$ and $v_Z^2$:
\beq
\half v_Z^2  =  \half 
v_{weak}^2 + \Pi_{33} - \Pi_{3Q}^T 
\eeq
and:
\beq
 \half {v}_W^2 =  
\half v_{weak}^2 + \Pi_{WW} - \Pi_{3Q}^T
\eeq
we see that:
\beq
\sigma = \frac{2S+U}{16\pi};\qquad 
\omega = \frac{U}{16\pi};\qquad
\tau = \frac{\sin^2\theta\cos^2\theta M_Z^2T}{4\pi v_r^2}
= \frac{\alpha}{2}T;\qquad 
\ee

It is straightforward to compute $S$, $T$ and $U$.  
One finds for the $S$ parameter:
\bea
S & =  & \frac{N_c}{6\pi}
\left[1 - Y
\log\left\{ m_b^2 / m_t^2\right\}\right]
\eea
$T$ is given by the usual $\rho$--parameter expression, or the
difference of the zero--momentum expressions for 
$v_W^2$ and  $v_Z^2$:
\bea
T &=&  \frac{N_c}{4\pi \sin^2\theta \cos^2\theta M_Z^2}
\left[(m_t^2+m_b^2)
-\frac{2m_t^2m_b^2}{(m_t^2-m_b^2)}\log(m_t^2/m_b^2)
\right.
\nonumber \\
& &\left. + \frac{M_W^2m_H^2}{m_H^2-M_W^2}\ln(m_H^2/M_W^2)
- \frac{M_Z^2 m_H^2}{m_H^2-M_Z^2}\ln(m_H^2/M_Z^2)
\right]
\eea
Similarly, for $U$ we can
obtain:
\bea
U & =  & \frac{N_c}{6\pi}
\left[-\frac{5m_t^4-22m_t^2m_b^2 + 5m_b^4}{3(m_t^2-m_b^2)^2} \right.
\nonumber \\
&  & +
\left. \frac{m_t^6 -3m_t^4m_b^2 -3m_t^2m_b^4 + m_b^6}{(m_t^2-m_b^2)^3} 
\log\left\{ m_t^2 / m_b^2\right\}\right]
\eea

Combining the inputs from different measurements, such as $G_F =
1/2\sqrt{2}v_{weak}(0)^2$,\ \  $M_Z^2 = \half (g_1^2(M_Z^2)+
g_2^2(M_Z^2))v_{weak}^2(M_Z^2)$, \ \  $M_W^2 = \half
g_2^2(M_W^2)v_{weak}^2(M_W^2)$,\ \  $\sin^2\theta_{Z-pole} =
g_1(M_Z^2)/(g_1(M_Z^2)+g_2(M_Z^2))$, and $m_{top}$, one identifies (at
some chosen confidence level) an ellipse-shaped region of the $S-T$
plane with which the data is most consistent.  The $S-T$ coordinate
system is chosen with arbitrary offsets so that some particular
Standard Model value of $M_H$ defines the origin. One can then overlay
the trajectories corresponding to varying the value of $M_H$ or the
effects of other new physics.  An example of an $S-T$ error ellipse plot
with theoretical overlays is shown in Figure (\ref{stnew}).


\section*{Appendix B: The Nambu-Jona-Lasinio Model}
       \renewcommand{\theequation}{B.\arabic{equation}}
       \setcounter{equation}{0}
\addcontentsline{toc}{section}{Appendix B: The Nambu-Jona-Lasinio Model }

Much of our intuition about dynamical symmetry breaking
comes from the BCS theory of superconductivity 
\cite{Bardeen:1957kj,Bardeen:1957mv}. 
This involves a fermion pairing dynamics,
leading to a rearrangement of the vacuum (ground state)
in the presence of a strong attractive
interactions. It was imported into elementary particle physics to
provide a successful picture of mass generation in
strong interactions, also known as chiral symmetry breaking
\cite{Nambu:1960xd,Gell-Mann:1960np}. 
A toy version of this, known as the Nambu--Jona-Lasinio (NJL) model,
\cite{Nambu:1961tp}
provides a simple physical picture of chiral symmetry breaking.
We present a synopsis of it here.

Consider an effective four--fermion vertex
with coefficient, $G= g^2/\Lambda^2$. 
The theory at the ``high energy physics'' scale $\Lambda$ is:
\beq
\label{ctheq1}
{\cal{L}}_\Lambda = \bar{\psi}^a_L i\slash{\partial }\psi^a_L + \bar{\psi}^a_R 
i\slash{\partial }\psi^a_R  + G(\bar{\psi}^{a}_L\psi_{Ra})(\bar{\psi}_{R}^b\psi_{Lb})
\eeq
where $(a,b)$ are global $SU(N)$  ``color''
indices.
Note that the four--fermion form of the interaction is contained in a Fierz
rearrangement of a single--coloron (massive gluon) exchange potential:
\beq
\half g^2 \left(\bar{\psi}
\gamma_\mu\frac{\lambda^A}{2}\psi\right) \frac{g^{\mu\nu}}{q^2- \Lambda^2} 
\left(\bar{\psi}\gamma_\nu\frac{\lambda^A}{2}\psi\right)
\eeq
thus $G=g^2/\Lambda^2$
at $q^2=0$.  Thus, in some sense we may view this an approximation
to what we believe is happening in QCD on scales $M^2 \sim \Lambda_{QCD}^2$,
ignoring the effects of confinement.  

Let us view eq. (\ref{ctheq1}) as an effective Lagrangian 
at the scale $\Lambda$; 
the coupling constant $g$ is thus the renormalized effective
coupling constant at that scale. Of course, in any
realistic strongly interacting theory this would seem to be a drastic
approximation \cite{Hasenfratz:1991it}, but it
underlies the reasonably successful chiral constituent
quark model (see \cite{Manohar:1984md,Bijnens:1993uz} and references
therein).
We consider the solution to the theory at lower energies
based upon the effects of the fermion loops,
\ie,
a fermion bubble approximation. This is equivalent to a
large--$N_{color}$ expansion. We will present the renormalization
group version of the solution to the NJL model, which is quite
physical and transparent.  The widely-used alternative 
older approach is
discussed in refs.\cite{Bardeen:1990ds,Hill:1992jc}.

We can rewrite eq.(\ref{ctheq1}), introducing an auxilliary field $H$,
as:
\beq
\label{ctheq2}
{\cal{L}} 
= {\cal{L}}_{kinetic} + (g\overline{\psi}_L\psi_RH + h.c.) - \Lambda^2 H^\dagger H
\eeq
If we integrate out the field $H$ we reproduce the four--fermion
vertex as an induced interaction with $G \equiv g^2/\Lambda^2$.  Note
that $G > 0$ implies an attractive interaction and permits the
factorization in this form.  More specifically, eq.(\ref{ctheq2}) is
the effective Lagrangian on a scale $\Lambda$.  To obtain the
effective Lagrangian on a scale $\mu < \Lambda$ in the fermion bubble
approximation we integrate out the fermion field components on scales
$\mu \leftrightarrow \Lambda$.  The full induced effective Lagrangian
at the new scale $\mu$ will then take the form:
\bea
{\cal{L}}_\mu & = & {\cal{L}}_{kinetic} 
+ g\overline{\psi}_L\psi_RH + h.c. 
\nonumber \\
  &   & + Z_H|\partial_\nu H|^2 - m_H^2H^\dagger H 
  -\frac{\lambda_0}{2}(H^\dagger H)^2 
-\xi_0 R H^\dagger H
\eea
Here $R$ is the geometric scalar curvature, and we see there is
an induced ``nonminimal'' coupling of the Higgs field to
gravity, $\xi$.
A direct evaluation of the induced parameters by
computing the relevant Feynman loops gives:
\bea
Z_H & = & \frac{g^2N_c}{(4\pi)^{2}}  \log(\Lambda^2/\mu^2);
\qquad
m^2_H  =  \Lambda^2- \frac{2g^2N_c}{(4\pi)^{2}} (\Lambda^2 - \mu^2)
\nonumber  \\
\lambda_0 & = & \frac{2g^4N_c}{(4\pi)^{2}} \log(\Lambda^2/\mu^2);
\qquad
\xi_0  =  \frac{1}{6}\frac{g^2N_c}{(4\pi)^{2}} \log(\Lambda^2/\mu^2).
\eea
(the parameter $g$ is unrenormalized 
at this stage in fermion loop approximation).
The induced low energy parameters, $Z_H$ and $\lambda_0$, and $\xi_0$
are determined in terms of $\Lambda$, and we are
interested in the $\mu \approx 0$ limit of the theory.

We emphasize that the effective theory applies in either the 
spontaneously broken or
unbroken phases.  The broken phase is selected by demanding that
$m_H^2 < 0$ for scales $\mu^2 \ll \Lambda^2$, thus  requiring
that $\Lambda^2(1- g^2N_c/8\pi^2) < 0$;
hence, $g^2 > 8\pi^2/N_c = g_c^2$ defines a critical
coupling.  On the other hand,
for positive $m_H^2$ as $\mu \rightarrow 0$ the
theory remains unbroken (this is equivalent to a subcritical four--fermion
coupling constant, $g^2 \leq g^2_c$).  

Let us bring the effective Lagrangian  into
a conventionally normalized form
by rescaling the field $H\rightarrow H/\sqrt{Z_H}$:
\bea
{\cal{L}} & = & {\cal{L}}_{kinetic} 
+ \tilde{g}\overline{\psi}_L\psi_RH + h.c.  
\nonumber \\
  &   & + |\partial_\nu H|^2 - \widetilde{m}_H^2H^\dagger H 
-\frac{\tilde{\lambda}}{2}(H^\dagger H)^2 
-\xi R H^\dagger H
\eea
Where we find:
\bea
\label{LIST}
\tilde{g}^2 & = & g^2/Z_H = \frac{16\pi^{2}}{N_c  \log(\Lambda^2/\mu^2)}
\nonumber \\
\widetilde{m}^2_H & = & m_H^2/Z_H 
\\
\tilde{\lambda} & = & \lambda_0/Z_H^2
= \frac{32 \pi^{2}}{N_c  \log(\Lambda^2/\mu^2)}
\nonumber \\
\xi & = & \xi_0/Z_H = 1/6
\nonumber
\eea
These are the physical renormalized coupling constants.
We tune the low energy value of $\widetilde{m}_H^2$ to
the desired value, and
the remaining predictions of the model are contained in $\tilde{g}$,
$\tilde{\lambda}$ (and $\xi$) as we will see below.
The compositeness of the $H$ boson is essentially contained
in the result that $g$ and $\lambda$ are singular as $\mu
\rightarrow \Lambda$ (while $\xi$ is constant and equal
to its conformal value of $1/6$).  We will refer to these as
the ``compositeness conditions.''

These results
are easily recovered directly from
the conventional differential renormalization group
equations, supplemented with ``compositeness
conditions'' as high energy boundary conditions.
We utilize
the approximate $\beta$-functions which reflect only the
presence of fermion loops:
\beq
16\pi^2~\frac{\partial }{\partial\ln\mu}g = 
N_c{g}^3
\eeq
\beq
\label{LAM}
16\pi^2~\frac{\partial }{\partial\ln\mu}\lambda 
= (-4N_cg^4 + 4N_cg^2\lambda)
\eeq
Solving the first RG equation gives:
\beq
\frac{1}{g^2(\mu)} - \frac{1}{g^2(\Lambda)} 
= \frac{N_c}{16\pi^2}\ln(\Lambda^2/\mu^2)
\eeq
If we now use the boundary condition, $1/g^2(\Lambda) = 0$
we see that we recover the above result $\tilde{g}^2 = g_c^2(\mu)$.
The second RG equation may then be solved by hypothesizing an 
anzatz of the form
$\lambda = cg^2$. Substituting  one finds:
\beq
16\pi^2~\frac{\partial }{\partial\ln\mu}g = 
\frac{1}{2c}(4c-4)N_c{g}^3
\eeq
and demand that this must be consistent with the other
RG equation. Thus one finds:
$c=2$
and:
\beq
\frac{1}{\lambda(\mu)} - \frac{1}{\lambda(\Lambda)} 
= \frac{N_c}{32\pi^2}\ln(\Lambda^2/\mu^2)
\eeq
and again ${\lambda(\Lambda)}^{-1} = 0 $ leads to the 
above result of $\tilde{\lambda} = \lambda(\mu)$.

To obtain the phenomenological
results of the NJL model
we examine the low energy Higgs potential from  the action
with $\mu = m$:
\beq
V(H) = -\widetilde{m}_H^2 H^\dagger H + \frac{\tilde{\lambda}}{2} (H^\dagger H)^2
-  (\tilde{g}\overline{\psi}_L\psi_RH + h.c.)
\eeq
Let us assume that $\widetilde{m}^2_H < 0$ so the neutral Higgs field develops
a VEV:
$Re(H^0) = (v + h)/\sqrt{2}$.
Therefore we find the dynamical fermion mass: 
\beq
 m = \tilde{g}v/\sqrt{2} ; 
\eeq
and
the $h$ mass is:
\beq
m^2_h = v^2\tilde{\lambda}
\eeq
and so:
\beq
m_h^2/m^2 = 2\tilde{\lambda}/\tilde{g^2} = 4  \qquad
\makebox{or:} \qquad m_h = 2 m
\eeq
using eqs.(B.7).
This is the familar NJL result, $m_h = 2m$.
We note that this result is subject to renormalizion
when effects of other interactions
are included \cite{Bardeen:1990ds,Hill:1992jc}.
It also does not imply that the Higgs is
loosely bound (see below)!
We also obtain presently 
the Pagels-Stokar relation:
\beq
\label{vweak17}
\half v^2 = m^2/\tilde{g}^2 = m^2\frac{N_c}{16\pi^2}\ln(\Lambda^2/m^2)
\eeq

It is also amusing to study the result $\xi = 1/6$
in the differential renormalization group \cite{Hill:1992jc}. One
finds that
the solution for the scalar coupling
to curvature,
\beq
\xi(\mu) = 1/6,
\eeq
is a RG constant for all scales.
More generally, as one descends toward the infrared,
$\xi = 1/6$ is an {\em attractive infra-red fixed point}. Therefore,
no matter what is the initial value for $\xi$ at
the large scale $\Lambda$, given enough RG running time
$\xi$ will eventually reach 1/6 for small $\mu$.
Of course, the RG running only occurs for scales
$\mu > m_H$.
We thus find 
in the usual fermion bubble approximation
that $H$ is conformally coupled to gravity, even though
scale breaking dynamics exists at high energies $\Lambda$.
Moreover, $\xi = 1/6$ is an attractive renormalization group
fixed point in the infrared in this approximation \cite{Hill:1992jc}. 

This physical particle $H$
is a bound state of $\bar{\psi}\psi$, arising by the attractive
four--fermion interaction
at the scale $\Lambda $. 
One might think that this is a loosely bound state, since it lies on
top of the threshold for open $\bar{\psi} \psi$
and apparently has vanishing binding energy
to this order. However, this is {\em not a nonrelativistic bound state},
and normal intuition does not apply. The state is built from
fully relativistic
Feynman loops with momenta that extend from $m$ to $\Lambda$.
The prediction $m_h = 2m$ 
cannot be viewed  as an exact one and is subject
to corrections from subleading $N$ effects and
other interactions.

The RG approach can also be used to go beyond
the leading order in large $N_c$.  Essentially, we
keep the compositeness boundary conditions
but include the full dynamics into the RG equations
\cite{Bardeen:1990ds}. In this case, $\tilde{g}$ and
$\tilde{\lambda}$ develop nontrivial
infra-red fixed points 
\cite{Hill:1981sq,Pendleton:1981as,Hill:1985tg,Hill:1981yr}. 
The NJL model is also readily generalized with the incorporation of
flavor to be a chiral constituent quark model (see
\cite{Manohar:1984md,Bijnens:1993uz} and references therein), or
techniquark model. As discussed in Section 4, it is also the basis of
Topcolor models.

The NJL model is readily adapted to QCD, or more
general strongly interacting theories with nontrivial
cboundstate or chiral dynamics. In that case the field
$H$ becomes a more general object, 
such as a linear $\sigma$-model
field (see, e.g., \cite{Bijnens:1993uz} or
the Appendix of \cite{Bardeen:1994ae}, which closely
follows the present discussion).

\newpage


\bibliographystyle{unsrt}

\addcontentsline{toc}{section}{Bibliography }
\bibliography{allbiblio}

\begin{thebibliography}{100}

\bibitem{Nambu:1960xd}
Y.~Nambu.
\newblock Axial vector current conservation in weak interactions.
\newblock {\em Phys. Rev. Lett.}, 4:380--382, 1960.

\bibitem{Goldstone:1961eq}
J.~Goldstone.
\newblock Field theories with 'superconductor' solutions.
\newblock {\em Nuovo Cim.}, 19:154--164, 1961.

\bibitem{Gell-Mann:1960np}
M~Gell-Mann and M~Levy.
\newblock The axial vector current in beta decay.
\newblock {\em Nuovo Cim.}, 16:705, 1960.

\bibitem{Gell-Mann:1953aa}
M.~Gell-Mann.
\newblock {\em Phys. Rev.}, 92:883, 1953.

\bibitem{Gell-Mann:1956aa}
M.~Gell-Mann.
\newblock {\em Nuovo Comento}, 4, Suppl. 2:848, 1956.

\bibitem{Zweig:1964jf}
G.~Zweig.
\newblock An su(3) model for strong interaction symmetry and its breaking. 2.
\newblock CERN-TH-412.

\bibitem{Bjorken:1969dy}
J.~D. Bjorken.
\newblock Asymptotic sum rules at infinite momentum.
\newblock {\em Phys. Rev.}, 179:1547--1553, 1969.

\bibitem{Bjorken:1969ja}
J.~D. Bjorken and Emmanuel~A. Paschos.
\newblock Inelastic electron proton and gamma proton scattering, and the
  structure of the nucleon.
\newblock {\em Phys. Rev.}, 185:1975--1982, 1969.

\bibitem{Feynman:1969ej}
Richard~P. Feynman.
\newblock Very high-energy collisions of hadrons.
\newblock {\em Phys. Rev. Lett.}, 23:1415--1417, 1969.

\bibitem{Fritzsch:1973pi}
H.~Fritzsch, M.~Gell-Mann, and H.~Leutwyler.
\newblock Advantages of the color octet gluon picture.
\newblock {\em Phys. Lett.}, B47:365, 1973.

\bibitem{Greenberg:1964pe}
O.~W. Greenberg.
\newblock Spin and unitary spin independence in a paraquark model of baryons
  and mesons.
\newblock {\em Phys. Rev. Lett.}, 13:598--602, 1964.

\bibitem{Yang:1954ek}
C.~N. Yang and R.~L. Mills.
\newblock Conservation of isotopic spin and isotopic gauge invariance.
\newblock {\em Phys. Rev.}, 96:191--195, 1954.

\bibitem{Gross:1973id}
D.~J. Gross and F.~Wilczek.
\newblock Ultraviolet behavior of nonabelian gauge theories.
\newblock {\em Phys. Rev. Lett.}, 30:1343--1346, 1973.

\bibitem{Politzer:1973fx}
H.~D. Politzer.
\newblock Reliable perturbative results for strong interactions?
\newblock {\em Phys. Rev. Lett.}, 30:1346--1349, 1973.

\bibitem{Bardeen:1957kj}
J.~Bardeen, L.~N. Cooper, and J.~R. Schrieffer.
\newblock Microscopic theory of superconductivity.
\newblock {\em Phys. Rev.}, 106:162, 1957.

\bibitem{Bardeen:1957mv}
J.~Bardeen, L.~N. Cooper, and J.~R. Schrieffer.
\newblock Theory of superconductivity.
\newblock {\em Phys. Rev.}, 108:1175--1204, 1957.

\bibitem{Coleman:1973jx}
S.~Coleman and E.~Weinberg.
\newblock Radiative corrections as the origin of spontaneous symmetry breaking.
\newblock {\em Phys. Rev.}, D7:1888--1910, 1973.

\bibitem{Fermi:1934hr}
E.~Fermi.
\newblock An attempt of a theory of beta radiation. 1.
\newblock {\em Z. Phys.}, 88:161--177, 1934.

\bibitem{Glashow:1961tr}
S.~L. Glashow.
\newblock Partial symmetries of weak interactions.
\newblock {\em Nucl. Phys.}, 22:579--588, 1961.

\bibitem{Weinberg:1967tq}
S.~Weinberg.
\newblock A model of leptons.
\newblock {\em Phys. Rev. Lett.}, 19:1264--1266, 1967.

\bibitem{Salam:1980jd}
Abdus Salam.
\newblock Gauge unification of fundamental forces.
\newblock {\em Rev. Mod. Phys.}, 52:525, 1980.

\bibitem{LEPEWWG:2001}
LEPEWWG.
\newblock 2001 summary of results presented at 2000 conferences.
\newblock 2001.

\bibitem{Chanowitz:2001bv}
Michael~S. Chanowitz.
\newblock The z --> anti-b b decay asymmetry: Lose-lose for the standard model.
\newblock {\em Phys. Rev. Lett.}, 87:231802, 2001.

\bibitem{Lee:1977eg}
Benjamin~W. Lee, C.~Quigg, and H.~B. Thacker.
\newblock Weak interactions at very high-energies: The role of the higgs boson
  mass.
\newblock {\em Phys. Rev.}, D16:1519, 1977.

\bibitem{'tHooft:1980xb}
(ed.~) G.~'t Hooft et~al.
\newblock Recent developments in gauge theories. proceedings, nato advanced
  study institute, cargese, france, august 26 - september 8, 1979.
\newblock New York, Usa: Plenum ( 1980) 438 P. ( Nato Advanced Study Institutes
  Series: Series B, Physics, 59).

\bibitem{Wilson:1971dh}
Kenneth~G. Wilson.
\newblock Renormalization group and critical phenomena. 2. phase space cell
  analysis of critical behavior.
\newblock {\em Phys. Rev.}, B4:3184--3205, 1971.

\bibitem{Wilson:1974jj}
K.~G. Wilson and J.~Kogut.
\newblock The renormalization group and the epsilon expansion.
\newblock {\em Phys. Rept.}, 12:75--200, 1974.

\bibitem{Georgi:1974sy}
H.~Georgi and S.~L. Glashow.
\newblock Unity of all elementary particle forces.
\newblock {\em Phys. Rev. Lett.}, 32:438--441, 1974.

\bibitem{Langacker:1991an}
Paul Langacker and Ming xing Luo.
\newblock Implications of precision electroweak experiments for m(t), rho(0),
  sin**2-theta(w) and grand unification.
\newblock {\em Phys. Rev.}, D44:817--822, 1991.

\bibitem{Amaldi:1991cn}
Ugo Amaldi, Wim de~Boer, and Hermann Furstenau.
\newblock Comparison of grand unified theories with electroweak and strong
  coupling constants measured at lep.
\newblock {\em Phys. Lett.}, B260:447--455, 1991.

\bibitem{Hill:1984xh}
Christopher~T. Hill.
\newblock Are there significant gravitational corrections to the unification
  scale?
\newblock {\em Phys. Lett.}, B135:47, 1984.

\bibitem{Shafi:1984gz}
Q.~Shafi and C.~Wetterich.
\newblock Modification of gut predictions in the presence of spontaneous
  compactification.
\newblock {\em Phys. Rev. Lett.}, 52:875, 1984.

\bibitem{Goldberger:1958tr}
M.~L. Goldberger and S.~B. Treiman.
\newblock Decay of the pi meson.
\newblock {\em Phys. Rev.}, 110:1178--1184, 1958.

\bibitem{Nambu:1961tp}
Y.~Nambu and G.~Jona-Lasinio.
\newblock Dynamical model of elementary particles based on an analogy with
  superconductivity. i.
\newblock {\em Phys. Rev.}, 122:345, 1961.

\bibitem{Nambu:1961fr}
Y.~Nambu and G.~Jona-Lasinio.
\newblock Dynamical model of elementary particles based on an analogy with
  superconductivity. ii.
\newblock {\em Phys. Rev.}, 124:246, 1961.

\bibitem{Bardeen:1995kv}
William~A. Bardeen.
\newblock On naturalness in the standard model.
\newblock Presented at the 1995 Ontake Summer Institute, Ontake Mountain,
  Japan, Aug 27 - Sep 2, 1995.

\bibitem{Weinberg:1976gm}
Steven Weinberg.
\newblock Implications of dynamical symmetry breaking.
\newblock {\em Phys. Rev.}, D13:974--996, 1976.

\bibitem{Susskind:1978ms}
Leonard Susskind.
\newblock Dynamics of spontaneous symmetry breaking in the weinberg- salam
  theory.
\newblock {\em Phys. Rev.}, D20:2619, 1979.

\bibitem{Georgi:1984af}
Howard Georgi and David~B. Kaplan.
\newblock Composite higgs and custodial su(2).
\newblock {\em Phys. Lett.}, B145:216, 1984.

\bibitem{Arkani:2001nc}
Nima Arkani-Hamed, Andrew~G. Cohen, and Howard Georgi.
\newblock Electroweak symmetry breaking from dimensional deconstruction.
\newblock {\em Phys. Lett.}, B513:232--240, 2001.

\bibitem{Hill:2000mu}
Christopher~T. Hill, Stefan Pokorski, and Jing Wang.
\newblock Gauge invariant effective lagrangian for kaluza-klein modes.
\newblock {\em Phys. Rev.}, D64:105005, 2001.

\bibitem{Arkani-Hamed:2001ca}
Nima Arkani-Hamed, Andrew~G. Cohen, and Howard Georgi.
\newblock (de)constructing dimensions.
\newblock {\em Phys. Rev. Lett.}, 86:4757--4761, 2001.

\bibitem{Farhi:1981xs}
Edward Farhi and Leonard Susskind.
\newblock Technicolor.
\newblock {\em Phys. Rept.}, 74:277, 1981.

\bibitem{Sikivie:1980fm}
P.~Sikivie.
\newblock An introduction to technicolor.
\newblock Lectures given at the Int. School of Physics, Enrico Fermi, Varenna,
  Italy, Jul 21 - Aug 2, 1980.

\bibitem{Dimopoulos:1980tn}
S.~Dimopoulos, S.~Raby, and P.~Sikivie.
\newblock Problems and virtues of scalarless theories of electroweak
  interactions.
\newblock {\em Nucl. Phys.}, B176:449, 1980.

\bibitem{Lane:1980wc}
Kenneth~D. Lane and Michael~E. Peskin.
\newblock An introduction to weak interaction theories with dynamical symmetry
  breaking.
\newblock Lectures given at 15th Rencontre de Moriond, Les Arcs, France, Mar
  9-21, 1980.

\bibitem{Kaul:1983uk}
Romesh~K. Kaul.
\newblock Technicolor.
\newblock {\em Rev. Mod. Phys.}, 55:449, 1983.

\bibitem{Kane:1981ee}
G.~L. Kane.
\newblock Generalized higgs physics and technicolor.
\newblock Based on lectures given at Les Houches Summer School in Theoretical
  Physics, Les Houches, France, Aug 3 - Sep 11, 1981.

\bibitem{Ellis:1981gx}
John Ellis.
\newblock Technicolor: Oasis or mirage?
\newblock Presented at SLAC Topical Conf. on Particle Physics of SLAC Summer
  Inst., Stanford, Calif., Jul 27 - Aug 7, 1981.

\bibitem{King:1989ec}
S.~F. King.
\newblock Technicolor today.
\newblock {\em Nucl. Phys. Proc. Suppl.}, 16:635, 1990.

\bibitem{Farhi:1982vt}
(ed.~) E.~Farhi and (ed.~) R.~Jackiw.
\newblock Dynamical gauge symmetry breaking. a collection of reprints.
\newblock Singapore, Singapore: World Scientific ( 1982) 403p.

\bibitem{Lane:2000pa}
Kenneth Lane.
\newblock Technicolor 2000.
\newblock 2000.

\bibitem{Lane:1993wz}
Kenneth Lane.
\newblock An introduction to technicolor.
\newblock 1993.

\bibitem{Chivukula:1996uy}
R.~Sekhar Chivukula.
\newblock An introduction to dynamical electroweak symmetry breaking.
\newblock 1996.

\bibitem{Chivukula:1993nj}
R.~S. Chivukula.
\newblock Beyond the standard model.
\newblock Prepared for 16th International Symposium on Lepton and Photon
  Interactions, Ithaca, NY, 10-15 Aug 1993.

\bibitem{Chivukula:1998if}
R.~Sekhar Chivukula.
\newblock Models of electroweak symmetry breaking.
\newblock 1998.

\bibitem{King:1995yr}
Stephen~F. King.
\newblock Dynamical electroweak symmetry breaking.
\newblock {\em Rept. Prog. Phys.}, 58:263--310, 1995.

\bibitem{Rubakov:1982vi}
V.~A. Rubakov and M.~E. Shaposhnikov.
\newblock Grand unification theory and technicolor. (in russian).
\newblock In *Dubna 1982, Proceedings, High Energy Physics For Young
  Scientists*, 5-142.

\bibitem{King:1985si}
Stephen~F. King.
\newblock Compositeness, unification and technicolor.
\newblock HUTP-85/A023.

\bibitem{Decker:1982ci}
R.~Decker.
\newblock Sin**2-theta-w and proton lifetime in unextended technicolor models.
  (talk, abstract only).
\newblock In *Hamburg 1982, Proceedings, Electroweak Interactions At High
  Energies*, 287-290.

\bibitem{Decker:1982ae}
R.~Decker.
\newblock Technicolor and some su(5) predictions.
\newblock DO-TH 82/17.

\bibitem{Anselm:1981pz}
A.~A. Anselm.
\newblock Evolution of theory after the 'standard model' (grand unification and
  technicolor). (in russian).
\newblock In *Leningrad 1981, Proceedings, Physics Of Elementary Particles*,
  3-21.

\bibitem{Elias:1980ej}
Victor Elias.
\newblock Some consequences of embedding technicolor in grand unified theories.
\newblock {\em Phys. Rev.}, D22:2879, 1980.

\bibitem{Chivukula:1990qb}
R.~Sekhar Chivukula and Terry~P. Walker.
\newblock Technicolor cosmology.
\newblock {\em Nucl. Phys.}, B329:445, 1990.

\bibitem{Witten:1981nf}
Edward Witten.
\newblock Dynamical breaking of supersymmetry.
\newblock {\em Nucl. Phys.}, B188:513, 1981.

\bibitem{Dine:1981za}
Michael Dine, Willy Fischler, and Mark Srednicki.
\newblock Supersymmetric technicolor.
\newblock {\em Nucl. Phys.}, B189:575--593, 1981.

\bibitem{Dine:1982qj}
Michael Dine and Mark Srednicki.
\newblock More supersymmetric technicolor.
\newblock {\em Nucl. Phys.}, B202:238, 1982.

\bibitem{Hill:1993ev}
Christopher~T. Hill, Dallas~C. Kennedy, Tetsuya Onogi, and Hoi-Lai Yu.
\newblock Spontaneously broken technicolor and the dynamics of virtual vector
  technimesons.
\newblock {\em Phys. Rev.}, D47:2940--2948, 1993.

\bibitem{Kolb:1984dn}
Edward~W. Kolb and L.~McLerran.
\newblock The structure of 'techni'jets.
\newblock {\em Phys. Lett.}, B143:505, 1984.

\bibitem{Bijnens:1993uz}
Johan Bijnens, Christophe Bruno, and Eduardo de~Rafael.
\newblock Nambu-jona-lasinio like models and the low-energy effective action of
  qcd.
\newblock {\em Nucl. Phys.}, B390:501--541, 1993.

\bibitem{Manohar:1984md}
Aneesh Manohar and Howard Georgi.
\newblock Chiral quarks and the nonrelativistic quark model.
\newblock {\em Nucl. Phys.}, B234:189, 1984.

\bibitem{Pagels:1979hd}
Heinz Pagels and Saul Stokar.
\newblock The pion decay constant, electromagnetic form-factor and quark
  electromagnetic selfenergy in qcd.
\newblock {\em Phys. Rev.}, D20:2947, 1979.

\bibitem{Appelquist:1988yc}
Thomas Appelquist, Kenneth Lane, and Uma Mahanta.
\newblock On the ladder approximation for spontaneous chiral symmetry breaking.
\newblock {\em Phys. Rev. Lett.}, 61:1553, 1988.

\bibitem{Lane:1988en}
Kenneth Lane.
\newblock Walking technicolor beyond the ladder approximation.
\newblock Presented at 1988 Division of Particles and Fields of the APS Mtg.,
  Storrs, CT, Aug 15-18, 1988.

\bibitem{Mahanta:1989sn}
Uma~Prasad Mahanta.
\newblock Higher order corrections in walking technicolor theories.
\newblock UMI-89-13676.

\bibitem{Kogut:1983sm}
J.~Kogut et~al.
\newblock Studies of chiral symmetry breaking in su(2) lattice gauge theory.
\newblock {\em Nucl. Phys.}, B225:326, 1983.

\bibitem{Kogut:1982fn}
J.~Kogut et~al.
\newblock The scales of chiral symmetry breaking in quantum chromodynamics.
\newblock {\em Phys. Rev. Lett.}, 48:1140, 1982.

\bibitem{Peskin:1980gc}
Michael~E. Peskin.
\newblock The alignment of the vacuum in theories of technicolor.
\newblock {\em Nucl. Phys.}, B175:197--233, 1980.

\bibitem{Preskill:1981mz}
John Preskill.
\newblock Subgroup alignment in hypercolor theories.
\newblock {\em Nucl. Phys.}, B177:21--59, 1981.

\bibitem{Binetruy:1982uf}
P.~Binetruy, S.~Chadha, and P.~Sikivie.
\newblock Vacuum alignment by broken gauge interactions.
\newblock {\em Nucl. Phys.}, B207:505, 1982.

\bibitem{Leon:1983dg}
J.~Leon and M.~Ramon-Medrano.
\newblock On the vacuum alignment in the breaking of g (etc).
\newblock {\em Phys. Rev.}, D28:915, 1983.

\bibitem{Luty:1992xz}
Markus~A. Luty.
\newblock Vacuum alignment in 'composite technicolor' models.
\newblock {\em Phys. Lett.}, B292:113--118, 1992.

\bibitem{Georgi:1994at}
Howard Georgi.
\newblock Physics from vacuum alignment in a technicolor model.
\newblock {\em Nucl. Phys.}, B416:699--721, 1994.

\bibitem{Chivukula:1998uf}
R.~Sekhar Chivukula and Howard Georgi.
\newblock Large-n and vacuum alignment in topcolor models.
\newblock {\em Phys. Rev.}, D58:075004, 1998.

\bibitem{Lane:2000es}
Kenneth Lane, Tonguc Rador, and Estia Eichten.
\newblock Vacuum alignment in technicolor theories. i: The technifermion
  sector.
\newblock {\em Phys. Rev.}, D62:015005, 2000.

\bibitem{'tHooft:1977am}
G.~'t~Hooft.
\newblock Can we make sense out of 'quantum chromodynamics'?
\newblock Lectures given at Int. School of Subnuclear Physics, Erice, Sicily,
  Jul 23 - Aug 10, 1977.

\bibitem{'tHooft:1978yv}
G.~'t~Hooft.
\newblock Nonabelian gauge theories and quark confinement.
\newblock Lecture given at Conf. on Current Trends in the Theory of Fields,
  Tallahassee, Fl., Apr 6-7, 1978.

\bibitem{'tHooft:1983wm}
Gerard 't~Hooft.
\newblock Planar diagram field theories.
\newblock Presented at Summer School 'Progress in Gauge Field Theory', Cargese,
  France, Sep 1-15, 1983.

\bibitem{Georgi:1986kr}
Howard Georgi and Lisa Randall.
\newblock Flavor conserving cp violation in invisible axion models.
\newblock {\em Nucl. Phys.}, B276:241, 1986.

\bibitem{Dimopoulos:1980sp}
Savas Dimopoulos.
\newblock Technicolored signatures.
\newblock {\em Nucl. Phys.}, B168:69--92, 1980.

\bibitem{Witten:1982fp}
E.~Witten.
\newblock An su(2) anomaly.
\newblock {\em Phys. Lett.}, B117:324--328, 1982.

\bibitem{DiVecchia:1980xq}
P.~Di Vecchia and G.~Veneziano.
\newblock Minimal composite higgs systems.
\newblock {\em Phys. Lett.}, B95:247, 1980.

\bibitem{Tandean:1995ci}
Jusak Tandean.
\newblock Observing the technieta at a photon linear collider.
\newblock {\em Phys. Rev.}, D52:1398--1403, 1995.

\bibitem{Witten:1979vv}
Edward Witten.
\newblock Current algebra theorems for the u(1) 'goldstone boson'.
\newblock {\em Nucl. Phys.}, B156:269, 1979.

\bibitem{Witten:1980sp}
E.~Witten.
\newblock Large n chiral dynamics.
\newblock {\em Ann. Phys.}, 128:363, 1980.

\bibitem{Coleman:1980mx}
Sidney Coleman and Edward Witten.
\newblock Chiral symmetry breakdown in large n chromodynamics.
\newblock {\em Phys. Rev. Lett.}, 45:100, 1980.

\bibitem{Dimopoulos:1979qi}
Savas Dimopoulos and Leonard Susskind.
\newblock A technicolored solution to the strong cp problem.
\newblock Print-79-0196 (COLUMBIA).

\bibitem{Chivukula:1990rn}
R.~Sekhar Chivukula and Mitchell Golden.
\newblock Observing the techniomega at the ssc.
\newblock {\em Phys. Rev.}, D41:2795, 1990.

\bibitem{Dimopoulos:1981yf}
S.~Dimopoulos, S.~Raby, and G.~L. Kane.
\newblock Experimental predictions from technicolor theories.
\newblock {\em Nucl. Phys.}, B182:77, 1981.

\bibitem{Dimopoulos:1982fj}
Savas Dimopoulos and John Ellis.
\newblock Challenges for extended technicolor theories.
\newblock {\em Nucl. Phys.}, B182:505--528, 1982.

\bibitem{Eichten:1986eq}
E.~Eichten, I.~Hinchliffe, K.~D. Lane, and C.~Quigg.
\newblock Signatures for technicolor.
\newblock {\em Phys. Rev.}, D34:1547, 1986.

\bibitem{Bardeen:1994ae}
William~A. Bardeen and Christopher~T. Hill.
\newblock Chiral dynamics and heavy quark symmetry in a solvable toy field
  theoretic model.
\newblock {\em Phys. Rev.}, D49:409--425, 1994.

\bibitem{Eichten:1984eu}
E.~Eichten, I.~Hinchliffe, K.~Lane, and C.~Quigg.
\newblock Super collider physics.
\newblock {\em Rev. Mod. Phys.}, 56:579--707, 1984.

\bibitem{Eichten:1994nc}
Estia Eichten and Kenneth Lane.
\newblock Multiscale technicolor and top production.
\newblock {\em Phys. Lett.}, B327:129--135, 1994.

\bibitem{Weinberg:1968de}
Steven Weinberg.
\newblock Nonlinear realizations of chiral symmetry.
\newblock {\em Phys. Rev.}, 166:1568--1577, 1968.

\bibitem{Coleman:1969sm}
S.~Coleman, J.~Wess, and Bruno Zumino.
\newblock Structure of phenomenological lagrangians. 1.
\newblock {\em Phys. Rev.}, 177:2239--2247, 1969.

\bibitem{Callan:1969sn}
Jr. Curtis G.~Callan, Sidney Coleman, J.~Wess, and Bruno Zumino.
\newblock Structure of phenomenological lagrangians. 2.
\newblock {\em Phys. Rev.}, 177:2247--2250, 1969.

\bibitem{Appelquist:1980vg}
Thomas Appelquist and Claude Bernard.
\newblock Strongly interacting higgs bosons.
\newblock {\em Phys. Rev.}, D22:200, 1980.

\bibitem{Longhitano:1980iz}
Anthony~C. Longhitano.
\newblock Heavy higgs bosons in the weinberg-salam model.
\newblock {\em Phys. Rev.}, D22:1166, 1980.

\bibitem{Longhitano:1981tm}
Anthony~C. Longhitano.
\newblock Low-energy impact of a heavy higgs boson sector.
\newblock {\em Nucl. Phys.}, B188:118, 1981.

\bibitem{Renken:1983ap}
Ray Renken and Michael~E. Peskin.
\newblock Corrections to weak interaction parameters in theories of
  technicolor.
\newblock {\em Nucl. Phys.}, B211:93, 1983.

\bibitem{Gasser:1984yg}
J.~Gasser and H.~Leutwyler.
\newblock Chiral perturbation theory to one loop.
\newblock {\em Ann. Phys.}, 158:142, 1984.

\bibitem{Gasser:1985gg}
J.~Gasser and H.~Leutwyler.
\newblock Chiral perturbation theory: Expansions in the mass of the strange
  quark.
\newblock {\em Nucl. Phys.}, B250:465, 1985.

\bibitem{Golden:1991ig}
Mitchell Golden and Lisa Randall.
\newblock Radiative corrections to electroweak parameters in technicolor
  theories.
\newblock {\em Nucl. Phys.}, B361:3--23, 1991.

\bibitem{Holdom:1990tc}
B.~Holdom and J.~Terning.
\newblock Large corrections to electroweak parameters in technicolor theories.
\newblock {\em Phys. Lett.}, B247:88--92, 1990.

\bibitem{Dobado:1991zh}
Antonio Dobado, Domenec Espriu, and Maria~J. Herrero.
\newblock Chiral lagrangians as a tool to probe the symmetry breaking sector of
  the sm at lep.
\newblock {\em Phys. Lett.}, B255:405--414, 1991.

\bibitem{Weinberg:1966kf}
Steven Weinberg.
\newblock Pion scattering lengths.
\newblock {\em Phys. Rev. Lett.}, 17:616--621, 1966.

\bibitem{Chanowitz:1998wi}
Michael~S. Chanowitz.
\newblock Strong w w scattering at the end of the 90's: Theory and experimental
  prospects.
\newblock 1998.

\bibitem{Chanowitz:1996si}
Michael~S. Chanowitz.
\newblock Gauge invariant formulation of strong w w scattering.
\newblock {\em Phys. Lett.}, B388:161--166, 1996.

\bibitem{Chanowitz:1987vj}
Michael Chanowitz, Mitchell Golden, and Howard Georgi.
\newblock Low-energy theorems for strongly interacting ws and zs.
\newblock {\em Phys. Rev.}, D36:1490, 1987.

\bibitem{Chanowitz:1986hu}
Michael Chanowitz, Mitchell Golden, and Howard Georgi.
\newblock Universal scattering theorems for strongly interacting ws and zs.
\newblock {\em Phys. Rev. Lett.}, 57:2344, 1986.

\bibitem{Chanowitz:1985hj}
M.~S. Chanowitz and Mary~K. Gaillard.
\newblock The tev physics of strongly interacting w's and z's.
\newblock {\em Nucl. Phys.}, B261:379, 1985.

\bibitem{Chanowitz:1984ne}
Michael~S. Chanowitz and Mary~K. Gaillard.
\newblock Multiple production of w and z as a signal of new strong
  interactions.
\newblock {\em Phys. Lett.}, B142:85, 1984.

\bibitem{Berger:1992tv}
Micheal~S. Berger and Michael~S. Chanowitz.
\newblock Probing w boson and top quark mass generation with strong z z
  scattering signals.
\newblock {\em Phys. Rev. Lett.}, 68:757--760, 1992.

\bibitem{Chanowitz:1994zh}
Michael~S. Chanowitz and William Kilgore.
\newblock Complementarity of resonant and nonresonant strong w w scattering at
  the lhc.
\newblock {\em Phys. Lett.}, B322:147--153, 1994.

\bibitem{Cornwall:1974km}
John~M. Cornwall, David~N. Levin, and George Tiktopoulos.
\newblock Derivation of gauge invariance from high-energy unitarity bounds on
  the s - matrix.
\newblock {\em Phys. Rev.}, D10:1145, 1974.

\bibitem{Yao:1988aj}
York-Peng Yao and C.~P. Yuan.
\newblock Modification of the equivalence theorem due to loop corrections.
\newblock {\em Phys. Rev.}, D38:2237, 1988.

\bibitem{Bagger:1990fc}
Jonathan Bagger and Carl Schmidt.
\newblock Equivalence theorem redux.
\newblock {\em Phys. Rev.}, D41:264, 1990.

\bibitem{He:1992ng}
Hong-Jian He, Yu-Ping Kuang, and Xiao-yuan Li.
\newblock On the precise formulation of equivalence theorem.
\newblock {\em Phys. Rev. Lett.}, 69:2619--2622, 1992.

\bibitem{He:1994yd}
Hong-Jian He, Yu-Ping Kuang, and Xiao-yuan Li.
\newblock Further investigation on the precise formulation of the equivalence
  theorem.
\newblock {\em Phys. Rev.}, D49:4842--4872, 1994.

\bibitem{He:1997cm}
Hong-Jian He and William~B. Kilgore.
\newblock The equivalence theorem and its radiative correction-free formulation
  for all r(xi) gauges.
\newblock {\em Phys. Rev.}, D55:1515--1532, 1997.

\bibitem{He:1995br}
Hong-Jian He, Yu-Ping Kuang, and C.~P. Yuan.
\newblock Equivalence theorem and probing the electroweak symmetry breaking
  sector.
\newblock {\em Phys. Rev.}, D51:6463--6473, 1995.

\bibitem{He:1994qa}
Hong-Jian He, Yu-Ping Kuang, and Xiao-yuan Li.
\newblock Proof of the equivalence theorem in the chiral lagrangian formalism.
\newblock {\em Phys. Lett.}, B329:278--284, 1994.

\bibitem{Golden:1995xv}
M.~Golden, T.~Han, and G.~Valencia.
\newblock Strongly-interacting electroweak sector: Model independent
  approaches.
\newblock 1995.

\bibitem{Georgi:1978gs}
H.~M. Georgi, S.~L. Glashow, M.~E. Machacek, and D.~V. Nanopoulos.
\newblock Higgs bosons from two gluon annihilation in proton-proton collisions.
\newblock {\em Phys. Rev. Lett.}, 40:692, 1978.

\bibitem{Dicus:1987dj}
Duane~A. Dicus, Chung Kao, and W.~W. Repko.
\newblock Gluon production of gauge bosons.
\newblock {\em Phys. Rev.}, D36:1570, 1987.

\bibitem{Glover:1989rg}
E.~W.~N. Glover and J.~J. van~der Bij.
\newblock Z boson pair production via gluon fusion.
\newblock {\em Nucl. Phys.}, B321:561, 1989.

\bibitem{Kao:1991tt}
Chung Kao and Duane~A. Dicus.
\newblock Production of w+ w- from gluon fusion.
\newblock {\em Phys. Rev.}, D43:1555--1559, 1991.

\bibitem{Jones:1979bq}
D.~R.~T. Jones and S.~T. Petcov.
\newblock Heavy higgs bosons at lep.
\newblock {\em Phys. Lett.}, B84:440, 1979.

\bibitem{Dicus:1985zg}
Duane~A. Dicus and Scott S.~D. Willenbrock.
\newblock Higgs bosons from vector boson fusion in e+ e-, e p and p p
  collisions.
\newblock {\em Phys. Rev.}, D32:1642, 1985.

\bibitem{Hioki:1983yz}
Zenro Hioki, Shoichi Midorikawa, and Hiroyuki Nishiura.
\newblock Higgs boson production in high-energy lepton - nucleon scattering.
\newblock {\em Prog. Theor. Phys.}, 69:1484, 1983.

\bibitem{Han:1985zn}
T.~Han and H.~C. Liu.
\newblock Production of charged and neutral higgs bosons in high- energy lepton
  nucleon interactions.
\newblock {\em Z. Phys.}, C28:295--301, 1985.

\bibitem{Cahn:1984ip}
R.~N. Cahn and Sally Dawson.
\newblock Production of very massive higgs bosons.
\newblock {\em Phys. Lett.}, B136:196, 1984.

\bibitem{Boos:1999kj}
E.~Boos et~al.
\newblock Strongly interacting vector bosons at tev e+- e- linear colliders.
  (addendum).
\newblock {\em Phys. Rev.}, D61:077901, 2000.

\bibitem{Boos:1998gw}
E.~Boos et~al.
\newblock Strongly interacting vector bosons at tev e+- e- linear colliders.
\newblock {\em Phys. Rev.}, D57:1553, 1998.

\bibitem{Accomando:1998wt}
E.~Accomando et~al.
\newblock Physics with e+ e- linear colliders.
\newblock {\em Phys. Rept.}, 299:1--78, 1998.

\bibitem{He:1997rb}
H.~J. He, Y.~P. Kuang, and C.~P. Yuan.
\newblock Estimating the sensitivity of lhc to electroweak symmetry breaking:
  Longitudinal/goldstone boson equivalence as a criterion.
\newblock {\em Phys. Rev.}, D55:3038--3067, 1997.

\bibitem{He:1996nm}
Hong-Jian He, Yu-Ping Kuang, and C.~P. Yuan.
\newblock Global power counting analysis on probing electroweak symmetry
  breaking mechanism at high energy colliders.
\newblock {\em Phys. Lett.}, B382:149--156, 1996.

\bibitem{Rosenfeld:1988th}
Rogerio Rosenfeld and Jonathan~L. Rosner.
\newblock Substructure of the strongly interacting higgs sector.
\newblock {\em Phys. Rev.}, D38:1530, 1988.

\bibitem{Rosenfeld:1989im}
Rogerio Rosenfeld.
\newblock Techniomega production in p p colliders.
\newblock {\em Phys. Rev.}, D39:971, 1989.

\bibitem{Carone:1994vg}
Christopher~D. Carone and Mitchell Golden.
\newblock Detecting the technirho in technicolor models with scalars.
\newblock {\em Phys. Rev.}, D49:6211--6219, 1994.

\bibitem{Casalbuoni:1995as}
R.~Casalbuoni et~al.
\newblock Pseudogoldstones at future colliders from the extended bess model.
\newblock {\em Z. Phys.}, C65:327--336, 1995.

\bibitem{Bagger:1994zf}
J.~Bagger et~al.
\newblock The strongly interacting w w system: Gold plated modes.
\newblock {\em Phys. Rev.}, D49:1246--1264, 1994.

\bibitem{Chanowitz:1995ap}
Michael~S. Chanowitz and William~B. Kilgore.
\newblock W+ z and w+ gamma* backgrounds to strong w+ w+ scattering at the lhc.
\newblock {\em Phys. Lett.}, B347:387--393, 1995.

\bibitem{Casalbuoni:1993su}
R.~Casalbuoni et~al.
\newblock Vector resonances from a strong electroweak sector at linear
  colliders.
\newblock {\em Z. Phys.}, C60:315--326, 1993.

\bibitem{Lane:1991qh}
Kenneth Lane and M.~V. Ramana.
\newblock Walking technicolor signatures at hadron colliders.
\newblock {\em Phys. Rev.}, D44:2678--2700, 1991.

\bibitem{Eichten:1998kn}
Estia Eichten, Kenneth Lane, and John Womersley.
\newblock Narrow technihadron production at the first muon collider.
\newblock {\em Phys. Rev. Lett.}, 80:5489--5492, 1998.

\bibitem{Zerwekh:2001uq}
Alfonso~R. Zerwekh and Rogerio Rosenfeld.
\newblock Gauge invariance, color-octet vector resonances and double technieta
  production at the tevatron.
\newblock {\em Phys. Lett.}, B503:325--330, 2001.

\bibitem{SekharChivukula:2001gv}
R.~Sekhar~Chivukula, Aaron Grant, and Elizabeth~H. Simmons.
\newblock Two-gluon coupling and collider phenomenology of color- octet
  technirho mesons.
\newblock 2001.

\bibitem{Sakurai:1960ju}
J.~J. Sakurai.
\newblock Theory of strong interactions.
\newblock {\em Annals Phys.}, 11:1--48, 1960.

\bibitem{Bando:1988br}
Masako Bando, Taichiro Kugo, and Koichi Yamawaki.
\newblock Nonlinear realization and hidden local symmetries.
\newblock {\em Phys. Rept.}, 164:217--314, 1988.

\bibitem{Bando:1988ym}
Masako Bando, Takanori Fujiwara, and Koichi Yamawaki.
\newblock Generalized hidden local symmetry and the a1 meson.
\newblock {\em Prog. Theor. Phys.}, 79:1140, 1988.

\bibitem{Bando:1985rf}
Masako Bando, Taichiro Kugo, and Koichi Yamawaki.
\newblock On the vector mesons as dynamical gauge bosons of hidden local
  symmetries.
\newblock {\em Nucl. Phys.}, B259:493, 1985.

\bibitem{Bando:1985ej}
M.~Bando, T.~Kugo, S.~Uehara, K.~Yamawaki, and T.~Yanagida.
\newblock Is rho meson a dynamical gauge boson of hidden local symmetry?
\newblock {\em Phys. Rev. Lett.}, 54:1215, 1985.

\bibitem{Farhi:1979zx}
E.~Farhi and Leonard Susskind.
\newblock A technicolored g.u.t.
\newblock {\em Phys. Rev.}, D20:3404--3411, 1979.

\bibitem{Eichten:1980ah}
Estia Eichten and Kenneth Lane.
\newblock Dynamical breaking of weak interaction symmetries.
\newblock {\em Phys. Lett.}, B90:125--130, 1980.

\bibitem{Chadha:1981yt}
S.~Chadha and M.~E. Peskin.
\newblock Implications of chiral dynamics in theories of technicolor. 2. the
  mass of the p+.
\newblock {\em Nucl. Phys.}, B187:541, 1981.

\bibitem{Chadha:1981rw}
S.~Chadha and M.~E. Peskin.
\newblock Implications of chiral dynamics in theories of technicolor. 1.
  elementary couplings.
\newblock {\em Nucl. Phys.}, B185:61, 1981.

\bibitem{Chivukula:1995dt}
R.~Sekhar Chivukula, Rogerio Rosenfeld, Elizabeth~H. Simmons, and John Terning.
\newblock Strongly coupled electroweak symmetry breaking: Implication of
  models.
\newblock 1995.

\bibitem{Casalbuoni:1998fs}
R.~Casalbuoni et~al.
\newblock Detecting and studying the lightest pseudo-goldstone boson at future
  p p, e+ e- and mu+ mu- colliders.
\newblock {\em Nucl. Phys.}, B555:3--52, 1999.

\bibitem{Casalbuoni:1992nw}
R.~Casalbuoni et~al.
\newblock The pseudogoldstones mass spectrum.
\newblock {\em Phys. Lett.}, B285:103--112, 1992.

\bibitem{Dimopoulos:1979es}
Savas Dimopoulos and Leonard Susskind.
\newblock Mass without scalars.
\newblock {\em Nucl. Phys.}, B155:237, 1979.

\bibitem{Eichten:1979ah}
Estia Eichten and Kenneth Lane.
\newblock Dynamical breaking of weak interaction symmetries.
\newblock {\em Phys. Lett.}, B90:125, 1980.

\bibitem{Skiba:1996ne}
Witold Skiba.
\newblock Signatures of technicolor models with the gim mechanism.
\newblock {\em Nucl. Phys.}, B470:84--112, 1996.

\bibitem{Devoto:1983sh}
A.~Devoto, D.~Ww. Duke, J.~f.~Owens, and R.~g.~Roberts.
\newblock Direct analysis of scaling violations in large q**2 deep inelastic
  neutrino and muon scattering.
\newblock {\em Phys. Rev.}, D27:508--522, 1983.

\bibitem{Hewett:1993ks}
J.~L. Hewett, T.~G. Rizzo, S.~Pakvasa, H.~E. Haber, and A.~Pomarol.
\newblock Vector leptoquark production at hadron colliders.
\newblock 1993.

\bibitem{Chivukula:1991zk}
R.~Sekhar Chivukula, Mitchell Golden, and Elizabeth~H. Simmons.
\newblock Multi - jet physics at hadron colliders.
\newblock {\em Nucl. Phys.}, B363:83--96, 1991.

\bibitem{Chivukula:1991di}
R.~Sekhar Chivukula, Mitchell Golden, and Elizabeth~H. Simmons.
\newblock Six jet signals of highly colored fermions.
\newblock {\em Phys. Lett.}, B257:403--408, 1991.

\bibitem{Parke:1986gb}
Stephen~J. Parke and T.~R. Taylor.
\newblock An amplitude for n gluon scattering.
\newblock {\em Phys. Rev. Lett.}, 56:2459, 1986.

\bibitem{Kunszt:1988it}
Z.~Kunszt and W.~J. Stirling.
\newblock Multi - jet cross-sections in hadronic collisions.
\newblock {\em Phys. Rev.}, D37:2439, 1988.

\bibitem{Mangano:1989rp}
Michelangelo Mangano and Stephen Parke.
\newblock Approximate multi - jet cross-sections in qcd.
\newblock {\em Phys. Rev.}, D39:758, 1989.

\bibitem{Mangano:1990by}
Michelangelo~L. Mangano and Stephen~J. Parke.
\newblock Multiparton amplitudes in gauge theories.
\newblock {\em Phys. Rept.}, 200:301, 1991.

\bibitem{Berends:1989ie}
F.~A. Berends, W.~T. Giele, and H.~Kuijf.
\newblock On six jet production at hadron colliders.
\newblock {\em Phys. Lett.}, B232:266, 1989.

\bibitem{Berends:1990hf}
F.~A. Berends, W.~T. Giele, and H.~Kuijf.
\newblock Exact and approximate expressions for multi - gluon scattering.
\newblock {\em Nucl. Phys.}, B333:120, 1990.

\bibitem{Belyaev:1999xe}
Alexander Belyaev, Rogerio Rosenfeld, and Alfonso~R. Zerwekh.
\newblock Tevatron potential for technicolor search with prompt photons.
\newblock {\em Phys. Lett.}, B462:150--157, 1999.

\bibitem{Hayot:1981gg}
F.~Hayot and O.~Napoly.
\newblock Detecting a heavy colored object at the fnal tevatron.
\newblock {\em Zeit. Phys.}, C7:229, 1981.

\bibitem{Hill:1994hs}
Christopher~T. Hill and Stephen~J. Parke.
\newblock Top production: Sensitivity to new physics.
\newblock {\em Phys. Rev.}, D49:4454--4462, 1994.

\bibitem{Bagger:1991qx}
J.~Bagger, S.~Dawson, and G.~Valencia.
\newblock Testing electroweak symmetry breaking through gluon fusion at p p
  colliders.
\newblock {\em Phys. Rev. Lett.}, 67:2256--2259, 1991.

\bibitem{Chivukula:1992bk}
R.~Sekhar Chivukula, Mitchell Golden, and M.~V. Ramana.
\newblock Colored pseudogoldstone bosons and gauge boson pairs.
\newblock {\em Phys. Rev. Lett.}, 68:2883--2886, 1992.

\bibitem{Soldate:1990fh}
Mark Soldate and Raman Sundrum.
\newblock Z couplings to pseudogoldstone bosons within extended technicolor.
\newblock {\em Nucl. Phys.}, B340:1--32, 1990.

\bibitem{Chivukula:1993gi}
R.~Sekhar Chivukula, Michael~J. Dugan, and Mitchell Golden.
\newblock Analyticity, crossing symmetry and the limits of chiral perturbation
  theory.
\newblock {\em Phys. Rev.}, D47:2930--2939, 1993.

\bibitem{Barklow:1994uf}
Timothy~L. Barklow.
\newblock Studies of strong electroweak symmetry breaking at future e+ e-
  linear colliders.
\newblock Presented at 1994 Meeting of the American Physical Society, Division
  of Particles and Fields (DPF 94), Albuquerque, NM, 2-6 Aug 1994.

\bibitem{Barklow:1995sk}
T.~L. Barklow.
\newblock Studies of w w gamma and w w z couplings at future e+ e- linear
  colliders.
\newblock In *Los Angeles 1995, Vector boson self-interactions* 307- 322.

\bibitem{Barklow:2002su}
Timothy~L. Barklow, R.~Sekhar Chivukula, Joel Goldstein, Tao Han, and et~al.
\newblock Electroweak symmetry breaking by strong dynamics and the collider
  phenomenology.
\newblock 2002.

\bibitem{Barger:1995cn}
V.~Barger, Kingman Cheung, T.~Han, and R.~J.~N. Phillips.
\newblock Probing strongly interacting electroweak dynamics through w+ w- / z z
  ratios at future e+ e- colliders.
\newblock {\em Phys. Rev.}, D52:3815--3825, 1995.

\bibitem{Kurihara:1993jv}
Y.~Kurihara and R.~Najima.
\newblock Heavy higgs search at tev e+ e- linear collider.
\newblock {\em Phys. Lett.}, B301:292--297, 1993.

\bibitem{Barger:1994wa}
V.~Barger, J.~F. Beacom, Kingman Cheung, and T.~Han.
\newblock Production of weak bosons and higgs bosons in e- e- collisions.
\newblock {\em Phys. Rev.}, D50:6704--6712, 1994.

\bibitem{Kurihara:1993jg}
Yoshimasa Kurihara and Ryuichi Najima.
\newblock Study of the electroweak symmetry breaking through a reaction e+ e-
  $\to$ electron-neutrino anti-electron- neutrino w+ w-.
\newblock KEK-PREPRINT-93-90.

\bibitem{Brodsky:1993xp}
Stanley~J. Brodsky.
\newblock Photon-photon collisions at the next linear collider: Theory.
\newblock Presented at 2nd International Workshop on Physics and Experiments
  with Linear e+ e- Colliders, Waikoloa, HI, 26- 30 Apr 1993.

\bibitem{Jikia:1995vz}
G.~Jikia.
\newblock Four weak gauge boson production at photon linear collider and heavy
  higgs signal.
\newblock {\em Nucl. Phys.}, B437:520--540, 1995.

\bibitem{Cheung:1994sa}
Kingman Cheung.
\newblock Studies of strong electroweak symmetry breaking at photon colliders.
\newblock {\em Phys. Rev.}, D50:4290--4298, 1994.

\bibitem{Bell:1969ts}
J.~S. Bell and R.~Jackiw.
\newblock A pcac puzzle: pi0 $\to$ gamma gamma in the sigma model.
\newblock {\em Nuovo Cim.}, A60:47--61, 1969.

\bibitem{Adler:1969gk}
S.~L. Adler.
\newblock Axial vector vertex in spinor electrodynamics.
\newblock {\em Phys. Rev.}, 177:2426--2438, 1969.

\bibitem{Manohar:1990eg}
Aneesh Manohar and Lisa Randall.
\newblock Searching for neutral pseudogoldstone bosons in z0 decays.
\newblock {\em Phys. Lett.}, B246:537--540, 1990.

\bibitem{Randall:1992gp}
Lisa Randall and Elizabeth~H. Simmons.
\newblock Signatures of neutral pseudogoldstone bosons from technicolor.
\newblock {\em Nucl. Phys.}, B380:3--21, 1992.

\bibitem{Rupak:1995kg}
Gautam Rupak and Elizabeth~H. Simmons.
\newblock Limits on pseudoscalar bosons from rare z decays at lep.
\newblock {\em Phys. Lett.}, B362:155--163, 1995.

\bibitem{Lubicz:1996xi}
Vittorio Lubicz and Pietro Santorelli.
\newblock Production of neutral pseudogoldstone bosons at lep-2 and nlc in
  multiscale walking technicolor models.
\newblock {\em Nucl. Phys.}, B460:3--36, 1996.

\bibitem{Ellis:1981hz}
John Ellis, Mary~K. Gaillard, Dimitri~V. Nanopoulos, and Pierre Sikivie.
\newblock Can one tell technicolor from a hole in the ground?
\newblock {\em Nucl. Phys. B}, 182:529--545, 1981.

\bibitem{Holdom:1981bg}
Bob Holdom.
\newblock A phenomenological lagrangian for hypercolor.
\newblock {\em Phys. Rev. D}, 24:157--183, 1981.

\bibitem{Lynch:2000hi}
Kevin~R. Lynch and Elizabeth~H. Simmons.
\newblock Z0 boson decays to composite scalars: Constraining technicolor
  theories.
\newblock {\em Phys. Rev.}, D64:035008, 2001.

\bibitem{Appelquist:1993gi}
Thomas Appelquist and John Terning.
\newblock Revenge of the one family technicolor models.
\newblock {\em Phys. Lett.}, B315:139--145, 1993.

\bibitem{Lane:1999uh}
Kenneth Lane.
\newblock Technihadron production and decay in low-scale technicolor.
\newblock {\em Phys. Rev.}, D60:075007, 1999.

\bibitem{Lane:1999uk}
Kenneth Lane.
\newblock Technihadron production and decay rates in the technicolor straw man
  model.
\newblock Companion to \cite{Lane:1999uh}, 1999.

\bibitem{Swartz:1996}
M.L. Swartz.
\newblock The search for pseudogoldstone bosons at a high energy linear
  collider.
\newblock {\em Proceedings, Snowmass 1996}, pages 956--960, 1996.

\bibitem{Ginzburg:1983vm}
I.~F. Ginzburg, G.~L. Kotkin, V.~G. Serbo, and V.~I. Telnov.
\newblock Colliding gamma e and gamma gamma beams based on the single pass
  accelerators (of vlepp type).
\newblock {\em Nucl. Instr. Meth.}, 205:47, 1983.

\bibitem{Wang:1998us}
Xue-Lei Wang, Yu-Ping Kuang, Hong-Yi Zhou, Hua Wang, and Ling Zhang.
\newblock Single top quark production in e gamma collisions and testing
  technicolor models.
\newblock {\em Phys. Rev.}, D60:014002, 1999.

\bibitem{Dimopoulos:1981kp}
S.~Dimopoulos and L.~Susskind.
\newblock Mass generation by nonstrong interactions.
\newblock {\em Nucl. Phys.}, B191:370, 1981.

\bibitem{Sikivie:1980hm}
P.~Sikivie, L.~Susskind, M.~Voloshin, and V.~Zakharov.
\newblock Isospin breaking in technicolor models.
\newblock {\em Nucl. Phys.}, B173:189, 1980.

\bibitem{Raby:1980my}
Stuart Raby, Savas Dimopoulos, and Leonard Susskind.
\newblock Tumbling gauge theories.
\newblock {\em Nucl. Phys.}, B169:373, 1980.

\bibitem{Georgi:1981mh}
Howard Georgi, Lawrence Hall, and Mark~B. Wise.
\newblock Remarks on mass hierarchies from tumbling gauge theories.
\newblock {\em Phys. Lett.}, B102:315, 1981.

\bibitem{Dimopoulos:1983gc}
Savas Dimopoulos, Howard Georgi, and Stuart Raby.
\newblock Technicolor gimnastycs.
\newblock {\em Phys. Lett.}, B127:101, 1983.

\bibitem{Veneziano:1981yz}
G.~Veneziano.
\newblock Tumbling and the strong anomaly.
\newblock {\em Phys. Lett.}, B102:139, 1981.

\bibitem{Eichten:1982mu}
E.~Eichten and F.~Feinberg.
\newblock Comment on tumbling gauge theories.
\newblock {\em Phys. Lett.}, B110:232, 1982.

\bibitem{Kobayashi:1986fz}
Tsunehiro Kobayashi.
\newblock Quark - lepton generations and tumbling complementarity in chiral
  preon models with su(n) hypercolor.
\newblock UTHEP-159.

\bibitem{Martin:1992aq}
Stephen~P. Martin.
\newblock A tumbling top quark condensate model.
\newblock {\em Phys. Rev.}, D46:2197--2202, 1992.

\bibitem{Georgi:1989pt}
Howard Georgi.
\newblock An extended technicolor model with custodial su(2) symmetry.
\newblock {\em Phys. Lett.}, B216:155, 1989.

\bibitem{Brown:2001mg}
H.~N. Brown et~al.
\newblock Precise measurement of the positive muon anomalous magnetic moment.
\newblock {\em Phys. Rev. Lett.}, 86:2227--2231, 2001.

\bibitem{Onderwater:2001ie}
Gerco Onderwater.
\newblock A new precise measurement of the muon anomalous magnetic moment.
\newblock 2001.

\bibitem{Xiong:2001rt}
Zhao-Hua Xiong and Jin~Min Yang.
\newblock Muon anomalous magnetic moment in technicolor models.
\newblock 2001.

\bibitem{Yue:2001db}
Chong xing Yue, Qing jun Xu, and Guo li~Liu.
\newblock Topcolor assisted technicolor models and muon anomalous magnetic
  moment.
\newblock 2001.

\bibitem{DiVecchia:1980wy}
P.~Di Vecchia.
\newblock Chiral dynamics for color and technicolor qcd like theories at large
  n.
\newblock WU B 80-36.

\bibitem{Chivukula:1992ap}
R.~Sekhar Chivukula, Stephen~B. Selipsky, and Elizabeth~H. Simmons.
\newblock Nonoblique effects in the z b anti-b vertex from etc dynamics.
\newblock {\em Phys. Rev. Lett.}, 69:575, 1992.

\bibitem{Lynn:1985fg}
B.~W. Lynn, Michael~E. Peskin, and R.~G. Stuart.
\newblock Radiative corrections in su(2) x u(1): Lep / slc.
\newblock Contribution to LEP Physics Study Group, March 1985: to appear in
  Proc. of LEP Physics Workshop, CERN Report 1985.

\bibitem{Peskin:1990zt}
Michael~E. Peskin and Tatsu Takeuchi.
\newblock A new constraint on a strongly interacting higgs sector.
\newblock {\em Phys. Rev. Lett.}, 65:964--967, 1990.

\bibitem{Peskin:1992sw}
Michael~E. Peskin and Tatsu Takeuchi.
\newblock Estimation of oblique electroweak corrections.
\newblock {\em Phys. Rev.}, D46:381--409, 1992.

\bibitem{Langacker:1994ah}
Paul Langacker.
\newblock Theoretical study of the electroweak interaction: Present and future.
\newblock 1994.

\bibitem{Blondel:1994ke}
Alain Blondel.
\newblock Precision electroweak physics at lep.
\newblock {\em Nucl. Phys. Proc. Suppl.}, 37B:3--22, 1994.

\bibitem{Yue:2000ay}
Chong-Xing Yue, Yu-Ping Kuang, Xue-Lei Wang, and Wei bin Li.
\newblock Restudy of the constraint on topcolor-assisted technicolor models
  from r(b).
\newblock {\em Phys. Rev.}, D62:055005, 2000.

\bibitem{Burdman:1997pf}
Gustavo Burdman and Dimitris Kominis.
\newblock Model-independent constraints on topcolor from r(b).
\newblock {\em Phys. Lett.}, B403:101--107, 1997.

\bibitem{Buras:1983ff}
A.~J. Buras, S.~Dawson, and A.~N. Schellekens.
\newblock Fermion masses, rare processes and cp violation in a class of
  extended technicolor models.
\newblock {\em Phys. Rev.}, D27:1171, 1983.

\bibitem{Giudice:1992sz}
Gian~F. Giudice and Stuart Raby.
\newblock A new paradigm for the revival of technicolor theories.
\newblock {\em Nucl. Phys.}, B368:221--247, 1992.

\bibitem{Raby:1990pq}
Stuart Raby and Gian~F. Giudice.
\newblock Jump starting stalled gutts.
\newblock Presented at 1990 Int. Workshop on Strong Coupling Gauge Theories and
  Beyond, Nagoya, Japan, Jul 28-31, 1990.

\bibitem{Peskin:2001rw}
Michael~E. Peskin and James~D. Wells.
\newblock How can a heavy higgs boson be consistent with the precision
  electroweak measurements?
\newblock 2001.

\bibitem{Swartz:1999xv}
Morris~L. Swartz.
\newblock Precision electroweak physics at the z.
\newblock {\em eConf}, C990809:307--332, 2000.

\bibitem{Dobrescu:1998nm}
Bogdan~A. Dobrescu and Christopher~T. Hill.
\newblock Electroweak symmetry breaking via top condensation seesaw.
\newblock {\em Phys. Rev. Lett.}, 81:2634--2637, 1998.

\bibitem{Chivukula:1998wd}
R.~Sekhar Chivukula, Bogdan~A. Dobrescu, Howard Georgi, and Christopher~T.
  Hill.
\newblock Top quark seesaw theory of electroweak symmetry breaking.
\newblock {\em Phys. Rev.}, D59:075003, 1999.

\bibitem{Chivukula:1992nw}
R.~Sekhar Chivukula, Michael~J. Dugan, and Mitchell Golden.
\newblock Electroweak corrections in technicolor reconsidered.
\newblock {\em Phys. Lett.}, B292:435--441, 1992.

\bibitem{Chivukula:1991bx}
R.~Sekhar Chivukula and Mitchell Golden.
\newblock Hiding the electroweak symmetry breaking sector.
\newblock {\em Phys. Lett.}, B267:233--239, 1991.

\bibitem{Chivukula:1999az}
R.~Sekhar Chivukula and Nick Evans.
\newblock Triviality and the precision bound on the higgs mass.
\newblock {\em Phys. Lett.}, B464:244--248, 1999.

\bibitem{Hall:1999fe}
Lawrence Hall and Christopher Kolda.
\newblock Electroweak symmetry breaking and large extra dimensions.
\newblock {\em Phys. Lett.}, B459:213--223, 1999.

\bibitem{Appelquist:1998xf}
Thomas Appelquist and Francesco Sannino.
\newblock The physical spectrum of conformal su(n) gauge theories.
\newblock {\em Phys. Rev.}, D59:067702, 1999.

\bibitem{Dugan:1991ck}
Michael~J. Dugan and Lisa Randall.
\newblock The sign of s from electroweak radiative corrections.
\newblock {\em Phys. Lett.}, B264:154--160, 1991.

\bibitem{Georgi:1991ci}
Howard Georgi.
\newblock Effective field theory and electroweak radiative corrections.
\newblock {\em Nucl. Phys.}, B363:301--325, 1991.

\bibitem{Gates:1991uu}
Evalyn Gates and John Terning.
\newblock Negative contributions to s from majorana particles.
\newblock {\em Phys. Rev. Lett.}, 67:1840--1843, 1991.

\bibitem{Luty:1992fe}
Markus~A. Luty and Raman Sundrum.
\newblock Technicolor theories with negative s.
\newblock 1992.

\bibitem{Maekawa:1995yd}
Nobuhiro Maekawa.
\newblock Vector - like strong coupling theory with small s and t parameters.
\newblock {\em Prog. Theor. Phys.}, 93:919--926, 1995.

\bibitem{Einhorn:1981cy}
M.~B. Einhorn, D.~R.~T. Jones, and M.~Veltman.
\newblock Heavy particles and the rho parameter in the standard model.
\newblock {\em Nucl. Phys.}, B191:146, 1981.

\bibitem{Simmons:1989pu}
Elizabeth~H. Simmons.
\newblock Separating electroweak symmetry breaking from flavor physics in an
  almost standard model.
\newblock {\em Nucl. Phys.}, B324:315, 1989.

\bibitem{Carone:1993rh}
Christopher~D. Carone and Elizabeth~H. Simmons.
\newblock Oblique corrections in technicolor with a scalar.
\newblock {\em Nucl. Phys.}, B397:591--615, 1993.

\bibitem{Carone:1994xc}
Christopher~D. Carone and Howard Georgi.
\newblock Technicolor with a massless scalar doublet.
\newblock {\em Phys. Rev.}, D49:1427--1436, 1994.

\bibitem{Rizzo:1999br}
Thomas~G. Rizzo and James~D. Wells.
\newblock Electroweak precision measurements and collider probes of the
  standard model with large extra dimensions.
\newblock {\em Phys. Rev.}, D61:016007, 2000.

\bibitem{Chankowski:2000an}
Piotr Chankowski et~al.
\newblock Do precision electroweak constraints guarantee e+ e- collider
  discovery of at least one higgs boson of a two- higgs-doublet model?
\newblock {\em Phys. Lett.}, B496:195--205, 2000.

\bibitem{Sundrum:1993xy}
Raman Sundrum.
\newblock A realistic technicolor model from 150-tev down.
\newblock {\em Nucl. Phys.}, B395:60--76, 1993.

\bibitem{Appelquist:1997fp}
Thomas Appelquist, John Terning, and L.~C.~R. Wijewardhana.
\newblock Postmodern technicolor.
\newblock {\em Phys. Rev. Lett.}, 79:2767--2770, 1997.

\bibitem{Randall:1993vt}
Lisa Randall.
\newblock Etc with a gim mechanism.
\newblock {\em Nucl. Phys.}, B403:122--140, 1993.

\bibitem{Chivukula:1987fw}
R.~Sekhar Chivukula, H.~Georgi, and L.~Randall.
\newblock A composite technicolor standard model of quarks.
\newblock {\em Nucl. Phys.}, B292:93--108, 1987.

\bibitem{Chivukula:1987py}
R.~Sekhar Chivukula and Howard Georgi.
\newblock Composite technicolor standard model.
\newblock {\em Phys. Lett.}, B188:99, 1987.

\bibitem{Glashow:1977nt}
Sheldon~L. Glashow and Steven Weinberg.
\newblock Natural conservation laws for neutral currents.
\newblock {\em Phys. Rev.}, D15:1958, 1977.

\bibitem{Abachi:1994hr}
S.~Abachi et~al.
\newblock First generation leptoquark search in p anti-p collisions at s**(1/2)
  = 1.8-tev.
\newblock {\em Phys. Rev. Lett.}, 72:965--969, 1994.

\bibitem{Abe:1995fj}
F.~Abe et~al.
\newblock A search for second generation leptoquarks in p anti-p collisions at
  s**(1/2) = 1.8-tev.
\newblock {\em Phys. Rev. Lett.}, 75:1012--1016, 1995.

\bibitem{Chivukula:1994mn}
R.~S. Chivukula, E.~H. Simmons, and J.~Terning.
\newblock A heavy top quark and the z b anti-b vertex in noncommuting extended
  technicolor.
\newblock {\em Phys. Lett.}, B331:383--389, 1994.

\bibitem{Lane:1989ej}
Kenneth Lane and Estia Eichten.
\newblock Two scale technicolor.
\newblock {\em Phys. Lett.}, B222:274, 1989.

\bibitem{Terning:1995sc}
John Terning.
\newblock Chiral technicolor and precision electroweak measurements.
\newblock {\em Phys. Lett.}, B344:279--286, 1995.

\bibitem{Hill:1995hp}
Christopher~T. Hill.
\newblock Topcolor assisted technicolor.
\newblock {\em Phys. Lett.}, B345:483--489, 1995.

\bibitem{Chivukula:1990bc}
R.~Sekhar Chivukula, Andrew~G. Cohen, and Kenneth Lane.
\newblock Aspects of dynamical electroweak symmetry breaking.
\newblock {\em Nucl. Phys.}, B343:554--570, 1990.

\bibitem{Appelquist:1991kn}
Thomas Appelquist, John Terning, and L.~C.~R. Wijewardhana.
\newblock Mass enhancement and critical behavior in technicolor theories.
\newblock {\em Phys. Rev.}, D44:871--877, 1991.

\bibitem{Chivukula:1996gu}
R.~S. Chivukula, E.~H. Simmons, and J.~Terning.
\newblock Limits on noncommuting extended technicolor.
\newblock {\em Phys. Rev.}, D53:5258--5267, 1996.

\bibitem{ehsrsc:2002}
R.S. Chivukula and E.H. Simmons.
\newblock Electroweak limits on non-universal weak bosons.
\newblock 2002.

\bibitem{Gusynin:1983kp}
V.~P. Gusynin, V.~A. Miransky, and Yu.~A. Sitenko.
\newblock On the dynamics of tumbling gauge theories.
\newblock {\em Phys. Lett.}, B123:407--412, 1983.

\bibitem{Kikukawa:1992ig}
Yoshio Kikukawa and Noriaki Kitazawa.
\newblock Tumbling and technicolor theory.
\newblock {\em Phys. Rev.}, D46:3117--3122, 1992.

\bibitem{Kitazawa:1994pd}
Noriaki Kitazawa.
\newblock Tumbling and technicolor theory.
\newblock 1994.

\bibitem{Sualp:1983bq}
Gonul Sualp and Sinan Kaptanoglu.
\newblock A systematic investigation of all su(n) tumbling gauge models.
\newblock {\em Ann. Phys.}, 147:460, 1983.

\bibitem{Appelquist:1994sg}
Thomas Appelquist and John Terning.
\newblock An extended technicolor model.
\newblock {\em Phys. Rev.}, D50:2116--2126, 1994.

\bibitem{Hill:1991ge}
Christopher~T. Hill, Markus~A. Luty, and Emmanuel~A. Paschos.
\newblock Electroweak symmetry breaking by fourth generation condensates and
  the neutrino spectrum.
\newblock {\em Phys. Rev.}, D43:3011--3025, 1991.

\bibitem{Gell-Mann:1978pg}
M.~Gell-Mann, P.~Ramond, and R.~Slansky.
\newblock Color embeddings, charge assignments, and proton stability in unified
  gauge theories.
\newblock {\em Rev. Mod. Phys.}, 50:721, 1978.

\bibitem{Hill:1990vn}
C.~T. Hill and E.~A. Paschos.
\newblock A naturally heavy fourth generation neutrino.
\newblock {\em Phys. Lett.}, B241:96, 1990.

\bibitem{Akhmedov:1996vm}
Eugeni Akhmedov, Manfred Lindner, Erhard Schnapka, and Jose W.~F. Valle.
\newblock Dynamical left-right symmetry breaking.
\newblock {\em Phys. Rev.}, D53:2752--2780, 1996.

\bibitem{Martin:1991xw}
Stephen~P. Martin.
\newblock Dynamical electroweak symmetry breaking with top quark and neutrino
  condensates.
\newblock {\em Phys. Rev.}, D44:2892--2898, 1991.

\bibitem{Evans:1993yj}
N.~J. Evans, S.~F. King, and D.~A. Ross.
\newblock Electroweak radiative corrections in dynamical models with majorana
  neutrinos.
\newblock {\em Phys. Lett.}, B303:295--302, 1993.

\bibitem{King:1992ev}
Stephen~F. King and Samjid~H. Mannan.
\newblock Quark and lepton masses in extended technicolor.
\newblock {\em Nucl. Phys.}, B369:119--138, 1992.

\bibitem{Holdom:1981rm}
Bob Holdom.
\newblock Raising the sideways scale.
\newblock {\em Phys. Rev.}, D24:1441, 1981.

\bibitem{Yamawaki:1986zg}
Koichi Yamawaki, Masako Bando, and Ken iti Matumoto.
\newblock Scale invariant technicolor model and a technidilaton.
\newblock {\em Phys. Rev. Lett.}, 56:1335, 1986.

\bibitem{Bando:1986bg}
Masako Bando, Ken iti Matumoto, and Koichi Yamawaki.
\newblock Technidilaton.
\newblock {\em Phys. Lett.}, B178:308, 1986.

\bibitem{Appelquist:1986an}
Thomas~W. Appelquist, Dimitra Karabali, and L.~C.~R. Wijewardhana.
\newblock Chiral hierarchies and the flavor changing neutral current problem in
  technicolor.
\newblock {\em Phys. Rev. Lett.}, 57:957, 1986.

\bibitem{Appelquist:1987tr}
Thomas Appelquist and L.~C.~R. Wijewardhana.
\newblock Chiral hierarchies and chiral perturbations in technicolor.
\newblock {\em Phys. Rev.}, D35:774, 1987.

\bibitem{Johnson:1964da}
K.~Johnson, M.~Baker, and R.~Willey.
\newblock Selfenergy of the electron.
\newblock {\em Phys. Rev.}, 136:B1111--B1119, 1964.

\bibitem{Jackiw:1973tr}
R.~Jackiw and K.~Johnson.
\newblock Dynamical model of spontaneously broken gauge symmetries.
\newblock {\em Phys. Rev.}, D8:2386, 1973.

\bibitem{Cohen:1989sq}
Andrew Cohen and Howard Georgi.
\newblock Walking beyond the rainbow.
\newblock {\em Nucl. Phys.}, B314:7, 1989.

\bibitem{Mahanta:1989rb}
U.~Mahanta.
\newblock Walking technicolor beyond the ladder approximation.
\newblock {\em Phys. Rev. Lett.}, 62:2349, 1989.

\bibitem{Holdom:1985sk}
Bob Holdom.
\newblock Techniodor.
\newblock {\em Phys. Lett.}, B150:301, 1985.

\bibitem{Appelquist:1987fc}
Thomas Appelquist and L.~C.~R. Wijewardhana.
\newblock Chiral hierarchies from slowly running couplings in technicolor
  theories.
\newblock {\em Phys. Rev.}, D36:568, 1987.

\bibitem{Appelquist:1996kp}
Thomas Appelquist, Nick Evans, and Stephen~B. Selipsky.
\newblock Phenomenology of the top mass in realistic extended technicolor
  models.
\newblock {\em Phys. Lett.}, B374:145--151, 1996.

\bibitem{Chivukula:1993tz}
R.~S. Chivukula, E.~Gates, E.~H. Simmons, and J.~Terning.
\newblock Walking technicolor and the z b anti-b vertex.
\newblock {\em Phys. Lett.}, B311:157--162, 1993.

\bibitem{Hill:1991at}
Christopher~T. Hill.
\newblock Topcolor: Top quark condensation in a gauge extension of the standard
  model.
\newblock {\em Phys. Lett.}, B266:419--424, 1991.

\bibitem{Baluni:1981ty}
Varouzhan Baluni.
\newblock Screening of instantons in broken gauge theories (origin of quark -
  lepton masses).
\newblock Print-81-0328 (IAS,PRINCETON).

\bibitem{Appelquist:1992is}
Thomas Appelquist and George Triantaphyllou.
\newblock Precision tests of technicolor.
\newblock {\em Phys. Lett.}, B278:345--350, 1992.

\bibitem{Sundrum:1993rf}
Raman Sundrum and Stephen D.~H. Hsu.
\newblock Walking technicolor and electroweak radiative corrections.
\newblock {\em Nucl. Phys.}, B391:127--146, 1993.

\bibitem{Intriligator:1996au}
K.~Intriligator and N.~Seiberg.
\newblock Lectures on supersymmetric gauge theories and electric- magnetic
  duality.
\newblock {\em Nucl. Phys. Proc. Suppl.}, 45BC:1--28, 1996.

\bibitem{Banks:1982nn}
T.~Banks and A.~Zaks.
\newblock On the phase structure of vector - like gauge theories with massless
  fermions.
\newblock {\em Nucl. Phys.}, B196:189, 1982.

\bibitem{Appelquist:1998rb}
Thomas Appelquist, Anuradha Ratnaweera, John Terning, and L.~C.~R.
  Wijewardhana.
\newblock The phase structure of an su(n) gauge theory with n(f) flavors.
\newblock {\em Phys. Rev.}, D58:105017, 1998.

\bibitem{Terning:1997jj}
John Terning.
\newblock Duals for su(n) susy gauge theories with an antisymmetric tensor:
  Five easy flavors.
\newblock {\em Phys. Lett.}, B422:149, 1998.

\bibitem{Eichten:1997yq}
Estia Eichten, Kenneth Lane, and John Womersley.
\newblock Finding low-scale technicolor at hadron colliders.
\newblock {\em Phys. Lett.}, B405:305--311, 1997.

\bibitem{Marciano:1980zf}
William~J. Marciano.
\newblock Exotic new quarks and dynamical symmetry breaking.
\newblock {\em Phys. Rev.}, D21:2425, 1980.

\bibitem{Balaji:1997va}
Bhashyam Balaji.
\newblock Top decay in topcolor-assisted technicolor.
\newblock {\em Phys. Lett.}, B393:89--93, 1997.

\bibitem{Lane:1995gw}
Kenneth Lane and Estia Eichten.
\newblock Natural topcolor assisted technicolor.
\newblock {\em Phys. Lett.}, B352:382--387, 1995.

\bibitem{Eichten:1996dx}
Estia Eichten and Kenneth Lane.
\newblock Low-scale technicolor at the tevatron.
\newblock {\em Phys. Lett.}, B388:803--807, 1996.

\bibitem{Groom:2000in}
D.~E. Groom et~al.
\newblock Review of particle physics.
\newblock {\em Eur. Phys. J.}, C15:1, 2000.

\bibitem{Affolder:1999du}
T.~Affolder et~al.
\newblock Search for technicolor particles in lepton plus two jets and multijet
  channels in p anti-p collisions at s**(1/2) = 1.8-tev.
\newblock FERMILAB-PUB-99-141-E.

\bibitem{Affolder:2000hp}
T.~Affolder et~al.
\newblock Search for color singlet technicolor particles in p anti-p collisions
  at s**(1/2) = 1.8-tev.
\newblock {\em Phys. Rev. Lett.}, 84:1110, 2000.

\bibitem{Handa:1999gx}
Takanobu Handa.
\newblock Technicolor limits at the tevatron.
\newblock To be published in the proceedings of 13th Topical Conference on
  Hadron Collider Physics, Mumbai, India, 14-20 Jan 1999.

\bibitem{Kounine:1999tp}
Andrei Kounine et~al.
\newblock Search for technicolour production at lep.
\newblock L3 Collaboration note 2428, contributed paper to the International
  Europhysics Conference on High Energy Physics 99, Tampere, Finland, 1999.

\bibitem{Borisov:2000tp}
G.~Borisov, F.~Richard, J.~Cuevas, and J.~Marco.
\newblock Search for technicolor with delphi.
\newblock DELPHI Collaboration note DELOHI2000-034 CONF 353, contributed paper
  to the XXXVth Rencontres de Moriond, Les Arcs, France, 11-25 March 2000.

\bibitem{opal:2001prelim}


\bibitem{Abazov:2001qd}
V.~M. Abazov et~al.
\newblock Search for heavy particles decaying into electron positron pairs in p
  anti-p collisions.
\newblock {\em Phys. Rev. Lett.}, 87:061802, 2001.

\bibitem{Abdallah:2001ft}
J.~Abdallah et~al.
\newblock Search for technicolor with delphi.
\newblock {\em Eur. Phys. J.}, C22:17--29, 2001.

\bibitem{Abe:1998jc}
F.~Abe et~al.
\newblock Search for a technicolor omega(t) particle in events with a photon
  and a b quark jet at cdf.
\newblock {\em Phys. Rev. Lett.}, 83:3124--3129, 1999.

\bibitem{ATLAS:TDR}
ATLAS Collaboration.
\newblock Atlas detector and physics performance technical design report.
\newblock CERN/LHCC/99-14.

\bibitem{Godfrey:1998pm}
Stephen Godfrey, Tao Han, and Pat Kalyniak.
\newblock Discovery limits for techniomega production in e gamma collisions.
\newblock {\em Phys. Rev.}, D59:095006, 1999.

\bibitem{Lynch:2000md}
Kevin~R. Lynch, Elizabeth~H. Simmons, Meenakshi Narain, and Stephen Mrenna.
\newblock Finding z' bosons coupled preferentially to the third family at lep
  and the tevatron.
\newblock {\em Phys. Rev.}, D63:035006, 2001.

\bibitem{Lane:1996ua}
Kenneth Lane.
\newblock Symmetry breaking and generational mixing in topcolor-- assisted
  technicolor.
\newblock {\em Phys. Rev.}, D54:2204--2212, 1996.

\bibitem{Lane:1998qi}
Kenneth Lane.
\newblock A new model of topcolor-assisted technicolor.
\newblock {\em Phys. Lett.}, B433:96--101, 1998.

\bibitem{Djouadi:1996gv}
A.~Djouadi, J.~Kalinowski, and P.~M. Zerwas.
\newblock Two- and three-body decay modes of susy higgs particles.
\newblock {\em Z. Phys.}, C70:435--448, 1996.

\bibitem{Ma:1998up}
Ernest Ma, D.~P Roy, and Jose Wudka.
\newblock Enhanced three-body decay of the charged higgs boson.
\newblock {\em Phys. Rev. Lett.}, 80:1162--1165, 1998.

\bibitem{higgshunt}
John~F. Gunion, Howard~E. Haber, Gordon Kane, and Sally Dawson.
\newblock {\em The Higgs Hunter's Guide}.
\newblock Ad\-di\-son-Wes\-ley Publishing Company, Reading, MA, 1990.

\bibitem{Alam:1995aw}
M.~S. Alam et~al.
\newblock First measurement of the rate for the inclusive radiative penguin
  decay b $\to$ s gamma.
\newblock {\em Phys. Rev. Lett.}, 74:2885--2889, 1995.

\bibitem{Ahmed:1999fh}
S.~Ahmed et~al.
\newblock b --> s gamma branching fraction and cp asymmetry.
\newblock 1999.

\bibitem{Ellis:1986xy}
R.~G. Ellis, G.~C. Joshi, and M.~Matsuda.
\newblock Hard gamma emission from b decay and the two higgs boson doublet
  model.
\newblock {\em Phys. Lett.}, B179:119, 1986.

\bibitem{Barger:1990fj}
V.~Barger, J.~L. Hewett, and R.~J.~N. Phillips.
\newblock New constraints on the charged higgs sector in two higgs doublet
  models.
\newblock {\em Phys. Rev.}, D41:3421, 1990.

\bibitem{Chetyrkin:1997vx}
Konstantin Chetyrkin, Mikolaj Misiak, and Manfred Munz.
\newblock Weak radiative b-meson decay beyond leading logarithms.
\newblock {\em Phys. Lett.}, B400:206--219, 1997.

\bibitem{Kagan:1998ym}
Alexander~L. Kagan and Matthias Neubert.
\newblock {QCD} anatomy of b --> x/s gamma decays.
\newblock {\em Eur. Phys. J.}, C7:5--27, 1999.

\bibitem{Adriani:1993yn}
O.~Adriani et~al.
\newblock Inclusive search for the charmless radiative decay of the b quark (b
  $\to$ s gamma).
\newblock {\em Phys. Lett.}, B317:637--646, 1993.

\bibitem{Buskulic:1995gj}
D.~Buskulic et~al.
\newblock Measurement of the b $\to$ tau- anti-tau-neutrino x branching ratio
  and an upper limit on b- $\to$ tau- anti- tau-neutrino.
\newblock {\em Phys. Lett.}, B343:444--452, 1995.

\bibitem{Adam:1996ts}
W.~Adam et~al.
\newblock Study of rare b decays with the delphi detector at lep.
\newblock {\em Z. Phys.}, C72:207--220, 1996.

\bibitem{Barate:1998vz}
R.~Barate et~al.
\newblock A measurement of the inclusive b --> s gamma branching ratio.
\newblock {\em Phys. Lett.}, B429:169--187, 1998.

\bibitem{Ackerstaff:1998yk}
K.~Ackerstaff et~al.
\newblock Measurement of the michel parameters in leptonic tau decays.
\newblock {\em Eur. Phys. J.}, C8:3--21, 1999.

\bibitem{Abe:1998uz}
F.~Abe et~al.
\newblock Search for new particles decaying to b anti-b in p anti-p collisions
  at s**(1/2) = 1.8-tev.
\newblock {\em Phys. Rev. Lett.}, 82:2038--2043, 1999.

\bibitem{Affolder:2000ny}
T.~Affolder et~al.
\newblock Search for second and third generation leptoquarks including
  production via technicolor interactions in p anti-p collisions at s**(1/2) =
  1.8-tev.
\newblock {\em Phys. Rev. Lett.}, 85:2056--2061, 2000.

\bibitem{Abe:1998iq}
F.~Abe et~al.
\newblock Search for third-generation leptoquarks from technicolor models in p
  anti-p collisions at s**(1/2) = 1.8-tev.
\newblock {\em Phys. Rev. Lett.}, 82:3206, 1999.

\bibitem{Abe:1997dn}
F.~Abe et~al.
\newblock Search for third generation leptoquarks in anti-p p collisions at
  s**(1/2) = 1.8-tev.
\newblock {\em Phys. Rev. Lett.}, 78:2906--2911, 1997.

\bibitem{Muller:1996dj}
David~J. Muller and Satyanarayan Nandi.
\newblock Topflavor: A separate su(2) for the third family.
\newblock {\em Phys. Lett.}, B383:345--350, 1996.

\bibitem{Malkawi:1996fs}
Ehab Malkawi, Tim Tait, and C.~P. Yuan.
\newblock A model of strong flavor dynamics for the top quark.
\newblock {\em Phys. Lett.}, B385:304--310, 1996.

\bibitem{Muller:1997eg}
D.~J. Muller and S.~Nandi.
\newblock A separate su(2) for the third family: Topflavor.
\newblock {\em Nucl. Phys. Proc. Suppl.}, 52A:192--194, 1997.

\bibitem{Muller:1996qs}
D.~J. Muller and S.~Nandi.
\newblock A separate su(2) for the third family: Topflavor.
\newblock 1996.

\bibitem{Simmons:1997ws}
Elizabeth~H. Simmons.
\newblock New gauge interactions and single top quark production.
\newblock {\em Phys. Rev.}, D55:5494--5500, 1997.

\bibitem{Smith:1996ij}
Martin~C. Smith and S.~Willenbrock.
\newblock Qcd and yukawa corrections to single-top-quark production via q qbar
  -> t bbar.
\newblock {\em Phys. Rev.}, D54:6696--6702, 1996.

\bibitem{Heinson:1996pi}
A.~P. Heinson.
\newblock Future top physics at the tevatron and lhc.
\newblock 1996.

\bibitem{Heinson:1997zm}
A.~P. Heinson, A.~S. Belyaev, and E.~E. Boos.
\newblock Single top quarks at the fermilab tevatron.
\newblock {\em Phys. Rev.}, D56:3114--3128, 1997.

\bibitem{Buras:1983nb}
A.~J. Buras and T.~Yanagida.
\newblock Raising the extended technicolor scale through supersymmetry.
\newblock {\em Phys. Lett.}, B121:316, 1983.

\bibitem{Murayama:2000dw}
Hitoshi Murayama.
\newblock Supersymmetry phenomenology.
\newblock 2000.

\bibitem{Dobrescu:1995gz}
Bogdan~A. Dobrescu.
\newblock Fermion masses without higgs: A supersymmetric technicolor model.
\newblock {\em Nucl. Phys.}, B449:462--482, 1995.

\bibitem{Buras:1983yg}
A.~J. Buras and W.~Slominski.
\newblock Gauge invariant effective lagrangians for goldstone like particles of
  supersymmetric technicolor models.
\newblock {\em Nucl. Phys.}, B223:157, 1983.

\bibitem{Nilles:1982my}
H.~Peter Nilles.
\newblock Is supersymmetry afraid of condensates?
\newblock {\em Phys. Lett.}, B112:455, 1982.

\bibitem{Liu:1999ah}
Chun Liu.
\newblock Topcolor-assisted supersymmetry.
\newblock {\em Phys. Rev.}, D61:115001, 2000.

\bibitem{Samuel:1990dq}
Stuart Samuel.
\newblock Bosonic technicolor.
\newblock {\em Nucl. Phys.}, B347:625--650, 1990.

\bibitem{Dine:1990jd}
Michael Dine, Alex Kagan, and Stuart Samuel.
\newblock Naturalness in supersymmetry, or raising the supersymmetry breaking
  scale.
\newblock {\em Phys. Lett.}, B243:250--256, 1990.

\bibitem{Kagan:1991gh}
Alex Kagan and Stuart Samuel.
\newblock Renormalization group aspects of bosonic technicolor.
\newblock {\em Phys. Lett.}, B270:37--44, 1991.

\bibitem{Kagan:1990az}
Alex Kagan and Stuart Samuel.
\newblock The family mass hierarchy problem in bosonic technicolor.
\newblock {\em Phys. Lett.}, B252:605--610, 1990.

\bibitem{Kagan:1992gi}
Alex Kagan and Stuart Samuel.
\newblock Multi - higgs systems in bosonic technicolor: A model for ssc
  physics.
\newblock {\em Int. J. Mod. Phys.}, A7:1123--1186, 1992.

\bibitem{Kagan:1991ng}
Alex Kagan.
\newblock Recent developments in bosonic technicolor.
\newblock to appear in Proc. of 15th Johns Hopkins Workshop on Current Problems
  in Particle Theory, Baltimore, MD, Aug 26-28, 1991.

\bibitem{Dobrescu:1998ci}
Bogdan~A. Dobrescu and Elizabeth~H. Simmons.
\newblock Top-bottom splitting in technicolor with composite scalars.
\newblock {\em Phys. Rev.}, D59:015014, 1999.

\bibitem{Carone:1995mx}
Christopher~D. Carone, Elizabeth~H. Simmons, and Yumian Su.
\newblock b $\to$ s gamma and z $\to$ b anti-b in technicolor with scalars.
\newblock {\em Phys. Lett.}, B344:287--292, 1995.

\bibitem{Hemmige:2001vq}
Vagish Hemmige and Elizabeth~H. Simmons.
\newblock Current bounds on technicolor with scalars.
\newblock {\em Phys. Lett.}, B518:72--78, 2001.

\bibitem{Kaplan:1991dc}
David~B. Kaplan.
\newblock Flavor at ssc energies: A new mechanism for dynamically generated
  fermion masses.
\newblock {\em Nucl. Phys.}, B365:259--278, 1991.

\bibitem{Kagan:1995qg}
Alexander~L. Kagan.
\newblock Implications of tev flavor physics for the delta i = 1/2 rule and
  br(l)(b).
\newblock {\em Phys. Rev.}, D51:6196--6220, 1995.

\bibitem{Dobrescu:1997kt}
Bogdan~A. Dobrescu and John Terning.
\newblock Negative contributions to s in an effective field theory.
\newblock {\em Phys. Lett.}, B416:129, 1998.

\bibitem{Atwood:1995vm}
D.~Atwood, A.~Kagan, and T.~G. Rizzo.
\newblock Constraining anomalous top quark couplings at the tevatron.
\newblock {\em Phys. Rev.}, D52:6264--6270, 1995.

\bibitem{Pendleton:1981as}
B.~Pendleton and G.~G. Ross.
\newblock Mass and mixing angle predictions from infrared fixed points.
\newblock {\em Phys. Lett.}, B98:291, 1981.

\bibitem{Hill:1981yr}
Christopher~T. Hill.
\newblock Fixed points; fermion mass predictions.
\newblock Talk delivered at the 2nd Workshop on Grand Unification, Ann Arbor,
  Michigan, Apr 24-26, 1981 (see also the concluding remarks of S. Weinberg in
  this conference volume).

\bibitem{Hill:1981sq}
Christopher~T. Hill.
\newblock Quark and lepton masses from renormalization group fixed points.
\newblock {\em Phys. Rev.}, D24:691, 1981.

\bibitem{Hill:1985tg}
Christopher~T. Hill, Chung~Ngoc Leung, and Sumathi Rao.
\newblock Renormalization group fixed points and the higgs boson spectrum.
\newblock {\em Nucl. Phys.}, B262:517, 1985.

\bibitem{Oehme:1985jy}
Reinhard Oehme, Klaus Sibold, and Wolfhart Zimmermann.
\newblock Construction of gauge theories with a single coupling parameter for
  yang-mills and matter fields.
\newblock {\em Phys. Lett.}, B153:142, 1985.

\bibitem{Bardeen:1994rv}
W.~A. Bardeen, M.~Carena, S.~Pokorski, and C.~E.~M. Wagner.
\newblock Infrared fixed point solution for the top quark mass and unification
  of couplings in the mssm.
\newblock {\em Phys. Lett.}, B320:110--116, 1994.

\bibitem{Bardeen:1990ds}
William~A. Bardeen, Christopher~T. Hill, and Manfred Lindner.
\newblock Minimal dynamical symmetry breaking of the standard model.
\newblock {\em Phys. Rev.}, D41:1647, 1990.

\bibitem{Nambu:1988mr}
Y.~Nambu.
\newblock Quasisupersymmetry, bootstrap symmetry breaking and fermion masses.
\newblock Invited talk to appear in Proc. of 1988 Int. Workshop New Trends in
  Strong Coupling Gauge Theories, Nagoya, Japan, Aug 24-27, 1988.

\bibitem{Miransky:1989xi}
V.~A. Miransky, Masaharu Tanabashi, and Koichi Yamawaki.
\newblock Dynamical electroweak symmetry breaking with large anomalous
  dimension and t quark condensate.
\newblock {\em Phys. Lett.}, B221:177, 1989.

\bibitem{Miransky:1989ds}
V.~A. Miransky, Masaharu Tanabashi, and Koichi Yamawaki.
\newblock Is the t quark responsible for the mass of w and z bosons?
\newblock {\em Mod. Phys. Lett.}, A4:1043, 1989.

\bibitem{Marciano:1990mj}
William~J. Marciano.
\newblock Dynamical symmetry breaking and the top quark mass.
\newblock {\em Phys. Rev.}, D41:219, 1990.

\bibitem{Marciano:1989xd}
W.~J. Marciano.
\newblock Heavy top quark mass predictions.
\newblock {\em Phys. Rev. Lett.}, 62:2793--2796, 1989.

\bibitem{Miransky:1991gz}
Vladimir~A. Miransky.
\newblock Electroweak symmetry breaking and dynamics of tight bound states.
\newblock {\em Int. J. Mod. Phys.}, A6:1641--1658, 1991.

\bibitem{Hasenfratz:1991it}
Anna Hasenfratz, Peter Hasenfratz, Karl Jansen, Julius Kuti, and Yue Shen.
\newblock The equivalence of the top quark condensate and the elementary higgs
  field.
\newblock {\em Nucl. Phys.}, B365:79--97, 1991.

\bibitem{Lindner:1993ah}
Manfred Lindner.
\newblock Top condensates as higgs substitute.
\newblock {\em Int. J. Mod. Phys.}, A8:2167--2240, 1993.

\bibitem{Lindner:1992bs}
Manfred Lindner and Douglas Ross.
\newblock Top condensation from very massive strongly coupled gauge bosons.
\newblock {\em Nucl. Phys.}, B370:30--50, 1992.

\bibitem{Blumhofer:1993kv}
A.~Blumhofer and M.~Lindner.
\newblock Custodial su(2) violation and the origin of fermion masses.
\newblock {\em Nucl. Phys.}, B407:173--190, 1993.

\bibitem{Carena:1992ky}
M.~Carena, T.~E. Clark, C.~E.~M. Wagner, W.~A. Bardeen, and K.~Sasaki.
\newblock Dynamical symmetry breaking and the top quark mass in the minimal
  supersymmetric standard model.
\newblock {\em Nucl. Phys.}, B369:33--53, 1992.

\bibitem{Clark:1990tq}
T.~E. Clark, S.~T. Love, and William~A. Bardeen.
\newblock The top quark mass in a supersymmetric standard model with dynamical
  symmetry breaking.
\newblock {\em Phys. Lett.}, B237:235, 1990.

\bibitem{Luty:1990bg}
Markus~A. Luty.
\newblock Dynamical electroweak symmetry breaking with two composite higgs
  doublets.
\newblock {\em Phys. Rev.}, D41:2893, 1990.

\bibitem{Andrianov:1999xd}
A.~A. Andrianov, V.~A. Andrianov, and R.~Rodenberg.
\newblock Composite two-higgs models and chiral symmetry restoration.
\newblock {\em JHEP}, 06:003, 1999.

\bibitem{Andrianov:1996ag}
A.~A. Andrianov, V.~A. Andrianov, V.~L. Yudichev, and R.~Rodenberg.
\newblock Composite two-higgs models.
\newblock {\em Int. J. Mod. Phys.}, A14:323, 1999.

\bibitem{Babu:1991vx}
K.~S. Babu and Rabindra~N. Mohapatra.
\newblock Top quark mass in a dynamical symmetry breaking scheme with radiative
  b quark and tau lepton masses.
\newblock {\em Phys. Rev. Lett.}, 66:556--559, 1991.

\bibitem{Froggatt:1990wa}
C.~D. Froggatt, I.~G. Knowles, and R.~G. Moorhouse.
\newblock Third generation masses from a two higgs model fixed point.
\newblock {\em Phys. Lett.}, B249:273--280, 1990.

\bibitem{Siegemund-Broka:1992ei}
Stephan Siegemund-Broka.
\newblock A leptoquark model of dynamical electroweak symmetry breaking.
\newblock {\em Phys. Rev.}, D46:1141--1147, 1992.

\bibitem{Lebed:1992qv}
Richard~F. Lebed and Mahiko Suzuki.
\newblock Making electroweak models of composite fermions realistic.
\newblock {\em Phys. Rev.}, D45:1744--1750, 1992.

\bibitem{Cvetic:1997eb}
G.~Cvetic.
\newblock Top quark condensation: A review.
\newblock {\em Rev. Mod. Phys.}, 71:513, 1999.

\bibitem{Hill:1990mt}
Christopher~T. Hill.
\newblock Minimal dynamical symmetry breaking of the electroweak interactions
  and m(top).
\newblock {\em Mod. Phys. Lett.}, A5:2675--2682, 1990.

\bibitem{Bonisch:1991hm}
Ralf Bonisch.
\newblock Gauge created top quark condensate and heavy top.
\newblock 1991.

\bibitem{'tHooft:1976up}
G.~'t~Hooft.
\newblock Symmetry breaking through bell-jackiw anomalies.
\newblock {\em Phys. Rev. Lett.}, 37:8--11, 1976.

\bibitem{Buchalla:1996dp}
Gerhard Buchalla, Gustavo Burdman, C.~T. Hill, and Dimitris Kominis.
\newblock Gim violation and new dynamics of the third generation.
\newblock {\em Phys. Rev.}, D53:5185--5200, 1996.

\bibitem{Kominis:1995fj}
Dimitris Kominis.
\newblock Flavor changing neutral current constraints in topcolor assisted
  technicolor.
\newblock {\em Phys. Lett.}, B358:312--317, 1995.

\bibitem{Burdman:2000in}
Gustavo Burdman, Kenneth~D. Lane, and Tonguc Rador.
\newblock Anti-b b mixing constrains topcolor-assisted technicolor.
\newblock {\em Phys. Lett.}, B514:41--46, 2001.

\bibitem{Simmons:2001va}
Elizabeth~H. Simmons.
\newblock Limitations of b meson mixing bounds on technicolor theories.
\newblock {\em Phys. Lett.}, B526:365--369, 2002.

\bibitem{Aranda:2000vk}
Alfredo Aranda and Christopher~D. Carone.
\newblock Bounds on bosonic topcolor.
\newblock {\em Phys. Lett.}, B488:351--358, 2000.

\bibitem{Aranda:2000xy}
Alfredo Aranda and Christopher~D. Carone.
\newblock Bosonic topcolor.
\newblock 2000.

\bibitem{Carone:1997kc}
Christopher~D. Carone, Lawrence~J. Hall, and Takeo Moroi.
\newblock New mechanism of flavor symmetry breaking from supersymmetric strong
  dynamics.
\newblock {\em Phys. Rev.}, D56:7183--7192, 1997.

\bibitem{Triantaphyllou:2000uu}
George Triantaphyllou and George Zoupanos.
\newblock Strongly interacting fermions from a higher-dimensional unified gauge
  theory.
\newblock {\em Phys. Lett.}, B489:420--426, 2000.

\bibitem{Lindner:1998er}
Manfred Lindner and George Triantaphyllou.
\newblock Mirror families in electro-weak symmetry breaking.
\newblock {\em Phys. Lett.}, B430:303--313, 1998.

\bibitem{Triantaphyllou:1998ke}
George Triantaphyllou.
\newblock New physics with mirror particles.
\newblock {\em J. Phys. G}, G26:99, 2000.

\bibitem{He:1999vp}
Hong-Jian He, Tim Tait, and C.~P. Yuan.
\newblock New topflavor models with seesaw mechanism.
\newblock {\em Phys. Rev.}, D62:011702, 2000.

\bibitem{Chivukula:1998vd}
R.~Sekhar Chivukula and Howard Georgi.
\newblock Effective field theory of vacuum tilting.
\newblock {\em Phys. Rev.}, D58:115009, 1998.

\bibitem{Martin:1993mj}
Stephen~P. Martin.
\newblock Selfbreaking technicolor.
\newblock {\em Nucl. Phys.}, B398:359--375, 1993.

\bibitem{Chivukula:1996yr}
R.~S. Chivukula, A.~G. Cohen, and E.~H. Simmons.
\newblock New strong interactons at the tevatron ?
\newblock {\em Phys. Lett.}, B380:92--98, 1996.

\bibitem{Popovic:1998vb}
Marko~B. Popovic and Elizabeth~H. Simmons.
\newblock A heavy top quark from flavor-universal colorons.
\newblock {\em Phys. Rev.}, D58:095007, 1998.

\bibitem{Bardeen:1986sm}
W.~A. Bardeen, C.~N. Leung, and S.~T. Love.
\newblock The dilaton and chiral symmetry breaking.
\newblock {\em Phys. Rev. Lett.}, 56:1230, 1986.

\bibitem{Leung:1986sn}
C.~N. Leung, S.~T. Love, and William~A. Bardeen.
\newblock Spontaneous symmetry breaking in scale invariant quantum
  electrodynamics.
\newblock {\em Nucl. Phys.}, B273:649--662, 1986.

\bibitem{Appelquist:1988fm}
Thomas Appelquist, Mark Soldate, Tatsu Takeuchi, and L.~C.~R. Wijewardhana.
\newblock Effective four fermion interactions and chiral symmetry breaking.
\newblock To be publ. in Proc. of 12th Johns Hopkins Workshop on Current
  Problems in Particle Theory, Baltimore, MD, Jun 8- 10, 1988.

\bibitem{Kondo:1989qd}
Kei-ichi Kondo, Hidetoshi Mino, and Koichi Yamawaki.
\newblock Critical line and dilaton in scale invariant qed.
\newblock {\em Phys. Rev.}, D39:2430, 1989.

\bibitem{Chivukula:1995dc}
R.~Sekhar Chivukula, B.~A. Dobrescu, and J.~Terning.
\newblock Isospin breaking and fine tuning in topcolor assisted technicolor.
\newblock {\em Phys. Lett.}, B353:289--294, 1995.

\bibitem{Burdman:1997vn}
Gustavo Burdman.
\newblock Constraints on strong dynamics from rare b and k decays.
\newblock 1997.

\bibitem{Burdman:1997qn}
Gustavo Burdman.
\newblock Effects of the electroweak symmetry breaking sector in rare b and k
  decays.
\newblock {\em Phys. Lett.}, B409:443--449, 1997.

\bibitem{Hill:1995di}
Christopher~T. Hill and Xin min Zhang.
\newblock Z $\to$ b anti-b versus dynamical electroweak symmetry breaking
  involving the top quark.
\newblock {\em Phys. Rev.}, D51:3563--3568, 1995.

\bibitem{Loinaz:1998jg}
Will Loinaz and Tatsu Takeuchi.
\newblock Constraints on topcolor assisted technicolor models from vertex
  corrections.
\newblock {\em Phys. Rev.}, D60:015005, 1999.

\bibitem{Burdman:1999sr}
Gustavo Burdman.
\newblock Scalars from top-condensation models at hadron colliders.
\newblock {\em Phys. Rev. Lett.}, 83:2888--2891, 1999.

\bibitem{Burdman:1996iv}
Gustavo Burdman.
\newblock Topcolor models and scalar spectrum.
\newblock 1996.

\bibitem{Cao:2002af}
Jun-jie Cao, Zhao-hua Xiong, and Jin~Min Yang.
\newblock Probing topcolor-assisted technicolor from top-charm associated
  production at lhc.
\newblock 2002.

\bibitem{Yue:2000fe}
Chong-xing Yue, Qing-jun Xu, Guo-li Liu, and Jian-tao Li.
\newblock Production and decay of the neutral top-pion in high energy e+ e-
  colliders.
\newblock {\em Phys. Rev.}, D63:115002, 2001.

\bibitem{Yue:2000xa}
Chong-xing Yue, Gong-ru Lu, Jun-jie Cao, Jian-tao Li, and Guoli Liu.
\newblock Neutral top-pion and top-charm production in high energy e+ e-
  collisions.
\newblock {\em Phys. Lett.}, B496:93--97, 2000.

\bibitem{Yue:2001uv}
Chong-xing Yue, Yuan-ben Dai, Qing-jun Xu, and Guo-li Liu.
\newblock The process e+ e- --> anti-t c in topcolor-assisted technicolor
  models.
\newblock {\em Phys. Lett.}, B525:301--307, 2002.

\bibitem{Huang:2001pm}
Jin-shu Huang, Zhao-hua Xiong, and Gong-ru Lu.
\newblock Top quark pair production at e+ e- colliders in the topcolor-assisted
  technicolor model.
\newblock 2001.

\bibitem{He:1998ie}
Hong-Jian He and C.~P. Yuan.
\newblock New method for detecting charged (pseudo-)scalars at colliders.
\newblock {\em Phys. Rev. Lett.}, 83:28--31, 1999.

\bibitem{Balazs:1998sb}
Csaba Balazs, Hong-Jian He, and C.~P. Yuan.
\newblock {QCD} corrections to scalar production via heavy quark fusion at
  hadron colliders.
\newblock {\em Phys. Rev.}, D60:114001, 1999.

\bibitem{He:2002fd}
Hong-Jian He, Shinya Kanemura, and C.~P. Yuan.
\newblock Determining the chirality of yukawa couplings via single charged
  higgs boson production in polarized photon collision.
\newblock {\em Phys. Rev. Lett.}, 89:101803, 2002.

\bibitem{Balazs:1998nt}
C.~Balazs, J.~L. Diaz-Cruz, H.~J. He, T.~Tait, and C.~P. Yuan.
\newblock Probing higgs bosons with large bottom yukawa coupling at hadron
  colliders.
\newblock {\em Phys. Rev.}, D59:055016, 1999.

\bibitem{Diaz-Cruz:1998qc}
J.~Lorenzo Diaz-Cruz, Hong-Jian He, Tim Tait, and C.~P. Yuan.
\newblock Higgs bosons with large bottom yukawa coupling at tevatron and lhc.
\newblock {\em Phys. Rev. Lett.}, 80:4641--4644, 1998.

\bibitem{Han:2003pu}
T.~Han, D.~Rainwater, and G.~Valencia.
\newblock Tev resonances in top physics at the lhc.
\newblock 2003.

\bibitem{tracking}
tracking down this reference!

\bibitem{Abachi:1997jv}
S.~Abachi et~al.
\newblock Direct measurement of the top quark mass.
\newblock {\em Phys. Rev. Lett.}, 79:1197--1202, 1997.

\bibitem{Abbott:1998fv}
B.~Abbott et~al.
\newblock Measurement of the top quark mass using dilepton events. d0
  collaboration.
\newblock {\em Phys. Rev. Lett.}, 80:2063--2068, 1998.

\bibitem{Abbott:1998dc}
B.~Abbott et~al.
\newblock Direct measurement of the top quark mass at d0.
\newblock {\em Phys. Rev.}, D58:052001, 1998.

\bibitem{Abbott:1998dn}
B.~Abbott et~al.
\newblock Measurement of the top quark mass in the dilepton channel.
\newblock {\em Phys. Rev.}, D60:052001, 1999.

\bibitem{Affolder:2000dt}
T.~Affolder et~al.
\newblock Measurement of the top quark p(t) distribution.
\newblock {\em Phys. Rev. Lett.}, 87:102001, 2001.

\bibitem{Abbott:1998wh}
B.~Abbott et~al.
\newblock The dijet mass spectrum and a search for quark compositeness in
  anti-p p collisions at s**(1/2) = 1.8- tev.
\newblock {\em Phys. Rev. Lett.}, 82:2457--2462, 1999.

\bibitem{Affolder:2000eu}
T.~Affolder et~al.
\newblock Search for new particles decaying to t anti-t in p anti-p collisions
  at s**(1/2) = 1.8-tev.
\newblock {\em Phys. Rev. Lett.}, 85:2062--2067, 2000.

\bibitem{Amidei:1996dt}
D.~Amidei et~al.
\newblock Future electroweak physics at the fermilab tevatron: Report of the
  tev-2000 study group.
\newblock FERMILAB-PUB-96-082.

\bibitem{Simmons:1997fz}
Elizabeth~H. Simmons.
\newblock Coloron phenomenology.
\newblock {\em Phys. Rev.}, D55:1678--1683, 1997.

\bibitem{Burdman:1999us}
Gustavo Burdman, R.~Sekhar Chivukula, and Nick Evans.
\newblock Precision bounds on flavor gauge bosons.
\newblock {\em Phys. Rev.}, D61:035009, 2000.

\bibitem{Bertram:1998wf}
Iain Bertram and Elizabeth~H. Simmons.
\newblock Dijet mass spectrum limits on flavor-universal colorons.
\newblock {\em Phys. Lett.}, B443:347, 1998.

\bibitem{Abbott:1998yy}
B.~Abbott et~al.
\newblock Coloron limits using the d0 dijet angular distribution.
\newblock 1998.

\bibitem{Chivukula:1996cc}
R.~S. Chivukula and J.~Terning.
\newblock Precision electroweak constraints on top-color assisted technicolor.
\newblock {\em Phys. Lett.}, B385:209--217, 1996.

\bibitem{kdl:temp}
KDL.
\newblock tc2 model with big lepton hypercharges made vectorial.
\newblock 1997.

\bibitem{Rador:1998is}
Tonguc Rador.
\newblock Lepton number violation in top-color assisted technicolor.
\newblock {\em Phys. Rev.}, D59:095012, 1999.

\bibitem{Yue:2000fh}
Chongxing Yue, Guoli Liu, and Jiantao Li.
\newblock The bound on the mass of the new gauge boson z' from the process mu
  --> 3e.
\newblock {\em Phys. Lett.}, B496:89--92, 2000.

\bibitem{Murakami:2001cs}
Brandon Murakami.
\newblock The impact of lepton-flavor violating z' bosons on muon g-2 and other
  muon observables.
\newblock 2001.

\bibitem{Su:1997nn}
Yumian Su, Gian~Franco Bonini, and Kenneth Lane.
\newblock Fermilab tevatron constraints on topcolor-assisted technicolor.
\newblock {\em Phys. Rev. Lett.}, 79:4075, 1997.

\bibitem{Harris:1999ya}
Robert~M. Harris, Christopher~T. Hill, and Stephen~J. Parke.
\newblock Cross section for topcolor z'(t) decaying to t anti-t.
\newblock 1999.

\bibitem{Collins:1999rz}
Hael Collins, Aaron Grant, and Howard Georgi.
\newblock The phenomenology of a top quark seesaw model.
\newblock {\em Phys. Rev.}, D61:055002, 2000.

\bibitem{Georgi:2000wt}
Howard Georgi and Aaron~K. Grant.
\newblock A topcolor jungle gym.
\newblock {\em Phys. Rev.}, D63:015001, 2001.

\bibitem{Burdman:1998vw}
Gustavo Burdman and Nick Evans.
\newblock Flavour universal dynamical electroweak symmetry breaking.
\newblock {\em Phys. Rev.}, D59:115005, 1999.

\bibitem{Popovic:2001cj}
Marko~B. Popovic.
\newblock Third generation seesaw mixing with new vector-like weak- doublet
  quarks.
\newblock {\em Phys. Rev.}, D64:035001, 2001.

\bibitem{He:2001fz}
Hong-Jian He, Christopher~T. Hill, and Tim M.~P. Tait.
\newblock Top quark seesaw, vacuum structure and electroweak precision
  constraints.
\newblock 2001.

\bibitem{Langacker:1988ur}
Paul Langacker and David London.
\newblock Mixing between ordinary and exotic fermions.
\newblock {\em Phys. Rev.}, D38:886, 1988.

\bibitem{Popovic:2000dx}
Marko~B. Popovic and Elizabeth~H. Simmons.
\newblock Weak-singlet fermions: Models and constraints.
\newblock {\em D62}, page 035002, 2000.

\bibitem{Abe:1998iz}
F.~Abe et~al.
\newblock Measurement of the top quark mass and t anti-t production cross
  section from dilepton events at the collider detector at fermilab.
\newblock {\em Phys. Rev. Lett.}, 80:2779--2784, 1998.

\bibitem{Abachi:1997re}
S.~Abachi et~al.
\newblock Measurement of the top quark pair production cross section in p
  anti-p collisions.
\newblock {\em Phys. Rev. Lett.}, 79:1203--1208, 1997.

\bibitem{Ackerstaff:1998cz}
K.~Ackerstaff et~al.
\newblock Search for unstable heavy and excited leptons in e+e- collisions at
  s**(1/2) = 170-gev to 172-gev.
\newblock {\em Eur. Phys. J.}, C1:45--64, 1998.

\bibitem{Abreu:1998jw}
P.~Abreu et~al.
\newblock Search for composite and exotic fermions at lep2.
\newblock {\em Eur. Phys. J.}, C8:41, 1999.

\bibitem{Affolder:1999bs}
T.~Affolder et~al.
\newblock Search for a fourth-generation quark more massive than the z0 boson
  in p anti-p collisions at s**(1/2) = 1.8-tev.
\newblock {\em Phys. Rev. Lett.}, 84:835, 2000.

\bibitem{King:1992qb}
S.~F. King.
\newblock Dynamical symmetry breaking solution to the problem of flavor.
\newblock {\em Phys. Rev.}, D45:990--992, 1992.

\bibitem{Georgi:1990ej}
H.~Georgi.
\newblock Technicolor and families.
\newblock Presented at Conf. SCGT 90, Nagoya, Japan, Jul 28-31, 1990.

\bibitem{Burdman:2000yq}
Gustavo Burdman, R.~Sekhar Chivukula, and Nick Evans.
\newblock Flavor gauge bosons at the tevatron.
\newblock {\em Phys. Rev.}, D62:075007, 2000.

\bibitem{Antoniadis:1990ew}
Ignatios Antoniadis.
\newblock A possible new dimension at a few tev.
\newblock {\em Phys. Lett.}, B246:377--384, 1990.

\bibitem{Antoniadis:1997hk}
Ignatios Antoniadis and M.~Quiros.
\newblock Large radii and string unification.
\newblock {\em Phys. Lett.}, B392:61--66, 1997.

\bibitem{Lykken:1996fj}
Joseph~D. Lykken.
\newblock Weak scale superstrings.
\newblock {\em Phys. Rev.}, D54:3693--3697, 1996.

\bibitem{Arkani-Hamed:1998rs}
Nima Arkani-Hamed, Savas Dimopoulos, and Gia Dvali.
\newblock The hierarchy problem and new dimensions at a millimeter.
\newblock {\em Phys. Lett.}, B429:263, 1998.

\bibitem{Antoniadis:1998ig}
Ignatios Antoniadis, Nima Arkani-Hamed, Savas Dimopoulos, and G.~R. Dvali.
\newblock New dimensions at a millimeter to a fermi and superstrings at a tev.
\newblock {\em Phys. Lett.}, B436:257--263, 1998.

\bibitem{Dienes:1998vh}
Keith~R. Dienes, Emilian Dudas, and Tony Gherghetta.
\newblock Extra spacetime dimensions and unification.
\newblock {\em Phys. Lett.}, B436:55--65, 1998.

\bibitem{Dienes:1998vg}
Keith~R. Dienes, Emilian Dudas, and Tony Gherghetta.
\newblock Grand unification at intermediate mass scales through extra
  dimensions.
\newblock {\em Nucl. Phys.}, B537:47--108, 1999.

\bibitem{Randall:1999ee}
Lisa Randall and Raman Sundrum.
\newblock A large mass hierarchy from a small extra dimension.
\newblock {\em Phys. Rev. Lett.}, 83:3370--3373, 1999.

\bibitem{Randall:1999vf}
Lisa Randall and Raman Sundrum.
\newblock An alternative to compactification.
\newblock {\em Phys. Rev. Lett.}, 83:4690--4693, 1999.

\bibitem{Horava:1996ma}
Petr Horava and Edward Witten.
\newblock Eleven-dimensional supergravity on a manifold with boundary.
\newblock {\em Nucl. Phys.}, B475:94--114, 1996.

\bibitem{Arkani-Hamed:1999dc}
Nima Arkani-Hamed and Martin Schmaltz.
\newblock Hierarchies without symmetries from extra dimensions.
\newblock {\em Phys. Rev.}, D61:033005, 2000.

\bibitem{Nussinov:2001ps}
Shmuel Nussinov and R.~Shrock.
\newblock Effects of gauge interactions on fermion masses in models with
  fermion wavefunctions separated in higher dimensions.
\newblock {\em Phys. Lett.}, B526:137--143, 2002.

\bibitem{Nussinov:2001rb}
Shmuel Nussinov and Robert Shrock.
\newblock n anti-n oscillations in models with large extra dimensions.
\newblock {\em Phys. Rev. Lett.}, 88:171601, 2002.

\bibitem{Alishahiha:2001nb}
Mohsen Alishahiha.
\newblock (de)constructing dimensions and non-commutative geometry.
\newblock {\em Phys. Lett.}, B517:406--414, 2001.

\bibitem{Dai:2001bx}
Jian Dai and Xing-Chang Song.
\newblock Spontaneous symmetry broken condition in (de)constructing dimensions
  from noncommutative geometry.
\newblock 2001.

\bibitem{Adams:2001ne}
Allan Adams and Michal Fabinger.
\newblock Deconstructing noncommutativity with a giant fuzzy moose.
\newblock 2001.

\bibitem{Nambu:1973qe}
Yoichiro Nambu.
\newblock Generalized hamiltonian dynamics.
\newblock {\em Phys. Rev.}, D7:2405--2414, 1973.

\bibitem{Dobrescu:1998dg}
Bogdan~A. Dobrescu.
\newblock Electroweak symmetry breaking as a consequence of compact dimensions.
\newblock {\em Phys. Lett.}, B461:99--104, 1999.

\bibitem{Cheng:1999rq}
Hsin-Chia Cheng, Bogdan~A. Dobrescu, and Christopher~T. Hill.
\newblock Electroweak symmetry breaking by extra dimensions.
\newblock 1999.

\bibitem{Cheng:1999bg}
Hsin-Chia Cheng, Bogdan~A. Dobrescu, and Christopher~T. Hill.
\newblock Electroweak symmetry breaking and extra dimensions.
\newblock {\em Nucl. Phys.}, B589:249--268, 2000.

\bibitem{Cheng:1999fu}
Hsin-Chia Cheng, Bogdan~A. Dobrescu, and Christopher~T. Hill.
\newblock Gauge coupling unification with extra dimensions and gravitational
  scale effects.
\newblock {\em Nucl. Phys.}, B573:597--616, 2000.

\bibitem{Arkani-Hamed:2000hv}
Nima Arkani-Hamed, Hsin-Chia Cheng, Bogdan~A. Dobrescu, and Lawrence~J. Hall.
\newblock Self-breaking of the standard model gauge symmetry.
\newblock {\em Phys. Rev.}, D62:096006, 2000.

\bibitem{Arkani-Hamed:2001nc}
Nima Arkani-Hamed, Andrew~G. Cohen, and Howard Georgi.
\newblock Electroweak symmetry breaking from dimensional deconstruction.
\newblock {\em Phys. Lett.}, B513:232--240, 2001.

\bibitem{Arkani-Hamed:2001vr}
Nima Arkani-Hamed, Andrew~G. Cohen, and Howard Georgi.
\newblock Accelerated unification.
\newblock 2001.

\bibitem{Cheng:2001vd}
Hsin-Chia Cheng, Christopher~T. Hill, Stefan Pokorski, and Jing Wang.
\newblock The standard model in the latticized bulk.
\newblock {\em Phys. Rev.}, D64:065007, 2001.

\bibitem{Cheng:2001nh}
Hsin-Chia Cheng, Christopher~T. Hill, and Jing Wang.
\newblock Dynamical electroweak breaking and latticized extra dimensions.
\newblock {\em Phys. Rev.}, D64:095003, 2001.

\bibitem{Bardeen:1980xx}
William~A. Bardeen, Robert~B. Pearson, and Eliezer Rabinovici.
\newblock Hadron masses in quantum chromodynamics on the transverse lattice.
\newblock {\em Phys. Rev.}, D21:1037, 1980.

\bibitem{Chivukula:2002ej}
R.~Sekhar Chivukula and Hong-Jian He.
\newblock Unitarity of deconstructed five-dimensional yang-mills theory.
\newblock 2002.

\bibitem{SehkarChivukula:2001hz}
R.~Sehkar~Chivukula, Duane~A. Dicus, and Hong-Jian He.
\newblock Unitarity of compactified five dimensional yang-mills theory.
\newblock {\em Phys. Lett.}, B525:175--182, 2002.

\bibitem{HillHeTait:2000xx}
Hong-Jian He, Christopher~T. Hill, and Tim M.~P. Tait.
\newblock Top quark seesaw, vacuum structure and electroweak precision
  constraints.
\newblock {\em Phys. Rev.}, D65:055006, 2002.

\bibitem{Hashimoto:2000uk}
Michio Hashimoto, Masaharu Tanabashi, and Koichi Yamawaki.
\newblock Top mode standard model with extra dimensions.
\newblock {\em Phys. Rev.}, D64:056003, 2001.

\bibitem{Gusynin:2002cu}
V.~Gusynin, M.~Hashimoto, M.~Tanabashi, and K.~Yamawaki.
\newblock Dynamical chiral symmetry breaking in gauge theories with extra
  dimensions.
\newblock 2002.

\bibitem{Muck:2001yv}
Alexander Muck, Apostolos Pilaftsis, and Reinhold Ruckl.
\newblock Minimal higher-dimensional extensions of the standard model and
  electroweak observables.
\newblock 2001.

\bibitem{Kim:2001gk}
Hyung~Do Kim.
\newblock To be (finite) or not to be, that is the question. 'kaluza- klein
  contribution to the higgs mass'.
\newblock 2001.

\bibitem{Chankowski:2001hz}
Piotr~H. Chankowski, Adam Falkowski, and Stefan Pokorski.
\newblock Unification in models with replicated gauge groups.
\newblock 2001.

\bibitem{Cheng:2001qp}
Hsin-Chia Cheng, Konstantin~T. Matchev, and Jing Wang.
\newblock Gut breaking on the lattice.
\newblock {\em Phys. Lett.}, B521:308--314, 2001.

\bibitem{Csaki:2001em}
Csaba Csaki, Joshua Erlich, Christophe Grojean, and Graham~D. Kribs.
\newblock 4d constructions of supersymmetric extra dimensions and gaugino
  mediation.
\newblock {\em Phys. Rev.}, D65:015003, 2002.

\bibitem{Cheng:2001an}
H.~C. Cheng, D.~E. Kaplan, M.~Schmaltz, and W.~Skiba.
\newblock Deconstructing gaugino mediation.
\newblock {\em Phys. Lett.}, B515:395--399, 2001.

\bibitem{Csaki:2001qm}
Csaba Csaki, Graham~D. Kribs, and John Terning.
\newblock 4d models of scherk-schwarz gut breaking via deconstruction.
\newblock {\em Phys. Rev.}, D65:015004, 2002.

\bibitem{Kobayashi:2001fr}
Tatsuo Kobayashi, Nobuhito Maru, and Koichi Yoshioka.
\newblock 4d construction of bulk supersymmetry breaking.
\newblock 2001.

\bibitem{Hill:2001bt}
Christopher~T. Hill.
\newblock Topological solitons from deconstructed extra dimensions.
\newblock {\em Phys. Rev. Lett.}, 88:041601, 2002.

\bibitem{Csaki:2001zx}
Csaba Csaki et~al.
\newblock Exact results in 5d from instantons and deconstruction.
\newblock 2001.

\bibitem{Arkani-Hamed:2001ie}
Nima Arkani-Hamed, Andrew~G. Cohen, David~B. Kaplan, Andreas Karch, and Lubos
  Motl.
\newblock Deconstructing (2,0) and little string theories.
\newblock 2001.

\bibitem{Bander:2001qk}
Myron Bander.
\newblock Gravity in dynamically generated dimensions.
\newblock {\em Phys. Rev.}, D64:105021, 2001.

\bibitem{Skiba:2002nx}
Witold Skiba and David Smith.
\newblock Localized fermions and anomaly inflow via deconstruction.
\newblock 2002.

\bibitem{Kaplan:2001ga}
David~Elazzar Kaplan and Tim M.~P. Tait.
\newblock New tools for fermion masses from extra dimensions.
\newblock {\em JHEP}, 11:051, 2001.

\bibitem{Kaplan:1984fs}
David~B. Kaplan and Howard Georgi.
\newblock Su(2) x u(1) breaking by vacuum misalignment.
\newblock {\em Phys. Lett.}, B136:183, 1984.

\bibitem{Georgi:1984ef}
Howard Georgi, David~B. Kaplan, and Peter Galison.
\newblock Calculation of the composite higgs mass.
\newblock {\em Phys. Lett.}, B143:152, 1984.

\bibitem{Kaplan:1984sm}
David~B. Kaplan, Howard Georgi, and Savas Dimopoulos.
\newblock Composite higgs scalars.
\newblock {\em Phys. Lett.}, B136:187, 1984.

\bibitem{Arkani-Hamed:2002qx}
N.~Arkani-Hamed et~al.
\newblock The minimal moose for a little higgs.
\newblock {\em JHEP}, 08:021, 2002.

\bibitem{Arkani-Hamed:2002qy}
N.~Arkani-Hamed, A.~G. Cohen, E.~Katz, and A.~E. Nelson.
\newblock The littlest higgs.
\newblock {\em JHEP}, 07:034, 2002.

\bibitem{Arkani-Hamed:2002pa}
Nima Arkani-Hamed, Andrew~G. Cohen, Thomas Gregoire, and Jay~G. Wacker.
\newblock Phenomenology of electroweak symmetry breaking from theory space.
\newblock 2002.

\bibitem{Low:2002ws}
Ian Low, Witold Skiba, and David Smith.
\newblock Little higgses from an antisymmetric condensate.
\newblock {\em Phys. Rev.}, D66:072001, 2002.

\bibitem{Eguchi:1982nm}
Tohru Eguchi and Hikaru Kawai.
\newblock Reduction of dynamical degrees of freedom in the large n gauge
  theory.
\newblock {\em Phys. Rev. Lett.}, 48:1063, 1982.

\bibitem{Lane:2002pe}
Kenneth Lane.
\newblock A case study in dimensional deconstruction.
\newblock 2002.

\bibitem{Csaki:2002qg}
Csaba Csaki, Jay Hubisz, Graham~D. Kribs, Patrick Meade, and John Terning.
\newblock Big corrections from a little higgs.
\newblock 2002.

\bibitem{Hewett:2002px}
JoAnne~L. Hewett, Frank~J. Petriello, and Thomas~G. Rizzo.
\newblock Constraining the littlest higgs.
\newblock 2002.

\bibitem{Burdman:2002ns}
Gustavo Burdman, Maxim Perelstein, and Aaron Pierce.
\newblock Collider tests of the little higgs model.
\newblock 2002.

\bibitem{Han:2003wu}
Tao Han, Heather~E. Logan, Bob McElrath, and Lian-Tao Wang.
\newblock Phenomenology of the little higgs model.
\newblock 2003.

\bibitem{Kennedy:1989sn}
D.~C. Kennedy and B.~W. Lynn.
\newblock Electroweak radiative corrections with an effective lagrangian: Four
  fermion processes.
\newblock {\em Nucl. Phys.}, B322:1, 1989.

\bibitem{Altarelli:1991zd}
Guido Altarelli and Riccardo Barbieri.
\newblock Vacuum polarization effects of new physics on electroweak processes.
\newblock {\em Phys. Lett.}, B253:161--167, 1991.

\bibitem{Hill:1992jc}
Christopher~T. Hill and David~S. Salopek.
\newblock Calculable nonminimal coupling of composite scalar bosons to gravity.
\newblock {\em Ann. Phys.}, 213:21--30, 1992.

\end{thebibliography}

\end{document}